\documentclass{book}
\usepackage[utf8]{inputenc}
\usepackage[LGR,T1]{fontenc}
\usepackage[british]{babel}
\usepackage{textgreek}
\usepackage{amsmath}
\usepackage{amssymb}
\usepackage{amsthm} 
\usepackage{amscd}
\usepackage{bm}
\usepackage{mathtools}
\usepackage{mathrsfs}
\usepackage{mathdots}
\usepackage{fancyhdr}
\usepackage{fancyvrb}
\usepackage{pifont}
\usepackage{lmodern}
\usepackage[top=4cm]{geometry}
\usepackage[shortlabels]{enumitem}
\usepackage[clock]{ifsym}
\setlist[enumerate,1]{
{1.},
ref={\arabic*}
}
\setlist[enumerate,2]{
{(a)},
ref={\theenumi{}(\alph*)}
}
\setlist[itemize]{
label={--}
}
\usepackage[a-2b,mathxmp]{pdfx}
\hypersetup{
    colorlinks=true,
    linkcolor=blue, 	
    anchorcolor=black, 	
    citecolor=green, 	
    filecolor=cyan, 	
    urlcolor=magenta
    }
\usepackage{graphicx}
\usepackage{tikz-cd}
\usetikzlibrary{positioning,arrows.meta,trees, shapes.geometric, fit, calc}
\usepackage{worldflags}
\allowdisplaybreaks
\theoremstyle{plain}
\newtheorem{thm}{Theorem}[section]
\newtheorem{corollary}[thm]{Corollary}
\newtheorem{lemma}[thm]{Lemma}
\newtheorem{proposition}[thm]{Proposition}
\theoremstyle{definition}
\newtheorem{definition}[thm]{Definition}
\newtheorem{remark}[thm]{Remark}
\newtheorem{example}[thm]{Example}
\numberwithin{equation}{chapter}
\newcommand{\mf}{\mathfrak}
\newcommand{\mc}{\mathcal}
\newcommand{\ms}{\mathscr}
\newcommand{\op}{\operatorname}

\newcommand{\id}{\operatorname{id}}
\newcommand{\im}{\operatorname{im}}

\newcommand{\ev}{\operatorname{ev}}

\newcommand{\R}{\mathbb R}
\newcommand{\Z}{\mathbb Z}
\newcommand{\N}{\mathbb N}
\newcommand{\Nast}{{\N^{\ast}}}
\newcommand{\Q}{\mathbb Q}

\newcommand{\D}{\mathbb D}

\newcommand{\F}{\mathbb F}

\renewcommand{\Pr}{\mathrm{Pr}}
\newcommand{\e}{\varepsilon}
\newcommand{\p}{\varphi}

\newcommand{\inj}{\hookrightarrow}
\newcommand{\surj}{\twoheadrightarrow}

\newcommand{\E}{\mathbb{E}}

\newcommand{\bigmid}{~ \Big |~}

\newcommand{\A}{\mathbb{A}}
\newcommand{\B}{\mathbb{B}}

\newcommand{\Pot}{{\mathcal P}}

\newcommand{\T}{\mathbb T}

\newcommand{\Out}{\op{Out}}

\renewcommand{\P}{{\mathbb P}}
\renewcommand{\d}{{\mathrm d}}
\newcommand{\X}{{\bm{\mathsf X}}}
\newcommand{\x}{{\bm{\mathsf x}}}

\newcommand{\Tr}{{\bm{\mathsf T}}}
\newcommand{\y}{{\bm{\mathsf y}}}
\newcommand{\w}{{\bm{\mathsf w}}}
\renewcommand{\H}{{\bm{\mathsf H}}}
\newcommand{\h}{{\bm{\mathsf h}}}

\renewcommand{\L}{{\mathbb L}}
\newcommand{\Today}{2nd June 2025}
\newcommand{\ovT}{{\overline{\T}}}
\newcommand{\Ta}{{\T_\alpha}}

\newcommand{\ovTa}{{\overline{\T_\alpha}}}

\newcommand{\Pos}{{\mathbf{Pos}}}
\newcommand{\PPos}{{P\text{-}\Pos}}
\newcommand{\DM}{\mathbf{DM}}
\newcommand{\bRp}{{\overline{\R_+}}}
\newcommand{\On}{\mathrm{On}}

\newcommand{\Prg}{{\mathrm{Prg}}}
\newcommand{\Prd}{{\mathrm{Prd}}}
\newcommand{\Opt}{{\mathrm{Opt}}}
\newcommand{\ovl}[1]{\overline{#1}}

\newcommand{\convT}{{\stackrel{\mathsf T}{\,\to\,}}}
\newcommand{\Plus}{\text{\ding{58}}}
\newcommand{\prj}{{\op{prj}}}
\newcommand\bluebf[1]{\textcolor{blue}{\textbf{#1}}}
\newcommand\blue[1]{\textcolor{blue}{{#1}}}
\newcommand\dredbf[1]{\textcolor{red}{\textbf{#1}}}
\newcommand\dred[1]{\textcolor{red}{{#1}}}

\newcommand{\silentchapter}[1]{
    \chapter*{#1}
    \markboth{#1}{#1}
    \addcontentsline{toc}{chapter}{#1}
}
\newcommand{\silentsection}[1]{
    \section*{#1}
    \addcontentsline{toc}{section}{#1}
}

\makeatletter
\def\@year{}
\renewcommand{\year}[1]{\def\@year{#1}}
\def\@currentdegree{}
\newcommand{\currentdegree}[1]{\def\@currentdegree{#1}}
\def\@ORCID{}
\newcommand{\ORCID}[1]{\def\@ORCID{#1}}
\def\@faculty{}
\newcommand{\faculty}[1]{\def\@faculty{#1}}
\def\@requesteddegreelong{}
\newcommand{\requesteddegreelong}[1]{\def\@requesteddegreelong{#1}}
\def\@requesteddegreeshort{}
\newcommand{\requesteddegreeshort}[1]{\def\@requesteddegreeshort{#1}}
\def\@president{}
\newcommand{\president}[1]{\def\@president{#1}}
\def\@firstreferee{}
\newcommand{\firstreferee}[1]{\def\@firstreferee{#1}}
\def\@secondreferee{}
\newcommand{\secondreferee}[1]{\def\@secondreferee{#1}}
\def\@dateofdefence{\Today}
\newcommand{\dateofdefence}[1]{\def\@dateofdefence{#1}}
\makeatother

\makeatletter
\def\maketitle{\begin{titlepage}%
  \let\footnotesize\small
  \let\footnoterule\relax
  \let \footnote \thanks
  \null\vfil
  \vskip 2em%
  \begin{center}%
  \let \footnote \thanks
    {\LARGE \@title \par}%
    \vskip 5em%
    {\large
      \lineskip .5em%
      \begin{tabular}[t]{c}%
        {vorgelegt von} \\
        \\
        \@author, \@currentdegree \\
        {ORCID: \@ORCID}
      \end{tabular}\par}%
    \vskip 5em%
    {\large
      \lineskip .5em%
      \begin{tabular}[t]{c}%
        {an der \@faculty} \\
        {der Technischen Universität Berlin} \\
        {zur Erlangung des adakemischen Grades}
      \end{tabular}\par}%
    \vskip 1em%
    {\large
      \lineskip .5em%
      \begin{tabular}[t]{c}%
        {\@requesteddegreelong} \\
        {-- \, \@requesteddegreeshort \, --} 
      \end{tabular}\par}%
    \vskip 1em%
    {\large genehmigte Dissertation}
  \end{center}%
  \par
  \vskip 3em%
  {\large
      \lineskip .5em%
      \begin{tabular}[t]{l}%
        {Promotionsausschuss:} \\
        \\
        {Vorsitzender: \@president} \\
        {Gutachter: \@firstreferee} \\
        {Gutachter: \@secondreferee} \\
        \\
        {Tag der wissenschaftlichen Aussprache: \@dateofdefence}
      \end{tabular}\par}%
  \vskip 3em%
  \begin{center}
    {\large Berlin \@year\par}
  \end{center}
  \vfil
  \end{titlepage}%
  \setcounter{footnote}{0}%
  \global\let\thanks\relax
  \global\let\maketitle\relax
  \global\let\@thanks\@empty
  \global\let\@author\@empty
  \global\let\@date\@empty
  \global\let\@title\@empty
  \global\let\title\relax
  \global\let\author\relax
  \global\let\date\relax
  \global\let\and\relax
}
\makeatother

\begin{document}
\pagenumbering{roman}

\title{On the Foundations of Dynamic Games and Probability: \\Decision Making in Stochastic Extensive Form}
\date{\Today}
\year{2025}
\author{Ernst Emanuel Rapsch}
\currentdegree{M.\ Sc.}
\ORCID{0009-0000-8921-3894}
\faculty{Fakultät II -- Mathematik und Naturwissenschaften}
\requesteddegreelong{Doktor der Naturwissenschaften}
\requesteddegreeshort{Dr.\ rer.\ nat.}
\president{Professor Dr.\ Max Klimm}
\firstreferee{Professor Dr.\ Christoph Knochenhauer}
\secondreferee{Professor Dr.\ Frank Riedel}
\dateofdefence{24.\ Juli 2025}

\maketitle

\newpage
\thispagestyle{empty} 

\vspace*{\fill}
\noindent On the Foundations of Dynamic Games and Probability: Decision Making in Stochastic Extensive Form © 2025 by E. Emanuel Rapsch is licensed under Creative Commons Attribution 4.0 International. To view a copy of this license, visit \href{https://creativecommons.org/licenses/by/4.0/}{https://creativecommons.org/licenses/by/4.0/}, or send a letter to Creative Commons, PO Box 1866, Mountain View, CA 94042, United States of America.\\
\\

\noindent 
The author's website is \href{https://erapsch.github.io/}{https://erapsch.github.io/}, with contact information and publication updates.

\silentchapter{Acknowledgements}

First and foremost, I would like to thank Christoph Knochenhauer for supervising my doctoral studies, for his suggestions and advice, the helpful discussions, as well as for his continued support. He granted me a substantial amount of freedom to define and develop my own research, and I express my sincere gratitude for this. I am also very grateful to Frank Riedel for agreeing to serve as a member of the committee, for the valuable discussions, and for his literature suggestions, which were highly relevant to the project. I would also like to thank Max Klimm for presiding over the committee. 
Furthermore, I would like to address my sincerest thanks to Peter Bank for his support and availability, and his helpful general advice. Being part of the research community in stochastic analysis in Berlin under his guidance was an invaluable asset for this project. Finally, I would like to express my special gratitude to Peter Tankov, who first introduced me to game-theoretic modelling with stochastic analysis, and who has given me much-appreciated support and good advice.

I would like to thank Klaus Ritzberger for the interest in my project, our stimulating discussion, and his particularly helpful suggestions. Moreover, I am also grateful to Elizabeth Baldwin, Miguel Ballester, Samuel N.\ Cohen, and Jan-Henrik Steg for detailed and very helpful discussions on this project. Finally, I wish to thank Daniel Andrei, Karolina Bassa, Gerrit Bauch, Peter E.\ Caines, Fanny Cartellier, Robert Denkert, Peter Paul Dimke, Roxana Dumitrescu, Sebastian Ertel, Jobst Heitzig, Ulrich Horst, Christoph Kühn, Ali Lazrak, Marcos Leutscher, Kristoffer Lindensjö, Edwin Lock, Christopher Lorenz, Berenice Anne Neumann, Louis Pape, Manos Perdikakis, Boy Schultz, Frank Seifried, Ludvic Sinander, Mehdi Talbi, Luca Taschini, and Jacco Thijssen.

I am indebted to several institutions, in particular the \href{https://gepris.dfg.de/gepris/projekt/410208580}{Berlin-Oxford IRTG 2544 Stochastic Analysis in Interaction},\footnote{Funded by Deutsche Forschungsgemeinschaft (DFG, German Research Foundation), Project ID: 410208580.} whose speaker is Peter Bank. His relentless engagement, together with the great work by the coordinators Tom Klose and later Thomas Wagenhofer made this project a success story. I also gratefully acknowledge partial funding by this institution as well as by \href{https://gepris.dfg.de/gepris/projekt/390685689}{EXC 2046/1 The Berlin Mathematics Research Center MATH+}.\footnote{Funded by Deutsche Forschungsgemeinschaft (DFG, German Research Foundation), Project ID: 390685689.} My thanks further go to the Berlin Mathematical School for its highly professional organisation and its rich scientific and pedagogical programme. I am grateful to Samuel N.\ Cohen and Elizabeth Baldwin, who supervised my studies at the University of Oxford during Michaelmas Term 2023. I also thank the organisers of the many stimulating conferences, workshops, seminars I attended and presented my work at. For a particularly inspiring course and two very instructive seminars, I am indebted to Ottmar Edenhofer, Terry Lyons, and Alex Stomper. 
I am especially grateful to my teaching colleagues at Technische Universität Berlin for the trustful and intellectually stimulating collaboration, and to both Technische Universität Berlin and the University of Oxford for the academic, administrative, and technical support. Last but not least, I thank Jean Downes in particular for her admirable engagement and professionalism, as well as her successor Anika Bartens for continuing this work with equal dedication.

I would also like to commemorate Jean-Pierre Demailly whom I am deeply indebted to for his generosity in sharing his profound intellectual strength and ethos. I am particularly grateful to Sebastian Goette, Sandra Rozensztajn, Jean-Claude Sikorav, and Martin Ziegler, and express my sincere gratitude toward Hans Jürgen Scheuer and Sebastian Schöttler. I acknowledge the important role played by Wolfram Bäurle, Peter Hübner, Volkmar Topp, and Hella Walter, whose pedagogic conviction and serious commitment to the subjects they taught cannot be overstated. As mathematics is speaking about formal expressions, I thank the excellent, keen, and inspiring conversation partners and friends I found along the way. I am deeply grateful to my parents for all they have done for me. This text has been written in memory of my grandparents and their confidence in the future. It is dedicated to Adeline, to whom I owe infinite gratitude and who yet merits even more.

\silentchapter{Preface}
\noindent\textbf{1. Dixitque alter ad proximum suum: Venite, faciamus lateres, et coquamus eos igni. Habueruntque lateres pro saxis, et bitumen pro caemento.} \smallskip

--- The Latin Bible (Biblia Sacra Vulgata), Gen 11:3.\footnote{Cf.\ \cite{Beriger2018Biblia}. Translation (by the author): And they said to one another, ``Come, let us make bricks and bake them with fire.'' And they had brick for stone, and bitumen for mortar.} \medskip

Game theory has become deeply embedded in the lexicon of contemporary social sciences, engineering, and mathematics. Its terminology — games, strategies, equilibria, types, signals, subgames — appears with great frequency, accompanied by an expanding array of refinements. The term equilibrium alone admits numerous qualifiers, including ``Nash,'' ``subgame-perfect,'' ``perfect,'' ``Bayesian,'' ``Markov perfect,'' ``correlated,'' and ``coarse-correlated,'' as well as compound forms such as ``perfect Bayesian equilibrium.'' However, these terms are context-dependent, with identical words employed across distinct formal frameworks lacking clear conceptual linkage and leading to a quasi-Babylonian confusion of tongues. This raises the question: do such labels reflect a coherent underlying concept, or are they largely products of ad-hoc usage and metaphorical extension?\bigskip

\noindent\textbf{2. Doch ein Begriff muss bei dem Worte sein.} \smallskip

--- Goethe, Faust I, said by the student.\footnote{Cf.\ \cite{Goethe2012Faust}. Translation (by the author): There must be a concept connected with the word.} \medskip

Whatever the devil’s advocate might argue, game-theoretic terms must ultimately be grounded in the first principles of decision theory. Yet as dynamics and uncertainty grow more complex, this grounding becomes increasingly obscure. This motivates the need for a foundational theory --- an abstract language for formally comparing usages and meanings. Here, foundations are not dogmata, but consistent elements enabling a comparative analysis of existing theory.\bigskip\newpage

\noindent\textbf{3. Ω ξειν', αγγελλειν Λακεδαιμoνιoις oτι τηδε \textbackslash{}\textbackslash{} κειμεθα, τoις κεινων ρημασι πειθoμενoι.} \smallskip

--- Simonides of Ceos, Epitaph at Thermopylae.\footnote{According to Herodotus \cite[Book V, Chapter 228]{Herodotus2015Herodoti}. Translation (Godley \cite{Herodotus1922Persian}): Go tell the Spartans, thou that passest by, \textbackslash{}\textbackslash{}
That here obedient to their words we lie.} \medskip

Rationality is one such fundamental principle, central to Western philosophy and going back to Ancient Greek ways of thought and policy making. The decades, centuries, even millennia-long tradition of skilful critique of rationality does not disprove this at all. To the contrary, the persistence and methods of its critics rather confirm the principle’s relevance. This remains true even though the principle is under strong pressure today, a pressure which is partly self-inflicted through conceptual overstretch. In its basic form, (economic) rationality simply means that outcomes are, to some extent, generated by agents deciding according to personal objectives. However, it does not imply that agents are human, intelligent, or omniscient; that macroscopic outcomes appear rational; or that outcomes are fully generated by agents’ decisions.\bigskip

\noindent\textbf{4. But men may construe things after their fashion \textbackslash{}\textbackslash{} Clean from the purpose of the things themselves.} \smallskip

--- Shakespeare, Julius Caesar, said by Cicero.\footnote{Cf.\ \cite{Shakespeare1998Julius}} \medskip

Decision making is the active process of reducing uncertainty; agency is the capacity to do so. This motivates the classical idea of formalising an act as a set-theoretic function from states to consequences. For example, 
\[ s \colon \{\text{Osteria}, \text{Sushi-Ramen bar}\} \to \{\text{eat pasta}, \text{eat fish}\}, \]
may specify a person’s choice of meal in whichever restaurant that person ends up tonight. Personal objectives are then formalised in terms of statements of the form ``I prefer act A over act B'', or more formally, as preference relations on the set of acts. As a rule, rationality --- together with assumptions about how agents reason about one another’s choices and beliefs --- implies that the profile of acts executed by the agents is in strategic equilibrium. That is, the acts taken together are self-enforcing: no agent has an incentive to deviate unilaterally.\footnote{This theory goes back to von Neumann--Morgenstern \cite{Neumann1944Theory}, Nash \cite{Nash1950Equilibrium, Nash1951Non}, Savage \cite{Savage1972Foundations}, and others.} \bigskip

\noindent\textbf{5. Ma sendo l’intenzione mia stata scrivere cosa che sia utile a chi la intende, mi è parso piú conveniente andare dreto alla verità effettuale della cosa che alla immaginazione di essa.}\smallskip

--- Machiavelli, Il Principe, Chapter XV.\footnote{Cf.\ \cite{Machiavelli1997Il}. Translation (by the author): But since my intention is to write something useful to those who understand it, it has seemed to me more fitting to pursue the effectual truth of the matter rather than its imagined form.} \medskip

In such an equilibrium, universally undesirable or inefficient outcomes can result from rational decision making. This ``effectual truth'' can be seen in such diverse examples as anthropogenic climate change (the tragedy of the commons), war (commitment and information problems), or oligopolistic market economies (oligopolistic surplus). Game theory thus provides a general and flexible language for describing interaction under anarchy, i.e., in the absence of rulers. Even where ``rulers'' exist, they are understood as agents pursuing objectives and engaging in equilibrated interaction with fellow agents. Understanding outcomes, therefore, depends not on interpreting a ruler’s will, but on analysing the rules --- objectives, information, and options of agents. It may be tempting to adopt the ruler’s perspective and assume omniscience and fantastic forms of ``rationality'', but it can also be deceptive and amount to wishful thinking.  \bigskip\newpage

\noindent\textbf{6. C’est la persuasion que, pour être bon praticien, il faut commencer par être savant.}\smallskip

--- Louis Pasteur, Lecture at Ecole des Beaux-Arts, Paris 1864.\footnote{Cf.\ \cite{Pasteur2025First}. Translation (by the author): It is the persuasion that, to be a good practitioner, one must first be learned.} \medskip

As I am writing these lines, the profound changes in international relations force us to sharpen our capacity to think strategically. Similar remarks apply to climate change mitigation and the ongoing economic transformation. This dissertation therefore focuses on developing the theory of games with complex dynamics and uncertainty. The classical model of dynamic choice under uncertainty is the strategy, i.e., a complete contingent plan of action, for example:
\begin{gather*}
    s \colon \{\text{\bluebf{Outside}},{\blue{\worldflag[length=0.25cm,width=0.25cm]{IT}}}, {\blue{\worldflag[length=0.25cm,width=0.25cm]{JP}}}\} \to \{{\dredbf{\text{enter }\worldflag[length=0.25cm,width=0.25cm]{IT}}}, {\dredbf{\text{enter }\worldflag[length=0.25cm,width=0.25cm]{JP}}}, \text{\dred{eat pasta}}, \text{\dred{eat fish}}\}.
\end{gather*}

\begin{figure}
    \centering
    \includegraphics[width=0.3\linewidth]{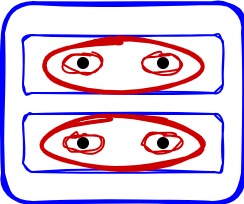}\hspace{5em}
    \includegraphics[width=0.3\linewidth]{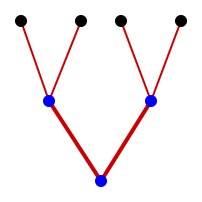}
    \caption{Choosing a meal, in the refined partitions and tree formulations, where: black bullets = outcomes, blue bullets / rectangular loops = states = moves, red edges / elliptic loops = consequences = choices.}
    \label{fig:refined_partitions_restaurant}
\end{figure}
Figure~\ref{fig:refined_partitions_restaurant} illustrates this in two equivalent ways: through refined partitions of the set of outcomes (left) and a graph-theoretic tree (right).\footnote{Many similar figures can be found in the foundational book \cite{Neumann1944Theory}, see, for instance, Figures~5 and~9 therein.} A classical representation of preference then consists in computing the expected utility of the outcome distribution:\footnote{In the following expression, $\mathrm{EU}$ denotes the expected utility, the real number $u(\bullet)$ denotes the utility of the outcome $\bullet$, and $p_1,p_2,p_3,p_4$ are probabilities, that is, positive real numbers that sum to one.}
\begin{align*}
    \mathrm{EU} = p_1 \cdot u(\worldflag[length=0.25cm,width=0.25cm]{IT}\,\text{pasta}) + p_2 \cdot u(\worldflag[length=0.25cm,width=0.25cm]{IT}\,\text{fish}) + p_3 \cdot u(\worldflag[length=0.25cm,width=0.25cm]{JP}\,\text{pasta}) + p_4 \cdot u(\worldflag[length=0.25cm,width=0.25cm]{JP}\,\text{fish}).
\end{align*}
However, as the tree grows complex, classical decision- and game-theoretic concepts cease to be well-defined. Typically, this problem is circumvented by a plethora of ad-hoc, case-specific solutions, creating the mentioned confusion of tongues. By contrast, I claim that, when formulated properly, these concepts remain meaningful within a more general, unifying framework, allowing comparison and evaluation of common ad-hoc usages. This requires, first, a suitable concept of exogenous uncertainty, explaining those parts of outcomes not attributable to agency, though still relevant to agents’ decision making. Second, it requires an abstract, graded language of dynamic games under exogenous uncertainty, compatible both with the traditional baseline model of dynamic games --- the extensive form --- and with an intrinsic model of ``limit games'', such as continuous-time stochastic games, which are central in many applications.

\newpage
According to a remark commonly attributed to Einstein, one should make things as simple as possible, but not simpler. Quite some mathematical work will be necessitated by the outlined non-trivial programme, and even more for the detailed development of its core ideas. Still, the reader who keeps in mind the small set of simple yet powerful concepts just introduced will not be distracted from the essence. On the contrary, such a reader will see the very simplicity of game theory, for so much can be understood with so few. And most welcome will be the reader willing to ``throw away the ladder, after he has climbed up on it.''\footnote{As suggested by Wittgenstein in Proposition~6.54 of the Tractatus logico-philosophicus: ``(Er muss sozusagen die Leiter wegwerfen, nachdem er auf ihr hinaufgestiegen ist.)'' Cf.\ \cite{Wittgenstein2006Tractatus,Wittgenstein1922Tractatus}.}

\vspace*{3em}
\hfill E.\ Emanuel Rapsch \qquad\quad Berlin, August 2025

\silentchapter{Abstract, in English and German}

\silentsection{Abstract}
In this dissertation, an abstract and general language for the fundamental objects underlying dynamic games under probabilistic uncertainty is developed. Combining the theory of decision trees by Alós-Ferrer--Ritzberger (2005) and a Harsanyian notion of exogenous uncertainty, the concept of \emph{stochastic decision forests} is introduced. Exogenous information is modelled via filtration-like objects providing dynamic updates on the ``realised tree'', and an abstract decision-theoretic model of \emph{adapted choice} is formulated. 

Based on this, a consistent model of ``rules'' is introduced, leading to the notion of \emph{stochastic extensive forms}, generalising the works of Alós-Ferrer--Ritzberger (2008, 2011). Well-posedness is completely characterised in terms of order-theoretic properties of the underlying forest. Moreover, the language of stochastic extensive forms addresses a vast class of dynamic decision problems formulated in terms of time-indexed paths of action --- a first step towards an approximation theory of continuous-time games based on stochastic processes. In this formulation, a well-posed theory obtains if and only if the time half-axis is essentially well-ordered.

Therefore, a relaxed game-theoretic model of ``extensive form characteristics'' is introduced: the \emph{stochastic process form}. Its action processes arise from well-posed action path stochastic extensive forms under tilting convergence, which is introduced in order to faithfully describe accumulating reaction behaviour. The problem of instantaneous reaction and information about it is tackled by introducing \emph{vertically extended continuous time}, for which a suitable stochastic analysis is developed. Stochastic process forms admit a natural notion of information sets, subgames, and equilibrium. The theory applies to stochastic differential and timing games, e.g., addressing open issues in the works by Fudenberg--Tirole (1985) and Riedel--Steg (2017).

\newpage

\silentsection{Kurzzusammenfassung}

In dieser Dissertation wird eine abstrakte und allgemeine Sprache für die fundamentalen Objekte entwickelt, die dynamischen Spielen unter probabilistischer Unsicherheit zugrunde liegen. Durch die Verbindung der Theorie der Entscheidungsbäume von Alós-Ferrer--Ritzberger (2005) mit einem Harsanyischen Begriff exogener Unsicherheit wird das Konzept der \emph{stochastischen Entscheidungs\-wälder} eingeführt. Exogene Information wird über filtrationsähnliche Objekte modelliert, die dy\-namische Lageberichte über den ``realisierten Baum'' bereitstellen, und ein abstraktes entscheidungs\-theoretisches Modell \emph{adaptierter Auswahl} wird formuliert.  

Darauf aufbauend wird ein konsistentes Modell von ``Regeln'' eingeführt, das zum Begriff der \emph{stochastischen Extensivformen} führt, was die Arbeiten von Alós-Ferrer--Ritzberger (2008, 2011) verallgemeinert. Wohlgestelltheit wird vollständig durch ordnungstheoretische Eigenschaften des zugrunde liegenden Waldes charakterisiert. Darüber hinaus erfasst die Sprache der stochastischen Extensivformen eine große Klasse dynamischer Entscheidungsprobleme, die über zeit\-indizierte Ak\-tionspfade formuliert sind --- ein erster Schritt hin zu einer Approximationstheorie zeitstetiger Spiele auf der Grundlage stochastischer Prozesse. In dieser Formulierung ergibt sich eine wohlgestellte Theorie genau dann, wenn die Zeit-Halbachse im Wesentlichen wohlgeordnet ist.  

Daher wird ein relaxiertes spieltheoretisches Modell von ``Extensivform-Charakteristiken'' eingeführt: die \emph{stochastische Prozessform}. Ihre Aktionsprozesse entstehen aus wohlgestellten Aktionspfad-basierten stochastischen Extensivformen unter Kippkonvergenz, die eingeführt wird, um akkumulierendes Reaktionsverhalten getreu zu beschreiben. Das Problem instantaner Reaktion und Information dazu wird durch die Einführung \emph{vertikal erweiterter stetiger Zeit} behandelt, für die eine geeignete stochastische Analysis entwickelt wird. Stochastische Prozessformen erlauben auf natürliche Weise Begriffe von Informationsmengen, Teilspielen und Gleichgewicht. Die Theorie findet Anwendung auf stochastische Differential- und Stoppspiele und adressiert dabei u.\,a.\ offene Fragen aus den Arbeiten von Fudenberg--Tirole (1985) sowie Riedel--Steg (2017).

\silentchapter{Further declarations}

In both its conception and its reality, this text is a monograph. However, preliminary versions of the first two chapters have been made publicly available in order to serve as a better basis for communi\-cation within the scientific community. 
The text of Chapter~\ref{chap:1-SDF_AC} is a modified version of a paper, which has been uploaded on arXiv and SSRN: \cite{Rapsch2024Decision}, DOI:  	
\href{https://doi.org/10.48550/arXiv.2404.12332}{https://doi.org/10.48550/arXiv.2404.12332}. The text of Chapter~\ref{chap:2-SEF_G} is a modified version of a paper, which has been uploaded on arXiv and SSRN: \cite{Rapsch2024Decisiona}, DOI: \href{https://doi.org/10.48550/arXiv.2411.17587}{https://doi.org/10.48550/arXiv.2411.17587}. The entire doctoral dissertation has also been uploaded on arXiv in the version submitted at TU Berlin in June 2025, up to minor corrections; the corresponding reference is \cite{Rapsch2025Foundationsa}, DOI: \href{https://doi.org/10.48550/arXiv.2508.04752}{https://doi.org/10.48550/arXiv.2508.04752}.

Other than that, the text of Chapter~\ref{chap:3-SPF_VECT} has not yet been made public, but there is a plan to use parts of it in joint work with Christoph Knochenhauer and to submit the resulting papers to appropriate journals. The introduction of this thesis is a combination of strongly modified versions of the introductions of the two cited papers and an additional introduction for Chapter~\ref{chap:3-SPF_VECT}. A similar remark applies regarding the conclusions. None of the mentioned papers has been submitted to a journal yet. However, it is planned to do so in the near- to mid-term future.\medskip

\begin{figure}
    \centering
    \includegraphics[width=0.8\linewidth]{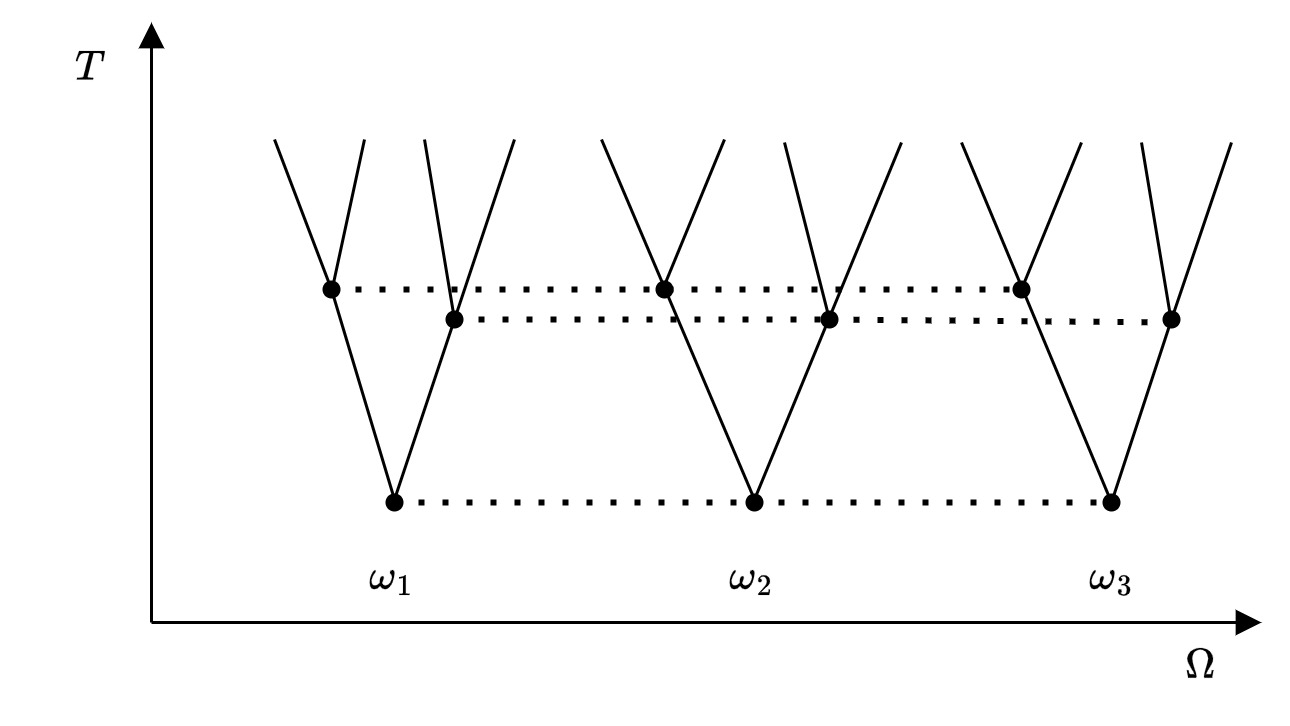}
    \caption{A quite simple stochastic decision forest. The horizontal dimension or axis describes ``exogenous'' scenarios, given by $\Omega$. The vertical dimension or axis describes ``endogenous'' decision making schematically, given by the connected components alias trees $T$ of $(F,\supseteq)$ and random moves $\x\in\X$. Moves are indicated by bullet points, random moves by dotted lines.}
    \label{fig:SDF}
\end{figure}

\begin{figure}
    \centering
    \includegraphics[width=0.95\linewidth]{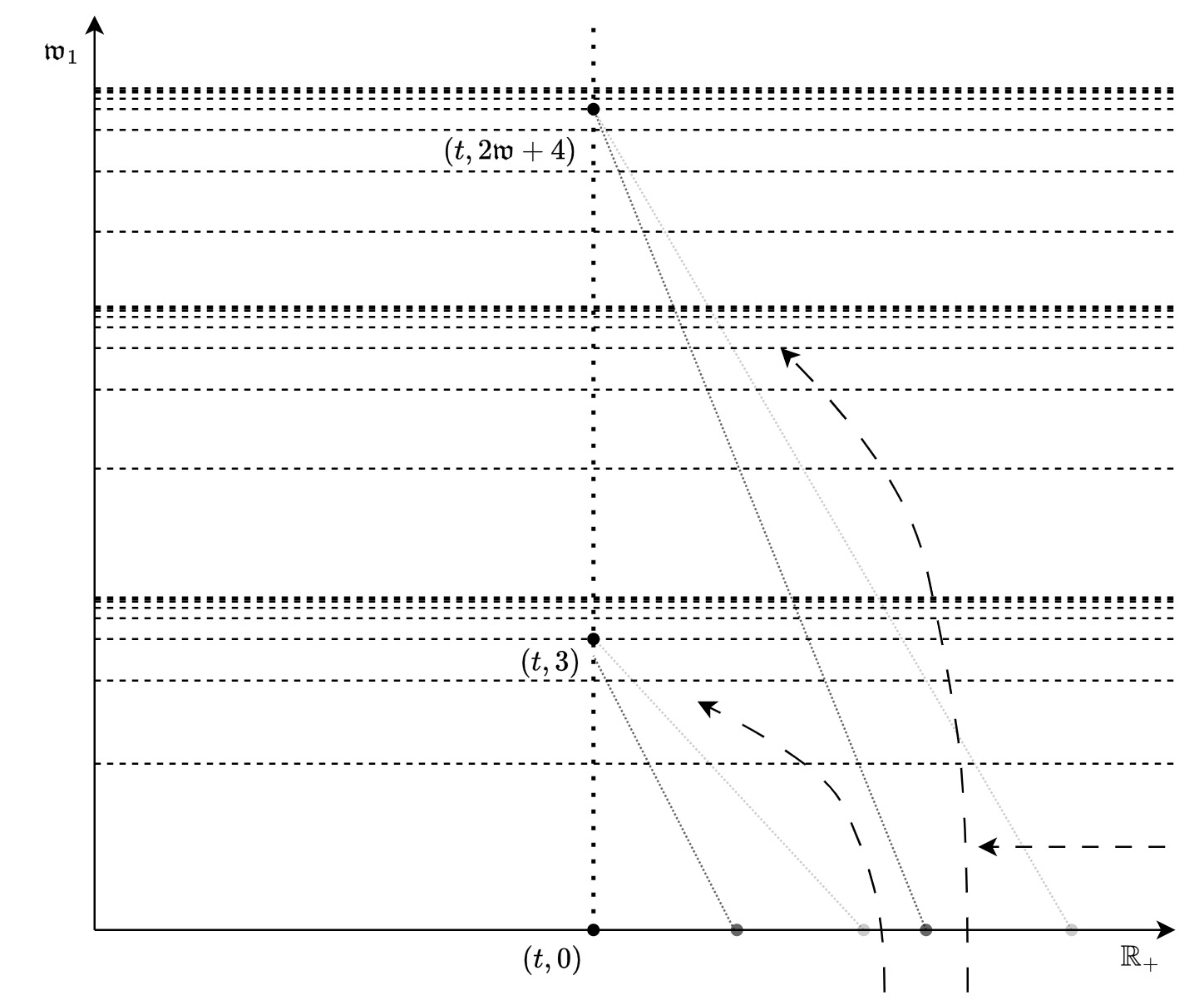}
    \caption{An illustration of vertically extended continuous time and tilting convergence: above any real time $t\in\R_+$ we attach a sufficiently large well-order. The action process consisting of three actions at times $(t,0)$, $(t,3)$, and $(t,2\mf w+4)$ can be obtained by a limit of sequences of classical continuous-time decision making on refining, convergent grids in $\R_+$, on which these actions occur at the grid points with indices $0$, $3$, and $2\mf w+4$, respectively. The limit procedure is illustrated by the dashed arrow pushing the classical continuous-time decision making to the left. They eventually accumulate at the real time $t$; but, via the special notion of convergence proposed here, they are tilted by $90^\circ$ counterclockwise and end up on the vertical axis above $t$. --- Notation: Roughly speaking, up to unique isomorphism, $\mf w$ is the ``smallest infinite well-order'', and $2\mf w+4$ is the twofold concatenation of this well-order concatenated with the four-element well-order. $\mf w_1$ can be seen as the set containing, up to unique isomorphisms, all well-orders embeddable into $\R_+$. For rigorous definitions and more details, see Section~\ref{3-SPF_VECT.sec:vERT}.}
    \label{fig:VECT_and_tilting_conv}
\end{figure}

In selected places and to a very limited extent, the software \emph{ChatGPT} in its current version (that is, versions~4 and~5) was used to check spelling and improve the English formulation of the author's thoughts. Any suggestion has been critically reviewed by the author and the author takes full responsibility for resulting modifications of the text. \LaTeX~has been used to generate this typescript technically. Figures~\ref{fig:refined_partitions_restaurant}, \ref{fig:SDF}, and~\ref{fig:VECT_and_tilting_conv} have been generated by the author using the graphics software \emph{draw.io}, version v27.0.9.\medskip

Compared to the version submitted at TU Berlin in June 2025, the front matter has been slightly revised --- a preface has been added, the abstract shortened, the acknowledgements compressed, and the order changed --- and minor misprints have been corrected throughout the document. One figure has been omitted from the main matter (Figure~2 in the version from June), since Figure~\ref{fig:refined_partitions_restaurant} has now been added in the preface and serves the same purpose.

\cleardoublepage
\phantomsection
\addcontentsline{toc}{chapter}{Contents}
\tableofcontents

\mainmatter
\pagenumbering{arabic}

\silentchapter{Introduction}
\silentsection{Extensive form characteristics}

This thesis aims at developing a unified language for dynamic games and decision problems under probabilistic uncertainty in terms of their \emph{extensive form characteristics}. In this text, extensive form characteristics are loosely defined as those properties characterising a decision problem or game by the flow of information about past choices and exogenous events, along with a set of adapted choices locally available to decision makers. The paradigmatic model of these extensive form characteristics is provided by classical \emph{extensive form} theory, as established by von Neumann and Morgenstern in \cite{Neumann1944Theory} and furthered by Kuhn in \cite{Kuhn1950Extensive,Kuhn1953Extensive}. This model has two variants. The first uses either refined partitions on a fixed set $W$ of outcomes; information is described by partitions of $W$, and choosing consists in refining them. The second is grounded on (decision) trees along with partitions on moves; information is described by these partitions, and choosing consists in selecting branches. For a graphical illustration of this, see Figure~\ref{fig:refined_partitions_restaurant} in the preface of this dissertation.\footnote{Many similar figures can be found in the foundational book \cite{Neumann1944Theory}, see, for instance, Figures~5 and~9 therein.} 
Both these formulations admit clear representations for information flow, outcomes, choices, and ``points'' where agents can choose. Canonical dynamic refinements of Nash equilibrium (e.g.\ subgame-perfect \cite{Selten1965Spieltheoretische} or perfect Bayesian equilibrium \cite{Fudenberg1991Perfect}) rely on this structure.

However, other relevant formulations of problems sharing extensive form characteristics exist. One important example for continuous-time models employing game-theoretic terminology is given by (stochastic) differential games. These are usually formulated in terms of a (stochastic) differential equation of the form
\begin{equation}\label{eq:SDG_SDE}
    \d \chi_t = V(\xi_t,\chi_t)\,\d \eta_t, \qquad t\in\R_+,
\end{equation}
where $\xi_t$ is the tuple of all agents' actions at time $t$, valued in, say, $\R^n$, $\eta$ is a suitable integrator (for instance, a function of bounded variation, or a continuous $\L^2$-martingale with respect to some probability measure $\P$), valued in, say, $\R^m$, $V\colon \R^{n+d}\to \R^{d\times m}$ is a sufficiently regular map so that this equation admits a unique solution in a reasonable sense, and $\chi$ describes the $\R^d$-valued solution, called ``state process'', which is the payoff-relevant quantity and which agents can partially condition future decisions on. In this context, information available to the agents is modelled via the state process's path $\chi$, filtrations, and the requirement that the processes, particularly $\xi$ and the strategy inducing $\xi$, are adapted to them. One typical definition of equilibrium would be, given a bounded continuous function $u$ from a suitable path space to $\R$:
\[ \E_x u(\chi^s) \ge \E_x u(\chi^{(\tilde s^i,s^{-i})}), \qquad \forall x\in \R^d, \]
for all players $i=1,\dots,n$ and all unilateral deviations $\tilde s^i$. In this example, $s = \xi$ is the strategy profile, directly determining the action process, $n$ is the number of players, and $s^i = \xi^i$ is its $i$-th component. $\chi^s$ describes the solution to Equation~\ref{eq:SDG_SDE} given $\xi = s$, with initial condition $x$ under the measure $\P_x$, whose expectation is denoted by $\E_x$. This is a stacked strategic form perspective: For any initial value of the state process, an own strategic form is defined.\footnote{See \cite{Isaacs1999Differential, Friedman1972Stochastic} for the initial accounts due to Isaacs and Friedman. For a more recent overview on differential games in general, see \cite{Dockner2000Differential}. Regarding stochastic differential games, see \cite{Carmona2018Probabilistic, Carmona2018Probabilistica} for a recent textbook focusing on stochastic differential games where the ``vector field'' $V$ depends also on the distribution of $(\xi,\chi)$, called ``mean field games'', introduced independently in \cite{Huang2006Large,Lasry2007Mean}.} 

As another natural example, timing games can be seen as the simplest non-trivial class of dynamic games. In these, at each point in time, agents can choose a number in $\{0,1\}$, the choice of $0$ being irreversible. These games are often formulated in terms of the first times the number~$0$ is chosen by the agents (``optional'' or ``stopping'' times). An alternative description is based on decreasing $\{0,1\}$-valued\footnote{Or $[0,1]$-valued, which is equivalent, up to taking some conditional expectation of the $\{0,1\}$-valued process in question, see \cite{Bismut1979Temps, Touzi2002Continuous}.} stochastic processes adapted to some information flow modelled by a filtration.\footnote{See, e.g., \cite{Dynkin1967Game, Dynkin1969Theorems} for the start of the literature on ``Dynkin games'', \cite{Touzi2002Continuous, Laraki2013Equilibrium, Hamadene2014Multiplayer} and the references therein for more recent works on this topic. For another, more abstract ``game-theoretic'' stream of the literature on timing games, see, for instance, \cite{Fudenberg1985Preemption, Laraki2005Continuous, Riedel2017Subgame, Steg2018Preemptive} and the references therein. For a review of the vast literature on economic applications via real options theory, see \cite{Azevedo2014Developing}.} Due to their relative simplicity, timing games allow to study conceptual problems in game theory and solutions to these like under a magnifying glass. For example, one classical conundrum in continuous-time preemption games is the difficulty of constructing a symmetric pre\-emption equilibrium, which is readily possible in discrete time. The problem is that complicated chains of randomised action and reaction on smaller and smaller discrete-time grids collapse in the continuous-time limit --- or seen from a different perspective, in continuous time, waiting for a tiny, but positive amount of time with positive probability still gives the opportunity to an opponent to preempt you. 

Decision trees and extensive forms based on them provide the canonical model of extensive form characteristics as defined above. Although classical extensive form theory following \cite{Neumann1944Theory, Kuhn1950Extensive, Kuhn1953Extensive} relies on strong finiteness assumptions, the concept itself is very general and broadly applicable. In a series of papers, including \cite{AlosFerrer2005Trees, AlosFerrer2008Trees, AlosFerrer2011Comment}, and in the monograph \cite{AlosFerrer2016Theory}, Alós-Ferrer and Ritzberger develop an abstract, highly general theory of extensive form games and decision problems and give a concrete order-theoretic characterisation of its own boundaries. Beyond these boundaries, the notions of strategy, outcome, and equilibrium lose rigorous decision-theoretic meaning when applied to the extensive form characteristics of a given decision problem. Accordingly, it is plausible to presume that any decision problem and game exhibiting extensive form characteristics lies within these boundaries or is, at least, some sort of limit of objects within these boundaries. The inclusion, or the precise meaning of this limit, allows us to rigorously determine the decision-theoretic meaning of strategies, outcomes, equilibrium, etc.\ for the given decision problem.

However, to the best of the author's knowledge, the existing literature leaves three important questions unanswered.
\begin{enumerate}
    \item From an abstract and general perspective, how can the ``extensive form characteristics'' of stochastic games, decision problems exposed to randomness, or those endowed with rando\-misation devices be modelled faithfully? 
    \item Given an abstract, general, and faithful model of these ``extensive form characteristics'' under probabilistic uncertainty, what are abstract and general consistency conditions giving rise to a complete game-theoretic model consisting of a stochastic extensive form (as we may call it) and an equilibrium concept for strategy profiles in this form?
    \item Given an answer to the two previous questions, what precise meaning can be given to continuous-time games under probabilistic uncertainty --- Bayesian and stochastic games, stochastic differential games and timing games, games with randomisation devices or ``mixed'' strategies --- as limits of extensive form games?
\end{enumerate}
Regarding the first of these questions, let us recall the abstract principle according to which a game or decision problem is specified by a complete set of rules --- since otherwise it could not be played or solved. In particular, there would be no point in analysing solution concepts. Yet, the interpretation of this principle is subtle in the context of randomness.

\silentsection{Probabilistic uncertainty in game theory}
The theories of games and of probability have a closely intertwined history. Probability theory evolved out of the study of games of chance. It originated in questions about the fair value of such games (Pascal, Fermat, Huygens) and eventually led to the development of an epistemic and mathematical concept of probability (notably in Jacob Bernoulli's \emph{Ars conjectandi}, followed by De Moivre and Laplace, among many others, and then the modern theory, following Kolmogorov and so many others, we teach today); see \cite{Shafer1993Early, Sylla2014Tercentenary} and the references therein. Game theory, on the other hand, is ``interactive decision theory'' (see \cite{Aumann2020Game}). Thus, as it matters essentially what players know, how probable they esteem those events they do not know, and what they know about what the others know (or believe), and so forth, probability is essential. It is not by coincidence that game theory motivated the von Neumann-Morgenstern theory of representing preferences in terms of expected utility, a theory pioneered by Daniel Bernoulli (cf.\ \cite{Bernoulli1954Exposition}); and it is not by coincidence either that the central result on equilibrium existence in finite games by Nash (cf.\ \cite{Nash1950Equilibrium,Nash1951Non}) relies on the possibility of players to randomise.

The foundational book by von Neumann and Morgenstern \cite{Neumann1944Theory}, then, adopted a particularly strict reading of the principle of complete rules to model probability in games. On this reading, randomness in decision problems and games is modelled by probability measures over actions called ``lotteries'', leading to the classical notion of mixed strategies. In this view, there is no difference between information about choices of agents and other sources of uncertainty. The latter is just interpreted as uncertainty about an additional, virtual agent's behaviour. That agent is called ``nature'', and the other agents receive the qualifier ``personal''.\footnote{This approach is described, for instance, in, \cite[Subsection 2.2.3]{AlosFerrer2005Trees} or \cite{Fudenberg1991Game, AlosFerrer2016Theory}. It is also adopted in the context of stochastic games following Shapley \cite{Shapley1953Stochastic}.} 

Yet, the approach of generating probabilistic uncertainty via classical mixed strategies is restrictive. First, it implicitly makes substantial assumptions on probability regarding objectivity and independence. Note that a classical mixed strategy does not describe the randomisation procedure explicitly. Although the analogy to lotteries on horse races is often made as a motivation, in that formalism there are no horses and no market for lottery tickets. Hence, the interaction of players in that market, previous individual experience on that market, and the players' commitment to the realisation of the horse race remain inside a black box (see, e.g., the discussion in \cite{Luce1989Games}). In fact, as Aumann pointed out in \cite{Aumann1974Subjectivity,Aumann1987Correlated}, this process can give rise to many kinds of correlated and subjective probabilities, and resulting behaviour may indeed constitute a certain type of equilibrium. In that view, classical mixed strategies are one possible distributional outcome of it, arising under special hypotheses on information and independence, but by far not the only one. Adapting \cite{Aumann1974Subjectivity,Aumann1987Correlated}, this information can be seen as part of the extensive form, the independence as a result of the agents' beliefs. 

Ultimately, the mentioned problem arises by setting too narrow a focus on the ``objective'' and ``frequentist'' aspects of probability. Decision-theoretic arguments favouring a detailed treatment of its ``subjective'' facets and an abstract theory thereof can be found in the literature initiated by De Finetti, Ramsey, Savage (see \cite{Savage1972Foundations} and the references therein). Moreover, the empirical observation underlying the Allais ``paradox'' provides further grounds to such a theory (see \cite[Example~2.32]{Foellmer2016Stochastic} for a textbook presentation, and the references therein). The branch of Bayesian statistics, going back to Laplace's ``inverse probabilities'' (cf.\ \cite{Shafer1993Early}), is based on this reading of probability.\footnote{The different aspects of probability (objective, subjective, Bayesian, frequentist, constructive) are not necessarily in contradiction, though; to the contrary, they are necessary complements. For instance, J.\ Bernoulli's law of large numbers can be read in a Bayesian or in a frequentist way. Analysing i.i.d.\ coin flips, as Bernoulli did, might seem as a rather objectivist exercise --- nevertheless, his book is entitled \emph{Ars conjectandi}, and not \emph{Scientia aestimandi} or the like. See \cite{Sylla2014Tercentenary} for further discussion on J.\ Bernoulli's understanding of probability.}

Second, as probability measures on actions, classical mixed strategies oblige one to specify $\sigma$-algebras on actions --- only for the purpose of randomisation. In the dynamic setting, however, there is a well-known trade-off between allowing for interesting $\sigma$-algebras on actions and measurability of the induced outcome map.\footnote{See the attacker--defender example in \cite{Aumann1964Mixed}, or the similar ultimatum bargaining example in \cite[Example~1]{AlosFerrer2016Characterizations} alias \cite[Example~6.7]{AlosFerrer2016Theory}).} Aumann (cf.\ \cite{Aumann1963Choosing, Aumann1964Mixed}) argues that the best one can do is using an external randomisation device, given by a fixed probability space $(\Omega,\ms E,\P)$, and restricting to maps from $\Omega$ to (complete contingent plans of) actions having sufficient measurability properties. Third, if uncertainty arises through probability measures on actions alone, then the conceivable real situation of agents not knowing payoffs or the set of admissible actions falls out of the game-theoretic framework, which demands completely specified rules. Still, this ``feels like a game'', whence the paradoxical term of ``games of incomplete information''.

This leads to Harsanyi's idea of ``insourcing'' such uncertainty to the ``nature agent's'' move at the game tree's root.\footnote{See \cite{Harsanyi1967Games, Harsanyi1968Games, Harsanyi1968Gamesa} for Harsanyi's foundational papers; see \cite{Mertens1985Formulation, Aumann1987Correlated, Aumann1995Epistemic} and the references therein for the further formal and conceptual development.} In this view, ``incomplete information'' emerges as a special kind of imperfect information. This information is exogenous in the sense that it is neither the result of nor influenced by decision making of ``personal agents'' and by respective equilibrium analysis.\footnote{The word ``exogenous'' has many meanings in many contexts, including economics, which the present meaning should not be strictly equated with.} According to Harsanyi's approach, this \emph{exogenous information} is described by a measurable space $(\Omega,\ms E)$ of ``states'', the ``uncertainty domain'', by measurable dependence of ``{personal agents'}'' choices on sub-$\sigma$-algebras $\ms F\subseteq \ms E$, or equivalently (via $\ms F = \sigma(\theta)$), random variables $\theta\colon (\Omega,\ms E)\to(\Theta,\ms T)$ modelling signals about ``types'' describing information about the ``true'' state $\omega\in\Omega$, and, on a possibly subjective level, by belief systems of personal agents, given in terms of probability measures $\P$ on states. 

A key observation is that, first, this modelling scheme is not only about ``incomplete information''. Namely, exogenous information can be any information that is, as written just above, neither the result of nor influenced by decision making of ``personal agents''. This information is, at least partially, revealed to ``personal agents'' by signals. That can be signals about some random event, about the realisation of a state process, about the result of a coin flip or roulette wheel --- all this can be considered as randomisation, and there is no need to go formally beyond ``pure'' strategies defined as complete contingent plans of action. The statistical dependence or correlation between signals --- and, \emph{a fortiori}, between strategies --- is described by the agents' beliefs. Second, it does not make an essential difference if, by extension, the only move of the ``nature agent'' is the game tree's root. The relevant structure is not the ``nature agent's'' choices, but the information about it available to ``personal agents'' when they move. Going one step further, we may simply drop the concept of a ``nature agent'' as well as the qualifier ``personal'', and restrict ourselves to an exogenous scenario space $(\Omega,\ms E)$ and a structure of exogenous information $\ms F_\x\subseteq\ms E$ available to agents at the their ``moves'' $\x$.

From this perspective, the outlined modelling scheme following Harsanyi's ideas is decision-theoretically more general and more flexible than the approach of generating probabilistic uncertainty via classical mixed strategies. It is indeed compatible with subjective probability (in the sense of \cite{Savage1972Foundations, Aumann1974Subjectivity, Aumann1987Correlated}, as discussed above), and with the idea of ``external randomisation devices'' as in \cite{Aumann1964Mixed}. Under the assumption of a common prior, a sufficiently rich uncertainty domain $(\Omega,\ms E)$, and suitable symmetry and independence properties of signals (translating into corresponding properties of subjective posterior beliefs), we can also retrieve the objective approach. Moreover, as strategies equal ``pure'' strategies in this setting, no $\sigma$-algebras on actions etc.\ have to be chosen for the sole purpose of randomisation. Ultimately, the measurability trade-off mentioned above does not concern the extensive form, but equilibrium analysis on it. That is, given a strategy profile $s$, given the induced outcome $\Out(s \mid .) \colon\Omega\to W$ mapping any exogenous scenario (alias state) $\omega\in\Omega$ to an outcome $\Out(s\mid\omega) = w\in W$, and given a, say, bounded, ``taste'' (alias ``payoff'' alias ``utility'') function $u\colon W\to \R$, an expected utility criterion requires measurability of the \emph{composition} $u\circ\Out(s \mid .)\colon \Omega\to \R$. No measurability condition need be formulated directly on $W$. Finally, the outlined modelling scheme yields a game-theoretic formalism under probabilistic uncertainty based on the principle of ``completely specified rules''.

\silentsection{Stochastic decision forests and adapted choices}

The extensive form is the basic game-theoretic formulation of the extensive form characteristics of a dynamic decision problem. The aforementioned two most traditional models of extensive forms are based on either \emph{refined partitions} on a fixed set of outcomes, or on \emph{decision trees} whose nodes describe moves and outcomes. This goes back to \cite{Neumann1944Theory, Kuhn1950Extensive, Kuhn1953Extensive}, and has been substantially generalised in \cite{AlosFerrer2005Trees, AlosFerrer2008Trees}. A third and also quite common formulation --- which we call \emph{action path} formulation --- is based on paths valued in some action space, indexed over time, as in repeated strategic form games (see e.g., \cite[Subsection~2.2]{AlosFerrer2005Trees}).

Arguably, the tree-based model is the standard formulation used in textbooks and much of the abstract literature.\footnote{See, e.g.\, \cite{Fudenberg1991Game, MasColell1995Microeconomic}.} Therein, choices are traditionally described by actions which uniquely determine a successor node at any node of a given information set; information sets, given by partitions on moves, are a primitive in that description. 
However, following an argument by \cite{AlosFerrer2005Trees}, the approach based on refined partitions has a particular decision-theoretic appeal because then strategies can be understood as maps from states to consequences (Savage acts, cf.\ \cite{Savage1972Foundations}) --- or, more precisely, from local states to local consequences --- with both local states and local consequences being described by certain sets of outcomes (= global consequences). The article \cite{AlosFerrer2005Trees} shows that both approaches are equivalent. Indeed, using the terms from \cite{AlosFerrer2005Trees},\footnote{For formal details, we refer to \cite{AlosFerrer2005Trees}, and to Chapter~\ref{chap:1-SDF_AC} of this thesis, where this is reviewed and further developed.} any decision tree $(T,\ge)$ can be represented by a system $\hat T$ of subsets of the set $W$ of all maximal chains (``plays'', or paths of local decisions) in $T$ such that $\hat T$ defines a decision tree with respect to the superset relation $\supseteq$. Moreover, if $(T,\ge)$ is already a set tree on a set $V$ (i.e.\ $T\subseteq \mc P(V)$ for some set $V$, and $\ge$ equals the superset relation $\supseteq$), then it is canonically isomorphic to this representation. Actually, the theory in \cite{AlosFerrer2005Trees} works out as mentioned without assuming connectedness of the graph $(T,\ge)$, and thus can be applied to disjoint unions of decision trees, that is, what we effectively call \emph{decision forests}.\footnote{This follows indeed graph-theoretical conventions, see \cite{Bollobas2013Modern}, as detailed in Chapter~\ref{chap:1-SDF_AC}.} 

Following Harsanyi's \emph{insourcing} of ``incomplete information'' into an uncertainty domain $(\Omega,\ms E)$,\footnote{Cf.\ \cite{Harsanyi1967Games, Harsanyi1968Games, Harsanyi1968Gamesa}.} this text suggests to \emph{outsource} uncertainty as uncertainty about the realised decision tree, or put equivalently, the realised connected component of the decision forest. In view of the preceding discussion on probabilistic uncertainty in game theory, $(\Omega,\ms E)$ admits several possible interpretations, e.g.\ as an external randomisation device, as a Bayesian uncertainty domain on that subjective probabilistic beliefs can be formed, or as a (rather objective) source of ``exogenous'' noise or randomness. The events in $\ms E$ are \emph{exogenous} in that players' choices do not affect them. No ``nature'' agent is needed in this view.

In order to describe the exogenous information dynamically revealed to agents in terms of sub-$\sigma$-algebras of $\ms E$, a structure identifying moves in different trees is needed --- since the moves are discrete objects and we may wish to allow for non-discrete sub-$\sigma$-algebras. Then, despite moves being discrete, agents may have non-discrete information about which of the identified moves across different trees is the realised one. This leads us to introduce the notion of \emph{stochastic decision forests}, around which Chapter~\ref{chap:1-SDF_AC} is centred. In a simplified version, a stochastic decision forest on $(\Omega,\ms E)$ is given by a decision forest $F$ on a set $W$ of outcomes, a surjection $\pi\colon F\to \Omega$ whose fibres $\pi^{-1}(\{\omega\}) = T_\omega$ are the connected components of $(F,\supseteq)$, and a set $\X$ such that, if $X$ denotes the set of $F$'s moves, first, any element $\x\in\X$ is a section of moves, that is, it is a map $\x\colon \Omega \to X$ satisfying $\pi\circ\x = \id_{\Omega}$, and, second, $\X$ induces a covering of $F$'s moves, that is, $\{\x(\omega) \mid \x\in\X,\,\omega\in \Omega\} = X$. The elements of $\X$ are called \emph{random moves}. This is illustrated in Figure~\ref{fig:SDF}.

In this approach, it may seem as if two forms of information, or uncertainty, are separated: 1)~information about the measurable space of scenarios $(\Omega,\ms E)$, 2)~information about (past) choices of agents. The first is not the result of the play of any decision maker; in that sense it is ``exogenous''. The second arises from decision making; in that sense it is ``endogenous''. For the second, a complete dynamic description of the process generating information is provided by the order structure given by the forest $(F,\supseteq)$; not so for the first. Players generate ``endogenous'' information by strategic behaviour and receive bits of it; regarding ``exogenous'' information, it is sufficient to regard players as receivers. As a side effect, the problem of whether a strategy profile induces a unique outcome (arising e.g.\ in continuous-time models) is a non-issue regarding exogenous noise: ``nature'', so to speak, can ``play'' Brownian motion although (the paths of) this presumed ``strategy'' do not lie within the maximal strategy set described by \cite{Stinchcombe1992Maximal} or cannot really be generated by ``playable'' (alias well-posed) extensive forms according to \cite{AlosFerrer2008Trees, AlosFerrer2011Comment}.

To any random move, we can associate a sub-$\sigma$-algebra of $\ms E$ describing exogenous information in the mentioned way. 
Then, adaptedness with respect to this (not necessarily discrete alias partition-generated) ``exogenous information'' can be formulated directly on the level of choices. In the refined partitions-approach following \cite{AlosFerrer2005Trees}, choices can be defined as unions of nodes, which are thus non-empty sets of outcomes. Consequently, the availability of a choice at a move --- including information about preceding choices of all agents --- is already encoded in the choice itself. This gives rise to a notion of information sets, which is further developed in Chapter~\ref{chap:2-SEF_G}. To any random move, then, we can attach a set of reference choices an agent can, in principle, make at that random move. A choice is \emph{adapted}, in this framework, iff, conditional on any available reference choice, its availability can be inferred from the given exogenous information. That way, a notion of choices obtains that is compatible both with general measurable spaces along the exogenous dimension, and with the complete discrete description of dynamic decision making along the tree dimension. As a side effect, any form of randomisation can be formulated in terms of $(\Omega,\ms E)$ and, so, all measurability issues can be understood in terms of $(\Omega,\ms E)$. 

As illustrated in the beginning, the fundamental motivation for this theory becomes apparent in a large class of concrete applications. Thus, its presentation is tightly accompanied by examples. This includes simple textbook examples of finite trees (and forests). Following \cite{Gilboa1997Comment}, it also includes a model of the absent-minded driver story by \cite{Piccione1997Interpretation} which is sometimes seen as a challenge to extensive form theory (so in \cite{Kuhn1953Extensive, AlosFerrer2005Trees}). 
Furthermore, in the deterministic model in \cite{AlosFerrer2005Trees}, action path-based models have been included and discussed in some examples. Herein, special attention was payed to the continuous-time case (as an attempt to model differential games). In this text, we strongly generalise these examples to stochastic decision forests based on a small set of readily verifiable axioms, allowing for many different time regimes, many different specifications of the outcomes (e.g.\ paths of timing games, also in the case of the scenario-dependent expiration of certain options for action), and general stochastic noise. Of course, it also includes the deterministic case. To the best of the author's knowledge, this theoretical unification of action path-based decision problems with extensive form characteristics within a single framework is another new contribution of this chapter in its own right.

\silentsection{Stochastic extensive forms and games}\label{0-intro.sec:SEF_G}

In order to obtain an extensive form model under probabilistic uncertainty, it remains to describe a consistent way of equipping stochastic decision forests with structures of exogenous information, and of reference and adapted choices. Introducing and studying this model --- called the \emph{stochastic extensive form} --- is the subject of the second chapter. Using that stochastic decision forests are, in particular, forests of decision trees, we do so by generalising the extensive form model from \cite{AlosFerrer2008Trees, AlosFerrer2011Comment}. 

Not all of these generalisations are evident. Based on the idea that the $\Omega$-component of strategic behaviour is fully contained in choices, we only allow for complete choices, that is, choices available at entire random moves. With this, it is reasonable to demand that the sets $P(c)$, where $c$ runs over all choices of a fixed agents, form a partition of that agent's moves. Here, $P(c)$ denotes the set of moves choice $c$ is available at. This gives rise to a --- partition-based --- notion of ``endogenous'' (i.e.\ decision-related) \emph{information sets}, reflecting that there is no \emph{a priori} choice of measurability structure on outcomes or choices. Then, a \emph{strategy} is simply a complete contingent plan of action, that is, a map from information sets to choices available at these. In this model, adaptedness is a feature of choices, not strategies.

The first question about a stochastic extensive form is whether it is well-posed, that is, whether, given any history, any strategy profile induces a unique outcome. Beyond the finite case, this is actually an intricate problem. Many revealing examples for this can be found in \cite{Simon1989Extensive,Stinchcombe1992Maximal} (see also \cite{AlosFerrer2016Theory}, and the references therein). To take (a simplified version of) one of these examples, consider the naive continuous-time two-player extensive form where outcomes are paths $\R_+ \to \{0,1\}^I$ describing the actions of both players over time, with $I = \{1,2\}$ being the set of players. A naive history at time $t\in\R_+$ is a truncated path $h_t\colon [0,t)_{\R_+}\to \{0,1\}^I$. Then, the map $s^i$, $i\in I$, associating to any history $h_t$, $t\in\R_+$, action $0$ if $h_t$ is constant equal to $0$ and $1$ else, is a complete contingent plan of action, a strategy. If both players choose that strategy, however, there is an infinity of outcomes compatible with it. Namely, for any $u\in\R_+$, the outcome $w\colon\R_+\ni t\mapsto (1\{u<t\},1\{u<t\})$ is compatible with $s = (s^1,s^2)$: For $t\in[0,u]_{\R_+}$, the history $w|_{[0,t)_{\R_+}}$ is constant equal to zero, hence both players select action $0$. For $t\in (u,\infty)_{\R_+}$, the history $w|_{[0,t)_{\R_+}}$ is not constant equal to zero, hence both players select action $1$. The restriction to right-continuous outcomes $\R_+\to\{0,1\}^I$ is not a solution either, since then the symmetric strategy profile given by ``choose $1$'' at time zero and ``choose $0$'' at any time $t>0$ does not admit any compatible outcome.\footnote{Note that we made a restriction on outcomes, not on strategies. A priori, there is no reason to restrict strategies; strategies derive from the primitive data --- such as outcomes, decision points, information --- as complete contingent plans of action.}
By considering a subset of strategy profiles satisfying certain regularity conditions (roughly, some form of right-continuity), one can restore existence of unique outcomes (cf.\ \cite{Stinchcombe1992Maximal}). The fact that this rules out instantaneous reaction can be compensated by adding infinitesimal reaction times (cf.\ \cite{Stinchcombe1992Maximal}). But then, in any case, not any complete contingent plan of action is accepted as a strategy which violates a fundamental principle of extensive form modelling. 

Understanding the general conditions under that an extensive form is well-posed \emph{a priori}, with its canonical set of strategies, is therefore crucial. This issue has been a main driver for the study of the deterministic model in \cite{AlosFerrer2008Trees}, and we need to understand it as well in the generalised stochastic setting. In Chapter~\ref{chap:2-SEF_G}, we show that restricting a stochastic extensive form to a fixed exogenous scenario induces an extensive form in the classical sense of \cite{AlosFerrer2011Comment} --- and that a stochastic extensive form is well-posed iff this holds true for the induced classical extensive form in any exogenous scenario. As a consequence, we can apply the fundamental characterisation results from \cite{AlosFerrer2008Trees, AlosFerrer2011Comment} and find that well-posedness can be characterised by basic order-theoretic properties of the underlying forest. An essential, necessary condition is up-discreteness, meaning that any non-empty chain of nodes has an earliest element; or put equivalently, that all maximal chains alias decision paths are well-orders with respect to $\supseteq$. 

This classification result shows in particular that, in order to obtain well-posedness of a stochastic extensive form, decision paths must be well-ordered --- a strong restriction, already known from \cite{AlosFerrer2008Trees, AlosFerrer2011Comment} in the deterministic case. If supposing an underlying time axis, modelled by a total order, and equating decision making with a (possible) change of action, this implies that the resulting action paths must be locally right-constant. 

In this text, we also develop an abstract and general model of stochastic extensive forms based on time-indexed paths of action. For these action path stochastic extensive forms we then infer the abstract and general result that well-posedness is equivalent to the (relevant part of) the time half-axis being well-ordered. This both illustrates the range and delimits the boundaries of action path stochastic extensive forms. 
Although this rules out typical stochastic differential and other continuous-time games, we can construct a vast class of well-posed stochastic extensive forms based on action paths on a fixed well-ordered time grid. This construction --- and (a selection of) the outcomes they induce --- form the basis for the ``approximate extensive form'' model introduced later in Chapter~\ref{chap:3-SPF_VECT}, providing a description of continuous-time games in terms of their extensive form characteristics. 
Besides action path stochastic extensive forms, we also continue the discussion of the key concepts using simple examples, including a full description of the absent-minded driver phenomenon.

In order to define a game, the last step consists in defining a solution, or equilibrium, concept. We propose such a notion, based on the general extensive form concept of perfect Bayesian equilibrium (see \cite{Fudenberg1991Perfect}), which can be naturally implemented given the stochastic extensive form model --- and in particular the model of histories, random moves, and information sets describing points ``where'' player can revise their plans. This also includes subgame-perfect equilibrium (cf.\ \cite{Selten1965Spieltheoretische}). The concept is illustrated by various examples, including the absent-minded driver model just mentioned.

We also discuss the model of randomisation in stochastic extensive forms. In this model, randomisation --- e.g. in the ``mixed strategy'' sense --- is explicitly described in terms of functional dependence of choices on exogenous scenarios, and corresponding correlation, dependence, and distributional properties of relevant bits of exogenous information under the agents' belief, much like in \cite{Aumann1974Subjectivity}. As we do not add measurability structure across sets of available choices or the like, the classical trade-off between relevant $\sigma$-algebras on actions and measurability of the induced outcome map (cf.\ \cite{Aumann1964Mixed, AlosFerrer2016Characterizations}), discussed above, remains, though in a different, and perhaps helpful, form.

\silentsection{Stochastic process forms in vertically extended continuous time}

The interplay of continuous time and complex uncertainty represents both a litmus test and a fundamental challenge for game theory. When time is modelled as a continuum and information is subject to intricate structures of uncertainty, even the most basic notions --- such as strategy and outcome, simultaneity and reaction, randomisation and beliefs, subgames and information sets --- become elusive. Ultimately, canonical meta-concepts of equilibrium and optimality become difficult to implement rigorously and in a natural way, raising the questions of their general decision-theoretic meaning and of their interpretation in concrete applications.

The reasons for continuous-time modelling are manifold. One is pragmatic. Working in continuous time unlocks the powerful toolbox of (real, stochastic, functional, ...) analysis and considerably helps pushing the frontiers of mathematical tractability. Relatedly, the numerics --- calculating optimisers, values, equilibria --- may become much more tractable and the convergence of numerical schemes easier to analyse. Another reason is decision-theoretic. Working in discrete time assumes that there is some entity or natural constraint that can enforce agents to act only at the times of the predefined grid --- though in many situations it is well conceivable that a real agent could try to act in between two grid points. This is a severe restriction for non-cooperative game theory, where, once the rules have been fixed, agents are otherwise anarchic. A fourth traditional argument is rather ``philosophical'', based on the statement that time really ``is'' continuous. If one accepts this point of view, any realistic model must employ continuous time.\footnote{We express no philosophical opinion regarding the validity of this argument; we just mention its existence. It has a prominent tradition in the debate about the interpretation of physics. Another example for its use may be volatility modelling in finance. At least when assuming the concept of volatility to represent reality in one way or the other as opposed to a mainly instrumental view on it, the claim that ``volatility is rough'' (cf.\ \cite{Gatheral2018Volatility}, see also \cite{Fukasawa2020Volatility} and there references therein, not least their titles) would make little sense without the implicit claim that ``time is continuous''.}

For examples, we refer to the discussion of stochastic differential and continuous-time timing games from the beginning. On this occasion we emphasise that in stochastic differential games, the action process $\xi$ and the state process $\chi$ do typically not have locally right-constant paths, and hence there is little hope of being able to formulate the problem directly within a well-posed action path stochastic extensive form, by the results of Chapter~\ref{chap:2-SEF_G}.\footnote{In \cite{AlosFerrer2005Trees}, game trees based on action paths are actually linked to ``differential games''. This neglects the fact that usually differential games are not formulated via trees, but via differential equations and filtrations. Be it as it may, the corresponding decision tree, or decision forest in the stochastic framework we propose in this text, is not such as to give rise to extensive forms or, even if so, to ensure their well-posedness, as we conclude in Chapter~\ref{chap:2-SEF_G}, thereby generalising findings from \cite{AlosFerrer2008Trees,AlosFerrer2011Comment}.} We also note that in timing games instantaneous reaction to new exogenous information, i.e.\ a Brownian motion hitting some boundary, may be a relevant optimiser or best response candidate which is also incompatible with the well-posed action path stochastic extensive form models constructed in Chapter~\ref{chap:2-SEF_G}.

The sheer size, breadth, and relevance of the literature on continuous-time games and control, which involve uncertainty (stochastic noise, randomisation, ``incomplete information'', ...) to varying, but often high degrees, makes it necessary to develop an abstract decision-theoretic understanding. Existing standard approaches typically fall into two extremes: either a stacked strategic form, as is common in stochastic differential games and control (see \cite{Friedman1972Stochastic, Riedel2017Subgame, Cohen2015Stochastic}), without any canonical way of implementing extensive form notions like subgame-perfect equilibrium (cf.\ \cite{Selten1965Spieltheoretische}), or a fully developed well-posed extensive form (cf.\ \cite{Neumann1944Theory, Kuhn1953Extensive, AlosFerrer2016Theory}), based on action paths indexed over time (cf.\ \cite{AlosFerrer2005Trees, AlosFerrer2008Trees} for the basic deterministic example, and Chapters~\ref{chap:1-SDF_AC} and~\ref{chap:2-SEF_G} for a general stochastic theory). One must further mention the product form (cf.\ \cite{Witsenhausen1971Information, Witsenhausen1975Intrinsic, Heymann2022Kuhns}) which leaves the extensive form terrain in a decent way in order to center on measurability, also with respect to information on past choices. Yet, to best of the author's knowledge, measurability along time (or more generally, instances of decision making, e.g.\ subgames) is not focal in this formal model, while it is critical for continuous-time theory and the application of (stochastic) analysis.

All these perspectives aim to capture the essence of dynamic strategic interaction, and all seek to be analysable through dynamic refinements of the Nash equilibrium concept. This thesis argues that these approaches, despite their differences, share underlying extensive form characteristics and can be unified through an abstract and general formal model. We show that this can be done in a way incorporating the necessary structure to implement dynamic equilibrium (like perfect Bayesian or subgame-perfect) and to be compatible with complex forms of probabilistic uncertainty in continuous time and, in particular, the general theory of stochastic processes (cf.\ \cite{Dellacherie1978Probabilities, ElKaroui1981Les, Jacod2003Limit, Protter2005Stochastic}).

In a temporal setup involving probabilistic uncertainty, focusing on extensive form characteris\-tics leads to the basic insight that the fundamental object describing outcomes of possible choices are not decision trees or forests (as in extensive form theory), nor mere collections of random variables over decision points or ``agents'' (as in the product form), but rather stochastic processes satisfying particular measurability properties with respect to a given filtration. Yet, strategies must be understandable as complete contingent plans of action, alias local choices, given these infor\-mational alias measurability constraints; it must be explained which outcome processes they induce; it must be explained what are the ``decision points'', ``subgames'', ``information sets'' (including counterfactuals) conditional on that outcome processes are induced and, ultimately, preferences are formulated. As in extensive form theory, the stacked strategic form ultimately used in equilibrium analysis must derive from these data.

Thus, we develop a framework that speaks the language of stochastic processes while syste\-matically preserving the structural hallmarks of the underlying extensive form characteristics. This abstract model is deliberately general: it subsumes a broad class of games and control problems formulated in terms of stochastic processes and offers a conceptual bridge to extensive form-based reasoning. It enables rigorous comparisons with stochastic extensive forms (in the sense of this thesis, Chapter~\ref{chap:2-SEF_G}, a generalisation of \cite{AlosFerrer2016Theory, Kuhn1953Extensive, Neumann1944Theory}) and captures key strategic properties without relying on case-by-case ad-hoc adaptations of extensive form theory.

The relevance of such a model lies not only in its capacity to represent strategic dynamics with greater clarity, but also in its ability to enable a general theory of a) reformulating classical continuous-time models like stochastic differential games and then approximating them using rigo\-rous action path extensive form models, and of b) conversely describing decision-theoretic limits of action path extensive form models as reaction lags shrink to zero. Indeed, as mentioned above, a very general class of action path stochastic extensive forms --- that is stochastic extensive forms wherein outcomes are given by pairs $(\omega,f)$ of an exogenous scenario and a time-indexed path $f$ of action --- can be constructed, as demonstrated in Chapter~\ref{chap:2-SEF_G}. It is shown in Theorem~\ref{2-SEF_G.thm:AP_sef_well-posed} of this thesis that, if restricting to a well-ordered time grid in $\R_+$, these stochastic extensive forms are well-posed, i.e.\ strategy profiles induced unique outcomes. Under this hypothesis, action paths can be seen as locally right-constant paths in continuous time whose jump times lie in a fixed, well-ordered grid.

With this approximation intuition in mind, we make a fundamental observation concerning the limit behaviour of outcomes: as intervals of interaction shrink to zero, information about potential reactions vanishes. Precisely, if on the grid $G_n$, $n\in\Z$, given by $G_n(k) = k 2^{-n}$ for all $k=0,1,\dots$, Alice acts according to the process $\xi^n = 1[0,2^{-n})_{\R_+}$ and Bob according to $\xi^{n-1}$, then Alice switches to zero strictly before Bob, and there is no difficulty in modelling Bob observing this until time $2^{1-n}$. Letting go $n\to\infty$, both $\xi^n$ and $\xi^{n-1}$ converge pointwise to the action process $1\{0\}$. The order of action and also Bob's observation of Alice's action is lost in the limit. Moreover, there is no instant of time at that the stopping really occurs: at time zero, the value is still $1$, while at any time $\e>0$, the stopping must have already happened before --- a paradox.

This problem essentially underlies the challenge described and analysed in \cite{Fudenberg1985Preemption, Riedel2017Subgame} in the context of continuous-time preemption games. Yet, the analysis of these articles does not formally describe the ``limit'' outcome processes, which yet are a crucial part of the extensive form characteristics of the problem. If we wish to do so in the language of stochastic processes, the described phenomenon suggests allowing for well-ordered chains of ``reaction nodes'' at a single real point in time. Interpreting these as instances of instantaneous reaction, it becomes natural to glue well-orders above every real time point --- leading to a vertically extended continuous-time structure. That is, we consider the vertically extended set of continuous time $\tilde\T = \R_+\times\alpha$ for some well-order $\alpha$, whose smallest three elements we call $0,1,2$, and we equip $\tilde\T$ with lexicographic order. Then, the intuitive limit as $n\to\infty$ of Alice's behaviour $\xi^n$ is $1\{(0,0)\}$, whereas that of Bob's $\xi^{n-1}$ is $1\{(0,0),(0,1)\}$. Then, there is no loss of information in the limit. The set $\tilde\T$ and the notion of ``tilting'' convergence just described are illustrated in Figure~\ref{fig:VECT_and_tilting_conv}.

This extension raises several foundational questions: What size and structure should such well-orders have in order to be consistent with the above-mentioned limit procedure? How are stochastic processes to be defined on such an extended time scale? What are the appropriate notions of order, topology, and measurability on the vertically extended time half-axis? How can we formalise ``points'' that agents can consider their options and revise their plans at, and what are counterfactuals? What concept to use in order to describe the corresponding instances of time, optional times, and, based on that, optional processes generated by these times and corresponding actions, adapted to the information flow? Finally, how can we formally describe the ``tilting'' limit procedure motivated above, and establish a link between grid-dependent decision making in continuous time and decision making in vertically extended time? These questions, rooted in game, decision, and control theory, lead to an extended theory of continuous time and stochastic analysis on that time half-axis, which is the subject of the first two sections of Chapter~\ref{chap:3-SPF_VECT}. In the third section, a rigorous abstract game-theoretic model of stochastic extensive form characteristics based on stochastic processes in vertically extended time is introduced, which we call \emph{stochastic process form}, and we argue that it responds to the above challenges. Within this model, we derive notions of strategy and outcomes, information sets and subgames, randomisation and beliefs, and of equilibrium. We examine the contrasts with classical extensive, product, and strategic form approaches and explore the consequences of our framework for dynamic equilibrium analysis. To demonstrate both the breadth and specificity of the theory, we apply it to stochastic differential games in general and provide a detailed treatment of a prototypical class: timing games. 

On the one hand, we discuss on a rather abstract level how and why stochastic differential games fall within the framework of stochastic process forms, which provides a stronger decision-theoretic footing than the traditional stacked strategic form framework. A detailed analysis is performed in the case of timing games. Concerning the latter, we recall that \cite{Fudenberg1985Preemption, Riedel2017Subgame, Steg2018Preemptive} give a solution to the continuous-time preemption problem in terms of a stacked strategic form approach, based on extended mixed strategies. There, payoffs are defined as direct functions of strategy profiles, using a ``discrete time with an infinitesimally fine grid'' limit consideration. This approach does not formally describe outcomes, nor does it describe how payoffs arise from outcomes. One side effect of this is that payoffs are rather hard to formulate, and it probably would be even harder for more than two players. Moreover, in the cited approach strategies are actually ``stacked strategies'' alias large families of stochastic processes, one for each subgame. Thus, strategies formally depend on subgames --- though this dependency is \emph{ex post} weakened by a dynamic consistency requirement. However, this consistency condition is not further justified and appears as a solution to a difficulty arising from the formalism rather than from the formalised problem itself. After all, a strategy is a complete contingent plan of action alias local choices. Further, a ``subgame'' is a ``point'' where agents can revise these local choices. Both notions are no primitives in any strong sense, but arise from the description of information flow and local choices, i.e., the extensive form characteristics. Provided well-posedness of the game-theoretic model, a strategy profile induces a unique outcome in any ``subgame''. In the stacked strategic form model from the cited literature, these decision-theoretically important steps are skipped. This raises the question whether the ``stacked strategies'' can be integrated into one strategy process, defining a complete contingent plan of action. Similarly, we ask how ``subgames'', ``decision points'', ``information sets'' arise from information flow and local choices. This would clearly strengthen the interpretation of subgame-perfect equilibrium, because an equilibrium must be a strategy profile in its own right, understandable in terms of the basic extensive form characteristics.

Responding to this, we propose a well-posed stochastic process form model for timing games. Aside from being more general (general finite number of players, asymmetric information, full closed-loop setting, including a larger class of subgames), it gives a systematic explanation of outcomes and strategies, information, subgames, and equilibrium, deriving from the basic principles of the stochastic process form model. We show that the expected symmetric preemption equilibrium obtains via one global strategy profile which induces the expected outcome of randomisation on the vertical half-axis above the preemption boundary. As a corollary of the theory of ``tilting convergence'', transforming reaction on smaller and smaller time lags into chains of infinitesimal reaction, we obtain a representation of this outcome in terms of a limit of discrete-time approximations. This extends a similar result on the deterministic two-player timing game from \cite{Steg2018Preemption}.

\silentsection{Organisation of the thesis}

This thesis is organised as follows. 

In Chapter~\ref{chap:1-SDF_AC}, the basic, tree-based model of the stochastic extensive form characteristics under probabilistic uncertainty is introduced, culminating in the notions of stochastic decision forests, exogenous information structures, and adapted choices. The theory is tightly accompanied with examples, including absent-mindedness. We also explain how action path-based models can be formulated in this language, in high generality.

In Chapter~\ref{chap:2-SEF_G}, the central notion of stochastic extensive forms is defined, its informational properties are analysed, and strategies are introduced. Moreover, we introduce and analyse a suitable notion of histories and classify well-posedness in terms of order-theoretic properties of the underlying forest. This general theory, and in particular the well-posedness results, are applied to action path stochastic decision forests and extensive forms. Furthermore, dynamic equilibrium is implemented and illustrated by examples.

In Chapter~\ref{chap:3-SPF_VECT}, vertically extended continuous time $\ovT$ is introduced as the smallest complete total order containing all countable accumulations of well-orders embedded into $\R_+$. Topology and measurable structures on $\ovT$ are studied. Then, stochastic processes and random times are investigated, and suitable notions of progressive measurability, optional and predictable times and processes are introduced and their basic properties analysed. The notion of tilting convergence is introduced and so the fundamental link to outcomes of well-posed action path stochastic extensive forms is established. In the last section, stochastic process forms are introduced in full generality and their information sets analysed. The section is concluded with a case study of continuous-time stochastic timing games and a short discussion of stochastic differential games.

All proofs can be found in the corresponding sections of Appendix~Chapter~\ref{chap:App1-Proofs}. A pedagogic, self-contained treatment of the Dedekind-MacNeille completion of partially ordered sets, which we use in Section~\ref{3-SPF_VECT.sec:vERT}, is found in Appendix~Chapter~\ref{chap:App2-Additional_mat}.

\silentsection{Notations}

We give a list of some notations or conventions used throughout the text some of which are not completely standard.

\begin{itemize}[label=--]
    \item $\N = \Z_+$ = the positive integers including zero, $\Nast = \N \setminus \{0\}$, $\Q$ = the rational numbers, $\R$ = the real numbers, $\R_+ = \{x\in\R \mid x \ge 0\}$, all of them understood to be equipped with the standard order and algebraic structure;
    \item a function $f\colon D \to V$ from a set $D$ called \emph{domain} to a set $V$ is a subset of $D\times V$ such that for all $x\in D$ there is a unique $y\in V$ with $(x,y)\in f$, and this unique $y$ is denoted by $f(x)$; in other words, $f$ is described through its graph; as an abbreviation, the constant map on $D$ with value $y\in V$ is denoted by $y_{D} = D \times \{y\}$;
    \item $\mc P(A) = \mc P A$ = the set of subsets of a given set $A$, $\mc P(f) = \mc P f$ = the function $\mc P A \to\mc P B,\, M \mapsto \{ f(m) \mid m\in M \}$ for a given function $f\colon A \to B$ between two sets $A$ and $B$;\footnote{$\mc P$ defines a covariant endofunctor on the category of sets.}
    \item $\im f = (\mc P f)(D)$ = the image of a set-theoretic function $f\colon D \to V$;
    \item $\bigcup M $ = the union of a set $M$ = the set of all $x$ that are the element of some $S\in M$, also written $\bigcup_{i\in I} S_i$ in case $M$ is the image of some function $I\ni i\mapsto S_i$, for some set $I$;
    \item $f\times g$ = the function $D_1\times D_2 \to V_1\times V_2,\, (x,y) \mapsto (f(x),g(y))$ for functions $f\colon D_1\to V_1$, $g\colon D_2\to V_2$;
    \item $\mf w$ = the smallest infinite ordinal, $\mf w_1$ = the smallest uncountable ordinal;
    \item $|M|$ = the cardinality of a set $M$;
    \item $x<y$ means ``$x\le y$ and $x\neq y$'', given a partial order $\le$ an a set $M$ and $x,y\in M$; similarly, $>$ denotes the strict partial order associated to a partial order $\ge$;
    \item $[x,y)_T = \{z\in T \mid x \le z < y\}$ for any poset (partially ordered set) $T$ and all $x,y\in T$; intervals $[x,y]_T$, $(x,y]_T$, $(x,y)_T$ are defined similarly according to usual conventions;
    \item $\ms B_T$ = the Borel $\sigma$-algebra of a given topological space $T$;
    \item $\ms E|_D = \{E \cap D \mid E\in\ms E\}$, for any measurable space $(\Omega,\ms E)$ and any subset $D\subseteq \Omega$;
    \item $[\![\sigma,\tau)\!) = \{(t,\omega) \in \ovT\times\Omega \mid \sigma(\omega) \le t < \tau(\omega)\}$ for the poset $\ovT$ introduced in Section~\ref{3-SPF_VECT.sec:vERT}, any set $\Omega$, and all maps $\sigma,\tau\colon \Omega\to \ovT$, known under the name \emph{stochastic interval} in probability theory; the intervals $[\![\sigma,\tau]\!]$, $(\!(\sigma,\tau]\!]$, and $(\!(\tau,\sigma)\!)$ are defined similarly according to usual conventions; moreover, $[\![\tau]\!] = [\![\tau,\tau]\!]$;
    \item $f_\ast \mu$ is the push-forward of a measure $\mu$ on a given measurable space $(\Omega,\ms E)$ by a map $f\colon \Omega\to Y$ into some set $Y$, defined on the $\sigma$-algebra $f_\ast\ms E = \{B\subseteq Y \mid f^{-1}(B) \in \ms E\}$ and given by $f_\ast \mu(B) = \mu(f^{-1}(B))$, $B\in f_\ast\ms E$;
    \item $f^\ast \ms Y = \{ f^{-1}(B) \mid B\in\ms Y\}$ is the pull-back of $\ms Y$ by $f$, for any map $f\colon \Omega\to Y$ and any $\sigma$-algebra $\ms Y$ on $Y$;
    \item $\mf P_{\ms E}$ is the set of probability measures on a measurable space $(\Omega,\ms E)$;
    \item $\ms E^{\mathrm u}$ denotes the universal completion of a $\sigma$-algebra $\ms E$, that is, the intersection of the completions of $\ms E$ with respect to all elements of $\mf P_{\ms E}$.
\end{itemize}

\chapter{Stochastic decision forests and adapted choices}\label{chap:1-SDF_AC}
\section{Decision forests}\label{1-SDF_AC.sec:def}

The basic object of classical extensive form decision and game theory is the decision tree. In the stochastic generalisation presented here, nature does not act as an agent taking decisions dynamically, but is simply replaced with a device ``randomly'' selecting the decision tree the ``personal'' agents follow during their decision making process. This implies that we consider \emph{decision forests}, rather than decision trees. Although this term is not new (as testified by \cite{Rokach2016Decision}), we consider it in the context of abstract decision theory, and more precisely within the --- decision-theoretically natural --- refined partitions framework, and in the aim of making this framework amenable to exogenous noise in the sense of general probability theory. This framework has originally been developed for trees, rather than forests, pioneered in \cite[Section~8]{Neumann1944Theory}, and developed in much more generality in \cite{AlosFerrer2005Trees,AlosFerrer2008Trees} and subsequent papers, much of which is covered in the monograph \cite{AlosFerrer2016Theory}. In this section, we fix some order-theoretic language, present a definition and interpretation of decision forests within the refined partitions approach, and show that decision forests are nothing but forests of decision trees. For a graphical illustration of both the tree and the refined-partitions approaches, and of their duality discussed in the following, we refer to Figure~\ref{fig:refined_partitions_restaurant}, printed in the preface. 

\subsection{Order- and graph-theoretic conventions}\label{1-SDF_AC.subs:conventions}
We first recall some basic definitions from graph and order theory, thereby fixing conventions used in this text, which combine those from \cite{Bollobas2013Modern,AlosFerrer2005Trees,Davey2002Introduction}. In a partially ordered set (in short, \emph{poset}) $(N,\ge)$ a \emph{chain} is a subset $c\subseteq N$ such that for all $x,y\in c$, $x\ge y$ or $y\ge x$ holds true. A \emph{maximal chain} is a chain that is maximal as a chain with respect to set inclusion in $\mc P(N)$. $x\in N$ is called a \emph{maximal element} iff there is no $y\in N$ other than $x$ such that $y\ge x$. $x\in N$ is called \emph{maximum} iff for all $y\in N$, $x \ge y$. For $x\in N$, the \emph{principal up-set} $\uparrow x$ and \emph{principal down-set} $\downarrow x$ are defined by
\[ \uparrow x = \{y\in N \mid y\ge x\}, \qquad \downarrow x = \{y\in N \mid x \ge y\}. \]

Moreover, in this text, a poset $(F,\ge)$ is called a \emph{forest} iff for every $x\in F$, $\uparrow x$ is a chain. A forest $(F,\ge)$ is called \emph{rooted} iff $F\neq\emptyset$ and for every $x\in F$, $\uparrow x$ contains a maximal element of $(F,\ge)$. A forest $(T,\ge)$ is called a \emph{tree} iff for every $x,y\in T$, $(\uparrow x) \cap (\uparrow y) \neq \emptyset$. Given a forest $(F,\ge)$, the elements $x\in F$ are called \emph{nodes}. Nodes $x\in F$ such that $\downarrow x = \{x\}$ are called \emph{terminal}. We state the following lemma, fundamental for what follows. It can actually be seen as an explicitly order-theoretic reformulation of a basic result from graph theory (see the discussion in \cite[Section~I.1]{Bollobas2013Modern}).\footnote{As the claim that it is a reformulation requires proof, and also for the reader's convenience, a proof can be found in the appendix.}

\begin{lemma}\label{1-SDF_AC.lemma:partion_of_forest}
    For any forest $(F, \ge)$ there exists a unique partition $\mc F$ of $F$ into trees such that for all $x, y \in F$ with $x \ge y$ there is $T \in \mc F$ with $x, y \in T$. If $(F,\ge)$ is rooted, then for any $T\in\mc F$, $(T,\ge)$ is a rooted tree and has a maximum.
\end{lemma}

The elements of $\mc F$ are called \emph{connected components} of $(F,\ge)$. The maximum of a rooted tree $(T,\ge)$ is called the \emph{root}. The \emph{roots} of a forest $(F,\ge)$ are the roots of its connected components. A \emph{decision forest} (\emph{decision tree}) is a rooted forest (tree, respectively) $(F,\ge)$ such that all $x,y\in F$ with $x\neq y$ can be separated by some maximal chain $c\subseteq F$, that is, $c\cap \{x,y\}$ is a singleton. A \emph{move} in a decision forest $(F,\ge)$ is a non-terminal node $x\in F$. In the present text, let us call a decision forest $(F,\ge)$ \emph{(everywhere) non-trivial} iff some (any, respectively) root is a move.

If $V$ is a set, a \emph{$V$-poset} is a subset $N\subseteq \mc P(V)$. The name derives from the fact that $(N,\supseteq)$ defines a poset of subsets of $V$ ordered by set inclusion.

\begin{remark}\label{1-SDF_AC.rmk:def_tree_AFK}
    Note that the definition of a tree used in this text is an order-theoretic transcription of a graph-theoretical concept. What is called forest here, corresponds indeed to a ``forest'', also named ``acyclic graph'' in graph theory (see \cite[Sections I.1, I.2]{Bollobas2013Modern}), but is called ``tree'' in \cite[Definition~1]{AlosFerrer2005Trees}. This latter terminology adapts the use of this term in order theory (see \cite{Davey2002Introduction} and compare the discussion in \cite[Remark~2]{AlosFerrer2005Trees}). In the present text the use of forests with multiple connected components, describing exogenous scenarios, is central, and hence using the term ``tree'' for this may be misleading. Hence, the definitions of the present text insist stronger on the graph-theoretical aspect as presented in \cite{Bollobas2013Modern}, and especially on the botanic metaphor of trees and forests, than do those in \cite{AlosFerrer2005Trees,Davey2002Introduction}, though without adding unnecessary discreteness assumptions. 
    
    Note, however, that a ``rooted tree'' in the sense of \cite[Definition~1]{AlosFerrer2005Trees} corresponds to a rooted tree in this text because the \cite{AlosFerrer2005Trees}-definition of ``rooted'' demands the existence of a maximum, not only of maximal elements for principal up-sets. But a rooted forest in the sense of this text need not be a ``rooted tree'' in the sense of \cite{AlosFerrer2005Trees}. 

    It should also be noted that we use the partial order $\ge$ rather than $\le$ because of the representation used in the subsequent subsection, following the decision- and game-theoretic texts \cite{AlosFerrer2005Trees,AlosFerrer2016Theory}, though this may differ from the usual convention in other contexts.
\end{remark}

\subsection{Decision forests}

The following definition is, formally, a transcription of the characterisation of (a subclass of) ``game trees'' in \cite[Theorem~3]{AlosFerrer2005Trees}. So, although the object is formally not new, we look at it from a slightly different perspective which is why it is recalled here. First, it emphasises the fact that there can be multiple connected components, but restricts the attention to those \cite{AlosFerrer2005Trees}-``game trees'' whose connected components are rooted. Second, the terminology in this text is different also in that it insists on the purely decision-theoretic aspect. What is called ``decision forest (or tree) over a set'' and ``decision path'' here, respectively, is called ``game tree`` and ``play'' in \cite{AlosFerrer2005Trees}. Third, while \cite{AlosFerrer2005Trees}-``game trees'' are defined via certain set-theoretic properties which can be rather easily verified in applications, and are then characterised via the so-called ``representation by plays'', the present text perceives the decision-theoretic essence of decision forests over sets rather as being the duality between outcomes and nodes expressed by that representation, and thus uses it as the definition.

\begin{definition}\label{1-SDF_AC.def:decision_forest}
    Let $V$ be a set. A \emph{decision forest on $V$} is a $V$-poset $F$ such that:
    \begin{enumerate}
        \item\label{1-SDF_AC.def:decision_forest.rooted_forest} $(F,\supseteq)$ is a rooted forest;
        \item\label{1-SDF_AC.def:decision_forest:repr_by_dec_paths} $F$ is \emph{its own representation by decision paths}, that is, if $W$ denotes the set of maximal chains in $(F,\supseteq)$, and for every $y\in F$, $W(y) = \{w\in W \mid y\in w\}$, then there is a bijection $f\colon V \to W$ such that for every $y\in F$, $(\mc P f)(y) = W(y)$.
    \end{enumerate}
    $F$ is called \emph{decision tree on $V$} iff, in addition, for all $x,y\in F$ there is $z\in F$ with $z\supseteq x\cup y$.

    The nodes, terminal nodes, and moves of $(F,\supseteq)$ are also called \emph{nodes}, \emph{terminal nodes}, and \emph{moves} of $F$, respectively, and the elements of $V$ are called \emph{outcomes}. The set of moves of $F$ is denoted by $X(F)$ or $X$ in short.
\end{definition}

Following \cite{AlosFerrer2005Trees}, $V$ can be seen as the set of possible \emph{outcomes} of the game, and the forest $(F,\supseteq)$ specifies how, as the decision problem is tackled dynamically, the set of realisable outcomes becomes smaller and smaller. Thus, elements of $F$ can be interpreted as nodes. See \cite{AlosFerrer2005Trees} for handy criteria on the set $F$ characterising the situation of it being a decision forest on $V$: essentially, this corresponds to $F$ being a ``game tree'' on $V$ in the sense of \cite[Definition~4]{AlosFerrer2005Trees} such that all connected components of $(F,\supseteq)$ have maximal elements.

Condition \ref{1-SDF_AC.def:decision_forest}.\ref{1-SDF_AC.def:decision_forest.rooted_forest} clarifies that the predecessors of any node $x\in F$ can be totally ordered and contain a root, and therefore constitute an account of the past with a beginning, a history. The set $W$ of maximal chains can be interpreted as the set of \emph{decision paths}. Condition \ref{1-SDF_AC.def:decision_forest}.\ref{1-SDF_AC.def:decision_forest:repr_by_dec_paths} says that possible outcomes and decision paths are in one-to-one correspondence, in such a way that a node is contained in some decision path iff the corresponding outcome is contained in that node. Even more is true: according to \cite[Theorem~3, Corollary~2]{AlosFerrer2005Trees} and Remark~\ref{1-SDF_AC.rmk:def_tree_AFK}, the bijection $f$ is uniquely determined. More precisely:

\begin{proposition}[\cite{AlosFerrer2005Trees}]\label{1-SDF_AC.prop:f(v)=uparrow v}
    Let $F$ be a decision forest on some set $V$ and $f\colon V \to W$ be a map as in Definition~\ref{1-SDF_AC.def:decision_forest}, where $W$ is the set of maximal chains in $(F,\supseteq)$. Then,
    \[ \forall v\in V\colon\quad f(v) = \uparrow \{v\} = \{x\in F \mid v\in x\}. \]
\end{proposition}
Thus $V$, which formally is the set of possible outcomes, and $W$, which formally is the set of possible decision paths, can be identified in a uniquely determined way, namely via the $f$ above. Under this identification, we have the following duality statement:
\[ x\in f(v) \qquad \Longleftrightarrow \qquad v\in x, \]
for all nodes $x\in F$ and all outcomes $v\in V$. As a consequence, in the remainder of this thesis, the set a decision forest is defined on is denoted by $W$ rather than by $V$, consistent with other pieces of the literature, notably with \cite{AlosFerrer2005Trees,AlosFerrer2016Theory}. \smallskip

There is a second duality, namely between two approaches to dynamic decision theory: graphs, based on tree-like objects, and refined partitions, based on a set of outcomes. This is made precise in the following two propositions, which are contained in \cite[Lemma 14 and Theorem~1]{AlosFerrer2005Trees}, following Remark~\ref{1-SDF_AC.rmk:def_tree_AFK}:

\begin{proposition}[\cite{AlosFerrer2005Trees}]\label{1-SDF_AC.prop:decision_forest_over_set_is_decision_forest}
    Let $F$ be a decision forest on a set $V$. Then the poset $(F,\supseteq)$ is a decision forest.
\end{proposition}

\begin{proposition}[\cite{AlosFerrer2005Trees}]\label{1-SDF_AC.prop:repr_by_dec_paths_of_decision_forest}
     Let $(F_0,\ge)$ be a decision forest. Let $V$ be the set of its maximal chains and for any $x_0\in F_0$ let $V(x_0)$ be the set of $v\in V$ with $x_0\in v$. Let \[F = \{V(x_0) \mid x_0\in F_0\}.\] Then $F$ defines a decision forest on $V$, and $(F,\supseteq)$ is order-isomorphic to $(F_0,\ge)$.
\end{proposition}

Rephrasing the ``representation by plays'' from \cite{AlosFerrer2005Trees}, we call $F$ the \emph{representation by decision paths} of $(F_0,\ge)$. The definition of a decision forest on a set requires essentially that the operations from the two preceding propositions are, up to isomorphism, inverse to each other. See \cite[Section~4]{AlosFerrer2005Trees} for more details on this.

Thus, there is a rigorous sense in that both of the mentioned descriptions are equivalent. While the graph-theoretical approach is graphically more convenient, the refined partitions approach is naturally aligned with Savage's acts as a model for choice under uncertainty (see the introduction of \cite{AlosFerrer2005Trees} for a discussion). We note that, although phrased differently and with a different aim, all three preceding propositions and the duality concepts constitute one of the essential innovations of \cite{AlosFerrer2005Trees}. 

\subsection{Forests of decision trees}

In the present text, it is crucial to deal with forests and not only with trees, in the sense of Subsection~\ref{1-SDF_AC.subs:conventions} and Remark~\ref{1-SDF_AC.rmk:def_tree_AFK}. Often, the study of forests can be reduced to the study of trees by considering the connected components separately. Is this true for decision forests as well, and in what sense? In other words, is the duality of nodes and outcomes compatible with the forest structure? This question gets fully answered in the following theorem. More precisely, we have:

\begin{thm}\label{1-SDF_AC.thm:decision_forest=forest_of_decision_trees}
	Let $V$ be a set and $F$ be a $V$-poset defining a rooted forest with respect to $\supseteq$. Let $\mc F$ denote the set of connected components of the rooted forest $(F,\supseteq)$. For every $T\in\mc F$, let $V_T$ denote the root of $(T,\supseteq)$.
	
	Then $F$ is a decision forest on $V$ iff 
    \begin{enumerate}
        \item\label{1-SDF_AC.thm:decision_forest=forest_of_decision_trees.partition} $\{V_T\mid T\in\mc F\}$ is a partition of $V$, and
        \item\label{1-SDF_AC.thm:decision_forest=forest_of_decision_trees.T_decision_tree} for every $T\in\mc F$, $(T,\supseteq)$ defines a decision tree on $V_{T}$.
    \end{enumerate}
\end{thm}

To verify the Propositions \ref{1-SDF_AC.prop:f(v)=uparrow v}, \ref{1-SDF_AC.prop:decision_forest_over_set_is_decision_forest}, and \ref{1-SDF_AC.prop:repr_by_dec_paths_of_decision_forest}, one can cite the results in \cite{AlosFerrer2005Trees} in combination with Remark~\ref{1-SDF_AC.rmk:def_tree_AFK}. In view of Theorem~\ref{1-SDF_AC.thm:decision_forest=forest_of_decision_trees}, it suffices, however, to cite the results for the case of rooted trees only. This procedure will be of interest at a later stage when analysing outcome existence and uniqueness for stochastic extensive forms.

\begin{remark}\label{1-SDF_AC.rmk:decision_forests_over_sets}
    Theorem~\ref{1-SDF_AC.thm:decision_forest=forest_of_decision_trees} offers a way of constructing decision forests from collections of decision trees. Let $\mc F_0$ be a non-empty set of decision trees on sets. For any $T_0\in\mc F_0$, let $V_{T_0}$ be its root. Then, let
    \[ V = \bigcup_{T_0\in\mc F_0} V_{T_0} \times \{T_0\} = \{ (v,T_0) \mid T_0 \in \mc F_0,\, v\in T_0\}, \]
    the disjoint union of all roots, and let
    \[ F = \big\{ x \times \{T_0\} \mid T_0 \in \mc F_0,\, x\in T_0\big\}. \]
    Then, $F$ is a decision forest on $V$. The set of connected components is given by
    \[ \mc F = \Big\{ \big\{ x \times \{T_0\} \mid x\in T_0\big\} \bigmid T_0 \in \mc F_0\Big\}. \]
    Moreover, Theorem~\ref{1-SDF_AC.thm:decision_forest=forest_of_decision_trees} states that all decision forests on sets can be represented in this form.
\end{remark}

For examples of decision forests on sets, the reader is referred to the following section, where decision forests whose connected components are indexed over the set of exogenous scenarios are considered.\footnote{Also note the examples of decision trees on sets in \cite{AlosFerrer2005Trees} and the more extensive version in \cite{AlosFerrer2016Theory} in combination with the preceding Remark~\ref{1-SDF_AC.rmk:decision_forests_over_sets}.}

\section{Stochastic decision forests}\label{1-SDF_AC.sec:sdf}

In this section, the central notion of this chapter --- stochastic decision forests --- is introduced. The main idea behind this is to weaken the traditional assumptions on exogenous information, which is no more assumed to arise through the dynamic decision making of a nature agent. Rather, an exogenous scenario $\omega$ is ``randomly'' realised within a given measurable space which determines the decision tree underlying the actual decision makers' problem. Therefore, before the main definition, measurable spaces are discussed as a model for exogenous scenarios and sets of scenarios whose probability can be measured. Subsequently, the definition of stochastic decision forests is given and analysed, simple examples are presented, and the class of action path stochastic decision forests is constructed. For a basic graphical illustration of the abstract objects introduced in this section, the reader is referred to Figure~\ref{fig:SDF} printed in the introduction.

\subsection{Exogenous scenario spaces}\label{1-SDF_AC.subs:Ex_sc_sp}

For the remainder of the thesis, an \emph{exogenous scenario space} is a measurable space $(\Omega,\ms E)$ such that $\Omega\neq\emptyset$.\footnote{For more information on measure and probability theory, the reader is referred to the introductory texts \cite{Bogachev2007Measure,Kallenberg2021Foundations}.} $\Omega$ describes the set of possible \emph{exogenous scenarios} which replace the outcomes possibly generated by the behaviour of a ``nature'' agent. $\ms E$ is the set of \emph{events} alias collections of exogenous scenarios that can be measured by all relevant decision makers. That is, we implictly suppose that decision makers or agents form beliefs about the probability of events, in the sense of probability theory, and here, $\ms E$ describes the set of collections of exogenous scenarios that agents assign probabilities to.\footnote{These beliefs can be interpreted as representation devices of agents' preferences over the random outcomes of strategic interaction in terms of expected utility --- ``random'' because of the dependence on exogenous scenarios, which in this particular context play the role of states while outcomes correspond to the consequences, put in Savage's words, see, e.g.\ \cite{Savage1972Foundations,Gilboa1989Maxmin,Hara2023Multiple}.}

Why does one suppose $\ms E$ to be a $\sigma$-algebra? As an algebra of sets, $\ms E$ can be seen as a Boolean algebra and can therefore model basic logical operations on events. Moreover, in the probabilistic approach, we are only interested in those functions $\P\colon\ms E\to [0,1]$ compatible with the algebra structure\footnote{The meaning of the phrase ``compatible with the algebra structure'' can be reduced to the statement: a) $\P(\Omega) = 1$, and b) for all disjoint $E_1,E_2\in\ms E$ we have $\P(E_1\cup E_2) = \P(E_1) + \P(E_2)$.} that are $\sigma$-additive, which can be thought of as a continuity property, and it is a fundamental measure-theoretic result, that, for such $\P$, without loss of generality, $\ms E$ can be supposed to be closed under countable unions, that is, to be a $\sigma$-algebra (see \cite[Section~1.5]{Bogachev2007Measure}).

If $\ms E$ can be generated by a countable partition $\ms P$ of $\Omega$, any probability measure on it is entirely described through the probabilities on these partitioning events. So, though all elements of $\ms E$ including unions of partition members are events, one can formally describe the situation by the set $\ms P$ and a countable family of positive numbers $(p_E)_{E\in \ms P}$ adding up to $1$ and indexed by that set. However, in general, $\ms E$ need not be generated by a partition of $\Omega$. For instance, it may be that all singletons are events and have probability zero, as in the case of the Lebesgue measure on the unit interval $[0,1]$. Hence, from a general perspective, scenarios may be neither a relevant nor a sufficient description of exogenous data for decision makers, but it is the ($\sigma$-)algebra $\ms E$ of events that these agents are concerned about. Moreover, as famously illustrated by the Lebesgue measure, in case $\Omega$ is uncountable, there is a non-trivial trade-off between the fineness\footnote{Recall that a $\sigma$-algebra $\ms E'$ on $\Omega$ is \emph{finer} than $\ms E$ iff $\ms E \subseteq \ms E'$.} 
of $\ms E$ and the number of probability measures on it (see \cite[Section~1.7]{Bogachev2007Measure}). Hence, to describe the agents' relation to exogenous data and admissible beliefs about their probability, it is crucial to specify $\ms E$. This is important not the least because many relevant applications require an uncountable $\Omega$ (with a $\sigma$-algebra $\ms E$ that cannot be generated by a countable partition), be it the uniform distribution on $[0,1]$, be it continuous martingales omnipresent in finance (see, e.g.\ \cite{Delbaen2006Mathematics,Bayer2023Rough}).

\subsection{Stochastic decision forests}
In light of the preceding argument, we model uncertainty about exogenous events by an exogenous scenario space $(\Omega,\ms E)$ and suppose that some $\omega\in\Omega$ is realised, determining the relevant decision tree without being directly communicated to the decision makers. So, the agents do not necessarily have complete information about exogenous events in $\ms E$, which tell in which tree they are while making choices. Note that events in $\ms E$ need not be generated by a partition of $\Omega$. Hence, there must be a structure of similarity among trees that can serve as a consistent basis for exogenous information revelation in an abstract and general probabilistic sense. This distinguishes the present approach from the traditional model of games or decision problems under incomplete information (cf.\ \cite{Harsanyi1967Games,Harsanyi1968Games,Harsanyi1968Gamesa}) or stochastic games (cf.\ \cite{Shapley1953Stochastic}, see also \cite[Subsection~2.2.2.5]{AlosFerrer2016Theory}), represented via a randomising nature agent and partition-based information sets with respect to that agent's past action.

\begin{definition}\label{1-SDF_AC.def:sdf}
    A \emph{stochastic decision forest}, in short \emph{\textsc{sdf}}, on an exogenous scenario space $(\Omega,\ms E)$ is a triple $(F,\pi,\X)$ consisting of:
    \begin{enumerate}
        \item\label{1-SDF_AC.def:sdf.df} a decision forest $F$ on some set $W$;
        \item\label{1-SDF_AC.def:sdf.conn_comp} a surjective map $\pi\colon F\to \Omega$ such that the set $\mc F$ of connected components of $(F,\supseteq)$ is given by the fibres of $\pi$, that is,
        \[ \mc F = \{\pi^{-1}(\{\omega\}) \mid \omega\in \Omega\}; \]
        \item\label{1-SDF_AC.def:sdf.X} a set $\X$ such that: 
        \begin{enumerate}
            \item\label{1-SDF_AC.def:sdf.X.section} any element $\x\in\X$ is a section of moves defined on some non-empty event, that is, it is a map $\x\colon D_\x \to X$ satisfying $\pi\circ\x = \id_{D_\x}$ for some $D_\x\in\ms E\setminus\{\emptyset\}$;
            \item\label{1-SDF_AC.def:sdf.X.cov} $\X$ induces a covering of $X$, that is, $\{\x(\omega) \mid \x\in\X,\,\omega\in D_\x\} = X$.
        \end{enumerate}
    \end{enumerate}
    The elements of $\X$ are called \emph{random moves}. For $\omega\in\Omega$, let $T_\omega = \pi^{-1}(\{\omega\})$ and $W_\omega$ be the root of $T_\omega$. For $E\subseteq\Omega$, let $W_E = \bigcup_{\omega\in E} W_\omega$ and $F_E = \bigcup_{\omega\in E} T_\omega$.
\end{definition}

In other words, a stochastic decision forest is a decision forest on a set (Axiom~\ref{1-SDF_AC.def:sdf.df}), whose connected components are indexed by $\Omega$ (Axiom~\ref{1-SDF_AC.def:sdf.conn_comp}), admitting a set of sections of moves (Axiom~\ref{1-SDF_AC.def:sdf.X.section}) called random moves that form a covering (Axiom~\ref{1-SDF_AC.def:sdf.X.cov}) of $X$. As discussed in the remainder of this thesis, random moves form a flexible basis for a general description of information revelation for different agents, under the innocent hypothesis that there is at least one active agent per move. 

If one visualises a stochastic decision forest in two dimensions, any tree growing along the vertical ``decision path'' axis, the forest's trees being placed along the horizontal ``$\Omega$'' axis, then random moves are --- roughly speaking --- horizontal or diagonal sections covering the set of moves in a decision-theoretically interpretable way. Hence, in a stochastic decision forest, the flow of information can be decomposed into the endogenous movement of agents along the set of random moves caused by decision making and the exogenous information revelation at those random moves. Note also that stochastic decision forests are allowed to vary across scenarios: the trees in $\mc F$ need not be isomorphic. The crowns may become shallower in some scenarios (because some options are no more available there). See Figure~\ref{fig:SDF} for a graphical illustration.

It is possible that random moves as such can be ordered in a way consistent with the ordering on $F$. This means that one can relate these points of exogenous information revelation uniformly by means of the words ``before'' and ``after''. This possibility is necessary for conceiving an agent's capacity to condition action on knowledge about exogenous information received ``earlier''. Note that this order consistency of random moves is not about information sets and, in particular, not to be confounded with the strong order on information sets as introduced in \cite{Ritzberger1999Recall}. An example of an order inconsistent stochastic decision forest is presented in the following subsection (namely, Gilboa's interpretation of the absent-minded driver formulated as a stochastic decision forest).
Furthermore, we introduce two non-triviality conditions.
In many contexts it seems reasonable to assume first that any root is a move --- a situation which (provided sufficient measurability) can always be obtained by eliminating those scenarios $\omega$ with singleton $T_\omega$ alias singleton $W_\omega$ --, and second that the set of random moves describes the basis of exogenous information most efficiently, in the sense that it cannot be extended without compromising its structure.

\begin{definition}\label{1-SDF_AC.def:sdf.addon}
    Given a stochastic decision forest $(F,\pi,\X)$ on an exogenous scenario space $(\Omega,\ms E)$, let $\ge_\X$ denote the partial order on $\X$ defined by
    \[ \x_1 \ge_\X \x_2 \quad \Longleftrightarrow \quad \Big[ D_{\x_1} \supseteq D_{\x_2}~ \text{ and }~\forall \omega\in D_{\x_2}\colon \x_1(\omega) \supseteq \x_2(\omega)\Big]. \]
    A set $\tilde\X\subseteq\X$ of random moves is said \emph{order consistent} iff for all $\x_1,\x_2\in\tilde\X$:
    \begin{equation*}
        \Big[\exists\omega\in D_{\x_1}\cap D_{\x_2}\colon~ \x_1(\omega) \supseteq \x_2(\omega)\Big] \qquad \Longrightarrow \qquad \x_1 \ge_\X \x_2.
    \end{equation*}

    A stochastic decision forest $(F,\pi,\X)$
    \begin{enumerate}[label=4(\alph*),ref=4(\alph*)]
        \item\label{1-SDF_AC.def:sdf.X.OC} is said \emph{order consistent} iff $\X$ is order consistent;
        \item\label{1-SDF_AC.def.sdf.X.surely_NT} is said \emph{surely non-trivial} iff $(F,\supseteq)$ is everywhere non-trivial;
        \item\label{1-SDF_AC.def:sdf.X.max} that is order consistent, is said \emph{maximal} iff for every set $\bar\X$ such that $(F,\pi,\bar\X)$ is an order consistent stochastic decision forest and that is \emph{refined by $\X$} in that for all $\bar\x\in\bar\X$ there is $P_{\bar\x}\subseteq \X$ with $\bar\x = \bigcup P_{\bar\x}$,\footnote{According to standard set-theoretic conventions, $\bar\x = \bigcup P_{\bar\x}$ means: $\bar\x$ is a map with domain $\bigcup_{\x\in P_{\bar\x}} D_\x$ and for all $\x\in P_{\bar\x}$ and $\omega\in D_\x$, $\bar\x(\omega) = \x(\omega)$.} we have $\bar\X = \X$.
    \end{enumerate}
\end{definition}

As is going to be shown in the sequel, some of these properties are actually stronger than the definition formally indicates. In the case of order consistency, random moves cannot cross each other, and in particular, the root property is compatible with random moves.

\begin{lemma}\label{1-SDF_AC.lemma:sdf.X.roots} 
    Let $(F,\pi,\X)$ be a stochastic decision forest. Let $D = \{\omega\in\Omega \mid W_\omega\in X\}$.
    \begin{enumerate}
        \item\label{1-SDF_AC.lemma:sdf.X.roots.OC}  If $(F,\pi,\X)$ is order consistent, then, for any $\x\in\X$, $\x(\omega)$ is a root in $(F,\supseteq)$ for all or no $\omega\in D_\x$. 
        \item\label{1-SDF_AC.lemma:sdf.X.roots.surely_non-trivial}  $D=\Omega$ iff $(F,\pi,\X)$ is surely non-trivial.
        \item\label{1-SDF_AC.lemma:sdf.X.roots.root_random_move} If $(F,\pi,\X)$ is order consistent and maximal, and $X\neq\emptyset$, then $\x_0\colon D \to X,\,\omega\mapsto W_\omega$ is a random move, i.e.\ $\x_0\in\X$.
    \end{enumerate}
\end{lemma}

Moreover, in the order consistent, surely non-trivial and maximal case, random moves can be seen as moves of a rooted decision tree in their own right. This makes the endogenous movement of agents along $\X$ particularly ordered: as said before, it becomes possible to relate two points of exogenous information revelation via ``before'' and ``after''. Let us make this precise. First, we might have to add terminal nodes, if they exist. For example, think of an ``infinite centipede'' which contains infinitely many maximal chains but only one maximal chain consisting only of moves. Hence, we extend the partial order $\ge_\X$ to a binary relation $\ge_\Tr$ on the disjoint union \[\Tr = \X \cup\Big \{\big\{(\omega,\{w\})\big\}\bigmid (\omega,w)\in\Omega\times W\colon \{w\} \in F,\,\pi(\{w\}) = \omega\Big\}\] by letting, for $\x\in\X$ and $(\omega,w),(\omega',w')\in\Omega\times W$ with $\{w\},\{w'\}\in F$ and $\pi(\{w\})=\omega$, $\pi(\{w'\})=\omega'$:
\begin{itemize}[label=--]
    \item $\x\ge_\Tr \big\{(\omega,\{w\})\big\}$ iff $\omega\in D_\x$ and $w\in\x(\omega)$;
    \item $\big\{(\omega,\{w\})\big\} \ge_\Tr \big\{(\omega',\{w'\})\big\}$ iff $(\omega,w)=(\omega',w')$;
    \item $\big\{(\omega,\{w\})\big\}\ngeq_\Tr \x$.
\end{itemize}
Note that a set $\big\{(\omega,\{w\})\big\}$ as above is nothing else than the set-theoretic function $\y\colon D_\y \to F$ with $D_\y = \{\omega\}$ and $\y(\omega)=\{w\}$. In analogy with random moves and since its image is a terminal node, we call such $\y$ a \emph{random terminal node}. The analogy is substantiated by the subsequent proposition which (seemingly) strengthens Axiom~\ref{1-SDF_AC.def:sdf.X.OC} in Definition~\ref{1-SDF_AC.def:sdf.addon} above.

\begin{proposition}\label{1-SDF_AC.prop:ev_on_Tr_is_iso}
    Let $(F,\pi,\X)$ be an order consistent stochastic decision forest on some exogenous scenario space $(\Omega,\ms E)$ and $\Tr \bullet \Omega = \{(\y,\omega) \in \Tr\times \Omega \mid \omega \in D_\y\}$. Then
    the evaluation map $\ev\colon \Tr\bullet \Omega \to F, (\y,\omega) \mapsto \y(\omega)$ is a bijection such that for all $(\y_1,\omega_1),(\y_2,\omega_2)\in\Tr \bullet \Omega$:
    \[ \Big[\y_1 \ge_\Tr \y_2 \text{ and } \omega_1 = \omega_2\Big] \quad \Longleftrightarrow \quad \y_1(\omega_1) \supseteq \y_2(\omega_2). \]
\end{proposition}

In order-theoretic terms, the evaluation map defines an order isomorphism between $\Tr\bullet\Omega$, equipped with the order induced by the product of $\ge_\Tr$ on $\Tr$ and equality on $\Omega$, and $(F,\supseteq)$. 
Now, as announced, we can make rigorous the sense in that random moves are the moves of a rooted decision tree, under the assumptions presented in Definition~\ref{1-SDF_AC.def:sdf.addon}.

\begin{thm}\label{1-SDF_AC.thm:Xrm_is_dec_tree}
    Let $(F,\pi,\X)$ be an order consistent stochastic decision forest on some exogenous scenario space $(\Omega,\ms E)$ such that $\x_0\colon \Omega\to F,\, \omega\mapsto W_\omega$ is a random move.\footnote{By Lemma~\ref{1-SDF_AC.lemma:sdf.X.roots}, order consistency, sure non-triviality, and maximality of $(F,\pi,\X)$ imply that $\x_0$ is a random move.} 
    Then, $(\Tr,\ge_\Tr)$ defines a rooted decision tree and $\X$ is the set of its moves.
\end{thm}

Note, however, that $F$ cannot be subsumed under this derived object. The latter as such does not faithfully account for the fact that there is a realised scenario. For instance, only those maximal chains $\mathbf c$ in $(\Tr,\ge_\Tr)$ correspond to outcomes in $W = \bigcup F$ that satisfy $\bigcap_{\y\in\mathbf c} D_\y \neq \emptyset$, which need not be the case. This is developed further in Chapter~\ref{chap:2-SEF_G}.

\subsection{Simple examples}\label{1-SDF_AC.subs:simple_sdf}
\begin{figure}
    \centering
    \begin{tikzpicture}[node distance={20mm}, thick, main/.style = {draw, circle}] 
    \node[main] (1) {$\x_0(\omega_1)$}; 
    \node[main] (2) [above left of=1] {$\x_1(\omega_1)$}; 
    \node[main] (3) [above right of=1] {$\x_2(\omega_1)$}; 
    \node[main] (4) [right of=3] {$\x_1(\omega_2)$}; 
    \node[main] (5) [below right of=4] {$\x_0(\omega_2)$};
    \node[main] (6) [above right of=5] {$\x_2(\omega_2)$};
    \node[] (8) [above=0.5cm of 2] {$\{w_{112}\}$};
    \node[] (7) [left=0.2cm of 8] {$\{w_{111}\}$};
    \node[] (10) [above=0.5cm of 3] {$\{w_{122}\}$};
    \node[] (9) [left=0.2cm of 10] {$\{w_{121}\}$};
    \node[] (11) [above=0.5cm of 4] {$\{w_{211}\}$};
    \node[] (12) [right=0.2cm of 11] {$\{w_{212}\}$};
    \node[] (13) [above=0.5cm of 6] {$\{w_{221}\}$};
    \node[] (14) [right=0.2cm of 13] {$\{w_{222}\}$};
    \draw[->] (1) -- (2); 
    \draw[->] (1) -- (3); 
    \draw[->] (5) -- (4); 
    \draw[->] (5) -- (6); 
    \draw[->] (2) -- (7); 
    \draw[->] (2) -- (8); 
    \draw[->] (3) -- (9); 
    \draw[->] (3) -- (10); 
    \draw[->] (4) -- (11); 
    \draw[->] (4) -- (12); 
    \draw[->] (6) -- (13); 
    \draw[->] (6) -- (14); 
    \end{tikzpicture} 
    \caption{A simple stochastic decision forest represented as a directed graph, with $w_{\ell km} = (\omega_\ell,k,m)$, for $(\ell,k,m)\in \{1,2\}^3$. Moves are indicated by circles.}
    \label{1-SDF_AC.fig:simple_sdf}
\end{figure}
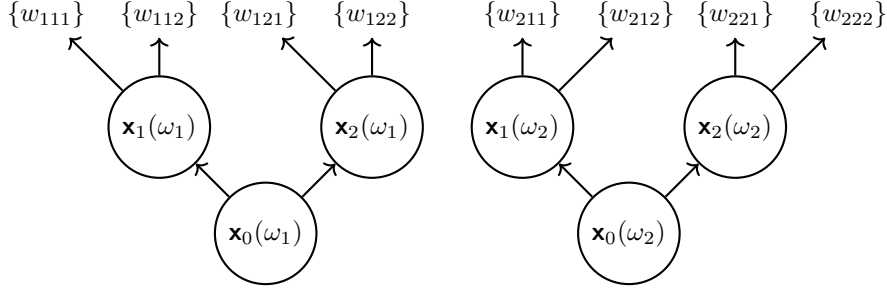

In the following we present three very simple examples of stochastic decision forests. The first example is illustrated in Figure~\ref{1-SDF_AC.fig:simple_sdf}. 
It indicates \emph{pars pro toto} how finite stochastic extensive form decision problems can be formalised. This forest can be used to describe a two-period stochastic control problem, e.g.\ that of a portfolio manager who in each period decides whether to buy or sell a share of some stock. Another example is a stochastic ultimatum bargaining. In a first round a first agent can make one out of two possible offers (``high'' or ``low'') --- e.g. how to split some amount of money --- which the second can accept or decline in the second round. The amount of money, or the amount offered, may depend on which of the two exogenous scenarios $\omega_1$ or $\omega_2$ are realised. Hence, an outcome is a triplet $(\omega,k,m)$, where $\omega$ encodes the exogenous scenario, $k$ the type of the offer (``high'' or ``low''), and $m$ the response of the second agent (``accept'' or ``reject'').

Formally, let $\Omega = \{\omega_1,\omega_2\}$ be some exogenous scenario space with two elements and $\ms E = \mc P \Omega$. Let $W = \Omega \times \{1,2\}^2$ and $\x_0,\x_1,\x_2\colon \Omega \to \mc P(W)$ given by $\x_0(\omega) = \{\omega\} \times \{1,2\}^2$ and $\x_k(\omega) = \{(\omega,k)\}\times \{1,2\}$, $F = \{\x_k(\omega) \mid \omega\in\Omega,~k=0,1,2\} \cup \{ \{w\} \mid w\in W\}$, $\pi\colon F \to\Omega$ be the map sending any node to the first entry of an arbitrary choice among its elements.

\begin{lemma}\label{1-SDF_AC.lemma:simple_sdf1}
    The tuple $(F,\pi,\X)$ defines an order consistent, surely non-trivial, and maximal stochastic decision forest.
\end{lemma}

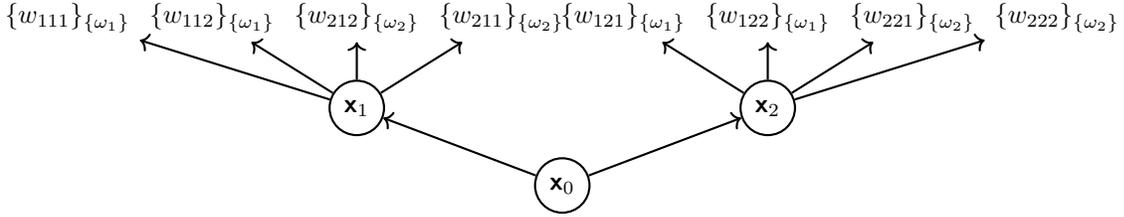
\begin{figure}
    \centering
    \begin{tikzpicture}[node distance={20mm}, thick, main/.style = {draw, circle}] 
    \node[main] (1) {$\x_0$}; 
    \node[main] (2) [above left=0.5cm and 2.2cm of 1] {$\x_1$}; 
    \node[main] (3) [above right=0.5cm and 2.2cm of 1] {$\x_2$};
    \node[] (9) [above=0.5cm of 2] {$\{w_{212}\}_{\{\omega_2\}}$};
    \node[] (8) [left=0cm of 9] {$\{w_{112}\}_{\{\omega_1\}}$};
    \node[] (7) [left=0cm of 8] {$\{w_{111}\}_{\{\omega_1\}}$};
    \node[] (10) [right=0cm of 9] {$\{w_{211}\}_{\{\omega_2\}}$};
    \node[] (12) [above=0.5cm of 3] {$\{w_{122}\}_{\{\omega_1\}}$};
    \node[] (13) [right=0cm of 12] {$\{w_{221}\}_{\{\omega_2\}}$};
    \node[] (11) [left=0cm of 12] {$\{w_{121}\}_{\{\omega_1\}}$};
    \node[] (14) [right=0cm of 13] {$\{w_{222}\}_{\{\omega_2\}}$};
    \draw[->] (1) -- (2); 
    \draw[->] (1) -- (3); 
    \draw[->] (2) -- (7); 
    \draw[->] (2) -- (8); 
    \draw[->] (2) -- (9); 
    \draw[->] (2) -- (10); 
    \draw[->] (3) -- (11); 
    \draw[->] (3) -- (12); 
    \draw[->] (3) -- (13); 
    \draw[->] (3) -- (14); 
    \end{tikzpicture} 
    \caption{The decision tree $(\Tr,\ge_\Tr)$ for the simple stochastic decision forest, with $w_{\ell km} = (\omega_\ell,k,m)$, for $(\ell,k,m)\in \{1,2\}^3$. (Random) moves are indicated by circles. Elements of $\Tr \setminus \X$, of the form $\{(\omega,\{w\})\}$ and seen as maps $\omega\mapsto \{w\}$, are denoted by $\{w\}_{\{\omega\}}$.}
    \label{1-SDF_AC.fig:simple_sdf_Tr}
\end{figure}

The corresponding decision tree $(\Tr,\ge_\Tr)$ is illustrated in Figure~\ref{1-SDF_AC.fig:simple_sdf_Tr}.
\smallskip

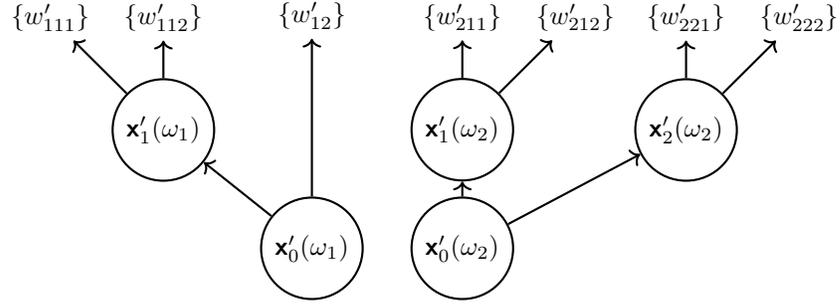
\begin{figure}
    \centering
    \begin{tikzpicture}[node distance={20mm}, thick, main/.style = {draw, circle}] 
    \node[main] (1) {$\x'_0(\omega_1)$}; 
    \node[main] (2) [above left=0.6cm and 1cm of 1] {$\x'_1(\omega_1)$}; 
    \node[] (3) [above=2.1cm of 1] {$\{w'_{12}\}$}; 
    \node[main] (5) [right of=1] {$\x'_0(\omega_2)$};
    \node[main] (4) [above=0.2cm of 5] {$\x'_1(\omega_2)$}; 
    \node[main] (6) [above right=0.6cm and 2cm of 5] {$\x'_2(\omega_2)$};
    \node[] (8) [above=0.5cm of 2] {$\{w'_{112}\}$};
    \node[] (7) [left=0.2cm of 8] {$\{w'_{111}\}$};
    \node[] (11) [above=0.5cm of 4] {$\{w'_{211}\}$};
    \node[] (12) [right=0.2cm of 11] {$\{w'_{212}\}$};
    \node[] (13) [above=0.5cm of 6] {$\{w'_{221}\}$};
    \node[] (14) [right=0.2cm of 13] {$\{w'_{222}\}$};
    \draw[->] (1) -- (2); 
    \draw[->] (1) -- (3); 
    \draw[->] (5) -- (4); 
    \draw[->] (5) -- (6);  
    \draw[->] (2) -- (7); 
    \draw[->] (2) -- (8); 
    \draw[->] (4) -- (11); 
    \draw[->] (4) -- (12); 
    \draw[->] (6) -- (13); 
    \draw[->] (6) -- (14); 
    \end{tikzpicture} 
    \caption{A variant of the simple stochastic decision forest in Figure~\ref{1-SDF_AC.fig:simple_sdf} represented as a directed graph, with $w'_{\ell km} = (\omega_\ell,k,m)$, for all triples $(\ell,k,m)\in\{1,2\}^3$ with $(\omega_\ell,k,m)\in W'$, and $w'_{12} = (\omega_1,2)$. Moves are indicated by circles.}
    \label{1-SDF_AC.fig:simple_sdf_variant}
\end{figure}

As a variant, identifying the elements $(\omega_1,2,1)$ and $(\omega_1,2,2)$ in $W$ is seen below to provide a stochastic decision forest with a random move that is not defined on all of $\Omega$, as illustrated in Figure~\ref{1-SDF_AC.fig:simple_sdf_variant}. 
In terms of the bargaining example, this means that in the exogenous scenario $\omega_1$, when confronted with the ``low'' offer ($k=2$), the second agent can only reject (perhaps his principal told him to do so, or for legal reasons). Formally, let $W' = W\setminus\{(\omega_1,2,1),(\omega_1,2,2)\} \cup \{(\omega_1,2)\}$ and let $\x'_0 = \x_0$, $\x'_1 = \x_1$, and $\x'_2\colon \{\omega_2\}) \to \mc P(W')$ be given by $\x'_2(\omega_2) = \{(\omega_2,2)\}\times \{1,2\}$. Let $\X' = \{\x'_0,\x'_1,\x'_2\}$. Let $D_{\x'_0}=\Omega$, $D_{\x'_1} = \Omega$, $D_{\x'_2}=\{\omega_2\}$ and $F'=\{\x'(\omega) \mid \x'\in\X',~\omega\in D_{\x'}\} \cup \{\{w'\}\mid w'\in W'\}$. Let $\pi'\colon F'\to\Omega$ be the map sending any node to the first entry of an arbitrary choice among its elements.

\begin{lemma}\label{1-SDF_AC.lemma:simple_sdf2}
    The tuple $(F',\pi',\X')$ defines an order consistent, surely non-trivial, and maximal stochastic decision forest.
\end{lemma}

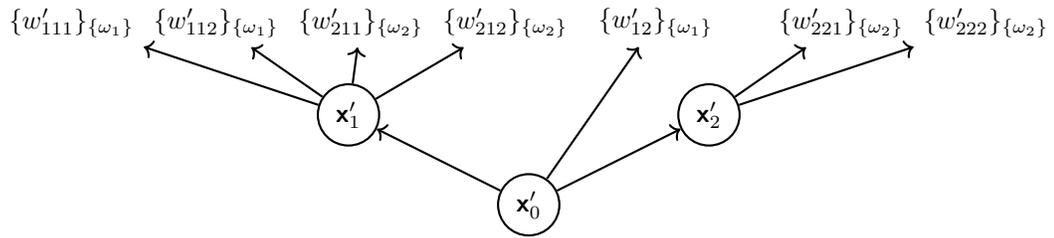
\begin{figure}
    \centering
    \begin{tikzpicture}[node distance={20mm}, thick, main/.style = {draw, circle}] 
    \node[main] (1) {$\x'_0$}; 
    \node[main] (2) [above left=0.6cm and 1.8cm of 1] {$\x'_1$}; 
    \node[] (3) [above right=1.8cm and 0.5cm of 1] {$\{w'_{12}\}_{\{\omega_1\}}$};  
    \node[main] (6) [above right=0.6cm and 1.8cm of 1] {$\x'_2$};
    \node[] (8) [above left=0.6cm and 0.5cm of 2] {$\{w'_{112}\}_{\{\omega_1\}}$};
    \node[] (7) [left=0cm of 8] {$\{w'_{111}\}_{\{\omega_1\}}$};
    \node[] (11) [right=0cm of 8] {$\{w'_{211}\}_{\{\omega_2\}}$};
    \node[] (12) [right=0cm of 11] {$\{w'_{212}\}_{\{\omega_2\}}$};
    \node[] (13) [above right=0.6cm and 0.5cm of 6] {$\{w'_{221}\}_{\{\omega_2\}}$};
    \node[] (14) [right=0cm of 13] {$\{w'_{222}\}_{\{\omega_2\}}$};
    \draw[->] (1) -- (2); 
    \draw[->] (1) -- (3); 
    \draw[->] (1) -- (6);  
    \draw[->] (2) -- (7); 
    \draw[->] (2) -- (8); 
    \draw[->] (2) -- (11); 
    \draw[->] (2) -- (12); 
    \draw[->] (6) -- (13); 
    \draw[->] (6) -- (14); 
    \end{tikzpicture} 
    \caption{The decision tree $(\Tr',\ge_{\Tr'})$ for the variant of the simple stochastic decision forest, with $w'_{\ell km} = (\omega_\ell,k,m)$, for all triples $(\ell,k,m)\in \{1,2\}^3$ with $(\omega_\ell,k,m) \in W'$, and $w'_{12} = (\omega_1,2)$. (Random) moves are indicated by circles. Elements of $\Tr' \setminus \X'$, of the form $\{(\omega,\{w'\})\}$ and seen as maps $\omega\mapsto \{w'\}$, are denoted by $\{w'\}_{\{\omega\}}$.}
    \label{1-SDF_AC.fig:simple_sdf_variant_Tr}
\end{figure}

The corresponding decision tree $(\Tr',\ge_{\Tr'})$ is illustrated in Figure~\ref{1-SDF_AC.fig:simple_sdf_variant_Tr}.
\smallskip

The third example is an \textsc{sdf} representation of Gilboa's model \cite{Gilboa1997Comment} of the absent-minded driver phenomenon first described in \cite{Piccione1997Interpretation}. Recall that in this story a driver thinks about her way home. She knows that she has to take the second exit of the motorway to get there. If she takes the first one, she will end up in a disastrous region. If, however, she misses the second one, she has to sleep in a motel. The problem is that, being absent-minded, she will not recognise the number of any intersection she passes by. As discussed in \cite{Piccione1997Interpretation, Aumann1997Absent, Gilboa1997Comment}, this is a challenge to tree-based game theory. In \cite{Gilboa1997Comment}, Gilboa proposes a model based on a decision tree with a nature agent, which we now reformulate using a stochastic decision forest.

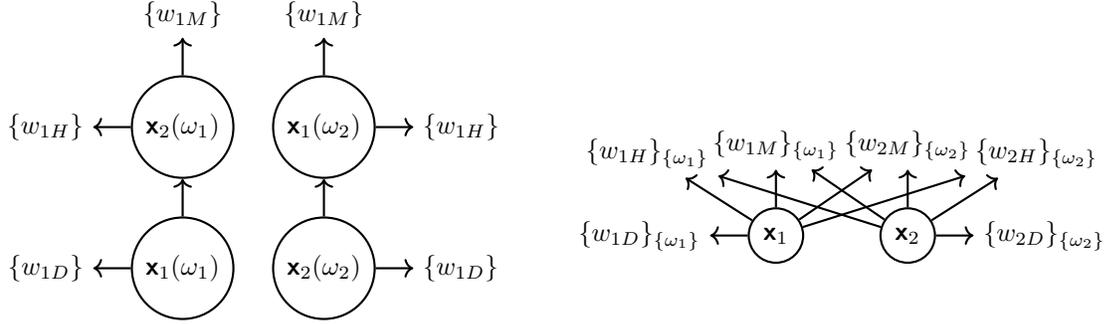
\begin{figure}
    \centering
    \begin{tikzpicture}[node distance={20mm}, thick, main/.style = {draw, circle}] 
    \node[main] (11) {$\x_1(\omega_1)$}; 
    \node[] (1D) [left=0.5 of 11] {$\{w_{1D}\}$};
    \node[main] (12) [above=0.5cm of 11] {$\x_2(\omega_1)$}; 
    \node[] (1H) [left=0.5 of 12] {$\{w_{1H}\}$};
    \node[] (1M) [above=0.5cm of 12] {$\{w_{1M}\}$}; 
    \node[main] (22) [right=0.5cm of 11] {$\x_2(\omega_2)$}; 
    \node[] (2D) [right=0.5 of 22] {$\{w_{1D}\}$};
    \node[main] (21) [above=0.5cm of 22] {$\x_1(\omega_2)$}; 
    \node[] (2H) [right=0.5 of 21] {$\{w_{1H}\}$};
    \node[] (2M) [above=0.5cm of 21] {$\{w_{1M}\}$}; 
    \draw[->] (11) -- (12); 
    \draw[->] (11) -- (1D); 
    \draw[->] (12) -- (1H); 
    \draw[->] (12) -- (1M);  
    \draw[->] (22) -- (21); 
    \draw[->] (22) -- (2D); 
    \draw[->] (21) -- (2H); 
    \draw[->] (21) -- (2M);  
    \end{tikzpicture} 
    \qquad
    \begin{tikzpicture}[node distance={20mm}, thick, main/.style = {draw, circle}] 
        \node[] (0) {};
        \node[main] (1) [above=0.5 of 0] {$\x_1$};
        \node[] (1D) [left=0.5cm of 1] {$\{w_{1D}\}_{\{\omega_1\}}$};
        \node[] (1H) [above left=0.5cm and 0.5cm of 1] {$\{w_{1H}\}_{\{\omega_1\}}$};
        \node[] (1M) [above=0.5cm of 1] {$\{w_{1M}\}_{\{\omega_1\}}$};  
        \node[main] (2) [right=1 of 1] {$\x_2$};
        \node[] (2D) [right=0.5cm of 2] {$\{w_{2D}\}_{\{\omega_2\}}$};
        \node[] (2H) [above right=0.5cm and 0.5cm of 2] {$\{w_{2H}\}_{\{\omega_2\}}$};
        \node[] (2M) [above=0.5cm of 2] {$\{w_{2M}\}_{\{\omega_2\}}$}; 
        \draw[->] (1) -- (1D);
        \draw[->] (1) -- (1H);
        \draw[->] (1) -- (1M);
        \draw[->] (1) -- (2M);
        \draw[->] (1) -- (2H);
        \draw[->] (2) -- (2M);
        \draw[->] (2) -- (1H);
        \draw[->] (2) -- (1M);
        \draw[->] (2) -- (2D);
        \draw[->] (2) -- (2H);
    \end{tikzpicture}
    \caption{The absent-minded driver \textsc{sdf}, following Gilboa: $(F,\supseteq)$ represented as a directed graph, in case $\rho$ is injective, with $\omega_\ell = \rho^{-1}(\ell)$, $w_{\ell S} = (\omega_\ell,S)$, for all $\ell\in \{1,2\}$ and symbols $S$ such that $(\omega,S)\in W$ (left), $(\Tr,\ge_\Tr)$ where elements of $\Tr \setminus \X$, of the form $\{(\omega,\{w\})\}$ and seen as maps $\omega\mapsto \{w\}$, are denoted by $\{w\}_{\{\omega\}}$ (right). Non-minimal elements are indicated by circles, respectively.}
    \label{1-SDF_AC.fig:absent_minded_driver_Gilboa_sdf}
\end{figure}

For this, let $(\Omega,\ms E)$ be some exogenous scenario space and $\rho\colon \Omega \to \{1,2\}$ an $\ms E$-$\mc P\{1,2\}$-measurable surjection. Further, let $D,H,M$ be three pairwise distinct symbols meaning ``disastrous region'', ``home'', and ``motel'' as in the original story from \cite{Piccione1997Interpretation}. Let $W = \Omega \times \{D,H,M\}$ and $\x_1,\x_2$ be maps defined on the whole of $\Omega$ given by
\[ \x_k(\omega) = \begin{cases} \{\omega\} \times \{D,H,M\}, &\text{ if } \rho(\omega) = k, \\ \{\omega\} \times \{H,M\}, & \text{ if } \rho(\omega) = 3-k, \end{cases} \qquad k\in\{1,2\}. \]
$F$ is defined as the union of the images of $\x_1$ and $\x_2$ and the set of all singleton sets in $W$. $\pi\colon F \to \Omega$ maps any element of $F$ to the first component of its elements. $(F,\supseteq)$ and $(\Tr,\ge_\Tr)$ are illustrated in Figure \ref{1-SDF_AC.fig:absent_minded_driver_Gilboa_sdf}. Note that $(\Tr,\ge_\Tr)$ is not even a forest, both $\x_1$ and $\x_2$ are maximal elements, and both $\im\x_1$ and $\im\x_2$ contain both a root and a move that is not a root, respectively.

\begin{lemma}\label{1-SDF_AC.lemma:absent_minded_driver_Gilboa_sdf}
    The tuple $(F,\pi,\X)$ just defined is a stochastic decision forest which is not order consistent.
\end{lemma}

\subsection{Action path stochastic decision forests}\label{1-SDF_AC.subs:APsdf}

In most pieces of the literature, dynamic games are defined by supposing a notion of time and specifying outcomes as certain paths of action at instants of time. \cite[Subsection~2.2]{AlosFerrer2005Trees} provides a broad overview for this, including classical textbook definitions as in \cite{Fudenberg1991Game}, infinite bilateral bargaining in discrete time as in \cite{Rubinstein1982Perfect}, repeated games, the long cheap talk game in \cite{Aumann2003Long}, and a decision-theoretic interpretation of differential games as in \cite{Dockner2000Differential}. We desire to add to this list stochastic control in both discrete and continuous time (see, e.g.\ \cite{Pham2009Continuous,Bertsekas1996Stochastic,Karatzas1998Methods}) and stochastic differential games (see, e.g.\ \cite{Carmona2018Probabilistic}) without restrictions on the noise in question, and the first concrete step in this direction is taken in what follows next.\footnote{As described in the introduction, this is a guiding theme of the whole thesis and its investigation culminates only in the third and final chapter, Chapter~\ref{chap:3-SPF_VECT}.}

In this subsection, this approach is formulated in one abstract and general framework. This framework is based on a specific structure pertaining to all of these examples, namely \emph{time}. Interestingly, time is not included in the abstract formulation of decision forests, and it serves as a particularly strong similarity structure for trees and even branches of one and the same tree. Of course, every decision path is totally ordered and therefore has its own time axis --- but in the following construction, the ``clocks'' of all decision paths are synchronised.\footnote{We do not address the question whether all ``relevant'' order consistent and maximal stochastic decision forests can be represented in this framework --- that is, admit a ``clock'' synchronising all their decision paths --- though it does not seem obvious to conceive a counterexample.} The second major point of the following framework is that it will allow for general exogenous stochastic noise, going strictly beyond the ``nature'' agent setting.\smallskip

Let $(\T,\le)$ be a total order admitting a minimum which we denote by $0$. Further, let $I$ be a non-empty finite set (which will represent decision makers alias agents), $(\A^i)_{i\in I}$ be a family of non-empty Polish spaces, and $\A = \prod_{i\in I}\A^i$ be their topological product, with canonical projections $p^i\colon\A \to \A^i$, $i\in I$. Of course, the case of singleton $I$ is included in this setting.

Let $(\Omega,\ms E)$ be an exogenous scenario space. Let $W \subseteq \Omega\times\A^\T$ be such that for all $\omega\in\Omega$, there is $f\in\A^\T$ with $(\omega,f)\in W$. An outcome will thus be a pair of an exogenous scenario and a path $f\colon \T \to \A$ of ``action'', and any scenario is required to admit at least one outcome.
For any $t\in\T$ and $\tilde w = (\omega,f)\in\Omega\times\A^\T$, let \[x_t(\tilde w) = x_t(\omega,f) = \{ (\omega',f')\in W \mid \omega' = \omega, \, f'|_{[0,t)_\T} = f|_{[0,t)_\T}\}.\] 
Let $F=\{x_t(w) \mid t\in\T, \,w\in W\} \cup \{\{w\} \mid w\in W\}$. Further, let $\pi\colon F\to \Omega$ be the unique function mapping any $x\in F$ to the first item of one choice of its elements. 

For $(t,f)\in\T\times\A^\T$, let $D_{t,f} = \{\omega \in\Omega\mid |x_t(\omega,f)| \ge 2\}$. This will turn out as the event that $x_t(.,f)$ is a move.
We consider the following assumptions:
\begin{itemize}[label=--]
    \item\hypertarget{1-SDF_AC.Ass:AP.SDF0}{\textbf{Assumption~AP.SDF0.}}~For all $t\in\T$ and $f\in\A^\T$, $D_{t,f}\in\ms E$. 
    \item\hypertarget{1-SDF_AC.Ass:AP.SDF1}{\textbf{Assumption~AP.SDF1.}}~For all $w\in W$ and all $t,u\in\T$ with $t\neq u$ and $x_t(w) = x_u(w)$, we have $x_t(w) = \{w\}$.
    \item\hypertarget{1-SDF_AC.Ass:AP.SDF2}{\textbf{Assumption~AP.SDF2.}}~For all $\omega\in\Omega$ and $\tilde f\in\A^\T$, and for all subsets $\T'\subseteq \T$ satisfying $x_t(\omega,\tilde f) \in F$ for all $t\in \T'$, there is $f\in\A^\T$ with $(\omega,f)\in W$ and $f|_{[0,t)_\T} = \tilde f|_{[0,t)_\T}$ for all $t\in \T'$.
    \item\hypertarget{1-SDF_AC.Ass:AP.SDF3}{\textbf{Assumption~AP.SDF3.}}~For all $t\in\T$ and $f,g\in\A^\T$ such that $D_{t,f},D_{t,g}\neq\emptyset$ and $D_{t,f} \cap D_{t,g} = \emptyset$, there is $u\in [0,t)_\T$ such that $D_{u,f} \cap D_{u,g} \neq\emptyset$ and $f|_{[0,u)_\T} \neq g|_{[0,u)_\T}$.
\end{itemize}
Assumption~\hyperlink{1-SDF_AC.Ass:AP.SDF0}{AP.SDF0} requires that $D_{t,f}$ is indeed an event in $(\Omega,\ms E)$. Assumption~\hyperlink{1-SDF_AC.Ass:AP.SDF1}{AP.SDF1} stipulates that any (presumptive) move has a unique time associated to it. Indeed, we have the following result. For any $x\in F$, let 
\[\T_x = \{t\in\T \mid \exists w\in x\colon x = x_t(w)\}. \]
It is evident that $\T_x$ is a non-empty convex set. We recall that convexity means that for all $t_0,t_1\in\T_x$, we have $[t_0,t_1]_\T \subseteq \T_x$. 

\begin{lemma}\label{1-SDF_AC.lemma:AP_sdf.AssmAP.SDF1}
    Assumption~\hyperlink{1-SDF_AC.Ass:AP.SDF1}{AP.SDF1} is satisfied iff for all $x\in F$ that are not singletons, $\T_x$ is a singleton.
\end{lemma}

Note that the following weak converse statement holds true independently of whether \hyperlink{1-SDF_AC.Ass:AP.SDF1}{AP.SDF1} is assumed or not:

\begin{lemma}\label{1-SDF_AC.lemma:AP_sdf.AssmAP.SDF1_addon}
    All $x\in F$ such that $\T_x = \{t\}$ for some $t\in\T$ that is not maximal in $\T$, are no singletons.
\end{lemma}

We immediately infer:

\begin{corollary}\label{1-SDF_AC.cor:AP_sdf.AssmAP.SDF1}
    If Assumption~\hyperlink{1-SDF_AC.Ass:AP.SDF1}{AP.SDF1} is satisfied and $\T$ has no maximum, then for all $x\in F$ the following statements are equivalent:
    \begin{itemize}[label=--]
        \item $x$ is a singleton.
        \item $\T_x$ is not a singleton.\qed
    \end{itemize}
\end{corollary}

Assumption~\hyperlink{1-SDF_AC.Ass:AP.SDF2}{AP.SDF2} corresponds to what is generally called ``boundedness'' in \cite[Subsection~3.4]{AlosFerrer2005Trees}. In the present context, this is the \emph{conditio sine qua non} demanding that any (presumptive) decision path, alias maximal chain of nodes, corresponds to a path of action $\T\to\A$. 
Assumption~\hyperlink{1-SDF_AC.Ass:AP.SDF3}{AP.SDF3} refers to the maximality in Axiom~\ref{1-SDF_AC.def:sdf.X.max} and ensures that different paths in different scenarios cannot be identified without violating the order consistency. If the assumption does not hold true, then it will become possible to identify the path $f$ on $D_{t,f}$ with the path $g$ on $D_{t,g}$, as for all $u<t$ until that (exclusively) $f$ and $g$ can be distinguished, $D_{u,f}$ and $D_{u,g}$ continue to be disjoint. 

Further, let $\X$ be the set of maps
\[ \x_t(f) \colon D_{t,f} \to F,\, \omega \mapsto \x_t(f)(\omega) = x_t(\omega,f), \]
ranging over all $t\in\T$, $f\in \A^\T$ with $D_{t,f} \neq \emptyset$.

\begin{definition}
    The tuple $(I,\A,\T,W)$ is called \emph{action path stochastic decision forest (\textsc{sdf}) data} on $(\Omega,\ms E)$ iff Assumptions \hyperlink{1-SDF_AC.Ass:AP.SDF0}{AP.SDF$k$}, $k=0,1,2$, are satisfied. The data is called \emph{maximal} iff, moreover, Assumption \hyperlink{1-SDF_AC.Ass:AP.SDF3}{AP.SDF3} is satisfied. 
    $(F,\pi,\X)$ is said to be the \emph{action path stochastic decision forest (\textsc{sdf}) candidate} (on $(\Omega,\ms E)$) \emph{induced} by these action path \textsc{sdf} data. 
    
    If $(F,\pi,\X)$ defines an \textsc{sdf} on $(\Omega,\ms E)$, the word ``candidate'' is dropped. $\A = \prod_{i\in I} \A^i$ is called \emph{action space}, $\T$ \emph{time axis}, $I$ the \emph{action index set} and $\A^i$ the \emph{$i$-th action space factor} of the action path \textsc{sdf} data, and \textit{a fortiori}, of the induced action path \textsc{sdf}.\footnote{Note that $\A$, $\T$, $I$ are part of the data only and, in contrast to $W$, not uniquely determined by the \textsc{sdf}, although ``minimal'' representatives might be inferred.}
\end{definition}

\begin{example}
    For $\T = \R_+$, singleton $\Omega$, $W = \Omega \times \A^\T$, the induced action path \textsc{sdf} candidate $(F,\pi,\X)$, more precisely, $F$ yields the decision tree of what is called ``differential game'' in \cite[Subsection~2.2.5]{AlosFerrer2005Trees}. Hence, the action path \textsc{sdf} is a generalisation of this example in several directions: first and foremost, by adding an exogenous scenario space, second, by generalising the time axis, and third, by allowing for much higher flexibility regarding the set of outcomes.
\end{example}

\begin{example}
    For $\T = \{0,\dots,N\}$, for $N\in\Nast$, or $\T = \N$ any corresponding action path \textsc{sdf} candidate can serve as a basis for describing the classical finite or infinite horizon discrete-time stochastic decision problem (see, e.g.\ \cite{Bertsekas1996Stochastic}). The admissible actions can be specified through the choice of $W$. The case $W=\Omega\times\A^\T$ corresponds to the case where at each move, the set of possible joint action is given by $\A$. It is discussed in Example~\ref{1-SDF_AC.ex:APsdf} that these action path \textsc{sdf} data induce an action path \textsc{sdf}, even for general $\T$.

    Note that in contrast to the traditional ``nature'' model of stochastic games and decision problems in discrete time, exogenous information cannot be deduced from the order-theoretic properties of the (random) moves. The ``nature'' player or agent is replaced with a forest of decision trees and a structure of random moves to that one may attach exogenous information in the form of $\sigma$-algebras as explained in the next section. 

    $\T$ can also equal more general well-orders: For $\T = \N \cup \{+\infty\}$ with canonical order, that is, essentially, the second smallest infinite ordinal, the long cheap talk game tree (see \cite{Aumann2003Long}) --- and general stochastic variants thereof in the sense of \textsc{sdf}s --- can be obtained. This example, in case of singleton $\Omega$, is treated in \cite{AlosFerrer2005Trees}.
\end{example}

An important result of this chapter, expressed in the next theorem, is that the preceding construction --- encompassing a large set of decision problems and games in their elementary decision-theoretic structure, including very general stochastic versions of them --- is well-defined and yields an \textsc{sdf}. This provides the basis for formulating a large class of stochastic decision problems in extensive form --- already in this chapter it is shown how exogenous information and choices adapted to it, and in the following chapters how decision problems can be modelled on that basis.

\begin{thm}\label{1-SDF_AC.thm:AP_sdf}
    Let $(I,\A,\T,W)$ be action path \textsc{sdf} data on an exogenous scenario space $(\Omega,\ms E)$. Then, the induced action path \textsc{sdf} candidate is an order consistent stochastic decision forest on $(\Omega,\ms E)$. It is maximal in case the data $(I,\A,\T,W)$ is maximal. It is surely non-trivial iff for all $\omega\in\Omega$, there are $f,f'\in\A^\T$ with $f\neq f'$ and $(\omega,f),(\omega,f')\in W$.
\end{thm}

As will be discussed in the second chapter, for well-ordered $\T$, well-posed decision problems can be defined on this action path \textsc{sdf}. This may not true in the general case. One approach to model the extensive form characteristics appropriately nevertheless, consists in approximating by means of outcomes from action-path \textsc{sdf} on well-ordered time grids. This will be discussed in the third chapter of this thesis, Chapter~\ref{chap:3-SPF_VECT}.\smallskip

Given an action path \textsc{sdf} as above, let $\mf t\colon X\to \T$ be the map assigning to any move $x\in X$ the unique element $\mf t(x)$ of $\T_x$, see Lemma \ref{1-SDF_AC.lemma:AP_sdf.AssmAP.SDF1}.

\begin{lemma}\label{1-SDF_AC.lemma:mf_t}
    Let $(F,\pi,\X)$ be the action path \textsc{sdf} induced by action path \textsc{sdf} data $(I,\A,\T,W)$ on an exogenous scenario space $(\Omega,\ms E)$. Then $\mf t$ is strictly decreasing, that is, for all $x,y\in X$ with $x\supsetneq y$ we have $\mf t(x) < \mf t(y)$. Moreover, for all $\x\in\X$ and all $\omega,\omega'\in D_{\x}$, $\mf t(\x(\omega)) = \mf t(\x(\omega'))$.
\end{lemma}

By this lemma, $\mf t$ induces a map $\X\to\T$ which we denote also by $\mf t$.\smallskip

We conclude this section with examples of action path \textsc{sdf}. This includes the illustrative examples from Subsection~\ref{1-SDF_AC.subs:simple_sdf} and several typical classes of well-known decision problems. 

\begin{lemma}\label{1-SDF_AC.lemma:simple_sdf_as_APsdf}
    The two simple stochastic decision forests from Subsection~\ref{1-SDF_AC.subs:simple_sdf} can be represented as a maximal action path \textsc{sdf} with time $\T = \{0,1\}$.
\end{lemma}

The proof of this lemma in the appendix shows that the representation of the variant $(F',\pi',\X')$ as an action path \textsc{sdf} has to specify a ``dummy'' action at the terminal node $\{w'_{12}\}$ without alternative, that is, an action meaning inaction. This is an artefact of the modelling decision to explicitly include a temporal dimension, as reflected in the action path formulation.\footnote{Assumption~\hyperlink{1-SDF_AC.Ass:AP.SDF1}{AP.SDF1} ensures that moves have a unique time associated to them, but terminal nodes need not. In many cases they do not, as made apparent by the Lemmata \ref{1-SDF_AC.lemma:AP_sdf.AssmAP.SDF1} and \ref{1-SDF_AC.lemma:AP_sdf.AssmAP.SDF1_addon} and, most strikingly, their Corollary~\ref{1-SDF_AC.cor:AP_sdf.AssmAP.SDF1}. But there may be an instant of time that certain branches still ``move'' at, and others do not and actually identify that instant with later points in time, which appears a bit artificial.} 
Stochastic decision forests are based on first principles and do no include such a dimension. Therefore, apart from being more general and flexible, their decision-theoretic interpretation is much clearer. At the same time, they include action path \textsc{sdf}s which model a structure present in many applications and which can therefore simplify the formal representation, at the cost of possibly introducing artificial phenomena like inactive activity. It may be noted that, in action path \textsc{sdf}, time plays a role complementary to that of random moves: it defines similarity of moves across branches, while random moves define similarity across trees. The random moves in action path \textsc{sdf}s are defined such as to be perfectly compatible with time.

\begin{example}\label{1-SDF_AC.ex:APsdf}
    \begin{itemize}[label=--]
        \item Without further assumptions on the data $(I,\A,\T,W)$, it holds true that if $W= \Omega \times \A^\T$, then they are maximal action path \textsc{sdf} data.
        
        \item Suppose that $\A^i = \{0,1\}$ for all $i\in I$. Let $W$ be the set of pairs $(\omega,f)$ where $f\colon\T \to \A$ is componentwise decreasing. Then, the quadruple $(I,\A,\T,W)$ defines maximal action path \textsc{sdf} data. The corresponding action path \textsc{sdf} is a natural (and at least approximate) candidate for describing stochastic decision problems of \emph{timing} (alias \emph{stopping}) (see, e.g.\ the monographs \cite{Shiryaev2007Optimal,Peskir2006Optimal,ElKaroui1981Les} for the mathematical theory, and \cite{Riedel2017Subgame,Karatzas1998Methods} for the link to applications in economics and finance).
        
        \item Let us consider an example of a forest that becomes shallower towards its crown in some areas of the exogenous scenario space. We use an example from finance, namely the exercise of an American up-and-out option (cf.\ \cite[Chapter~26]{Hull2018Options}). Let $P = (P_t)_{t\in\R_+}$ be a continuous stochastic process on $(\Omega,\ms E)$ with strictly positive real values describing the price of a financial security. The initial price is $P_0 = 1$. As long as the price has not reached $2$, the holder of the option can exercise it, but once the price reaches $2$, the option expires irreversibly. This problem can be modelled using the action path stochastic decision forest associated to the following data. 
        
        Let $I$ be a singleton, $\T=\R_+$, and $\A = \{0,1\}$. Take $W$ to be the set of all $(\omega,f)$ where $\omega\in\Omega$ and $f\colon\R_+ \to \{0,1\}$ is decreasing such that, if $f$ takes the value $0$, then $t^\ast_f = \inf\{t\in\R_+ \mid f(t) = 0\}$ satisfies $\max_{t\in[0,t^\ast_f]} P_{t}(\omega) < 2$. Then, $W$ induces a maximal action path \textsc{sdf}, and 
        \[D_{t,f} = \{\omega\in\Omega \mid \max_{u\in[0,t]} P_u(\omega) < 2\},\]
        for all $t\in\R_+$ and decreasing $f\in\{0,1\}^{\R_+}$ with $f(t-) = 1$.
    \end{itemize}
    The proofs of these claims can be found in the corresponding part of the appendix.
\end{example}

\section{Exogenous information}\label{1-SDF_AC.sec:exogenous_information}

As observed in the previous section, exogenous information is not contained in the order-theoretic structure of random moves. The approach put forward in this thesis consists in attaching exogenous information to random moves in the form of $\sigma$-algebras, in a way analogous to (and in some sense, more general than) the use of filtrations in probability theory. Therefore, this latter approach is analysed and interpreted in the first subsection, and based on that, in the following subsections, a concept of exogenous information revelation on stochastic decision forests is introduced and explained, as well as illustrated by examples for the stochastic decision forests encountered in the previous section.

\subsection{Filtrations in probability theory}\label{1-SDF_AC.subs:filtr_in_prob_th}

Recall that for a measurable space $(\Omega,\ms E)$, the usual interpretation of $\ms E$ sees its elements as being those subsets of $\Omega$ that are measurable for a given observer or agent. In the context of what we call exogenous scenario spaces, this has essentially the meaning that for all admissible beliefs (alias probability measures) on $\Omega$, probabilities can be computed for these subsets (see Subsection~\ref{1-SDF_AC.subs:Ex_sc_sp}). However, there is a second meaning to ``measurable'', ubiquitous in probability theory and many of its applications.

In probability theory, a filtration is a monotone map $\ms F$ from a non-empty totally ordered set $\T$ modelling time, typically a suborder of the two-sided compactification of $\R$, to the set of sub-$\sigma$-algebras of $\ms E$, thus mapping any $t\in\T$ to a sub-$\sigma$-algebra $\ms F_t \subseteq \ms E$ such that for all $t,u\in\T$ with $t\le u$, $\ms F_t \subseteq \ms F_u$ (see, e.g.\ \cite[Chapter~9]{Kallenberg2021Foundations} or \cite[Chapter~3]{Cohen2015Stochastic}). The customary interpretation refers to an agent being equipped with that filtration $\ms F$ such that, for all times $t\in\T$, $\ms F_t$ describes the information the agent has at time $t$, or put differently, $\ms F_t$ is the set of events measurable for that agent at time $t$. Of course, this agent is able to compute probabilities for all $E\in\ms E$, but the elements of $\ms F_t$ are measurable in an even stronger sense. 

Namely, one tacitly supposes that there is one realised scenario $\omega\in\Omega$, drawn ``at random'', so to speak, but information about it is only revealed to the agent progressively via $\ms F$, in the following way: For all instants of time $t\in\T$ and each $E\in\ms F_t$ the agent can discern at time $t$ whether it contains the realised scenario or not. In particular, the agent when equipped with a belief in form of a probability measure on $\ms E$ can derive updated probabilities and expectations by evaluating conditional probabilities and expectations on $E$. Yet, this notion of measurement must be read with some caution. Actually, this measurement capacity of the agent includes logical operations given by the $\sigma$-algebra property, in particular countable logical disjunctions alias unions. Uncountable operations might be excluded however. Moreover, under the belief of the agent, $E$ may have probability zero and hence the evaluation of conditional probabilities and expectations on $E$ may be meaningless (cf.\ \cite{Kallenberg2021Foundations,Cohen2015Stochastic}). This can be illustrated by the following examples.

In the discrete setting, where $\ms E$ is generated by a countable partition of $\Omega$, the interpretation is evident. Then all $\ms F_t$, $t\in\T$, are also generated by countable partitions, and the partition of later instants of time refine the preceding partitions. The members of the partition generating $\ms F_t$ can be thought of as the agent's information sets regarding the ``nature'' agent's past choices at time $t$.\footnote{This will be discussed in detail in Chapter~\ref{chap:2-SEF_G}.} 
In that discrete case, thus, filtrations are formally and conceptually a special case of order-theoretic rooted forests on the set $\Omega$, as discussed in Subsection~\ref{1-SDF_AC.sec:def}.

But in general, $\ms F_t$, for some $t\in\T$, may, for instance, contain all the singletons subsets of $\Omega$, without containing all elements of $\ms E$, let alone all subsets of $\Omega$ --- think of $\Omega$ being the unit interval with Borel $\sigma$-algebra $\ms E$ and $\ms F_t$ being the set of subsets $E\subseteq [0,1]$ that are countable or have countable complement. So the agent would actually be able to measure the realised scenario without discerning some $E\in\ms E$, say the interval $[0,\frac 12]$, although it is a union of measurable singletons. The point is that this union is uncountable, and we do not assume agents to be able to perform uncountable logical operations. Hence, the ability to discern the realised scenario does not necessarily imply the ability to discern all events, that is elements of $\ms E$. Moreover, all but countably many (and in the case of the Lebesgue measure on $[0,1]$ even all) singletons must have probability zero, and so the ability to discern the realised scenario may not be relevant from a decision-theoretic perspective: If the conditional expectation of some interesting quantity, e.g.\ a payoff, given $\ms F_t$ is computed, it is completely inconclusive to evaluate it on some $E\in\ms F_t$ whose probability is zero, because the conditional expectation is defined almost surely. 

Once again we note that the full descriptor of exogenous information at time $t$ is the set of ``measurable'' events $\ms F_t$, where ``measurable'' is used in the second, stronger sense, not solely the set of scenarios $\Omega$ (compare the discussion in Subsection~\ref{1-SDF_AC.subs:Ex_sc_sp}). In particular, this discussion suggests that, in general, filtrations cannot be subsumed under the theory of decision forests and trees, which is ``discrete'' in that it can be represented via refined partitions, and thus filtrations provide a means to truly extend the scope of stochastic game and decision theory beyond those types of stochastic noise a dynamically choosing ``nature'' agent can simulate.

\subsection{Exogenous information structures}

Stochastic decision forests are built such as to provide instances at that exogenous information can be revealed. In particular, these instances must be ``uniform'' across trees or ``neutral'' such as to reveal no essential exogenous information by themselves. These instances are the random moves and exogenous information is encoded as information about the realised decision tree. In view of the preceding discussion, a general probabilistic model must be based on sub-$\sigma$-algebras of the exogenous scenario space $(\Omega,\ms E)$, rather than on refined partitions generated by some nature agent's dynamic behaviour. By the order structure on both random moves and subsets of $\Omega$, a straightforward notion of learning, forgetting, recalling information obtains as well.

\begin{definition}\label{1-SDF_AC.def:EIS}
    Let $(F,\pi,\X)$ be a stochastic decision forest on an exogenous scenario space $(\Omega,\ms E)$ and let $\tilde\X\subseteq\X$. An \emph{exogenous information structure on $\tilde\X$} is a family $\ms F = (\ms F_\x)_{\x\in\tilde\X}$ such that for all $\x\in\tilde\X$, $\ms F_\x$ is a sigma-algebra on $D_\x$ with $\ms F_\x\subseteq\ms E$. An exogenous information structure $\ms F$ on $\tilde\X$ is said to admit \emph{recall} iff for all $\x,\x'\in\tilde\X$ with $\x \ge_\X \x'$ and every $E\in \ms F_\x$, we have $E\cap D_{\x'} \in \ms F_{\x'}$.
\end{definition}

We think of all decision makers alias agents as being equipped with some exogenous information structure on the set of their random moves $\tilde\X$. For any of a given agent's random moves $\x\in\tilde\X$, $\ms F_\x$ is interpreted as the set of events $E\in\ms E$ relevant at $\x$, meaning that $E\subseteq D_\x$, and that the agent can measure at $\x$, meaning that the agent can discern whether $E$ contains the realised scenario or not. This second property is to be read in the sense discussed in the previous Subsection~\ref{1-SDF_AC.subs:filtr_in_prob_th}. $\ms F$ can be thought of as a dynamic oracle: As the agent passes through the stochastic decision forest, random move per random move, the oracle reveals partial information on the realised exogenous scenario. 
Using the language of ``types'' (following \cite{Harsanyi1967Games, Harsanyi1968Games, Harsanyi1968Gamesa, Mertens1985Formulation}), for each position alias random move $\x\in\tilde\X$, $\ms F_\x$ describes the private information on all ``states of the world'', including the type of his own and that of other agents.

The more straightforward condition of recall\footnote{Note that recall of an exogenous information structure is a weak adaptation of the notion of ``weak recall'' in \cite{Ritzberger1999Recall} along the exogenous dimension.} ensures that the agent does not forget exogenous information: if the agent is at $\x'$ and has been at $\x$ already, and was able to measure $E$ at $\x$, then this agent can measure $E$ given the domain of $\x'$ also at $\x'$.
Note that, if for all $\x\in\X$, we have $D_\x = \Omega$, then an exogenous information structure on $\X$ admitting recall is nothing but a monotone decreasing map $\ms F\colon\x\mapsto \ms F_\x$ from $\X$ into the set of sub-sigma-algebras of $\ms E$, ``monotone decreasing'' meaning that for all $\x,\x'\in\X$ with $\x \ge_\X \x'$, we have $\ms F_\x \subseteq \ms F_{\x'}$. 

It should be noted that what we have at hand here is a generalised adaptation of the notion of filtrations from probability theory to decision trees. In both cases, the index set is partially ordered. However, in the former case the order is total, whereas in the latter quite the opposite case is true, since it is a decision tree. Moreover, the underlying forest can ``thin out'' towards the crowns non-uniformly across trees alias exogenous scenarios, so that, in general, domains of random moves must be taken into account.

\subsection{Simple examples}\label{1-SDF_AC.subs:simple_sdf_EIS}
In the following we discuss all exogenous information structures admitting recall for the simple examples from the previous section, Subsection~\ref{1-SDF_AC.subs:simple_sdf}.\smallskip

First, consider the simple stochastic decision forest $(F,\pi,\X)$ from Subsection~\ref{1-SDF_AC.subs:simple_sdf}, illustrated in Figure~\ref{1-SDF_AC.fig:simple_sdf}. Recall that this can model , modelling a stochastic ultimatum bargaining problem or a two-period stochastic control problem. Consider the five following families $\ms F = (\ms F_\x)_{\x\in\X}$:
\begin{enumerate}
    \item $\ms F_\x = \{\Omega,\emptyset\}$ for all $\x\in\X$: at all moves, it is unknown which scenario is realised;
    \item $\ms F_{\x_0} = \{\Omega,\emptyset\}$ and one of the following three cases is true:
    \begin{enumerate}
        \item $\ms F_{\x_1} = \ms F_{\x_2} = \mc P \Omega$: only at the second move, it becomes known which scenario is realised, irrespective of which one is the second move;
        \item $\ms F_{\x_1} = \mc P \Omega$, $\ms F_{\x_2} = \{\Omega,\emptyset\}$: $\x_1$ is the only move at that the realised scenario is revealed; in the control proble, an agent with this exogenous information may have interest in choosing (if possible) $\x_1$ rather than $\x_2$ in order to learn, modelling the trade-off \emph{exploration vs.\ exploitation}; that way, problems with partial information, dynamic learning, adaptive control can be modelled;
        \item $\ms F_{\x_2} = \mc P \Omega$, $\ms F_{\x_1} = \{\Omega,\emptyset\}$: analogous to the preceding situation;  in the ultimatum bargaining problem, this amounts to the situation where the second agent knows the exogenous scenario iff the first agents makes the ``low'' offer, providing a strategic ``trade-off'' to the first agent between exploiting the risk attitude of he second agent and making a low (and possibly profitable) offer;
    \end{enumerate}
    \item $\ms F_\x = \mc P \Omega$ for all $\x\in\X$: at all moves, the realised scenario is known.
\end{enumerate}

\begin{lemma}\label{1-SDF_AC.lemma:simple_sdf1_EIS}
    For the simple stochastic decision forest $(F,\pi,\X)$ from Subsection~\ref{1-SDF_AC.subs:simple_sdf} there are exactly five exogenous information structures on $\X$ admitting recall, and these are given by the families $\ms F$ considered just above.
\end{lemma}

Now, consider the variant $(F',\pi',\X')$ from Subsection~\ref{1-SDF_AC.subs:simple_sdf}, as illustrated in Figure~\ref{1-SDF_AC.fig:simple_sdf_variant}. Consider the three following families $(\ms F'_{\x'})_{\x'\in\X'}$.
In all three cases, let $\ms F'_{\x'_2} = \{\{\omega_2\},\emptyset\}$. Moreover, separate the following three cases.
\begin{enumerate}
    \item $\ms F'_{\x'_0} = \ms F'_{\x'_1} = \{\Omega,\emptyset\}$: again, there could be an exploitation vs.\ exploration trade-off (compare the cases 2(c) above): if the stock trader chooses option $2$ in the first period, she will learn the exogenous scenario in the second;
    \item $\ms F'_{\x'_0} = \{\Omega,\emptyset\}$, $\ms F'_{\x'_1} =\mc P(\Omega)$: this is similar to case 2(a) above;
    \item $\ms F'_{\x'_0} = \ms F'_{\x'_1} = \mc P(\Omega)$: the realised scenario is known at any move (similar to case 3 above).
\end{enumerate}

\begin{lemma}\label{1-SDF_AC.lemma:simple_sdf2_EIS}
    For the stochastic decision forest $(F',\pi',\X')$ from Subsection~\ref{1-SDF_AC.subs:simple_sdf} there are exactly three exogenous information structures on $\X'$ admitting recall, and these are given by the families $\ms F'$ considered just above.
\end{lemma}

We conclude this subsection with a note on the absent-minded driver story as modelled by Gilboa in \cite{Gilboa1997Comment}. Let $I = \{1,2\}$ and $\xi^1,\xi^2$ real-valued random variables on $(\Omega,\ms E)$ such that there is a probability measure $\P$ on it making $(\rho,\xi^1,\xi^2)$ independent. On the basis of the \textsc{sdf} $(F,\pi,\X)$ proposed in Subsection~\ref{1-SDF_AC.subs:simple_sdf}, let $\ms F^i_{\x_i} = \sigma(\xi^i)$ for both $i\in I$, that is, the agent active at $\x_i$ has no non-trivial information about $\rho$ if both agents have posterior beliefs derived from $\P$. These give trivially rise to exogenous information structures on $\X$ admitting recall.

\subsection{Action path stochastic decision forests}\label{1-SDF_AC.subs:APsdf_EIS}

In this subsection, we discuss examples of exogenous information structures for action path \textsc{sdf} data $(I,\A,\T,W)$ and the induced action path \textsc{sdf} $(F,\pi,\X)$ as in Subsection~\ref{1-SDF_AC.subs:APsdf}, defined on an exogenous scenario space $(\Omega,\ms E)$. First, we observe a necessary condition on these in the case of recall: When evaluated along maximal chains of random moves in $(\Tr,\ge_\Tr)$, we obtain a filtration in the sense of classical probability theory. More precisely:

\begin{lemma}\label{1-SDF_AC.lemma:APsdf_EIS_induces_filtration}
    Let $(F,\pi,\X)$ be the action path \textsc{sdf} induced by action path \textsc{sdf} data $(I,\A,\T,W)$ on an exogenous scenario space $(\Omega,\ms E)$. 
    Let $\ms F$ be an exogenous information structure on $\X$ admitting recall, and let $f\in\A^\T$. Let $\T_f = \{t\in\T \mid D_{t,f} \neq \emptyset\}$. 
    
    Then, for all $t,u\in\T_f$ with $t<u$ and $E\in\ms F_{\x_t(f)}$, $E\cap D_{u,f} \in \ms F_{\x_u(f)}$. In particular, if $D_{t,f} = \Omega$ for all $t\in\T_f$, then $(\ms F_{\x_t(f)})_{t\in\T_f}$ defines a filtration with time index set $\T_f$.
\end{lemma}

In the already discussed special case that $\A$ has at least two elements and $W=\Omega \times \A^\T$, $(\ms F_{\x_t(f)})_{t\in\T}$ is a filtration.\smallskip

Now we turn to sufficient conditions, or more precisely, to the construction of exogenous information structures for the action path \textsc{sdf}. Let $\ms G = (\ms G_t)_{t\in\T}$ be a filtration on $(\Omega,\ms E)$. For example, $\ms G$ could be generated by some stochastic process, possibly augmented with nullsets with respect to some set of probability measures on $(\Omega,\ms E)$.

\begin{enumerate}
    \item The basic setting in stochastic control of exogenous noise revealed over time independently of the agents' behaviour (see, e.g.\ \cite{Pham2009Continuous}) corresponds to letting, for all $\x\in\X$, \[\ms F_{\x} = \ms G_{\mf t(\x)} \mid_{D_{\x}}, \] and $\ms F = (\ms F_\x)_{\x\in\X}$. This exogenous information structure admits recall.
    \item We can also consider a more general case that allows for exogenous information depending on previous behaviour --- which can serve as a basis for describing problems with partial information involving dynamic learning and stochastic filtering, the trade-off exploration vs.\ exploitation (see, e.g.\ \cite{Bain2009Fundamentals,Cohen2015Stochastic,Cohen2023Optimal}). Again, let $\ms G = (\ms G_t)_{t\in\T}$ be a filtration on $(\Omega,\ms E)$. Let $(Y_\x)_{\x\in\X}$ be a family of random variables with values in some Polish space $\B$. Let, for any $\x\in\X$:
    \[ \ms F_\x =\Big( \sigma(Y_{\x'} \mid \x' \ge_\X \x) \vee \ms G_{\mf t(\x)} \Big)  \bigmid_{D_{\x}}. \]
    Again, let $\ms F = (\ms F_\x)_{\x\in\X}$, yielding an exogenous information structure admitting recall.
    \item Taking $\ms F_\x = \sigma(Y_\x)$, for all $\x\in\X$, can and does often provide an exogenous information structure not admitting recall.
\end{enumerate}

In particular, these cases can be used to model ``open loop'' and ``closed loop'' controls, respectively, covering the definitions in \cite[Chapter~2]{Carmona2018Probabilistic}, for example. However, as will be discussed in the Subsection~\ref{1-SDF_AC.subs:APsdf_AC} about adapted choices, the crux lies in that local decisions not only depend on the filtration-like object representing exogenous information at the current random move, but also on the current random move itself. Hence, when comparing a counterfactual random move to a reference random move revealing the same exogenous information, the same strategy may demand a different choice, because the random move is different.

\begin{example}\label{1-SDF_AC.ex:APsdf_EIS}
    We give a typical example of such a family $(Y_\x)_{\x\in\X}$ in the case $\B = \R$, $\A = \R$, and $W = \Omega \times C(\R)$. Consider an auxiliary family of real-valued random variables $(\tilde Y_\x)_{\x\in\X}$ satisfying the stochastic differential equation
    \[ \tilde Y_{\x_t(f)} = \int_0^t \tilde b(f(u),\tilde Y_{\x_u(f)})~\d \tilde Z_{u}, \]
    for all $t\in\T$ and $f\in C(\R)$, and bounded continuous $\tilde b\colon \R\times \R \to \R$ and a suitable $\ms G$-adapted stochastic integrator $\tilde Z$, say Brownian motion.\footnote{... with respect to a probability measure on $(\Omega,\ms E)$} Then let $(Y_\x)_{\x\in\X}$ solve the stochastic differential equation
    \[ Y_{\x_t(f)} = \int_0^t b(\tilde Y_{\x_u(f)},Y_{\x_u(f)})~\d u + \int_0^t \sigma(Y_{\x_u(f)})~\d Z_{u} \]
    for all $t\in\T$ and $f\in C(\R)$, and bounded continuous $b\colon \R\times \R \to \R$ and $\sigma \colon \R \to \R$, and a suitable $\ms G$-adapted stochastic integrator $Z$, say Brownian motion. 

    $\tilde Y$ can be interpreted as a noisy and not directly observable signal controlled by an agent through $f$, while $Y$ can be seen as an observation depending on $\tilde Y$. This is a typical setting in control theory involving stochastic filtering (see, for instance, \cite{Bain2009Fundamentals,Cohen2023Optimal}, \cite[Part~V]{Cohen2015Stochastic}).
\end{example}

\begin{thm}\label{1-SDF_AC.thm:AP_sdf_EIS}
    Let $(F,\pi,\X)$ be the action path \textsc{sdf} induced by action path \textsc{sdf} data $(I,\A,\T,W)$ on an exogenous scenario space $(\Omega,\ms E)$. Consider the family $\ms F$ from above, in its general version as in point (2). Then, $\ms F$ defines an exogenous information structure on $\X$ admitting recall.
\end{thm}

\section{Adapted choices}\label{1-SDF_AC.sec:adapted_choices}

In this section, a concept of choices is introduced that aims at reconciling the classical decision-theoretic model of choice under uncertainty and the probabilistic setting of adapted processes. First, these two concepts, their differences and intersections are discussed. Then, the concept of adapted choices on stochastic decision forests is introduced and explained. The section is completed with examples for the stochastic decision forests and exogenous information structures introduced in the two preceding sections.

\subsection{Choice under uncertainty and adapted processes}

In the refined partitions approach classical in sequential decision theory and developed in high generality in \cite{AlosFerrer2005Trees}, a choice consists in selecting a member of a given partition of the set of outcomes that are still possible according to the information the agent has at the current move. This is illustrated in Figure~\ref{fig:refined_partitions_restaurant} in the preface. The partition members can be seen as ``local consequences'', pertaining to the current information. Upon relabelling moves as ``local states'', this is a direct application of Savage's concept of acts (cf.\ \cite{Savage1972Foundations}) which assign a (local) consequence to each (local) state. If the overall problem is well-posed this happens in such a way that the set of possible outcomes (which can be seen as global consequences) is successively reduced to a singleton.\footnote{This point is discussed in Chapter~\ref{chap:2-SEF_G}.}

In probabilistic models based on filtrations which describe exogenous information on an abstract measurable space $(\Omega,\ms E)$, choices are modelled differently. If $\ms F = (\ms F_t)_{t\in\T}$ is a filtration on $(\Omega,\ms E)$ with some totally ordered time index set $\T$, the choice of agent $i$ equipped with that information and capable of action described by a Polish space $\A^i$ at time $t$ is typically modelled by an $\ms F_t$-measurable function $g_t\colon\Omega\to \A^i$, modulo regularity conditions in $t$ (leading to optional or predictable processes, for example), see, e.g., \cite{Pham2009Continuous,Cohen2015Stochastic}. $\ms F_t$-measurability of $g_t$ means that for any measurable set of actions $A_t^i \subseteq \A^i$, the event that $i$ chooses an action in $A_t^i$ is measurable to agent $i$ at time $t$, that is, $(g_t)^{-1}(A_t^i)\in \ms F_t$. Actually, $A_t^i$ can be understood as a set-valued reference choice and $(g_t)^{-1}(A_t^i)$ is just the event that both choices, $g_t$ and $A^i_t$, are compatible.

In the discrete case, this approach can be subsumed under the refined partitions framework going back to \cite{Neumann1944Theory} and presented in high generality in \cite{AlosFerrer2016Theory}. Indeed, if $\ms E$ is generated by a countable partition, then so is $\ms F_t$ for each $t\in\T$, and $\ms F_t$-measurability is another way of expressing the requirement that $g_t$ is constant on each member of the partition generating $\ms F_t$. Recall from Subsection~\ref{1-SDF_AC.subs:filtr_in_prob_th} that each realised partition member is thought of as the agent's information set regarding the ``nature'' agent's past choices. In that sense, $g_t$ corresponds to a map from moves at time $t$ to actions such that at moves from the same information set the same action is selected. A family of such maps $g_t$, ranging over all times $t$, also called adapted process in the language of stochastic processes, then defines a complete contingent plan of action for this agent $i$ provided $i$ has only exogenous information.

On general measurable spaces, however, the $\ms F_t$-measurability cannot be rephrased like this in terms of partitions. Moreover, agents may have endogenous information about their own or other agents' past behaviour and condition their own behaviour on this information, that is, they may condition on what they know about their current random move. This suggests that one may adapt the more general measure-theoretic concept of measurable functions to stochastic decision forests as regards exogenous information about the horizontal $\Omega$ axis, while maintaining the refined partitions logic along the vertical tree axis in order to rigorously explain the interactive decision making in extensive form. The aim of this section is to introduce such a concept, bringing together the measure-theoretic probabilistic and the refined partitions based tree-like approaches to information.\footnote{In this section we do however not make formally precise which axioms the set of choices an agent is equipped with should satisfy in order to consistently define a stochastic extensive form following the refined partitions paradigm and how exactly information sets can be modelled, nor do we discuss how and when this general stochastic approach could be represented using a ``nature'' agent: this is the aim of the second part of this thesis, see Chapter~\ref{chap:2-SEF_G}.}

\subsection{Choices in stochastic decision forests}

A basic principle in extensive form theory is that at any ``move'' it is ``known'' to decision makers whether a given ``choice'' is available to them or not. One of the important facts the refined partitions approach formalised in \cite{AlosFerrer2005Trees} elucidates, is that the availability of a choice at a given move can be completely described in terms of the underlying set-theoretic structure: A choice is available at a move iff the latter is an immediate predecessor of the former. More precisely, if $(F,\pi,\X)$ is a stochastic decision forest on an exogenous scenario space $(\Omega,\ms E)$, $W=\bigcup F$ and $c\subseteq W$ is some subset (for instance, a union of nodes representing a choice), then, with
\[ \downarrow c = \{ x\in F \mid c \supseteq x \}, \]
let, as in the classical setting of \cite{AlosFerrer2005Trees}, the set of \emph{immediate predecessors of $c$} be defined by:
\[ P(c) = \{x\in F \mid \exists y \in \downarrow c\colon \uparrow x = \uparrow y \setminus \downarrow c \}. \]
Recall that, for any set $E\subseteq\Omega$, $F_E = \bigcup_{\omega\in\Omega} T_\omega$ is the forest of all trees in $F$ corresponding to elements of $E$ via $\pi$, and $W_E = \bigcup_{\omega\in E} W_\omega$ is the union of all their roots, alias the set of all outcomes feasible in scenarios in $E$.

\begin{lemma}\label{1-SDF_AC.lemma:P(c)_compatible_with_conn_comp}
    For each subset $E\subseteq\Omega$, and each subset $c\subseteq W$, 
    \[ P(c \cap W_E) = P(c) \cap F_E. \]
\end{lemma}

As in \cite{AlosFerrer2005Trees}, a \emph{choice} is a non-empty union $c$ of nodes. For $x\in X$, $c$ is said to be \emph{available at $x$} iff $x\in P(c)$, as in \cite{AlosFerrer2005Trees}. The sets $P(c)$ are a model for information sets in \cite{AlosFerrer2005Trees} and \cite{AlosFerrer2008Trees,AlosFerrer2011Comment} and will turn out to have, in a weaker sense, a similar role in the setting of stochastic decision forests, as discussed in Chapter~\ref{chap:2-SEF_G}.

While in a discrete setting the description of availability of choices in terms of predecessors makes it relatively easy to interpret and thus adhere to the above-mentioned basic principle, this becomes less evident in the present general measure-theoretic context, because of the tension between the discreteness of choices and the general measure-theoretic description of exogenous information in terms of systems of ``measurable'' sets rather than using partitions.
We therefore explicitly assume a locally defined structure of reference choices like the $A_t^i$ above, that can be measured by agents at this random move. Choices made by agents must be adapted to their individual exogenous information structure and that structure of reference choices. Hence, on the one hand, choices can be ``discrete'' in the sense of constituting partitions and thus eliminating a sufficient amount of alternatives so that progressive choosing results in a unique outcome. And on the other hand, choices can be adapted in the sense that the availability of its restriction to any reference choice is a measurable event with respect to the current exogenous information.

\begin{definition}
    Let $(F,\pi,\X)$ be a stochastic decision forest on an exogenous scenario space $(\Omega,\ms E)$, let $\tilde \X\subseteq\X$, and let $\ms F$ be an exogenous information structure on $\tilde \X$.
    \begin{enumerate}
        \item A choice $c$ is said
        \begin{enumerate}
            \item \emph{non-redundant} iff for any $\omega\in\Omega$ with $P(c) \cap T_\omega = \emptyset$, we have $c\cap W_\omega = \emptyset$;
            \item \emph{$\tilde \X$-complete} iff for every random move $\x\in \tilde\X$, $\x^{-1}(P(c))$ is either empty or equal to $D_\x$;
            \item \emph{complete} iff it is $\X$-complete.
        \end{enumerate}
        \item For any random move $\x\in\X$, a choice $c$ is said \emph{available at $\x$} iff $\x^{-1}(P(c)) = D_\x$.
        \item A \emph{reference choice structure on $\tilde\X$} is a family $\ms C = (\ms C_\x)_{\x\in\tilde\X}$ of sets $\ms C_\x$ of non-redundant and $\tilde\X$-complete choices available at $\x$, $\x\in\tilde\X$.
        \item Let $\ms C$ be a reference choice structure on $\tilde\X$. An \emph{$\ms F$-$\ms C$-adapted choice} is a non-redundant and $\tilde\X$-complete choice $c$ such that for all $\x\in\tilde\X$ that $c$ is available at and all $c'\in\ms C_\x$:
        \[ \x^{-1}(P(c \cap c')) = \{\omega\in D_\x \mid \x(\omega) \in P(c\cap c') \} \in \ms F_\x. \]
    \end{enumerate}
\end{definition}

From the an agent's perspective, both for exogenous information revelation and adapted choices, the relevant order (or even tree-like) structure is the partial order (or even decision tree) $(\Tr,\ge_\Tr)$, and more precisely some subset $\tilde\X$ describing the moves of that agent. In that respect, choices are made on that tree with respect to the exogenous information revealed along it. Regarding outcomes and outcome generation, however, the $\Omega$ dimension and thus the forest $F$ are crucial. $\X$ builds the link between both, and exogenous information as well as adapted choices are defined with respect to $\X$ and such as to be compatible with each other. It is more general than usual continuous-time stochastic control and differential games formulations (as in \cite{Pham2009Continuous,Cohen2015Stochastic,Carmona2018Probabilistic}) because of the dependence on $\X$, rather than on the more rigid notion of time. This point, among others, will be clarified in the following examples.


\subsection{Simple examples}\label{1-SDF_AC.subs:simple_sdf_AC}

For the simple examples from Subsection~\ref{1-SDF_AC.subs:simple_sdf}, both the basic version and its variation, and the absent-minded driver model, we provide a reference choice structure, and a list of adapted choices, one for each exogenous information structure from Example~\ref{1-SDF_AC.subs:simple_sdf_EIS}.\smallskip

First, consider the basic version $(F,\pi,\X)$. Let $M$ be the set of maps $\Omega\to \{1,2\}$, modelling a contingent action at a given move. For $k\in \{1,2\}$ and $f,g\in M$, let
\begin{align*}
    c_{f\bullet} =&~ \{(\omega,k',m')\in W \mid k' = f(\omega)\}, \\
    c_{k g} =&~  \{(\omega,k',m')\in W \mid k' = k, ~ m' = g(\omega)\}, \\
    c_{\bullet g} =&~\{(\omega,k',m')\in W \mid  m' = g(\omega)\}.
\end{align*}
The first line describes the choice to act according to $f$ in the first step; the second line describes the choice to act according to $g$ in the second step provided action $k$ has been chosen in the first; the third line describes the choice to act according to $g$ in the second step independently of what has happened before. 
Define $c_{k\bullet}$, $c_{\bullet m}$, and $c_{km}$ by identifying $k,m\in\{1,2\}$ with the constant maps on $\Omega$ with values $k$ and $m$, respectively.  
Let $\ms C_{\x_0} = \{c_{1 \bullet},c_{2 \bullet}\}$, $\ms C_{\x_1} = \ms C_{\x_2} = \{c_{\bullet1},c_{\bullet2}\}$. Note the partitioned structure of these sets, reflecting the discreteness of the situation. 

\begin{lemma}\label{1-SDF_AC.lemma:simple_sdf1_RCS}
    $\ms C = (\ms C_\x)_{\x\in\X}$ defines a reference choice structure on $\X$.
\end{lemma} 

Next, consider the following table. It reads as follows: Each line specifies a set of subsets of $W$ for each of the five exogenous information structures (\textsc{eis}) listed in Subsection~\ref{1-SDF_AC.subs:simple_sdf_EIS}, first part; these subsets are classified according to whether they will correspond to choices at the beginning of the ``first period'' (at time $0$) or of the ''second period'' (at time $1$), if perceived as action path \textsc{sdf} according to Lemma \ref{1-SDF_AC.lemma:simple_sdf_as_APsdf}:
\begin{center}
    \begin{tabular}{r| c c}
     \textsc{eis}& 1st period & 2nd period  \\
     \hline
      1. & $c_{k \bullet}$ : $k\in\{1,2\}$ &$c_{k m}, c_{\bullet m}$ : $k,m\in\{1,2\}$ \\
      2.(a) &$c_{k \bullet}$ : $k\in\{1,2\}$ & $c_{k g}, c_{\bullet g}$ : $k\in\{1,2\},\, g\in M$ \\
      2.(b) & $c_{k \bullet}$ : $k\in\{1,2\}$ & $c_{1g}, c_{2m}, c_{\bullet m}$ : $m\in\{1,2\},\, g\in M$ \\
      2.(c) & $c_{k \bullet}$ : $k\in\{1,2\}$ & $c_{1m}, c_{2g}, c_{\bullet m}$ : $m\in\{1,2\},\, g\in M$ \\
      3. &  $c_{f \bullet}$ : $f\in M$ & $c_{k g}, c_{\bullet g}$ : $k\in\{1,2\},\, g\in M$ \\
    \end{tabular}
\end{center}
For example, consider the stock trader control problem with exogenous information according to case 2.(b). In the first period, the agent can choose ``buy'' or ``sell'' but cannot make the choice dependent on the exogenous scenario (for example, some insider information regarding future price development). In the second, the agent receives this insider information only if she chose to ``sell'' (= action $2$) before, and only then she can condition her choice on that information by selecting an arbitrary contingent action $g\in M$ via $c_{2g}$. If she chose to ``buy'' (= action $1$) in the first period, she can only choose $c_{1m}$, $m\in\{1,2\}$. The choice $c_{\bullet m}$, $m\in\{1,2\}$, --- that is, choose action $m$ at the second move, whatever happened before --- is also adapted. However, when formulating consistency criteria for a game-theoretic model based on adapted choices (which is the subject of Chapter~\ref{chap:2-SEF_G}), this choice will be eliminated here: Indeed, this choice supposes that the agent cannot recall what she did in the first period. However, she must be able to recall this because the exogenous information revealed at both second-period moves is different.

\begin{lemma}\label{1-SDF_AC.lemma:simple_sdf1_AC}
    For each exogenous information structure $\ms F$, the subsets of $W$ given in the corresponding line of the preceding table are $\ms F$-$\ms C$-adapted choices on $(F,\pi,\X)$.
\end{lemma}

Second, consider the variant $(F',\pi',\X')$ of the simple stochastic decision forest. Let, again, $M$ be the set of maps $\Omega\to \{1,2\}$ (contingent actions). For $k\in \{1,2\}$ and $f,g\in M$, let
\begin{align*}
    c'_{f\bullet} =&~ \{(\omega,k',m'), (\omega,k')\in W \mid k' = f(\omega)\}, \\
    c'_{k g} =&~ \{(\omega,k',m')\in W \mid k' = k, \, m' = g(\omega)\}, \\
    c'_{\bullet g} =&~ \{(\omega,k',m') \in W \mid  m' = g(\omega)\}.
\end{align*}
The interpretations are similar to those from the basic model above. Define $c'_{k\bullet}$, $c'_{\bullet m}$, and $c'_{km}$ by identifying $k,m\in\{1,2\}$ with the constant maps on $\Omega$ with values $k$ and $m$, respectively.  
Let $\ms C'_{\x'_0} = \{ c'_{1 \bullet},c'_{2 \bullet}\}$, $\ms C'_{\x'_1} = \ms C'_{\x'_2} = \{c'_{\bullet1},c'_{\bullet2}\}$. 

\begin{lemma}\label{1-SDF_AC.lemma:simple_sdf2_RCS}
    $\ms C' = (\ms C'_{\x'})_{\x'\in\X'}$ defines a reference choice structure on $\X'$.
\end{lemma}

Next, consider the following table. It reads as above.
\begin{center}
    \begin{tabular}{r| c c}
     \textsc{eis}& 1st period & 2nd period  \\
     \hline
      1. & $c'_{k \bullet}$ : $k\in\{1,2\}$ & $c'_{k m}, c'_{\bullet m}$ : $k,m\in\{1,2\}$ \\
      2. & $c'_{k \bullet}$ : $k\in\{1,2\}$ & $c'_{k g}, c'_{\bullet g}$ : $k\in\{1,2\},\, g\in M$ \\
      3. & $c'_{f \bullet}$ : $f\in M$ & $c'_{k g}, c'_{\bullet g}$ : $k\in\{1,2\},\, g\in M$ \\
    \end{tabular}
\end{center}   
For example, in the ultimatum bargaining example, in case 1., the agents can not at all condition their action on exogenous events. Usually, the second agent can observe what the first offered, so the choice $c'_{km}$ is available at the random move $\x'_k$, for all $k,m\in\{1,2\}$. If the offer is not communicated and the second agent has to accept or reject ``blindly'', then the second agent only has the choices $c'_{\bullet m}$, $m\in\{1,2\}$.

\begin{lemma}\label{1-SDF_AC.lemma:simple_sdf2_AC}
    For each exogenous information structure $\ms F'$, the subsets of $W'$ given in the corresponding line of the preceding table are $\ms F'$-$\ms C'$-adapted choices on $(F',\pi',\X')$.
\end{lemma}

Note that the adaptedness of choices can be rephrased, namely by requiring that $f$ and $g$ be measurable with respect to the $\sigma$-algebra of exogenous information at the random move the choice is available at, respectively. This more convenient language is used in the context of action paths in the next subsection which contains the previous two examples following up on Lemma \ref{1-SDF_AC.lemma:simple_sdf_as_APsdf}.

Let us resume that, in general, a choice like $c_{kg}$ is conditional on the knowledge that $\x_k$ has been chosen at the root of $\X$, while $c_{\bullet g}$ is independent of the initial choice. This is reflected by the respective sets of immediate predecessors, see Lemma \ref{1-SDF_AC.lemma:simple_sdf1_choices}. A similar remark is true for the variant, see Lemma \ref{1-SDF_AC.lemma:simple_sdf2_choices}. Hence, as in \cite[Section~5]{AlosFerrer2005Trees}, choices can reflect the endogenous information, that is, the information about the position in the decision tree $(\Tr,\ge_\Tr)$, agents have. While $c_{kg}$ is a choice of perfect (endogenous) information, $c_{\bullet g}$ is not. The discussion on this theme will be continued in Chapter~\ref{chap:2-SEF_G}.\smallskip

Concerning the \textsc{sdf} model $(F,\pi,\X)$ of the absent-minded driver story following Gilboa, with the exogenous information structure $\ms F$ as discussed in Subsection~\ref{1-SDF_AC.subs:simple_sdf_EIS}, let, for $i\in I = \{1,2\}$:
\[ \op{Ex}_i = \underbrace{[\rho^{-1}(i)\times\{D\}] \cup [\rho^{-1}(3-i)\times\{H\}]\}}_{=\text{``exit''}}, \qquad \op{Ct}_i = \underbrace{[\rho^{-1}(i)\times\{H,M\}] \cup [\rho^{-1}(3-i) \times \{M\}]}_{\text{=``continue''}},\]
and $\ms C_{\x_i}^i = \{ \op{Ex}_i,\op{Ct}_i\}$.
Futhermore, for any $E\in\ms F^i_{\x_i}$, let
\[ c_i(E) = (W_E \cap \op{Ex}_i)\cup(W_{E^\complement} \cap \op{Ct}_i), \]
the choice of agent $i$ to exit in the event $E$ and to continue in the opposite event $E^\complement$. $E$ might be thought about as an event independent of $\rho$, allowing for individual ``randomisation''. Let $C^i = \{c_i(E) \mid E\in\ms F^i_{\x_i}\}$. That is, at both random moves $\x_i$, $i\in I=\{1,2\}$, the active agent $i$ has two basic choices: ``exit'' and ``continue'', between that $i$ can randomise depending on $i$'s exogenous information, that is, in an $\ms F_{\x_i}^i$-measurable way. It is easily seen that $\ms C^i$ defines a reference choice structure on $\{\x_i\}$ and that $C^i$ is a set of $\ms F^i$-$\ms C^i$-adapted choices, for both $i\in I$.

\subsection{Action path stochastic decision forests}\label{1-SDF_AC.subs:APsdf_AC}

Finally, we consider the action path \textsc{sdf} $(F,\pi,\X)$ from Subsection~\ref{1-SDF_AC.subs:APsdf}, induced by action path \textsc{sdf} data $(I,\A,\T,W)$ on an exogenous scenario space $(\Omega,\ms E)$. \smallskip

Let $t\in\T$. For any set $A_{<t}\subseteq \A^{[0,t)_\T}$ and any family $A_t = (A_{t,\omega})_{\omega\in\Omega}\in\mc P(\A)^\Omega$ of subsets of $\A$, let
\[ c(A_{<t},A_t) = \{(\omega,f)\in W \mid f|_{[0,t)_\T} \in A_{<t},\, f(t) \in A_{t,\omega} \}. \]
For $t\in\T$, let $\ms C_t$ be the set of all $c(A_{<t},A_t)$ ranging over all $A_{<t}\subseteq\A^{[0,t)_\T}$ and all families $A_t = (A_{t,\omega})_{\omega\in\Omega}\in \mc P(\A)^\Omega$ of subsets of $\A$ satisfying the following assumptions:
\begin{itemize}[label=--]
    \item \hypertarget{1-SDF_AC.Ass:AP.C0}{\textbf{Assumption~AP.C0.}}~$c(A_{<t},A_t)\neq\emptyset$.
    \item \hypertarget{1-SDF_AC.Ass:AP.C1}{\textbf{Assumption~AP.C1.}}~For all $w\in c(A_{<t},A_t)$, there is $w'\in x_t(w)\setminus c(A_{<t},A_t)$. 
    \item \hypertarget{1-SDF_AC.Ass:AP.C2}{\textbf{Assumption~AP.C2.}}~For all $f\in\A^\T$ with $f|_{[0,t)_\T} \in A_{<t}$, we have \[x_t(\omega,f)\cap c(A_{<t},A_t) \neq \emptyset\] for all or for no $\omega\in D_{t,f}$.
\end{itemize}
So we consider choices that correspond to actions at a predefined time $t$. Again, there is some sort of duality here: action paths are the result of progressive choosing; choices are collections of action paths, essentially. More precisely, $c(A_{<t},A_t)\in\ms C_t$ describes the choice of an action in $A_{t,\omega}$ in scenario $\omega$ and at time $t$, given the history is contained in $A_{<t}$. We assume that such an action is really possible for at least some scenario (Assumption~\hyperlink{1-SDF_AC.Ass:AP.C0}{AP.C0}), that it really constitutes a choice in that there is an alternative (Assumption~\hyperlink{1-SDF_AC.Ass:AP.C1}{AP.C1}), and that it is complete in the sense that it only trivially intersects with random moves (Assumption~\hyperlink{1-SDF_AC.Ass:AP.C2}{AP.C2}).

The principal down-sets and sets of predecessors of such choices take the expected form, as affirmed by the following lemmata. In particular, such a choice $c(A_{<t},A_t)$ is available exactly at all those moves $x$, whose time $\mf t(x)$ is $t$, and that contain an outcome compatible with the choice.

\begin{lemma}\label{1-SDF_AC.lemma:AP_downarrow_c}
    For all $t\in\T$ and $c \in\ms C_t$, we have:
    \[ \downarrow c = \{ x_u(w) \mid w \in c,\,u\in\T\colon t<u\} \cup \{\{w\} \mid w\in c\}.  \]
\end{lemma}

\begin{lemma}\label{1-SDF_AC.lemma:AP_P(c)}
    For all $t\in\T$ and $c \in\ms C_t$, we have:
    \[ P(c) = \{ x_t(w) \mid w \in c\}.  \]
\end{lemma}

As a result, we obtain a large class of non-redundant and complete choices for action path \textsc{sdf}s.

\begin{lemma}\label{1-SDF_AC.lemma:C_t_non-redundant_complete}
    Let $t\in\T$ and $c\in \ms C_t$. Then, $c$ defines a non-redundant and complete choice. Moreover, for all $\x\in\X$ that $c$ is available at, there is $(\omega,f)\in c$ such that $\omega\in D_{t,f} = D_\x$ and $\x = \x_t(f)$.
\end{lemma}

We fix an action index $i\in I$ (in perspective, modelling an agent), a set of random moves $\tilde\X^i$ (of that agent), and an exogenous information structure $\ms F^i = (\ms F^i_\x)_{\x\in\tilde\X^i}$ (this agent is equipped with). We denote the canonical projection $\A\to\A^i$ by $p^i$.
For all $t\in\T$ and $\x\in\tilde\X^i$ with $\mf t(\x) = t$, let $\ms C^i_{\x}$ be \emph{a} set of sets $c(A_{<t},A_t)$ as above such that
\begin{enumerate}
    \item\label{1-SDF_AC.def:msC.1} $A_{<t}\subseteq\A^{[0,t)_\T}$;
    \item\label{1-SDF_AC.def:msC.2} $A_t = (A_{t,\omega})_{\omega\in\Omega}$ such that there is $A_t^i\in \ms B(\A^i)$ satisfying, for all $\omega\in \Omega$, \[A_{t,\omega} = \begin{cases} (p^i)^{-1}(A^i_t), &\quad \omega\in D_\x, \\ \emptyset, &\quad \omega\notin D_\x; \end{cases}\]
    \item\label{1-SDF_AC.def:msC.3} $c(A_{<t},A_t)\in\ms C_t$; and
    \item\label{1-SDF_AC.def:msC.4} for all $\omega\in D_{\x}$, $\x(\omega) \cap c(A_{<t},A_t) \neq\emptyset$.
\end{enumerate}
These properties are referenced as (AP-RCS.\ref{1-SDF_AC.def:msC.1}) etc. Hence, a choice $c(A_{<t},A_t)\in\ms C^i_\x$ allows for choosing a measurable set of individual actions for ``agent'' $i$ at the random move $\x$ given the endogenous past $A_{<t}$.

\begin{proposition}\label{1-SDF_AC.prop:APsdf_RCS}
    $\ms C^i = (\ms C^i_\x)_{\x\in\tilde\X^i}$ defines a reference choice structure on $\tilde\X^i$.
\end{proposition}

Let $t\in\T$, $A_{<t}\subseteq \A^{[0,t)_\T}$, $D\in\ms E$, and $g\colon D\to\A^i$. Let $A_t^{i,g} = (A_{t,\omega}^{i,g})_{\omega\in\Omega}$ be given by
\[ A_{t,\omega}^{i,g} = \begin{cases} \{a\in \A \mid p^i(a) = g(\omega) \}, &\quad\omega\in D, \\ \emptyset, &\quad \omega\notin D. \end{cases} \]
Let $c(A_{<t},i,g) = c(A_{<t},A_t^{i,g})$. Provided it is an element of $\ms C_t$, this models the choice of ``agent'' $i$, given an endogenous history in $A_{<t}$, to take the random action $g$, that is, the action $g(\omega)$ in scenario $\omega\in D$, at time $t$.
The perfect (endogenous) information case corresponds to $A_{<t} = \{f|_{[0,t)_\T}\}$ for some $f\in\A^\T$ such that $D_{t,f}\neq\emptyset$.

\begin{example}
    For $\T=\R_+$, singleton $\Omega$, singleton $A_{<t} = \{f|_{[0,t)_\T}\}$ for $t\in\T$ and $f\in\A^\T$, and $W = \Omega \times \A^\T \cong \A^\T$, $c(A_{<t},i,g)$ corresponds to the so-called ``differential game'' choice in \cite[Section~2]{AlosFerrer2011Comment} (see also \cite[Example~4.14]{AlosFerrer2016Theory}), where agent $i$ chooses the value of $g$ given the history $f|_{[0,t)_\T}$.
\end{example}

\begin{example}\label{1-SDF_AC.ex:adapted_choice_vs_adapted_process}
    If $\T = \{0,1,\dots,N\}$ for some $N\in\N$ or $\T = \N$, then these $c(A_{<t},i,g)$ reflect the usual specification of choices in terms of adapted processes from decision problems in discrete time (see, e.g.\ \cite{Bertsekas1996Stochastic}) if $g$ is supposed to be $\ms F_\x^i$-adapted at all $\x\in\tilde\X^i$ that $c(A_{<t},i,g)$ is available at. For the case of continuous-time, $\T=\R_+$ (see, e.g.\, \cite{Pham2009Continuous}), a similar remark can be made (note, however, that additional regularity conditions along time --- progressive measurability, optionality, predictability, for example --- are imposed for complete contingent plans of action, which is discussed in the third part).
\end{example}

The next and final theorem of this chapter is concerned with the following question, which can be motivated by the preceding Example~\ref{1-SDF_AC.ex:adapted_choice_vs_adapted_process}. Provided $c = c(A_{<t},i,g) \in \ms C_t$, what is the link between the adaptedness of $c$ and the measurability of the function $g|_{D_\x}$ with respect to $\ms F^i_\x$, for all $\x\in\tilde\X^i$ that $c$ is available at?\footnote{As stated in the theorem below, the availability of $c$ at $\x$ implies $D_\x \subseteq D$.} Indeed, in stochastic control and game theory, action path specifications are often implicitly used and choices are defined by assigning an action $g(\omega)$ to any scenario $\omega$, for $g$ measurable with respect to the value of the underlying filtration at the current time (see, e.g.\ \cite[Chapter~8]{Bertsekas1996Stochastic}, \cite{Pham2009Continuous,Cohen2015Stochastic,Carmona2018Probabilistic}). Hence, if we can answer the question above by providing a tight link, then this customary modelling paradigm in stochastic control and game theory can be explained as a derivative of adapted choices in stochastic decision forests and can therefore be interpreted in terms of traditional decision theory.

The theorem below affirms that the $\ms F^i_\x$-measurability of $g|_{D_\x}$ for relevant $\x\in\tilde\X^i$ is sufficient for the adaptedness of $c$. Moreover, it is also necessary, provided the reference choice structure sufficiently reflects the Borel $\sigma$-algebra on $\A^i$. 
For making this precise, let us call a set $\mc M$ \emph{stable under non-trivial intersections} iff for all $A,B\in\mc M$ with $A\cap B\neq\emptyset$, we have $A\cap B\in\mc M$. This property is equivalent to saying that $\mc M \cup \{\emptyset\}$ is stable under intersections. For example, any partition has this property. 
Now, given some $t\in\T$, $A_{<t}\subseteq \A^{[0,t)_\T}$, $i\in I$, $D\in\ms E$, and $g\colon D \to \A^i$ such that $\tilde c = c(A_{<t},i,g)\in\ms C_t$, we consider the following assumption:
\begin{itemize}[label=--]
    \item \hypertarget{1-SDF_AC.Ass:AP.C3}{\textbf{Assumption~AP.C3}} on the triple $(A_{<t},i,g)$ and on $\ms C^i$.~For all $\x\in\tilde\X^i$ that $\tilde c$ is available at, there is
    a generator $\ms G(\A^i)$ of the Borel $\sigma$-algebra of $\A^i$, stable under non-trivial intersections, such that for all $G\in\ms G(\A^i)$, upon letting $A^{i,G}_t = (A^{i,G}_{t,\omega})_{\omega\in\Omega}$ be given by $A^{i,G}_{t,\omega} = (p^i)^{-1}(G)$ for $\omega\in D_\x$ and $A^{i,G}_{t,\omega} = \emptyset$ for $\omega\notin D_\x$, we have \[c(A_{<t},A^{i,G}_t)\in \ms C^i_{\x}.\]
\end{itemize}

\begin{thm}\label{1-SDF_AC.thm:APsdf_AC}
    Let $(F,\pi,\X)$ be the action path \textsc{sdf} induced by action path \textsc{sdf} data $(I,\A,\T,W)$ on an exogenous scenario space $(\Omega,\ms E)$. Further, let $i\in I$ be an action index, $\tilde\X^i\subseteq\X$ be a set of random moves, $\ms F^i$ be an exogenous information structure on $\tilde\X^i$ and $\ms C^i$ be a reference choice structure on $\tilde\X^i$ as above satisfying Axioms (AP-RCS.$k$), $k=1,\dots,4$. 
    Let $t\in\T$, $A_{<t}\subseteq \A^{[0,t)_\T}$, $D\in\ms E$, and $g\colon D\to\A^i$ such that $c(A_{<t},i,g)\in\ms C_t$. 
    
    Then, we have:
    \begin{enumerate}
        \item\label{1-SDF_AC.thm:APsdf_AC.non_red_and_compl} $c(A_{<t},i,g)$ is a non-redundant and $\tilde\X^i$-complete choice. 
        \item\label{1-SDF_AC.thm:APsdf_AC.Dx_subset_D} For all $\x\in\tilde\X^i$ that $c(A_{<t},i,g)$ is available at, we have $D_\x \subseteq D$.
        \item\label{1-SDF_AC.thm:APsdf_AC.Fx_mb_=>_adapted} If for all $\x\in\tilde\X^i$ that $c(A_{<t},i,g)$ is available at $g|_{D_\x}$ is $\ms F^i_{\x}$-measurable, then $c(A_{<t},i,g)$ is $\ms F^i$-$\ms C^i$-adapted.
        \item\label{1-SDF_AC.thm:APsdf_AC.adapted_=>_Fx_mb} If Assumption~\hyperlink{1-SDF_AC.Ass:AP.C3}{AP.C3} is satisfied for $(A_{<t},i,g)$ and $\ms C^i$, and if moreover $c(A_{<t},i,g)$ is $\ms F^i$-$\ms C^i$-adapted, then $g|_{D_\x}$ is $\ms F^i_{\x}$-measurable for all $\x\in\tilde\X^i$ that $c(A_{<t},i,g)$ is available at.
    \end{enumerate}
\end{thm}

\begin{example}\label{1-SDF_AC.ex:APsdf_AC}
    Recall that we consider the action path \textsc{sdf} $(F,\pi,\X)$ induced by action path \textsc{sdf} data $(I,\A,\T,W)$ on an exogenous scenario space $(\Omega,\ms E)$. 
    \begin{itemize}[label=--]
        \item If we consider the simple \textsc{sdf}s from the preceding subsection in action path formulation, according to Lemma \ref{1-SDF_AC.lemma:simple_sdf_as_APsdf}, then the preceding construction yields exactly the adapted choices from the preceding subsection.
        
        \item Suppose that $W=\Omega\times\A^\T$. Fix some $i\in I$ and suppose that $\A^i$ has at least two elements. Let $t\in\T$ and $A_{<t}\subseteq \A^{[0,t)_\T}$.
        Then, for any map $g\colon\Omega\to\A^i$ we have $c(A_{<t},i,g) \in \ms C_t$, that is, it satisfies Assumptions \hyperlink{1-SDF_AC.Ass:AP.C0}{AP.C0}, \hyperlink{1-SDF_AC.Ass:AP.C1}{AP.C1}, and \hyperlink{1-SDF_AC.Ass:AP.C2}{AP.C2}. Moreover, if for all $\x\in\tilde\X^i$ with $\mf t(\x) = t$, $\ms C^i_\x$ contains all $c(A_{<t},A_t)$ ranging over all $A_t$ satisfying (AP-RCS.$k$), $k=2,3,4$, then Assumption~\hyperlink{1-SDF_AC.Ass:AP.C3}{AP.C3} is satisfied for $(A_{<t},i,g)$ and $\ms C^i$.
        
        \item In case of the timing problem (see Example~\ref{1-SDF_AC.ex:APsdf}, and, e.g.\ \cite{Shiryaev2007Optimal,Karatzas1998Methods}), we have $\A^i = \{0,1\}$ for all $i\in I$. Fix $i\in I$. Let $t\in\T$ and $A_{<t}\subseteq \A^{[0,t)_\T}$ be a non-empty set of componentwise decreasing paths $f_t$ such that $p^i\circ f_t = 1_{[0,t)_\T}$. 
        Then, for all $g\colon\Omega\to\{0,1\}$, we have $c(A_{<t},i,g)\in\ms C_t$, that is, it satisfies Assumptions \hyperlink{1-SDF_AC.Ass:AP.C0}{AP.C0}, \hyperlink{1-SDF_AC.Ass:AP.C1}{AP.C1}, \hyperlink{1-SDF_AC.Ass:AP.C2}{AP.C2}. Moreover, if for all $\x\in\tilde\X^i$ with $\mf t(\x) = t$, $\ms C^i_\x$ contains all $c(A_{<t},A_t)$ ranging over all $A_t$ satisfying (AP-RCS.$k$), $k=2,3,4$, then Assumption~\hyperlink{1-SDF_AC.Ass:AP.C3}{AP.C3} is satisfied for $(A_{<t},i,g)$ and $\ms C^i$.
        
        \item An analogous statement holds true in the case of the up-and-out option control problem (see Example~\ref{1-SDF_AC.ex:APsdf}, and \cite{Hull2018Options}). Here $I$ is a singleton, $\T = \R_+$, and $\A^i = \A = \{0,1\}$ for the unique $i\in I$. Let $t\in\R_+$ such that, with  
        \[ D = \{ \omega\in \Omega \mid \max_{u\in[0,t]} P_u(\omega) < 2\}, \]
        we have $D\neq\emptyset$. Further, let $g\colon D\to\{0,1\}$ be a map and let $A_{<t} = \{1\}^{[0,t)}$. 
        Then, we have $c(A_{<t},i,g)\in\ms C_t$, that is, it satisfies Assumptions \hyperlink{1-SDF_AC.Ass:AP.C0}{AP.C0}, \hyperlink{1-SDF_AC.Ass:AP.C1}{AP.C1}, \hyperlink{1-SDF_AC.Ass:AP.C2}{AP.C2}. Moreover, if for all $\x\in\tilde\X^i$ with $\mf t(\x) = t$, $\ms C^i_\x$ contains all $c(A_{<t},A_t)$ ranging over all $A_t$ satisfying (AP-RCS.$k$), $k=2,3,4$, Assumption~\hyperlink{1-SDF_AC.Ass:AP.C3}{AP.C3} is satisfied for $(A_{<t},i,g)$ and $\ms C^i$.
    \end{itemize}
    The proofs of these claims can be found in the corresponding part of the appendix.
\end{example}

It is important to note that a choice $c = c(A_{<t},i,g) \in \ms C_t$ as above, available at a random move $\x$, with $A_{<t}\subseteq \A^{[0,t)_\T}$ and $\ms F^i_\x$-measurable $g\colon D \to \A^i$, $D\in\ms E$, can condition on two distinct things: first, on the endogenous information $A_{<t}$, and second, on the exogenous information via the $\ms F^i_\x$-measurability of $g$. For instance, one can imagine the case that there are $\x'\in\tilde\X^i \setminus \{\x\}$ with $\ms F^i_\x = \ms F^i_{\x'}$ and $\mf t(\x) = t = \mf t(\x')$, $\ms F^i_{\x}$-measurable $g'\colon D \to \A^i$ with $g\neq g'$ and $A'_{<t}\subseteq \A^{[0,t)_\T}$ such that $c(A'_{<t},i,g')$ is available at $\x'$ and an agent's complete contingent plan of action specifies $c(A_{<t},i,g)$ at $\x$, but $c(A'_{<t},i,g')$ at $\x'$.
In the context of Example~\ref{1-SDF_AC.ex:APsdf_EIS}, with $\x = \x_t(f)$ and $\x' = \x_t(f')$ for suitable $f,f'\in \A^\T$, this would mean that, even if $Y_{\x_u(f)}$, $u\le t$, and $Y_{\x_u(f')}$, $u\le t$, generate the same $\sigma$-algebra, i.e.\ $\ms F^i_{\x_t(f)} = \ms F^i_{\x_t(f')}$, an agent need not make the very choice the agent makes at $\x_t(f)$ at the counterfactual random move $\x_t(f')$ as well.

Let us close on a note about this specific example for that we consider an agent only having choices of the form $c = c(A_{<t},i,g)$ as in the preceding paragraph. In case $A_{<t} = \A^{[0,t)_\T}$ for all choices available to this agent, then the complete contingent plans of action this agent can build model what is described by the term ``open loop'' in the control- and game-theoretic literature. If, conversely, only choices with singleton $A_{<t}$ are available to the agent, then we obtain what is subsumed under the term ``closed loop'' (see, e.g.\ \cite{Fudenberg1988Open,Fudenberg1991Game}). On the other hand, the alternative between the case a) $\ms F^i_{\x_t(f)} = \ms G_t$ for all $f\in\A^\T$ and $t\in D_{t,f}$, and a fixed filtration $\ms G = (\ms G_t)_{t\in\T}$ (case 1 in Subsection~\ref{1-SDF_AC.subs:APsdf_EIS}), and the case b) of $\ms F^i_{\x_t(f)}$ depending on a stochastic state controlled by $f$, for $f\in\A^\T$ and $t\in D_{t,f}$, in the sense of case 2 of Subsection~\ref{1-SDF_AC.subs:APsdf_EIS}, possibly in combination with Example~\ref{1-SDF_AC.ex:APsdf_EIS}, marks the difference between a) ``control-independent'' knowledge about the realised scenario and b) strategic dynamic learning of exogenous information (as, for instance, in adaptive control or exploration--vs.--exploitation problems).

Note that there are intermediate regimes in both (endogenous, exogenous) dimensions, and that these can be combined. However, within a consistent framework, as discussed in Chapter~\ref{chap:2-SEF_G}, ``large'' alias uninformative $A_{<t}$ and informative $\ms F^i$ like in case 2 of Subsection~\ref{1-SDF_AC.subs:APsdf_EIS} can be combined only to a limited extent: poor endogenous information restricts the granularity of exogenous information since otherwise the latter could reveal parts of the former. In any case, the situation of informative $\ms F^i$ is not to be confounded with ``closed loop'' controls (see, e.g., the corresponding discussion in \cite[p.\ 72--76]{Carmona2018Probabilistic}). Actually, the agent can only react to the knowledge on the noise, controlled or uncontrolled, but not explicitly to the employed controls or the actual paths of the controlled noise themselves --- which can make a crucial difference when evaluating counterfactuals, especially in case several agents interact. 
On the other hand, ``small'' alias informative $A_{<t}$ may well coexist with case 1 of Subsection~\ref{1-SDF_AC.subs:APsdf_EIS}, that is, ``$\ms F^i_{\x_t(f)} = \ms G_t$'', which is near, or in the extreme case of singleton $A_{<t}$ equal to closed loops with respect to endogenous information, but is equivalent, on the exogenous information side, to the absence of strategic learning of the realised exogenous scenario. In any case, we conclude, that the terminology of ``closed'' and ``open'' loops must be used with caution in the mixed regimes just discussed, and that specifying the underlying extensive form characteristics in terms of a stochastic decision forest, exogenous information structures, and adapted choices can be helpful in using it.

\chapter{Stochastic extensive forms and games}\label{chap:2-SEF_G}
\section{Stochastic extensive forms}\label{2-SEF_G.sec:Stochastic extensive forms}

The fundamental object of classical extensive form decision and game theory is the decision tree (cf.~\cite{Neumann1944Theory,Kuhn1950Extensive,Kuhn1953Extensive,AlosFerrer2016Theory}). The stochastic generalisation constructed in the present thesis replaces the traditional ``nature'' agent with a device ``randomly'' selecting a decision tree. As noted in Chapter~\ref{chap:1-SDF_AC}, this implies that we consider decision forests rather than decision trees. In that chapter, the former have been formalised using the refined partitions framework and made amenable to exogenous noise in the sense of general probability theory, and it has been shown that these are in a strong sense decision-theoretically natural. The present chapter is not self-contained as it refers to the notions introduced in Chapter~\ref{chap:1-SDF_AC}. Based on this, the central concept of the current chapter and arguably of the whole thesis, that of \emph{stochastic extensive forms}, is introduced in the first subsection below. This makes it possible, in turn, to give a definition of strategies in the sense of Savage acts (cf.~\cite{Savage1972Foundations}) and their extension due to Anscombe and Aumann in \cite{Anscombe1963Definition} and Aumann in \cite{Aumann1974Subjectivity}, respectively. Towards the end of this section, the new concepts are illustrated by simple examples, including the absent-minded driver story. For graphical representations of the refined-partitions approach to dynamic decision making and of stochastic decision forests, we refer to Figures~\ref{fig:refined_partitions_restaurant} and~\ref{fig:SDF} printed in the preface and the introduction, respectively.

\subsection{Definition: Stochastic extensive forms}\label{2-SEF_G.subs:SEF}

Chapter~\ref{chap:1-SDF_AC} is about a model of stochastic decision forests that consistently combines exogenous and endogenous information flow and allows for a notion of choices adapted to exogenous information. While exogenous information is modelled through an additional structure, endogenous information is given in terms of the sets of immediate predecessors. In that sense, choices are ``adapted to'' endogenous information by construction. Given a stochastic decision forest $(F,\pi,\X)$, a set $I$ of agents, and families $\ms F = (\ms F^i)_{i\in I}$, $\ms C = (\ms C^i)_{i\in I}$ and $C = (C^i)_{i\in I}$ of exogenous information structures, reference choice structures, and sets of adapted choices, respectively, additional criteria are required in order to define an extensive form. First, choices must be partition refining along the trees: alternative future nodes must be separable by alternative choices and alternative choices must be disjoint, in any scenario; endogenous information sets must be disjoint; any possible future node must be compatible with some choice. Second, endogenous and exogenous information must be compatible: exogenous information must be identical across endogenous information sets. Finally, as we allow for multiple agents to act ``simultaneously'' at the same move (as in \cite{AlosFerrer2005Trees, AlosFerrer2008Trees, AlosFerrer2011Comment}), any admissible combination of choices must be compatible with at least one outcome.

The notation and definition that follow are generalisations of those from \cite{AlosFerrer2005Trees, AlosFerrer2008Trees, AlosFerrer2011Comment}. While the latter are based on (what we call) decision trees, the former are based on (more general) stochastic decision forests.

Let $I$ be a set. If $C = (C^i)_{i\in I}$ is a family of sets of choices on some stochastic decision forest $(F,\pi,\X)$, then for any move $x\in X$, any random move $\x\in \X$, and any $i\in I$,
$$ A^i(x) = \{c\in C^i \mid x\in P(c) \}, \qquad A^i(\x) = \{c\in C^i \mid \x^{-1}(P(c)) \neq \emptyset \} $$
are the sets of choices in $C^i$ \emph{available at} $x$, $\x$, respectively.  The notation $A$, commonly used and linked to the term ``action'', is discussed later in this subsection.

For any move $x\in X$, any random move $\x\in\X$, let
$$ J(x) = \{ i\in I \mid A^i(x) \neq \emptyset\}, \qquad J(\x) = \{i\in I \mid A^i(\x) \neq \emptyset\}. $$
For any $i\in I$, let 
\[ X^i = \{x\in X \mid i \in J(x) \}, \qquad \X^i = \{\x\in\X \mid i \in J(\x)\}. \]
Let $\X^i \bullet \Omega = \{(\x,\omega)\in\X^i\times \Omega\mid \omega\in D_\x\}$. 

Clearly, if all $c\in C^i$ are $\X^i$-complete, then we have $A^i(\x) = \{c\in C^i \mid \x^{-1}(P(c)) = D_\x\} = A^i(\x(\omega))$ and $J(\x) = J(\x(\omega))$ for all $\x\in\X^i$ and $\omega\in D_\x$. It is also clear that if $c\in C^i$ is $\X^i$-complete, then it is already complete.

\begin{definition}\label{2-SEF_G.def:SEF}
    Let $(\Omega,\ms E)$ be an exogenous scenario space. A \emph{stochastic pseudo-extensive form}, in short: \emph{$\psi$-\textsc{sef}}, \emph{on $(\Omega,\ms E)$}, is a tuple $\F = (F,\pi,\X,I,\ms F, \ms C,C)$ such that $(F,\pi,\X)$ is a stochastic decision forest on $(\Omega,\ms E)$, $I$ is a set, $C=(C^i)_{i\in I}$ is a family of sets of choices, $\ms F = (\ms F^i)_{i\in I}$ is a family of exogenous information structures on $\X^i$, $i\in I$, $\ms C = (\ms C^i)_{i\in I}$ is a family of reference choice structures on $\X^i$, $i\in I$, such that all elements of $C^i$ are $\ms F^i$-$\ms C^i$-adapted and the evaluation map $\X^i\bullet\Omega\to X$ is injective, for all $i\in I$, and satisfying the following axioms:
    \begin{enumerate}
        \item\label{2-SEF_G.def:SEF.P(c)} For all $i\in I$, all $c,c'\in C^i$ such that $P(c) \cap P(c') \neq \emptyset$ we have $P(c) = P(c')$, and for all $\omega\in\Omega$, we have either $c \cap W_\omega = c' \cap W_\omega$ or $c\cap c' \cap W_\omega = \emptyset$.
        \item\label{2-SEF_G.def:SEF.outcomes_faithful} For all $x\in X$ and all $(c^i)_{i\in J(x)} \in \bigtimes_{i\in J(x)} C^i$, we have
        $$ x\cap\bigcap_{i\in J(x)} c^i \neq \emptyset. $$
        \item\label{2-SEF_G.def:SEF.weak_separation} For all $y,y'\in F$ with $\pi(y) = \pi(y')$ and $y\cap y' = \emptyset$ there are $i\in I$ and $c,c'\in C^i$ such that $y\subseteq c$, $y\subseteq c'$ and $c\cap c' \cap W_{\pi(y)} = \emptyset$.
        \item\label{2-SEF_G.def:SEF.enough_choices} For all $x\in X$, all $i\in J(x)$, all $y\in \downarrow x\setminus\{x\}$, there is $c\in A^i(x)$ with $c\supseteq y$.
        \item\label{2-SEF_G.def:SEF.endo_exo_compatible} For all $i\in I$, $\x,\x'\in \X^i$ such that $A^i(\x) \cap A^i(\x') \neq \emptyset$, we have $\ms F^i_\x = \ms F^i_{\x'}$ and $\ms C^i_\x = \ms C^i_{\x'}$.
        \item\label{2-SEF_G.def:SEF.choice_completeness} For all $i\in I$, all $\ms F^i$-$\ms C^i$-adapted choices $c'$ such that
        \begin{enumerate}[label=(\roman*),ref=\theenumi{}(\roman*)]
            \item\label{2-SEF_G.def:SEF.choice_completeness.i} any $\omega\in\Omega$ with $c'\cap W_\omega \neq \emptyset$ admits $c\in C^i$ with $c'\cap W_\omega = c \cap W_\omega$,
            \item\label{2-SEF_G.def:SEF.choice_completeness.ii} and there is $c\in C^i$ with $P(c') = P(c)$,
        \end{enumerate}
         are choices for $i$, that is, satisfy $c'\in C^i$.
    \end{enumerate}
    The elements of $I$ are called \emph{agents} or \emph{decision makers}. For each agent $i\in I$, $\ms F^i$ is called \emph{$i$'s exogenous information structure}, $\ms C^i$ is called \emph{$i$'s reference choice structure}, and the elements of $C^i$ are called \emph{$i$'s choices}. For $i\in I$ and $x\in X$ ($\x\in \X$) $i$ is said \emph{active at $x$ ($\x$, respectively)} iff $i\in J(x)$ ($i\in J(\x)$, respectively). For any agent $i\in I$, the sets $P(c)$, $c\in C^i$, are called \emph{immediate predecessor sets of $i$'s choices}. For any agent $i\in I$, the elements of $X^i$ and $\X^i$ are called \emph{$i$'s moves} and \emph{$i$'s random moves}, respectively.

    If $\F$ is a stochastic pseudo-extensive form on $(\Omega,\ms E)$, then its items are typically denoted by
    \[ \F = (F,\pi,\X,I,\ms F,\ms C,C). \]

    A \emph{stochastic extensive form}, in short: \emph{\textsc{sef}}, \emph{on $(\Omega,\ms E)$}, is a stochastic pseudo-extensive form satisfying the following stronger separation axiom:
    \begin{enumerate}[label=3'.,ref=3']
        \item\label{2-SEF_G.def:SEF.separation} For all $y,y'\in F$ with $\pi(y) = \pi(y')$ and $y\cap y'= \emptyset$, there are $x\in X$, $i\in I$ and $c,c'\in C^i$ with $x\cap c\supseteq y$, $x\cap c'\supseteq y'$, $c\cap c' \cap W_{\pi(y)} = \emptyset$ and $x\in P(c) \cap P(c') \cap T_{\pi(y)}$. 
    \end{enumerate}

    A \emph{classical (pseudo-) extensive form} is the data $(F,I,C)$ for a stochastic (pseudo-) extensive form $\F$ on the singleton exogenous scenario space, respectively. 
\end{definition}

In other words, a stochastic pseudo-extensive form specifies a stochastic decision forest $(F,\pi,\X)$, a set of agents $I$, and for each agent $i\in I$, a ``dynamically updating oracle'' (exogenous information structure) $\ms F^i$ along $i$'s random moves, a set of reference choices $\ms C^i$ describing how $i$ can measure choices locally at each of $i$'s random moves, and a set $C^i$ of choices adapted to this data satisfying six axioms. Five of these axioms have already been motivated roughly, and all of them are discussed in more detail in the sequel. 

At this point, we make a first comparison with the notion of the ``extensive decision problem'' in the sense of \cite{AlosFerrer2005Trees} and of the ``extensive form'' in the sense of \cite{AlosFerrer2011Comment}, reproduced and further developed in the monographic version in \cite[Definition 5.2, p.\ 118]{AlosFerrer2016Theory}. Indeed, upon consulting the latter reference it becomes evident that, respectively, a triple $(T,I,C)$ is a classical (pseudo-) extensive form in the sense of the preceding definition iff $(T,C)$ is an (``extensive decision problem'') ``extensive form'' with set of ``players'' $I$ according to (cf.~\cite[Definition 4.1]{AlosFerrer2016Theory}) \cite[Definition 5.2, p.\ 118]{AlosFerrer2016Theory}, the tree $(T, \supseteq)$ is rooted, and $C = (C^i)_{i\in I}$ is such that for all $i\in I$ and all $c\in C^i$, $P(c) \neq \emptyset$.
In that sense, the concept of \cite[Definition 5.2]{AlosFerrer2016Theory}-``extensive forms'' is naturally equivalent to the concept of classical extensive forms, which is naturally embedded into the concept of stochastic extensive forms. An analogous statement holds true for \cite[Definition 4.1]{AlosFerrer2016Theory}-``extensive decision problems'' and classical pseudo-extensive forms. The term ``pseudo'' is used because in a stochastic pseudo-extensive form, separating two disjoint nodes from the same tree is possible but not necessarily at one single move --- and therefore may be effectless on the level of strategies. This is exactly the problem about the continuous-time example discussed in the introduction where an infinity of outcomes is compatible with a given strategy profile.

In the present text, the term ``pseudo-extensive forms'' is preferred over that of ``extensive decision problems'' because the latter, just as the term ``extensive (form) game'', can easily be understood to include a given preference relation on outcomes. This, however, is not the case and we wish to avoid any confusion about this. In this framing, pseudo-extensive forms as defined above, whether stochastic or classical, provide a form describing all possibilities of evolution, the information agents have about it, and what they can do. But to make a decision problem or game out of it, individual preferences about outcomes must be added. While the stronger separability property of extensive forms makes them more relevant in the end, their relaxed version (whatever its name) has been found to be important for understanding the problem of well-posedness in the classical case in \cite{AlosFerrer2008Trees,AlosFerrer2011Comment}. This is why here as well both versions are introduced, but the naming is chosen such as to underline the importance of the stronger version.

The term ``classical'' is used instead of ``deterministic'' which, at first sight, might seem to be more compelling. This is because a certain class of stochastic (pseudo-) extensive forms can be represented as classical (pseudo-) extensive forms, as is discussed in Section~\ref{2-SEF_G.sec:well-posedness_equilibrium}. In the case of classical extensive forms, Axioms \ref{2-SEF_G.def:SEF.endo_exo_compatible} and~\ref{2-SEF_G.def:SEF.choice_completeness} become redundant, and the formulation of the remaining four axioms slightly shorter. The interpretation of the first four axioms in the general stochastic case, as discussed next, therefore resembles the discussion in \cite{AlosFerrer2008Trees,AlosFerrer2011Comment,AlosFerrer2005Trees} (see \cite{AlosFerrer2016Theory} for a monographic treatment). 

Axiom \ref{2-SEF_G.def:SEF.enough_choices} means that at any move and for any possible future node, any active agent can choose not to discard it. Put less rigorously, it ensures that at any move any future node is compatible with some choice which is the last part of the refined partitions model. Axiom \ref{2-SEF_G.def:SEF.weak_separation} means that any pair of non-consecutive nodes from the same exogenous scenario $\omega$ can be separated by a pair of choices disjoint on $W_\omega$ and jointly available to one and the same agent, while Axiom \ref{2-SEF_G.def:SEF.separation} in addition requires this to be possible jointly at one and the same, and preceding move in $T_\omega$. Put less rigorously, both axioms ensure that alternative future nodes are separable by alternative choices, but to different extents. Axiom \ref{2-SEF_G.def:SEF.outcomes_faithful} means that at any move, all profiles of admissible choices by active agents are compatible with at least one outcome. This is a minimal requirement on the decision forest $F$ to be a faithful descriptor of outcomes, and is linked to the possibility that multiple agents can choose at once. This modelling ansatz, though non-standard compared to the historic literature, has been pursued in \cite{AlosFerrer2005Trees,AlosFerrer2008Trees,AlosFerrer2011Comment,AlosFerrer2016Theory}. We adopt this convention because it simplifies the presentation of interactive settings.

\subsection{Information sets}

For the understanding of Axioms~\ref{2-SEF_G.def:SEF.P(c)} and~\ref{2-SEF_G.def:SEF.endo_exo_compatible}, let us note the following.
\begin{proposition}\label{2-SEF_G.prop:information_sets}
    Let $\F$ be a stochastic pseudo-extensive form and $i\in I$ an agent.
    \begin{enumerate}
        \item\label{2-SEF_G.prop:information_sets.P(c)_partition} $\{P(c) \mid c\in C^i\}$ is a partition of $X^i$.
        \item\label{2-SEF_G.prop:information_sets.A(x)_partition} $\{A^i(x) \mid x\in X^i\}$ is equal to $\{ A^i(\x) \mid \x \in \X^i\}$ and is a partition of $C^i$.
        \item\label{2-SEF_G.prop:information_sets.P(c)=P(c')} For all $c,c'\in C^i$, we have $P(c) = P(c')$ iff there is $x\in X$ with $c,c'\in A^i(x)$.
        \item\label{2-SEF_G.prop:information_sets.A(x)=A(x')} For all $x,x'\in X$, we have $A^i(x) = A^i(x')$ iff there is $c\in C^i$ with $x,x'\in P(c)$.
        \item\label{2-SEF_G.prop:information_sets.exists_mfP} There is a unique partition ${\mf P^i}$ of $\X^i$ such that for all $\x,\x'\in \X^i$ we have $A^i(\x) = A^i(\x')$ iff there is ${\mf p}\in{\mf P^i}$ such that $\x,\x'\in{\mf p}$.
        \item\label{2-SEF_G.prop:information_sets.Bij_mfP_P(c)} The assignment \[{\mf P^i} \ni {\mf p} \mapsto \bigcup_{\x\in\mf p} \im \x\] defines a bijection ${\mf P^i} \to \{P(c) \mid c\in C^i\}$.
        \item\label{2-SEF_G.prop:information_sets.msC_msF_const_on_mfP} For all ${\mf p}\in{\mf P^i}$, all $\x,\x'\in{\mf p}$, we have $D_\x = D_{\x'}$, $\ms C^i_\x = \ms C^i_{\x'}$, and $\ms F^i_\x = \ms F^i_{\x'}$.
    \end{enumerate}
\end{proposition}

\begin{remark}\label{2-SEF_G.rmk:prop_information_sets}
    In the proof of the preceding proposition in Subsection~\ref{2-SEF_G.subsec:appendix.proofs.1}, a bit more is shown. Namely, let $\F = (F,\pi,\X,I,\ms F,\ms F,C)$ be a tuple as in Definition~\ref{2-SEF_G.def:SEF} satisfying Axioms~\ref{2-SEF_G.def:SEF}.$k$, $k=1,\dots,5$, but not necessarily Axiom~\ref{2-SEF_G.def:SEF}.\ref{2-SEF_G.def:SEF.choice_completeness}. Then the conclusions of Proposition~\ref{2-SEF_G.prop:information_sets} hold true.
\end{remark}

Thus, the immediate predecessor sets partition $i$'s moves, and the sets of available choices partition $i$'s choices. Two choices have identical immediate predecessor sets iff they are available at a common move; and two moves $i$ is active at have identical available choices iff they are jointly immediate predecessors to one and the same choice. Moreover, immediate predecessor sets as well as the preceding statements can equivalently be formulated on the level of random moves, which gives rise to a model of endogenous information sets for stochastic extensive forms. This model is consistent with respect to exogenous information in that at two random moves belonging to the same endogenous information set ${\mf p}$ the same exogenous information is revealed. 

In light of Proposition~\ref{2-SEF_G.prop:information_sets}, let, for ${\mf p}\in{\mf P^i}$ and $\x\in{\mf p}$:
$$ A^i({\mf p}) = A^i(\x), \qquad D_{\mf p} = D_\x, \qquad \ms F^i_{\mf p} = \ms F^i_\x, \qquad \ms C^i_{\mf p} = \ms C^i_\x. $$
Previous definitions about moves can be lifted accordingly. For instance, we call $c$ \emph{available} at an endogenous information set $\mf p$ iff $c \in A^i(\mf p)$. 

To fix names for the discussed concept of information, let us note that in stochastic pseudo-extensive forms, information is revealed along two different channels: exogenous information is revealed via $\ms F$, endogenous information (about agents' behaviour) is revealed via the position in $(\X,\ge_\X)$. Therefore, it is natural to decompose the property of perfect recall accordingly.

\begin{definition}
    Let $\F$ be a stochastic pseudo-extensive form and $i\in I$ be an agent. 
    \begin{enumerate}
        \item The elements of $\mf P^i$ are called \emph{$i$'s endogenous information sets}. 

        \item The set of choices of agent $i$ is said to admit and agent $i$ itself is said to have \emph{perfect endogenous information} iff all $\mf p\in \mf P^i$ are singletons and for all $j\in I\setminus\{i\}$, all $\x\in\X^i$, $\x'\in\X^j$, we have $\im\x \cap\im\x' = \emptyset$. Agent $i$ is said to have \emph{perfect exogenous information} iff $\ms F^i_\x = \ms E|_{D_\x}$ for all $\x\in\X^i$. Agent $i$ is said to have \emph{perfect information} iff $i$ has both perfect endogenous and exogenous information, and $\F$ is said to be of \emph{perfect information} iff this holds true for all $i\in I$.

        \item The set of choices of agent $i$ and agent $i$ itself are said to admit \emph{perfect endogenous recall} iff all $c,c'\in C^i$ and $\omega\in\Omega$ with $c\cap c'\cap W_\omega \neq \emptyset$ satisfy $c\cap W_\omega \supseteq c'\cap W_\omega$ or $c\cap W_\omega \subseteq c'\cap W_\omega$. Agent $i$ is said to admit \emph{perfect exogenous recall} iff ${\ms F}^i$ admits recall. Agent $i$ is said to admit \emph{perfect recall} iff $i$ admits both perfect endogenous and exogenous recall, and $\F$ is said so iff this holds true for all $i\in I$.
    \end{enumerate}
\end{definition}

Perfect recall with respect to endogenous information is defined by the requirement that for two choices available at moves along a given decision path the earlier one cannot condition on less endogenous information (this criterion is compatible with many classical definitions of perfect recall for a large class of classical extensive forms, see \cite[Subsections~6.4.1, 6.4.2]{AlosFerrer2016Theory}). Perfect information with respect to endogenous information is essentially defined in the classical way, namely, by requiring information sets to be minimally small, among and across agents. Note, however, that we formulate the notion both for individual agents and the stochastic (pseudo-) extensive form as a whole. Perfect information with respect to exogenous information is defined analogously. However, it is not a very interesting case as no ``nature'' agent is supposed to act dynamically. The stochastic component of stochastic (pseudo-) extensive forms is relevant just because there may be agents having imperfect exogenous information (about the realised scenario). For the sake of an illustration of the above-defined notions it is shown in Lemma~\ref{2-SEF_G.lemma:perfect_endo_information_implies_perfect_endo_recall} in the appendix that, as to be expected, perfect information implies perfect recall.

Perfect endogenous recall can be analysed along the lines of \cite{Ritzberger1999Recall}. Perfect recall, however, has a richer structure in the present setting with general sigma-algebras.
We also note that in stochastic extensive forms, no information set can be visited twice by a given decision path (as in the classical case discussed in \cite{AlosFerrer2005Trees} and \cite{AlosFerrer2016Theory}), which the author proposes to call the \emph{Heraclitus property}:

\begin{lemma}[Heraclitus Property]\label{2-SEF_G.lemma:Heraclitus_property}
   Let $(F,\pi,\X,I,\ms F,\ms C,C)$ be a stochastic pseudo-extensive form on an exogenous scenario space $(\Omega,\ms E)$. Then we have for all agents $i\in I$:
   \begin{enumerate}
       \item\label{2-SEF_G.lemma:Heraclitus_property.X} for all $x,x'\in X$ with $A^i(x) \cap A^i(x') \neq \emptyset$ and $x\supseteq x'$ we have $x=x'$;
       \item\label{2-SEF_G.lemma:Heraclitus_property.rmX} for all $\x,\x'\in\X$ with $A^i(\x) \cap A^i(\x') \neq \emptyset$ and $\x \ge_\X \x'$ we have $\x = \x'$.
   \end{enumerate}
\end{lemma}

The possibility of crossing an information set twice is typically referred to as ``absent-mindedness'', as a response to the absent-minded driver story and the subsequent modelling idea both due to Piccione and Rubinstein in \cite{Piccione1997Interpretation}. Together with the preceding result, we might thus be tempted to reject either the hypothesis that ``absent-mindedness'' as a concept is compatible with classical decision theory, as brought up in \cite{Piccione1997Interpretation}, or the claim of the generality of extensive form modelling (compare the corresponding discussion in \cite{AlosFerrer2016Theory}). The author believes that to resolve this seeming dilemma it might be helpful to distinguish between a phenomenon (such as the absent-minded driver story) and a model or formal attempt to analyse or describe a phenomenon.

Note that from a point of view of classical decision theory and in particular of refined partitions-based (stochastic) extensive forms, the \emph{phenomenon} of absent-mindedness as expressed in this story is not contradictory in itself. 
This has been discussed, for instance, in \cite{Aumann1997Absent,Gilboa1997Comment}. Gilboa, for instance, has proposed an alternative description which can be succinctly formulated in stochastic extensive form (see the third of the simple examples in Chapter~\ref{chap:1-SDF_AC}, rediscussed in Subsection~\ref{2-SEF_G.subs:SEF_simple-examples}). Moreover, both \cite{Aumann1997Absent} and \cite{Gilboa1997Comment} make disappear much of the paradoxical conclusions from \cite{Piccione1997Interpretation} which suggests that the latter arise rather from the model than from the phenomenon.

In that sense, the \emph{phenomenon} of absent-mindedness is not at odds with the refined-partitions theory of (stochastic) extensive forms. There is such a model describing the strategic phenomenon convincingly and concisely. To make that clear we have used the term ``Heraclitus property'' for the above-described \emph{formal property} as opposed to the word ``absent-mindedness'' which we use here only for the phenomenon (without denying its use in purely graph-based models that one cannot always make sense of from the rigorous decision-theoretic refined partitions-based perspective). This point of view also distinguishes the present treatment from the one in \cite{AlosFerrer2016Theory}.\smallskip

Let us conclude this subsection with two remarks. The first is about the actual information flow perceived by a given agent. If an agent $i\in I$ is at random move $\x\in\X^i$, the three pieces of information the agent has are $A^i(\x)$, $\ms F_\x^i$, and $\ms C_\x^i$. From this, the agent can infer the current information set ${\mf p}\in{\mf P^i}$ alias $P(c)$, for $c\in A^i(\x)$, the fact that the realised scenario $\omega$ is an element of $D_\x$, and for any $E\in\ms F_\x^i$ the fact whether $\omega\in E$ or not, including $E = \x^{-1}(P(c \cap c'))$ for all $c'\in\ms C_\x^i$. But the consistency conditions imply that all of this does not reveal more information along the vertical tree axis (e.g.\ about which $\x'\in{\mf p}$ is the actual one) or about the horizontal scenario axis (e.g.\ about events not contained in $\ms F_\x^i$).

Also note that the refined partitions approach implemented through stochastic (pseudo-) extensive forms does not require action labels. Choices already implicitly contain the data specifying conditional on which endogenous information they can be made --- this point has been made in \cite{AlosFerrer2005Trees,AlosFerrer2008Trees,AlosFerrer2011Comment} already (for classical (pseudo-) extensive forms). This implies that there is no necessity to add action labels because they are already implicit in the definition of choices. In the present framework, a choice does not only tell whether to go left at one particular move, but also at which set of moves. So for instance, a choice can consist in going left at move $x_0$; but it can also consist in going left at any move at time $2$; or it can consists in going left if agent $j\neq i$ has gone left at time $1$. Moreover, the fact whether these choices are available or not determines the endogenous information the given agent has: knowing whether you are at move $x_0$ or not when you are actually there; knowing nothing about what agents did before time $2$; knowing whether $j$ has gone left at time $1$, respectively. This has partly been discussed in \ref{1-SDF_AC.sec:adapted_choices} and is further detailed in the upcoming examples. Consequently, for a stochastic (pseudo-) extensive form denoted as above, one could call the elements $c\in C^i$ \emph{actions} of agent $i$. Although this is avoided for reasons of simplicity, the standard notion $A^i(x)$ for the set of actions at move $x\in X$ is retained.

Moreover, note that Axiom \ref{2-SEF_G.def:SEF.P(c)} also includes the statement that for any agent $i\in I$, any pair of choices $c,c'\in C^i$ with $P(c) = P(c')$ when ``evaluated'' in a particular scenario $\omega\in\Omega$ is either equal or disjoint. Along any tree $T_\omega$, two choices available at the same move are either identical or disjoint (alias strict alternatives). 

\subsection{Completeness}

Finally, let us consider Axiom~\ref{2-SEF_G.def:SEF.choice_completeness}. This is a completeness axiom. Indeed, any tuple of the form $(F,\pi,\X,I,\ms F,\ms C,C)$ satisfying the conditions from Definition~\ref{2-SEF_G.def:SEF} except Axiom~\ref{2-SEF_G.def:SEF.choice_completeness} can be modified by extending the set of choices for any agent such that the result satisfies Axiom~\ref{2-SEF_G.def:SEF.choice_completeness} and is equivalent to $(F,\pi,\X,I,\ms F,\ms C,C)$. More precisely:

\begin{lemma}\label{2-SEF_G.lemma:completeness}
    Let $\F = (F,\pi,\X,I,\ms F,\ms C,C)$ be a tuple satisfying the conditions defining a stochastic extensive form on some exogenous scenario space $(\Omega,\ms E)$ possibly except Axiom~\ref{2-SEF_G.def:SEF.choice_completeness}, according to Definition~\ref{2-SEF_G.def:SEF}. For any $i\in I$, let $\hat C^i$ be the set of all $\ms F^i$-$\ms C^i$-adapted choices $\hat c$ such that 
    \begin{enumerate}[label=(\roman*),ref=(\roman*)]
        \item\label{2-SEF_G.lemma:completeness.def:hatC.i} any $\omega\in\Omega$ with $\hat c\cap W_\omega\neq \emptyset$ admits $c\in C^i$ satisfying $\hat c\cap W_\omega = c \cap W_\omega$,
        \item\label{2-SEF_G.lemma:completeness.def:hatC.ii} there is $c\in C^i$ such that $P(\hat c) \subseteq P(c)$.
    \end{enumerate}
    Let $\hat C = (\hat C^i)_{i\in I}$. \smallskip

    Then, $\hat \F = (F,\pi,\X,I,\ms F,\ms C,\hat C)$ defines a stochastic extensive form on $(\Omega,\ms E)$ such that
    \begin{enumerate}
        \item\label{2-SEF_G.lemma:completeness.property_1} for all $i\in I$, all $\omega\in\Omega$,
        \[ \{ \hat c \cap W_\omega \mid \hat c\in\hat C^i\} \setminus \{\emptyset\} = \{c \cap W_\omega \mid c\in C^i\} \setminus \{\emptyset\}; \]
        \item\label{2-SEF_G.lemma:completeness.property_2} for all $i\in I$, 
        \[ \{P(\hat c) \mid \hat c\in \hat C^i\} = \{ P(c) \mid c\in C^i\}. \]
    \end{enumerate}

    The lemma remains true if ``stochastic extensive form'' is replaced with ``stochastic pseudo-extensive form'' everywhere.
\end{lemma}

By this lemma, completeness implies that choices are fully determined by their structure along any tree $T_\omega$, $\omega\in\Omega$, their immediate predecessor sets, and the informational structure along $\Omega$ given by $\ms F$ and $\ms C$. More precisely, any $C^i$, $i\in I$, is determined by the set of all $c\cap W_\omega$, $c\in C^i$, $\omega\in\Omega$, the family of sets $P(c)$, $c\in C^i$, and the families $\ms F^i$ and $\ms C^i$.

\subsection{Strategies}\label{2-SEF_G.subs:strategies}

In classical game and decision theory, a strategy is an agent's complete contingent plan of action (see e.g.\ \cite{MasColell1995Microeconomic}). As explained in \cite{AlosFerrer2005Trees}, this can be viewed as a sequential version of acts in the theory of choice under uncertainty going back to Savage (cf.~\cite{Savage1972Foundations}). Acts map (given) ``states'' to (chosen) ``consequences''. 

\cite{AlosFerrer2005Trees} takes the view that states are given by moves and consequences by available choices. Concerning ``states'', there is, however, a point of possible confusion which implies that there are different ways acts are formulated in the dynamic setting. One may argue that the point is not really the set of moves, but the information an agent has about them. In \cite{AlosFerrer2005Trees}, as ``states'' are identified with moves, acts are required to assign identical consequences on information endogenous information sets. Although the present text formally takes the perspective that endogenous information sets appear as the most precise interpretation of states, it perceives both viewpoints as equally convincing and equivalent.

On the other side, the refined partitions approach sees (local) ``consequences'' at a state as members of a partition of the set of (global) consequences (alias outcomes), that is, as choices in the sense of the present text. We see again that action is already implicitly described by choices, and no further structure of action labels or the like is needed, compare the discussion in the previous subsection. 

In order to ensure consistency (alias adaptedness) with respect to endogenous information in the mentioned sense, a strategy needs to assign to any of the agent's endogenous information sets a choice that is available at it. In addition, by restricting to adapted choices the local compatibility, or measurability, of acts with respect to exogenous information can be assured. This leads to the following definition.

\begin{definition}\label{2-SEF_G.def:strategy}
    Let $(F,\pi,\X,I,\ms F,\ms C,C)$ be a stochastic pseudo-extensive form and $i\in I$ an agent. 
    A \emph{strategy for $i$} is a map $s^i\colon {\mf P^i} \to C^i$ such that for all ${\mf p}\in{\mf P^i}$, $s^i({\mf p}) \in A^i({\mf p})$. 
    
    Let $S^i$ denote the set of strategies for $i$, and $S = \bigtimes_{i\in I} S^i$. A \emph{strategy profile} is an element of $S$.
\end{definition}

As mentioned before, there are other ways of formally defining strategies which moreover seem more traditional. These interpret moves as states and impose restrictions on strategies in terms of (endogenous) information sets. As the setting of stochastic pseudo-extensive forms includes both moves and random moves, we obtain the following two temporary definitions. Let $i\in I$. 

An \emph{$X$-strategy for $i$} is a map $s^i\colon X^i \to C^i$ such that:
\begin{enumerate}
    \item for all $x\in X^i$, we have $s^i(x) \in A^i(x)$;
    \item for all $x,x'\in X^i$ with $A^i(x) = A^i(x')$, $s^i(x) = s^i(x')$.
\end{enumerate}
Clearly, this is equivalent to saying that for all $c\in C^i$, 
$$ \{x\in X^i \mid s^i(x) = c\} \in \{\emptyset, P(c)\}. $$
This definition is a direct formal generalisation of the corresponding definition in \cite[Subsection~5.2]{AlosFerrer2005Trees} and exactly coincides with it in case of singleton $\Omega$.

An \emph{$\X$-strategy for $i$} is a map $s^i\colon \X^i \to C^i$ such that:
\begin{enumerate}
    \item for all $\x\in\X^i$, we have $s^i(\x) \in A^i(\x)$;
    \item for all $\x,\x'\in\X^i$ with $A^i(\x) = A^i(\x')$, $s^i(\x) = s^i(\x')$.
\end{enumerate}
Denote the set of $X$-strategies for $i$ by $S_X^i$ and the set of $\X$-strategies for $i$ by $S_\X^i$.

Furthermore, consider the natural surjections 
\begin{equation}\label{2-SEF_G.eq:Xi_surj}
    X^i \surj \X^i \surj {\mf P^i},
\end{equation}
with respect to which any map with domain ${\mf P^i}$ induces a map with domain $\X^i$ and any map with domain $\X^i$ induces a map with domain $X^i$, respectively. Then we have the following result.

\begin{proposition}\label{2-SEF_G.prop:strategies}
    Let $(F,\pi,\X,I,\ms F,\ms C,C)$ be a stochastic pseudo-extensive form, and $i\in I$ an agent. Then the maps in Equation~\ref{2-SEF_G.eq:Xi_surj} induce bijections
    $$ S^i \quad \stackrel{\cong}{\longrightarrow}\quad S_\X^i  \quad\stackrel{\cong}{\longrightarrow}\quad  S_X^i. $$  
\end{proposition}

Therefore, in the following, if $s^i$ is a strategy, both its corresponding $\X$- and $X$-strategy are denoted by $s^i$ as well.

\begin{remark}\label{2-SEF_G.rmk:strategies_adapted}
    Let $(F,\pi,\X,I,\ms F,\ms C,C)$ be a stochastic pseudo-extensive form and $i\in I$ an agent. Any strategy for agent $i$ is adapted in the sense that for all $\x\in\X^i$ the choice $s^i(\x)$ is adapted, that is, for all reference choices $c'\in\ms C_\x$, and all $\x'\in\X^i$ that $s^i(\x)$ is available at: 
    $$ \x'^{-1}(P(s^i(\x)\cap c')) \in \ms F_{\x'}^i. $$
    This abstract adaptedness is a consequence of the structure of the underlying stochastic pseudo-extensive form. It is not part of the definition of a strategy and, in that sense, not a primitive of the theory. 

    In the case of action path stochastic pseudo-extensive forms this property is seen to correspond exactly to the adaptedness in the language of the theory of stochastic processes (see Subsection~\ref{2-SEF_G.subs:AP_SEF.information_histories_adapted_choices} and Subsection~\ref{1-SDF_AC.subs:APsdf_AC}).
\end{remark}

It is a standard procedure in game and decision theory to extend acts (alias strategies) so that they become maps from states to lotteries over consequences. As discussed in \cite{Anscombe1963Definition}, this is consistent with the theory of subjective probability. It is an elementary insight of game theory that, contrary to what one may naively infer from the single-agent situation, additional randomisation may improve coordination. In the dynamic setting, where states can be seen as endogenous information sets and consequences as available choices, this extension leads to the abstract notion of behaviour strategies. From an abstract point of view, a behaviour strategy is a complete contingent plan of lotteries over action. That is, the agent can draw an action at random at any endogenous information set. In contrast to this, an abstract mixed strategy is a lottery over complete contingent plans of action, that is, over strategies (which are then said ``pure'', in order to distinguish). That is, the agent can draw a ``pure'' strategy at random before the start and then commits to it from the beginning until the end. While for mixed strategies correlation over different endogenous information sets is possible, the lottery draws of a behaviour strategy at distinct endogenous information sets are independent.

Following von Neumann and Morgenstern, lotteries are interpreted as probability measures. The term ``lottery'' describes a typically unpredictable procedure of determining a consequence, for instance, the procedure consisting of observing the winner of a horse race. Yet, a probability measure describes only the statistical distribution of such a procedure's result whatever the meaning of ``statistical'' (frequentist, subjectivist, or other). It thus needs a way of transforming this abstract distribution into a procedure of the above-mentioned sort. In the literature, there are conflicting ways of doing so.\footnote{About the interpretation of randomised strategies, see the detailed discussion in \cite{Luce1989Games}.} 

The basic interpretation starts as follows: go to the horse race in question and observe the result. But then, one might either act blindly according to it or revise the own strategy in view of that new information. The point here is whether the horse race's result is stochastically independent of all exogenous information the agent has at that moment, and in particular, whether the agent has to commit to its result (as if another agent behaved within a mandate given by the original agent, or as if the horse race were some unconscious cognitive process). 

The second interpretation which does not require commitment can be implemented using an exogenous scenario space, as discussed in \cite{Aumann1974Subjectivity,Aumann1987Correlated}. In particular, it can be implemented in stochastic extensive forms by using strategies as defined above and profiles thereof. The randomisation is given in terms of exogenous scenarios, beliefs on these, and the dependence alias correlation structure across endogenous information sets and agents.\footnote{This is discussed in more detail in Section~\ref{2-SEF_G.sec:well-posedness_equilibrium}, once beliefs have been introduced.} In that sense, strategy profiles in stochastic extensive forms are profiles of correlated strategies whose correlation device and information structure are given by the exogenous scenario space, the exogenous information structures of the agents and beliefs. Also note that for the fundamental solution concept of Nash equilibrium (and its derivatives) the difference between the two interpretations above evaporates (there, commitment to the result of randomisation conforms to personal interest); hence, restricting on the second interpretation does not seem to imply a restriction.

\subsection{Simple examples}\label{2-SEF_G.subs:SEF_simple-examples}

In this subsection, we recall the three simple examples of stochastic decision forests, including the exogenous information structures admitting recall, reference choices structures, and adapted choices for them introduced in Subsections~\ref{1-SDF_AC.subs:simple_sdf}, \ref{1-SDF_AC.subs:simple_sdf_EIS}, and~\ref{1-SDF_AC.subs:simple_sdf_AC} in Chapter~\ref{chap:1-SDF_AC}. As we demonstrate here, these data do indeed yield stochastic extensive forms. 

We begin with the basic simple \textsc{sdf}. It is illustrated in Figure~\ref{1-SDF_AC.fig:simple_sdf}. Recall that it indicates in a stylised fashion how to model finite dynamic decision problems using stochastic extensive forms. The basic example can be illustrated with the problem of a stock trader, who can buy or sell a share of a given stock in two periods. The prices depend on the realised exogenous scenario, and in our case there are only two exogenous scenarios.\footnote{This is weakly related to the classical one-period binomial model from mathematical finance, see, for example, \cite{Foellmer2016Stochastic}. However, the agent acts in two periods and the information and choice structures of the agent clarify how both ``dimensions'' relate.} An outcome is a triplet $(\omega,k,m)$ of an exogenous scenario $\omega$, and the symbols indicating buying or selling decisions $k$ and $m$ in the first and second period, respectively. For formal details, we refer to Subsection~\ref{1-SDF_AC.subs:simple_sdf}. The corresponding decision tree $(\Tr,\ge_\Tr)$ is illustrated in Figure~\ref{1-SDF_AC.fig:simple_sdf_Tr}. 
For the sake of simplicity, we consider only one agent $i$ --- for example, the stock trader as mentioned above ---  i.e.\ $I = \{i\}$.

Regarding exogenous information, it is shown in Lemma~\ref{1-SDF_AC.lemma:simple_sdf1_EIS} that there are exactly five exogenous information structure admitting recall and these are listed in Subsection~\ref{1-SDF_AC.subs:simple_sdf_EIS}. 
Concerning choices, we refer to definitions made in Subsection~\ref{1-SDF_AC.subs:simple_sdf_AC}, Chapter~\ref{chap:1-SDF_AC}; here, we only recall that $M$ denotes the set of random actions alias maps $\Omega\to\{1,2\}$. 
We then consider the following table which is slightly different from the one in Subsection~\ref{1-SDF_AC.subs:simple_sdf_AC} because we are now interested in the consistency requirements defining stochastic extensive form, rather than just in providing a list of adapted choices. The table here reads as follows: Each line specifies a set of subsets of $W$ for one of the five exogenous information structures (\textsc{eis}) from Lemma~\ref{1-SDF_AC.lemma:simple_sdf1_EIS} and recalled above; these subsets are classified according to whether they will correspond to choices at the beginning of the ``first period'' (at time $0$) or of the ''second period'' (at time $1$), if perceived as action path \textsc{sdf} according to Lemma~\ref{1-SDF_AC.lemma:simple_sdf_as_APsdf}:
\begin{center}
    \begin{tabular}{r| c c}
     \textsc{eis}& 1st period & 2nd period  \\
     \hline
      1. & $c_{k \bullet}$ : $k\in\{1,2\}$ &$c_{k m}$ : $k,m\in\{1,2\}$ \\
      1. & $c_{k \bullet}$ : $k\in\{1,2\}$ &$c_{\bullet m}$ : $m\in\{1,2\}$ \\
      2.(a) &$c_{k \bullet}$ : $k\in\{1,2\}$ & $c_{k g}$ : $k\in\{1,2\}, g\in M$ \\
      2.(a) &$c_{k \bullet}$ : $k\in\{1,2\}$ & $c_{\bullet g}$ : $g\in M$ \\
      2.(b) & $c_{k \bullet}$ : $k\in\{1,2\}$ & $c_{1g}, c_{2m}$ : $m\in\{1,2\}, g\in M$ \\
      2.(c) & $c_{k \bullet}$ : $k\in\{1,2\}$ & $c_{1m}, c_{2g}$ : $m\in\{1,2\}, g\in M$ \\
      3. &  $c_{f \bullet}$ : $f\in M$ & $c_{k g}$ : $k\in\{1,2\}, g\in M$ \\
      3. &  $c_{f \bullet}$ : $f\in M$ & $c_{\bullet g}$ : $g\in M$ \\
    \end{tabular}
\end{center}
For example, consider the line corresponding to the exogenous information structure 2.(c). As described in Chapter~\ref{chap:1-SDF_AC}, Subsection~\ref{1-SDF_AC.subs:APsdf_AC}, in ``the first period, the agent can choose `buy' or `sell' but cannot make the choice dependent on the exogenous scenario (for example, some insider information regarding future price development). In the second, the agent receives this insider information only if she chose to `sell' (= action $2$) before, and only then she can condition her choice on that information by selecting an arbitrary contingent action $g\in M$ via $c_{2g}$.''

\begin{lemma}\label{2-SEF_G.lemma:simple_sef1}
    Let $I$ be a singleton, $i\in I$, $\ms F^i$ be any of the five families from Subsection~\ref{1-SDF_AC.lemma:simple_sdf1_EIS}, $\ms F = (\ms F^i)$, $\ms C^i$ be as defined in Subsection~\ref{1-SDF_AC.lemma:simple_sdf1_AC}, $\ms C = (\ms C^i)$, and $C^i$ be a set of choices corresponding to it via the preceding table, $C = (C^i)$. Then, the tuple $\F = (F,\pi,\X,I,\ms F,\ms C,C)$ defines a stochastic extensive form on $(\Omega,\ms E)$.
\end{lemma}

\begin{remark}
    Note that for the exogenous information structure 2.(b), the set
    \[ \tilde C^i = \Big\{c_{k \bullet} \mid k\in\{1,2\}\Big \} \cup \Big\{ c_{\bullet m} \mid m\in\{1,2\}\Big\} \]
    gives not rise to a stochastic extensive form because Axiom \ref{2-SEF_G.def:SEF.endo_exo_compatible} is violated. A similar remark can be made regarding 2.(c). In other words, the exogenous information available at time $1$ would reveal the current random move though the endogenous information at that node would not do so (the endogenous information set would contain both random moves at time $1$). Axiom \ref{2-SEF_G.def:SEF.endo_exo_compatible} stipulates that such an inconsistency must not arise. 
\end{remark}

Next, we consider the variant of the simple example from Chapter~\ref{chap:1-SDF_AC}. It starts from the preceding example with two-element exogenous scenario space with scenarios denoted by $\omega_1$ and $\omega_2$, but identifies the elements $(\omega_1,2,1)$ and $(\omega_1,2,2)$ in $W$ which provides a stochastic decision forest with a random move that is not defined on all of $\Omega$, as illustrated in Figure~\ref{1-SDF_AC.fig:simple_sdf_variant}. In terms of our financial example, if the agent chose to ``buy''  in the first period and scenario $\omega_1$ is realised, she cannot choose anymore in the second period (perhaps because of liquidity or legal constraints). For the formal description of the stochastic decision forest, the reader is referred to Subsection~\ref{1-SDF_AC.subs:simple_sdf}. The induced decision tree $(\Tr',\ge_{\Tr'})$ is illustrated in Figure~\ref{1-SDF_AC.fig:simple_sdf_variant_Tr}.

It is shown in Lemma~\ref{1-SDF_AC.lemma:simple_sdf2_EIS} that there are exactly three exogenous information structures admitting recall, described in detail in Subsection~\ref{1-SDF_AC.subs:simple_sdf_EIS}. 
Concerning choices, the reader is referred to Subsection~\ref{1-SDF_AC.subs:simple_sdf_AC}. We illustrate the consistency properties of \textsc{sef} by the following table. Its interpretation is completely analogous to the preceding one from the basic version of the simple example.
\begin{center}
    \begin{tabular}{r| c c}
     \textsc{eis}& 1st period & 2nd period  \\
     \hline
      1. & $c'_{k \bullet}$ : $k\in\{1,2\}$ & $c'_{k m}$ : $k,m\in\{1,2\}$ \\
      2. & $c'_{k \bullet}$ : $k\in\{1,2\}$ & $c'_{k g}$ : $k\in\{1,2\}, g\in M$ \\
      3. & $c'_{f \bullet}$ : $f\in M$ & $c'_{k g}$ : $k\in\{1,2\}, g\in M$ \\
    \end{tabular}
\end{center}   
For example, in the trading example, in case of exogenous information described by line 2., the agent can recall her choice from the first period when in the second, and she can condition her choice in the second period on the realised scenario --- but not so in the first period.

\begin{lemma}\label{2-SEF_G.lemma:simple_sef2}
    Let $I'$ be a singleton, $i\in I'$, $\ms F^{\prime i}$ be any of the three preceding families, $\ms F' = (\ms F^{\prime i})$, $\ms C^{\prime i}$ as defined above, $\ms C = (\ms C^{\prime i})$, and ${C^{\prime i}}$ be the set of choices corresponding to it via the preceding table, $C' = (C^{\prime i})$. Then, the tuple $\F' = (F',\pi',\X',I',\ms F',\ms C',C')$ defines a stochastic extensive form on $(\Omega,\ms E)$.
\end{lemma}

\begin{remark}
    Note that for all three exogenous information structures and any $g\in M$, $c'_{\bullet g}$ cannot be the choice of an agent for some stochastic extensive form on $(F',\pi',\X')$. This is because the exogenous information available at time $1$ necessarily reveals the current random move, while the endogenous information at any move at time $1$ does not for $c'_{\bullet g}$ (the endogenous information set of that choice would contain both random moves at time $1$). Axiom \ref{2-SEF_G.def:SEF.endo_exo_compatible} stipulates that such an inconsistency must not arise. 
\end{remark}

The third example is a representation of Gilboa's interpretation in \cite{Gilboa1997Comment} of the absent-minded driver phenomenon in stochastic extensive form. For the motivating story, we refer to \cite{Piccione1997Interpretation}; it has also been recalled in Subsection~\ref{1-SDF_AC.subs:simple_sdf}. The corresponding \textsc{sdf}, exogenous information and reference choice structures, and adapted choices have been introduced in --- and for all notational details, the reader is referred to --- Chapter~\ref{chap:1-SDF_AC}, Subsections~\ref{1-SDF_AC.subs:simple_sdf}, \ref{1-SDF_AC.subs:simple_sdf_EIS}, and~\ref{1-SDF_AC.subs:simple_sdf_AC}, respectively. 

We nevertheless recall the main aspects of the definitions from these subsections in the following, and while doing so, we put them together in a suitable way. In this model, $(\Omega,\ms E)$ is an exogenous scenario space and $\rho\colon \Omega \to \{1,2\}$ is an $\ms E$-$\mc P\{1,2\}$-measurable surjection. Further, let $\xi^1,\xi^2$ be $[0,1]$-valued random variables. Suppose that $(\Omega,\ms E)$ is rich enough to admit a probability measure under which $(\rho,\xi^1,\xi^2)$ is independent and whose marginals are uniformly distributed on $\{1,2\}$ and $[0,1]$, respectively. 
Further, let $D,H,M$ be three different symbols meaning ``disastrous region'', ``home'' and ``motel'' as in the original story from \cite{Piccione1997Interpretation}. We have $W = \Omega \times \{D,H,M\}$ and the only two random moves $\x_1,\x_2$ are defined on the whole of $\Omega$ by
\[ \x_k(\omega) = \begin{cases} \{\omega\} \times \{D,H,M\}, &\text{ if } \rho(\omega) = k, \\ \{\omega\} \times \{H,M\}, & \text{ if } \rho(\omega) = {3-k}, \end{cases} \qquad k\in\{1,2\}. \]
$F$ has been defined as the union of the images of $\x_1$ and $\x_2$ and the set of all singleton sets in $W$. $\pi\colon F \to \Omega$ maps any element of $F$ to the first component of its elements. The decision forest $(F,\supseteq)$ and the induced poset of random nodes $(\Tr,\ge_\Tr)$ are illustrated in Figure \ref{1-SDF_AC.fig:absent_minded_driver_Gilboa_sdf}. 

Then, let $I =\{1,2\}$ and $\ms F^i_{\x_i} = \sigma(\xi^i)$, that is, the agents have no information other than their private signals $\xi^i$. 
For $i\in I $, we have defined:
\[ \op{Ex}_i = \underbrace{[\rho^{-1}(i)\times\{D\}] \cup [\rho^{-1}(3-i)\times\{H\}]\}}_{=\text{``exit''}}, \qquad \op{Ct}_i = \underbrace{[\rho^{-1}(i)\times\{H,M\}] \cup [\rho^{-1}(3-i) \times \{M\}]}_{\text{=``continue''}},\]
and $\ms C_{\x_i} = \{ \op{Ex}_i,\op{Ct}_i\}$.
Futhermore, for any $E\in\ms F^i_{\x_i}$, we have defined
\[ c_i(E) = (W_E \cap \op{Ex}_i)\cup(W_{E^\complement} \cap \op{Ct}_i), \]
to be the choice of ``agent'' to exit in the event $E$ and to continue in the opposite event $E^\complement$. As discussed in Subsection~\ref{1-SDF_AC.subs:simple_sdf_AC}, $E$ can be interpreted ``as an event independent of $\rho$, allowing for individual {`randomisation'}''. We have defined $C^i = \{c_i(E) \mid E\in\ms F^i_{\x_i}\}$. As discussed, this means that, at both random moves $\x_i$, $i\in I=\{1,2\}$, the corresponding active agent $i$ has two ``pure'' choices, which are ``exit'' and ``continue''. Between these, agent $i$ can randomise as a function of private information. It is easy to see that, for both $i\in I$, $\ms C^i$ defines a reference choice structure on $\{\x_i\}$ and that $C^i$ is a set of $\ms F^i$-$\ms C^i$-adapted choices.\footnote{See Subsection~\ref{1-SDF_AC.subs:simple_sdf_AC}.} Let $\ms F = (\ms F^i)_{i\in I}$, $\ms C = (\ms C^i)_{i\in I}$, and $C = (C^i)_{i\in I}$. 

Now, these data do indeed take a consistent decision-theoretic form:

\begin{thm}\label{2-SEF_G.thm:absent_minded_driver_Gilboa_sef}
    $(F,\pi,\X,I,\ms F,\ms C,C)$ defines a stochastic extensive form with perfect recall and imperfect information. 
\end{thm}

\section{Action path stochastic extensive forms}\label{2-SEF_G.sec:AP_SEF}

In most pieces of the literature, dynamic games are defined by supposing a notion of \emph{time} and specifying outcomes as certain paths of action at instants of time. \cite[Subsection~2.2]{AlosFerrer2005Trees} provides a broad overview for this, including classical textbook definitions as in \cite{Fudenberg1991Game}, infinite bilateral bargaining in discrete time as in \cite{Rubinstein1982Perfect}, repeated games, the long cheap talk game in \cite{Aumann2003Long}, and a decision-theoretic interpretation of differential games as in \cite{Dockner2000Differential}. In this thesis the author desires to generalise this by, first, allowing for a large class of outcome paths and, second, adding a truly stochastic dimension. As an application, studied in Chapter~\ref{chap:3-SPF_VECT}, this will be used to provide an ``asymptotic'' model of stochastic control in not only in discrete, but also in continuous time (see, e.g.\ \cite{Pham2009Continuous,Bertsekas1996Stochastic,Karatzas1998Methods}) and stochastic differential games (see, e.g.\ \cite{Carmona2018Probabilistic}) without restrictions on the noise in question.

In Subsections~\ref{1-SDF_AC.subs:APsdf}, \ref{1-SDF_AC.subs:APsdf_EIS}, and~\ref{1-SDF_AC.subs:APsdf_AC} of Chapter~\ref{chap:1-SDF_AC}, a first part of this approach has been formalised in one abstract and general framework in terms of stochastic decision forests, that allows for general exogenous stochastic noise, therefore going strictly beyond the ``nature'' agent setting. The author insists on what has been said there, namely that ``this framework is based on a specific structure pertaining to all of these examples, namely \emph{time}. Interestingly, time is not included in the abstract formulation of decision forests, and it serves as a particularly strong similarity structure for trees and even branches of one and the same tree'' (Subsection~\ref{1-SDF_AC.subs:APsdf}). In this section, it is shown under which conditions, how, and in what sense the construction from Subsections~\ref{1-SDF_AC.subs:APsdf}, \ref{1-SDF_AC.subs:APsdf_EIS}, and~\ref{1-SDF_AC.subs:APsdf_AC} of Chapter~\ref{chap:1-SDF_AC} gives rise to a stochastic extensive form.\smallskip

\subsection{Information, history structures, and adapted choices}\label{2-SEF_G.subs:AP_SEF.information_histories_adapted_choices}

Let us fix action path \textsc{sdf} data $(I,\A,\T,W)$ on a given exogenous scenario space $(\Omega,\ms E)$. The first steps consist in determining reference choice structures and adapted choices for any $i\in I$. While reference choice structures and adapted choices in Subsection~\ref{1-SDF_AC.subs:APsdf_AC} have been studied with respect to their consistency with respect to exogenous information, it remains to clarify their consistency with endogenous information. In action path \textsc{sdf}s, endogenous information at time $t$ can be modelled via partitions of histories prior to that $t$, as is clarified in the sequel.

For this, we introduce new notation. For $i\in I$, $t\in\T$, any subset $\T_t\subseteq \T$ with $[0,t)_\T \subseteq \T_t$, and all $f\in\A^{\T_t}$, let $D_{t,f}^i$ be the set of $\omega\in \Omega$ such that there are $f',f''\in\A^\T$ with $(\omega,f'), (\omega,f'')\in W$, $f'|_{[0,t)_\T} = f|_{[0,t)_\T} = f''|_{[0,t)_\T}$ and $p^i \circ f'(t) \neq p^i \circ f''(t)$.  This set will turn out as the event that, given the historic path $f|_{[0,t)_\T}$, agent $i$ can choose. Clearly, $D_{t,f}^i \subseteq D_{t,f}$, if $\T = \T_t$. In Proposition~\ref{2-SEF_G.prop:Dfti} this is stated in higher generality and it is seen that equality holds true unless $D_{t,f}^i = \emptyset$. Given action path \textsc{sdf} data $(I,\A,\T,W)$ and $i\in I$, let $\tilde\X^i = \{\x_t(f) \mid t\in\T,\,f\in\A^\T\colon D_{t,f}^i \neq \emptyset\}$.

\begin{definition}\label{2-SEF_G.def:mcH}
    Let $(I,\A,\T,W)$ be action path \textsc{sdf} data on an exogenous scenario space $(\Omega,\ms E)$ as before and $\ms F = (\ms F^i)_{i\in I}$ be a family of exogenous information structures $\ms F^i = (\ms F^i_\x)_{\x\in\tilde\X^i}$ on $\tilde\X^i$, $i\in I$. A \emph{history structure for $i$, given $(I,\A,\T,W,\ms F)$} is a family $\mc H^i = (\mc H^i_t)_{t\in\T}$ such that:
    \begin{enumerate}
        \item\label{2-SEF_G.def:mcH.partition} for each $t\in\T$, $\mc H^i_t$ is a partition of the set of all $f\in \A^{[0,t)_\T}$ such that $D_{t,f}^i \neq \emptyset$, 
        \item\label{2-SEF_G.def:mcH.msF_compatible} for all $t\in\T$, $A_{<t}\in \mc H^i_t$, and $f,f'\in A_{<t}$ we have $\ms F_{\x_t(f)}^i = \ms F_{\x_t(f')}^i$.
    \end{enumerate}
\end{definition}

The elements of $\mc H^i$ serve as a basis for the model of endogenous information sets in terms of partitions of histories, consistent with exogenous information. Axiom \ref{2-SEF_G.def:mcH}.\ref{2-SEF_G.def:mcH.partition} is meant to ensure that for any $i\in I$ and $t\in\T$, any element of $\mc H^i_t$ partitions the set of possible histories for moves of $i$ at time $t$. Axiom \ref{2-SEF_G.def:mcH}.\ref{2-SEF_G.def:mcH.msF_compatible} postulates the consistency with respect to exogenous information. Note that $\ms F_{\x_t(f)}^i = \ms F_{\x_t(f')}^i$ implies $D_{t,f} = D_{t,f'}$.

In the following, we fix, for any $i\in I$, an exogenous information structure $\ms F^i = (\ms F^i_\x)_{\x\in\tilde\X^i}$ and let $\ms F = (\ms F^i)_{i\in I}$. In addition, we fix a family $\mc H = (\mc H^i)_{i\in I}$ of history structures $\mc H^i$ for $i$, given $(I,\A,\T,W,\ms F)$, $i\in I$. For all $t\in\T$, $i\in I$, and $\x\in\tilde\X^i$ with $\mf t(\x) = t$, let $\ms C^i_{\x}$ the set of \emph{all} sets $c(A_{<t},A_t)$ as above such that
\begin{enumerate}
    \item\label{2-SEF_G.def:msC.1} $A_{<t}\in\mc H^i_t$;
    \item\label{2-SEF_G.def:msC.2} $A_t = (A_{t,\omega})_{\omega\in\Omega}$ such that there is $A_t^i\in \ms B(\A^i)$ satisfying, for all $\omega\in \Omega$, \[A_{t,\omega} = \begin{cases} (p^i)^{-1}(A^i_t), &\quad \omega\in D_\x, \\ \emptyset, &\quad \omega\notin D_\x; \end{cases}\]
    \item\label{2-SEF_G.def:msC.3} $c(A_{<t},A_t)\in\ms C_t$; and
    \item\label{2-SEF_G.def:msC.4} for all $\omega\in D_{\x}$, $\x(\omega) \cap c(A_{<t},A_t) \neq\emptyset$.
\end{enumerate}
These properties are referenced as (AP-RCS*.\ref{2-SEF_G.def:msC.1}), (AP-RCS*.\ref{2-SEF_G.def:msC.2}) etc. In contrast to Subsection~\ref{1-SDF_AC.subs:APsdf}, we have now specified the history structure implicit in $\ms C^i_\x$. $c(A_{<t},A_t)$ allows for choosing a measurable set of actions in the $i$-th action space factor at the random move $\x$ given an endogenous past $A_{<t}$ at time $t$, as specified by $\mc H$. By Proposition~\ref{1-SDF_AC.prop:APsdf_RCS}, $\ms C^i = (\ms C_\x^i)_{\x\in\tilde\X^i}$ defines a reference choice structure for any $i\in I$. 

\begin{remark}
    By Lemma~\ref{1-SDF_AC.lemma:simple_sdf_as_APsdf}, the ``simple examples'' for stochastic decision forests discussed also in Subsection~\ref{2-SEF_G.subs:SEF_simple-examples} can be represented as action path \textsc{sdf}. However, the reference choice structures considered in Lemmata~\ref{1-SDF_AC.lemma:simple_sdf1_RCS} and~\ref{1-SDF_AC.lemma:simple_sdf2_RCS}, and recalled in Subsection~\ref{2-SEF_G.subs:SEF_simple-examples}, are not those obtained by the previous construction. Actually, the previous construction is obtained from the former by elementwise intersecting with the known endogenous past. 
    
    Let us spell this out for the basic version. Denote by $\ms C^{\text{AP}}$ the action path construction and by $\ms C^{\text{orig}}$ the ``original'' definition from Chapter~\ref{chap:1-SDF_AC}, recalled in Subsection~\ref{2-SEF_G.subs:SEF_simple-examples}. Then, recalling the table in Subsection~\ref{2-SEF_G.subs:SEF_simple-examples}, describing the different combinations of exogenous information structures, reference choice structures and sets of choices:
    \begin{enumerate}
        \item $\ms C^{\text{AP}}_{\x_0} = \ms C^{\text{orig}}_{\x_0}$, and
        \item for both $k=1,2$:
        \begin{enumerate}
            \item if there is a bullet in the second period, that is, for the second, fourth or eighth line of the table, then: $\ms C^{\text{AP}} = \ms C^{\text{orig}}$ 
            \item else, that is, for the first, third, fifth, sixth, seventh line, then: 
            \[ \ms C^{\text{AP}}_{\x_k} = \{ c \cap \im \x_k \mid c\in \ms C^{\text{orig}}_{\x_k} \} = \{ c \cap c_{k\bullet} \mid c\in \ms C^{\text{orig}}_{\x_k} \}. \]
        \end{enumerate}
    \end{enumerate}

    While the action path construction is more accurate in that it precisely conditions on the piece of endogenous information the agent really has, the sets of adapted choices are identical under both specifications of the reference choice structure. After all, reference choice structures are about $\ms B_{\A^i}$ rather than about $\A^{[0,t)_\T}$. In Chapter~\ref{chap:1-SDF_AC}, therefore, it was not relevant to make that distinction, and for the simple examples it does not really matter. In general, however, the general construction given for action path \textsc{sdf} data given above is preferable, because available actions may vary across different pasts $A_{<t}$ so that a reference choice for a given past may no more be a choice for another past.
\end{remark}

Next, the actual choices prospective agents make are introduced. Here, we follow Subsection~\ref{1-SDF_AC.subs:APsdf_AC}, but, again, with the difference of explicitly referring to $\mc H$ in doing so. Let $i\in I$, $A_{<t}\in\mc H^i_t$, $D\in\ms E$, and $g\colon D\to\A^i$. Let $A_t^{i,g} = (A_{t,\omega}^{i,g})_{\omega\in\Omega}$ be given by
\[ A_{t,\omega}^{i,g} = \begin{cases} \{a\in \A \mid p^i(a) = g(\omega) \}, &\quad\omega\in D, \\ \emptyset, &\quad \omega\notin D. \end{cases} \]
Let $c(A_{<t},i,g) = c(A_{<t},A_t^{i,g})$. Very similarly to what has been said in Subsection~\ref{1-SDF_AC.subs:APsdf_AC}, if this set is an element of $\ms C_t$, it models the choice to take, given an endogenous history in $A_{<t}$, the action $g(\omega)$ in the $i$-th action space factor in scenario $\omega\in D$ at time $t$.

Further assumptions are needed to construct a stochastic (pseudo-) extensive form out of all this. First, as we wish to interpret action indices as agents, simultaneous actions in different action space factors should be independent from each other. Furthermore, we require \emph{in fine} that scenariowise choices available at a move $x = \x(\omega)$, for some $(\x,\omega) \in \X\bullet\Omega$ can be extended a) to a representative class of elements of $\ms C^i_\x$ and b) to adapted choices of the form $c(A_{<t},i,g)$. In this sense, this requires $W$ to be measurable with respect to the Borel $\sigma$-algebra on $\A$. In addition, separation assumptions with respect to $W$ are proposed that makes it possible to separate outcomes via choices in the relevant way.

In the following, let us call a set $\mc M$ \emph{stable under non-trivial intersections} iff for all $A,B\in\mc M$ with $A\cap B\neq\emptyset$, we have $A\cap B\in\mc M$. This property is equivalent to saying that $\mc M \cup \{\emptyset\}$ is stable under intersections. For example, any partition has this property. 

\begin{itemize}[label=--]
    \item\hypertarget{2-SEF_G.Ass:AP.SEF0}{\textbf{Assumption AP.SEF0.}}~For all subsets $J\subseteq I$, all $t\in\T$, all $(f_j)_{j\in J} \in(\A^{\T})^J$ and $\omega\in \Omega$ with $(\omega,f_j)\in W$ for all $j\in J$ and $f_j|_{[0,t)_\T} = f_{j'}|_{[0,t)_\T}$ for all $j,j'\in J$, there is $f\in \A^\T$ such that 
    \[ \forall j\in J\colon \quad \Big[ p^j \circ f(t) = p^j \circ f_j(t), \quad (\omega,f) \in x_t(\omega,f_j)\Big]~. \]
    \item\hypertarget{2-SEF_G.Ass:AP.SEF1}{\textbf{Assumption AP.SEF1.}}~For all $i\in I$, $t\in\T$, $f\in\A^\T$ such that $D_{t,f}^i \neq \emptyset$ and the unique $A_{<t}\in\mc H^i_t$ with $f|_{[0,t)_\T}\in A_{<t}$, there is a generator $\ms G(\A^i)$ of $\ms B(\A^i)$ stable under non-trivial intersections such that for all $G\in\ms G(\A^i)$, upon letting $A_t^{i,G} = (A_{t,\omega}^{i,G})_{\omega\in\Omega}$ be given by $A_{t,\omega}^{i,G} = (p^i)^{-1}(G)$ for $\omega\in D_{t,f}$ and $A_{t,\omega}^{i,G} = \emptyset$ for $\omega\notin D_{t,f}$, we have
    \[ c(A_{<t},A_t^{i,G}) \in \ms C_{\x_t(f)}^i. \]
    \item\hypertarget{2-SEF_G.Ass:AP.SEF2}{\textbf{Assumption AP.SEF2.}}~For all $i\in I$, $t\in\T$, $f\in\A^\T$, $\omega\in D_{t,f}^i$ with $(\omega,f)\in W$, and the unique $A_{<t}\in\mc H^i_t$ satisfying $f|_{[0,t)_\T}\in A_{<t}$, there is a map $g\colon D_{t,f} \to \A^i$ such that $p^i\circ f(t) = g(\omega)$, $c(A_{<t},i,g)\in \ms C_t$, $c(A_{<t},i,g)$ is $\ms F^i$-$\ms C^i$-adapted, and for all $(\omega',{f'}_{<t})\in D_{t,f}\times A_{<t}$ there is $f'\in\A^\T$ satisfying $(\omega',f')\in c(A_{<t},i,g)$ and $f'|_{[0,t)_\T} = {f'}_{<t}$.
    \item\hypertarget{2-SEF_G.Ass:AP.psi-SEF3}{\textbf{Assumption AP.$\psi$-SEF3.}}~ For all $f,f'\in\A^\T$ and $t_0\in\T$ with $f(t_0)\neq f'(t_0)$ and $\omega\in\Omega$ such that $(\omega,f),(\omega,f')\in W$, there are $t\in\T$ and $i\in I$ such that $t\le t_0$, $p^i\circ f(t) \neq p^i \circ f'(t)$ and $\omega\in D_{t,f}^i \cap D_{t,f'}^i$.
    \item\hypertarget{2-SEF_G.Ass:AP.SEF3}{\textbf{Assumption AP.SEF3.}}~ For all $f,f'\in\A^\T$ with $f\neq f'$ and $\omega\in\Omega$ such that $(\omega,f),(\omega,f')\in W$, the set
    \[ \{ t\in\T \mid f(t) \neq f'(t)\}\]
    has a minimum.
\end{itemize}

Assumption \hyperlink{2-SEF_G.Ass:AP.SEF1}{AP.SEF1} is tightly linked to Assumption AP.C3 in Subsection~\ref{1-SDF_AC.subs:APsdf} (see Theorem~\ref{2-SEF_G.thm:adapted_choices_mb_functions} later in this section). Hence, if we assume \hyperlink{2-SEF_G.Ass:AP.SEF1}{AP.SEF1}, then by Theorem~\ref{1-SDF_AC.thm:APsdf_AC}, provided that $c = c(A_{<t},i,g)\in\ms C_t$, the $\ms F^i$-$\ms C^i$-adaptedness of the choice $c(A_{<t},i,g)$ is equivalent to the $\ms F_\x^i$-measurability of $g|_{D_\x}$ for all $\x\in\tilde\X^i$ that $c$ is available at. Note that Assumption \hyperlink{2-SEF_G.Ass:AP.SEF3}{AP.SEF3} is satisfied if all $f\in\A^\T$ with $(\omega,f)\in W$ for some $\omega\in\Omega$ are locally right-constant, with respect to the order topology on $\T$.\footnote{By definition, $f$ is locally right-constant iff for every non-maximal $t\in\T$ there is $u\in\T$ with $t<u$ such that $f|_{[t,u)_\T}$ is constant.} This is necessarily true if $(\T,\le)$ is a well-order.

It has been discussed in some detail in Subsection~\ref{1-SDF_AC.subs:APsdf} that these choices can model dependence on exogenous and endogenous information independently from each other. For instance, if $A_{<t} = \{f|_{[0,t)_\T}\}$ for some $f\in\A^\T$ such that $D_{t,f}^i\neq\emptyset$, for an action index $i\in I$, the prospective agent corresponding to $i$ who can choose $c(A_{<t},i,g)$ can make her choice dependent on whether the past actions are described by $f$ or not --- while $\ms F_{\x_t(f)}$ could well be very coarse allowing only a small amount of functions $g$, that is, a weak dependence on the exogenous scenario $\omega$. Similar remarks can be made in the opposite case and in mixed regimes. This means that ``open'' and ``closed loop'' decision making and control can be understood with respect to endogenous and exogenous information independently: in the example just mentioned, the loop could be closed with respect to endogenous information, but need not be with respect to exogenous information. See Subsection~\ref{1-SDF_AC.subs:APsdf} for more details, Proposition~\ref{2-SEF_G.thm:link_endogenous_information_H} below for the link between $\mc H$ and the endogenous information structure, and Section~\ref{1-SDF_AC.sec:exogenous_information} for more details regarding exogenous information structures.\smallskip

For each $i\in I$, let $C^i$ be the set of all sets of the form $c(A_{<t},i,g)$ where $t\in\T$, $A_{<t} \in \mc H^i_t$, $D\in\ms E$, $g\colon D\to \A^i$ such that $c(A_{<t},i,g)\in\ms C_t$, $c(A_{<t},i,g)$ is $\ms F^i$-$\ms C^i$-adapted, and for all $(\omega,f_{<t})\in D \times A_{<t}$ there is $f\in \A^\T$ with $(\omega,f)\in c(A_{<t},i,g)$ and $f|_{[0,t)_\T} = f_{<t}$. Let $C=(C^i)_{i\in I}$. This concludes the construction of the necessary data.

\subsection{Action path stochastic extensive form data}

In this subsection, we show that the data introduced just before yield a stochastic extensive form based on paths of action. Moreover, we show that the traditional models of choices --- given by measurable functions --- and of endogenous information --- effectively given by history structures --- can be embedded into and provide a vast class of examples for the stochastic extensive form. 
We start with fixing some names for the objects just constructed, before showing that they provide stochastic (pseudo-) extensive forms whose properties are directly related to corresponding properties of the data.

\begin{definition}\label{2-SEF_G.def:SEF_data}
    Let $(\Omega,\ms E)$ be an exogenous scenario space.
    \begin{enumerate}
        \item \emph{Action path stochastic pseudo-extensive form ($\psi$-\textsc{sef}) data} on $(\Omega,\ms E)$ are defined to be a tuple
        \[ (\ast) \qquad \D = \big (I,\A,\T,W,\ms F,\mc H\big) \]
        such that $(I,\A,\T,W)$ is action path \textsc{sdf} data on $(\Omega,\ms E)$, $\ms F = (\ms F^i)_{i\in I}$ is a family of exogenous information structures $\ms F^i$ on $\tilde\X^i$, $i\in I$, and $\mc H = (\mc H^i)_{i\in I}$ is a family of history structures $\mc H^i$ for $i\in I$, such that Assumptions \hyperlink{2-SEF_G.Ass:AP.SEF0}{AP.SEF$k$}, $k=0,1,2$, and \hyperlink{2-SEF_G.Ass:AP.psi-SEF3}{AP.$\psi$-SEF3} are satisfied.
        If $\D$ is action path $\psi$-\textsc{sef} data on $(\Omega,\ms E)$, then its entries are denoted as in $(\ast)$.
        \item \emph{Action path stochastic extensive form (\textsc{sef}) data} on $(\Omega,\ms E)$ are action path $\psi$-\textsc{sef} data satisfying Assumption \hyperlink{2-SEF_G.Ass:AP.SEF3}{AP.SEF3}.
        \item Suppose that $\D$ is action path $\psi$-\textsc{sef} data on $(\Omega,\ms E)$. Associate to it $\ms C = (\ms C^i)_{i\in I}$ and $C = (C^i)_{i\in I}$ in the way defined above. The tuple
            \[ \F = (F,\pi,\X,I,\ms F,\ms C,C) \]
        is said to be \emph{the \textsc{sef} candidate induced by $\D$}. The term ``\textsc{sef} candidate'' can be replaced with ``(pseudo-) stochastic extensive form'' if $\F$ satisfies the respective property.
    \end{enumerate}
\end{definition}

Before clarifying whether \textsc{sef} data induce \textsc{sef}, whether pseudo or not, respectively, we explain the meaning of the sets $D_{t,f}^i$. For this define, for all $t\in\T$, all sets of time $\T_t \subseteq \T$ with $[0,t)_\T\subseteq \T_t$, all $f\in\A^{\T_t}$, the set
\[ \hat D_{t,f} = \{ \omega\in\Omega \mid \exists f',f'' \in\A^\T\colon (\omega,f'), (\omega,f'')\in W,~ f'|_{[0,t)_\T} = f|_{[0,t)_\T} = f''|_{[0,t)_\T},~ f'(t) \neq f''(t)\}. \]
\begin{proposition}\label{2-SEF_G.prop:Dfti}
    Let $(\Omega,\ms E)$ be an exogenous scenario space, $\D$ be action path $\psi$-\textsc{sef} data on it and $\F$ be the induced action path $\psi$-\textsc{sef} candidate. Then, the following statements hold true.
    \begin{enumerate}
        \item\label{2-SEF_G.prop:Dfti.hatDft=unionDfti} For all $t\in\T$, sets of time $\T_t \subseteq \T$ with $[0,t)_\T\subseteq \T_t$,  $f\in\A^{\T_t}$, we have
    \[ \hat D_{t,f} = \bigcup_{i\in I} D_{t,f}^i. \]
        \item\label{2-SEF_G.prop:Dfti.hatDft_subseteq_Dft} For all $t\in\T$ and $f\in\A^\T$, we have $\hat D_{t,f} \subseteq D_{t,f}$.
        \item\label{2-SEF_G.prop:Dfti.Dfti_nonempty} For all $t\in\T$, $f\in\A^\T$, and $i\in I$, we have:
        \[ \Big[ D_{t,f} \neq \emptyset \text{ and } \x_t(f) \in \X^i \Big] \quad \Longleftrightarrow \quad D_{t,f}^i \neq \emptyset, \]
        and if either side of the equivalence holds true, then $D_{t,f} = D_{t,f}^i$.
    \end{enumerate}
\end{proposition}

\begin{corollary}\label{2-SEF_G.cor:tildeXi=Xi}
    Let $(\Omega,\ms E)$ be an exogenous scenario space, $\D$ be action path $\psi$-\textsc{sef} data on it and $\F$ be the induced action path $\psi$-\textsc{sef} candidate. Then, for all $i\in I$, $\tilde\X^i = \X^i$. \hfill \qed
\end{corollary}

Note, however, that it is easy to construct action path \textsc{sef} data with $\hat D_{t,f} \subsetneq D_{t,f}$ for some $(t,f)\in\T\times\A^\T$, see Example \ref{2-SEF_G.ex:SEF_with_Dtf_neq_hatDtf}. Hence, the induced \textsc{sef} candidate satisfies $\bigcup_{i\in I} \X^i \subsetneq \X$, that is, provided it defines an \textsc{sef} (which it indeed does, by the following theorem), there are random moves no agent is active at. 

After these preparations, we can now state the central result of this section.

\begin{thm}\label{2-SEF_G.thm:AP_sef}
    Let $(\Omega,\ms E)$ be an exogenous scenario space and consider action path stochastic (pseudo-) extensive form data $\D$ on it and let $\F$ be the \textsc{sef} candidate induced by $\D$ as in Definition~\ref{2-SEF_G.def:SEF_data}. Then $\F$ is a stochastic (pseudo-) extensive form.
\end{thm}

\begin{example}
    The reader is referred to Examples~\ref{1-SDF_AC.ex:APsdf}, \ref{1-SDF_AC.ex:APsdf_EIS}, and \ref{1-SDF_AC.thm:APsdf_AC} for concrete examples of stochastic decision forests, exogenous information structures, and adapted choices in the action path case. In view of Theorem~\ref{2-SEF_G.thm:AP_sef}, both the ``$W = \Omega\times\A^\T$'' example and the timing game example can be easily used to construct stochastic extensive forms if $\T$ is well-ordered. 

    Beyond that setting, not all of these examples give rise to stochastic extensive forms however. For example, building on the third \textsc{sdf} from the mentioned family of examples (the American up-and-out option exercise problem), will cause problems regarding Axiom~\ref{2-SEF_G.def:SEF.separation} alias \hyperlink{2-SEF_G.Ass:AP.SEF3}{{Assumption AP.SEF3.}}, the strong separation axiom, because paths need not be right-continuous. This is a light version of the example discussed in the introduction.\footnote{See Introduction, Stochastic extensive forms and games.} 
\end{example}

\begin{remark}
    This tension already indicates that going beyond a well-ordered setting may not be possible within the setting of (action path) stochastic extensive forms. Nevertheless, an approximation theory for game-theoretic language in a continuous-time setting can be based on a large class of action path stochastic extensive forms on well-ordered time grids. This is further studied in Section~\ref{2-SEF_G.sec:well-posedness_equilibrium} (especially, in Theorem~\ref{2-SEF_G.thm:AP_sef_well-posed}), and then, on a larger scale, in Chapter~\ref{chap:3-SPF_VECT}.
\end{remark}

A traditional model of choices in game-theoretic models based on action paths (or stochastic processes) identifies choices with random actions $g$ and information sets with collections $A_{<t}$ of past truncations of paths. In the following proposition, we explicitly compute the sets of available choices and information sets for action path $\psi$-\textsc{sef}, and see that it is compatible with the traditional model.
\begin{proposition}\label{2-SEF_G.prop:compute_Ai(x)_mfp}
    Let $(\Omega,\ms E)$ be an exogenous scenario space, $\D$ be action path $\psi$-\textsc{sef} data on it, $\F$ be the induced action path $\psi$-\textsc{sef}, and $i\in I$.
    \begin{enumerate}
        \item\label{2-SEF_G.prop:compute_Ai(x)_mfp.Ai(x)} For all $\x\in\X$ and $t\in\T$, $f\in\A^\T$ with $\x = \x_t(f)$, we have:
        \[ A^i(\x) = \{c\in C^i \mid \exists A_{<t}\in \mc H^i_t~\exists g\colon D_{t,f}\to \A^i\colon c = c(A_{<t},i,g), \, f|_{[0,t)_\T}\in A_{<t}\}. \]
        \item\label{2-SEF_G.prop:compute_Ai(x)_mfp.mfp} The function mapping any pair $(t,A_{<t})$, where $t\in\T$ and $A_{<t}\in\mc H^i_t$, to
        \[ \mf p = \{\x_t(f) \mid f\in\A^\T\colon f|_{[0,t)_\T} \in A_{<t} \} \]
        is well-defined, injective and has image $\mf P^i$.
    \end{enumerate}
\end{proposition}

In the next theorem, which can be seen as a direct consequence of Theorem~\ref{1-SDF_AC.thm:AP_sdf}, we clarify how the structural assumption of adaptedness of choices relates to the measure-theoretic concept of measurability of random actions. As the latter is used in traditional game-theoretic models based on action paths (or stochastic processes) in order to describe adaptedness to information, the theorem shows that the game forms of traditional models can be understood in the language of stochastic extensive forms. The point of this theorem is to explain the usual measurability assumption on random action in terms of the decision-theoretic concept of adapted choices.\footnote{Of course, this measurability requirement is not sufficient to imply $c\in C^i$ in general, because, in addition, the latter requires that for all $(\omega,f_{<t})\in D \times A_{<t}$ there is $f\in \A^\T$ with $(\omega,f)\in c(A_{<t},i,g)$ and $f|_{[0,t)_\T} = f_{<t}$. Without this requirement, first, $c$ could be available at $\x_t(f)$, but not at $\x_t(f')$, for some $f,f'\in\A^\T$ with $f|_{[0,t)_\T},f'|_{[0,t)_\T}\in A_{<t}$, and, second, $D$ could be too large.}
\begin{thm}\label{2-SEF_G.thm:adapted_choices_mb_functions}
    Let $(\Omega,\ms E)$ be an exogenous scenario space, $\D$ be action path $\psi$-\textsc{sef} data on it, $\F$ be the induced action path $\psi$-\textsc{sef}, and $i\in I$. Further, let $t\in\T$, $A_{<t}\in\mc H^i_t$, $D\in\ms E$ and $g\colon D\to \A^i$ be a map such that $c = c(A_{<t},i,g)\in\ms C_t$. 
    Then we have:
    \begin{enumerate}
        \item\label{2-SEF_G.thm:adapted_choices_mb_functions.non_red_compl} $c$ is a non-redundant and $\X^i$-complete choice.
        \item\label{2-SEF_G.thm:adapted_choices_mb_functions.Dx_subseteq_D} For all $\x\in\X^i$ that $c$ is available at, we have $D_\x \subseteq D$.
        \item\label{2-SEF_G.thm:adapted_choices_mb_functions.mb} $c$ is $\ms F^i$-$\ms C^i$-adapted iff for all $\x\in\X^i$ that $c$ is available at, $g|_{D_\x}$ is $\ms F^i_\x$-measurable.
    \end{enumerate}
\end{thm}

We conclude this subsection (and section) by classifying the endogenous information structure of an action path \textsc{sef} in terms of $\mc H$.
For $t,u\in\T$ with $t<u$ let $p_{u,t}$ be the restriction
\[ \A^{[0,u)_\T} \to \A^{[0,t)_\T}, ~ f\mapsto f|_{[0,t)_\T}. \]
More precisely, perfect endogenous recall and information can be formulated on the level of the history structure $\mc H$, that is, on the level of action paths. Hence, also on that level, action path-based models of perfect recall can be formulated using the language of stochastic extensive forms.

\begin{thm}\label{2-SEF_G.thm:link_endogenous_information_H}
    Let $(\Omega,\ms E)$ be an exogenous scenario space, $\D$ be action path $\psi$-\textsc{sef} data on it, $\F$ be the induced action path $\psi$-\textsc{sef}, and $i\in I$. Then the following statements hold true:
    \begin{enumerate}
        \item\label{2-SEF_G.thm:link_endogenous_information_H.perfect_recall} $i$ admits perfect endogenous recall iff for all $t,u\in\T$ with $t<u$, all $A_{<t}\in\mc H^i_t$, all $A_{<u}\in\mc H^i_u$, we have 
        \begin{itemize}[label=--]
            \item either: a) $(\mc P p_{u,t})(A_{<u}) \subseteq A_{<t}$ and b) all $f,f'\in A_{<u}$ satisfy $p^i \circ f(t) = p^i\circ f'(t)$,
            \item or: $(\mc P p_{u,t})(A_{<u}) \cap A_{<t} = \emptyset$.
        \end{itemize}
        \item\label{2-SEF_G.thm:link_endogenous_information_H.perfect_information} $i$ has perfect endogenous information iff for all $t\in\T$, all $A\in\mc H^i_t$ are singletons such that $A\cap A' = \emptyset$ for all $j\in I\setminus\{i\}$ and $A'\in \mc H^j_t$.
    \end{enumerate}
\end{thm}

That is, the finer $\mc H^i$ is, the more precise endogenous information agent $i$ has, and vice versa; and $i$ admits perfect recall iff, under the identification introduced by the projection operator $p$, $\mc H^i_t$ is essentially a flow of refined partitions in $t$ and $i$ recalls past decisions.

\section{Well-posedness and equilibrium}\label{2-SEF_G.sec:well-posedness_equilibrium}

A crucial property of extensive forms is what the author suggests calling ``well-posedness'', namely, that conditional on any history, any strategy profile induces a unique outcome and that any outcome can be attained this way. In this section, the concept of histories from \cite[p.\ 219]{AlosFerrer2008Trees} is reformulated within the setting of decision forests. We relate this to the new concept of random histories in the case of stochastic decision forests $(F,\pi,\X)$ which, in the order consistent, surely non-trivial, and maximal case, turn out as the histories of the associated decision tree $(\Tr,\ge_\Tr)$ and corresponds to the scenario-wise principal up-sets of random moves. Then, well-posedness is formulated following \cite{AlosFerrer2008Trees} and characterised in terms of well-posedness of the scenario-wise classical extensive forms. As a result, notions of preferences on strategy profiles and equilibrium concepts can be defined in well-posed stochastic extensive forms because consequences of strategy profiles always exist and are unique.

In the three final subsections, examples of well- and ill-posed stochastic extensive forms are discussed, first of rather pedagogic nature, second in the case of action path outcomes. Moreover, it is emphasised using examples why the nature representation of dynamic noise is not sufficient in general, an issue that the theory of games in stochastic extensive form can resolve.

\subsection{Histories}

Let us start with extending the definition of histories from \cite[p.\ 219]{AlosFerrer2008Trees} to decision forests, following \cite[Section~4.4]{AlosFerrer2016Theory}. For this, let us remind the reader that a subset $h\subseteq N$ of a poset $(N,\ge)$ is \emph{upward closed} iff for all $x\in h$, we have $\uparrow x \subseteq h$.

\begin{definition}[\cite{AlosFerrer2016Theory}]\label{2-SEF_G.def:history}
    Let $(F,\ge)$ be a decision forest. A \emph{history in $(F,\ge)$} is a non-empty, non-maximal and upward closed chain. The set of histories in $(F,\ge)$ is generically denoted by $H$.
\end{definition}

Actually, two histories may be equivalent in that they have the same sets of maximal chains containing them, respectively. This motivates the following lemma and definition.

\begin{lemma}\label{2-SEF_G.lemma:closed_history}
    Let $(F,\ge)$ be a decision forest and let $h\in H$ be a history.
    \begin{enumerate}
        \item\label{2-SEF_G.lemma:closed_history.1} There is a unique upward closed chain $\overline h$ in $(F,\ge)$ a) satisfying $\overline h \subseteq w$ for all maximal chains $w$ in $(F,\ge)$ with $h\subseteq w$ and b) that is maximal with respect to set inclusion among all chains in $(F,\ge)$ satisfying Property a).
        \item\label{2-SEF_G.lemma:closed_history.2} We have
        \[ \overline h = \bigcap \{ \uparrow x\mid x\in F\colon h\subseteq \uparrow x\}. \]
        \item\label{2-SEF_G.lemma:closed_history.3} For any further history $h'\in H$, we have $\overline {h'} = \overline h$ iff for any maximal chain $w$ in $(F,\ge)$ we have
        \[ h'\subseteq w \quad \Longleftrightarrow\quad h\subseteq w. \]
        \item\label{2-SEF_G.lemma:closed_history.4} We have $\overline h = h \cup\{\inf h\}$ if $h$ admits an infimum, and $\overline h = h$ otherwise.
    \end{enumerate}
\end{lemma}

\begin{definition}\label{2-SEF_G.def:closed_history}
    Let $(F,\ge)$ be a decision forest and let $h\in H$ be a history. $\overline h$ is said the \emph{closure of $h$}. $h$ is said to be \emph{closed} iff $h=\overline h$.
\end{definition}

From the perspective of outcome generation, it seems to suffice to describe historical dependence in terms of closed histories. This is confirmed by Lemma~\ref{2-SEF_G.lemma:R_continuous_in_h} in the following subsection.\smallskip

Within the realm of stochastic decision forests the exogenous information agents can condition on is described based on random moves rather than on moves. This is particularly true in the order consistent case which motivates the following concept of histories compatible with random moves.

\begin{definition}\label{2-SEF_G.def:random_history}
    Let $(F,\pi,\X)$ be an order consistent stochastic decision forest on an exogenous scenario space $(\Omega,\ms E)$. A \emph{(closed) random history in $(F,\pi,\X)$ }is a map $\h$ with domain $D_\h\in\ms E \setminus \{\emptyset\}$ such that $\h(\omega)$ is a (closed) history in $(T_\omega,\supseteq)$ for all $\omega\in D_\h$ and such that for all $\x\in\X$ admitting $\omega\in D_\x\cap D_\h$ with $\x(\omega)\in\h(\omega)$, we have $D_\x \supseteq D_\h$ and $\x(\omega') \in \h(\omega')$ for all $\omega'\in D_\h$. The set of random histories in $(F,\pi,\X)$ is generically denoted by $\H$.
\end{definition} 

\begin{lemma}\label{2-SEF_G.lemma:histories_canonical_inj}
    Let $(F,\pi,\X)$ be a stochastic decision forest on an exogenous scenario space $(\Omega,\ms E)$. Then, there is an injection $X\inj H$ associating to any move $x\in X$ the closed history $\uparrow x$. Moreover, if $(F,\pi,\X)$ is order consistent, there is an injection $\X\inj \H$ associating to any random move $\x\in\X$ the closed random history 
    \[ \h \colon D_\x \to H, \, \omega \to \uparrow \x(\omega). \]
\end{lemma}

Via these injections, we consider $X$ as a subset of $H$ and, if applicable, $\X$ as a subset of $\H$. Moreover, if $(F,\pi,\X)$ is surely non-trivial, then, via the natural injection $\Omega\to X,\,\omega\mapsto W_\omega$, $\Omega$ can be seen as a subset of $X$ and thus of $H$, too.\smallskip

The following proposition establishes that, in an order consistent \textsc{sdf} such that $\x_0 \colon \Omega\to F,\, \omega\mapsto W_\omega$ is a random move, random histories and histories in the induced decision tree compatible with at least one scenario naturally correspond to each other. 

\begin{proposition}\label{2-SEF_G.prop:random_histories}
    Let $(F,\pi,\X)$ be an order consistent stochastic decision forest on an exogenous scenario space $(\Omega,\ms E)$ such that $\x_0 \colon \Omega\to F,\, \omega\mapsto W_\omega$ is a random move and $(\Tr,\ge_\Tr)$ be the induced decision tree. Denote the set of histories in $(\Tr,\ge_\Tr)$ by $H_\Tr$. Let $f$ be the map with domain $\H$ associating to any $\h\in\H$ the set of $\x\in\X$ such that there is $\omega\in D_\x\cap D_\h$ with $\x(\omega)\in \h(\omega)$. Then, the following statements hold true.
    \begin{enumerate}
        \item\label{2-SEF_G.prop:random_histories.image_H_Tr} The image of $f$ is given by the set of a) all $h_\Tr\in H_\Tr$ that are non-maximal chains in $(\X,\ge_\X)$, and b) all maximal chains $h_\Tr$ in $(\X,\ge_\X)$ admitting non-empty $D\in\ms E\setminus\{\emptyset\}$ with $D\subseteq \bigcap_{\x\in h_\Tr} D_\x$ such that for any $\omega\in D$ there is $w\in W_\omega$ with $w\in\bigcap_{\x\in h_\Tr} \x(\omega)$. 
        \item\label{2-SEF_G.prop:random_histories.faithful} $f$ is faithful in that for all $\h_1,\h_2\in\H$ with $f(\h_1) = f(\h_2)$ there is $\h\in\H$ with $D_{\h_1}\cup D_{\h_2}\subseteq D_\h$ and $\h|_{D_{\h_k}} = \h_k$ for both $k=1,2$.
        \item\label{2-SEF_G.prop:random_histories.closed} For any closed random history $\h\in\H$, $f(\h)$ is a closed history in $(\Tr,\ge_\Tr)$. 
    \end{enumerate}
\end{proposition}

\subsection{Induced outcomes and well-posedness}

In extensive form theory, the outcome of strategic interaction is defined as the consequence of local decision making compatible with the rules defined by the tree and choice structures. More precisely, ``the'' outcome $w$ compatible with a strategy profile $s$, given a history $h$, is characterised by the fact that it is not discarded by any of the strategy profile's choices at moves $x$ succeeding the history $h$ and containing $w$. Yet, beyond the realm of standard finite or discrete-time situations, such an outcome need not exist, or there may be several of them, and there may even be outcomes that can never be attained that way (see, i.a.\ \cite{Simon1989Extensive,Stinchcombe1992Maximal,AlosFerrer2008Trees}). In a well-posed extensive form model of decision making, these three problems must not arise.

From a decision-theoretic point of view, this implies two things. First, it is relevant to formally define well-posedness of (stochastic) extensive forms. This is done in the following definition. As the theory exposed here is based on the refined partitions approach developed by Alós-Ferrer and Ritzberger (see, e.g.\ \cite{AlosFerrer2016Theory}), this definition here is necessarily structurally similar to the definitions in \cite[p.\ 228]{AlosFerrer2008Trees} and \cite[pp.\ 102, 105, 106]{AlosFerrer2016Theory}. For the sake of accessability, a notation similar to that used in these texts is used here as well. 

Second, additional assumptions are required to make an extensive form well-posed. This is discussed below the following definition.

\begin{definition}\label{2-SEF_G.def:sef_well-posed}
    Let $\F = (F,\pi,\X,I,\ms F,\ms C,C)$ be a stochastic pseudo-extensive form on an exogenous scenario space $(\Omega,\ms E)$. 

    \begin{enumerate}
        \item\label{2-SEF_G.def:sef_well-posed.induced_outcome} Let $R = R(\F)$ be the map assigning to all triples $w\in W$, $s\in S$, $h\in H$ with $w\in \bigcap h$ the set
        \[ R(w,s \mid h) = \bigcap \Big\{ s^i(x) \mid x \in X,\,i\in J(x)\colon~ w\in x \subseteq \bigcap h \Big\}. \]

        Given such data, $w$ is said \emph{compatible with $s$ given $h$} iff $ w \in R(s,w \mid h) $.
        \item\label{2-SEF_G.def:sef_well-posed.well_posed} $\F$ is called \emph{well-posed} iff 
        \begin{enumerate}
            \item\label{2-SEF_G.def:sef_well-posed.well_posed.onto_outcomes} for all $h\in H$ and all $w\in\bigcap h$, there is $s\in S$ such that $w$ is compatible with $s$ given $h$;
            \item\label{2-SEF_G.def:sef_well-posed.well_posed.existence} for all $s\in S$ and all $h\in H$, there is $w\in \bigcap h$ that is compatible with $s$ given $h$;
            \item\label{2-SEF_G.def:sef_well-posed.well_posed.uniqueness} for all $s\in S$ and all $h\in H$, there is at most one $w\in \bigcap h$ that is compatible with $s$ given $h$, and in this case, $R(w,s \mid h) = \{w\}$.
        \end{enumerate}

        \item\label{2-SEF_G.def:sef_well-posed.well_posed.outcome_map} Suppose that $\F$ is well-posed. Then, for any strategy profile $s\in S$ and history $h\in H$, the unique outcome $w\in \bigcap h$ compatible with $s$ given $h$ is said \emph{outcome induced by $s$ given $h$}, and denoted by $w = \Out(s\mid h)$. The corresponding map $\Out(.\mid .)\colon S\times H\to W$ is called \emph{(induced) outcome map}. 
    \end{enumerate}
\end{definition}

Note that $R(w,s \mid h)$ describes the set of outcomes that are compatible with both the outcome --- or decision path associated to --- $w$ and the strategy profile $s$ given the history $h$. By definition, $w$ is compatible with $s$ given $h$ iff it is compatible with itself and $s$ given $h$. A minimum requirement on the result of strategic interaction in terms of $s$ and given history $h$ is not to be discarded by $s$, given $h$, that is, to be compatible with $s$ given $h$.

Property \ref{2-SEF_G.def:sef_well-posed.well_posed.onto_outcomes} means that given any history $h$, any outcome yet undiscarded (by $h$) can be compatible with some strategy profile. If the other two existence and uniqueness properties are satisfied so that the outcome map can be defined at all, Property \ref{2-SEF_G.def:sef_well-posed.well_posed.onto_outcomes} is equivalent to that for any history $h\in H$ we have $\im\Out(. \mid h) = \bigcap h$. Property \ref{2-SEF_G.def:sef_well-posed.well_posed.existence} describes existence of compatible outcomes for all strategy profiles given any history. Property \ref{2-SEF_G.def:sef_well-posed.well_posed.uniqueness} describes uniqueness of compatible outcomes $w$ for all strategy profiles $s$ given any history $h$ and, moreover, of outcomes compatible with such triplets $(w,s,h)$.

Note that in the special case of classical extensive forms this corresponds to the definition of ``induced'' outcomes and conditional versions of conditions (A0), (A1), (A2) in \cite[Chapter~5, pp.\ 102, 105, 106]{AlosFerrer2016Theory} though the term ``well-posed'' is not used therein, to the best of the author's knowledge. As explained above, it is of utmost importance to understand when a stochastic extensive form is well-posed, and a main contribution of \cite{AlosFerrer2008Trees,AlosFerrer2011Comment}, comprehensively presented in \cite{AlosFerrer2016Theory}, is to characterise this in the case of classical extensive forms, namely in terms of order-theoretic properties of the underlying decision tree.

The next lemma confirms formally that, as claimed in the preceding subsection, it is sufficient to restrict to closed histories in what concerns outcome generation.

\begin{lemma}\label{2-SEF_G.lemma:R_continuous_in_h}
    Let $\F = (F,\pi,\X,I,\ms F,\ms C,C)$ be a stochastic pseudo-extensive form on an exogenous scenario space $(\Omega,\ms E)$. Furthermore, let $s\in S$ be a strategy profile, $h\in H$ be a history, and $w\in W=\bigcup F$ be an outcome. Then,
    \[ R(w,s \mid h) = R(w,s \mid \overline h).\]
\end{lemma}

Next, we discuss under which conditions an extensive form is well-posed. First, we consider the property whether any undiscarded outcome can be attained.
\begin{thm}\label{2-SEF_G.thm:well_posed.onto_outcomes}
    For any stochastic extensive form $\F$, Property \ref{2-SEF_G.def:sef_well-posed}.\ref{2-SEF_G.def:sef_well-posed.well_posed.onto_outcomes} is satisfied.
\end{thm}
In case of classical extensive forms and the unconditional version of Property \ref{2-SEF_G.def:sef_well-posed}.\ref{2-SEF_G.def:sef_well-posed.well_posed.onto_outcomes}, this basic fact has already been established in \cite[Theorem~5.1]{AlosFerrer2016Theory}.

Furthermore, we classify existence and uniqueness. For this, the classification from the classical case in \cite{AlosFerrer2008Trees,AlosFerrer2011Comment} and \cite{AlosFerrer2016Theory} is applied. Therefore, it remains to describe existence and uniqueness in terms of adequate existence and uniqueness properties for suitable classical extensive forms. In fact, these classical pseudo-extensive forms are given by the scenario-wise truncations of the given stochastic pseudo-extensive form:

\begin{proposition}\label{2-SEF_G.prop:scwise_SEF}
    Let $(F,\pi,\X,I,\ms F,\ms C,C)$ be a stochastic (pseudo-) extensive form on some exogenous scenario space $(\Omega,\ms E)$ and $\omega\in\Omega$. Then, $(T_\omega,I,C_\omega)$ is a classical (pseudo-) extensive form, respectively, where $C_\omega = (C_\omega^i)_{i\in I}$ and
    $$ C_\omega^i = \{ c\cap W_\omega \mid c\in C^i\} \setminus \{\emptyset\}, \qquad i\in I. $$
\end{proposition}

We turn to the central theorem of this subsection translating well-posedness properties of stochastic pseudo-extensive forms into the language of classical extensive forms.

\begin{thm}\label{2-SEF_G.thm:SEF_well-posed}
    Let $\F = (F,\pi,\X,I,\ms F,\ms C,C)$ be a stochastic pseudo-extensive form on an exogenous scenario space $(\Omega,\ms E)$. Then $\F$ satisfies Properties \ref{2-SEF_G.def:sef_well-posed}.\ref{2-SEF_G.def:sef_well-posed.well_posed.onto_outcomes}, \ref{2-SEF_G.def:sef_well-posed}.\ref{2-SEF_G.def:sef_well-posed.well_posed.existence}, \ref{2-SEF_G.def:sef_well-posed}.\ref{2-SEF_G.def:sef_well-posed.well_posed.uniqueness}, well-posedness, iff for all $\omega\in\Omega$, the classical pseudo-extensive form $(T_\omega,I,C_\omega)$ does so,\footnote{Precisely, some and any stochastic pseudo-extensive form with set of nodes $T_\omega$, set of agents $I$, and family of sets of individual choices $C_\omega = (C^i_\omega)_{i\in I}$ \emph{does so}. Clearly, these properties do only depend on $T_\omega$, $I$, and $C_\omega$ in the singleton-$\Omega$ case.} respectively.
\end{thm}

This theorem makes it possible to apply the classification results from \cite{AlosFerrer2008Trees,AlosFerrer2011Comment}, see \cite{AlosFerrer2016Theory} for a comprehensive monographic treatment. For this, we recall four important notions from these works (see, e.g.\ \cite[Definitions~4.2, 5.1]{AlosFerrer2016Theory}).\footnote{Without loss of generality, we reformulate them for decision forests.} Let $(F,\ge)$ be a decision forest. For any history $h$ in $(F,\ge)$, a \emph{continuation} is the complement $w\setminus h$ of $h$ in a maximal chain alias decision path $w$ containing $h$. $(F,\ge)$ is said
\begin{enumerate}
    \item \emph{weakly up-discrete} iff for all non-terminal nodes $x\in F$ any maximal chain in $\downarrow x \setminus \{x\}$ has a maximum;
    \item \emph{up-discrete} iff any non-empty chain has a maximum;
    \item \emph{coherent} iff every history without minimum has at least one continuation with a maximum;
    \item \emph{regular} iff for all non-maximal $x\in F$, the history $\uparrow x \setminus \{x\}$ has an infimum.
\end{enumerate}
See \cite{AlosFerrer2008Trees,AlosFerrer2011Comment} and \cite{AlosFerrer2016Theory} for a discussion and examples. We only give a brief overview here. Weak up-discreteness requires the existence of successor nodes, while up-discreteness even demands all chains to be well-ordered with respect to $\ge$.\footnote{This might be confusing because typically well-orders are considered with respect to $\le$, rather than $\ge$.} The latter is equivalent to the existence of a maximum for every continuation of every history (cf.~\cite[Lemma 5.4]{AlosFerrer2016Theory}). Roughly speaking, coherence ensures the existence of (inductive) limit candidates and regularity requires such a limit to be uniquely identifiable. Weak up-discreteness, coherence, and regularity are mutually independent of each other. See \cite{AlosFerrer2008Trees} and \cite{AlosFerrer2016Theory} for more details.

We obtain the following two corollaries:

\begin{corollary}\label{2-SEF_G.cor:SEF_existence=~order-theoretic_properties}
    Let $F$ be a decision forest over a set $W$ such that for any $x\in X$ and $w\in x$, we have
    \begin{equation}\label{2-SEF_G.eq:available_choices}
        x \supsetneq \bigcup \Big\{ y\in \downarrow x \setminus \{x\} \bigmid w\in y\Big\}. 
    \end{equation}
    Then, the following statements are equivalent:
    \begin{enumerate}
        \item\label{2-SEF_G.cor:SEF_existence=~order-theoretic_properties.order_properties.existence} Every stochastic pseudo-extensive form $\F$ with decision forest $F$ satisfies Property \ref{2-SEF_G.def:sef_well-posed}.\ref{2-SEF_G.def:sef_well-posed.well_posed.existence}.
        \item\label{2-SEF_G.cor:SEF_existence=~order-theoretic_properties.order_properties} $(F,\supseteq)$ is weakly up-discrete and coherent.
    \end{enumerate}
\end{corollary}

In the context of classical extensive forms in \cite{AlosFerrer2008Trees,AlosFerrer2016Theory}, the fact that for all $x\in X$ and $w\in x$ Equation~\ref{2-SEF_G.eq:available_choices} is described by the sentence that $F$ has \emph{available choices}. It characterises the mere possibility of defining a stochastic pseudo-extensive form on it. See \cite[Subsection~4.5]{AlosFerrer2016Theory} for more details. By \cite[Corollary 1]{AlosFerrer2008Trees} (or identically, \cite[Corollary 4.1]{AlosFerrer2016Theory}), weak up-discreteness of $(F,\supseteq)$ implies that property. In particular, the implication ``\ref{2-SEF_G.cor:SEF_existence=~order-theoretic_properties.order_properties} $\Rightarrow$ \ref{2-SEF_G.cor:SEF_existence=~order-theoretic_properties.order_properties.existence}'' holds true without additionally requiring available choices.

\begin{corollary}\label{2-SEF_G.cor:SEF_well-posedness=~order-theoretic_properties}
    Let $\F$ be a stochastic extensive form on an exogenous scenario space $(\Omega,\ms E)$. Then, the following statements are equivalent:
    \begin{enumerate}
        \item\label{2-SEF_G.cor:SEF_well-posedness=~order-theoretic_properties.well-posed} $\F$ is well-posed.
        \item\label{2-SEF_G.cor:SEF_well-posedness=~order-theoretic_properties.order_properties} $(F,\supseteq)$ is weakly up-discrete, coherent, and regular.
        \item\label{2-SEF_G.cor:SEF_well-posedness=~order-theoretic_properties.order_properties2} $(F,\supseteq)$ is up-discrete and regular.
    \end{enumerate}
\end{corollary}

Hence, the consistency requirements of stochastic extensive forms suffice to fully characterise well-posedness in terms of easily verifiable order-theoretic properties of the underlying decision forest $F$. One the hand, this allows to obtain a vast class of well-posed stochastic extensive forms; on the other, we can clearly delimit the decision-theoretic boundaries of stochastic extensive form theory.

First, recall that any outcome naturally corresponds to a maximal chain of nodes alias decision path, by definition of decision forests (Definition~\ref{1-SDF_AC.def:decision_forest}). By the preceding corollary, in any well-posed stochastic extensive form, any decision path is well-ordered. That is, if thinking an outcome as a path of actions indexed over some fixed time half-axis $\T$, this path must be locally right-constant.\footnote{A map $f\colon\T\to S$ from a total order $\T$ into a set $S$ is locally right-constant iff for any non-maximal $t\in\T$ there is $u>t$ such that $f|_{[t,u)_\T}$ is constant.}

As shown in \cite{AlosFerrer2015Repeated}, conversely, well-posed classical extensive forms can be constructed whose ``main'' outcomes are given by locally right-constant paths by adding inertia moves after any original ``action'' move, where agents must choose a strictly positive time lag of inaction. In view of Theorem~\ref{2-SEF_G.thm:SEF_well-posed}, it is not daring to suppose that this can be fully generalised to the setting of stochastic extensive forms.\footnote{Details on this topic are omitted here in the interest of brevity. The author would be happy to provide further details on request.} However, the framework in \cite{AlosFerrer2015Repeated} is not compatible with the action path extensive form theory of the present text, because \hyperlink{1-SDF_AC.Ass:AP.SDF1}{{Assumption~AP.SDF1.}} is violated: Once all the finitely many players have chosen their inertia lag, a positive amount of time elapses without reducing the set of possible outcomes. In that sense, it remains questionable whether and, if so, in what precise meaning the framework in \cite{AlosFerrer2015Repeated} describes strategic interaction in \emph{continuous time}. In any case, it is not an action path extensive form with time half-axis $\R_+$.

If we consider any action path stochastic extensive form, then well-posedness is equivalent to the well-order property of the relevant part of the time half-axis, as shows the following theorem.

\begin{thm}\label{2-SEF_G.thm:AP_sef_well-posed}
    Let $\F$ be the stochastic extensive form induced by action path \textsc{sef} data with time $\T$. If $\T$ is well-ordered, then $\F$ is well-posed. Conversely, if $\F$ is well-posed and for any $t\in\T$, there is $f\in\A^\T$ with $D_{t,f}\neq\emptyset$, then $\T$ is well-ordered.
\end{thm}

We conclude from our discussion that --- in the sense of action path stochastic extensive forms --- continuous-time decision making, whether in a ``stochastic'' or ``deterministic'' environment, is confined to well-ordered time grids. Therefore, describing extensive form characteristics of games and decision problems in continuous time in a well-posed way requires a notion of approximating general continuous-time outcomes by those arising on well-ordered time grids. Based on the theory of well-posed action path stochastic extensive forms, we can generate such approximators, in a general probabilistic setting. The approximand and an adequate notion of approximation is constructed and studied in Chapter~\ref{chap:3-SPF_VECT}. This is tightly related to the important question of the decision-theoretic meaning continuous-time games under probabilistic uncertainty, including ``stochastic differential games'' and preemption games.
\smallskip

While in classical extensive forms subgames are defined in terms of moves, the obvious analogon in stochastic extensive form is given by random moves, that is, the sections of moves that, potentially, exogenous information is revealed and choices are available at. 
From this perspective, it is important to understand the link between random moves and histories, similarly to the classical theory (as discussed also in, e.g.\ \cite[p.\ 106]{AlosFerrer2016Theory}). Moreover, in the order consistent case, the link between random moves and random histories is interesting since the latter correspond to the relevant histories in the induced decision tree $(\Tr,\ge_\Tr)$, by Proposition~\ref{2-SEF_G.prop:random_histories}. 

\begin{proposition}\label{2-SEF_G.prop:random_histories=~random_moves}
    Let $\F = (F,\pi,\X,I,\ms F,\ms C,C)$ be a well-posed stochastic extensive form on an exogenous scenario space $(\Omega,\ms E)$. Then for any closed history $h\in H$ there is $x\in X$ with $h = \uparrow x$. Moreover, if $(F,\pi,\X)$ is order consistent, then for any closed random history $\h$ there is $\x\in\X$ such that $D_\h \subseteq D_\x$ and for all $\omega\in D_\x$, $\h(\omega) = \uparrow \x(\omega)$.
\end{proposition}

We see that in well-posed stochastic extensive forms, closed histories always have minima. From the perspective of outcome generation, it suffices to consider closed histories, by Lemma \ref{2-SEF_G.lemma:R_continuous_in_h}. Hence, sections of closed histories whose minima constitute random moves are the crucial object for representing subgames. In the order consistent case, moreover, this corresponds exactly to closed random histories.

\subsection{Expected utility preferences}\label{2-SEF_G.subs:EU_preferences}

In a well-posed stochastic extensive form $\F$, dynamic decision making can be analysed by comparing consequences of strategy profiles, given any random move $\x$. By Proposition~\ref{2-SEF_G.prop:random_histories=~random_moves} and Lemma~\ref{2-SEF_G.lemma:R_continuous_in_h}, this is equivalent to conditioning on collections of closed histories whose infima constitute random moves, and in the order consistent case, this is equivalent to conditioning on closed random histories. Actually, an agent can only condition on its information set. Hence, agent $i\in I$ is ultimately interested in the maps
\[ \Out(s\mid .)\colon \mf p\times D_\mf p \to W,\, (\x,\omega) \mapsto \Out(s\mid \x(\omega)), \]
ranging over all strategy profiles $s\in S$ and all endogenous information sets $\mf p\in\mf P^i$.

In the following, a general concept of analysis is presented. We begin with a concept of comparing consequences. This concept is only an example, though a quite central and general one, and not generic. It implements the idea of expected utility in its basic form (which itself exhibits strong links to preference-based rational choice under uncertainty, \cite{Savage1972Foundations,Anscombe1963Definition,Gilboa1989Maxmin, Hara2023Multiple}). On it, as is discussed later, a theoretical equilibrium concept can be based, that covers (subgame-) perfect versions of Nash, correlated and Bayesian equilibrium. For this, dynamic consistency of ``comparison'' is central which is why this is discussed in some detail in the following. 

\begin{definition}\label{2-SEF_G.def:EU_pref_str}
    Let $\F$ be a well-posed stochastic extensive form on an exogenous scenario space $(\Omega,\ms E)$.
    \begin{enumerate}
        \item A \emph{belief system on $\F$} is a family $\Pi = (p_{i,\mf p},\ms P_{i,\mf p},\P_{i,\mf p})_{i\in I,\,\mf p\in\mf P^i}$ such that, for any $i\in I$ and any $\mf p\in\mf P^i$, $\ms P_{i,\mf p}$ is a $\sigma$-algebra on $\mf p$, $\P_{i,\mf p}$ is a probability measure on $\ms E|_{D_\mf p}$, and $p_{i,\mf p} \colon D_{\mf p} \to \mf p$ is an $\ms E|_{D_\mf p}$-$\ms P_{i,\mf p}$-measurable map.
        \item A \emph{taste system on $\F$} is a family $U = (u_{i,\mf p})_{i\in I,\,\mf p\in\mf P^i}$ of maps $u_{i,\mf p} \colon W \to \R$.
        \item An \emph{expected utility (\textsc{eu}) preference structure on $\F$} is a tuple $\Pr = (\Pi,U,\ms W)$ where
        \begin{itemize}[label=--]
            \item $\Pi$ is a belief system on $\F$,
            \item $U$ is a taste system on $\F$, and
            \item $\ms W$ is a $\sigma$-algebra on $W$,
        \end{itemize}
        such that, 
        we have, for all $i\in I$ and $\mf p\in\mf P^i$:
        \begin{enumerate}
            \item\label{2-SEF_G.def:EU_pref_str.uip_mb} $u_{i,\mf p}$ is $\ms W$-Borel-measurable;
            \item\label{2-SEF_G.def:EU_pref_str.Out_mb} $\Out_{i,\mf p}^s\colon D_\mf p \to W,\, \omega \mapsto \Out(s \mid p_{i,\mf p}(\omega),\omega)$ is $ \ms E|_{D_\mf p}$-$\ms W$-measurable for all $s\in S$;
            \item\label{2-SEF_G.def:EU_pref_str.uip_Out_intb} $u_{i,\mf p} \circ \Out_{i,\mf p}^s$ is Lebesgue-quasi-integrable\footnote{That is, the negative part or the positive part is Lebesgue-integrable.} with respect to $\P_{i,\mf p}$ for all $s\in S$;
            \item\label{2-SEF_G.def:EU_pref_str.W_rich_enough} the map $\psi_\mf p\colon W\to \mf p \cup \{\emptyset\}$ assigning to any $w\in W$ the unique\footnote{If two immediate predecessors in $X$ of a choice can be compared, then they are equal, see Lemma~\ref{2-SEF_G.lemma:Heraclitus_property}. As the evaluation on $\X^i \bullet \Omega$ is injective, $\x$ must be unique.} random move $\x\in \mf p$ such that $w\in \bigcup \im\x$ if it exists and $\emptyset$ else, is measurable with respect to $\ms W$ and the $\sigma$-algebra on $\mf p \cup \{\emptyset\}$ generated by $\ms P_{i,\mf p} $.
        \end{enumerate}
    \end{enumerate}
\end{definition}
In other words, an \textsc{eu} preference structure fixes beliefs (probability measures and random draws $p_{i,\mf p}$ on endogenous information sets $\mf p$, inducing probabilities $(p_{i,\mf p})_\ast \P_{i,\mf p}$ on $\mf p$), tastes (utility functions, or payoffs), and a measurability structure on outcomes, such that tastes and outcome generation is measurable (\ref{2-SEF_G.def:EU_pref_str.uip_mb}, \ref{2-SEF_G.def:EU_pref_str.Out_mb}), expected utility can be computed (\ref{2-SEF_G.def:EU_pref_str.uip_Out_intb}, which is always satisfied if $u_{i,\mf p}$ is bounded below), and the sets of outcomes in $W$ describing endogenous information sets are measurable (\ref{2-SEF_G.def:EU_pref_str.W_rich_enough}). Moreover, these data are given conditional on all agents and all of their endogenous information sets because beginning at any of these, a well-posed stochastic extensive form is induced, defining a decision situation in its own right.\footnote{We omit the formal argument behind this statement, in the interest of brevity.} In the perfect information case, these exactly correspond to classical subgames. An analysis of the original stochastic extensive form might necessitate an analysis of these induced \textsc{sef}, in order to rule out non-credible threats, for instance (which motivated the concept of subgame-perfect equilibrium in \cite{Selten1965Spieltheoretische}).

Once fixed, an \textsc{eu} preference structure is considered common knowledge among all agents. For a rational agent whose rationality is common knowledge, then, it seems plausible that beliefs on two different information sets are consistent with respect to the outcome generation map induced by the given strategy profile. This is a version of the Harsanyi doctrine (cf.~\cite{Harsanyi1967Games, Harsanyi1968Games, Harsanyi1968Gamesa, Aumann1974Subjectivity}), according to which differences in posterior beliefs should only result from differences in information. We suggest using a dynamic variant asking for common priors ``locally'' between any two given information sets, therefore taking into account consistency off the equilibrium path.\footnote{Here, we explicitly refrain from building an epistemic theory of types and higher-order beliefs, of rationality and equilibrium, following \cite{Mertens1985Formulation, Aumann1995Epistemic}; this is beyond the scope of the present text.} 

One can also argue that for rational agents whose rationality is common knowledge, tastes are necessarily consistent for any individual agent across endogenous information sets. Indeed, consider an agent $i$ performing equilibrium analysis. Why should $i$ make contingent plans for action at two different, but possibly related endogenous information sets at which tastes differ meaning that the consequences of action are evaluated differently? Put equivalently, supposing a specific form of rationality might lead us to split dynamically inconsistent ``agents'' into ``multiple selves'' (see \cite{Gilboa1997Comment,Strotz1955Myopia}). These notions of consistency are formalised as follows.

\begin{definition}\label{2-SEF_G.def:consistent_EU_pref_str}
    Let $\F$ be a well-posed stochastic extensive form on an exogenous scenario space $(\Omega,\ms E)$, $\Pi$ be a belief system and $U$ be a taste system on $\F$. Moreover, let $\ms W$ be a $\sigma$-algebra on $W$.
    \begin{enumerate}
        \item Let $\Pi$ and $\ms W$ satisfy Assumptions~\ref{2-SEF_G.def:EU_pref_str.Out_mb} and~\ref{2-SEF_G.def:EU_pref_str.W_rich_enough} in Definition~\ref{2-SEF_G.def:EU_pref_str}, and let $s\in S$ be a strategy profile. The pair $(\Pi,s)$ is said \emph{dynamically consistent} iff for all sets $J \subseteq \bigcup_{i\in I} \{i\} \times \mf P^i$ with at most two elements,
        \begin{enumerate}
            \item\label{2-SEF_G.def:consistent_EU_pref_str.beliefs.p} for all $(i,\mf p_i),(j,\mf p_j)\in J$, all $\omega\in D_{\mf p_i}$ with
            \[ (\ast) \qquad \psi_{j,\mf p_j} \circ \Out^s_{i,\mf p_i}(\omega) \neq\emptyset, ~\text{ and }~ p_{i,\mf p_i}(\omega)(\omega) \supseteq \big[\psi_{j,\mf p_j} \circ \Out^s_{i,\mf p_i}(\omega)\big](\omega) \]
            satisfy
            \[ p_{j,\mf p_j}(\omega) = \psi_{j,\mf p_j} \circ \Out^s_{i,\mf p_i}(\omega); \]
            \item for all $(i,\mf p_i),(j,\mf p_j)\in J$ the set 
            \[ E_{i,j} = \{ \omega\in D_{\mf p_i} \cap D_{\mf p_j} \mid p_{i,\mf p_i}(\omega)(\omega) \supseteq  p_{j,\mf p_j}(\omega)(\omega) \} \]
            is an event, i.e.\ satisfies $E_{i,j} \in \ms E$;
            \item\label{2-SEF_G.def:consistent_EU_pref_str.beliefs.common_prior} there is a \emph{common prior for $J$}, that is, a probability measure $\P$ on $(\Omega,\ms E)$ such that for all $((i,\mf p_i),(j,\mf p_j))\in J^2$ with $(i,\mf p_i)\neq(j,\mf p_j)$ we have $\P(D_{\mf p_i}\cup D_{\mf p_j}) = 1$, and the events
            \[  E_{\neg i,j} = D_{\mf p_j} \setminus E_{i,j}, \quad E_{i,j}^s = \{ \omega \in E_{i,j} \mid \psi_{j,\mf p_j} \circ \Out^s_{i,\mf p_i}(\omega) \neq \emptyset \} \]
            and any $E\in \ms E|_{D_{\mf p_j}}$ satisfy
            \[ \P_{j,\mf p_j}(E) \cdot \P(E_{\neg i,j} \cup E_{i,j}^s) = \P(( E_{\neg i,j} \cup E_{i,j}^s)\cap E). \]
        \end{enumerate}
        \item $U$ is said \emph{dynamically consistent} iff for all $i\in I$, there is a map $u_i\colon W \to \R$ such that for any $\mf p\in\mf P^i$, we have $u_{i,\mf p} = u_i$.
        \item Provided $\Pr = (\Pi,U,\ms W)$ is an \textsc{eu} preference structure, and given a strategy profile $s\in S$, $(\Pr,s)$ it is said \emph{dynamically consistent} iff both $(\Pi,s)$ and $U$ are dynamically consistent.
    \end{enumerate}
\end{definition}

Some of the technical details of the definition may require some verbal explanation. Given $i, j\in I$, $\mf p_i\in\mf P^{i}$, $\mf p_j\in\mf P^j$ as above, then $\psi_{j,\mf p_j}\circ \Out^s_{i,\mf p_i}(\omega)$ equals a) the random move in $\mf p_j$ reached by strategy profile $s$ when starting in information set $\mf p_i$ in scenario $\omega$ and random move $p_{i,\mf p_i}(\omega)$, if it exists (which is the case precisely iff $(\ast)$ holds true); b) the unique $\x\in\X^{j}$ such that $\x(\omega)$ strictly precedes $p_{i,\mf p_i}(\omega)(\omega)$ in case it exists (which is the case of the left-hand side in $(\ast)$ being true, but not the right-hand side); c) the empty set meaning that neither of the preceding two things exist, that is, in scenario $\omega$, $\mf p_j$ does not precede $\mf p_i$ nor is $\mf p_j$ reached through $s$ starting from $\mf p_i$, given the belief of being at move $p_{i,\mf p_i}(\omega)$ in $\mf p_i$. 

Moreover, $E_{i,j}$ is the set of scenarios at that information set $\mf p_i$ precedes information set $\mf p_j$, according to the ``random move beliefs'' $p_{i,\mf p_i}$ and $p_{j,\mf p_j}$; $E_{\neg i,j}$ is the set of scenarios in $D_{\mf p_j}$ at that $\mf p_i$ does not precede $\mf p_j$ in that sense, and thus, where probabilities cannot be explained by outcome generation given $\mf p_i$; and $E_{i,j}^s$ is the set of scenarios at that $\mf p_j$ is actually reached via $s$ given $\mf p_i$, according to $p_{i,\mf p_i}$. Note that, if Property \ref{2-SEF_G.def:consistent_EU_pref_str.beliefs.p} is satisfied, then, by the Heraclitus Property in Lemma \ref{2-SEF_G.lemma:Heraclitus_property},
\[ E_{i,j}^s = \{\omega\in D_{\mf p_i} \cap D_{\mf p_j} \mid \exists \x\in\mf p_j\colon\, p_{i,\mf p_i}(\omega)(\omega) \supseteq \x(\omega), ~ \psi_{j,\mf p_j} \circ \Out^s_{i,\mf p_i}(\omega) \neq \emptyset\}. \]

Thus, dynamic consistency of $(\Pi,s)$ requires that for all two information sets of some agents, which can but need not be identical, first, in all scenarios $\omega$ where the second is reached via $s$ starting from the first and given the belief about actual position within the first, the belief at the second evaluated in $\omega$ equals exactly the attained random move; second, the set $E_{i,j}$ of scenarios at that information set $\mf p_i$, according to the ``belief'' $p_{i,\mf p_i}$, precedes information set $\mf p_j$ in any way, is an event; third, there is a common prior $\P$ on $(\Omega,\ms E)$ such that for both distinct pairs $((i,\mf p_i),(j,\mf p_j))$ the posterior $\P_{j,\mf p_j}$ is equal to the conditional probability given the event that $\mf p_i$ does not precede $\mf p_j$, or it does and $s$ leads from $\mf p_i$ to $\mf p_j$, always according to the belief $p_{i,\mf p_i}$ on the realised random move in $\mf p_i$. Note that this can be a trivially void statement for one of the two pairs, but not for both because
\[ E_{\neg i,j} \cup E_{i,j}^s \cup E_{\neg j,i} \cup E_{j,i}^s = D_{\mf p_i} \cup D_{\mf p_j},\]
and $\P$ is concentrated on $D_{\mf p_i} \cup D_{\mf p_j}$. However, Part \ref{2-SEF_G.def:consistent_EU_pref_str.beliefs.common_prior} can be a void statement in some situations, for instance, if
\[ \P_{i,\mf p_i}(E_{i,j}) = 1, \qquad \P_{i,\mf p_i}(E_{i,j}^s) = 0. \]
In that case, $\P = \iota_\ast \P_{i,\mf p_i}$, where $\iota$ is the inclusion map $D_{\mf p_i} \inj \Omega$, does the job.
This corresponds to the case where, essentially, at $\mf p_i$, $\mf p_i$ is believed to precede $\mf p_j$ although the outcome of $s$ is believed to almost never reach $\mf p_j$ out of $\mf p_i$. Then, the relation between the beliefs $\P_{i,\mf p_i}$ and $\P_{j,\mf p_j}$ is unaffected by condition \ref{2-SEF_G.def:consistent_EU_pref_str.beliefs.common_prior} applied to $\{(i,\mf p_i),(j,\mf p_j)\}$.

The definition above is very general, and it simplifies in specific situations. For instance, in many situations endogenous information sets can be ordered in the weak sense that for all $i,j\in I$ and $\mf p_i\in \mf P^i$ and $\mf p_j\in\mf P^j$ admitting $\omega\in D_{\mf p_i} \cap D_{\mf p_j}$, $\hat\x_i\in\mf p_i$ and $\hat\x_j\in\mf p_j$ such that $\hat\x_i(\omega) \supseteq \hat\x_j(\omega)$, then $D_{\mf p_i}\supseteq D_{\mf p_j}$ and for all $\omega\in D_{\mf p_j}$ and $\x_j\in\mf p_j$, there is $\x_i\in\mf p_i$ with $\x_i(\omega) \supseteq \x_j(\omega)$. In that case, for distinct $(i,\mf p_i),(j,\mf p_j)$ as in the hypothesis of \ref{2-SEF_G.def:consistent_EU_pref_str.beliefs.common_prior}, we have for $(i,\mf p_i)$ preceding $(j,\mf p_j)$ as in the preceding sentence that $E_{\neg j,i} \cup E_{j,i}^s = D_{\mf p_i}$. Hence, for any $E\in\ms E|_{D_{\mf p_i}}$, 
\[ \P(E) = \P_{i,\mf p_i}(E) \cdot \P(D_{\mf p_i}) = \P_{i,\mf p_i}(E). \]
Thus, condition \ref{2-SEF_G.def:consistent_EU_pref_str.beliefs.common_prior} reduces to the unidirectional Bayesian updating rule given by
\[ \P_{j,\mf p_j}(E) \cdot \P_{i,\mf p_i}(E_{\neg i,j} \cup E_{i,j}^s) = \P_{i,\mf p_i}(( E_{\neg i,j} \cup E_{i,j}^s)\cap E), \qquad E\in\ms E|_{D_{\mf p_j}}. \]
If, moreover, perfect endogenous information is given, and $\mf p_i = \{\x_i\}$, $\mf p_j = \{\x_j\}$ with $\x_i >_\X \x_j$, we have $E_{i,j} = D_{\x_j}$, $E_{\neg i,j} = \emptyset$, so that condition \ref{2-SEF_G.def:consistent_EU_pref_str.beliefs.common_prior} reduces to the statement:
\[ \P_{j,\mf p_j}(E) \cdot \P_{i,\mf p_i}(\psi_{j,\mf p_j} \circ \Out^s_{i,\mf p_i}\neq \emptyset) = \P_{i,\mf p_i}(\{\psi_{j,\mf p_j} \circ \Out^s_{i,\mf p_i}(\omega) \neq \emptyset\}\cap E), \qquad E\in\ms E|_{D_{\mf p_j}}. \]

From this discussion it becomes obvious that dynamic consistency of belief systems is not a very strong notion if endogenous information sets are ``small'' relative to the amount of random moves eligible for forming an endogenous information set, provided the exogenous scenario space is large and belief probability measures are non-atomic, because then $\P_{i,\mf p_i}(\psi_{j,\mf p_j} \circ \Out^s_{i,\mf p_i}\neq \emptyset)$ is likely to be zero in most of the cases. This problem can be alleviated when giving up a tree-based formulation in favour of one focusing on time --- as typical in game-theoretic formulations based on stochastic processes. This will be addressed in Chapter~\ref{chap:3-SPF_VECT}.

However, the problem is actually even deeper: for an arbitrary strategy profile $s$, there may be essentially no non-trivial way for making $\Out_{i,\mf p_i}^s$ measurable. Consider action path \textsc{sef} data with $\A = [0,1]$, $\T = \{0,1\}$, and perfect endogenous information on the exogenous scenario space given by the unit interval $\Omega = [0,1]$ with Borel $\sigma$-algebra $\ms E = \ms B([0,1])$. Suppose that the agent $i$ active at the root $\x_0$ has full information available, i.e.\ $\ms F^i_{\x_0} = \ms E$. Let $g\colon \Omega \to \A$ be the identity on $[0,1]$ and $h\colon \A\to\A$ be a non-measurable function. Then, the strategy profile $s$ mapping $\x_0$ to the random action $g$ and the move $\x_1(f)$, where $f\in\A^\T$ with $f(0) = a$, to the deterministic (alias constant ``random'') action $h(a)$, for any $a\in[0,1]$, has a problematic outcome map: $p_{i,\{\x_0\}}$ maps any $\omega\in\Omega$ to $\x_0$, and hence $\Out^s_{i,\{\x_0\}}(\omega) = (\omega,(0,\omega),(1,h(\omega)))$. If we wished to allow for utility functions on $W$ depending in a non-trivial, Borel-measurable on the third component of an outcome $w\in W = \Omega\times \A^\T$, for instance, the map $u\colon W\to [0,1]$ mapping any $w = (\omega,f)$ to $f(1)$, then $h = u\circ \Out^s_{i,\{\x_0\}}$ would have to be measurable --- which it is not.\footnote{From the purely mathematical perspective, this problem has been known for long from the perspective of randomisation. Mathematically, the preceding example is largely similar to Aumann's ``attacker--defender'' example in \cite{Aumann1964Mixed} and its version in \cite[Example 6.7]{AlosFerrer2016Theory}. We refer to the Subsection~\ref{2-SEF_G.subs:randomisation} for a discussion of this aspect and the related literature. However, we take a somewhat different view on the problem on the level of decision theory, as explained in the following paragraph.} One way out of this problem is to require strategies to have non-trivial measurability properties with respect to endogenous information which cannot be explained on the basis of partitions, or equivalently of the tree structure. Although from a purely decision-theoretic perspective, this is an \emph{ex-post} restriction of strategies contradicting basic principles from non-cooperative game-theory, this is an approach proposed in different contexts. This includes the product form (see \cite{Witsenhausen1971Information, Witsenhausen1975Intrinsic, Heymann2022Kuhns}) and games using the theory of stochastic processes (as stochastic differential games, where closed-loop feedback functions are supposed to be measurable on the state space, see e.g.\ \cite{Carmona2018Probabilistic}). Here, as well time serves as a second-best substitute for the nodes and edges of a decision tree. This topic is discussed in Chapter~\ref{chap:3-SPF_VECT}.

We thus clearly see that in well-posed stochastic extensive forms there is a tension between non-triviality of endogenous information (agents having information about past behaviour) and non-triviality of \textsc{eu} preference structures (non-atomic beliefs thereby requiring uncountable $\Omega$, and complex utility functions). The main point, however, is not about the existence of such a tension, but the fact that it is precisely described in terms of formal conditions and existing non-trivial examples, like action path \textsc{sef} data. Precisely, the tension does not imply that discrete structures are necessary for extensive form theory to be sensible and non-empty. On the contrary, from the preceding discussion and the remainder of the present section it becomes clear that relevant well-posed action path \textsc{sef} data with coarse endogenous information sets may well admit non-trivial \textsc{eu} preference structures, with non-atomic beliefs and interesting utility functions. Although this eventually breaks down if endogenous information becomes finer, these observations suggests an approximation theory, by approximating \textsc{sef} with fine endogenous information with refining sequences of \textsc{sef} with coarse endogenous information. 

\subsection{Dynamic rationality and equilibrium}\label{2-SEF_G.subs:equilibrium}

Having provided a model of beliefs and tastes with natural and sufficient measurability properties, a general concept of ``dynamic rationality'' for well-posed stochastic extensive forms can be formulated (see, e.g.\ \cite[Section~9.C]{MasColell1995Microeconomic}).\footnote{Note that the term ``sequential'' is replaced with the more suitable and general one of ``dynamic''.} It is based on the classical idea of best response to correct beliefs about other agents' strategies, conditional on all induced decision situations given through endogenous information sets, given the specified beliefs and tastes about endogenous and exogenous events which are common knowledge. As mentioned earlier, the conditioning on induced stochastic extensive forms is supposed to model the dynamic aspect of decision making (decisions are taken at any information set). As a prime example of this, this may rule out empty threats. 
As outlined in the motivation of dynamic consistency, by adding the latter, an equilibrium concept of general scope obtains.

\begin{definition}\label{2-SEF_G.def:dynamic_rationality}
    Let $\F$ be a well-posed stochastic extensive form on an exogenous scenario space $(\Omega,\ms E)$, and let $\Pr = (\Pi,U,\ms W)$ be an \textsc{eu} preference structure on $\F$. 
    For any $i\in I$, any $\mf p\in\mf P^i$, let $\E_{i,\mf p}$ denote the conditional expectation operator with respect to $\P_{i,\mf p}$ given $\ms F_\mf p^i$, that is, 
    \[ \E_{i,\mf p} = \E^{\P_{i,\mf p}}[. \mid {\ms F}_{\mf p}^i], \]
    where $\E^\mu$ denotes the Lebesgue integral operator with respect to a given measure $\mu$. For any strategy profile $s\in S$, $i\in I$ and $\mf p\in\mf P^i$, let $\pi_{i,\mf p}(s) = \E_{i,\mf p}[u_{i,\mf p}\circ\Out_{i,\mf p}^s]. $
    
    A strategy profile $s\in S$ is said \emph{dynamically rational given $\Pr$} iff for all $i\in I$, all $\mf p\in\mf P^i$, and all $\tilde s\in S$ with $\tilde s^{-i} = s^{-i}$, we have
    \[ \pi_{i,\mf p}(s) \ge \pi_{i,\mf p}(\tilde s). \]

    Let $s\in S$ be a strategy profile. Then, $(s,\Pr)$ is said \emph{in equilibrium} iff it is dynamically rational given $\Pr$ and $(\Pr,s)$ is dynamically consistent. 
\end{definition}

As discussed in the following remark, this equilibrium concept is a generalised model of perfect Bayesian equilibrium, including several other well-known classical concepts within one extensive form framework.
\begin{remark}[Perfect Bayesian, subgame-perfect Nash and correlated equilibrium]
    Let $\F$ be a well-posed stochastic extensive form on an exogenous scenario space $(\Omega,\ms E)$ and $\Pr = (\Pi,U,\ms W)$ be an \textsc{eu} preference structure. Let $s\in S$.
    \begin{enumerate}
        \item Let us consider the static case, that is, $\X$ is a singleton. Denote its unique element by $\x$. Dynamic consistency and perfect recall is trivially satisfied, of course. In Harsanyi's setting of \cite{Harsanyi1968Games}, $s$ is a ``Bayesian equilibrium point'' iff $(s,\Pr)$ is in equilibrium according to the preceding definition. In the current game-theoretic language, $(\F,\Pr)$ yields a generalisation of Bayesian games and the equilibrium property corresponds to a generalisation of Bayesian Nash equilibrium\footnote{According to \cite{MasColell1995Microeconomic,Escude2023Lecture}. In Aumann's setting of \cite{Aumann1974Subjectivity}, this is simply called ``equilibrium point''. Following the framework of \cite{Aumann1987Correlated}, this might be called ``Bayes rational at almost all states of the world''.} with respect to the ``information structure'' given by the exogenous scenario space $(\Omega,\ms E)$, with common prior $\P = \P_{i,\{\x\}}$ shared by all agents $i\in I$, and the individual ``signal'' $\sigma$-algebras $(\ms F_\x^i)_{i\in I}$. By weakening the dynamical consistency requirement on $(\Pi,s)$ to hold only if $i=j$, see Definition~\ref{2-SEF_G.def:consistent_EU_pref_str}, one obtains the more general case with subjective priors $\P_i = \P_{i,\{x\}}$, $i\in I$. In particular, with $(\Omega,\ms E)$ serving as a correlation device, $s$ can be seen as a correlated equilibrium, with respect to the subjective priors $\P_i$ --- though this equilibrium is typically formulated with respect to a common prior (see the discussion in \cite{Aumann1987Correlated}). As a special case, we obtain the Nash equilibrium (cf.~\cite{Nash1950Equilibrium}, in mixed and in pure strategies, see Subsection~\ref{2-SEF_G.subs:randomisation} for a possible meaning of this).
        \item The preceding point generalises to the dynamic setting. Then, we obtain a non-trivial generalisation of perfect Bayesian equilibrium for a generalised model of dynamic Bayesian games (see, e.g.\ \cite{Fudenberg1991Perfect,MasColell1995Microeconomic}). It is a non-trivial generalisation of Bayesian games and perfect Bayesian equilibrium, because general exogenous dynamic uncertainty can be handled that can not be seen as the outcome of a nature agent's decision making (for instance, Brownian noise in continuous time). 
        \item Under the hypothesis of perfect information, we obtain the subgame-perfect equilibrium (cf.~\cite{Selten1965Spieltheoretische}). If only perfect endogenous information and perfect recall are supposed instead of perfect information, we obtain a non-trivial generalisation of subgame-perfect equilibrium to the case of imperfect exogenous information.    
    \end{enumerate}
\end{remark}

Next, we comment on the multiple-selves approach to dynamic decision making.

\begin{remark}[Multiple selves]\label{2-SEF_G.rmk:multiple_selves_sef}
    Dynamic rationality rules out empty threats, eventually corresponding to certain plans that are optimal today, but suboptimal tomorrow.\footnote{More precisely, suboptimal in some possible future which the threat --- non-credibly --- attempts to deter others from bringing about.}  
    In the case of a dynamically inconsistent taste system, it may also rule out plans that are suboptimal today, but optimal tomorrow.
    
    A classical attempt for resolving both points is the so-called \emph{multiple selves} approach (going back to \cite{Strotz1955Myopia}, at least; see \cite{Piccione1997Interpretation,Gilboa1997Comment} in the context of the absent-minded driver story). One idea behind this says that strategically it does not make much sense to suppose an agent making contingent plans for the future, for its future self has the ability to revise it. In the inconsistent case, he might really do this. From that perspective, by definition, an agent should only act once. According to this principle, the following analysis can be performed in our setting.

    Let $\F$ be a well-posed stochastic extensive form on an exogenous scenario space $(\Omega,\ms E)$ and $\Pr = (\Pi,U,\ms W)$ be an \textsc{eu} preference structure. Let $s\in S$.
    
    Replace $\F$ with the stochastic extensive form 
    \[ \F = (F,\pi,\X,\hat I,\hat{\ms F},\hat{\ms C},\hat C),\]
    and $\Pr = (\Pi,U,\ms W)$ with $\hat{\Pr} = (\hat \Pi,\hat U,\ms W)$,
    where 
    \begin{itemize}[label=--]
        \item $\hat I =\{(i,\mf p) \mid i\in I,\, \mf p\in\mf P^i\}$;
        \item $\hat{\ms F}^{(i,\mf p)}_\x = \ms F^i_{\mf p}$ for all $i\in I$, $\mf p\in\mf P^i$, $\x\in\mf p$;\footnote{One can show that with the choices $\hat C$ defined below, the set of random moves for $(i,\mf p)$ is given by $\hat \X^{(i,\mf p)} = \mf p$. We recall that it does not matter how $\ms F$ and $\ms C$ are chosen outside of this set. This footnote hence also applies to the following point.}
        \item $\hat{\ms C}^{(i,\mf p)}_\x = \ms C^i_{\mf p}$, for all $i\in I$, $\mf p\in\mf P^i$, $\x\in\mf p$;
        \item $\hat C^{(i,\mf p)} = A^i(\mf p)$, for all $i\in I$, $\mf p\in\mf P^i$;
        \item $\hat\Pi$ is the family assigning to any $(i,\mf p)\in \hat I$ and the only endogenous information set of that agent, $\mf p$, the triple $(p_{i,\mf p},\ms P_{i,\mf p},\P_{i,\mf p})$;
        \item $\hat U$ is the family assigning to any $(i,\mf p)\in \hat I$ and the only endogenous information set of that agent, $\mf p$, the function $u_{i,\mf p}$.
    \end{itemize}
    That is, each agent is split into separate ``incarnations'' of itself at all of its endogenous information set. We do not bother the reader with the verification of the claim that this yields again a well-posed stochastic extensive form on $(\Omega,\ms E)$ and an \textsc{eu} preference structure on it. Note that the new structure trivially satisfies perfect recall and dynamic consistency of the taste system. 

    Now, dynamic rationality and equilibrium as defined above can be analysed in $(\hat\F,\hat\Pr)$.
\end{remark}

The expected utility framework, as implemented above, is attractive for several reasons, including its equivalence to a class of well-chosen rationality axioms on decision making. Nevertheless, it has been challenged on diverse grounds: in view of explicit experimental evidence, for its theoretical lack of aversion of Knightian uncertainty, for the assumption of complete preferences, for its tendency to overlook gain-loss-asymmetry, and other things (cf.~\cite{Ellsberg1961Risk, Aumann1962Utility, Kahneman1979Prospect, Gilboa1989Maxmin, Schmeidler1989Subjective, Bewley2002Knightian, Hara2023Multiple}, for overviews see \cite{Gilboa2009Theory, Etner2012Decision}). Stochastic extensive form theory is not made for expected utility, and alternative preference structures or ways of modelling decision making may be defined on it as well. The author has chosen the classical expected utility framework because of its historic importance, its current use and the fact that the proposed alternatives often depart from it.

At this point, the author wants to mention that the terms ``decision problem'' and ``game'' are not introduced formally because their use in the literature is ambiguous, and probably for good reasons. Sometimes, these terms just describe a phenomenon of strategic interaction or a non-formalised description of it (e.g.\ one can know the absent-minded driver's decision problem without knowing its extensive form description using graphs as proposed in \cite{Piccione1997Interpretation}). Sometimes, these terms are meant as a formal description of states, consequences, agents and choices (as in \cite{AlosFerrer2005Trees} where the term ``extensive decision problem'' is used for the classical pseudo extensive forms of the present text). Sometimes, these terms are used only if, in addition to the formal description of states, consequences, agents and choices, payoffs, utilities, beliefs, preferences or the like are provided (as typically the case in stochastic control theory). In other contexts, one might be even more radical and argue that in order to be a game it must be played, that the play must be in equilibrium, and that there cannot be any other equilibrium (because the model could not explain why the latter is not played). Implicitly, such an argument also requires well-posedness. From this perspective, a game (with stochastic extensive form characteristics) would be a well-posed stochastic extensive form equipped with an equilibrium concept such there exists a unique equilibrium. The present text does not suggest any of these views being superior compared to others, and its concepts are compatible with all of them.

\subsection{Randomisation}\label{2-SEF_G.subs:randomisation}

Randomisation, that is, extending the exogenous scenario space and the information structure on it, is a common procedure in many fields. It underlies Nash's idea of equilibrium existence by introducing lotteries over strategies (the latter called ``pure'' because deterministic, that is, essentially defined on a singleton exogenous scenario space); it is a fundamental way of representing choice under uncertainty through the theory of expected utility following \cite{Neumann1944Theory, Savage1972Foundations, Anscombe1963Definition, Aumann1974Subjectivity}; it underlies the implementation of decision rules as Bayesian equilibria and the revelation principle in mechanism design; it underlies the weak solution concept for stochastic differential equations, stochastic control problems and stochastic differential games. Mixed (and then also behaviour and pure strategies) can be defined in the spirit of \cite{Aumann1974Subjectivity}.
Certainly, this is not the right place to develop a general theory of randomisation of stochastic extensive forms. Let us restrict to some comments.

In the non-stochastic setting of finite games à la Nash, randomness enters only for ``mixing'' individual strategies and not as a correlation device. This typically happens under the common prior assumption, and so all strategies are objective. When one assumes that under the common prior the exogenous information structures of different agents are mutually independent, all strategy profiles are necessarily mixed. If, moreover, one assumes that the exogenous information available at different endogenous information sets is mutually independent, then any strategy profile is behavioural. 

By contrast, the point of most stochastic games lies in correlation given, for instance, by what one calls a state process that agents may have partial information about. Then pure, behavioural and mixed strategies are rather the exception than the rule. This is even more so if one does not assume common priors (see again the discussion in \cite{Aumann1974Subjectivity,Aumann1987Correlated}). Yet, similarly to the content of \cite[Assumption II]{Aumann1974Subjectivity}, one might wish $\ms E$ and $\Pi$ such as to assure reasonable options of behavioural strategy profiles $s$ (for instance, such that $\P_{i,\mf p}$ is non-atomic on $\ms F_{\mf p}^{i,s}$ for all $i\in I$, $\mf p\in\mf P^i$). 

Yet, the ``more'' randomisation is possible, the finer must be $\ms E$, hence, the less probability measures on it. 
Actually, this problem is not only related to randomisation, but also to tastes: the less trivial taste shall be, the finer must be $\ms W$, thus the finer must be $\ms P_{i,\mf p}$ and $\ms E$, and again, the smaller must be the set of admissible beliefs $\P_{i,\mf p}$. 
It is well-known from the case of classical extensive forms that without finiteness (or countability) assumptions, this creates a non-trivial trade-off between the richness of both randomisation procedures and tastes and this trade-off is well understood (Aumann in \cite{Aumann1961Borel,Aumann1963Choosing,Aumann1964Mixed}, also see the discussion by Alós-Ferrer and Ritzberger in \cite[Subsection~6.4.3]{AlosFerrer2016Theory}, in particular Example 6.7 therein). In Subsection~\ref{2-SEF_G.subs:EU_preferences}, we have re-interpreted this trade-off in terms of a tension between non-triviality of endogenous information and non-triviality of \textsc{eu} preference structures.

\subsection{Simple examples}

Let us first consider some equilibria for a randomised version of the stochastic extensive forms $\F$ from Subsection~\ref{2-SEF_G.subs:SEF_simple-examples}, Lemma~\ref{2-SEF_G.lemma:simple_sef1}. Obviously, already in this simple case, there is a large range of possibilities, which moreover are well-known and analysed in other formalisations. Thus, the following selection can only be illustrative of the generality of our approach, and attempts to clarify its mechanics.

\begin{example}
    We consider a randomised version in that we let $(\Omega,\ms E)$ be a general exogenous scenario space and $\rho\colon \Omega \to \{1,2\}$ an $\ms E$-$\mc P\{1,2\}$-measurable surjection. We suppose it rich enough to admit real-valued random variables $\xi^k$, $k=0,1,2$ and a probability measure under which these four random variables are independent, $\rho$ is $\frac 12$-Bernoulli-distributed, and each $\xi^k$ is uniformly distributed on $[0,1]$. The $\xi^k$ represent randomisation devices available at the random moves $\x_k$, respectively.
    
    We let $I$ be a singleton, $\T = \{0,1\}$, $\A = \{1,2\}$, and $W = \Omega \times \A^\T$, and consider the associated action path \textsc{sdf} $(F,\pi,\X)$ which gives us the \textsc{sdf} from the first simple example (see Lemma~\ref{1-SDF_AC.lemma:simple_sdf_as_APsdf}) if $\rho$ is also injective and $\omega_k$ denotes the preimage of $k\in\{1,2\}$ under $\rho$. Note that an element of $W$ is essentially a triple $(\omega,k,m) \in \Omega \times \{1,2\}\times\{1,2\}$. In reminiscence of the notation used for the simple example, denote the root of $(\X,\ge_\X)$ by $\x_0$ and let, for $k\in\{1,2\}$, $\x_k$ be a shorthand for the random move mapping any $\omega\in\Omega$ to $\x_k(\omega) = \Omega \times \{k\} \times\{1,2\}$ (which is nothing else than $\x_1(f)$ for any $f\colon \T\to \A$ with $f(0) = k$ when written in action path notation). 
    
    In view of Remark \ref{2-SEF_G.rmk:multiple_selves_sef} on the multiple selves approach, we let $\hat I = \{i,j\}$ be a set with two distinct elements, representing two agents, $i$ acting at time $0$, and $j$ at time $1$ --- so to speak, two independent copies of the unique action index $\in I$ above. Consider the map $u\colon W\to \R$ given by $u(\omega,f) = (-1)^{\rho(\omega) + f(0) + f(1)}$, which is a generalised form of the payoff known from ``matching pennies'', with two pennies whose two sides show $1$ and $2$, and where the fact whether matching means win or lose depends on the realisation of the random variable $\rho$. 
    
    Suppose that the taste system $U$ satisfies with $u_i = -u$ and $u_j = u$. We fix a prior belief, alias a probability measure on $(\Omega,\ms E)$, $\P$ such that $\rho$ and $\xi^k$, $k=0,1,2$, are $\P$-independent, and $\P(\rho = 1) = p$ for given $p\in [0,1]$. Given a strategy profile $s\in S$, let us call an \textsc{eu} preference structure $\Pr$ \emph{suitable} iff its taste system is $U$, $\P_{i,\{\x_0\}} = \P$ and $(\Pi,s)$ is dynamically consistent. Note that such a $\Pi$ can be constructed out of the data $\P$ and $s$ in any case, though not necessarily in a unique way. Also note that for any $s\in S$ and $\x\in\X$, $s^i(\x)$ can be identified with its random action alias an $\ms F^i(\x)$-measurable, $\A$-valued random variable. The independence of the signals means that we only consider ``mixed'' strategies in the traditional sense (no correlation) which is not a much of a restriction here anyway because of the zero-sum structure.

    \begin{enumerate}
        \item First suppose that the agents' information and choices are essentially given by the first line of the table defining $C$ in Subsection~\ref{2-SEF_G.subs:SEF_simple-examples}, so that, in particular, $\ms F$ essentially corresponds to case 1, i.e.\ no information about $\rho$, and, at time $1$, agent $j$ recalls $i$'s decision made at time $0$. That is, $\ms F^i_{\x_0} = \sigma(\xi^0)$, $\ms F^j_{\x_k} = \sigma(\xi^k)$, for $k=1,2$, and $\mc H^j_1 = \{\{(0,1)\},\{(0,2)\}\}$. 
    
        Let $s\in S$ be a strategy profile. 
        If $p=\frac 12$, then for any $s\in S$ and any suitable \textsc{eu} preference structure $\Pr$, $(s,\Pr)$ is in equilibrium, with expected utility zero. If $p>\frac 12$, then for $s\in S$ and any suitable \textsc{eu} preference structure $\Pr$, $(s,\Pr)$ is in equilibrium iff
        \[ \P_{j,\{\x_1\}}[s^j(\x_1) = 2] = \P_{j,\{\x_2\}}[s^j(\x_2) = 1] = 1. \]
        Its expected utility is $1-2p$ for $i$ and $2p-1$ for $j$, since $j$ can fully react to $i$'s action whatever the latter is (even if $i$ did randomise and even though $j$ cannot observe $i$'s personal randomisation signal $\xi^0$). 
        If $p<\frac 12$, conversely, then for $s\in S$ and any suitable \textsc{eu} preference structure $\Pr$, $(s,\Pr)$ is in equilibrium iff
        \[ \P_{j,\{\x_1\}}[s^j(\x_1) = 1] = \P_{j,\{\x_2\}}[s^j(\x_2) = 2] = 1. \]
        Its expected utility is $2p-1$ for $i$ and $1-2p$ for $j$.
        
        \item If we follow the second line of the table, the exogenous information structure remains similar, but agent $j$ cannot observe the choice made by agent $i$ --- which is more similar to the original ``matching pennies'' game. Then, again, $\ms F^i_{\x_0} = \sigma(\xi^0)$, but $\ms F^j_{\x_1} = \ms F^j_{\x_2} = \sigma(\xi^1,\xi^2)$ and $\mc H^j_1 = \{\{0\} \times\A\}$, as the agent receives the same signal at $\x_1$ and $\x_2$. Let $\mf p = \{\x_1,\x_2\}$.
        
        Let $s\in S$ and $\Pr$ be a suitable \textsc{eu} preference structure. If $p=\frac 12$, then $(s,\Pr)$ is in equilibrium without further restriction. If $p\neq\frac 12$, then $(s,\Pr)$ is in equilibrium iff
        \[ \P[s^i(\x_0) = 1] = \P_{j,\mf p}[s^j(\mf p) = 1] = \frac 12. \]
        In any case, equilibrium expected utility is equal to $0$ for both agents.

        \item Let us next consider the situation where agent $j$ has a further advantage by having full exogenous information, which corresponds to line 4, and in particular \textsc{eis} 2.(a). That is, $\ms F^i_{\x_0} = \sigma(\xi^0)$, but $\ms F^j_{\x_1} = \ms F^j_{\x_2} = \sigma(\rho,\xi^1,\xi^2)$ and $\mc H^j_1 = \{\{0\} \times\A\}$. Let again $\mf p = \{\x_1,\x_2\}$.

        Let $s\in S$ and $\Pr$ be a suitable \textsc{eu} preference structure. Let $\alpha = \P[s^i(\x_0) = 1]$ and for $k=1,2$ let $\beta_k\in [0,1]$ a number such that $\P[s^j(\mf p) = 1 \mid \rho = k] = \beta_k$, $\P$-almost surely. Then the expected utility of $i$, given $s^i(\x_0)$ is equal to
        \[ \pi_{i,\{\x_0\}}(s) = -(-1)^{s^i(\x_0)}\,\big(p(2\beta_1-1)+(1-p)(1-2\beta_2)\big), \]
        and the expected utility of $j$ given $\rho$ and $s^j(\mf p)$ is equal to
        \[ \pi_{j,\mf p}(s) = (1-2\alpha) (-1)^{\rho + s^j(\mf p)}. \]
        
        We conclude that $(s,\Pr)$ is in equilibrium iff 
        \[ \alpha = \frac 12, \qquad p(2\beta_1-1)+(1-p)(1-2\beta_2) = 0. \]
        For if $\alpha = \frac 12$, then any strategy for $j$ is a best response, but mixing is a best response for $i$ only if the right-hand side equality is satisfied. If $\alpha >\frac 12$, then $s^j$ is a best response iff $s^j(\mf p) = 3-\rho$, $\P$-almost surely. In this case, $\beta_1 = 0$ and $\beta_2 = 1$, so that $p(2\beta_1-1)+(1-p)(1-2\beta_2) = -1$. But then any best response $s^i$ by $i$ satisfies $s^i(\x_0) = 2$ $\P$-almost surely, whence $\alpha = 0$, a contradiction. A similar contradiction arises when assuming equilibrium with $\alpha < \frac 12$.

        Expected utility in equilibrium is almost surely zero for both agents. Hence, the informational advantage for $j$ (as compared to the preceding point) does not pay out.

        \item Let us now vary the preceding examples and consider a case where agent $i$ can determine what exogenous information agent $j$ obtains, which is a variation on the theme of exploration and exploitation. For this, we transcribe the fifth line with \textsc{eis} 2.(b). That is, $\ms F^i_{\x_0} = \sigma(\xi^0)$, but $\ms F^j_{\x_1} = \sigma(\rho,\xi^1)$, $\ms F^j_{\x_2} = \sigma(\xi^2)$ and $\mc H^j_1 = \{\{(0,1)\},\{(0,2)\}\}$.

        Let $s\in S$ and $\Pr$ be a suitable \textsc{eu} preference structure. Let $\alpha = \P[s^i(\x_0) = 1]$, and for $k=1,2$ let $\beta_k\in [0,1]$ a number such that $\P_{j,\{\x_1\}}[s^j(\x_1) = 1 \mid \rho = k] = \beta_k$, $\P_{j,\{\x_1\}}$-almost surely, and let $\gamma = \P_{j,\{\x_2\}}[s^j(\x_2) = 1]$. Expected utilities are given by
        \begin{align*}
            \pi_{i,\{\x_0\}}(s) =  - (-1)^{s^i(\x_0)}\, \Big( &1\{s^i(\x_0)=1\}\,\big(p(2\beta_1-1)+(1-p)(1-2\beta_2)\big) \\
            &+ 1\{s^i(\x_0)=2\}\,(1-2\gamma)(1-2p)\Big)
        \end{align*}
        for $i$, and
        \[ \pi_{j,\{\x_1\}}(s) = -(-1)^{\rho+s^j(\x_1)}, \qquad \pi_{j,\{\x_2\}}(s) = (1-2p)(-1)^{s^j(\x_2)} \]
        for $j$.

        We conclude that $(s,\Pr)$ is in equilibrium iff
        \[  \alpha \cdot p(1-p) = 0, \qquad \beta_1=0,~\beta_2=1, \qquad \begin{cases}
            \gamma = 0, &\text{ if } p < \frac 12, \\
            \gamma = 1, &\text{ if } p > \frac 12.
        \end{cases}. \]
        Note that $\alpha \cdot p(1-p) = 0$ means nothing else than: if $\rho$ is believed to be truly random, then $\alpha = 0$. In the deterministic case ($p=0$ or $p=1$), no restriction on $\alpha$ needs to be imposed; agent $i$ is completely indifferent in equilibrium.
        
        The proof is straightforward by backwards induction. The best response property of $s$ for agent $j$ implies that $\beta_1=0$ and $\beta_2=1$, and
        \[ \begin{cases}
            \gamma = 0, &\text{ if } p < \frac 12, \\
            \gamma = 1, &\text{ if } p > \frac 12.
        \end{cases}.\]
        Then, given this strategy of agent $j$, the expected utility of agent $i$ writes as
        \[  \pi_{i,\{\x_0\}}(s) =  - (-1)^{s^i(\x_0)}\, \big( -1\{s^i(\x_0)=1\}\,+ 1\{s^i(\x_0)=2\}\,|1-2p|\big) = -|1-2p|^{s^i(\x_0)-1},  \]
        and the best responses of $i$ are obviously the ones claimed above.

        Expected utility in equilibrium is almost surely $-|1-2p|$ for $i$ and $|1-2p|$ for $j$. If $j$ had access to full information about $\rho$ at $\x_1$ as well, it would be easy to see, $j$ could realise the expected utility of $1$ in equilibrium. In the present asymmetric case, however, provided $0<p<1$, agent $i$ can force $j$ to the random move with less exogenous information, reducing $j$'s expected equilibrium utility. Hence, $j$ cannot realise more expected utility in equilibrium than in the situation with no relevant information on $\rho$ at all (see the first case above). 
        \end{enumerate}
\end{example}

To close this subsection, let us consider the absent-minded driver stochastic extensive form discussed earlier, see Theorem~\ref{2-SEF_G.thm:absent_minded_driver_Gilboa_sef}. Following \cite{Piccione1997Interpretation}, we let taste $u$ of both agents by identically given by
\[ (\ast) \qquad u(\omega,D) = 0, \quad u(\omega,H) = 4, \quad u(\omega,M) = 1, \]
for $\omega\in\Omega$, and in equilibrium we suppose some form of dynamic consistency of strategy profile and belief. Here, we use the equilibrium notion introduced in the present text, restricting to those belief systems $\Pi$ whose common prior for $\{(1,\{\x_1\}),(2,\{\x_2\})\}$ is given by a fixed probability measure $\P$ with respect to that $(\rho,\xi^1,\xi^2)$ are independent and $\xi^1,\xi^2$ are both uniformly distributed on $[0,1]$.

Note that for any such $\P$ and any strategy profile $s\in S$, there is a belief system $\Pi$ such that $(\Pi,s)$ is dynamically consistent and $\P$ is a common prior for $\{(1,\{\x_1\}),(2,\{\x_2\})\}$. Moreover, $\Pi$ then necessarily satisfies $(\xi^i)_\ast\P_{i,\{\x_i\}} = (\xi^i)_\ast \P$, for both $i\in I$; or in other words:
\[ \P_{i,\{\x_i\}}(E) = \P(E), \qquad \text{ for all } E\in \ms F^i_{\x_i}. \]
$\Pi$ is essentially uniquely determined (i.e.\ ``along the equilibrium path'').

Also note that a strategy profile $s\in S$ corresponds to a pair $(E_1,E_2)$ of events $E_i\in\ms F_{\x_i}^i$ via $s^i(\x_i) = c_i(E_i)$, $i\in I$, and this correspondence is a bijection $S \to \ms F_{\x_1}^1 \times \ms F_{\x_2}^2$. In the sequel, we thus identify strategy profiles with their respective images under this bijection. $E_i$ is the event of agent $i$ exiting, $i\in I$.

Hence, fix $u$ and $\P$ as above. Given $s\in S$, represented by $(E_1,E_2)$ as above, and $p\in[0,1]$, call an \textsc{eu} preference structure \emph{suitable} iff $u_{i,\{\x_i\}} = u$ for both $i\in I$, $(\Pi,s)$ is dynamically consistent, and
\[ \P(\rho = 1) = \frac 12, \qquad \P(E_1) = \P(E_2) = 1-p. \]
Thus, $\frac 12$ is the prior probability of $\{\rho = 1\}$, and $p$ is the prior probability of continuing at either intersection, given the intersection. In particular, we deliberately focus on equilibria with 1) prior $\P$ under that $\rho$ is uniformly distributed, and 2) $\P(E_1) = \P(E_2)$. This selection of equilibria is natural in view of the symmetry of the problem, all the more in the case of a physical person split into two abstract agents. The indifference principle (a.k.a.\ principle of insufficient reason) further underpins that choice.

\begin{thm}\label{2-SEF_G.thm:absent_minded_driver_Gilboa_sef_well-posed}
    Consider the absent-minded driver \textsc{sef} introduced previously, defined on exogenous scenario space $(\Omega,\ms E)$. Let $p\in[0,1]$, let $s\in S$ be a strategy profile and let $\Pr$ be a suitable \textsc{eu} preference structure. Then, $(s,\Pr)$ is in equilibrium iff $p = \frac 23$.
\end{thm}

Hence, we find that the absent-minded driver paradox from \cite{Piccione1997Interpretation} evaporates, confirming the \emph{ex ante} optimal strategy of continuing independently with probability $\frac 23$ at each intersection, which is in a similar way the conclusion in \cite{Aumann1997Absent,Gilboa1997Comment} as well.

\subsection{Discussion on nature representations}

Traditionally, stochastic games in discrete time are modelled in what the author suggests calling \emph{nature representation}. That is, in a traditional extensive form model, using an additional agent called ``nature'' with perfect recall using a fixed mixed strategy. In the terminology of this thesis, this would entail an action path stochastic extensive form, in order to allow for that randomisation, and an agent with perfect recall. Provided the action space is regular enough (a Borel space), in that setting, any probability measure on discrete-time path space can be represented as the outcome of ``nature's'' strategy. This follows from classical disintegration results in probability theory (see \cite[Chapter~3]{Kallenberg2021Foundations}).

In continuous time, however, this representation can fail because the nature representation does not always yield a well-posed stochastic extensive form. There are exceptions: for instance, locally right-constant jump processes can be implemented like this (based on the deterministic model action-reaction extensive form framework in \cite{AlosFerrer2015Repeated}). An example of failure, however, is Brownian motion. Brownian motion is supported on paths of low regularity, paths that are far from being locally right-constant. Hence, a straightforward nature representation would suggest to take the action path pseudo-extensive form data $W=\Omega\times\A^\T$, for a suitable exogenous scenario space $(\Omega,\ms E)$, $\A=\R$, $\T=\R_+$, and singleton $I = \{\text{``nature''}\}$, such that ``nature'' admits perfect endogenous recall. By Theorem \ref{2-SEF_G.thm:link_endogenous_information_H}, perfect (endogenous) recall requires $\mc H^{\text{``nature''}}_t$ to contain only singletons, $t\in\T$. The corresponding decision forest is not weakly up-discrete, and, indeed, using well-known counterexamples from \cite{Simon1989Extensive, Stinchcombe1992Maximal, AlosFerrer2016Theory}, one can easily see that the induced action path pseudo-\textsc{sef} is not well-posed.

This issue might be circumvented by accepting the ``nature'' agent to admit imperfect recall. For instance, in the just-mentioned example one may instead consider the other extreme, i.e.\ assume $\mc H^{\text{``nature''}}_t$ to be a singleton for all $t\in\T$. Then, we easily obtain well-posedness (compare \cite[Example~5.5]{AlosFerrer2016Theory}, which can be directly adapted for a single agent). However, while modellers might fix ``nature's'' strategy, they cannot fix the personal agents' ones. Strategies would be functions of information sets, and therefore, adaptedness of ``personal'' agents' strategies would lack an explanation in the case of the non-discrete ``nature'' action space $\A$. While it seems already odd that the ``nature'' agent is forced to be forgetful on the grounds of extensive form well-posedness, we argue that the category of perfect recall is irrelevant for generating exogenous randomness or information in the first place. Moreover, the nature representation does not help to explain adaptedness of the actual, ``personal'' agents' strategic behaviour. We conclude that stochastic extensive forms provide a strict generalisation of existing extensive form theory: it contains classical extensive forms, and although many stochastic extensive forms can be nature-represented in classical extensive form, this is not the case for an important class of extensive forms with interesting noise, such as Brownian motion.

\chapter{Stochastic process forms in vertically extended continuous time}\label{chap:3-SPF_VECT}
\chaptermark{Stochastic process forms}
\section{Vertically extended continuous time}\label{3-SPF_VECT.sec:vERT}

In Chapter~\ref{chap:2-SEF_G}, we have shown the well-posedness of action path stochastic extensive forms on well-ordered time grids. We have also seen that, essentially, one cannot go beyond this within a rigorous action path stochastic extensive form setting. Thus, as described in the introduction to this thesis, we use the outcomes of these action path stochastic extensive forms defined on well-ordered time grids and let the grids become finer and finer to obtain asymptotic action processes, as illustrated in Figure~\ref{fig:VECT_and_tilting_conv} printed in the introduction. The corresponding notion of convergence is discussed later in Section~\ref{3-SPF_VECT.sec:sto_proc_in_vERT}. First, we need to vertically extend the continuous-time half-axis in order to faithfully represent relevant patterns of reaction on these smaller and smaller time grids in the limit.

The aim of this subsection is to introduce this vertically extended continuous time, that is, the smallest complete total order containing $\R_+$ and any tilted well-order embeddable into $\R_+$. We equip it with a suitable topology and with suitable $\sigma$-algebras. The order topology being too large, the interesting $\sigma$-algebra is not given by the Borel sets. Hence, we are led to studying the problems of the measurability of continuous functions and of measurable projection and section. 

\subsection{Preliminaries on order theory}\label{3-SPF_VECT.subs:Prel_order_theory}
We start with recalling some basic, well-known facts from order theory and thereby fix notation and conventions.

\subsubsection{Completions}
Basic notions from order theory are recalled in the appendix, see Section~\ref{3-SPF_VECT.sec:Dedekind-MacNeille_completion}. Here, we only recall the following special notions which are introduced in that appendix. Let $P$ be a poset and $\p\colon P \inj L$ be a completion. Then,
\begin{enumerate}
    \item we call $\p$ \emph{dense} iff $\im\p$ is both join- and meet-dense in $L$;
    \item we call $\p$ \emph{small} iff for any completion $\psi\colon P \inj M$, there is an embedding $f\colon L \to M$ with $\psi = f \circ \p$.
\end{enumerate}
It can be shown that a completion is dense iff it is small, for any poset $P$ there is a small completion $(\p,L)$, for any small completion $(\psi,M)$ there is a unique isomorphism $f\colon L \to M$ such that $\psi = f\circ\p$, and the small completion of $P$ can be represented by the Dedekind-MacNeille completion. For details, see Section~\ref{3-SPF_VECT.sec:Dedekind-MacNeille_completion}.

\subsubsection{Topology on total orders}

Without further mention, we equip any total order $T$ with the order topology $\ms O_T$. Namely, this is the topology generated by all sets of the form $\uparrow t \setminus \{t\} = \{u\in T \mid t < u\}$ and $\downarrow t \setminus \{t\} = \{u\in T \mid u<t\}$, $t\in T$. In the following definitions, let $T$ be a total order.

Recall that in a topological space $Y$, a neighbourhood of a point $y\in Y$ is a set $V\subseteq Y$ containing an open set $V'$ with $y\in V'\subseteq V$. 
A point $t\in T$ is a \emph{left-limit point} iff for every neighbourhood $U$ of $t$ we have $U\cap (\downarrow t \setminus \{t\})\neq\emptyset$. A point $t\in T$ is a \emph{right-limit point} iff for every neighbourhood $U$ of $t$ we have $U\cap (\uparrow t \setminus \{t\}) \neq\emptyset$. Denote the sets of left-limit (right-limit) points by ${T}_\nearrow$ (${T}_\swarrow$), respectively.

If $f\colon{T} \to S$ is a function on ${T}$ into some set $S$ and $t\in{T}$ is a left-limit (right-limit) point, then we call $f$ \emph{left-constant at $t$} (\emph{right-constant at $t$}) iff there is a neighbourhood $U$ of $t$ such that $f|_{U\cap\downarrow t}$ ($f|_{U\cap\uparrow t}$) is constant, respectively.\footnote{We choose this convention; a weaker alternative would have been to ask for $f|_{U\cap(\downarrow t\setminus\{t\})}$ ($f|_{U\cap(\uparrow t\setminus\{t\})}$, respectively) to be constant.} 

If $f\colon{T} \to Y$ is a function on ${T}$ into some topological space $Y$ and $t\in{T}$ is a left-limit (right-limit) point, then $y\in Y$ is said \emph{left-limit (right-limit) of $f$ at $t$} iff for every neighbourhood $V\subseteq Y$ of $y$, there is a neighbourhood $U\subseteq{T}$ of $t$ such that $U\cap(\downarrow t \setminus \{t\}) \subseteq f^{-1}(V)$ ($U\cap(\uparrow t \setminus \{t\}) \subseteq f^{-1}(V)$, respectively). Provided $Y$ is Hausdorff, if existent such $y$ is unique, and denoted by $y = \lim_{u\nearrow t} f(u)$ ($y = \lim_{u\searrow t} f(u)$, respectively). We call $f$ \emph{left-continuous at $t$} (\emph{right-continuous at $t$} iff, provided $t$ is a left-limit (right-limit) point, $f(t)$ is a left-limit (right-limit) of $f$ at $t$, respectively. With these definitions, $f$ is continuous at $t$ iff it is both left- and right-continuous at $t$; and $f$ is left-continuous (right-continuous) at $t$ if it is right-constant (left-constant) at $t$, respectively. If $Y$ is a total order as well, and $t\in T$ is a right-limit point, $f$ is said \emph{lower semicontinuous from the right at $t$} iff for any $y\in Y$ with $y < f(t)$, there is $v\in T$ with $t<v$ such that for all $u\in(t,v)_T$, we have $y < f(u)$.

A function $f\colon{T} \to S$ into some set $S$ is said \emph{locally right-constant} (\emph{locally left-constant}) iff it is right-constant (left-constant) at any left-limit (right-limit) point $t\in T$, respectively. In direct generalisation of the standard framework, we call a function $f\colon{T} \to Y$ into a topological space $Y$ \emph{làg} (\emph{làd}) iff it has left-limits (right-limits) at all left-limit (right-limit) points, respectively. We call a function $f\colon{T} \to Y$ into a topological space $Y$ \emph{left-continuous}, or \emph{càg} (\emph{right-continuous}, or \emph{càd}) iff it is left-continuous (right-continuous) at any $t\in{T}$, respectively. Agglutinations, like \emph{càdlàg}, stand for the conjunction, like ``{càd} and {làg}''. So, for example, $f$ is continuous iff it is càdcàg.\footnote{In principle, to all these definitions concerning the entire function $f$ one must add the qualifier ``locally''. As continuity is a local property, this makes no difference however. By contrast, ``constant'' is not a local property: for example, the identity on $\{0,1\}$ is locally right- and left-constant, but not constant.} A function $f\colon T\to Y$ from a totally ordered set $T$ into a further totally ordered set $Y$ is said \emph{lower semicontinuous from the right} iff it lowersemicontinuous from the right at all right-limit points in $T$.  

\subsubsection{Well-orders and ordinals}

We recall basic notions from the theory of well-orders within ZFC, see \cite[Chapter~2]{Ziegler2017Mathematische}. A \emph{well-order} is a poset $S$ such that any non-empty subset $M\subseteq S$ has a minimum. An \emph{ordinal (number)} is a set $\alpha$ such that all $y\in\alpha$ satisfy $y\subseteq\alpha$ and the element relation $\in$ defines a strict partial order on $\alpha$. The class of ordinals is denoted by $\On$. Then (see \cite[Section~9]{Ziegler2017Mathematische}),
\begin{enumerate}
    \item any set of ordinals is strictly totally ordered by the element relation $\in$,
    \item any non-empty set of ordinals has a minimum with respect to that partial order,
    \item for every $\alpha\in\On$, $\alpha = \{\beta\in\On \mid \beta < \alpha\}$, and
    \item $\On$ is not a set.
\end{enumerate}
Ordinals are important to us because any well-order is isomorphic to a unique ordinal via a unique isomorphism (cf.~\cite[Section~9]{Ziegler2017Mathematische}). 

The empty set is an ordinal, i.e.\ $\emptyset\in\On$, which is called \emph{zero} and denoted by $0=\emptyset$, any $\alpha\in\On$ has a unique successor, given by $\alpha + 1 = \alpha\cup\{\alpha\}$, and any downward closed set of ordinals is itself an ordinal. The successor of zero is \emph{one}, denoted by $1 = 0 + 1$ --- and so on. A \emph{successor ordinal} is an ordinal $\beta$ such that there is $\alpha\in\On$ with $\beta = \alpha+1$. A \emph{limit ordinal} is a non-zero, non-successor ordinal. For any set $S \subseteq\On$ of ordinals, there is a smallest ordinal $\beta$ such that all $\alpha\in S$ satisfy $\alpha\le\beta$, given by $\bigcup S$. Hence, if $S\subseteq\alpha+1$ for some $\alpha\in\On$, then $\sup S = \bigcup S$ in the poset $\alpha+1$. Based on the zero ordinal, the successor operation $+1$, and the supremum operation $\sup$, one can recursively define ordinal arithmetic in terms of an associative addition, a left-distributive multiplication, and a multiplicative exponentiation on ordinals. Care must be taken, however, since the algebraic structure of these operations is relatively limited in general, see \cite[p.\ 80]{Ziegler2017Mathematische} for details.

As a direct consequence of the infinity axiom, there exists a smallest infinite ordinal, denoted by $\mf w$. Equipped with ordinal addition and multiplication, and the respective neutral elements $0$ and $1$, $\mf w$ is identical to the algebraic structure $\N = \Z_+$ of natural numbers. One can show the existence of a smallest (alias first) uncountable ordinal, denoted by ${\mf w_1}$.\footnote{$\mf w$ is typically denoted by $\omega$, and $\mf w_1$ by $\omega_1$, but in view of the dominant role of probability in this text, and the typical notation ``$\omega$'' for scenarios, we choose this unusual notation.} Its successor is ${\mf w_1}+1 = \mf w_1 \cup \{\mf w_1\}$. By the results cited above, ${\mf w_1}$ is the set of all countable ordinal numbers. An ordinal can be embedded into $\R_+$ iff it is countable. For any countable set $S\subseteq{\mf w_1}$ of countable ordinals, $\bigcup S$ is countable, hence it admits a supremum in ${\mf w_1}$. Recall that, without further mention, we equip any totally ordered set with the order topology. ${\mf w_1}$ is sequentially compact, but not compact. As a consequence, any uncountable ordinal is not metrisable. Hence, an ordinal is Polish iff it is countable.

\subsection{The complete total order $\ovT$}\label{3-SPF_VECT.subs:vERT}
In view of the theory developed in the two previous chapters, and by Theorem~\ref{2-SEF_G.thm:AP_sef_well-posed} in particular, one can construct a broad class well-posed stochastic extensive forms based on paths of action indexed over well-ordered subsets of $\R_+$. 
The theory of this chapter is based on the intuition of eventually accumulating these time well-orders from the ``right''. Here, we graphically represent continuous, or ``real'', time $\R_+$ as a \emph{horizontal} half-axis oriented towards the right. As discussed later on in Subsections~\ref{3-SPF_VECT.subs:tilting_convergence}, in the limit, such well-orders can collapse form a purely horizontal point of view. As we wish to keep track of this order structure of decisions in the limit, we extend real time on a well-ordered \emph{vertical} half-axis.

Recalling that well-orders embedded in $\R_+$ are always countable, the time half-axis we require is
\begin{equation}\label{3-SPF_VECT.eq:T}
    \T = \R_+ \times {\mf w_1}. 
\end{equation}
We equip $\T$ with lexicographic order, that is, $(t,\beta), (u,\gamma)\in\T$ satisfy 
\begin{equation}\label{3-SPF_VECT.eq:lexicographic_order}
    (t,\beta) \le (u,\gamma) \qquad \Longleftrightarrow \qquad t < u, \text{ or } [t=u \text{ and } \beta\le \gamma].
\end{equation}
A stochastic analysis based on this time half-axis requires understanding two things: first, the small completion of $\T$ (for taking infima and suprema); second, suitable topologies and $\sigma$-algebras on that completion (for convergence, probability, and integration). 

In this subsection, we deal with the first question, but also prepare our later treatment of the second one. For that, the approach will consist in exhausting $\T$ by sufficiently ``small'' extensions. Namely, fix some $\alpha\in{\mf w_1}+1$, and let
\begin{equation}\label{3-SPF_VECT.eq:Ta}
    \Ta = \R_+ \times \alpha = \{t\in\T \mid \pi(t) < \alpha\}.
\end{equation}
Clearly, we have $\Ta = \T$ if $\alpha = {\mf w_1}$. More importantly, $\T = \bigcup_{\alpha\in{\mf w_1}} \Ta$.

Let us note that, via set inclusion, $\Ta$ is embedded into $\bRp\times({\mf w_1}+1)$ endowed with lexicographic order, i.e.\ all $(t,\beta), (u,\gamma)\in\bRp \times ({\mf w_1}+1)$ satisfy Statement~\ref{3-SPF_VECT.eq:lexicographic_order}. Here, $\bRp$ is the small completion of the poset $\R_+$, given by $\bRp = \R_+ \cup \{\infty\}$, with $\infty = \sup \R_+$. To obtain a (candidate for a) small completion of $\Ta$, let
\begin{equation}\label{3-SPF_VECT.eq:ovTa}
    \ovTa = \begin{cases} \T_{\alpha+1} \cup \{\infty\}, &\text{if }\alpha\text{ is a limit ordinal,}\\ \T_\alpha \cup \{\infty\}, &\text{else,} \end{cases}
\end{equation}
equipped with induced order.
If $\alpha = {\mf w_1}$, we simply write $\ovT = \ovTa$, that is, we let
\begin{equation}\label{3-SPF_VECT.eq:ovT}
    \ovT = [\R_+ \times ({\mf w_1}+1)] \cup\{\infty\}.
\end{equation}

Note that $\overline{\T_0} = \{\infty\}$ and, if $\alpha>0$,
\begin{equation}\label{3-SPF_VECT.eq:ovT=union(T,tops)}
    \ovTa = \Ta \cup \{(t,\sup\alpha)\mid t\in\R_+\} \cup \{\infty\},
\end{equation}
where $\sup\alpha$ is the supremum of the set $\alpha$ in ${\mf w_1}+1$. This union is disjoint iff $\alpha$ is a limit ordinal.

There are embeddings $\bRp \to \ovT \to \bRp \times ({\mf w_1}+1)$ mapping $t\mapsto (t,0)$ and $(t,\alpha) \mapsto (t,\alpha)$, by means of which we treat $\bRp$ as a subset of $\ovT$, and $\ovT$ as a subset of $\bRp \times ({\mf w_1}+1)$. Moreover, let $p\colon  \bRp \times ({\mf w_1}+1)\to \bRp$ and $\pi\colon \bRp \times ({\mf w_1}+1)\to ({\mf w_1}+1)$ be the canonical projections of the set-theoretic product. Clearly, $p$ is monotone and $\pi$ is not. 

We now answer the first question above. 

\begin{proposition}\label{3-SPF_VECT.prop:ovT_small_completion_of_T}
    Via set inclusion, $\ovTa$ is a small completion of $\Ta$. In particular, $\ovT$ is a small completion of $\T$.
\end{proposition}

The proof is based on the following lemma which is of independent interest.

\begin{lemma}\label{3-SPF_VECT.lemma:suprema_and_infima_in_ovT}
    Let $\alpha\in{\mf w_1}+1$ and $S\subseteq\ovTa$ be a subset. Furthermore, let
    \[ a = \inf \mc Pp(S), \qquad b = \sup \mc Pp(S), \qquad \text{in }\bRp. \]
    Then, $S$ has both an infimum and a supremum in $\ovTa$, given by
    \[ 
        \inf S = \begin{cases} (a,\gamma), & \text{if } a\in\mc Pp(S) \text{ and } \gamma = \inf \mc P\pi(S \cap [\{a\} \times (\sup\alpha+1)]) \text{ in }\sup\alpha+1, \\ (a,\sup\alpha), & \text{if } a\in\R_+\setminus \mc Pp(S), \\ \infty, &\text{else,} \end{cases}
    \]
    and
    \[ 
        \sup S = \begin{cases} (b,\gamma), & \text{if } b\in\mc Pp(S) \text{ and } \gamma = \sup \mc P\pi(S \cap [\{b\} \times (\sup\alpha+1)]) \text{ in }\sup\alpha+1, \\ b, & \text{if } b\notin\mc Pp(S). \end{cases}
    \]
\end{lemma}

\begin{definition}
    The complete and totally ordered lattice $\ovT$ is said \emph{vertically extended continuous, or real, time}, or shorter \emph{vertically extended time}, or even shorter, if the context permits, \emph{time}.
\end{definition}

As detailed above, the elements of $\T$ have a direct interpretation in terms of ``accumulated'' well-orders embedded into $\R_+$; this point is further detailed by the notion of tilting convergence, introduced in Subsection~\ref{3-SPF_VECT.subs:tilting_convergence}. The other elements of $\ovT$ arise by taking suprema and infima of elements of $\T$, as described in Proposition~\ref{3-SPF_VECT.prop:ovT_small_completion_of_T} and, more explicitly, in Lemma~\ref{3-SPF_VECT.lemma:suprema_and_infima_in_ovT}. In most decision-theoretic contexts, the element $\infty$ can be interpreted as ``never'' or as a terminal time. A natural interpretation of $(t,\mf w_1)$, for $t\in\R_+$, is ``never at real time $t$'' -- a contradiction, of course. Yet, this contradiction is later resolved by the fact that the relevant objects describing dynamic decision making, introduced in Section~\ref{3-SPF_VECT.sec:sto_proc_in_vERT}, --- optional times and optional processes --- overlook these vertical endpoints. 

\subsection{Topology and $\sigma$-algebras on $\ovT$}
In the remainder of this and the following section, we study suitable $\sigma$-algebras on the vertically extended continuous time half-axis $\ovT$, the small completion of $\T$. First, we note that $\ovT$ is equipped with the order topology $\ms O_\ovT$. For any subset $S\subseteq\ovT$, let $\ms G_\ovT(S)$ be the set of all down- and up-sets of the form $[0,t)_{\ovT}$ and $(t,\infty]_\ovT$, $t\in S$. Then, by definition, $\ms G_\ovT(\ovT)$ is a subbase of the topology $\ms O_\ovT$. This is slightly strengthened by the following simple result.

\begin{lemma}\label{3-SPF_VECT.lemma:msO_ovT.subbase_base}
    The set
    \begin{equation}\label{3-SPF_VECT.eq:lemma:msO_ovT.subbase_base.msU(T)}
        \ms U_\ovT(\T) = \{[0,u)_\ovT \mid u\in\T\} \cup \{(t,\infty]_\ovT \mid t\in\T\} \cup \{ (t,u)_\ovT \mid t,u\in\T\}
    \end{equation}
    is a base and $\ms G_\ovT(\T)$ is a subbase of the topology $\ms O_\ovT$ on $\ovT$.
\end{lemma}

Let $\ms B_\ovT = \sigma(\ms O_\ovT)$ be the Borel $\sigma$-algebra on $(\ovT,\ms O_\ovT)$. By definition, basically, $\ms B_\ovT$ 
\begin{enumerate}
    \item is the smallest $\sigma$-algebra $\ms B'$ on $\ovT$ such that for all topological spaces $Y$, all continuous maps $\ovT \to Y$ on are $\ms B'$-$\ms B_Y$-measurable,
    \item and is the largest $\sigma$-algebra $\ms B'$ on $\ovT$ such that for all topological spaces $X$, all continuous maps $X\to\ovT$ are $\ms B_X$-$\ms B'$-measurable.
\end{enumerate}

Deleting sets from $\ms B_\ovT$ removes some continuous maps $\ovT\to Y$ -- in probabilistic terms, \emph{a fortiori}, also some stochastic processes with continuous paths -- from our reach. Adding sets to $\ms B_\ovT$ removes some continuous maps $X\to\ovT$ -- in probabilistic terms, some random times -- from our reach. In that sense, $\ms B_\ovT$ is quite natural. 
However, despite being Hausdorff the topology $\ms O_\ovT$ on $\ovT$ is non-metrisable, because ${\mf w_1}$ is not metrisable. This makes standard methods from stochastic analysis hard to apply. Moreover, it is shown further below that $\ms B_\ovT$ is neither generated by the (compact) class of (closed) intervals, respectively, nor by that of products of compacts in $\bRp \times \alpha$, $\alpha\in{\mf w_1}$, and basic results from stochastic analysis, such as measurable projection, do not generalise to that $\sigma$-algebra.

Therefore, we consider the $\sigma$-algebras generated by intervals on the one hand, and by the products of compacts $\bRp \times \alpha$, $\alpha\in{\mf w_1}$ on the other. This defines the following programme which we are concerned with in most of Sections~\ref{3-SPF_VECT.sec:vERT} and~\ref{3-SPF_VECT.sec:sto_proc_in_vERT}:
\begin{enumerate}
    \item\label{3-SPF_VECT.step:sigma_alg_on_ovT.introduce_sigma_algebras} Introduce these $\sigma$-algebras formally, precisely describe their relevant generators, and establish their relationship;
    \item\label{3-SPF_VECT.step:sigma_alg_on_ovT.continuous_ovT_to_E_mb?} Determine a relevant class of topological spaces $Y$ such that continuous maps $\ovT\to Y$ are measurable;
    \item\label{3-SPF_VECT.step:sigma_alg_on_ovT.mb_projection_and_section?} Study whether the theorem of measurable projection and section holds true for these $\sigma$-algebras;
    \item\label{3-SPF_VECT.step:sigma_alg_on_ovT.mb_of_stochastic_processes} Study stochastic processes and random times with time $\ovT$ regarding relevant measurability properties.
\end{enumerate}

We start with Step~\ref{3-SPF_VECT.step:sigma_alg_on_ovT.introduce_sigma_algebras}. Let $\ms I_\ovT(\T) = \sigma(\ms G_\ovT(\T))$ be the $\sigma$-algebra generated by $\ms G_\ovT(\T)$, the complements of principal up- and down-sets of elements of $\T$ in $\ovT$. Next, for any $\alpha\in{\mf w_1}$, equip $\bRp$, $\alpha+1$, and their product $\bRp\times(\alpha+1)$ with their Polish topologies $\ms O_\bRp$, $\ms O_{\alpha+1}$, and $\ms O_{\bRp\times(\alpha+1)} = \ms O_\bRp\otimes\ms O_{\alpha+1}$, respectively, let
\begin{equation}\label{3-SPF_VECT.eq:def_pi_alpha}
    \rho^\alpha\colon\ovT \to \bRp\times(\alpha+1), \, t\mapsto \big( p(t), \sup [(\pi(t)+1) \cap (\alpha+1)] \big),
\end{equation}
and let $\ms P_\ovT^\alpha$ be the $\sigma$-algebra on $\ovT$ generated by $\rho^\alpha$ and $\ms B_{\bRp\times(\alpha+1)} = \sigma(\ms O_{\bRp\times(\alpha+1)})$, called the \emph{projection $\sigma$-algebra of rank $\alpha$}. Let $\ms P_\ovT$ be the $\sigma$-algebra on $\ovT$ generated by the set of pairs $\rho^\alpha$, $\ms B_{\bRp\times(\alpha+1)} = \sigma(\ms O_{\bRp\times(\alpha+1)})$, ranging over $\alpha\in{\mf w_1}$, called the \emph{projection $\sigma$-algebra}.\footnote{One may interpret the notation ``$\ms P$'' also by the words ``product'', ``Polish'', ``preimage''.} In formulae,
\begin{equation}\label{3-SPF_VECT.eq:msP_ovT^alpha_def}
    \ms P_\ovT^\alpha = \big\{(\rho^\alpha)^{-1}(B) \mid B\in\ms B_{\bRp\times(\alpha+1)} \big\}, \qquad \alpha\in\mf w_1,
\end{equation}
and $\ms P_\ovT = \bigvee_{\alpha\in\mf w_1} \ms P_\ovT^\alpha$.

We now describe several generators of $\ms I_\ovT(\T)$, making precise the statement that ``$\ms I_\ovT(\T)$ is the $\sigma$-algebra generated by the intervals''. For this, let
\[ \ms K_\ovT(\T) = \{ [t,u]_{\ovT} \mid t,u\in \T \}. \]

\begin{proposition}\label{3-SPF_VECT.prop:msK_compact_class_generating_msI}
    For any $t,u\in\ovT$, $[t,u]_\ovT$ is compact. $\ms K_\ovT(\T)$ is an intersection-stable compact class satisfying
    \[ \ms I_\ovT(\T) = \sigma(\ms G_\ovT(\T)) = \sigma(\ms K_\ovT(\T)). \]
\end{proposition}

\begin{corollary}\label{3-SPF_VECT.cor:msI_is_generated_by_all_kinds_of_principal_up/down-sets}
    The following sets of intervals are intersection-stable and generate the $\sigma$-algebra $\ms I_\ovT(\T)$:
    \[ \{ [0,u]_\ovT\mid u\in\T\}, \quad \{ [0,u)_\ovT \mid u\in\T\}, \quad \{ [t,\infty]_\ovT\mid t\in\T\}, \quad \{ (t,\infty]_\ovT \mid t\in\T\}. \]
\end{corollary}

Hence, we call $\ms I_\ovT(\T)$ the \emph{$\T$-interval $\sigma$-algebra on $\ovT$}, and abbreviate it by $\ms I_\ovT$. We continue with discussing the projection $\sigma$-algebras. 
Generators of $\ms P_\ovT^\alpha$, for $\alpha\in{\mf w_1}$, and $\ms P_\ovT$ are spelled out in the following lemma.
\begin{lemma}\label{3-SPF_VECT.lemma:generator_msP}
    For any $Y\in\{\bRp\} \cup {\mf w_1}$, let $\ms G_Y$ be an intersection-stable generator of the $\sigma$-algebra $\ms B_Y$ such that $Y$ is the countable union of elements of $\ms G_Y$. For $\alpha\in{\mf w_1}$, let
    \[ \ms G_{\ovT,\times}^\alpha = \big\{(\rho^\alpha)^{-1}(B\times C) \mid B\in\ms G_\bRp,\,C\in\ms G_{\alpha+1}\big\},  \]
    and let
    \begin{equation}\label{3-SPF_VECT.eq:msG_ovT,times}
        \ms G_{\ovT,\times} = \bigcup_{\alpha\in{\mf w_1}}\ms G_{\ovT,\times}^\alpha.
    \end{equation}
    Then,
    \begin{enumerate}
        \item\label{3-SPF_VECT.lemma:generator_msP.alpha} for all $\alpha\in{\mf w_1}$, $\ms G_{\ovT,\times}^\alpha$ is an intersection-stable generator of the $\sigma$-algebra $\ms P_\ovT^\alpha$ on $\ovT$;
        \item\label{3-SPF_VECT.lemma:generator_msP.mfw1} $\ms G_{\ovT,\times}$ is a generator of the $\sigma$-algebra $\ms P_\ovT$ on $\ovT$; moreover, if, for all $\alpha,\beta\in\mf w_1$ with $\alpha < \beta$ and all $C\in\ms G_{\alpha+1}$, we have either $\alpha\in C$ and $C\cup (\alpha,\beta]_{\mf w_1} \in \ms G_{\beta+1}$, or $\alpha\notin C$ and $C\in\ms G_{\beta+1}$, then the union in Equation~\ref{3-SPF_VECT.eq:msG_ovT,times} is increasing in $\alpha\in\mf w_1$ and $\ms G_{\ovT,\times}$ is intersection-stable;
        \item\label{3-SPF_VECT.lemma:generator_msP.compact_class} if $\ms G_Y$ consists of compact sets in $Y$ for all $Y\in\{\bRp\} \cup {\mf w_1}$ and the only element $B$ of $\ms G_\bRp$ with $\infty\in B$ is $B = \{\infty\}$, then the preceding generators are compact classes, respectively.
    \end{enumerate}
\end{lemma}

\begin{example}\label{3-SPF_VECT.ex:generator_msP}
    A typical example for the generators $\ms G_{\alpha+1}$, $\alpha\in\mf w_1$, is given by $\ms G_{\alpha+1} = \alpha+2 = \{\beta\in\mf w_1 \mid \beta\subseteq\alpha+1\}$. This provides an intersection-stable generator of $\alpha+1$, $\alpha\in\mf w_1$. Moreover, the hypothesis in the second sentence of Part~\ref{3-SPF_VECT.lemma:generator_msP.mfw1} is satisfied, i.e.\ for all $\alpha,\beta\in\mf w_1$ with $\alpha < \beta$ and all $C\in\ms G_{\alpha+1}$, we have either $\alpha\in C$ and $C\cup (\alpha,\beta]_{\mf w_1} \in \ms G_{\beta+1}$, or $\alpha\notin C$ and $C\in\ms G_{\beta+1}$. In order to obtain compact classes as in Part~\ref{3-SPF_VECT.lemma:generator_msP.compact_class}, we could restrict to those $\beta$ that are not limit ordinals, leading to $\{0\} \cup \{\gamma+1 \mid \gamma\in\alpha\} \subseteq \alpha+1$.
\end{example}
The $\sigma$-algebras $\ms P_\ovT^\alpha$, for $\alpha\in\mf w_1$, are defined by projecting down on the vertical levels $\alpha$ and below. As the following lemma indicates, they are insensitive to ``new information'' above these levels, and --- a result which is important in the context of stochastic processes --- this result is compatible with taking the product with a measurable space.
\begin{lemma}\label{3-SPF_VECT.lemma:sets_in_msPovT^alpha_otimes_msE_are_inactive_beyond_alpha}
    Let $\alpha\in{\mf w_1}$ be a countable ordinal and $(\Omega,\ms E)$ be a measurable space. Then, for any $M\in\ms P_\ovT^\alpha\otimes \ms E$, any $t\in\R_+$, and any $\omega\in\Omega$, we have
    \[ (u,\omega) \in M \]
    for all or no $u\in p^{-1}(t)\cap (\ovT\setminus \Ta)$.
\end{lemma}

Next, we ask the following: Is there an easy representation of the ``vertical limit'' $\sigma$-algebra $\ms P_\ovT = \bigvee_{\alpha\in\mf w_1} \ms P_\ovT^\alpha$? And, if so, is it compatible with taking the product with a measurable space? The next result gives a simple answer to these questions. It illustrates the interplay of the uncountable vertical half-axis and the countable additivity of $\sigma$-algebras.
\begin{proposition}\label{3-SPF_VECT.prop:exhaust_msP_ovT_otimes_msE}
    Let $(\Omega,\ms E)$ be a measurable space. Then,\begin{equation}\label{3-SPF_VECT.eq:msT_otimes_msA=union_over_alpha(preimages_of_B_bRp_times_alpha+1_otimes_msA)}
        \ms P_\ovT \otimes\ms E = \bigcup_{\alpha\in{\mf w_1}} \ms P_\ovT^\alpha \otimes\ms E = \big\{(\rho^\alpha\times\id_\Omega)^{-1}(S) \mid \alpha\in{\mf w_1},\, S\in\ms B_{\bRp\times(\alpha+1)}\otimes\ms E \big\}.
    \end{equation}
    The union is increasing in $\alpha\in{\mf w_1}$.
\end{proposition}
In particular, taking singleton $\Omega$, and recalling Equation~\ref{3-SPF_VECT.eq:msP_ovT^alpha_def}, we get the representation
\begin{equation}\label{3-SPF_VECT.eq:msP_ovT}
    \ms P_\ovT = \big\{(\rho^\alpha)^{-1}(B) \mid \alpha\in{\mf w_1},\, B\in\ms B_{\bRp\times(\alpha+1)} \big\}.
\end{equation}
Furthermore, we obtain the following corollary.
\begin{corollary}\label{3-SPF_VECT.cor:msP_ovT_quasi_subsetneq_msB_ovT}
    We have $\ms P_\ovT \subseteq \ms B_\ovT \vee \sigma(B\times\{0\} \mid B\in\ms B_\bRp)$ and $\ms P_\ovT \neq \ms B_\ovT$.
\end{corollary}

We continue with a result linking $\ms I_\ovT$ and $\ms P_\ovT$.
\begin{lemma}\label{3-SPF_VECT.lemma:msI_ovT_subseteq_msP_ovT}
    We have $\ms I_\ovT \subseteq \ms P_\ovT$.
\end{lemma}

We close this subsections with confirming that the relevant structural maps are measurable with respect to the $\sigma$-algebras under scrutiny. Let us denote, for any $\alpha\in{\mf w_1}$, the set-theoretic inclusion map $\ovT_{\alpha+1} \to \ovT$ by $\iota_\alpha$. Note that, as follows from Equation~\ref{3-SPF_VECT.eq:ovTa}, we have $\ovT_{\alpha+1}\in\ms B_{\bRp\times(\alpha+1)}$. 
\begin{lemma}\label{3-SPF_VECT.lemma:iota_alpha_&_p_mb}
    Let $\alpha\in{\mf w_1}$. Then, $\iota_\alpha$ is $\ms B_{\bRp\times(\alpha+1)}|_{\ovT_{\alpha+1}}$-$\ms P_\ovT$-measurable and $p$ is both $\ms I_\ovT$-$\ms B_\bRp$- and $\ms P_\ovT^\alpha$-$\ms B_\bRp$-measurable.
\end{lemma}
Note that, by Lemma~\ref{3-SPF_VECT.lemma:msI_ovT_subseteq_msP_ovT}, $\iota_\alpha$ is also $\ms B_{\bRp\times(\alpha+1)}|_{\ovT_{\alpha+1}}$-$\ms I_\ovT$-measurable, and $p$ is also $\ms P_\ovT$-$\ms B_\bRp$-measurable.

\subsection{Continuous functions on $\ovT$}

We continue with Step~\ref{3-SPF_VECT.step:sigma_alg_on_ovT.continuous_ovT_to_E_mb?}. Actually, a more granular answer is of interest, because it is natural in stochastic analysis to consider processes whose paths satisfy the weaker requirement of having left- and right-limits. For this, recall the definitions of limit points and continuity in Subsection~\ref{3-SPF_VECT.subs:Prel_order_theory}, which require some additional care when dealing with $\ovT$ instead of $\bRp$. To start, we determine the left- and right-limit points in $\ovT$.
\begin{lemma}\label{3-SPF_VECT.lemma:left/right-limit_points_in_ovT}
    The set $\ovT_\nearrow$ of left-limit points in $\ovT$ is given by all $t\in\ovT \setminus \{0\}$ such that $\pi(t)$ is not a successor ordinal. The set $\ovT_\swarrow$ of right-limit points in $\ovT$ is given by all $(t,{\mf w_1})$, $t\in\R_+$.
\end{lemma}

We now answer the main question of this subsection. In a first step, we note the special role of the points $t\in\ovT$ with $\pi(t) = \mf w_1$.

\begin{lemma}\label{3-SPF_VECT.lemma:Q1.mb_fcts_are_left-cont_on_pi=mfw1}
    Let $Y$ be a metrisable topological space and $f\colon\ovT\to Y$ be $\ms P_\ovT$-$\ms B_Y$-measurable. Then, for all $t\in\R_+$, $f$ is left-constant and, in particular, left-continuous at $(t,{\mf w_1})$.
\end{lemma}

This left-continuity requirement at the ``never at $t$''-instants is thus necessary for $\ms P_\ovT$-$\ms B_Y$-measurability (and, in particular, for $\ms I_\ovT$-$\ms B_Y$-measurability), but it is not a restriction as it may seem at first sight. First, for all topological spaces $Y$, any function $g\colon\bRp\to Y$ induces a function $f\colon\ovT\to Y$ that is left-continuous at $(t,{\mf w_1})$, for all $t\in\R_+$, namely $f = g\circ p$. $f$ inherits relevant properties from $g$, like làg, continuity, measurability.\footnote{The latter in view of Lemma~\ref{3-SPF_VECT.lemma:iota_alpha_&_p_mb}.} Moreover, note that a làg function $\ovT\to\R$ that is left-continuous at $(t,{\mf w_1})$, for all $t\in\R_+$, is essentially a làg function $\T\cup\{\infty\}\to\R$ that can be continuously extended from below, or the left, to the point at infinity of any vertical half-axis. That is, asymptotic behaviour of the function along the vertical half-axis can be explained by countably many values along each vertical half-axis. Speaking game-theoretically, the formally uncountable chains of vertical, infinitesimal (randomised or not) reaction are actually described by countably many ones.

Provided this unproblematic regularity assumption in $t\in\ovT$ with $\pi(t) = \mf w_1$, measurability can be assured under mild regularity conditions, as the following proposition clarifies.

\begin{proposition}\label{3-SPF_VECT.prop:Q1.cont_fcts_are_mb}
    Let $Y$ be a metrisable topological space. Any làg function $\ovT\to Y$ that is left-continuous at $(t,{\mf w_1})$, for all $t\in\R_+$, is $\ms I_\ovT$-$\ms B_Y$-measurable. In particular, it is $\ms P_\ovT$-$\ms B_Y$-measurable. 
\end{proposition}

We emphasise some of the special cases covered by the preceding proposition.

\begin{corollary}
    Let $Y$ be a metrisable topological space. The following functions $\ovT\to Y$ are $\ms I_\ovT$-$\ms B_Y$-measurable, and, in particular, $\ms P_\ovT$-$\ms B_Y$-measurable:
    \begin{enumerate}
        \item provided $Y=\R$, any monotone function that is left-continuous at $(t,{\mf w_1})$, for all $t\in\R_+$,
        \item any làdlàg function that is left-continuous at $(t,{\mf w_1})$, for all $t\in\R_+$,
        \item any càdlàg function that is left-continuous at $(t,{\mf w_1})$, for all $t\in\R_+$,
        \item any continuous function.\qed
    \end{enumerate}
\end{corollary}

\subsection{Measurable projection and section}

Next, we discuss Step~\ref{3-SPF_VECT.step:sigma_alg_on_ovT.mb_projection_and_section?}. The basic idea of the proof is already expressed in Proposition~\ref{3-SPF_VECT.prop:exhaust_msP_ovT_otimes_msE} and Lemma~\ref{3-SPF_VECT.lemma:sets_in_msPovT^alpha_otimes_msE_are_inactive_beyond_alpha}. 
As a consequence, the following proposition obtains. 

\begin{proposition}\label{3-SPF_VECT.prop:Q2.random_sets_of_time}
    Let $(\Omega,\ms E)$ be a measurable space, $\prj_\Omega\colon\ovT\times\Omega\to\Omega$ denote the canonical projection onto $\Omega$, and $M\in\ms P_\ovT \otimes\ms E$. Then, there is $\alpha\in{\mf w_1}$ such that
    \[ M_\alpha = M \cap (\ovT_{\alpha+1}\times\Omega)~ \in \ms B_{\bRp\times(\alpha+1)}\otimes\ms E, \qquad\text{and}\qquad  \mc P\prj_\Omega(M) = \mc P\prj_\Omega(M_\alpha). \]
\end{proposition}

We now state the theorem of measurable projection and section for vertically extended real time. For the setup, let us introduce some notation. For a measurable space $(\Omega,\ms E)$, let ${\ms E}^{\mathrm u}$ be the universal completion of $\ms E$, that is, the intersection of the completions of $\ms E$ with respect to all probability measures on $(\Omega,\ms E)$. Moreover, given a measurable space $(\Omega,\ms E)$, for any map $\tau\colon\Omega\to\ovT$, we let 
\begin{equation}\label{3-SPF_VECT.eq:converse_graph_tau}
    [\![\tau]\!] = \{(\tau(\omega),\omega) \mid \omega\in\Omega\}
\end{equation}
be its converse graph. Then, we have the following result.

\begin{thm}[Measurable Projection and Section]\label{3-SPF_VECT.thm:mb_proj_section}
    Let $(\Omega,\ms E)$ be a measurable space, $M\in\ms P_\ovT\otimes\ms E$ and $\prj_\Omega\colon\ovT\times\Omega\to\Omega$ be the projection onto $\Omega$. Then, 
    \[ \mc P\prj_\Omega(M) = \{\omega\in\Omega\mid \exists t\in\ovT\colon (t,\omega)\in M\} \in {\ms E}^{\mathrm u}. \]
    Moreover, there is an ${\ms E}^{\mathrm u}|_{\mc P\prj_\Omega(M)}$-$\ms P_\ovT$-measurable map $\sigma\colon\mc P\prj_\Omega(M)\to\overline\T$ such that $[\![\sigma]\!] \subseteq M$.
\end{thm}

\section{Stochastic processes in vertically extended continuous time}\label{3-SPF_VECT.sec:sto_proc_in_vERT}

In this section, we tackle Step~\ref{3-SPF_VECT.step:sigma_alg_on_ovT.mb_of_stochastic_processes}. For this, and for the entire Section~\ref{3-SPF_VECT.sec:sto_proc_in_vERT}, we fix a measurable space $(\Omega,\ms E)$ and a filtration $\ms F = (\ms F_t)_{t\in\ovT}$ on it with time index set $\ovT$, and a measurable space $(Y,\ms Y)$. Moreover, we will introduce a model of instants of vertically extended continuous time that options can be evaluated or decisions be made at, called optional times. Based on this, we are led to a theory of optional and predictable processes, yielding a natural model for decision making in vertically extended continuous time. We will see that these notions non-trivially extend the classical notions from stochastic calculus on $\R_+$. Finally, we show that these optional processes arise naturally by accumulating discrete-time decision making in classical continuous time $\R_+$, using the new concept of tilting convergence.

\subsection{Augmentation and right-limits of information flow}

As the theory of stochastic processes and stopping times involves both approximations from the right and projections, we have to discuss right-continuity and completeness assumption on $\ms F$. 

Regarding right-continuity, approximation from the right in the Polish spaces $\bRp\times(\alpha+1)$, $\alpha\in{\mf w_1}$, may require the \emph{strong right-continuous extension} $\ms F_\Plus = (\ms F_{t\Plus})_{t\in\ovT}$ which we define as follows: 
\[ \ms F_{\infty\Plus} = \ms F_\infty, \qquad \ms F_{(t,\beta)\Plus} = \bigcap_{u\in (t,\infty)_\bRp} \ms F_{(u,\beta)}, \quad t\in\R_+, \, \beta\in{\mf w_1}+1. \]
As $\ms F$ is a filtration, this implies that for all $t\in\ovT$, we have $\ms F_{t\Plus} = \ms F_{p(t)\Plus}$ --- a strong property. Namely, this is equivalent to saying that there is a filtration $\ms G = (\ms G_t)_{t\in\bRp}$ with time index set $\bRp$ such that $\ms F_\Plus = \ms G \circ p$. We note that $(\ms F_\Plus)_\Plus = \ms F_\Plus$ and we call $\ms F$, idem $(\Omega,\ms E,\ms F)$, \emph{strongly right-continuous} iff $\ms F = \ms F_\Plus$.

Note that the right-continuous extension defined above is much larger than the one with respect to the order topology on $\ovT$. With respect to the order topology, the \emph{right-continuous extension} is $\ms F_{+} = (\ms F_{t+})_{t\in\ovT}$, given by $\ms F_{t+} = \bigcap_{u\in (t,\infty]_\ovT} \ms F_u = \ms F_{t\Plus}$ if $t\in\ovT_\swarrow = \pi^{-1}(\{{\mf w_1}\})$,\footnote{See Lemma~\ref{3-SPF_VECT.lemma:left/right-limit_points_in_ovT}.} and $\ms F_{t+} = \ms F_t$ else. $\ms F$, idem $(\Omega,\ms E,\ms F)$, is said \emph{right-continuous} iff $\ms F = \ms F_+$. In the strong right-continuous extension information about events collapses along the vertical axes --- for any $x\in\R_+$, all information about the realised scenario included in $\ms F_{x}$ suffices to describe randomisation along the vertical half-axis $p = x$.

The second issue is completion. As beliefs are treated as part of agents' preferences, we do not fix probability measures at this stage. Hence, we make use of universal completions. We fix our conventions for this and recall basic properties.\footnote{For the essence of these notions and their properties, the methods of proof from the classical case can be directly adapted to the present setting. For the classical case, see, for instance, \cite[Chapter 1]{Sharpe1988General}.}

Let $\mf P_{\ms E}$ denote the set of probability measures on $(\Omega,\ms E)$ and, for any $\P\in\mf P_{\ms E}$, $\ms N_{\P} = \{M\subseteq\Omega\mid \exists N\in\ms E\colon M\subseteq N,\, \P(N) = 0\}$. Then, for any sub-$\sigma$-algebra $\ms A\subseteq\ms E$, call \emph{universal augmentation of $\ms A$ in $\ms E$} the $\sigma$-algebra
\begin{equation}\label{3-SPF_VECT.eq:universal_augmentation_in_msE}
    \ovl{\ms A} = \bigcap_{\P\in\mf P_{\ms E}} (\ms A \vee \ms N_\P).
\end{equation}
Moreover, we have 
\[ {\ms A}^{\mathrm u} \subseteq  \ovl{\ms A} \subseteq \ovl{\ms E}= {\ms E}^{\mathrm u}, \qquad \ovl{\ovl{\ms A}} = \ovl{\ms A}, \]
and the inequalities can be strict. $\ms A$ is said \emph{universally augmented in $\ms E$} iff $\ovl{\ms A} = \ms A$. We recall that, with this convention, $\ms E$ is universally complete iff it is universally augmented in itself.

The \emph{universal augmentation of $\ms F$ in $\ms E$} is the filtration $\ovl{\ms F} = (\ovl{\ms F_t})_{t\in\ovT}$. The filtration $\ms F$ is said \emph{universally augmented in $\ms E$} iff $\ms F = \ovl{\ms F}$. We call the filtered measurable space $(\Omega,\ms E,\ms F)$ \emph{universally complete} iff $\ms E = \ovl{\ms E}$ and $\ms F = \ovl{\ms F}$. We finally note that universal augmentation and right-continuous extension commute, i.e.\footnote{This follows from the analogous classical result. For the reader's convenience, the appendix contains a proof nevertheless.}
\begin{equation}\label{3-SPF_VECT.eq:(ovl_msF)_+=ovl_(msF_+)}
    \ovl{\ms F_\Plus} = \ovl{\ms F}_\Plus. 
\end{equation}
As a direct consequence, the same equation holds true with ``$\Plus$'' replaced by ``$+$''.

\subsection{Progressively measurable processes}

We continue with recalling some basic notions and thereby fixing notation. A \emph{stochastic process}, with time $\ovT$ and valued in $(Y,\ms Y)$, is a map $\xi\colon \ovT \times\Omega\to Y$ such that the maps $\xi_t = \xi(t,.)$, $t\in\ovT$, are $\ms E$-$\ms Y$-measurable. This is equivalent to the map $\Omega\to Y^{\ovT},\, \omega\mapsto (t\mapsto \xi(t,\omega))$, also denoted by $\xi$, being $\ms E$-$\ms Y^{\otimes\ovT}$-measurable. A stochastic process $\xi\colon \ovT \times\Omega\to Y$ is said \emph{(strictly) $\ms F$-adapted} iff, for all $t\in\ovT$, $\xi_t$ is even $\ms F_t$-$\ms Y$-measurable. We emphasise the following generalisation, which is less evident, because it depends on the choice of the $\sigma$-algebra on $\ovT$.

\begin{definition}\label{3-SPF_VECT.def:prog_mb}
    Let $\ms T_\ovT$ be a $\sigma$-algebra on $\ovT$ containing the $\T$-intervals, i.e.\ with $\ms I_\ovT(\T)\subseteq\ms T_\ovT$. 
    \begin{enumerate}
        \item A subset $M\subseteq\ovT\times\Omega$ is said \emph{$\ms F$-progressively measurable with respect to $\ms T_\ovT$} iff, for all $t\in\ovT$, 
        \begin{equation}\label{3-SPF_VECT.eq:prog_mb}
            M \cap ([0,t]_\ovT \times \Omega) \in \ms T_\ovT\otimes\ms F_t. 
        \end{equation}
        The set of $\ms F$-progressively measurable subsets of $\ovT\times\Omega$ with respect to $\ms T_\ovT$ is denoted by $\Prg(\ms T_\ovT,\ms F)$. 
        \item A stochastic process $\xi\colon \ovT \times\Omega\to Y$ is said \emph{$\ms F$-progressively measurable} with respect to $\ms T_\ovT$ iff $\xi^{-1}(B)$ is $\ms F$-progressively measurable with respect to $\ms T_\ovT$ for any $B\in\ms Y$.
        \item If the qualifier ``with respect to $\ms T_\ovT$'' is omitted, then $\ms T_\ovT = \ms P_\ovT$. Moreover, let $\Prg(\ms F) = \Prg(\ms P_\ovT,\ms F)$.
    \end{enumerate}
\end{definition}

\begin{remark}\label{3-SPF_VECT.rmk:prog_mb_basic_properties}
    The following facts are easily shown. Let the objects and notation be given as in the definition. Then:
    \begin{enumerate}
        \item\label{3-SPF_VECT.rmk:prog_mb_basic_properties.classical_theory} Standard real time notions of adapted and progressively measurable processes are essentially retrieved by considering stochastic processes $\xi\colon\ovT\times\Omega\to Y$ and filtrations $\ms F$ such that $\xi_t = \xi_{p(t)}$ and $\ms F_t = \ms F_{p(t)}$ for all $t\in\ovT$.
        \item\label{3-SPF_VECT.rmk:prog_mb_basic_properties.sigma_algebra_prog_mb} $\Prg(\ms T_\ovT,\ms F)$ defines a $\sigma$-algebra, and a stochastic process $\xi\colon\ovT\times\Omega\to Y$ is $\ms F$-progressively measurable with respect to $\ms T_\ovT$ iff it is $\Prg(\ms T_\ovT,\ms F)$-$\ms Y$-measurable. 
        \item\label{3-SPF_VECT.rmk:prog_mb_basic_properties.monotonicity_in_msT_ovT} If $\ms T_\ovT,\ms T'_\ovT$ are two $\sigma$-algebras on $\ovT$ containing the $\T$-intervals and such that $\ms T_\ovT\subseteq \ms T'_\ovT$, then $\Prg(\ms T_\ovT,\ms G) \subseteq \Prg(\ms T'_\ovT,\ms G)$, i.e.\ $\ms G$-progressively measurability with respect to $\ms T_\ovT$ implies that with respect to $\ms T'_\ovT$.
        \item\label{3-SPF_VECT.rmk:prog_mb_basic_properties.examples_prog_mb_sets} We have $\ms T_\ovT\otimes\ms F_0 \subseteq \Prg(\ms T_\ovT,\ms F) \subseteq \ms T_\ovT\otimes\ms F_\infty$. Hence, with respect to $\ms T_\ovT$, any set of the form $T\times E$, with $T\in\ms T_\ovT$, $A\in\ms F_0$, is $\ms F$-progressively measurable, and any progressively measurable set is $\ms T_\ovT\otimes\ms F_\infty$-measurable.
        \item\label{3-SPF_VECT.rmk:prog_mb_basic_properties.scwise_left-constant_at_(t,mfw1)} If $\ms T_\ovT\subseteq\ms P_\ovT$ and $Y$ is a metrisable topological space, then for any, with respect to $\ms T_\ovT$, $\ms F$-progressively measurable $\xi\colon\ovT\times\Omega\to Y$, all $\omega\in\Omega$ and all $t\in\R_+$, $\xi(.,\omega)$ is left-constant at $(t,\mf w_1)$. This follows from the fact that $\ovT\to\ovT\times\Omega,\, t\mapsto(t,\omega)$ is $\ms P_\ovT$-$\ms T_\ovT\otimes\ms E$-measurable, and Lemma~\ref{3-SPF_VECT.lemma:Q1.mb_fcts_are_left-cont_on_pi=mfw1}.\footnote{As $\mf w_1$ is uncountable, this does not have clear implications on the random variable $\xi_{(t,\mf w_1)}$ in general.}
        \item\label{3-SPF_VECT.rmk:prog_mb_basic_properties.prog_mg_=>_relaxed_adapted} Any, with respect to $\ms T_\ovT$, $\ms F$-progressively measurable stochastic process $\xi\colon\ovT\times\Omega\to Y$ as in Definition~\ref{3-SPF_VECT.def:prog_mb} is adapted in a relaxed sense. Namely, $\xi_t$ is $\ms F_t$-$\ms Y$-measurable (at least) for any $t\in\ovT$ with $\pi(t) < {\mf w_1}$, alias $t\in\T\cup\{\infty\}$. In view of the discussion in Subsection~\ref{3-SPF_VECT.subs:vERT}, these can be interpreted as (deterministic) optional times. The result follows from the previous item with $T=\{t\}$ and $E=\Omega$. 
    \end{enumerate}
\end{remark}

The simplest non-trivial progressively measurable processes are those generated by \emph{certain} random times -- making the emphasised qualifier precise, is our concern for the remainder of this and the following subsection. A \emph{random time} with respect to a $\sigma$-algebra $\ms T_\ovT$ on $\ovT$ containing the intervals is an $\ms E$-$\ms T_\ovT$-measurable map $\tau\colon\Omega\to\ovT$. These are in one-to-one correspondence to certain subsets of $\ovT\times\Omega$, by considering its converse graph $[\![\tau]\!]$ as defined in Equation~\ref{3-SPF_VECT.eq:converse_graph_tau}, its converse epi- or its converse hypograph. These sets can be defined in terms of stochastic intervals as well, which in turn are defined in complete analogy to the real-time case.\footnote{For instance, given two maps $\sigma,\tau\colon\Omega\to\ovT$, we have \[ [\![\sigma,\tau)\!) = \{(t,\omega)\in\ovT\times\Omega \mid \sigma(\omega) \le t < \tau(\omega) \}. \] That way, the converse graph of $\tau$ is given by $[\![\tau]\!] = [\![\tau,\tau]\!]$, the converse weak and strict epigraphs of $\tau$ are given by $[\![\tau,\infty]\!]$ and $(\!(\tau,\infty]\!]$, and the converse weak and strict hypographs of $\tau$ are given by $[\![0,\tau]\!]$ and $[\![0,\tau)\!)$.} By assigning to any $t\in\ovT$ the constant map $\Omega\to\ovT$ with value $t$, we obtain an injection of $\ovT$ into the set of random times, and we identify $\ovT$ with its image under this injection. Note that, hence, any set $M\subseteq \ovT\times\Omega$ is $\ms F$-progressively measurable with respect to $\ms T_\ovT$ as in the definition iff, for any $t\in\ovT$, $M\cap[\![0,t]\!]\in\ms T_\ovT\otimes\ms F_t$.

As in the classical theory, the notion of adaptedness for stochastic processes alias compatibility with the filtration $\ms F$ motivates defining the following subclass of random times (with respect to $\ms I_\ovT(\T)$). An \emph{$\ms F$-stopping time} is a map $\tau\colon\Omega\to\ovT$ with $\{\tau\le t\}\in\ms F_t$ for all $t\in\ovT$. In other words, $\tau\colon\Omega\to\ovT$ is an $\ms F$-stopping time iff the process $1[\![0,\tau)\!)$ is $\ms F$-adapted. That is, by Corollary~\ref{3-SPF_VECT.cor:msI_is_generated_by_all_kinds_of_principal_up/down-sets}, $\tau$ is a random time with respect to $\ms I_\ovT(\T)$ such that at any (order-completed) time $t\in\ovT$, the given information flow $\ms F$ can tell whether $\tau$ lies not in the future. It describes information about whether a fixed event in $\ms E$ has already happened --- including the present ---, or not. To any $\ovT$-valued stopping time $\tau$, we can associate the set
\begin{equation}
    \ms F_\tau = \{ E\in\ms E \mid \forall t\in\ovT \colon E \cap \{\tau \le t\} \in \ms F_t\}.
\end{equation}

\begin{remark}\label{3-SPF_VECT.rmk:tau_&_msF_tau} 
    Standard arguments, combined with Corollary~\ref{3-SPF_VECT.cor:msI_is_generated_by_all_kinds_of_principal_up/down-sets}, show that, for any sequence $(\tau_n)_{n\in\N}$ of $\ms F$-stopping times and $\tau = \tau_0$ we have:
    \begin{enumerate}
        \item\label{3-SPF_VECT.rmk:tau_&_msF_tau.sup_of_tau_n} Scenariowise supremum $\sup_{n\in\N} \tau_n$ is an $\ms F$-stopping time;
        \item\label{3-SPF_VECT.rmk:tau_&_msF_tau.inf_of_tau_n} Scenariowise infimum $\sigma = \inf_{n\in\N} \tau_n$ is an $\ms F_+$-stopping time, and even an $\ms F$-stopping time in case $\bigcup_{n\in\N} \{\sigma = \tau_n\} = \Omega$;
        \item\label{3-SPF_VECT.rmk:tau_&_msF_tau.tau<t_in_msG_t} For all $t\in\ovT$ with $\pi(t) < {\mf w_1}$, we have $\{\tau < t\},\,\{\tau = t\} \in \ms F_t$;
        \item\label{3-SPF_VECT.rmk:tau_&_msF_tau.F_tau_sigma-alg} $\ms F_\tau$ is a sub-$\sigma$-algebra of $\ms E$ and moreover, if $(\Omega,\ms E,\ms F)$ is universally complete, then it is universally augmented in $\ms E$;
        \item\label{3-SPF_VECT.rmk:tau_&_msF_tau.msF_tau=msG_t_if_tau=t} If $\tau = t$ holds true, for some $t\in\ovT$, then $\ms F_\tau = \ms F_t$;
        \item\label{3-SPF_VECT.rmk:tau_&_msF_tau.F_tau_subseteq_F_sigma_if_tau_leq_sigma} If $\tau_0\le\tau_1$ holds true, then $\ms F_{\tau_0}\subseteq\ms F_{\tau_1}$;
        \item\label{3-SPF_VECT.rmk:tau_&_msF_tau.tau_is_msF_tau_msI_mb} $\tau$ is $\ms F_\tau$-$\ms I_\ovT(\T)$-measurable.
    \end{enumerate}
\end{remark}

The following counterexample shows that other classical results about stopping times do not generalise. Detailed verifications can be found in the appendix.

\begin{example}\label{3-SPF_VECT.ex:counterex_stopping_times}
    Suppose that $\Omega=\R_+$, equipped with Lebesgue $\sigma$-algebra $\ms E$ and the exponential distribution $\P$ with parameter $1$, i.e.\ the identity $\sigma$ on $\R_+$ satisfies $\P(\sigma > t) = \op e^{-t}$, $t\in\R_+$. We can see $\sigma$ as the set-theoretic inclusion map $\R_+ \to \ovT$.
    Let $\ms F = (\ms F_t)_{t\in\ovT}$ be the $\P$-augmented filtration generated by $\sigma$, i.e.\ the collections $\ms F_t$ of sets $\{\sigma\le s\}$, $s\le t$, ranging over $t\in\ovT$.
    Let $V\subseteq \R_+$. Define $\tau\colon\Omega\to\ovT$ by letting $\tau(\omega) = \sigma(\omega)$ if $\omega\in V$, and $\tau(\omega) = (\sigma(\omega),1)$ else. Then the following statements hold true:
    \begin{enumerate}
        \item\label{3-SPF_VECT.ex:counterex_stopping_times.stopping_time} $\tau$ is an $\ms F$-stopping time;
        \item\label{3-SPF_VECT.ex:counterex_stopping_times.distribution} $\P_\tau = \P_\sigma$ on $\ms I_\ovT(\T)$;
        \item\label{3-SPF_VECT.ex:counterex_stopping_times.order} $\sigma\le\tau$ and $\{\sigma = \tau\} = V$;
        \item\label{3-SPF_VECT.ex:counterex_stopping_times.V_non_Lebesgue} If $V\notin\ms E$, then:
        \begin{enumerate}
            \item\label{3-SPF_VECT.ex:counterex_stopping_times.V_non_Lebesgue.tau_is_not_msF_tau_msT_mb} $\tau$ is not $\ms F_\tau$-$\ms P_\ovT$-measurable;
            \item\label{3-SPF_VECT.ex:counterex_stopping_times.V_non_Lebesgue.[[tau]]} $[\![\tau]\!]$, $[\![0,\tau)\!)$, and $(\!(\tau,\infty]\!]$ are not $\ms F$-progressively measurable (i.e.\ not even with respect to $\ms P_\ovT$).
        \end{enumerate} 
    \end{enumerate}
    Note that Part~\ref{3-SPF_VECT.ex:counterex_stopping_times.V_non_Lebesgue} is not void, since $\ms E \subsetneq \mc P\R_+$ (Vitali).
\end{example}

We conclude that there are $\ms F$-adapted processes with càg paths that are not $\ms F$-progressively measurable (i.e.\ not even with respect to $\ms P_\ovT$), for instance, $1(\!(\tau,\infty]\!]$ from Example~\ref{3-SPF_VECT.ex:counterex_stopping_times}, Part~\ref{3-SPF_VECT.ex:counterex_stopping_times.V_non_Lebesgue}. Moreover, the example demonstrates that two stopping times can have the same distribution, be ordered and still be different in any scenario (case $V=\emptyset$). We can also infer from Item~\ref{3-SPF_VECT.ex:counterex_stopping_times.V_non_Lebesgue.tau_is_not_msF_tau_msT_mb}, combined with Remark~\ref{3-SPF_VECT.rmk:tau_&_msF_tau}, Item~\ref{3-SPF_VECT.rmk:tau_&_msF_tau.tau_is_msF_tau_msI_mb}, that the inclusion in Lemma~\ref{3-SPF_VECT.lemma:msI_ovT_subseteq_msP_ovT} is strict:
\begin{corollary}
    We have $\ms I_\ovT(\T)\subsetneq\ms P_\ovT$.\qed
\end{corollary}

\subsection{Optional times}

The preceding discussion raises the question whether there is a natural subclass of stopping times exhibiting stronger measurability properties. This can mean different things: the measurability of the stopping time itself and the progressive measurability of its converse graph, epi- and hypograph. Making the $\sigma$-algebra on $\ovT$ larger works against the first, but in favour of the second requirement. Thus, we may rephrase the questions as whether there is a $\sigma$-algebra that solves this trade-off. Indeed, there is one, namely the projection $\sigma$-algebra, as the following theorem affirms.

\begin{thm}\label{3-SPF_VECT.thm:optional_times}
    For a map $\tau\colon\Omega\to\ovT$ consider the following seven statements.
    \begin{enumerate}
        \item\label{3-SPF_VECT.thm:optional_times.[tau]} The converse graph $[\![\tau]\!]$ is $\ms F$-progressively measurable.
        \item\label{3-SPF_VECT.thm:optional_times.[0,tau)} The converse strict hypograph $[\![0,\tau)\!)$ is $\ms F$-progressively measurable.
        \item\label{3-SPF_VECT.thm:optional_times.[tau,infty]} The converse weak epigraph $[\![\tau,\infty]\!]$ is $\ms F$-progressively measurable.
        \item\label{3-SPF_VECT.thm:optional_times.(tau,infty]} The converse strict epigraph $(\!(\tau,\infty]\!]$ is $\ms F$-progressively measurable and $\pi\circ\tau < {\mf w_1}$.
        \item\label{3-SPF_VECT.thm:optional_times.[0,tau]} The converse weak hypograph $[\![0,\tau]\!]$ is $\ms F$-progressively measurable and $\pi\circ\tau < {\mf w_1}$.
        \item\label{3-SPF_VECT.thm:optional_times.pi_bounded,pi_tau_tau-mb} There is $\alpha\in{\mf w_1}$ such that $\pi\circ\tau\le\alpha$ and, for all $\beta\in\alpha+1$ and $t\in\ovT$:
        \[ \{\pi\circ\tau = \beta,\, \tau\le t\} \,\in\ms F_t. \]
        \item\label{3-SPF_VECT.thm:optional_times.stopping_times,pi_bounded,tau_tau-mb} $\tau$ is an $\ms F$-stopping time, there is $\alpha\in{\mf w_1}$ such that $\pi\circ\tau \le \alpha$, and $\tau$ is $\ms F_\tau$-$\ms P_\ovT$-measurable.
    \end{enumerate}
    The following holds true:
    \begin{itemize}
        \item Statements~\ref{3-SPF_VECT.thm:optional_times.[0,tau)} and~\ref{3-SPF_VECT.thm:optional_times.[tau,infty]} are equivalent, and statements ~\ref{3-SPF_VECT.thm:optional_times.(tau,infty]} and~\ref{3-SPF_VECT.thm:optional_times.[0,tau]} are equivalent.
        \item Statements~\ref{3-SPF_VECT.thm:optional_times.pi_bounded,pi_tau_tau-mb} and~\ref{3-SPF_VECT.thm:optional_times.stopping_times,pi_bounded,tau_tau-mb} are equivalent, and they imply all the others.
        \item If $(\Omega,\ms E,\ms F)$ is universally complete, then all statements are equivalent.
    \end{itemize}
\end{thm}

\begin{definition}
    An \emph{$\ms F$-optional time} is an $\ms F$-stopping time satisfying $\pi\circ\tau \le \alpha$ for some $\alpha\in{\mf w_1}$ and being $\ms F_\tau$-$\ms P_\ovT$-measurable.
\end{definition}

\begin{remark}\label{3-SPF_VECT.rmk:real_valued_optional_times}
    A map $\tau\colon \Omega\to\bRp$ is an $(\ms F_t)_{t\in\R_+}$-optional time, or equivalently, an $(\ms F_t)_{t\in\R_+}$-stopping time, both in the classical sense (see e.g.\ Dellacherie--Meyer \cite{Dellacherie1978Probabilities}, Kallenberg \cite{Kallenberg2021Foundations}), iff it is an $\ms F$-stopping time in the extended sense, iff it is an $\ms F$-optional time. In the extended setting, the notion of optional times is stronger than that of stopping times. 
    
    Example~\ref{3-SPF_VECT.ex:counterex_stopping_times} together with Theorem~\ref{3-SPF_VECT.thm:optional_times} already illustrates the difference. On a more abstract level, a stopping time is a random time such that one can tell whether it is already over or not; this time can indeed be thought of as the time some given process on $(\Omega,\ms E)$ stops.
    
    In addition, an optional time lies in $\T$, unless it takes the value $\infty$, meaning ``never''; more precisely, even on $\T_{\alpha+1}$ for some countable $\alpha\in\mf w_1$. Recalling that the vertical axis stands for accumulating embedding of well-orders into $\R_+$ and that any well-order embedded into $\R_+$ must be countable, we see that a fixed optional time must be thinkable in terms of one such an embedding (or at most countably many) --- the discussion in Subsection~\ref{3-SPF_VECT.subs:tilting_convergence} and, in particular,  Proposition~\ref{3-SPF_VECT.prop:optional_times_are_tilting_limits_of_classical_very_simple_optional_procs} will make this precise. Moreover, the time itself must be measurable at the time of its realisation, with respect to the $\sigma$-algebra $\ms P_\ovT$, which takes into account information about the vertical time coordinate as well. In total, optional times describe times where one can revise options and make decisions in a way compatible with the given information flow. This is applied and further elaborated on in Section~\ref{3-SPF_VECT.sec:SPF} to formulate a general notion of ``information sets'' and ``subgames'' in a stochastic process-based game-theoretic model.
\end{remark}

\begin{corollary}\label{3-SPF_VECT.cor:xi_tau_is_F_tau_mb}
    If $(\Omega,\ms E,\ms F)$ is universally complete, $\xi\colon\ovT\times\Omega\to Y$ is $\ms F$-progressively measurable, and $\tau$ is an $\ms F$-optional time, then 
    \[ \xi_\tau\colon\Omega\to Y,\, \omega\mapsto \xi_{\tau(\omega)}(\omega) \]
    is $\ms F_\tau$-$\ms Y$-measurable.
\end{corollary}

the following propositions underline the importance of optional times, illustrate the utility of the previous theorem, and moreover prove useful in some of the upcoming proofs.
\begin{proposition}\label{3-SPF_VECT.prop:basic_properties_of_optional_times}
    Let $(\tau_n)_{n\in\N}$ be a sequence of $\ms F$-optional times and $\tau\colon\Omega\to\ovT$ be a map. The following statements hold true:
    \begin{enumerate}
        \item\label{3-SPF_VECT.prop:basic_properties_of_optional_times.translated_stopping_times} If $\tau$ is an $\ms F$-stopping time and $\e\ge0$ is real, then $p\circ\tau + \e$ is an $\ms F_\Plus$-optional time, and provided $\e>0$, it is even an $\ms F$-optional time.
        \item\label{3-SPF_VECT.prop:basic_properties_of_optional_times.lim} If $\alpha\in{\mf w_1}$ is such that, for all $n\in\N$, $\pi\circ\tau_n \le \alpha$, and 
        \[ \tau(\omega) = 
            \begin{cases}
                \lim_{n\to\infty} \tau_n(\omega), &\text{ if this limit exists,} \\
                \infty, &\text{ else, }
            \end{cases}
            \qquad \omega\in\Omega,
        \]
        the limit being taken in the Polish space $\bRp \times (\alpha+1)$, then $\tau$ is an $\ms F_\Plus$-optional time.
        \item\label{3-SPF_VECT.prop:basic_properties_of_optional_times.comparable} We have $\{\tau_0\le\tau_1\}\in \ms F_{\tau_0}\cap \ms F_{\tau_1}$.
        \item\label{3-SPF_VECT.prop:basic_properties_of_optional_times.max_min} $\tau_0\wedge\tau_1$ and $\tau_0\vee\tau_1$ are $\ms F$-optional times.
    \end{enumerate}      
\end{proposition}
In general, the proofs are a bit tedious. In the universally complete setting, they can be substantially simplified. This is discussed in the appendix, together with the proofs.\smallskip

As an illustration, we present the following result about approximating optional times with respect to the augmented filtration by optional times with respect to the right-limit of the original one --- the case of $\bRp$-valued $\ovl\tau$ is well-known (see, e.g., \cite{Sharpe1988General}).
\begin{proposition}\label{3-SPF_VECT.prop:any_ovl_msF-optional_time_has_msF-optional_time_version}
    Suppose that $\ms E$ is universally complete and let $\ovl{\ms F} = (\ovl{\ms F_t})_{t\in\ovT}$. Then, for any $\P\in\mf P_{\ms E}$ and any $\ovl{\ms F}$-optional time $\ovl\tau$ there is an $\ms F_\Plus$-optional time $\tau$ with $\P(\tau = \ovl\tau) = 1$.
\end{proposition}

A large class of stopping and optional times can be constructed as follows. For a set $M\subseteq\ovT\times\Omega$, let
\[ D_M\colon \Omega\to\ovT,\,\omega\mapsto \inf\{t\in\ovT \mid (t,\omega)\in M\} \]
be its \emph{d\'ebut} alias entry time. For example, $M = \{\xi\in B\}$ for a $Y$-valued stochastic process $\xi$ and a measurable set $B\in\ms Y$. Recall that $\xi$ is $\ms F$-progressively measurable iff $M$ is for any $B\in\ms Y$.  

\begin{thm}[Début]\label{3-SPF_VECT.thm:debut}
    Suppose that $\ms E$ is universally complete. Let $M\subseteq\ovT\times\Omega$ be $\ms F$-progressively measurable. Then, $D_M$ is an $\ovl{\ms F}_+$-stopping time. Moreover, the following conditions are equivalent:
    \begin{enumerate}
        \item\label{3-SPF_VECT.thm:debut.DM_optional_time} $D_M$ is an $\ovl{\ms F}$-optional time.
        \item\label{3-SPF_VECT.thm:debut.pi_circ_DM<omega1} $\pi\circ D_M < {\mf w_1}$.
        \item\label{3-SPF_VECT.thm:debut.[DM]_cap_[0,infty)_subseteq_M} $[\![D_M]\!]\cap[\![0,\infty)\!)\subseteq M$.
    \end{enumerate}
\end{thm}

\begin{remark}\label{3-SPF_VECT.rmk:optional_times_and_debuts}
    Note that, by Theorem~\ref{3-SPF_VECT.thm:optional_times}, conversely, any optional time $\tau$ is of the form $D_M$ with progressively measurable $M$, namely, with $M = [\![\tau,\infty]\!]$. This sheds further light on the interpretation of optional times. Namely, $D_M$ being an $\ovl{\ms F}$-optional time means that the option-revising or decision-making agent really does so \emph{at} time $\tau$, and not only in the infinitesimal future after $\tau$. In classical continuous time $\R_+$, an analogous statement is not true for optional times. For instance, the début of Brownian motion $\xi$ into an open set $U$ is the smallest time $\tau$ such that infinitesimally after this the Brownian motion enters $U$ --- but, typically,\footnote{That is, if $\xi$ does not already start in $U$, i.e.\ $\xi_0\notin U$, ...} $\xi_\tau\notin U$! Hence, the notion of the first entry time is problematic: it actually does not exist in this situation, there is only a greatest lower bound on all times $t$ with $\xi_t\in U$. 

    Not so for optional times in vertically extended continuous time! Here, Property~\ref{3-SPF_VECT.thm:debut.[DM]_cap_[0,infty)_subseteq_M} in Theorem~\ref{3-SPF_VECT.thm:debut} essentially expresses the fact that if in scenario $\omega\in\Omega$ the set $M$ is reached at some (finite or infinite) time, then it is reached at the infimal time $D_M$, i.e.\ $(D_M(\omega),\omega)\in M$.
\end{remark}

\subsection{Optional processes}

So far we have focused on the basic case of stopping times and corresponding processes. We continue with discussing a classes of progressively measurable processes describing decision making at optional times. By Theorem~\ref{3-SPF_VECT.thm:optional_times}, the $\sigma$-algebras on $\ovT\times\Omega$
\begin{equation*} 
    \begin{aligned}
        \Prd(\ms F) = &~ \big\{\{0\} \times E \mid E\in\ms F_0\big\} \vee \sigma\Big([\![0,\tau]\!] \mid \tau~\ms F\text{-optional time}\Big), \\
        \Opt(\ms F) = &~ \big\{\{\infty\} \times E \mid E\in\ms F_\infty\big\} \vee\sigma\Big([\![0,\tau)\!) \mid \tau~\ms F\text{-optional time}\Big),
    \end{aligned}
\end{equation*}
are contained in $\Prg(\ms F)$. In analogy with the classical case, call a subset $M\subseteq\ovT\times\Omega$ \emph{$\ms F$-predictable} (\emph{$\ms F$-optional}) iff $M\in \Prd(\ms F)$ ($M\in \Opt(\ms F)$, respectively); and idem for a stochastic process $\xi\colon\ovT\times\Omega\to Y$ iff this holds true for any $M$ of the form $M = \xi^{-1}(B)$, $B\in\ms Y$. As in the classical case, predictability implies optionality:
\begin{lemma}\label{3-SPF_VECT.lemma:Prd_subseteq_Opt}
    We have~ $\Prd(\ms F) \subseteq \Opt(\ms F)\subseteq\Prg(\ms F)$.
\end{lemma}

An important and illustrative example of optional processes is the following. Let us call \emph{real-valued simple $\ms F$-optional process} any map $\xi\colon\ovT\times\Omega\to \R$ of the form 
\begin{equation}\label{3-SPF_VECT.eq:Opt_proc_generated_by_xi*1[tau,sigma)}
    \xi = \xi^\alpha \circ \prj_\Omega\, 1[\![\tau_\alpha]\!] + \sum_{\beta\in\alpha} \xi^{\beta}\circ\prj_\Omega\, 1[\![\tau_{\beta},\tau_{\beta+1})\!)\,,
\end{equation}
for a countable ordinal $\alpha\in{\mf w_1}$, a family $(\tau_\beta)_{\beta\in\alpha+1}$ of $\ms F$-optional times with $\tau_0=0$, $\tau_\alpha=\infty$, $\tau_{\beta}\le\tau_{\gamma}$ for all $\beta,\gamma\in\alpha+1$ with $\beta\le\gamma$, and $\tau_\gamma = \sup_{\beta\in\gamma} \tau_\beta$ for all limit ordinals $\gamma\in\alpha+1$, and a family $(\xi^\beta)_{\beta\in\alpha+1}$ of real-valued $\ms F_{\tau_\beta}$-measurable $\xi^\beta$. 
Given a Polish space $Y$, a \emph{simple $\ms F$-optional process} is a map $\xi\colon\ovT\times\Omega\to Y$ such that for any measurable map $\p\colon Y \to \R$, $\p\circ\xi$ is a real-valued simple $\ms F$-optional process.

Let us briefly note that, as a special case, real-valued simple $\ms F$-optional processes contain what, in view of the classical stochastic analysis literature (see, e.g., \cite{Dellacherie1978Probabilities,Kallenberg2021Foundations}), we might call \emph{real-valued simple $\ms F$-predictable processes}. That is the case where $\tau_1 = (0,1)$, $\alpha = \gamma+1$ for some $\gamma\in\mf w_1$, $\xi^\gamma = \xi^\alpha$, and for all $\beta\in\alpha\setminus\{0\}$, there is an $\ms F$-optional time $\sigma_\beta$ with $p\circ\tau_\beta = p\circ\sigma_\beta$ and $\pi\circ\tau_\beta = \pi\circ\sigma_\beta + 1$, implying $[\![\tau_0,\tau_1)\!) = [\![0]\!]$, $[\![\tau_\beta,\tau_{\beta+1})\!) = (\!(\sigma_{\beta},\sigma_{\beta+1}]\!]$ for $\beta\in(0,\gamma)_{\mf w_1}$, and $\xi^\alpha\circ\prj_\Omega\, 1[\![\tau_\alpha]\!] + \xi^\gamma\circ\prj_\Omega\,1[\![\tau_\gamma,\tau_{\gamma+1})\!) = \xi^\gamma\circ\prj_\Omega\, 1(\!(\sigma_\gamma,\infty]\!]$.

\begin{lemma}\label{3-SPF_VECT.lemma:Opt_proc_generated_by_xi*1[tau,sigma)}
    $\Opt(\ms F)$ is the smallest $\sigma$-algebra $\ms M$ on $\ovT\times\Omega$ such that all real-valued simple $\ms F$-optional processes are $\ms M$-measurable.
\end{lemma}
\begin{lemma}\label{3-SPF_VECT.lemma:Opt_proc_generated_by_xi*1[tau,sigma)_2}
    The set of real-valued $\ms F$-optional processes is the smallest set of maps $\ovT\times\Omega\to\R$ a) containing $1[\![0,\tau)\!)$ for all $\ms F$-optional times and $1(\{\infty\} \times E)$ for all $E\in\ms F_\infty$, b) closed under pointwise addition and real scalar multiplication, and c) closed under pointwise convergence.
\end{lemma}

Similar results may be obtained for predictable processes. By definition, optional (and predictable) processes can be seen as a limit object of locally right-constant sequential decision making, progressively measurable with respect to the information flow. Note that we further elaborate on this in the following Subsection~\ref{3-SPF_VECT.subs:tilting_convergence}. Here, for predictable processes, the approximators are such that any action can be predicted (see the discussion above about simple predictable processes). Generalising these concepts to vertically extended real time allows to fully and consistently model an agent's capacity of sequential instantaneous re- or proaction with respect to information flow, including information progressively revealed only during this ``instantaneous'' process (of course, it is only instantaneous in the $\bRp$-coordinate). 

\begin{remark}
    \begin{enumerate}
        \item In view of the preceding discussion and Lemma~\ref{3-SPF_VECT.lemma:Prd_subseteq_Opt}, one can further develop the subtleness of re- or proaction with respect to information flow in the framework of vertically extended time, by considering generalised \emph{Meyer $\sigma$-algebras}, i.e.\ $\sigma$-algebras $\ms M$ on $\ovT\times\Omega$ satisfying
        \[ \Prd(\ms F) \subseteq \ms M \subseteq \Opt(\ms F). \]
        These objects have been introduced in classical continuous times by \cite{Lenglart1980Tribus}, further developed in both theory and applications in \cite{Bank2019Lenglarts, Bank2020Modelling, Bank2024Optimal, Bank2025How}.
        A detailed mathematical development of the related stochastic analysis in vertically extended continuous time is beyond the scope of the present work. It seems worth further inquiry, because it naturally arises in control- and game-theoretic models in order to mix predictable and optional decision making with respect to different sources of information (see Subsection~\ref{3-SPF_VECT.subs:SPF}).
        \item The notions of predictability and optionality correspond to the classical notions when supposing $\ms F$ and $\xi$ such that $\ms F_t = \ms F_{p(t)}$ and $\xi_t = \xi_{p(t)}$ for all $t\in\ovT$.
    \end{enumerate}
\end{remark}

The following proposition provides a hierarchical description of optional processes in the spirit of descriptive set theory (see \cite{Kechris1995Classical}). This allows to better understand the measurability of optional processes at all $t\in\ovT$ with $\pi(t) = {\mf w_1}$. For this, let us introduce the following notation. Let $\mc V_0$ denote the real vector space, with pointwise addition and scalar multiplication, generated by all maps of the form
\begin{equation}\label{3-SPF_VECT.eq:def.mcV0}
    1[\![0,\tau)\!), \quad 1\{\infty\} \times E, \qquad\qquad \tau ~\ms F\text{-optional time},\, E\in\ms F_\infty.
\end{equation}
For any ordinal $\alpha\in{\mf w_1}$, let $\mc V_{\alpha+1}$ be the set of pointwise limits of $\mc V_\alpha$-valued sequences. For any limit ordinal $\alpha\in{\mf w_1}$, let $\mc V_\alpha = \bigcup_{\beta \in\alpha}\mc V_\beta$.

By transfinite induction, one shows that for all $\alpha,\beta\in{\mf w_1}+1$ with $\alpha\le\beta$, we have $\mc V_\alpha \subseteq \mc V_\beta$. Moreover, for all $\alpha\in{\mf w_1}$, $\mc V_\alpha$ is an $\R$-vector space --- for if not, there would be a smallest $\alpha\in{\mf w_1}+1$ without that property, which is impossible.\footnote{The arguments are similar to those used in the context of the Borel hierarchy, see, e.g.\ \cite{Kechris1995Classical}.} We can now state and prove the announced proposition.

\begin{proposition}\label{3-SPF_VECT.prop:optional_processes=mcV_omega1}
    The set of real-valued $\ms F$-optional processes equals $\mc V_{{\mf w_1}}$.
\end{proposition}

\begin{corollary}\label{3-SPF_VECT.cor:Opt.uniform_upper_bound_on_vertical_activity}
    For any $\ms F$-optional process $\xi$ valued in a Polish space $Y$, there is $\alpha\in{\mf w_1}$ such that for all $t\in\ovT$ with $\pi(t) \ge \alpha$ and all $\omega\in\Omega$, $\xi(t,\omega) = \xi((p(t),\alpha),\omega)$ holds true. 
\end{corollary}

Note that, in view of Remark~\ref{3-SPF_VECT.rmk:prog_mb_basic_properties}, Part~\ref{3-SPF_VECT.rmk:prog_mb_basic_properties.scwise_left-constant_at_(t,mfw1)}, we already knew that for any progressively measurable process $\xi$ valued in a metrisable topological space, any $u\in\R_+$ and any $\omega\in\Omega$, there is $\alpha\in\mf{w_1}$ such that all $\beta\in\mf w_1+1$ with $\alpha\le\beta$ satisfy $\xi((u,\beta),\omega) = \xi((u,\alpha),\omega)$. By the preceding corollary, this can be considerably strengthened for optional processes: $\alpha$ can be chosen independent of both $u$ and $\omega$.

\begin{definition}\label{3-SPF_VECT.def:level_of_Opt}
    For any $\ms F$-optional process $\xi$ valued in a Polish space $Y$, its \emph{vertical level} is the smallest $\alpha\in{\mf w_1}$ such that for all $t\in\ovT$ with $\pi(t) \ge \alpha$ and all $\omega\in\Omega$, $\xi(t,\omega) = \xi((p(t),\alpha),\omega)$ holds true. The \emph{upper vertical level} of $\xi$ is the smallest $\beta\in\mf w_1$ satisfying the following property: For all $x\in\R_+$ and all $\omega\in\Omega$, there is $\alpha\in\beta$ such that for all $t\in\ovT$ with $p(t) = x$ and $\pi(t) \ge \alpha$, we have $\xi(t,\omega) = \xi((x,\alpha),\omega)$.
\end{definition}

\begin{remark}
    Note that the vertical level is inferior to the upper vertical level. Let $\alpha\in\mf w_1$. If $\alpha$ is the vertical level of $\xi$, then its upper vertical level is either $\alpha$ or $\alpha+1$. If $\alpha+1$ is the upper vertical level of $\xi$, then $\alpha$ is the vertical level of $\xi$. So, the upper vertical level is a more general and flexible concept than that of the vertical level, while the latter is a bit more accessible.
\end{remark}

\subsection{Tilting convergence}\label{3-SPF_VECT.subs:tilting_convergence}
In this subsection, we show that optional processes in vertically extended time arise by closing the set of classical, very simple optional processes with respect to binary continuous operations and \emph{natural} limit procedures. Given a Polish space $Y$, a \emph{very simple $\ms F$-optional process} is a map $\xi\colon\ovT\times\Omega\to Y$ such that for any measurable map $\p\colon Y \to \R$, $\p\circ\xi$ is a real-valued simple $\ms F$-optional process admitting a representation according to Equation~\ref{3-SPF_VECT.eq:Opt_proc_generated_by_xi*1[tau,sigma)} with deterministic $\tau_\beta$, for all $\beta\in\alpha+1$. We recall that a process $\xi\colon\ovT\times\Omega\to Y$ is said \emph{classical} iff $\xi = \xi\circ (p\times\id_\Omega)$. A classical, very simple $\ms F$-optional process can be represented as just mentioned with deterministic, $\bRp$-valued $\tau_\beta$.

The subtle point is to clarify what ``natural'' limit procedure means in our context and to formulate a suitable notion of convergence. Pointwise convergence has already been considered in the preceding section --- by Lemma \ref{3-SPF_VECT.lemma:Opt_proc_generated_by_xi*1[tau,sigma)_2}, optional processes are essentially generated by simple processes via pointwise limits. Stability under pointwise convergence is a requirement that emanates from basic measure and integration theory: Any $\sigma$-additive measure on the sample space $\ovT\times\Omega$ that can integrate (bounded measurable) functions of all simple $\ms F$-optional processes, can also integrate (bounded measurable) functions of general $\ms F$-optional processes.

In order to obtain general simple optional processes out of classical very simple ones --- and therefore give an interpretation of (simple, then general) optional processes in vertically extended time ---, one needs an additional notion of convergence, namely one capturing infinitesimally accumulating information. With the game-theoretic background of this text, mostly developed in Section~\ref{3-SPF_VECT.sec:SPF}, this is in particular information on decisions which, under pointwise convergence, may collapse in the limit. To give a visual description of the process, imagine an infinitely long sentence written in one horizontal half-axis (think of $\R_+$). A schematic representation of this is given in Figure~\ref{fig:VECT_and_tilting_conv}, printed in the introduction. Words are indicated by points, and the position of the words at the beginning of the convergence procedure are indicated by light gray. 

Now, there is a person sitting at $+\infty$, requesting an executive summary. The person therefore pushes the sentence with infinite strength in the direction of the start of the sentence (that is, zero) --- this is the dashed arrow pointing to the left in Figure~\ref{fig:VECT_and_tilting_conv}, moving the words (indicated by points) towards the left. The movement is indicated by the points being printed in darker gray. Now, at any point $t\in\R_+$ on the half-axis, there are parts of the sentence that accumulate in its right-hand neighbourhood. Things turn out such that these parts are \emph{tilted} by 90 degrees counterclockwise as to build a vertical strip just above $t$, to be read starting from below. In Figure~\ref{fig:VECT_and_tilting_conv}, the tilting is indicated by the curved dashed arrows. The accumulation process is most easily imagined at zero; but, given a sufficient kind of periodicity of the sentence, the asymptotics work out at any $t\in\R_+$ as well. 

Note that the sentence, although written on a continuous paper roll, is well-ordered and so are the vertical strips. On the paper roll, severely reshaped by the person, a continuous structure may yet arise. This should remind us of various pointwise approximation procedure in analysis, e.g.\ the approximation of measurable functions valued in $\R_+$ by simple ones, or even by step functions in certain cases. What we see in addition here, is the preservation of information on order about ``words'' (or more generally objects) that collapse at a single position $t$ on the paper roll within the limit. The game-theoretic importance of this construction seems rather obvious then: the order of actions, whose execution times collapse to one identical ``real'' time $t\in\R_+$ in the limit, should be preserved. 

Before getting to a formal description of this concept, let us note by means of an example that: a) pointwise convergence is far too restrictive, and b) convergence of the sequence $(\xi^n)_{n\in\N}$ alone does not lead to a satisfactory notion, i.e.\ the ``convergence'' depends on the sequence of grids. 
\begin{example}\label{3-SPF_VECT.ex:tilting_conv_1}
    For this, consider the real-valued processes indexed over $n\in\Z$ given by $\xi^n = 1[\![0,2^{-n})\!)$. Imagine that the process describes the actions of some agent called Alice on the dyadic grid $G_n$ given by $k2^{-n}$, $k\in\N$. What is a limit of $(\xi^n)_{n\in\N}$? 
    The pointwise limit, as $n\to +\infty$, is $1[\![0,(0,\mf w_1)]\!]$ --- but this gives no interesting information about the (vertically extended) time at that the agent switches to value zero (in short: stops). On the other hand, for any optional time $\tau$ with $p\circ \tau = 0$ and $\pi\circ\tau > 0$, $\xi = 1[\![0,\tau)\!)$ is a limit horizontally, i.e.\ $\xi^n(t) \to \xi(t)$ at any $t\in\bRp$. As only Alice is involved, the exact choice of $\tau$ does not seem to reflect much essential information.
    
    Now, add another agent called Bob acting according to $\xi^{n-1}$ in grid $G_n$. Horizontally, the same limit obtains, and for Bob alone the choice of the vertical stopping time seems irrelevant. However, in any grid $G_n$, Alice switches to zero strictly before Bob: Alice uses the second, Bob the third opportunity to stop. In that context, the adequate limit outcome for Alice would be $\xi^A = 1[\![0,\tau^A)\!)$ with $\tau^A = (0,1)$ and, similarly, for Bob $\xi^B = 1[\![0,\tau^B)\!)$ with $\tau^B = (0,2)$.\footnote{Note that, in the von Neumann hierarchy, the second ordinal is $1$, and the third ordinal is $2$.} Hence, the game-theoretically plausible limit depends on the chosen grid sequence and not only on the sequence $(\xi^n)_{n\in\N}$. It also depends on the fact that both agents use the same grid sequence. 
    Next, suppose that, in any grid $G_n$, $n\in\N$, a third agent called Carol acts according to $\xi^{n-2}$ if $n$ is even and according to $\xi^{n}$ if $n$ is odd. Then, comparing with Bob, in any grid both agents do not stop simultaneously, but any susceptible limits $\xi^B$ for Bob and $\xi^C$ for Carol, in any scenario $\omega\in\Omega$, must either let them stop --- also vertically --- instantaneously or gives one of them the priority --- which both would remain unexplained by the approximating sequences. A similar comparison with Alice is equally inconclusive. Thus, on the vertical half-axis, there may be no plausible limit at all. This holds true despite the facts that a) all sequences do converge in the strongest possible sense (pointwise on the full domain $\ovT\times\Omega$, and they are uniformly bounded) and b) the horizontal limit is plausible. 
    Hence, for understanding ``limit behaviour along the vertical half-axis'', looking at convergence of the sequence $(\xi^n)_{n\in\N}$ alone is insufficient.
\end{example}

Based on the theoretical motivation outlined in the beginning, underlined by the preceding example, we give the following definitions.
\begin{definition}
    An \emph{$\ms F$-adapted grid} is a map $G\colon(\alpha+1)\times\Omega\to\ovT$ for some $\alpha\in\mf w_1$ such that:
    \begin{enumerate}
        \item for each $\beta\in\alpha$, $\tau^G_\beta \colon\Omega\to\ovT,\, \omega\mapsto G(\beta,\omega)$ is an $\ms F$-optional time;
        \item $\tau^G_0 = 0$, $\tau^G_{\alpha} = \infty$;
        \item for all $\beta,\gamma\in\alpha$ with $\beta < \gamma$, we have $\tau^G_\beta(\omega) < \tau^G_\gamma(\omega)$ for all $\omega\in\{\tau^G_\beta < \infty\}$;
        \item for all limit ordinals $\gamma\in\alpha+1$ and all $\omega\in\Omega$, we have $G(\gamma,\omega) = \sup_{\beta\in\gamma} G(\beta,\omega)$.
    \end{enumerate}
    An $\ms F$-adapted grid is said \emph{classical} iff it is $\bRp$-valued. A \emph{(deterministic) grid} is an $\ms F$-random grid as above such that, for all $\beta\in\alpha$, $\tau^G_\beta$ is deterministic (i.e.\ constant).

    Given two $\ms F$-adapted grids $G\colon(\alpha+1)\times\Omega\to\ovT$, $G'\colon(\alpha'+1)\times\Omega\to\ovT$, $G'$ is said to \emph{refine} or to be a \emph{refinement of} $G$ iff there is an order-embedding $j\colon \alpha+1\inj\alpha'+1$ such that $G = G'\circ (j\times\id_\Omega)$. For any $\ms F$-adapted grid $G\colon(\alpha+1)\times\Omega\to\ovT$, let the \emph{grid size} at $\omega\in\Omega$ be given by
    \[ \Delta(G,\omega) = \begin{cases} \infty, & \text{ if } \sup_{\beta\in\alpha} G(\beta,\omega) < \infty, \\ \sup_{\substack{\beta\in\alpha\colon\\ \tau^G_\beta(\omega) < \infty}} \big(p\circ G(\beta+1,\omega) - p\circ G(\beta,\omega)\big), & \text{ else.} \end{cases} \]
    
    Let $(G_n)_{n\in\N}$ be a sequence of $\ms F$-adapted grids. It is said \emph{refining} iff for all $n\in\N$, $G_{n+1}$ refines $G_n$. It is said \emph{(pointwise uniformly) convergent} if, for every $\omega\in\Omega$, $\Delta(G_n,\omega) \to 0$ as $n\to\infty$.
\end{definition}

\begin{remark}
    Any $\ms F$-adapted grid $G\colon(\alpha+1)\times\Omega\to\ovT$ can be seen as a map $\mf w_1 \times \Omega\to\ovT$, by letting $G(\beta,\omega) = \infty$ for all arguments $(\beta,\omega)\in [\mf w_1 \setminus (\alpha+1)] \times\Omega$. We use this convention in the following.
\end{remark}

Simple / very simple (classical) $\ms F$-optional processes are defined via $\ms F$-adapted / deterministic (classical) grids, respectively. Let us fix a name for that.
\begin{definition}
    Let $Y$ be a Polish space and $\xi'$ be a simple $\ms F$-optional process valued in $Y$. An $\ms F$-adapted grid $G\colon(\alpha+1)\times\Omega\to \ovT$ is said \emph{compatible with $\xi'$} iff there are a real-valued simple $\ms F$-optional process $\xi$, given by Equation~\ref{3-SPF_VECT.eq:Opt_proc_generated_by_xi*1[tau,sigma)}, with $\tau_\beta = G(\beta,.)$ for all $\beta\in\alpha+1$, and measurable $\p\colon\R\to Y$ such that $\xi' = \p\circ \xi$.
\end{definition}

In other words: Compatibility essentially means that all jump times of $\xi$ are part of grid. By definition, for any simple (very simple) $\ms F$-optional process, there is a compatible $\ms F$-adapted (deterministic) grid. For simple $\ms F$-optional processes, there is even a smallest such grid, provided measure-theoretic completeness:

\begin{lemma}\label{3-SPF_VECT.lemma:simple_opt_proc_induced_grid}
    Suppose that $(\Omega,\ms E,\ms F)$ is universally complete. Let $Y$ be a Polish space and $\xi$ be a simple $\ms F$-optional process valued in $Y$. Then, there is an $\ms F$-adapted grid $G\colon(\alpha+1)\times\Omega\to\ovT$ satisfying, for all $\beta\in\alpha$ and $\omega\in\Omega$:
    \begin{equation}\label{3-SPF_VECT.eq:compatibility_simple_proc_grid}
        G(\beta+1,\omega) = \inf \{t\in [G(\beta,\omega),\infty]_\ovT \mid \xi_t(\omega) \neq \xi_{G(\beta,\omega)}(\omega) \}.
    \end{equation}
    Moreover, $\xi$ can be represented as in Equation~\ref{3-SPF_VECT.eq:Opt_proc_generated_by_xi*1[tau,sigma)} with $\tau_\beta = G(\beta,.)$ and $\xi^\beta = \xi_\beta$ for all $\beta\in\alpha+1$. If $\xi$ is classical, then so is $G$. 
\end{lemma}

It is easily shown using transfinite induction that if there is an order embedding $\alpha\inj\alpha'$ of one ordinal $\alpha$ into another $\alpha'$, then $\alpha\le\alpha'$. As an immediate consequence of this, if $G,G'$ are as in the definition such that $G'$ refines $G$, then $\alpha\le\alpha'$. As another consequence, we obtain the following lemma.

\begin{lemma}\label{3-SPF_VECT.lemma:refining,convergent_grid->psi,delta,gamma}
    Let $(G_n)_{n\in\N}$ be a refining, convergent sequence of $\ms F$-adapted grids $G_n\colon(\alpha_n+1)\times\Omega\to\ovT$ and let $(t,\omega)\in\bRp\times\Omega$. Then, 
    \begin{enumerate}
        \item\label{3-SPF_VECT.lemma:refining,convergent_grid->psi,delta,gamma.exists_delta^n_and_psi^n} for any $n\in\N$, there is a unique ordinal $\delta^n(t,\omega)$ admitting an order isomorphism \[ \psi^n(t,\omega)\colon \delta^n(t,\omega)+1 \to \{ \beta\in\alpha_n+1 \mid G_n(\beta,\omega) \ge t \}, \] and, moreover, this order isomorphism is unique and given by 
        \[ \beta'\mapsto \inf \{ \beta\in\alpha_n+1 \mid G_n(\beta,\omega) \ge t \} +\beta';\]
        \item\label{3-SPF_VECT.lemma:refining,convergent_grid->psi,delta,gamma.delta^n_incr_in_n} for all $n\in\N$, $\delta^n(t,\omega) \le \delta^{n+1}(t,\omega)$, and $\delta(t,\omega) = \sup_{n\in\N}(\delta^n(t,\omega)+1)$ is non-zero and countable.
    \end{enumerate}
\end{lemma}

Hence, to any refining, convergent sequence of $\ms F$-adapted grids $G_n\colon(\alpha_n+1)\times\Omega\to\ovT$, $n\in\N$, we can assign maps with domain $\bRp\times\Omega$ denoted by $\psi^n$, $\delta^n$, $\delta$, and $\gamma$ and given as in the lemma and by the following formula\footnote{The following infimum is computed in the complete lattice $\delta(t,\omega)+1$.}
\begin{equation}\label{3-SPF_VECT.eq:def.gamma}
    \gamma(t,\omega) = \inf \{ \beta\in\delta(t,\omega) \mid \limsup_{n\to\infty} p\circ G_n(\psi^n(t,\omega)(\beta),\omega) > t \}, \quad (t,\omega)\in\bRp\times\Omega.
\end{equation}
As $(G_n)_{n\in\N}$ is convergent, $\gamma(t,\omega)$ is a limit ordinal for all $(t,\omega)\in\R_+\times\Omega$.

\begin{definition}\label{3-SPF_VECT.def:tilting_convergence}
    Let $Y$ be a Polish space, $\xi$ and $\xi^n$, $n\in\N$, be stochastic processes $\ovT\times\Omega\to Y$, and $(G_n)_{n\in\N}$ be a refining, convergent sequence of $\ms F$-adapted grids $G_n\colon(\alpha_n+1)\times\Omega\to\ovT$.

    $(\xi^n \mid G_n)_{n\in\N}$ \emph{converges tiltingly to} $\xi$, or $(\xi^n)_{n\in\N}$ \emph{converges tiltingly along $(G_n)_{n\in\N}$} or $(\xi^n \mid G_n) \convT \xi$ as $n\to\infty$, iff, for all $(t,\beta,\omega)\in\ovT\times\Omega$, we have the following convergence in $Y$:
    \begin{equation}\label{3-SPF_VECT.eq:tilting_convergence}
        \xi(t,\beta,\omega) = \begin{cases} \lim_{n\to\infty} \xi^n\Big(G_n(\psi^n(t,\omega)(\beta),\omega),\omega\Big), & \text{if } \beta\in\gamma(t,\omega), \\ \lim_{\beta' \nearrow \gamma(t,\omega)} \xi(t,\beta',\omega), &\text{else.} \end{cases}
    \end{equation}
\end{definition}

Note that, for a fixed refining, convergent sequence of $\ms F$-adapted grids $G_n$, $n\in\N$, the tilting convergence $(\xi^n\mid G_n)\convT\xi$ as $n\to\infty$ determines $\xi$ uniquely at all arguments of the form $(t,\beta,\omega)\in\ovT\times\Omega$ with $\beta\in\gamma(t,\omega)$, by the first case in Equation~\ref{3-SPF_VECT.eq:tilting_convergence}. This includes $\bRp$, but not all of $\ovT$. The values on the (uncountable) remainder the vertical half-axis above $t$ are determined by extending $\xi(.,\omega)$ left-continuously at $(t,\gamma(t,\omega))$ and then constantly until $(t,\mf w_1)$. Indeed, using the metaphor from the subsections' beginning, these arguments are not attained by the ``infinitely strong push'' initiated by the person at $+\infty$. In other words, they do not contain relevant information about the asymptotics of $(\xi^n)_{n\in\N}$ along $(G_n)_{n\in\N}$. Note that, for this to work, an asymptotic limit must exist at the right-hand end of the information that accumulates near $t$ -- formally, left-continuity of $\xi(.,\omega)$ at $(t,\gamma(t,\omega))$ is necessary.

\begin{remark}\label{3-SPF_VECT.rmk:basic_properties_tilting_convergence}
    Let $Y$ be Polish spaces, $\xi$ and $\xi^n$, $n\in\N$, be stochastic processes $\ovT\times\Omega\to Y$, and $(G_n)_{n\in\N}$ be a refining, convergent sequence of $\ms F$-adapted grids $G_n\colon(\alpha_n+1)\times\Omega\to\ovT$ such that $(\xi^n\mid G_n)\convT \xi$. 
    \begin{enumerate}
        \item If $\hat\xi\colon\ovT\times\Omega\to Y$ is another stochastic process such that $(\xi^n\mid G_n) \convT \hat\xi$, then $\hat\xi = \xi$.
        \item Let $Z$ be another Polish space and $f\colon Y\to Z$ continuous. Then, $(f\circ \xi^n \mid G_n)\convT f\circ \xi$.
        \item Let $Y'$ be another Polish space, and $\xi'$ and ${\xi'}^n$, $n\in\N$, be further stochastic processes $\ovT\times\Omega\to Y'$ such that $({\xi'}^n\mid G_n)\convT \xi'$. Then, $((\xi^n,{\xi'}^n) \mid G_n) \convT (\xi,\xi')$, with respect to the topological product $Y\times Y'$.
        \item As a consequence, if $Y'$, $\xi'$, $({\xi'}^n)_{n\in\N}$ are given as in the preceding item, and if $Y = Y'$ is also a topological vector space on $\R$ and $(a_n)_{n\in\N}$ a sequence of scalars converging to $a\in\R$, then $(\xi^n + a_n {\xi'}^n \mid G_n) \convT \xi + a \xi'$ as $n\to\infty$.
    \end{enumerate}
    Note that we refrain from embedding tilting convergence into the language of general topology; though an interesting question, this is clearly beyond the scope of this text.
\end{remark}

To the best of the author's knowledge, the notion of ``tilting convergence'' is a new contribution to both the literature on stochastic analysis and control and that on limits in continuous-time games. Stochastic analysis in vertically extended time needs a notion of convergence that is adapted to both information flow given by the filtration $\ms F$ and to the vertical extension of time. Direct extensions of classical notions (such as pointwise / almost sure, measure-, or $\L^p$-convergence) seem inappropriate for this. Regarding game theory, in \cite{Fudenberg1986Limit}, Fudenberg and Levine study approximations of outcomes in continuous-time games in terms of outcomes generated by embedded refining, convergent sequences of grids, but they do not consider instantaneous reaction. The ``discrete time with an infinitesimally fine grid'' approximation by Simon and Stinchcombe (cf.~\cite{Simon1989Extensive}) does so, but it restricts to a deterministic setting, to approximators $\xi^n$ with a finite number of jumps and with stationary actions,\footnote{That is, for large $n$, the action at the $\beta$th point of grid $G_n$ does not depend on $n$, for a fixed grid index $\beta\in\alpha_n$.} and with piecewise constant $\xi$. Having said that, formally, tilting convergence can be seen a (though broad) generalisation of the convergence implied by the metric in \cite[Section~4, p.~1185]{Simon1989Extensive}.

\begin{example}
    Reconsider Example~\ref{3-SPF_VECT.ex:tilting_conv_1}. Let $\xi^n = 1[\![0,2^{-n})\!)$, and let $G_n\colon(\mf w+1)\times\Omega \to \ovT,\ (k,\omega) \mapsto k 2^{-n}$, for any $n\in\Z$, with the understanding $\mf w 2^{-n} = \infty$.     
    Then, Alice's and Bob's behaviour on $(G_n)_{n\in\N}$ converge tiltingly: $(\xi^n \mid G_n) \convT 1[\![0]\!] = 1[\![0,(0,1))\!)$ and $(\xi^{n-1} \mid G_{n})\convT 1[\![0,(0,1)]\!] = 1[\![0,(0,2))\!)$. Carol's, however, does not. Indeed, let, for $n\in\Z$,
    \[ \tilde \xi^n = \begin{cases} \xi^{n-2}, & \text{if } 2 \mid n, \\ \xi^n, &\text{else.} \end{cases} \]
    For all $n\in\N$ and $\omega\in\Omega$, $\psi^n(0,\omega) = \id_{\mf w+1}$; whence
    \[ \tilde\xi^n\Big(G_n(\psi^n(t,\omega)(1),\omega),\omega\Big) = \tilde\xi^n(2^{-n},\omega) = \begin{cases} 1, & \text{if } 2 \mid n, \\ 0 &\text{else,} \end{cases} \]
    which does not converge in $\R$ as $n\to+\infty$.
\end{example}

\begin{example}
    We again consider Example~\ref{3-SPF_VECT.ex:tilting_conv_1} in order to illustrate the grid-dependence. Let $\xi^n = 1[\![0,2^{-n})\!)$, and let $G_n\colon(\mf w^2+1)\times\Omega \to \ovT,\ (k\mf w + m,\omega) \mapsto (k+1-2^{-m}) 2^{-n}$, for any $n\in\Z$, with the understanding $\mf w 2^{-n} = \infty$ (case $k\mf w + m = \mf w^2$ alias $(k,m) = (\mf w,0)$). 
    Then, Alice's and Bob's behaviour on $(G_n)_{n\in\N}$ converge tiltingly, but to other limits: $(\xi^n \mid G_n) \convT 1[\![0]\!] = 1[\![0,(0,\mf w))\!)$ and $(\xi^{n-1} \mid G_{n})\convT 1[\![0,(0,2\mf w))\!)$. Alice only switches at the $\mf w$th moment, Bob only at the $2\mf w$th moment. Carol's behaviour does not converge on $(G_n)_{n\in\N}$ for similar reason as those from the previous example.
\end{example}

To some extent, it is possible to represent tilting convergence in terms of pointwise convergence:
\begin{lemma}\label{3-SPF_VECT.lemma:tilting_conv_as_ptw_conv?}
    Let $\xi^n$, $n\in\N$, be real-valued stochastic processes $\ovT\times\Omega\to \R$, and $(G_n)_{n\in\N}$ be a refining, convergent sequence of $\ms F$-adapted grids $G_n\colon(\alpha_n+1)\times\Omega\to\ovT$. Let $\mathrm{L}(\On)$ denote the class of limit ordinals. Then, for all $(t,\beta,\omega)\in\ovT\times\Omega$ with $\beta\in\delta^n(t,\omega)+1$, we have:
    \begin{equation}\label{3-SPF_VECT.eq:tilde_xi^n}
    \begin{aligned}
        \xi^n\Big(G_n(\psi^n(t,\omega)(\beta),\omega),\omega\Big) &= \xi^n_{\tau^{G_n}_{\beta}}(\omega)\,  1[\![0]\!](t,0,\omega) \\
        &\quad + \sum_{\beta_0\in\alpha_n+1} \xi^n_{\tau^{G_n}_{\beta_0+1+\beta}}(\omega)\, 1(\!(\tau^{G_n}_{\beta_0},\tau^{G_n}_{\beta_0+1}]\!](t,0,\omega) \\
        &\quad + \sum_{\beta_0\in(\alpha_n+1)\cap \mathrm{L}(\On)} \xi^n_{\tau^{G_n}_{\beta_0+\beta}}(\omega)\, 1[\![\tau^{G_n}_{\beta_0}]\!](t,0,\omega).
    \end{aligned}
    \end{equation}
\end{lemma}
With this representation, we see directly that tilting convergence ``accumulates information from the future''. In particular, it is -- in general -- not correct that any sequence of $\ms F$-optional processes converging tiltingly along a sequence of refining, convergent $\ms F$-adapted grids has an $\ms F$-optional (tilting) limit. Provided the grid converges sufficiently strongly, it appears natural to hope for an $\ms F_\Plus$-optional tilting limit. Making this precise is beyond the scope of the present work.

Here, the more relevant question for us is the following: What processes can be generated out of classical, very simple $\ms F$-optional processes via tilting and pointwise convergence? 

\begin{proposition}\label{3-SPF_VECT.prop:optional_times_are_tilting_limits_of_classical_very_simple_optional_procs}
    Let $\tau$ be an $\ms F$-optional time. Then, there are a sequence $(\xi^n)_{n\in\N}$ of classical, very simple $\ms F$-optional processes and a refining, convergent sequence $(G_n)_{n\in\N}$ of classical, deterministic grids $G_n$ compatible with $\xi^n$, for all $n\in\N$, such that $(\xi^n\mid G_n) \convT 1[\![0,\tau)\!)$ as $n\to\infty$.
\end{proposition}

For the following result, let us adopt the following conventions. A set $S$ of maps $\ovT\times\Omega\to Y$ is said \emph{optionally closed under tilting convergence} iff, for all $S$-valued sequences $(\xi^n)_{n\in\N}$ of $\ms F$-optional processes, all maps $\xi\colon\ovT\times\Omega\to Y$, and all refining, convergent sequences of $\ms F$-adapted grids $(G_n)_{n\in\N}$ such that a) $(\xi^n\mid G_n)\convT \xi$ as $n\to\infty$ and b) $\xi$ is $\ms F$-optional, it necessarily holds true that $\xi\in S$. Further, a set $S$ of maps $\ovT\times\Omega\to Y$ is said \emph{closed under continuous binary operations} iff for all $\xi,\xi'\in S$ and all continuous $f\colon Y\times Y\to Y$, the map $\xi''\colon\ovT\times\Omega\to Y,\, (t,\omega)\mapsto f(\xi_t(\omega),\xi'_t(\omega))$ satisfies $\xi''\in S$.

\begin{thm}\label{3-SPF_VECT.thm:density_classical_very_simple_opt_proc}
    Let $Y$ be a Polish space. The set of $Y$-valued $\ms F$-optional processes equals the smallest set of maps $\ovT\times\Omega\to Y$ a) containing all $Y$-valued classical, very simple $\ms F$-optional processes, b) closed under continuous binary operation, c) closed under pointwise convergence, and d) optionally closed under tilting convergence.
\end{thm}

We conclude that --- in the sense made precise in this subsection --- $\ms F$-optional processes are the processes generated by all classical simple $\ms F$-optional processes defined on deterministic grids (= very simple), by means of ``continuous completion'' (continuous binary operations), ``measurable completion'' (pointwise convergence), and ``decision-theoretic completion'' (tilting convergence).

\section{Stochastic process forms}\label{3-SPF_VECT.sec:SPF}

In this section, we introduce the abstract game- and decision-theoretic model of stochastic process forms. These implement extensive form characteristics using the language of stochastic processes, giving rise to a model that encompasses much of the continuous-time stochastic control literature, including stochastic differential games and timing games, but comes as close as arguably possible to an extensive form. The stochastic process form comes with a subtle model of information flow and information sets, or ``subgames'', using techniques from stochastic analysis. In stochastic process forms, strategies are complete contingent plans of action given by one stochastic process. A minimal requirement is well-posedness, i.e.\ any strategy profile induces a unique outcome compatible with it. This gives rise to a canonical way of implementing abstract concepts of dynamic equilibrium, including perfect Bayesian and subgame-perfect equilibrium. These point are illustrated concretely by a case study of the stochastic timing game and further discussed in the context of stochastic differential games.

\subsection{Introduction of stochastic process forms}\label{3-SPF_VECT.subs:SPF}

This chapter studies game-, decision-, control-theoretic models in that action is described by stochastic processes evolving in (possibly) continuous time. As shown in Chapters~\ref{chap:1-SDF_AC} and~\ref{chap:2-SEF_G} --- and in particular in Theorem~\ref{2-SEF_G.thm:AP_sef_well-posed} --- a large class of well-posed stochastic extensive forms based on paths of action indexed over well-ordered subsets of $\R_+$ can be constructed. The construction of action path stochastic extensive forms in Chapters~\ref{chap:1-SDF_AC} and~\ref{chap:2-SEF_G} moreover reveals that the induced outcomes generate adapted processes with respect to the exogenous information flow. These adapted processes on a well-ordered grid in $\R_+$ can be equivalently seen as locally right-constant adapted processes with time index set $\R_+$.\footnote{More precisely, in the language and notation of the previous chapters, given a well-ordered subset $\tilde\T\subseteq\R_+$ with $0\in\tilde\T$, the collection of induced outcomes $(\omega,f')\in W\subseteq\Omega\times\A^{\tilde \T}$ of a strategy profile $s$, given a random move $\x = \x_t(f)$ with domain $D_\x = D_{t,f}$, $(t,f)\in\tilde\T\times\A^\T$, and given scenario $\omega\in\Omega$, can be seen as map $\R_+\times D_\x\to\A$ with locally right-constant paths jumping only at times $\tilde \T$.} Thus, we see that there is an extensive form footing to stochastic games with locally right-constant continuous-time paths of action, given a fixed grid of admissible action times.

By the classical results due to \cite{Simon1989Extensive,Stinchcombe1992Maximal} and \cite{AlosFerrer2008Trees,AlosFerrer2011Comment}, going beyond such a locally right-constant setting while remaining strictly within extensive form theory is doomed to failure. However, when starting out of well-posed stochastic extensive forms with adapted (and thus optional) locally right-constant action process with time index set $\R_+$, continuous, measurable and decision-theoretic completion yields exactly the class of general optional processes in vertically extended continuous time, by Theorem~\ref{3-SPF_VECT.thm:density_classical_very_simple_opt_proc}. This implies two things. First, in that limit sense, defining a game-theoretic form on the basis of action processes with these properties has a footing on well-posed action path stochastic extensive forms. Second, decision-theoretic generality requires to work in vertically extended continuous time, based on the stochastic analysis developed so far in this text.

Therefore, for abstract game-theoretic reasons, the question arises what game-theoretic structures obtain when we describe action by stochastic (and in particular optional) processes in vertically extended time. This is moreover motivated by the existence of a huge literature on games, decision and control problems in continuous time using stochastic processes, including timing games and differential games, in various formulations.\footnote{For a recent textbook with many examples focusing on ``mean field games'', see \cite{Carmona2018Probabilistic}.} A third reason, linked to the two previously mentioned ones, is that explaining these games in terms of their extensive form characteristics also suggests a limit theory using action path stochastic extensive forms as approximators. In a first step, however, a susceptible limit must be identified, if we are interested in more than mere existence of it. Indeed, we wish to provide an abstract and general model \emph{a priori} of the extensive form characteristics of games based on stochastic processes. As this formulation is not an extensive form and the basic structure of it are not decision trees, but stochastic processes, it receives the name \emph{stochastic process form}. 

We motivate main parts of the following definition of the stochastic process form beforehand, and continue the detailed discussion afterwards. How do we model the ``extensive form characteristics'' in a stochastic process form, defined as ``the flow of information about past choices and exogenous events, along with a set of adapted choices locally available to decision makers'' in the introduction? The flow of an agent $i$'s information about past choices and exogenous events is given by a stochastic process $\chi$ on the one hand and on the other a pair consisting of a filtration $\ms H^i$ and a $\sigma$-algebra $\ms M^i$, respectively. This stochastic process is called state process, valued in some state space $\B$, as in the control-theoretic literature. The filtration is defined on the configuration space $W = \Omega\times\B^\ovT$, which is the product of the set of exogenous scenarios $\Omega$ and the path space for the state process $\B^\ovT$, as in the Witsenhausen product form (cf.~\cite{Witsenhausen1971Information, Witsenhausen1975Intrinsic, Heymann2022Kuhns}). Departing from a product form setting, $\ms M^i$ is a $\sigma$-algebra $\ovT\times W$ with $\Prd(\ms H^i) \subseteq \ms M^i \subseteq \Opt(\ms H^i)$, describing in flexible way what pieces of information revealed at time $\tau^i$ the agent $i$ can condition her action on (roughly speaking). These measurability conditions are also a clear departure from the extensive form setting because they use properties of entire processes and not of their evaluations at fixed deterministic times.

The condition defining $\ms M^i$ reveals it as what it well-known in stochastic analysis as \emph{Meyer $\sigma$-algebra} (a.k.a.\ $\sigma$-field) with respect to $\ms H^i$, introduced and developed in \cite{Lenglart1980Tribus,ElKaroui1981Les,Bank2019Lenglarts} in the classical setting; for recent work on applications to stochastic control, see, e.g.\ \cite{Bank2025How,Bank2024Optimal,Bank2020Modelling}. Following this literature, $\ms M^i$ allows to express the amount of information revealed ``at time $\tau^i$'' agent $i$ can use for action at time $\tau^i$, as opposed to action at times succeeding $\tau^i$. The presented framework of Meyer $\sigma$-algebras in vertically extended time permits to use the power of classical Meyer $\sigma$-algebras in describing action with respect to information at a given time in settings where longer well-ordered chains of instantaneous pro- and reaction are relevant, including games.

The set of choices locally available to decision makers is described by stochastic processes $s^i\colon\ovT\times W \to \A^i$, valued in some personal action space $\A^i$, and called strategies. These are thus formal primitives, but they are required to be $\ms H^i$-progressively measurable and ``locally'' $\ms M^i$-measurable processes. The meaning of ``locally'' is subtle: what are these loci alias decision points? When equipped with the $\sigma$-algebra $\ms M^i$, agent $i$ can check her options at any $\ms H^i$-optional time $\tau^i$ such that $[\![0,\tau^i)\!)\in\ms M^i$; for exactly in that case, she can really follow the abstract strategy $1[\![0,\tau^i)\!)$ of opting for value zero at time $\tau^i$. At time $\tau^i$, agent $i$ can observe the state process $\tilde\chi$ up to time $\tau^i$, that is everything she can see of $\ms M^i$-measurable functions (thus, possible strategies) of it up to time $\tau^i$. However, already optional times equipped with pointwise order do clearly not define a tree or forest; this is a clear departure from the extensive form. Still, we obtain a notion modelling instances $(\tau^i,\tilde\chi)$ at that choices alias options are available to agents, and strategies can be seen as complete contingent plans of action at all these instances, compatible with the information structure $\ms M^i$. We nevertheless insist on the difference to extensive forms where strategies are all complete contingent plans of locally available choices, without any condition on measurability along the time axis, or more precisely, over option-revision instances $(\tau^i,\tilde\chi)$. In the stochastic process form setting, it is the $\ms H^i$-progressive measurability and ``local'' $\ms M^i$-measurability that imply such a condition.

Based on this, one can also develop a notion of outcome, or actually, state processes induced by strategy profiles, given a starting point $(\tau^i,\tilde\chi)$ as above. Then, a minimal requirement for a stochastic process form in order to give rise to a proper game-theoretic model is well-posedness: that any strategy profile induces a unique state process, given any starting point. 

With these preparations, we introduce the formal definition. In what follows, $\ovT$ denotes vertically extended time as introduced in Subsection~\ref{3-SPF_VECT.subs:vERT}. In addition, we fix a measurable space $(\Omega,\ms E)$ with $\Omega\neq\emptyset$. The elements of $\Omega$ represent \emph{exogenous scenarios}, those of $\ms E$ \emph{events}.\footnote{In the language of the previous chapters, $(\Omega,\ms E)$ is an exogenous scenario space.} Moreover, fix a $\sigma$-ideal $\ms N$ on $\ms E$, that is a non-empty and strict subset of $\ms E$, stable under both intersection with elements of $\ms E$ and countable union. The relevant example for this is $\ms N = \ms E \cap \bigcap_{\P\in\mf P} \ms N_\P$ for a non-empty set $\mf P\subseteq \mf P_{\ms E}$ of prior beliefs alias probability measures on $\ms E$. This also includes the case $\ms N = \{\emptyset\}$.
Given this $\sigma$-ideal $\ms N$, we say that a property holds for \emph{$\ms N$-almost all $\omega\in\Omega$} or \emph{$\ms N$-almost surely} iff there is $N\in\ms N$ such that the property holds for all $\omega\in N^\complement$. For any set $S\subseteq\ovT\times\Omega$, two maps $\chi,\chi'\colon S\to Y$ are said \emph{$\ms N$-indistinguishable}, denoted by $\chi\cong_\ms N\chi'$ if $\ms N$-almost all $\omega\in\Omega$ satisfy the following property: for all $t\in\ovT$ with $(t,\omega)\in S$ we have $\chi_t(\omega) = \chi'_t(\omega)$. If $\ms N = \ms E \cap \ms N_\P$ for some $\P\in\mf P_{\ms E}$, then ``$\ms N$'' is replaced by ``$\P$'' in these phrases, as usual.
\begin{definition}\label{3-SPF_VECT.def:spf}
    \textsc{(Part A)}:~ For fixed $(\Omega,\ms E,\ms N)$, consider the data 
    \[ \mathbf F = (I,\A,\B,W,\mc W,\ms H,\ms M,\mc S), \]
    where:
    \begin{itemize}
        \item $I$ is a non-empty, finite set --- its elements are called \emph{agents};
        \item $\A = \prod_{i\in I} \A^i$ is the topological product of Polish spaces $\A^i$, $i\in I$ --- the elements of $\A^i$ are called \emph{$i$'s actions} and the elements of $\A$ \emph{action profiles};
        \item $\B$ is a Polish space --- it is called \emph{state space}, its elements are called \emph{states};
        \item $W\subseteq\Omega\times\B^\ovT$ is a subset --- its elements are called \emph{configurations};
        \item $\mc W$ is a set of pairs $\zeta = (\xi,\chi)$ of maps $\xi\colon \ovT\times\Omega\to\A$ and $\chi\colon\ovT\times\Omega\to\B$ such that, for all $\omega\in\Omega$, $(\omega,\chi(\omega))\in W$ --- an \emph{action process} is a $\xi$ such that there is $\chi$ with $(\xi,\chi)\in\mc W$, a \emph{state process} is a $\chi$ such that there is $\xi$ with $(\xi,\chi)\in\mc W$, an \emph{outcome process} is an element of $\mc W$, seen as a map $\ovT\times\Omega\to\A\times\B$;
        \item $\ms H = (\ms H^i)_{i\in I}$ is a family of filtrations $\ms H^i = (\ms H^i_t)_{t\in\ovT}$ on the sample space $W$, $i\in I$ --- for any $i\in I$, $\ms H^i$ is called \emph{basic information structure for $i$};
        \item $\ms M = (\ms M^i)_{i\in I}$ is a family of $\sigma$-algebras $\ms M^i$ on $\ovT\times W$ satisfying $\Prd(\ms H^i) \subseteq \ms M^i \subseteq \Opt(\ms H^i)$, $i\in I$ --- for any $i\in I$, $\ms M^i$ is called \emph{Meyer information structure for $i$};
        \item $\mc S = \bigtimes_{i\in I} \mc S^i$ is the set-theoretic product of sets $\mc S^i$ of {$\ms H^i$-progressively measurable} maps $s^i\colon \ovT\times W \to \A^i$ --- for any $i\in I$, a \emph{strategy process for $i$} is an element of $\mc S^i$.
    \end{itemize}
    \smallskip

    \textsc{(Part B)}:~ Let $i\in I$. An \emph{optional time for $i$} is an $\ms H^i$-optional time such that $[\![0,\tau^i)\!)\in\ms M^i$.\footnote{By Theorem~\ref{3-SPF_VECT.thm:optional_times}, $[\![0,\tau^i)\!)\in\Opt(\ms H^i)$ for any $\ms H^i$-optional time $\tau^i$. If $\ms H^i$ is augmented, the converse is true as well, by the same theorem. Note that this is a generalisation of stopping times with respect to Meyer-$\sigma$-algebras, going back to~\cite{Lenglart1980Tribus, ElKaroui1981Les}, see also~\cite[Subsection~2.1]{Bank2019Lenglarts}.} 
    A \emph{history for $i$} is a pair $(\tau^i,\chi)$ consisting of an optional time $\tau^i$ for $i$ and a state process $\chi$. Let $\chi,\chi'$ be state processes and $\tau^i$ be an optional time for $i$. Then, we say that \emph{$\chi'$ cannot be distinguished from $\chi$ until $\tau^i$ by $i$}, or that \emph{$(\tau^i,\chi')$ cannot be distinguished from $(\tau^i,\chi)$}, in symbols $\chi'\approx_{i,\tau^i} \chi$, iff we have, for $\ms N$-almost all $\omega\in\Omega$ and all $t\in [0,\tau^i(\omega,\chi(\omega))]_\ovT$, and for all real-valued $\ms M^i$-measurable maps $f\colon\ovT\times W\to\R$:
    \begin{equation}\label{3-SPF_VECT.eq:f(chi)=f(chi')}
        f(t,\omega,\chi(\omega)) = f(t,\omega,\chi'(\omega)).
    \end{equation}
    An \emph{(endogenous) information set for $i$} is a pair $\mf p = (\tau^i,\mf x)$ for an optional time $\tau^i$ for $i$ and an equivalence class $\mf x$ with respect to $\approx_{i,\tau^i}$ on the set of state processes.\footnote{It follows from the definition that $\approx_{i,\tau^i}$ is an equivalence relation. For the proof, take $f=1[\![0,\tau^i)\!)$. Inserting $t = \tau^i(\omega,\chi(\omega))$ yields $\tau^i(\omega,\chi'(\omega)) \le \tau^i(\omega,\chi(\omega))$. Hence, we can insert, in a second step, $t = \tau^i(\omega,\chi'(\omega))$ which yields $\tau^i(\omega,\chi(\omega)) \le \tau^i(\omega,\chi'(\omega))$. See Proposition~\ref{3-SPF_VECT.prop:information_sets} for further discussion.} $\tau^i$ is said the \emph{time} of the information set. 
    The set of information sets for $i$ is denoted by $\mf P^i$, and its subset of information sets with time $\tau^i$ is denoted by $\mf P^i(\tau^i)$, for any optional time $\tau^i$ for $i$. 

    Let, for any stochastic process $s\colon\ovT\times W\to\A$ and any stochastic process $\chi\colon\ovT\times\Omega\to\B$, the stochastic process $s\llcorner\upsilon$ be given by
    \[ s\llcorner\chi\colon\ovT\times\Omega\to\A,\, (t,\omega) \mapsto (s^i(t,\omega,\chi(\omega)))_{i\in I}. \] 
    We call $s$ \emph{admissible} iff for all $i\in I$, all  
    optional times $\tau^i$ for $i$, all state processes $\tilde\chi$, there is an, up to $\ms N$-indistinguishability, unique 
    state process $\chi$ extending $\tilde\chi$, i.e.\ satisfying $\chi|_{[\![0,\tau^i\circ(\id_\Omega\star\chi))\!)} \cong_\ms N \tilde\chi|_{[\![0,\tau^i\circ(\id_\Omega\star\chi))\!)}$, and indistinguishable from it until $\tau^i$, i.e.\ satisfying $\chi\approx_{i,\tau^i} \tilde\chi$,\footnote{Both properties are not necessarily equivalent. See Proposition~\ref{3-SPF_VECT.prop:information_sets} for a discussion of this.} 
    that admits an action process $\xi$ with $(\xi,\chi)\in\mc W$ satisfying
    \begin{equation}\label{3-SPF_VECT.eq:s_admissible}
        (s\llcorner\chi)|_{[\![\tau^i\circ (\id_\Omega\star\chi),\infty]\!]} \cong_\ms N \xi|_{[\![\tau^i\circ (\id_\Omega\star\chi),\infty]\!]}.
    \end{equation}  
    We call the --- up to $\ms N$-indistinguishability uniquely determined --- processs $\chi$ the \emph{state process induced by $s$ given $(\tau^i,\tilde\chi)$}, respectively, and use the notation $\chi = \Out^\star(s \mid \tau^i,\tilde\chi)$.\smallskip
    
    \textsc{(Part C)}:~ A \emph{stochastic process form} is given by data $\mathbf F$ as above such that
    \begin{enumerate}
        \item\label{3-SPF_VECT.def:spf.msH_non_anticipative} for all $i\in I$, $\ms H^i$ is non-anticipative, that is, there is a family of $\sigma$-algebras $\tilde{\ms H}^i_t$ on $\Omega\times\B^{[0,t]_\ovT}$, ranging over $t\in\ovT$, such that, with $\op{proj}_{[0,t]_\ovT}\colon \B^\ovT\to\B^{[0,t]_\ovT},\ f\mapsto f|_{[0,t]_\ovT}$, we have
        \[ {\ms H}^{i}_t = \{(\id_\Omega\times\op{proj}_{[0,t]_\ovT})^{-1}(H) \cap W \mid H\in\tilde{\ms H}^i_t\} ;\]
        \item\label{3-SPF_VECT.def:spf.msF^i_zeta_subseteq_msE} for all $\zeta=(\xi,\chi)\in\mc W$, all $i\in I$,  the map 
        \[ \id_\Omega \star \chi\colon \Omega\to W, \, \omega\mapsto (\omega,\chi(\omega)) \] 
        is $\ms E$-$\ms H^i_\infty$-measurable;
        \item\label{3-SPF_VECT.def:spf.nu_Obs(xi)} for all outcome processes $\zeta=(\xi,\chi),\, \zeta'=(\xi',\chi')\in\mc W$, for all $i\in I$, all optional times $\tau^i$ for $i$, such that, with $\hat\tau^i = \tau^i \circ (\id_\Omega\star\chi)$, 
        $\xi|_{[\![0,\hat\tau^i]\!]} \cong_{\ms N} \xi'|_{[\![0,\hat\tau^i]\!]}$ holds true, we have $\chi|_{[\![0,\hat\tau^i]\!]} \cong_{\ms N} \chi'|_{[\![0,\hat\tau^i]\!]}$;
        \item\label{3-SPF_VECT.def:spf.msMi-measurability} for any $i\in I$, any optional time $\tau^i$ for $i$, any $\beta\in\mf w_1$, any $s^i\in\mc S^i$, there is an $\ms M^i$-measurable process $\tilde s^i\colon\ovT\times W\to\A^i$ such that $\tilde s^i\in\mc S^i$ and $s^i_{\tau^i} = \tilde s^i_{\tau^i}$ on $\{\pi\circ\tau^i = \beta\}$.\footnote{Following the setting of \cite{Lenglart1980Tribus,ElKaroui1981Les}, this means nothing else than the measurability of $s^i_{\tau^i}$ with respect to the filtration associated with $\ms M^i$ evaluated at $\tau^i$, where all this is considered on the measurable space given by $\{\pi\circ\tau^i=\beta\}$ with induced $\sigma$-algebra.}
    \end{enumerate}
    \smallskip
    
    \textsc{(Part D)}:~ An \textsc{spf} $\mathbf F$ is said \emph{well-posed} iff all $s\in\mc S$ are admissible. 
\end{definition}

We make some additional remarks. Let us first note that, since everything is encoded as processes in time, information is non-anticipative (Axiom~\ref{3-SPF_VECT.def:spf.msH_non_anticipative}) and the state is a non-anticipative function of action (Axiom~\ref{3-SPF_VECT.def:spf.nu_Obs(xi)}), we obtain a basic structure for ``causality'' from the beginning. In that sense, the stochastic process form is more similar to the stochastic extensive form than to the product form. The stochastic process form therefore merges different concepts of information in order to provide a general and tractable setting for problems with uncountably many decision situations (``agents'' in the language of Witsenhausen; information sets in the setting of this text), as discussed in the sequel.

Furthermore, outcome processes are pairs of action and state processes, where the former determine the latter in a non-anticipative way compatible with optional times (Axiom~\ref{3-SPF_VECT.def:spf.nu_Obs(xi)}). However, this mapping need not be scenariowise; it can be purely ``statistical''. This is a further departure from the strict stochastic extensive form setting, but is common in many contexts. For instance, the state may arise through stochastic integration of a function of the action process with respect to a semimartingale. ``Endogenous'' information (that is, information about agents' behaviour) is only transmitted via the state process, and we may assume that only the state process is payoff-relevant --- both without loss of generality, because the state could include a copy of action.

Moreover, let us insist on the fact that the term ``information set'' does, of course, not have the same \emph{formal} meaning as in extensive forms. An information set $\mf p = (\tau^i,\mf x)$ describes the time $\tau^i$ at that an agent $i\in I$ currently considers her options and which set $\mf x$ of state processes is still possible to be realised. This explains in particular the information the agent has about the behaviour of all agents up to time $\tau^i$ (whence the qualifier ``endogenous''). Agent $i$'s information about $W$, including realisations of exogenous scenario $\omega\in\Omega$ and state processes $\chi(\omega)$, and including information ``at'' time $\tau^i$, can then be derived from $\ms M^i$. $\tau^i$ itself is compatible with this information because $[\![0,\tau^i)\!)\in\ms M^i$. In a considerable generalisation of \cite{Selten1965Spieltheoretische}, another name for information sets would be \emph{subgame} because information sets are the instances that agents consider options or revise decisions at. We refrain from this usage in general because the present setting goes beyond the situation of perfect information.

Finally, let us close the bracket opened in the beginning of this subsection by noting that in a well-posed stochastic extensive form, the action process $\xi$ with $\xi = s\llcorner \chi$, for the state process $\chi$ induced by $s$ given some history, is progressively measurable with respect to the filtration induced by all $\ms H^i$, $i\in I$, and the map $\id_\Omega\star\chi$. This follows from Axiom~\ref{3-SPF_VECT.def:spf.msF^i_zeta_subseteq_msE} and the optionality of $s^i$, for any $i\in I$. The proof is elementary; it is given in the special case of timing games later in the text. Moreover, by Axiom~\ref{3-SPF_VECT.def:spf.msMi-measurability}, at any joint optional time $\tau$ for all $i\in I$ (for example, elements of $\T$), for any $\beta\in\mf w_1$, there is $\tilde s\in\mc S$ with $\tilde s_\tau = s_\tau$ on $\{\pi\circ\tau=\beta\}$ whose $i$-th component is $\ms M^i$-measurable and therefore $\ms H^i$-optional, for all $i\in I$. Then, the action process $\tilde\xi$ with $\tilde\xi = \tilde s\llcorner \tilde\chi$, for the state process $\tilde\chi$ induced by $\tilde s$ given some history, is even optional with respect to the filtration induced by all $\ms H^i$, $i\in I$, and the map $\id_\Omega\star\tilde\chi$. We conclude that ``induced'' action processes are progressively measurable and ``locally'' optional, and if strategies for $i\in I$ are $\ms M^i$-measurable, even optional.

\subsection{Information sets, counterfactuals, and equilibrium}

In this subsection, we further discuss the ``problem of information'' \cite{Kuhn1953Extensive} in stochastic process forms. We treat the role of $\ms H$ and $\ms M$, and we analyse information sets. We first give some examples for classical choices of $\ms H$. 
\begin{example}\label{3-SPF_VECT.ex:msHit}
    Fix a stochastic process form $\mathbf F$ and an agent $i\in I$. We consider the separable, symmetric case where $\ms H^i_t$ equals $\ms F^i_t \otimes {\ms B}^{\prime i}_t|_W$, $i\in I$, $t\in\ovT$, for some filtration $\ms F^i$ on $(\Omega,\ms E)$ and some sub-$\sigma$-algebra ${\ms B}^{\prime i}_t$ of $(\ms B_\B)^{\otimes\ovT}$. One can take $(\Omega,\ms E,\ms F^i)$ to be universally complete. The following cases, classical in control theory,\footnote{See, e.g., the textbooks \cite{Cohen2015Stochastic,Carmona2018Probabilistic}.} obtain:
    \begin{enumerate}
        \item The case of \emph{state-independent exogenous information} obtains iff every state process $\chi$ is $\ms F^i$-progressively measurable.
        \item The case of \emph{dynamic learning of exogenous information} is the opposite. It can be formulated by fixing some filtration $\ms G^i$ with $\ms F^i_t\subseteq \ms G^i_t$ for all $t\in\ovT$ and considering state processes $\chi$ that are $\ms G^i$-adapted but not necessarily $\ms F^i$-adapted. Given two outcome processes $\zeta = (\xi,\chi)$ and $\zeta' = (\xi',\chi')$, the filtrations on $\Omega$ describing $i$'s information flow for either outcome, which are those induced by $\ms H^i$ and $\id_\Omega\star\chi$ and $\id_\Omega\star\chi'$, respectively, may therefore be strictly larger than $\ms F^i$ and differ.
        \item \emph{Open-loop} strategies (or controls) obtain iff ${\ms B}^{\prime i}_t = \{\emptyset, \B^\ovT\}$.
        \item \emph{Closed-loop} strategies (or controls) obtain iff, for all $t\in\ovT$, $(\ms B_\B)^{\otimes[0,t)_\ovT}\otimes \{\emptyset, \B^{\otimes[t,\infty]_\ovT}\} \subseteq {\ms B}^{\prime i}_t \subseteq (\ms B_\B)^{\otimes[0,t]_\ovT}\otimes \{\emptyset, \B^{\otimes(t,\infty]_\ovT}\}$ (for $t=0$, this reads: ${\ms B}^{\prime i}_0 \subseteq (\ms B_\B)^{\otimes\{0\}} \otimes \{\emptyset, \B^{\otimes(0,\infty]_\ovT}\}$).\footnote{We could, without loss of generality, ask for ${\ms B}^{\prime i}_t = (\ms B_\B)^{\otimes[0,t]_\ovT}\otimes \{\emptyset, \B^{\otimes(t,\infty]_\ovT}\}$ if $t>0$ since this does not necessarily imply that strategies can condition at time $t$ on the state at time $t$. Indeed, in that case, we could still have $\ms M^i = \Opt(\hat{\ms H}^i)$, for $\hat{\ms H}^i_t = \ms F^i_t \otimes (\ms B_\B)^{\otimes[0,t)_\ovT}\otimes \{\emptyset, \B^{\otimes[t,\infty]_\ovT}\}$, $t>0$, ${\ms B}^{\prime i}_0 = \{\emptyset, \B^{\otimes[0,\infty]_\ovT}\}$, and $\hat{\ms H}^i_0 = \ms H^i_0$.}
    \end{enumerate}
    These are stylised special cases, of course, and mixed regimes obtain quite easily. For instance, in a game, some dynamic learning of exogenous information may already come in by other players' using private randomisation devices.
    
    Out of the separable case one can construct more complicated information structures, intertwining information on exogenous and endogenous events. For example, $\ms H^i_t$ ($i\in I$, $t\in\ovT$) could also be given by $(\ms F^i_t \otimes {\ms B}^{\prime i}_t) \vee (\ms G^i_t \otimes {\ms B}^{\prime\prime i}_t)|_W$ for some filtration $\ms G^i$ on $(\Omega,\ms E)$ and some suitable sub-$\sigma$-algebra ${\ms B}^{\prime\prime i}_t$ of $(\ms B_\B)^{\otimes\ovT}$. 
    We also note that $\B$ could be a product of Polish spaces, and the information structures could depend in different ways on the different factors of that product. For example, one factor could describe a partially observable signal the agent cannot condition on (open-loop) and a second factor could describe an observation process the agent can condition on (closed-loop) and whose underlying exogenous randomness the agent try to learn. Moreover, information can be asymmetric in that not all agents can observe all factors in the same way. For example, any agent could have its own observation process, described by a corresponding component of $\chi$.
\end{example}

Next, we discuss an example for the choice of $\ms M$ and elaborate on the combination of Meyer $\sigma$-algebras and vertically extended time.
\begin{example}
    Meyer $\sigma$-algebra have been used in the financial literature recently in order to model information about imminent trade signals (see, for instance, \cite{Bank2024Optimal}). So far the focus has been set mainly on the single-agent setting however.\footnote{Yet, see \cite{Bank2025How} for a multi-agent model.} Reinterpreting this modelling approach in the language of the present setting, and adding vertically extended time in particular, yields a stochastic process form where $\ms M$ is given by
    \[ \ms M^i = \Prd(\ms H^i) \vee \sigma(Z^i), \qquad i\in I, \]
    for a fixed $\ms H^i$-optional process $Z^i$, for any $i\in I$, describing the signal observed by agent $i$. In \cite{Bank2024Optimal}, only a single agent is considered and this agent can essentially act twice per instance of time. 
    
    In the interactive setting or settings involving non-trivial consecutive, infinitesimal randomisation, e.g.\ in the framework of preemption as in \cite{Riedel2017Subgame}, one might wish to extend this. This requires adding additional virtual instants of time along the vertical half-axis, in order to keep track of the chain of reactions. Moreover, the signals $Z^i$ would no more be sufficient for describing the information agent $i$ has at the different vertical versions of the given real instant $\tau^i$ of time: if $Z^i$ contains sufficient information about agents' behaviour at $\tau^i$, there may be more than one (``I do what you do'') or no (``I do not do what I do'') outcome process $\zeta = (\xi,\chi)$ compatible with a given strategy profile (in the sense of satisfying Equation~\ref{3-SPF_VECT.eq:s_admissible}), therefore destroying well-posedness. In the well-posed setting, $Z^i$ will mainly be useful for describing \emph{exogenous} signal observations and pro- and reaction with respect to these, while vertically extended time accounts for \emph{endogenous} signal observations and pro- and reaction with respect to the latter. The formulation in terms of stochastic analysis in vertically extended time assures that both sorts of observation and action are consistent.
\end{example}

In stochastic dynamic games, it is important to analyse strategy profiles given counterfactual ``histories'' in a way compatible with the agents' information. By construction, the extensive form offers natural concepts for this, based on moves and, more generally, information sets. These concepts extend naturally to stochastic extensive forms, as introduced and studied in Chapter~\ref{chap:2-SEF_G}. In stochastic process forms these concepts are no more available in a strict sense; we have mimicked them in the definition above.

In a stochastic process form, an information set for an agent $i\in I$ is given by an optional time $\tau^i$ with respect to $i$'s information structure and an equivalence class of state processes $\chi$, where $\chi$ is identified with all other state processes $\chi'$ that $i$ cannot distinguish from it given the information $\ms M^i$ and time $\tau^i$. Here, $\chi$ and $\chi'$ cannot be distinguished given $\ms M^i$ and time $\tau^i$ iff any strategy based on information $\ms M^i$ (= $\ms M^i$-measurable) yields the same result up to time $\tau^i$ inclusively. It is clear that if the amount of information is too large --- implying too much future knowledge, and in particular times $\tau^i$ that depend on agents' action at $\tau^i$, i.e.\ arbitrary $\ms H^i$-optional times --- too many state processes can be distinguished which may prevent well-posedness. At the same time it may seem natural to permit ``subgames'' starting at the first jump of an exogenous Poisson process, for example. Hence, restricting to predictable $\ms H^i$-optional times is not a convincing solution either. Here, Meyer information structures provide a subtle device for managing this trade-off. This point is illustrated by the following proposition. 

Before stating it, we note that the mathematical tractability of this analysis is ensured by the fact that in stochastic process forms information is modelled on a ``universal'' configuration space $W\subseteq\Omega\times\B^\ovT$, not only on $\Omega$. This namely allows for a description of information that is independent of the choice of a concrete state process, and from which the state-dependent exogenous information --- that is, given a state process $\chi$, the filtration on $(\Omega,\ms E)$ generated by $\ms H^i$ and $\id_\Omega\star\chi$ --- can be derived. However, our approach is not primarily chosen for mathematical convenience, but because we esteem it natural in a stochastic process-based game-theoretic setting.\footnote{Compare the paper \cite{Jacka2018Informational} which treats a general adaptive stochastic control framework in a classical continuous-time setting.}

\begin{proposition}\label{3-SPF_VECT.prop:information_sets}
    Let $\mathbf F$ be a stochastic process form on $(\Omega,\ms E,\ms N)$ and $i\in I$. Consider the following additional assumptions:
    \begin{enumerate}[label=(\Alph*),ref=(\Alph*)]
        \item\label{3-SPF_VECT.prop:information_sets.Ass.Hi0_open-loop} We have $\ms H^i_0 \subseteq \ms E\otimes \{\emptyset,\B^\ovT\}$;
        \item\label{3-SPF_VECT.prop:information_sets.Ass.Hit_closed-loop} For all $t\in\ovT\setminus\{0\}$, $\{\emptyset,\Omega\}\otimes\B^{[0,t]_\ovT}\otimes\{\emptyset,\B^{(t,\infty]_\ovT}\} \subseteq \ms H^i_t$.
        \item\label{3-SPF_VECT.prop:information_sets.Ass.msMi_endogenously_predictable} We have $\ms M^i \subseteq \Opt(\ms E\otimes \{\emptyset,\B^\ovT\}) \vee \Prd(\ms H^i)$, where $\ms E\otimes \{\emptyset,\B^\ovT\}$ denotes the filtration equal to that $\sigma$-algebra at any time $t\in\ovT$. 
    \end{enumerate}
    Suppose that Assumptions~\ref{3-SPF_VECT.prop:information_sets.Ass.Hi0_open-loop} and~\ref{3-SPF_VECT.prop:information_sets.Ass.msMi_endogenously_predictable} are satisfied. Then, we have:
    \begin{enumerate}
        \item\label{3-SPF_VECT.prop:information_sets.f(t,omega,h)=f(t,omega,h')} For any $\ms M^i$-measurable $f\colon\ovT\times W\to \R$ and all $t\in\ovT$, $\omega\in\Omega$, and $h,h'\in\B^\ovT$ with $h|_{[0,t)_\ovT} = h'|_{[0,t)_\ovT}$:
        \[ f(t,\omega,h) = f(t,\omega,h'). \]
        \item\label{3-SPF_VECT.prop:information_sets.chi|<taui=chi|<taui_=>_chi_=~_chi'} For any optional time $\tau^i$ for $i$ and all state processes $\chi,\chi'$ with 
        \[ \chi|_{[\![0,\tau^i\circ(\id_\Omega\star\chi))\!)} \cong_{\ms N} \chi'|_{[\![0,\tau^i\circ(\id_\Omega\star\chi))\!)}, \] 
        we have $\chi\approx_{i,\tau^i} \chi'$.
        \item\label{3-SPF_VECT.prop:information_sets.chi_=~_chi'_=>_chi|<taui=chi|<taui} If in addition Assumption~\ref{3-SPF_VECT.prop:information_sets.Ass.Hit_closed-loop} is satisfied, then, for any optional time $\tau^i$ for $i$ and all state processes $\chi,\chi'$ that are left-continuous\footnote{Note that the pathwise left-continuity at times $u\in\ovT$ with $\pi(u) = \mf w_1$ is not at all a strong requirement; it suffices for $\chi$ to be progressively measurable with respect to some filtration (see Remark~\ref{3-SPF_VECT.rmk:prog_mb_basic_properties}, Part~\ref{3-SPF_VECT.rmk:prog_mb_basic_properties.scwise_left-constant_at_(t,mfw1)}). } at all $u\in\ovT$ with $\pi(u) = \mf w_1$ and satisfy $\chi\approx_{i,\tau^i} \chi'$, we have $\chi|_{[\![0,\tau^i\circ(\id_\Omega\star\chi))\!)} \cong_{\ms N} \chi'|_{[\![0,\tau^i\circ(\id_\Omega\star\chi))\!)}$.
    \end{enumerate}
    In particular, if all three assumptions above are satisfied, then, for all optional times $\tau^i$ for $i$ and all state processes $\chi,\chi'$ that are left-continuous at all $u\in\ovT$ with $\pi(u) = \mf w_1$, we have the equivalence:
    \[ \chi\approx_{i,\tau^i} \chi' \qquad \Longleftrightarrow \qquad \chi|_{[\![0,\tau^i\circ(\id_\Omega\star\chi))\!)} \cong_{\ms N} \chi'|_{[\![0,\tau^i\circ(\id_\Omega\star\chi))\!)}. \]
\end{proposition}

\begin{remark}[``Nodes'']\label{3-SPF_VECT.rmk:nodes}
    As a consequence of this proposition, if Assumptions~\ref{3-SPF_VECT.prop:information_sets.Ass.Hi0_open-loop} and~\ref{3-SPF_VECT.prop:information_sets.Ass.msMi_endogenously_predictable} are satisfied, then, for any $i\in I$, one could call \emph{nodes for $i$} all sets of the form
    \[ x_{\tau^i}(\tilde \chi) = \{\chi\, \text{ state processes} \mid \chi|_{[\![0,\tau^i\circ(\id_\Omega\star\chi))\!)} \cong_{\ms N} \chi'|_{[\![0,\tau^i\circ(\id_\Omega\star\chi))\!)} \}, \]
    ranging over all optional times $\tau^i$ for $i\in I$ and all state processes $\tilde\chi$. This is analogous to the definition of nodes in action path stochastic decision forests in Section~\ref{1-SDF_AC.sec:sdf} (similarly, in a deterministic and very special setting, under the name ``differential games'', in \cite{AlosFerrer2005Trees}). However, it is clear that the set of all these $x_{\tau^i}(\tilde\chi)$ does not at all define a tree or forest. This yields no extensive form in any strict sense.
\end{remark}

In this \textsc{spf} language for describing extensive form characteristics based on stochastic processes, including a model of information sets alias ``subgames'' alias instances of decision revision, we can conclude this subsection with a definition of equilibrium. It is a refinement of the classical Nash equilibrium concept in two ways: 1) the best-response condition must also be satisfied given counterfactual histories (``off the equilibrium path''), in the spirit of subgame-perfect equilibria (cf.~\cite{Selten1965Spieltheoretische,Selten1975Reexamination}); 2) the beliefs agents form about exogenous information and, in the case of imperfect information, the current ``move'' must be consistent, in the spirit of perfect Bayesian equilibria. As the stochastic process form is still very similar to stochastic extensive forms, it is unsurprising that the following definition is an adaptation of Definitions~\ref{2-SEF_G.def:EU_pref_str}, \ref{2-SEF_G.def:consistent_EU_pref_str}, and~\ref{2-SEF_G.def:dynamic_rationality} to the setting of stochastic process forms. 

Yet, there is an important difference here since in \textsc{spf} time provides a ``uniformising structure'' among information sets (making possible a refined meaning of conditional probability). Moreover, calculating expectations with respect to the posterior at a given endogenous information set requires to determine an adequate $\sigma$-algebra to condition on. The natural concepts for this, once again, derive from the general theory of Meyer-$\sigma$-algebras because at an information set $\mf p = (\tau^i,\mf x)$ for agent $i$, the agent can condition on all $\ms M^i$-measurable maps evaluated at $\tau^i$, at the realised scenario $\omega$, and at the belief $\chi_{i,\mf p}\in\mf x$ about the actual endogenous ``history''. We propose a  relaxed version of dynamic rationality and equilibrium --- allowing for a selection of information sets (and thus counterfactuals) to be checked\footnote{In the static Nash equilibrium case, one would only check the best response condition given the information set at time zero, provided it is unique. In a rigorous dynamic setting, on the contrary, one would wish to check given all information sets, of course.} --- because this a restriction typically made in the literature, for reasons that will become clearer in the case study of the timing game. For instance, one may focus on information sets $\mf p = (\tau^i,\mf x)$ such that $\mf x$ contains a deterministic (if not even constant) state process.
\begin{definition}\label{3-SPF_VECT.def:equilibrium}
    Let $\mathbf F$ be a well-posed \textsc{spf} on $(\Omega,\ms E,\ms N)$. 
    \begin{enumerate}
        \item A \emph{belief system on $\mathbf F$} is a family $\Pi = (\ms P^{i,\tau^i},\kappa^{i,\tau^i},p_{i,\mf p},\ms P_{i,\mf p})_{\mf p = (\tau^i,\mf x)\in\mf P^i,\,i\in I}$ such that, for any $i\in I$ and any optional time $\tau^i$ for $i$, $\ms P^{i,\tau^i}$ is a $\sigma$-algebra on $\mf P^i(\tau^i)$, $\kappa^{i,\tau^i}$ is a Markov kernel from $(\mf P^i(\tau^i),\ms P^{i,\tau^i})$ to $(\Omega,\ms E)$ with $\ms N\subseteq \ms N_{\kappa^{i,\tau^i}(.,\mf p)}$ for all $\mf p\in\mf P^i(\tau^i)$, and, moreover, for any information set $\mf p = (\tau^i,\mf x)\in\mf P^i(\tau^i)$ for $i$ with time $\tau^i$, $\ms P_{i,\mf p}$ is a $\sigma$-algebra on $\mf x$,\footnote{According to a classical choice of $\ms P_{i,\mf p}$, any $\chi,\chi'\in\mf x$, that coincide $\ms N$-almost surely on $[\![0,\tau^i]\!]$, would have to be inseparable by $\ms P_{i,\mf p}$, that is, for all $P\in\ms P_{i,\mf p}$, $\chi\in P$ iff $\chi'\in P$.}
         and $p_{i,\mf p} \colon \Omega \to \mf x$ is an $\ms E$-$\ms P_{i,\mf p}$-measurable map.
        \item A \emph{taste system on $\mathbf F$} is a family $U = (u_{i,\mf p})_{\mf p\in\mf P^i,\,i\in I}$ of maps $u_{i,\mf p} \colon W \to \R$.
        \item An \emph{expected utility (\textsc{eu}) preference structure on $\mathbf F$} is a tuple $\Pr = (\Pi,U,\ms W)$ where
        \begin{itemize}[label=--]
            \item $\Pi$ is a belief system on $\mathbf F$,
            \item $U$ is a taste system on $\mathbf F$, and
            \item $\ms W$ is a $\sigma$-algebra on $W$,
        \end{itemize}
        such that, 
        we have, for all $i\in I$, all optional times $\tau^i$ for $i$ and information sets $\mf p=(\tau^i,\mf x)\in\mf P^i(\tau^i)$:
        \begin{enumerate}
            \item\label{3-SPF_VECT.def:EU_pref_str.uip_mb} $u_{i,\mf p}$ is bounded and $\ms W$-Borel-measurable;
            \item\label{3-SPF_VECT.def:EU_pref_str.Out_mb} $\Out_{i,\mf p}^s\colon \Omega \to W,\, \omega \mapsto (\omega,\Out^\star(s \mid \tau^i,p_{i,\mf p}(\omega))(\omega))$ is $\ms E$-$\ms W$-measurable for all $s\in \mc S$;
            \item\label{3-SPF_VECT.def:EU_pref_str.phi_mb} for any $s\in\mc S$ and any optional time $\sigma^i$ for $i$ with $\tau^i \le \sigma^i$, the map \[\p^s_{i,\mf p,\sigma^i} \colon \mf x \to \mf P^i(\sigma^i),\] assigning to any $\chi\in\mf x$ the unique $\mf p' = (\sigma^i,\mf x')\in\mf P^i(\sigma^i)$ with $\Out^\star(s \mid \tau^i,\chi)\in\mf x'$, is $\mf P_{i,\mf p}$-$\mf P^{i,\sigma^i}$-measurable;
            \item\label{3-SPF_VECT.def:EU_pref_str.chi_i,mfp_state-proc} the map
            \[ \chi_{i,\mf p}\colon \Omega\to\B^\ovT,\, \omega\mapsto p_{i,\mf p}(\omega)(\omega) \]
            is a state process.
        \end{enumerate}
        \item Let $\Pr = (\Pi,U,\ms W)$ be an \textsc{eu} preference structure on $\mathbf F$ and $s\in\mc S$ a strategy profile. $(\Pr,s)$ is said \emph{dynamically consistent} iff:
        \begin{enumerate}
            \item\label{3-SPF_VECT.def:EU_pref_str.consistency.ui=uip} for all $i\in I$, there is $u_i$ with $u_{i,\mf p} = u_i$ for all $\mf p\in\mf P^i$;
            \item\label{3-SPF_VECT.def:EU_pref_str.consistency.p} for all $i\in I$, all optional times $\tau^i,\sigma^i$ for $i$ with $\tau^i\le\sigma^i$, all $\mf p\in\mf P^i(\tau^i)$, all $\omega\in\Omega$ satisfy\footnote{Note that for $\tau^i = \sigma^i$ this implies the equality $p_{i,\mf p}(\omega) = \Out^\star(s \mid p_{i,\mf p}(\omega))$. This reflects the fact that in \textsc{spf}, formally, information sets actually partition all possible state processes, rather than ``moves''. In \textsc{sef} already, moves contained in an information sets form a partition of attainable outcomes, and information sets partition moves, see Section~\ref{2-SEF_G.sec:Stochastic extensive forms}. However, note that there is no rigorous meaning to ``moves'' in \textsc{spf}.} 
            \[ p_{i,\p^s_{i,\mf p,\sigma^i}(p_{i,\mf p}(\omega))}(\omega) = \Out^\star(s \mid \tau^i,p_{i,\mf p}(\omega)); \]
            \item\label{3-SPF_VECT.def:EU_pref_str.consistency.P} for all $i\in I$, all optional times $\tau^i,\sigma^i$ for $i$ with $\tau^i\le\sigma^i$, all $\mf p\in\mf P^i(\tau^i)$ and the measure $\P_{i,\mf p} = \kappa^{i,\tau^i}(.,\mf p)$, and all $E\in\ms E$, we have, $\P_{i,\mf p}$-almost surely:\footnote{We recall that Equation~\ref{3-SPF_VECT.eq:dynamic_consistency_kappa} is, by definition of conditional expectation (and ``probability''), equivalent to the following statement: for all bounded, $\ms P^{i,\sigma^i}$-Borel-measurable maps $f\colon \mf P^i(\sigma^i)\to\R$, we have
            \[ \E^{\P_{i,\mf p}}\Big(f(\p^s_{i,\mf p,\sigma^i} \circ p_{i,\mf p})\cdot 1_E\Big) = \E^{\P_{i,\mf p}}\Big(f(\p^s_{i,\mf p,\sigma^i} \circ p_{i,\mf p}) \cdot\kappa^{i,\tau^i}(E,\p^s_{i,\mf p,\sigma^i} \circ p_{i,\mf p})\Big). \]}
            \begin{equation}\label{3-SPF_VECT.eq:dynamic_consistency_kappa}
                \kappa^{i,\sigma^i}(E,\p^s_{i,\mf p,\sigma^i} \circ p_{i,\mf p}) = {\P_{i,\mf p}}\Big(E \mid \p^s_{i,\mf p,\sigma^i} \circ p_{i,\mf p}\Big).
            \end{equation}
        \end{enumerate}
        \item Let $\Pr = (\Pi,U,\ms W)$ be an \textsc{eu} preference structure on $\mathbf F$. For any $i\in I$, any optional time $\tau^i$ for $i$, and any information set $\mf p\in\mf P^i(\tau^i)$, let $\P_{i,\mf p} = \kappa^{i,\tau^i}(.,\mf p)$ and 
        \[ 
            \ms F_{i,\mf p} = \Big\{ \{\omega\in\Omega \mid f_{\tau^i(\omega,\chi_{i,\mf p}(\omega))}(\omega,\chi_{i,\mf p}(\omega)) \in B\}  \mid f\colon\ovT\times W\to \R~\ms M^i\text{-measurable},\, B\in\ms B_\R\Big\}.\footnote{This is nothing else than the $\sigma$-algebra of the $\tau^i$-past associated with the Meyer-$\sigma$-algebra $\ms M^i$, pulled back onto $\Omega$ by $\id_\Omega\star\chi_{i,\mf p}$; introduced in \cite{Lenglart1980Tribus}, see also \cite{ElKaroui1981Les}, and \cite[Subsection~2.1]{Bank2019Lenglarts}. Using the notation from the latter survey, and of the pullback of $\sigma$-algebras, we thus have: $\ms F_{i,\mf p} = (\id_\Omega\star\chi_{i,\mf p})^\ast \ms F^{\ms M^i}_{\tau^i} = \{(\id_\Omega\star\chi_{i,\mf p})^{-1}(M) \mid M\in \ms F^{\ms M^i}_{\tau^i}\}$.}
        \] 
        Further, for any strategy profile $s\in\mc S$, let $\pi_{i,\mf p}(s)$ denote the conditional expectation of $u_{i,\mf p}\circ\Out^s_{i,\mf p}$ with respect to $\P_{i,\mf p}$ given $\ms F_{i,\mf p}$, that is,\footnote{Note that the following expression is well-defined in view of Equation~\ref{3-SPF_VECT.eq:dynamic_consistency_kappa}, because $\Out^s_{i,\mf p}\colon\Omega\to W$ is uniquely defined up to $\ms N$-indistinguishability, and $\ms N\subseteq\ms N_{\P_{i,\mf p}}$ by assumption.}
        \[ \pi_{i,\mf p}(s) = \E^{\P_{i,\mf p}}\Big(u_{i,\mf p}\circ \Out^s_{i,\mf p} \mid\ms F_{i,\mf p}\Big) . \]
        
        For any $i\in I$, fix a set of information sets $\tilde {\mf P}^i\subseteq \mf P^i$, and let $\tilde{\mf P} = (\tilde{\mf P}^i)_{i\in I}$. A strategy profile $s\in \mc S$ is said \emph{dynamically rational on $\tilde{\mf P}$ given $\Pr$} iff for all $i\in I$, all $\mf p\in\tilde{\mf P}^i$, and all $\tilde s\in \mc S$ with $\tilde s^{-i} = s^{-i}$, we have, $\P_{i,\mf p}$-almost surely,
        \[ \pi_{i,\mf p}(s) \ge \pi_{i,\mf p}(\tilde s). \]
        Let $s\in \mc S$ be a strategy profile. Then, $(s,\Pr)$ is said \emph{in equilibrium on $\tilde{\mf P}$} iff it is dynamically rational on $\tilde{\mf P}$ given $\Pr$ and $(\Pr,s)$ is dynamically consistent. The qualifier ``on $\tilde{\mf P}$'' can be dropped iff $\tilde{\mf P}^i = \mf P^i$ for all $i\in I$.
    \end{enumerate}
\end{definition}

\begin{remark}\label{3-SPF_VECT.rmk:prior_and_dynamic_consistency}
    Consider a stochastic process form $\mathbf F$ satisfying Assumption~\ref{3-SPF_VECT.prop:information_sets.Ass.Hi0_open-loop} in Proposition~\ref{3-SPF_VECT.prop:information_sets} (i.e.\ at time zero, agents have no information about the state process), and an \textsc{eu} preference structure $\Pr$ as in the definition. Then, it follows directly from the definition that, for any $i\in I$, there is a unique (endogenous!) information set at time zero. Denote this information set by $\mf p_0 = (0,\mf x_0)$. It clearly does not depend on the agent $i$. 

    Then, for any $s\in\mc S$ such that $(\Pr,s)$ is dynamically consistent, $p_{i,\mf p_0}$ is constant with value $\chi = \Out^\star(s\mid 0,\tilde\chi)$ (where $\tilde\chi$ can be any state process). More generally, for any information set $\mf p = (\tau^i,\mf x)$ for $i$ with $\chi\in\mf x$, $p_{i,\mf p}$ is constant with value $\chi$, and, in particular, $\P_{i,\mf p} = \P_{i,\mf p_0}$. Off the equilibrium path (or without Assumption~\ref{3-SPF_VECT.prop:information_sets.Ass.Hi0_open-loop}), this need not hold true.
    
    Moreover, in Bayesian language, for any $i\in I$, $\P_{i,\mf p_0}$ is the \emph{prior} of agent $i$. The \emph{common prior assumption}, alias \emph{Harsanyi doctrine}, says that $\P_{i,\mf p_0} = \P_{j,\mf p_0}$ for all $i,j\in I$.\footnote{See \cite{Harsanyi1967Games, Harsanyi1968Games, Harsanyi1968Gamesa, Aumann1987Correlated, Aumann1995Epistemic}.} The \emph{posterior for $i$ at information set $\mf p\in\mf P^i$} is the ``conditional probability'' $\P_{i,\mf p}(.\mid \ms F_{i,\mf p})$, though this need not admit a representation via a Markov kernel in the spirit of regular conditional probabilities. 
\end{remark}

\begin{example}\label{3-SPF_VECT.ex:equilibrium_for_quasi_perfect_information}
    Consider, for example, the special case of ``quasi''-perfect information as expressed by the Assumptions~\ref{3-SPF_VECT.prop:information_sets.Ass.Hi0_open-loop}, \ref{3-SPF_VECT.prop:information_sets.Ass.msMi_endogenously_predictable}, and \ref{3-SPF_VECT.prop:information_sets.Ass.Hit_closed-loop}. Then, with $\kappa^{i,\tau^i}$ being constant in the second component and independent of $i$ and $\tau^i$, the equilibrium definition above implements a generalised stochastic version of the concept of subgame-perfect equilibrium (cf.~\cite{Selten1965Spieltheoretische}). More generally, the definition above implements the concept of perfect Bayesian equilibrium (cf.~\cite{Harsanyi1968Gamesa, Fudenberg1991Perfect}) in stochastic process forms. In Subsections~\ref{3-SPF_VECT.subs:timing} and~\ref{3-SPF_VECT.subs:SDG}, we discuss this definition in the context of stochastic timing and differential games, with a concrete detailed example for the former.
\end{example}

\subsection{Timing games}\label{3-SPF_VECT.subs:timing}
We apply the developed theory to the simplest non-trivial dynamic game-theoretic model. That is, we introduce the general continuous-time stochastic timing game in stochastic process form, and illustrate it by proving the existence of the symmetric preemption equilibrium in the grab-the-dollar game. While natural in discrete time, it is hard to justify in continuous time. Yet, preemption is both an interesting theoretical example of subgame-perfection and a crucial phenomenon in many economic applications. In seminal papers, \cite{Fudenberg1985Preemption} and, in a general stochastic setting, \cite{Riedel2017Subgame, Steg2018Preemptive} provide a theory explaining the symmetric preemption equilibrium in continuous time using a stacked strategic form. 

However, in such a formulation, strategies necessarily depend on subgames. Although consistency across subgames may be assumed \emph{ex post}, the question remains on what grounds this happens. Moreover, the cited literature formally explains neither outcomes, that is, track records of the players' action, nor how payoffs derive from outcomes. Instead, payoffs are a direct function of strategy profiles, without factorising over (induced) outcomes. Hence, these models do not explain what happens, but what abstract strategies are chosen in each strategic form and what the payoffs are. This is reflected by the fact that payoffs are derived via a discrete-time approximation, see \cite{Steg2018Preemption}. As discussed in the introduction of \cite{Riedel2017Subgame}, and shown in the already-mentioned literature \cite{Stinchcombe1992Maximal, AlosFerrer2008Trees, AlosFerrer2011Comment}, a well-posed extensive form formulation based solely on paths with time index set $\R_+$ requires locally right-constant paths, so that reaction can only occur with a positive time lag of delay --- such a model is not very conclusive regarding preemption. The action-reaction model in \cite{AlosFerrer2015Repeated} provides a well-posed extensive form model making instantaneous reaction, and even $\mf w$ many of them, possible;\footnote{The model in \cite{AlosFerrer2015Repeated} is formulated with one reaction node per action node (i.e.\ roughly per instant of real time); the extension to $\mf w$ many reaction nodes is immediate.} however, as this model leaves the question of randomised strategies open, this theory is not sufficiently applicable to the preemption problem as well. This explains why the stacked strategic form approach has been chosen in \cite{Fudenberg1985Preemption} and \emph{idem} in \cite{Riedel2017Subgame}.

One problem about passing from discrete-time preemption to continuous-time preemption lies in the collapse of complicated patterns action and reaction near the preemption boundary to action at the preemption boundary with probability one. In classical continuous time, this implies a dramatic loss of information on the action process containing information about the action-reaction behaviour. However, this is a result of looking at action process convergence in terms of pointwise convergence. By contrast, tilting convergence preserves the detailed information by writing it on the vertical axis above the preemption boundary. Hence, stochastic process forms in vertically extended continuous time can bridge the gap between a) the need of a faithful model of the extensive form characteristics, based on an explicit description of outcomes, choices and information flow and including randomisation, and b) the desire to formally describe action and reaction behaviour, e.g.\ in preemption games, arising in continuous time via game-theoretic equilibrium analysis.

The mentioned kind of preemption is a very interesting example from the literature on timing games, but only one among many. Many other of these timing game models also use stacked strategic forms and ad-hoc variants of Nash equilibrium, which makes it difficult to understand the dynamic aspect of strategy and equilibrium. This includes the rich economic literature on real option games as well as the mathematical theory of Dynkin games --- we refer to the introduction of \cite{Riedel2017Subgame} for an overview on the literature. Another example would be timing games of asymmetric information, e.g.\ about price signals. Here, one agent can trade proactively thereby making profit and another one can only react (instantaneously in real time, one level higher on the vertical half-axis). But this other person's reaction may impact the price, and other players may pro- or react with respect to this.\footnote{For related models, see, e.g.\ \cite{Bank2025How, Bank2024Optimal}.}  The stochastic process form and the included abstract, dynamic equilibrium concept allows formulating a very general stochastic timing game model, which can shed light on this problem in general and is therefore of general interest.  

Let us start with introducing our formal model of timing problems. For convenience, we focus on the case of ``full'' endogenous information alias closed-loop controls (more precisely: at any time, all players know what the other players have done up to, exclusively, that time). This is the comparatively complicated case, weaker informational settings can be analysed similarly without essential additional effort. For this subsection's purpose, let $(\Omega,\ms E)$ be a universally complete measurable space. Suppose that $(\Omega,\ms E)$ is large enough to support a probability measure $\P$ and a, with respect to $\P$, $[0,1]$-uniformly distributed random variable. Let $\ms N = \{\emptyset\}$.

Let us fix an important technical convention. To any decreasing map $h\colon\ovT\to\{0,1\}$ we assign the map $h_- = h(.-)\colon\ovT\to\{0,1\}$, defined in any $t\in\ovT$ as follows. We let $h_-(0) = h(0-) = 1$ if $t=0$; $h(t-) = \lim_{u\nearrow t} h(u)$ if $t$ is a left-limit point; and, else, that is, if $t = (p(t),\beta+1)$ for some $\beta\in{\mf w_1}$, $h(t-) = h(p(t),\beta)$. We extend this convention to componentwise decreasing maps valued in $\{0,1\}^I$ in a componentwise manner, and to stochastic processes with decreasing paths valued in $\{0,1\}^I$.

Let the data $\mathbf F = (I,\A,\B,W,\mc W,\ms H,\ms M,\mc S)$ and $(\alpha,\upsilon,\ms G,\ms F,\ms F^\vee,(\tau_{\bm b},\tau^-_{\bm b})_{\bm b\in\B},z)$ be given as follows:
\begin{itemize}
    \item $I$ is a non-empty, finite set;
    \item $\A^i = \{0,1\}$ for any $i\in I$, $\A = \prod_{i\in I} \A^i$, equipped with discrete topology and the product order;
    \item $\B = \A$, thus also equipped with discrete topology and the product order, let $\bm 1 = (1,\dots,1)\in \B$ and $\bm 0 =(0,\dots,0)\in\B$ denote the constant functions $I\to\{0,1\}$ with value $1$ and $0$, respectively;
    \item $\alpha\in\mf w_1\setminus\{0\}$ is a countable non-zero ordinal;
    \item $W$ is the set of pairs $(\omega,h)\in\Omega\times \B^\ovT$ such that $h$ is right-continuous, decreasing, and has upper vertical level smaller than or equal to $\alpha$, and satisfies $h(\infty) = \bm 0$;
    \item $\ms G = (\ms G^i)_{i\in I}$ is a family of universally augmented filtrations on $(\Omega,\ms E)$ with time index set $\ovT$;
    \item $\upsilon\cong (\upsilon^i)_{i\in I} \colon\Omega\to[0,1]^I$ is $\ms E$-Borel-measurable, such that there is probability measure $\P$ on $(\Omega,\ms E)$ making $\upsilon$ uniformly distributed and independent from $\ms G$;
    \item $\ms F = (\ms F^i)_{i\in I}$, and for any $i\in I$, $\ms F^i = (\ms F^i_t)_{t\in\ovT}$ is the filtration given by $\ms F^i_t = \ovl{\ms G^i_t \vee \sigma(\upsilon^i)}$, the universal augmentation being taken in $\ms E$;
    \item $\ms F^\vee = (\ms F^\vee_t)_{t\in\ovT}$ is the in $\ms E$ universally augmented filtration generated by this family, i.e.\ $\ms F^\vee_t = \ovl{\bigvee_{i\in I} \ms F^i_t}$;
    \item $\mc W$ is the set of pairs $\zeta = (\xi,\chi)$ where $\xi\colon\ovT\times\Omega\to\A$ is an $\ms F^\vee$-optional, right-continuous, componentwise decreasing process with upper vertical level smaller than or equal to $\alpha$ satisfying $\xi_\infty = 0$, and $\chi = \xi$;\footnote{The condition $\chi = \xi$, for instance, could be relaxed; $\chi$ could be the solution to some (stochastic) differential equation depending non-anticipatively on $\xi$.}
    \item let
    \begin{equation}\label{3-SPF_VECT.def:z}
        z\colon \ovT\times W\to \B, \,(t,\omega,h) \mapsto \begin{cases} \bm 1, &\text{if } t= 0, \\ h(t), &\text{if } t> 0, \end{cases} 
    \end{equation}
    and, for any $\bm b\in\B$, 
    \[  \tau_{\bm b} = \inf\{u\in\ovT \mid z_u \le \bm b\}, \qquad \tau_{\bm b}^- = \inf\{u\in\ovT \mid z_{u-} \le \bm b\}; \]
    \item for any $i\in I$ and $t\in\ovT$, let ${\ms H}^{i,0}_t$ be the smallest $\sigma$-algebra on $W$ containing $\ms F^i_t \otimes \ms B_{\B}^{[0,t]_{\ovT}}\otimes\{\emptyset,\B^{(t,\infty]_\ovT}\}|_W$ and such that, for all $u\in[0,t]_\ovT$ and $\beta\in\mf w_1$, 
    \[ \{\tau_{\bm b} \le u,\, \pi\circ\tau_{\bm b} = \beta\} \, \in {\ms H}^{i,0}_t, \]
    let $\ms H^{i,1}_t$ be the universal augmentation of $\ms H^{i,0}_t$ in $[\ms H^{i,0}_\infty]^{\mathrm u}$; then, let, if $t>0$,
    \[ \ms H^i_t = \{ H\in \ms H^{i,1}_t \mid \exists \tilde H\subseteq \Omega\times\B^{[0,t]_\ovT}\colon H = \op{proj}_{[0,t]_\ovT}^{-1}(\tilde H) \cap W\},\]
    and, else,
    \[ \ms H^{i}_0 = \{(E \times \B^\ovT) \cap W \mid E\in\ms F^i_0\}; \]
    \item for any $i\in I$, $\ms M^i = \Opt(\ms F^i\otimes \{\emptyset,\B^\ovT\}) \vee \Prd(\ms H^i)$, where $\ms F^i\otimes \{\emptyset,\B^\ovT\}$ denotes the filtration given by $\ms F^i_t \otimes \{\emptyset,\B^\ovT\}$, $t\in\ovT$;
    \item for any $i\in I$, let $\mc S^i_0$ be the set of $\ms H^i$-progressively measurable, $\A^i$-valued processes $s^i\colon\ovT\times W\to\A^i$, lower semicontinuous from the right, with upper vertical level smaller than or equal to $\alpha$, and satisfying both $s^i_t(\omega,h) \le h^i(t-)$, for all $(t,\omega,h)\in\ovT\times W$, and $s^i_\infty = 0$; then, let $\mc S^i$ be the set of $s^i\in\mc S^i_0$ such that any optional time $\tau^i$ for $i$ admits $\ms M^i$-measurable $\tilde s^i\colon\ovT\times W \to \A^i$ with $\tilde s^i\in\mc S^i_0$ and $s^i_{\tau^i} = \tilde s^i_{\tau^i}$.
\end{itemize}

The basis of this model are the outcomes, like in extensive form models, and in contrast to the stacked strategic form model. Outcomes are exactly the outcomes of a timing game which, by definition, is a game with two actions, one of them being irreversible: that is, collections of decreasing $\{0,1\}$-valued paths for any player. There are no grounds for considering additional stopping intensities as in \cite{Fudenberg1985Preemption, Riedel2017Subgame}, nor for acting on the whole unit interval (as in almost all of the timing game literature with randomisation). The assumption of optionality of action processes is not even strictly required, since in the proof of well-posedness it can be seen that only optional action processes can be generated by strategic decision making according to $\mc S$. We have a countable uniform bound on activity; in the perspective of approximation of action processes via tilting convergence this corresponds to an upper bound on the well-order type of approximating grids. The assumption that the state process induced by a given action process equals the action process can be relaxed; it is made for simplicity here. 
Regarding the information structure $\ms H^i$ and $\ms M^i$ of player $i\in I$, we are in the setting of Proposition~\ref{3-SPF_VECT.prop:information_sets}. Here, $\ms H^i$ has been sufficiently enlarged to make the stochastic analysis of relevant débuts on $W$ possible, without violating the non-anticipativity axiom, Axiom~\ref{3-SPF_VECT.def:spf.msH_non_anticipative}, in the definition of \textsc{spf}. 

Let us discuss strategies in a bit more detail. In the \textsc{spf} setting, strategies are complete contingent plans of action -- contrasting stacked strategic form frameworks. They are globally defined objects, assembling local decisions based on available information (though this also requires measurability along ``nodes'', see Remark~\ref{3-SPF_VECT.rmk:nodes}). They must be globally progressively measurable, and, moreover, at any optional time for $i$, representable by an $\ms M^i$-measurable strategy. 
Lower semicontinuity from the right implies that the choice to remain at $1$ at the upper end of the vertical half-axis above some real time $t\in\R_+$ implies the agent to choose $1$ as well on some positive interval $((t,\mf w_1),t+\e)_\ovT$, for some $\e>0$, depending on $(\omega,h)$. On a scenariowise level, this reflects the ``identifiability'' axiom in \cite{Stinchcombe1992Maximal} and the inertia time lags in \cite{AlosFerrer2015Repeated}; yet, it is only a weak restriction because the player has the whole of $\{u\in\ovT \mid p(u) = t,\, \pi(u) < \alpha\}$ to react infinitesimally. However, as a difference to (the natural stochastic generalisation of) \cite{AlosFerrer2015Repeated}, the inertia time lag can depend on information revealed only at time $t+\e$. The inequality ``$s^i_t(\omega,h) \le h^i(t-)$'' simply expresses that at time $t$ and conditional on a history $h$ according to that $i$ has already chosen $0$, there is no other choice than $0$ left. Note, however, that it is absolutely possible that $s^i_t(\omega,h) = 1$ while $s^i_u(\omega,h) = 0$ for $t,u\in\ovT$ with $u<t$: $s^i(\omega,h)$ need not be decreasing! For instance, consider the strategy choosing $0$ whenever some real-valued Markov process $\eta$ stays in a closed set $C\subseteq\R$, and $1$ otherwise. Starting in a subgame $(\tau^i,\tilde\chi)$ with $\eta_{\tau^i\circ(\id_\Omega\star\tilde\chi)} \notin C$, the agent will stay at $1$ for a positive amount of classical real time. Yet, $\eta$ may have hit $C$ strictly before $\tau^i\circ(\id_\Omega\star\tilde\chi)$. Dropping this monotonicity assumption is crucial in the endeavour of formulating strategies as complete contingent plans of action, and independently of subgames. At the same time, it requires a consistent and careful application of stochastic analysis.\smallskip

Our first aim is to show well-posedness of $\mathbf F$. For this, we start with the stochastic analysis of the data introduced before. The first step is about the processes $z$ and $z_-$ and its débuts $\tau_{\bm b}$ and $\tau_{\bm b}^-$. They are crucial in constructing the induced outcome map; hence, we must verify their measurability properties beforehand.

\begin{lemma}\label{3-SPF_VECT.lemma:z_optional}
    For any $\bm b\in\B$, $\tau_{\bm b}$ and $\tau_{\bm b}^-$ are $\ms H^i$-optional times satisfying $\tau_{\bm 1} = \tau_{\bm 1}^- = 0$, and, if $\bm b \neq \bm 1$, $\pi\circ\tau_{\bm b} < \alpha$, $\pi\circ\tau_{\bm b}^- = \pi\circ\tau_{\bm b} + 1$, and $[\![0,\tau_{\bm b}^-)\!) = [\![0,\tau_{\bm b}]\!]$. $z$ is $\ms H^i$-optional and has the upper vertical level $\alpha$. $z_-$ is $\ms H^i$-predictable and has the vertical level $\alpha$, and even upper vertical level $\alpha$ if $\alpha$ is a limit ordinal.
\end{lemma}

In what follows, let $\ms H^\vee$ denote the augmented filtration generated by the family of $\ms H^j$, $j\in I$. That is, with $\ms H^{\vee,0}_t = \bigvee_{j\in I} \ms H^j_t$ for all $t\in\ovT$, $\ms H^\vee_t = \ovl{\ms H^{\vee,0}_t}$ is the augmentation of $\ms H^{\vee,0}_t$ in $[\ms H^{\vee,0}_\infty]^{\mathrm u}$.\smallskip

The next two lemmata are concerned with the questions whether optionality and progressive measurability are preserved under natural operations on the path space of the state process.

\begin{lemma}\label{3-SPF_VECT.lemma:fsharp_preserves_optionality}
    Let $f\colon\T\times W \to \B$ be $\ms H^\vee$-optional with right-continuous, decreasing paths, and upper vertical level smaller than or equal to $\alpha$ satisfying $f_\infty = \bm 0$. Then, with $f$ seen as a map $W \to \B^\ovT$, there is a map
    \begin{equation}\label{3-SPF_VECT.eq:fsharp}
        f^\#\colon \ovT\times W\to\ovT\times W,\, (t,\omega,h) \mapsto (t,\omega,f(\omega,h)) 
    \end{equation}
    which is both $\Opt(\ms H^\vee)$-$\Opt(\ms H^\vee)$- and $\Prg(\ms H^\vee)$-$\Prg(\ms H^\vee)$-measurable.
\end{lemma}

\begin{lemma}\label{3-SPF_VECT.lemma:eta_circ(id_Omega_star_chi)_inherits_optionality}
    Let $\eta\colon\T\times W\to\B$ be $\ms H^\vee$-optional and $\chi$ be a state process. Then, $\eta\circ[\id_\ovT\times(\id_\Omega\star\chi)]$ is $\ms F^\vee$-optional.
\end{lemma}

Now, we can state the well-posedness theorem.

\begin{thm}\label{3-SPF_VECT.thm:timing_SPF_well-posed}
    $\mathbf F$ is a well-posed stochastic process form on $(\Omega,\ms E,\{\emptyset\})$.
\end{thm}

We call $\mathbf F$ the \emph{timing \textsc{spf} of upper vertical level $\alpha$}. Combined with the equilibrium concept in Definition~\ref{3-SPF_VECT.def:equilibrium}, it provides a general continuous-time timing game model for finitely many players $i\in I$ with possibly asymmetric exogenous information $\ms G^i$, and augmented with private randomisation devices $\upsilon^i$. Players can react instantaneously to new information $\alpha$ times vertically above any real time $t\in\R_+$. By the choice of $\ms M^i$, at any optional time $\tau^i$ for $i$, a decision can be based on information $\ms H^i$ in a predictable way in general, but in addition on exogenous information $\ms F^i$ in a fully optional way. Loosely speaking, the player $i\in I$ can base a decision at time $\tau^i$ on exogenous information $\ms F^i$ until $\tau^i$ inclusively, but only on endogenous information that can be explained by endogenous information accumulated over previous instants of time.\footnote{A formally precise description of this can be given in terms of the $\sigma$-algebras $\ms F^{\ms M^i}_{\tau^i}$, in the sense of \cite[Subsection~2.1]{Bank2019Lenglarts}, following~\cite{Lenglart1980Tribus, ElKaroui1981Les}.} Note that ``indistinguishability up to an optional time $\tau^i$ for $i$'' and information sets can be easily characterised in this setting, see Proposition~\ref{3-SPF_VECT.prop:information_sets}.

Theorem~\ref{3-SPF_VECT.thm:timing_SPF_well-posed} is remarkable because already in the relatively simple case of timing games, counterexamples to well-posedness are well-known (see, in particular, \cite{Simon1989Extensive,Stinchcombe1992Maximal}). The analysis of a similar, but deterministic setting in \cite{Stinchcombe1992Maximal} concludes that well-posedness can only hold true for a specific subset of strategies, including a) a restriction of the number of simultaneous action, and b) an ``identifiability'' requirement regarding accumulating action from the right.\footnote{As already noted, the inertia nodes in \cite{AlosFerrer2015Repeated} play a similar role. However, this latter model is a rigorous extensive form model, in contrast to Stinchcombe's \emph{ex post} strategy set restriction.} 
An analogue to a) is given by the assumption of a uniform upper bound on the upper vertical level of outcomes processes and strategies. 
Note, however, that --- in contrast to the assumptions and conclusions in \cite{Stinchcombe1992Maximal} --- infinitely many jumps at a given real time are possible without risking well-posedness, as long as there is a uniform upper bound $\alpha$.\footnote{A similar, but finite structural requirement is contained in \cite[Assumption~F.1]{Simon1989Extensive}, also in a deterministic setting.} 
On the other hand, b) is addressed by the regularity requirement --- namely lower semicontinuity from the right --- of strategies.\footnote{A similar, but finite structural requirement is contained in \cite[Assumption~F.3]{Simon1989Extensive}. An analogue to \cite[Assumption~F.2]{Simon1989Extensive} (piecewise continuity) is automatically satisfied for the timing game.} By the inclusion of vertically extended real time, this requirement does not preclude instantaneous reaction.

\begin{remark}
    Theorem~\ref{3-SPF_VECT.thm:timing_SPF_well-posed} also obtains if we do not ask for a uniform upper bound on upper vertical levels. Formally, such an \textsc{spf} obtains by formally taking $\alpha = \mf w_1$ in the definition of the \textsc{spf}'s data. One can indeed show that the corresponding \textsc{spf} is well-posed, but we omit this result and the proof here because in many relevant applications, it appears, there is a uniform upper bound on the vertical level. 
\end{remark}

We next illustrate $\mathbf F$ by constructing an equilibrium in a specific toy example. 
This particular example is motivated by the literature on --- deterministic and stochastic --- games of preemption in continuous time, see \cite{Steg2018Preemptive, Steg2018Preemption, Riedel2017Subgame, Fudenberg1985Preemption}. As already said before, the existence and nature of symmetric equilibria crucially depends on the way instantaneous pro- and reaction is modelled and combined with randomisation. While the just-cited works take a ``discrete time with an infinitesimally fine grid'' perspective (see, e.g.\ \cite{Riedel2017Subgame, Simon1989Extensive}), we propose a formulation using abstract stochastic process forms. This is for the following reasons. First, this theory directly addresses the extensive form characteristics of the problem and provides canonical concepts of equilibrium in a general dynamic and Bayesian setting, including the here-relevant version of subgame-perfect equilibrium. Second, we insist on the fact that exogenous randomness, a Bayesian uncertainty domain, or randomisation can all be taken care of using the generalised stochastic calculus for vertically extended time proposed in Section~\ref{3-SPF_VECT.sec:sto_proc_in_vERT}. Third, the concrete definition of the game, including payoffs, in stochastic process form given below is ``intrinsic'' and does --- even not implicitly --- rely on approximation arguments, though approximation is an important interpretation device. As a consequence, the definition of the game in general, and of the payoffs in particular, can seem a little simpler while being more general than that in \cite[Subsection~4.B]{Fudenberg1985Preemption} or \cite[Definition~2.9]{Riedel2017Subgame} --- provided the abstract theory of stochastic process forms in vertically extended continuous time has been accepted. Fourth, by reformulating the stochastic timing game in the language of stochastic process forms we indicate how much more general games in continuous time, which critically involve instantaneous pro- and reaction, can be analysed in an abstract and tractable framework.

We consider a very stylised example illustrating key structures of the theory, and appearing quite similarly in the cited literature. It is a stochastic version of the ``grab-the-dollar game'' as described in the deterministic setting in \cite{Fudenberg1985Preemption}. Let us recall the basic facts about this example, following, e.g., \cite{Riedel2017Subgame} and the references therein. The story behind is about two players sitting in front of a one dollar bill. At any time, they can decide to (try to) grab it. The player grabbing first gets the dollar. If both grab at the same time, however, they both have to pay a fine. Clearly, this is a toy model for the modelling of preemption. The discrete-time version of the game admits a symmetric equilibrium, given by the behaviour strategy of grabbing with probability $1/2$ at any feasible time. In a standard continuous-time version such an equilibrium does not exist, as is well-known. Indeed, simultaneous grabbing at time zero with probability one is not in equilibrium. Hence, in any symmetric equilibrium in classical continuous time, both players have not stopped with probability one at some time $\e>0$. Given this behaviour of one player, the other player can do strictly better by stopping before $\e$ with probability one, in a way that reduces simultaneous grabbing.\footnote{The latter can be achieved by a distribution of stopping that is absolutely continuous with respect to Lebesgue measure on $(0,\e]_{\R_+}$.} Therefore, there is no evident symmetric continuous-time equilibrium. 

As a solution, Fudenberg and Tirole \cite{Fudenberg1985Preemption}, as well as Riedel and Steg \cite{Riedel2017Subgame}, argue within a ``discrete time with an infinitesimally fine grid''-approach, and introduce ``extended mixed strategies'' (cf.~\cite[Definition~2.7]{Riedel2017Subgame}) consisting not only of (mixed) stopping decisions along the time axis, but also of processes describing the ``conditional stopping probabilities'' of reaction on the infinitesimally fine grid (cf.~\cite{Riedel2017Subgame}). At least in the deterministic case, these have a precise interpretation as limits of grabbing probabilities in behaviour strategies on refining, convergent sequences of discrete-time grids (cf.~\cite{Steg2018Preemption}). They are motivated as a device to effectively control the order of stopping of the two agents if they happen to stop simultaneously, and this becomes clear in the definition of the payoffs in \cite[Definitions~2.9 and~2.11]{Riedel2017Subgame}. Hence, that approach contains an idea of different instances of time attached to one real point in time --- however, in the cited texts, this notion is not formally spelled out. As a consequence, the definition of payoffs is relatively hard to state, and the definition of subgame-perfect equilibria (cf.~\cite[Definition~2.14]{Riedel2017Subgame}) is not tightly linked to the abstract game-theoretic concept of subgame-perfect equilibrium, which is based on extensive forms --- or, at least, extensive form characteristics. This is, of course, linked to the mentioned general difficulties of formulating continuous-time and stochastic games and to the decision of these articles' authors to work with a stacked family of strategic forms rather than with an ``approximate'' extensive form. The latter did simply not exist and so it could not be applied. 

In the present chapter, we have developed such a theory out of abstract principles underlying extensive forms and decision making under probabilistic uncertainty. The notion of ``approximate extensive form'' we suggest is the stochastic process form. We note that it combines extensive form and stochastic aspects --- without being an extensive form, but based on outcomes that derive from extensive forms. Below, we see that the use of stochastic calculus in vertically extended time can give a meaning to a) non-simultaneous and ordered action at the same time that is b) measurable with respect to information given by $\sigma$-algebras. Moreover, c) it provides further fundamental insights regarding the definition of ``subgames'' in the stochastic setting, supporting and extending one key innovation of \cite{Riedel2017Subgame}.

We now discuss the model formally. As in the cited literature, we focus on the two-player case, $I = \{1,2\}$, with symmetric exogenous information, $\ms G^1 = \ms G^2$. Fix the level $\alpha = \mf w+1$ and a probability measure $\P$ on $(\Omega,\ms E)$ making $\upsilon$ uniformly distributed and independent from $\ms G^i$, $i\in I$. Slightly developing \cite[Example~2.4]{Riedel2017Subgame}, the stochastic component of the ``grab-the-dollar game'' now consists in two things. First, player $2$ is not American, and at time $t\in\R_+$, one dollar is worth $\eta_t$ in $2$'s local currency, where $\eta = (\eta_t)_{t\in\R_+}$ is $\ms G^1$-adapted, takes values in $(0,\infty)_\R$, and has continuous paths. Second, the dollar is only released at some exogenously given classical optional (alias stopping) time, for example, the time a neutral referee whistles. If a player acts before that time, both players are fined. All fines are payed in local currency. 
Let $\tau$ be an $\bRp$-valued $\ms G$-stopping time modelling the ``whistle''. 
For $i,j\in I$ with $i\neq j$, we consider the tastes $u_i,u_j\colon W\to\R$, given as follows. Let $(\omega,h)\in W$, and, for all $k\in I$, define $\sigma_k(h) = \inf\{t\in\ovT \mid h^k(t) = 0\}$. Further, let $\eta^1$ denote the constant process with value $1$, and $\eta^2 = \eta$. Let $a_i(h) = 1\{\pi \circ \sigma_i(h) < \mf w\}$ and
\begin{equation}\label{3-SPF_VECT.eq:def.u_i_timing_games}
\begin{aligned}
\begin{array}{lcll}
    u_i(\omega,h) &= u_j(\omega,h) &= -1,\qquad  &\text{if }\sigma_i(h) < \tau(\omega); \\
    u_i(\omega,h) &= \eta^i_{p(\sigma_i(h))}a_i(h),\quad\,u_j(\omega,h) &= 0, \qquad &\text{if }\tau(\omega)\le \sigma_i(h) < \sigma_j(h); \\
    u_i(\omega,h) &= u_j(\omega,h) &= - a_i(h), \qquad&\text{if }\tau(\omega) \le \sigma_i(h) = \sigma_j(h) < \infty; \\
    u_i(\omega,h) &= u_j(\omega,h) &= 0, \qquad &\text{if }\sigma_i(h) = \sigma_j(h) = \infty.
\end{array}
\end{aligned}
\end{equation}
Now, we define the strategy profile $s = (s^i)_{i\in I}$. Let $\phi \cong (\phi^n)_{n\in\mf w}\colon [0,1]\to[0,1]^{\mf w}$ be measurable such that, for each $i\in I$, $\phi(\upsilon^i) = (\phi^n(\upsilon^{i}))_{n\in\mf w} = (\upsilon^{i,n})_{n\in\mf w}$ is an i.i.d.\ sequence of $[0,1]$-uniformly distributed random variables, according to $\P$. Let $i\in I$, $j = 3-i$, and $(t,\omega,h)\in\ovT\times W$. If $h^i(t-) = 0$, or if $t=\infty$, let $s^i_t(\omega,h) = 0$. Else, let
\[ s^i_t(\omega,h) = 
    \begin{cases}
        1, &\text{if } t < \tau(\omega), \\ &\text{ or } \Big( \tau(\omega) \le t,\,\pi(t) < \mf w,\, \upsilon^{i,\pi(t)}(\omega) \ge \frac{\eta^j_{p(t)}}{1+\eta^j_{p(t)}}(\omega)
        ,\\
        0, &\text{else.} 
    \end{cases}
\]
That is, an agent $i$ pursuing strategy $s^i$, starting at an optional time $\tau^i$ of her's does the following, provided $\chi = \Out^\star(s \mid\tau^i,\tilde\chi)$ denotes the state process induced by $s$ given the process $\tilde\chi$ constant to $\bm 1$ (all players are still active exclusively until $\tau^i$). If $\tau$ is not yet reached, i.e.\ on $\{\tau^i\circ(\id_\Omega\star\chi) < \tau\}$, player $i$ waits until $\tau$, and then ``stops'' (i.e.\ switches to zero) with probability $\frac{\eta^j_{\tau}}{1+\eta^j_{\tau}}(\omega)$ at any $t\in\ovT$ with $\tau = p(t)$ and $\pi(t) < \mf w$, independently along the vertical axis below level $\mf w$, until player $i$ manages to actually stop. This implies that $i$ reaches state zero with probability one before time $(p\circ\tau,\mf w)$.

If $\tau$ has already been reached, two cases arise. Let $\hat\tau^i = \tau^i\circ(\id_\Omega\star\chi)$.  On $\{\hat\tau^i \ge \tau,\, \pi\circ\hat\tau^i < \mf w\}$, analogously at any instance of time on the vertical $\mf w$-axis above $\hat\tau^i$ the player stops with probability $\frac{\eta^j_{\hat\tau^i}}{1+\eta^j_{\hat\tau^i}}(\omega)$, until stopping occurs, reaching state zero almost surely before level $\mf w$. In the remaining event, on $\{\hat\tau^i \ge \tau,\, \pi\circ\hat\tau^i \ge \mf w\}$, $i$ stops immediately (and so does the opponent); given the payoffs, they are indifferent about different options at this time, and if $i$ wishes to stop ``in the near future'', then by lower semicontinuity from the right and the fixed bound $\alpha = \mf w+1$ on the upper vertical level, then she even has to stop immediately. Anyway, this latter case is reached with probability zero from the ``classical'' part of time, the real time axis $\R_+$. 

The outcome induced by $s$ can be obtained directly by tilting convergence, as states the following theorem. For simplicity, we restrict the statement to the unique information set at time $0$.\footnote{See Remark~\ref{3-SPF_VECT.rmk:prior_and_dynamic_consistency}.} The theorem can be seen as a stochastic variant of a similar result in \cite{Steg2018Preemption}, and as a representation of it on the level of outcomes. Conversely, this shows that tilting convergence allows for a substantial generalisation of the ``discrete time with an infinitesimally fine grid'' approach (see Subsection~\ref{3-SPF_VECT.subs:tilting_convergence} for further discussion). The theorem follows directly from Proposition~\ref{3-SPF_VECT.prop:optional_times_are_tilting_limits_of_classical_very_simple_optional_procs}. By the very definition of tilting convergence, the approximating sequence $\xi^n$ locally equals the outcome of the classical discrete-time symmetric equilibrium, written onto the grid $G_n$, for any $n\in\N$.

\begin{thm}\label{3-SPF_VECT.thm:timing_game-Out(s)_is_tilting_limit}
    Let $\chi$ be the state process induced by $s$ given the unique information set at time $0$, and $\xi = \chi$ be the corresponding action process. Then, there is a sequence $(\xi^n)_{n\in\N}$ of classical, very simple $\ms F^\vee$-optional processes and a refining, convergent sequence $(G_n)_{n\in\N}$ of classical, deterministic grids $G_n$ compatible with $\xi^n$, for all $n\in\N$, such that $(\xi^n \mid G_n) \convT \xi$ as $n\to\infty$.
\end{thm}

It is shown in the following that $s$ defines an equilibrium on some relatively large $\tilde{\mf P}$ and with respect to the payoffs $u_i$, $i\in I$, and the \textsc{eu} preference structure arising from $\P$ in the most consistent way possible. Precisely, we let $\tilde{\mf P} = (\tilde{\mf P}^i)_{i\in I}$, where for each $i\in I$, $\tilde{\mf P}^i$ is the set of information sets $\mf p = (\tau^i,\mf x)$ such that $\mf x$ contains an $\ms F^i$-progressively measurable state process. That is, counterfactual histories must be independent of the opponent's randomisation device. This restriction appears necessary to the author because the agents cannot use ``new'' independent randomisation devices at each real time $t$ --- it is well known from stochastic analysis that this is incompatible with measurability in the time variable, and in particular, it obstructs progressive measurability and $\ms M^i$-measurability.\footnote{We also refer to the proof of the theorem which would not work for general information sets because, in general, the history of the state process could correlate with the future randomisation of the opponent.} If we wish to obtain i.i.d.\ randomisation over all instants of real time such as to obtain a random distribution in the Schwartz sense, we obtain ``white noise'' which, in that sense, cannot be represented as a function, or process, valued in the original state space. Hence, in order to go beyond $\tilde{\mf P}$ as above, one would have to model strategies by random distributions rather than processes on action spaces. This is deliberately left for future research.

Nevertheless, the setting with $\tilde{\mf P}$ as above is already more general than the setting in \cite{Fudenberg1985Preemption} and that, more general one, in \cite{Riedel2017Subgame}. In the latter, dependence on endogenous histories is encoded by the dependence of each strategic form on \emph{modes} describing which players have already stopped and plans are only revised at real instants of time. In our setting, this would correspond to information sets $\mf p = (\tau^i,\mf x)$ such that $\mf x$ contains a deterministic path and $\tau^i$ is real-valued; and to strategies that can only condition on the left-limit of this path at the current instant of time. 

By contrast, in the model of the present text, $s^i$ is a best response to $s^j$, for both $i,j\in I$, $i\neq j$, with respect to a larger class of (generalised) ``subgames'', given by a general class of optional times $\sigma = \sigma^i$ possibly taking values on higher levels of the vertical half-axis, by our definition of equilibrium and information sets. Note that the model of subgame-perfection in \cite{Riedel2017Subgame,Fudenberg1985Preemption} does not contain that feature: Once the ``atomic'' randomisation procedure for ``extended mixed strategies'' has been started, players can no more revise their plan. By contrast, the stochastic process form in vertically extended time model allows for this. Agents can revise plans at all those cross-sections --- given by information sets --- through ``two-dimensional'' time measurable with respect to information and below level $\alpha = \mf w + 1$. Agents can even perform different actions, with different probabilities, on different levels within the same vertical strip.  

Furthermore, we do not restrict to strategies that are Markovian in the ``mode'', and find that the strategy profile $s$ (which is ``horizontally Markovian'') is an equilibrium even within the larger and more natural strategy space defined here. We also note that most timing game formulations, including the just-cited ones, integrate out the randomisation of both players, including the opponent, by considering the action space $[0,1]$ rather than $\{0,1\}$. In such an approach, the above-mentioned problem becomes invisible by construction yet remains unsolved.

Due to the specific information structure, especially the closed-loop information on all players' states and the symmetric exogenous information, we may interpret the equilibrium property as a stochastic version of subgame-perfect equilibrium (see Remark~\ref{3-SPF_VECT.ex:equilibrium_for_quasi_perfect_information}). This is indeed the term used for this equilibrium in the stochastic stacked strategic form setting by Riedel and Steg in \cite{Riedel2017Subgame}, and by Fudenberg and Tirole in \cite{Fudenberg1985Preemption}.

\begin{thm}\label{3-SPF_VECT.thm:Riedel-Steg_equilibrium}
    There is an \textsc{eu} preference structure $\Pr = (\Pi,U,{\ms W})$ such that:
    \begin{itemize}
        \item $(s,\Pr)$ is in equilibrium on $\tilde{\mf P}$,
        \item $U$ is given by $(u_i)_{i\in I}$ defined above,
        \item ${\ms W} = \ms H^\vee_\infty$.
    \end{itemize}
\end{thm}

In this equilibrium, no stopping occurs strictly before $\tau$. At all $\R_+$-valued $\ms G$-stopping times $\sigma$ not earlier than $\tau$, both players stop with $\P$-probability one on the vertical half-axis above $\sigma$. More precisely, they do so on the initial leg $\{\pi\in\mf w,\, p = \sigma\}$. Simultaneous stopping, sole stopping by player $i$, sole stopping by player $j$ (these terms referring to the extended time half-axis $\ovT$) --- all these three events have the probabilities known from \cite{Riedel2017Subgame}. If $\eta = 1$ is constant equal to one, then all these probabilities are $1/3$ under $\P$. This confirms the findings from \cite{Riedel2017Subgame,Fudenberg1985Preemption}.

\subsection{Stochastic differential games and control}\label{3-SPF_VECT.subs:SDG}

We conclude our study with discussing how stochastic differential games and control problems based on differential equations can be formulated using the language of stochastic process forms. The basic idea is as follows. On a complete probability space $(\Omega,\ms E,\P)$, with $\ms N = \ms N_\P$, we consider the system of abstract stochastic differential equations
\begin{equation}\label{3-SPF_VECT.eq:stochastic_differential_games.1}
    \d \chi^\beta_t = V_{(t,\beta)}(\chi_{[0,(t,\beta))_\ovT},\xi_{[0,(t,\beta))_\ovT}) \cdot\,\d\eta_t, 
    \qquad \beta\in{\mf w_1}+1,
\end{equation}
on the stochastic intervals consisting of $(t,\omega)\in\bRp\times\Omega$ with $\hat\tau(\omega) \le (t,\beta) \le \infty$,\footnote{In an alternative ``localised'' setting, we would only consider the stochastic interval consisting of $(t,\omega)\in\R_+\times\Omega$ with $\hat\tau(\omega) \le (t,\beta)$ and fix a value of $\chi$ at infinity, thereby restricting our attention to $\ovT \setminus \{\infty\}$. The difference between both settings is smaller than one might suppose at first sight, because of the stopping property of stochastic integrals and because $[\![0,1]\!] \subseteq [\![0,\infty)\!) \subseteq [\![0,\infty]\!]$, and $[\![0,1]\!] \to [\![0,\infty]\!], \, (t,\omega) \mapsto (-\log(1-t),\omega)$ is a map preserving many relevant structures.} with initial condition $\chi|_{[\![0,\hat\tau]\!]} = \hat\chi$, where: 
\begin{itemize}
    \item $I$ is a non-empty, finite set of \emph{agents}, or \emph{players};
    \item $\xi\colon\ovT\times\Omega\to\A$ is a stochastic processes describing the agents' \emph{action}, $\A = \prod_{i\in I} \A^i$ and $\A^i = \R^{a_i}$ for $a_i\in\Nast$ and $i\in I$;
    \item $\chi\colon\ovT\times\Omega\to\R^d$ is a stochastic process describing a \emph{state} or \emph{signal}, not fully observable by the agents, for $d\in\Nast$, for any $(t,\beta)\in\ovT$, $\chi^\beta_t = \chi_{(t,\beta)}$, and $\hat\chi\colon{[\![0,\hat\tau]\!]}\to\R^{d}$ is (the restriction of) a stochastic process describing a given initial state;
    \item $\hat\tau\colon\Omega\to\ovT$ is a random time with respect to $\ms P_\ovT$ such that $\pi\circ\hat\tau\le \alpha$ for some $\alpha\in{\mf w_1}$, describing the initial time;
    \item $\eta\colon\R_+\times\Omega\to\R^{m}$ describes an exogenous ``random'' perturbation of the state, where $m\in\Nast$;
    \item the maps $V_t\colon (\R^{d})^{[0,t)_\ovT}\times (\R^a)^{[0,t)_\ovT} \to \R^{d \times m}$, where $t\in\ovT$, describes the infinitesimal linear effect of these perturbations on the state process $\chi$ at time $t$, where $m\in\Nast$ and $a = \sum_{i\in I} a_i$.
\end{itemize}
Stochastic analysis treats the meaning and further properties of such equations, understood as integral equations with respect to the measure $\P$.\footnote{Going back to \cite{Ito1944Stochastic, Ito1951Stochastic}, see, for example, the textbooks \cite{Dellacherie1978Probabilities, Jacod2003Limit, Protter2005Stochastic, Friz2020Course}.} The dependence on the measure may be crucial, of course. For example, $\eta$ could be a continuous $\R^m$-valued $\L^2$-semimartingale with respect to $\P$ and the integral be understood in the sense of $\L^2(\P)$-convergence, following Ito. We only formulate one abstract non-anticipativity assumption on System~\ref{3-SPF_VECT.eq:stochastic_differential_games.1}, which is satisfied by the usual integration concepts:
\begin{itemize}
    \item \hypertarget{3-SPF_VECT.Ass:SDG}{\textbf{Assumption SDG.}}~ Suppose that, with the notation just introduced, $\chi$ solves System~\ref{3-SPF_VECT.eq:stochastic_differential_games.1} for $\xi$ and initial condition $\chi|_{[\![0,\hat\tau]\!]} = \hat\chi$ and initial time $\hat\tau$. Further, let $\xi'\colon\ovT\times\Omega\to\A$ and $\chi'\colon\ovT\times\Omega\to\R^d$ constitute another pair of stochastic processes satisfying $\xi'|_{[\![0,\hat\tau]\!]} = \xi|_{[\![0,\hat\tau]\!]}$ and $\chi'|_{[\![0,\hat\tau]\!]} = \hat\chi$ and such that $\chi'$ solves System~\ref{3-SPF_VECT.eq:stochastic_differential_games.1} for $\xi'$ and the initial condition $\chi'|_{[\![0]\!]} = \hat\chi|_{[\![0]\!]}$ and initial time $0$. Then, $\chi$ solves System~\ref{3-SPF_VECT.eq:stochastic_differential_games.1} for $\xi$ and the initial condition $\chi|_{[\![0]\!]} = \hat\chi|_{[\![0]\!]}$ and initial time $0$.
\end{itemize}

Imitating the stochastic differential games and control literature, System~\ref{3-SPF_VECT.eq:stochastic_differential_games.1} can be used to construct the set of outcomes $\mc W$. For this, we select a subset $W\subseteq\Omega\times\B^\ovT$ and fix the relevant information structures. Let, for any $i\in I$, $\ms H^i$ be a filtration on $W$ satisfying Axiom~\ref{3-SPF_VECT.def:spf.msH_non_anticipative} in Definition~\ref{3-SPF_VECT.def:spf} above, and $\ms M^i$ be a $\sigma$-algebra on $\ovT\times W$ satisfying $\Prd(\ms H^i) \subseteq\ms M^i\subseteq \Opt(\ms H^i)$. Let $\ms H = (\ms H^i)_{i\in I}$, $\ms M = (\ms M^i)_{i\in I}$. Then, let $\mc W$ be \emph{a} non-empty set of pairs $\zeta = (\xi,\chi)$ satisfying the following properties:
\begin{enumerate}
    \item for all $\zeta = (\xi,\chi),\,\zeta' = (\xi',\chi') \in \mc W$, $\chi_0 = \chi'_0$;
    \item for all $\zeta = (\xi,\chi)$ and all $\omega\in\Omega$, $(\omega,\chi(\omega))\in W$;
    \item for all $\zeta = (\xi,\chi)\in \mc W$, $\chi$ is the, up to $\ms N$-indistinguishability, unique stochastic process $\chi$ satisfying System~\ref{3-SPF_VECT.eq:stochastic_differential_games.1} for $\xi$ and initial data $(0,\chi|_{[\![0]\!]})$;
    \item for all $i\in I$, all $\ms H^i$-optional times $\tau^i$ with $[\![0,\tau^i)\!)\in\ms M^i$, all $\tilde\zeta = (\tilde\xi,\tilde\chi),\, \zeta' = (\xi',\chi') \in \mc W$, for $\xi = \tilde\xi$ and the initial data $(\hat\tau,\hat\chi) = (\tau^i \circ (\id_\Omega\star\chi'),\chi'|_{[\![0,\hat\tau]\!]})$, there is an, up to $\P$-indistinguishability, unique stochastic process $\chi$ satisfying System~\ref{3-SPF_VECT.eq:stochastic_differential_games.1}, and (for at least one representative thereof with respect to $\P$-indistinguishability) we have $(\xi,\chi)\in\mc W$.
\end{enumerate}
Let $\hat\chi^0 = \chi|_{[\![0]\!]}$ for some (and all) $(\xi,\chi)\in\mc W$. Moreover, let $\mc I$ denote the set of pairs $(\hat\tau,\hat\chi) = (\tau^i \circ (\id_\Omega\star\chi),\chi|_{[\![0,\hat\tau]\!]})$, where $i\in I$, $\tau^i$ is an $\ms H^i$-optional time with $[\![0,\tau^i)\!)\in\ms M^i$, $\chi$ is such that there is $\xi$ with $(\xi,\chi)\in\mc W$. 
For any $i\in I$, fix a set $\mc S^i$ of $\ms M^i$-measurable maps $s^i\colon\ovT\times W\to \A^i$, such that all elements of the product $\mc S = \bigtimes_{i\in I} \mc S^i$ are admissible.

\begin{proposition}\label{3-SPF_VECT.prop:SDG_well-posed_SPF}
    The data $\mathbf F = (I,\A,\B,W,\mc W,\ms H,\ms M,\mc S)$ defines a well-posed stochastic process form. 
\end{proposition}

\begin{remark}
    Note that, by well-posedness, for every $s\in\mc S$, the ansatz $\xi = s\llcorner\chi$ plugged into System~\ref{3-SPF_VECT.eq:stochastic_differential_games.1} yields the system of abstract stochastic differential equations 
    \begin{equation}\label{3-SPF_VECT.eq:stochastic_differential_games.2}
        \d \chi^\beta_t = V_{(t,\beta)}(\chi_{[0,(t,\beta))_\ovT},(s\llcorner \chi)_{[0,(t,\beta))_\ovT}) \cdot\,\d\eta_t,  
        \qquad \beta\in{\mf w_1}+1,
    \end{equation}
    on the stochastic intervals\footnote{Idem.} consisting of $(t,\omega)\in\bRp\times\Omega$ satisfying $(t,\beta)\le\infty$ and with initial condition $\chi|_{[\![0]\!]} = \hat\chi^0$. If $\chi = \Out^\star(s \mid 0,\tilde\chi)$ and $\xi = s\llcorner \chi$, where $\tilde\chi$ is a state process with $\tilde\chi|_{[\![0]\!]} = \hat\chi^0$, then $\chi$ solves System~\ref{3-SPF_VECT.eq:stochastic_differential_games.2}. If this system is, up to $\P$-indistinguishability, uniquely solvable, then the induced outcome processes of all strategy profiles, given the information set at time $0$, can be characterised by it. Counterfactual induced outcome processes are less handily characterised because they involve conditioning on counterfactual information sets.
\end{remark}

\begin{remark}
    We note that we have covered stochastic differential games and control problems in the so-called ``strong formulation'' here. What about the so-called ``weak formulation''? As a side remark, we note that often there is a way to translate a weakly formulated problem into the language of the strong formulation, e.g.\ by change-of-measure techniques (then, the agents' action consists in determining a density process).\footnote{See, for instance, the \cite[Chapter~21]{Cohen2015Stochastic} for a textbook account on that.} Relatedly, by the Yamada-Watanabe theorem, under certain conditions, weak existence of solutions for stochastic differential equations already implies strong existence.\footnote{See, for instance, the \cite[Theorem~32.14]{Kallenberg2021Foundations} for a textbook account on that.} So, in these cases, there is not much to worry about. However, what about an untranslated, and possibly untranslatable, weakly formulated problem?
    
    It can be seen as a relaxation of the above-discussed formalism. Indeed, this is by accepting that the probability space $(\Omega,\ms E,\P)$ and the information structure $\ms H$ need no more be fixed, but may vary depending on the outcome and strategy process. Moreover, the random perturbation and the initial states are only fixed in distribution, not (almost surely) pathwise. This situation can be interpreted in the sense of an outcome-dependent extension of the exogenous scenario space, adding sufficient randomisation devices in order make sense of the state dynamics in distribution. 
    
    The fundamental decision-theoretic problem with this model lies in that the mere existence of certain scenarios then depends on agents' strategies (creating ``unknown unknowns''). This can be compensated for by directly working on the path space for outcome and strategy processes, thereby fixing $(\Omega,\ms E)$ and a canonical outcome and strategy process on it, an approach often adopted in the stochastic analysis and control literature, already for reasons of mathematical convenience. Then, the decision making of any agent no more consists in choosing a strategy, but in selecting a ``non-anticipative'' probability measure on this path space $(\Omega,\ms E)$. In our approach,\footnote{Note that, in the present thesis, a ``mixed strategy'' in the von Neumann, Morgenstern and Nash sense is modelled simply by a strategy (perhaps only conditionally) independent from the exogenous information of other agents and inducing the same distributions under all agent's beliefs (``secret'' and ``objective'' in the sense of \cite{Aumann1974Subjectivity}). The randomising nature of a strategy is entirely based on its dependence on $(\Omega,\ms E)$. This is discussed in more detail in Subsections~\ref{2-SEF_G.subs:strategies} and~\ref{2-SEF_G.subs:randomisation}.} that is, essentially, a belief on the realised own strategy --- a true paradox which highlights that this weak approach does really relax the basic modelling principles of the present text.
\end{remark}

\begin{remark}
    We note that, without loss of generality, we have restricted our attention to a particular type of stochastic differential equations in this subsection. Instead of controlling the ``velocity'' (like drift rate or volatility) with respect to some exogenous random driver as in Equation~\ref{3-SPF_VECT.eq:stochastic_differential_games.1}, one might also control the driver itself, as, for instance, done in stochastic singular control or optimal stopping (or timing), and related multiple-agent (games) models. One could also make the field $V$ dependent on the distribution of $\chi$, leading to so-called McKean-Vlasov dynamics with applications in non-atomic stochastic or mean-field games (introduced in \cite{Lasry2007Mean,Huang2006Large}, see, e.g.\ \cite{Carmona2018Probabilistic,Carmona2018Probabilistica,Cardaliaguet2019Master}). The discussion of this subsection can be adapted to these cases as well.
\end{remark}

\begin{remark}
    We briefly comment --- on a very informal level, without going into technical details --- on the equilibrium concept for stochastic process forms in the context of this subsection. Stochastic differential games and control problems are often formulated in stacked strategic form. For instance, the optimality criterion in a strongly Markovian setting with a single agent $i$ may be of the form
    \[ (\ast) \qquad \text{Maximise }J(x;s) = \E_x [ u_i(\chi_\infty^s) ] \text{ over } s = s^i, \qquad \forall x\in \R^d, \]
    supposing $\chi^s$ to be a strong Markov process, given by $\chi^s = \Out^\star(s \mid 0,\tilde\chi)$ for some fixed state process $\tilde\chi$ corresponding to an initial condition, with continuous $u_i\colon\R^m \to [-4,2]$. Very roughly speaking, one may (try to) apply the strong Markov property in order to rewrite this as a maximisation problem of conditional expectations given a suitable filtration evaluated at relevant optional times. This in turn lies within the framework of the equilibrium (in the single-agent case, say, optimiser) concept for stochastic process forms in Definition~\ref{3-SPF_VECT.def:equilibrium}. If all this works out, then the stacked strategic forms thus turn out as a special case of a stochastic process form, also from the perspective of equilibria and optimisers.

    This perspective on dynamic optimisers (and equilibria) is also compatible with and explains standard methods from control theory. The representation in $(\ast)$ is used because it gives rise to a function in $x$, the value function $V(x) = \sup_s J(x;s)$. The dynamic programming (alias Bellman) principle exploits the optimality of $s$ at different ``information sets'' or ``subgames'', given sufficient regularity of $V$. It can be used to study local properties of $V$. This can be a powerful method to characterise, verify, or compute optimisers or equilibria using partial differential equations (called Hamilton-Jacobi-Bellman equations), thereby further justifying the approach using stochastic process forms in (possibly vertically extended) continuous time.\footnote{For the purely control-theoretic aspects, we refer, for instance, to the textbooks \cite{Cohen2015Stochastic, Pham2009Continuous}.}
\end{remark}

\silentchapter{Conclusion}
An abstract and general language of the extensive form characteristics of dynamic games under probabilistic uncertainty has been introduced. This language is compatible with the a) refined partitions approach, b) the decision tree approach, c) the action path approach along the endogenous (``decision making'') dimension; moreover, it allows for general probabilistic noise along the exogenous (``nature'', ``uncertainty domain'', ``external randomisation device'') dimension, described by a measurable space and families of sub-$\sigma$-algebras modelling dynamic updates.

In particular, it is possible to implement general stochastic processes as background noise on refined partitions-based decision forests without encountering outcome generation problems for a ``nature'' agent, while allowing for a rigorous decision-theoretic interpretation of the relationship between endogenous and exogenous information and choices, and in particular of dynamic equilibria. This represents an improvement over the existing state of the art in both refined partitions-based decision and game theory (e.g.\ in \cite{AlosFerrer2016Theory,AlosFerrer2015Repeated}) and stochastic control and differential games theory (e.g.\ \cite{Pham2009Continuous,Karatzas1998Methods,Carmona2018Probabilistic,Cohen2015Stochastic}). 

This has been achieved by abandoning the assumption of a ``nature'' agent and instead constructing a theory of \emph{stochastic decision forests}, where exogenous uncertainty relates to knowledge about the realised tree. These decision forests satisfy duality between outcomes and nodes and are therefore amenable to sequential decision theory. They can be represented as forests of decision trees, so that the theory in \cite{AlosFerrer2016Theory} applies on a scenariowise level. They allow for a notion of similarity across moves on different trees, given by random moves, which, under weak hypotheses on the underlying order-theoretic structure, form the moves of a decision tree in its own right. These serve as the basis for both a \emph{structure of exogenous information} revelation similar to filtrations in probability theory and a concept of \emph{adapted choices} compatible with exogenous information structures. Adapted choices implement the refined-partitions notion of choice under uncertainty in sequential decision theory while capturing the measurability assumptions on choices typically made in the literature.

Consistency criteria defined on suitable combinations of a stochastic decision forest, exogenous information structures, and adapted choices give rise to the notion of \emph{stochastic extensive forms}. These provide an implicit model of information sets and strategies as maps from information sets to choices available at these --- that is, complete contingent plans of action in the classical decision-theoretic sense. Well-posedness of stochastic extensive forms can be characterised in terms of basic order-theoretic properties of the underlying forest. In particular, well-posedness of a stochastic extensive form implies that decision paths are well-ordered. For well-posed stochastic extensive forms, dynamic equilibrium can be defined in a canonical way, implementing the meta-concepts of perfect Bayesian equilibrium and subgame-perfect equilibrium, and giving rise to a complete model of dynamic \emph{games} in stochastic extensive form. The problem of randomisation admits an alternative representation in terms of the exogenous scenario space.

In addition, a general model of action path stochastic decision forests and extensive forms has been constructed, based on a small number of easily verifiable conditions. This model can serve as a unified decision-theoretic foundation for a large class of stochastic decision problems. By the aforementioned general classification result, well-posedness essentially depends on the well-order property of the time half-axis. Thus, on the one hand, stochastic differential games and continuous-time timing games cannot be modelled that way. On the other hand, a large class of well-posed stochastic extensive forms based on action paths indexed over well-ordered time grids, for example those embedded into $\R_+$, obtains.

Taking limits of the corresponding outcomes, in a way that preserves accumulating reaction behaviour, leads to the notions of tilting convergence and of \emph{vertically extended continuous time}. On this extended time half-axis, a suitable stochastic analysis --- with consistent notions of progressive measurability, optional and decision times and processes, a Début Theorem etc.\ --- can be defined. The resulting relaxed game-theoretic model based on the \emph{stochastic process form}, which avoids a) well-posedness problems by reducing the set of strategies and b) measurability problems by supposing strategies to be progressively measurable (a strengthening, or, depending on the perspective, weakening, of the product form approach \cite{Witsenhausen1971Information,Witsenhausen1975Intrinsic,Heymann2022Kuhns}), can be justified on the grounds of tilting approximation of outcomes, but, at the same time, encompasses a vast class of applications: stochastic differential games, continuous-time timing games, continuous-time Bayesian games (e.g.\ principal-agent problems). In a case study of the timing game, we see that the symmetric, rando\-mised preemption equilibrium predicted by \cite{Fudenberg1985Preemption,Riedel2017Subgame} obtains also in this setting, conditional on a vast class of subgames.\smallskip

The language proposed in this dissertation is developed out of first principles of game and probability theory, which thus encompass different disciplines. Despite their strong conceptual and historical links these disciplines sometimes speak quite different languages, which blurs the view on those principles. The dissertation has pointed to some of the key difficulties, which makes it understandable that one often recurs to specific ad-hoc formulations. Yet, the dissertation also demonstrates that it is possible to make the link in a certain sense, given sufficient mathematical effort, and that this effort may improve the conceptual understanding. 
For example, the symmetric preemption equilibrium arises rather naturally from the general theory; no specific theory of ``stopping intensities'' alias conditional stopping probabilities as in \cite{Fudenberg1985Preemption,Riedel2017Subgame}, which heavily uses the two-player and timing games structure, is needed. Using mathematics, a convincing but see\-mingly ad-hoc solution in ``economic'' game theory can be represented by a canonical ``economic'' game-theoretic solution concept.

The present text focuses on abstract and general theory, illustrated via simple examples, and on the general link to stochastic process-based game and control theory. Moreover, we have seen that this theory yields a well-posed model of timing games, compatible with and providing further footing to the existing theory on preemption games. Other things remain to be addressed. For example, we argue that stochastic differential games are an important class of problems covered by stochastic process forms. It is beyond the scope of this text to provide another detailed case study of a differential game in stochastic process form within the necessary detail. We think that future research on stochastic differential games with preemption features, or asymmetric or partial information, with applications in economics and finance (as in \cite{Bank2024Optimal, Bank2025How}), for example, could benefit from and draw upon the game-theoretic formalism introduced in this thesis. 

Furthermore, using tilting convergence we have provided a general approximation mechanism on the level of outcomes. A game-theoretically very relevant question would be how this can be lifted to equilibria (see, e.g., \cite{Simon1989Extensive,Fudenberg1986Limit,Steg2018Preemption} for related literature). However, an approximation on the level of equilibria is more dependent on specific assumptions on the concrete problems at hand, e.g.\ regularity assumptions on tastes alias payoffs function. This is beyond the scope of this text, but could be analysed in more specific situations, under specific regularity assumptions, using the language from the present thesis.

This thesis also provides a contribution to stochastic analysis, which is formally independent from the decision-theoretic motivation underlying the present text. It might be useful at all places where accumulation creates discontinuities that become invisible in the usual pointwise limit. In that sense, this relates to the literature on stochastic integration and stochastic differential equations, which is a theory about limits of simple integrals as the time grids become arbitrarily fine. The question of finding an adequate notion for this in situations involving jumps arises in applications (see, e.g., \cite{Bank2024Optimal} and the references therein) and has motivated abstract theory (see, e.g., \cite{Chevyrev2024Superdiffusive, Marcus1981Modeling} and the references therein). Stochastic analysis in vertically extended continuous time provides an alternative candidate for this. A next step would necessarily involve attempts to formulate stochastic integration intrinsically in this setting, based on the notions of optional and predictable times and processes etc.\ introduced and studied in the present text.\footnote{Thanks are due to Peter Bank for a discussion on the subject.} \smallskip

To conclude, the theory presented in this text can help economists to better understand certain economics using mathematics, and it can help mathematicians to relate their ``own'' objects (such as differential games) more tightly to relevant economic questions (e.g.\ relevant dynamic equilibrium concepts). On that note, we end this text with Arnold Sommerfeld recalling Einstein to have ``said on one occasion: `Since the mathematicians have invaded the theory of relativity, I do not understand it myself any more {[sic!]'}'', which could leave us skeptical about the use and intentions of mathematicians. Yet, Sommerfeld continues that ``soon thereafter, at the time of the conception of the general theory of relativity, he readily acknowledged the indispensability of the four-dimensional scheme of Minkowski.''\footnote{Cf.\ \cite{Sommerfeld1949Albert}} Needless to say, extensive forms are certainly no Minkowski spacetime --- nevertheless, if Einstein changed his mind regarding the link of abstract mathematics and applications, there must be something true about it.

\cleardoublepage
\phantomsection
\addcontentsline{toc}{chapter}{Bibliography}
\bibliography{DSEF}
\bibliographystyle{plain}

\newpage

\appendix
\chapter{Proofs}\label{chap:App1-Proofs}
\section{Chapter~\ref{chap:1-SDF_AC}}
\subsection{Section~\ref{1-SDF_AC.sec:def}}

We first prove Lemma~\ref{1-SDF_AC.lemma:partion_of_forest} and Theorem~\ref{1-SDF_AC.thm:decision_forest=forest_of_decision_trees}, without using the Propositions \ref{1-SDF_AC.prop:f(v)=uparrow v}, \ref{1-SDF_AC.prop:decision_forest_over_set_is_decision_forest}, and \ref{1-SDF_AC.prop:repr_by_dec_paths_of_decision_forest}.

\begin{proof}
[Proof of Lemma~\ref{1-SDF_AC.lemma:partion_of_forest}] 
    \footnote{We recall that this lemma can actually be seen as an explicitly order-theoretic reformulation of a basic result from graph theory (see the discussion in \cite[Section~I.1]{Bollobas2013Modern}). As the claim that it is a reformulation requires proof, and also for the reader's convenience, a proof is given nonetheless.}
    For $x,y\in F$, let $x\sim y$ iff there is $z\in F$ such that $z\ge x$ and $z\ge y$. $\sim$ is clearly reflexive and symmetric. It is also transitive, because any principal up-set in $(F,\ge)$ is a chain and $\ge$ is transitive. The equivalence classes of $\sim$ define trees by construction of $\sim$.
	If there are $x,y\in F$ such that $x\ge y$, then (by taking $z=x$) we obtain that $x\sim y$. Hence, the equivalence classes of $\sim$ define a partition with the claimed property.
	
	Let $\mc F$ be an arbitrary partition with the claimed property. Then, for $x,y\in F$, $x$ and $y$ belong to the same partition member iff $x\sim y$. Indeed, if they belong to the same partition member $T$, which by assumption is a tree, then $x\sim y$ by definition of a tree. If conversely $x\sim y$, then there is $z\in F$ with $z\ge x$ and $z\ge y$. Hence, there are $T, T'\in\mc F$ such that $x,z\in T$ and $y,z\in T'$. Thus $T\cap T' \neq \emptyset$, whence $T=T'$. Hence, $\mc F$ is uniquely determined.

    We now prove the statement of the second sentence. Suppose the forest $(F,\ge)$ to be rooted, and let $T\in\mc F$. By definition of $\mc F$, $T$ is a non-empty tree. Let $x\in T$. Then the principal up-set $\uparrow x = \{y\in F \mid y \ge x\}$ in $F$ contains a maximal element $y$. By definition of $T$, $y\in\uparrow x$ implies $y\in T$. $\uparrow x$ is a chain, because $(F,\ge)$ is a forest. Hence, $y$ is even a maximum of $\uparrow x$ with respect to the induced order. Let $z\in T$ be an arbitrary element of the tree $T$. Then $T\cap\uparrow x \cap \uparrow z \neq \emptyset$. Take an element $u$ in this intersection. Then $y\ge u \ge z$, whence $y\ge z$. Thus $y$ is a maximum of $(T,\ge)$. In particular, $y$ is maximal in $(T,\ge)$. Therefore, $(T,\ge)$ is a rooted tree according to our definition and has a maximum.
\end{proof}

We prepare the proof of Theorem~\ref{1-SDF_AC.thm:decision_forest=forest_of_decision_trees} with two lemmata.

\begin{lemma}\label{1-SDF_AC.lemma:set_forest}
	Let $V$ be a set and $F$ be a $V$-poset such that $(F,\supseteq)$ is a rooted forest. Let $\mc F$ be the set of its connected components and for any $T\in\mc F$, let $V_T$ be the root of $(T,\supseteq)$. Then, for any $T\in\mc F$, we have:
	\begin{enumerate}
		\item\label{1-SDF_AC.lemma:set_forest.T=downarrow_V_T} $T = \{x\in F \mid V_T \supseteq x\}$;
		\item\label{1-SDF_AC.lemma:set_forest.W(V_T)=w_subseteq_T} $W(V_T) = \{w\in W \mid w\subseteq T\}$;
		\item\label{1-SDF_AC.lemma:set_forest.W(x)} for any $x\in T$, $W(x)$ is equal to the set of maximal chains $w$ in $(T,\supseteq)$ with $x\in w$; in particular, $W(V_T)$ is equal to the set of maximal chains in $(T,\supseteq)$.
	\end{enumerate}
\end{lemma}

\begin{proof}
	(Ad 1):~ If $x\in T$, then $V_T\supseteq x$, because $V_T$ is the root of $T$. Conversely, let $x\in F$ such that $V_T\supseteq x$. In combination with $V_T\in T$, we obtain $x\in T$, because $T\in\mc F$.\smallskip
	
	(Ad 2):~ Let $w\in W(V_T)$ and $x\in w$, then $V_T\supseteq x$ or $x\supseteq V_T$, whence $x\in T$, by definition of $V_T$ and $\mc F$. Thus $w\subseteq T$. Conversely, let $w\in W$ such that $w\subseteq T$. By definition of $V_T$, any $x\in w$ satisfies $V_T \supseteq x$. Because of maximality of the chain $w$, we must have $V_T\in w$.\smallskip
	
	(Ad 3):~ Let $x\in T$.
	
	Let $w\in W(x)$. Then $w\subseteq T$ since, for any $y\in w$ we have $x\supseteq y$ or $y\supseteq x$, whence $y\in T$. The chain $w$ is also maximal in $(T,\supseteq)$ because $T$ is a subset of $F$ and $w$ is maximal in $(F,\supseteq)$ by assumption.
	
	Now let $w$ be a maximal chain in $(T,\supseteq)$ with $x\in w$. By definition of $V_T$, we have $V_T\supseteq y$ for all $y\in w$, and maximality of the chain $w$ implies that $V_T\in w$. Let $y\in F$ be such that $w\cup\{y\}$ is a chain in $(F,\supseteq)$. Hence, $V_T\supseteq y$ or $y\supseteq V_T$, in particular we get $y\in T$. As $w$ is maximal in $(T,\supseteq)$, $y\in w$. Hence $w$ is already maximal in $(F,\supseteq)$.
	
	The second part follows from the first, since all maximal chains $w$ in $(T,\supseteq)$ satisfy $V_T\in w$. Indeed, as a maximal chain in the rooted tree $(T,\supseteq)$, $w$ is non-empty, and any element $x\in w\subseteq T$ satisfies $V_T\supseteq x$ by definition of $V_T$. By maximality of the chain $w$, we infer $V_T\in w$.
\end{proof}

\begin{lemma}\label{1-SDF_AC.lemma:decision_forest}
	Let $F$ be a decision forest on a set $V$ with $f$ as in Definition~\ref{1-SDF_AC.def:decision_forest}. Let $\mc F$ be the set of the connected components of $(F,\supseteq)$. Then, for any $v\in V$ and any $T\in\mc F$, the following statements are equivalent:
    \begin{enumerate}
        \item $f(v)\subseteq T$;
        \item $v\in V_T$;
        \item $V_T\in f(v)$.
    \end{enumerate}
\end{lemma}

\begin{proof}
	Let $v\in V$ and $T\in\mc F$. By Lemma~\ref{1-SDF_AC.lemma:set_forest}, Part~\ref{1-SDF_AC.lemma:set_forest.W(V_T)=w_subseteq_T}, the first and third statements are equivalent. Thus, it remains to show that $v\in V_T$ iff $V_T\in f(v)$.

    As $f$ is a bijection, $v\in V_T$ holds true iff $f(v) \in \mc P f(V_T)$. By definition of $f$, $\mc P f(V_T) = W(V_T)$. Hence, by definition of $W(V_T)$ and as $f(v)\in W$, $f(v)\in \mc P f(V_T)$ is equivalent to $V_T\in f(v)$. This proves the lemma.
\end{proof}

\begin{proof}[Proof of Theorem~\ref{1-SDF_AC.thm:decision_forest=forest_of_decision_trees}]
	(Ad necessity):~ Suppose $F$ to be a decision forest on $V$. Let $W$ be the set of maximal chains in $(F,\supseteq)$, and $f\colon V \rightarrow W$ a bijection such that, for every $x\in F$, $(\Pot f)(x) = W(x)$, just as in Definition~\ref{1-SDF_AC.def:decision_forest}.
	
	We claim that for all $v\in V$ there is a unique $T\in\mc F$ such that $f(v)\subseteq T$. Property \ref{1-SDF_AC.thm:decision_forest=forest_of_decision_trees.partition} then follows from Lemma~\ref{1-SDF_AC.lemma:decision_forest}.
	
	For the proof, let $v\in V$. As s maximal chain in the rooted forest $(F,\supseteq)$, $f(v)$ is non-empty. Hence, there is $x\in f(v)$.
	As $\mc F$ is a partition of $F$, there is a unique $T\in\mc F$ with $x\in T$. For arbitrary $y\in f(v)$, we have $x\subseteq y$ or $y\subseteq x$, since $f(v)$ is a chain. Hence, $x\sim y$, or in other words $y\in T$. Thus $T$ is the unique element of $\mc F$ with $f(v)\subseteq T$.
	
	For Property \ref{1-SDF_AC.thm:decision_forest=forest_of_decision_trees.T_decision_tree}, let $T\in\mc F$. Then $(T,\supseteq)$ defines a rooted tree with root $V_T$, by definition of $\mc F$. Moreover, for all $v\in V$, we have $f(v)\in W(V_T)$ iff $v\in V_T$, by Lemma~\ref{1-SDF_AC.lemma:decision_forest}. Hence, $f_T:=f|_{V_T}$ yields a bijection $V_T \rightarrow W(V_T)$. 
	
	By Lemma~\ref{1-SDF_AC.lemma:set_forest}, Part~\ref{1-SDF_AC.lemma:set_forest.W(x)}, second sentence, $W(V_T)$ equals the set of maximal chains in $(T,\supseteq)$. By definition of $V_T$, all $x\in T$ satisfy $V_T\supseteq x$ so that the definition of $f$ yields:
    \[ (\mc P f|_T)(x) = (\mc P f)(x) = W(x). \]
    By Lemma~\ref{1-SDF_AC.lemma:set_forest}, Part~\ref{1-SDF_AC.lemma:set_forest.W(x)}, first sentence, $W(x)$ consists exactly of all maximal chains in $(T,\supseteq)$ containing $x$. Hence, $T$ is its own representation by decision paths.\smallskip
	
	(Ad sufficiency):~ Let $F$ satisfy Properties \ref{1-SDF_AC.thm:decision_forest=forest_of_decision_trees.partition} and \ref{1-SDF_AC.thm:decision_forest=forest_of_decision_trees.T_decision_tree} from the theorem. It remains to show that $F$ is its own representation by decision paths, i.e.\ Part~\ref{1-SDF_AC.def:decision_forest:repr_by_dec_paths} in Definition~\ref{1-SDF_AC.def:decision_forest}.
	
	By Property \ref{1-SDF_AC.thm:decision_forest=forest_of_decision_trees.T_decision_tree} and Lemma~\ref{1-SDF_AC.lemma:set_forest}, Part~\ref{1-SDF_AC.lemma:set_forest.W(x)}, for each $T\in\mc F$, there is a bijection $f_T\colon V_T\rightarrow W(V_T)$ such that for all $x\in T$, $(\mc P f_T)(x)$ equals the set of maximal chains in $(T,\supseteq)$ containing $x$. By Part~\ref{1-SDF_AC.lemma:set_forest.W(x)} of Lemma~\ref{1-SDF_AC.lemma:set_forest}, this set is equal to $W(x)$. By Property \ref{1-SDF_AC.thm:decision_forest=forest_of_decision_trees.partition}, we obtain a map $f\colon V \rightarrow W$ satisfying, for any $T\in\mc F$ and $v\in V_T$, $f(v) = f_T(v)$. $f$ is injective by Lemma~\ref{1-SDF_AC.lemma:set_forest}, Part~\ref{1-SDF_AC.lemma:set_forest.W(V_T)=w_subseteq_T}, since $\mc F$ is a partition. $f$ is also surjective, hence bijective, since $F\neq\emptyset$ and so every maximal chain $w\in W$ contains an element $x$ for which there is $T\in\mc F$ with $x\in T$. Hence, $w\in W(x)$ which is a subset of $W(V_T)$ by Lemma~\ref{1-SDF_AC.lemma:set_forest}, Part~\ref{1-SDF_AC.lemma:set_forest.W(x)}. Hence, there is $v\in V_T$ with $f(v) = f_T(v) = w$.
	
	Let $x\in F$. Then there is $T\in\mc F$ with $x\in T$. Then $V_T \supseteq x$. Hence, $(\Pot f)(x) = (\Pot f_T)(x)$. By definition of $f_T$ and Part~\ref{1-SDF_AC.lemma:set_forest.W(x)} of Lemma~\ref{1-SDF_AC.lemma:set_forest}, we have  
	\[ (\Pot f_T)(x) = \{ w\in W(V_T) \mid x\in w \}. \]
	Lemma~\ref{1-SDF_AC.lemma:set_forest}, Part~\ref{1-SDF_AC.lemma:set_forest.W(x)}, implies that the latter set equals $W(x)$. Thus, $(\mc P f)(x) = W(x)$.
\end{proof}

\subsection{Section~\ref{1-SDF_AC.sec:sdf}}

\begin{proof}[Proof of Lemma~\ref{1-SDF_AC.lemma:sdf.X.roots}]
    (Ad Part~\ref{1-SDF_AC.lemma:sdf.X.roots.OC}):~ Suppose that $(F,\pi,\X)$ is order consistent. Let $\x\in\X$. Suppose there is $\omega\in D_\x$ such that $\x(\omega) = W_\omega$. Let $\omega'\in D_\x$. It suffices to show that $\x(\omega') = W_{\omega'}$.
    
    By the covering property in Axiom~\ref{1-SDF_AC.def:sdf.X.cov}, there is $\x'\in\X$ such that $\x'(\omega') = W_{\omega'}$. Then $\x'(\omega') \supseteq \x(\omega')$. Hence, $\x' \ge_\X \x$ by Axiom~\ref{1-SDF_AC.def:sdf.X.OC}. Thus, by definition of $\ge_\X$, $\omega\in D_{\x'}$ and $W_\omega \supseteq \x'(\omega) \supseteq \x(\omega) = W_\omega$, whence $\x'(\omega) = \x(\omega)$ which in turn implies $\x' = \x$ by order consistency, Axiom~\ref{1-SDF_AC.def:sdf.X.OC}. Thus, $\x(\omega') = \x'(\omega') = W_{\omega'}$.\smallskip

    (Ad Part~\ref{1-SDF_AC.lemma:sdf.X.roots.surely_non-trivial}):~ $D=\Omega$ iff for all $\omega\in\Omega$, $W_\omega\in X$. In view of Axiom~\ref{1-SDF_AC.def:sdf.conn_comp}, the set of roots of $(F,\supseteq)$ is given by precisely all $W_\omega$, $\omega\in\Omega$. The claimed equivalence follows.\smallskip

    (Ad Part~\ref{1-SDF_AC.lemma:sdf.X.roots.root_random_move}):~ Suppose that $(F,\pi,\X)$ is order consistent and maximal, and such that $X\neq\emptyset$. It suffices to show the claim $(\ast)$ that there is exactly one $\x\in\X$ whose image contains a root of $(F,\supseteq)$. Indeed, then, by Axiom~\ref{1-SDF_AC.def:sdf.X.cov}, $D\subseteq D_\x$ and $\x(\omega) = W_\omega = \x_0(\omega)$ for all $\omega\in D$, and by Part~\ref{1-SDF_AC.lemma:sdf.X.roots.OC} of this lemma, proven just before, $D=D_\x$, whence $\x_0 = \x\in\X$.

    Regarding the proof of $(\ast)$, note that by hypothesis there is $x\in X$. As $W_{\pi(x)}\supseteq x$, $W_{\pi(x)}\in X$. By Axiom~\ref{1-SDF_AC.def:sdf.X.cov}, there is $\x\in\X$ with $x\in\im\x$. Hence, if $(\ast)$ did not hold true, there would have to be at least two distinct random moves $\x_1,\x_2\in\X$ whose images contain roots of $F$. By Part~\ref{1-SDF_AC.lemma:sdf.X.roots.OC}, the images of both $\x_1$ and $\x_2$ then would contain roots only. As $\x_1\neq\x_2$, by Axiom~\ref{1-SDF_AC.def:sdf.X.OC}, we would have $D_{\x_1}\cap D_{\x_2} = \emptyset$. Then, let $\bar\x = \x_1 \cup \x_2$,\footnote{That is, $\bar\x\colon D_{\x_1} \cup D_{\x_2}\to X$ such that $\bar\x(\omega) = \x_k(\omega)$ for $\omega\in D_{\x_k}$, for both $k=0,1$.} 
    and $\bar\X = (\X \setminus \{\x_1,\x_2\}) \cup \{\bar\x\}$. 
    
    Clearly, $D_{\bar\x} = D_{\x_1} \cup D_{\x_2}$ would be a non-empty element of $\ms E$ and $\bar\x$ would be a section of $\pi$ on $D_{\bar\x}$. It is evident that $\bar\X$ would be order consistent and induce a covering of $X$. By construction, $\X$ would refine $\bar\X$. Thus, maximality (see Axiom~\ref{1-SDF_AC.def:sdf.X.max}) would imply the false statement $\bar\X = \X$. Hence, $(\ast)$ obtains and the proof is complete.
\end{proof}

\begin{proof}[Proof of Proposition~\ref{1-SDF_AC.prop:ev_on_Tr_is_iso}]
    (Ad ``order embedding''):~ Let $(\y_1,\omega_1),(\y_2,\omega_2)\in\Tr\bullet\Omega$. If both $\y_1,\y_2$ are elements of $\X$, the implication ``$\Rightarrow$'' is a consequence of the definition of $\ge_\X$, and the other implication ``$\Leftarrow$'' can be shown as follows: If $\y_1(\omega_1) \supseteq \y_2(\omega_2)$, then both moves belong to the same connected component of the rooted forest $(F,\supseteq)$ (whose existence Axiom~\ref{1-SDF_AC.def:sdf.df} ensures). By Axioms~\ref{1-SDF_AC.def:sdf.conn_comp} and \ref{1-SDF_AC.def:sdf.X.section}, $\omega_1 = \omega_2$. Then, Axiom~\ref{1-SDF_AC.def:sdf.X.OC} implies that $\y_1 \ge_\X \y_2$, whence $\y_1\ge_\Tr \y_2$. 
    
    Else, for some $k=1,2$, $\y_k$ is a random terminal node so that $D_{\y_k} = \{\omega_k\}$ and $\y_k(\omega_k) = \{w_k\}$ for some $w_k\in W$.    
    If $k=1$, then first, $\y_1 \ge_\Tr \y_2$ is equivalent to $\y_1 = \y_2$. Second, $\y_1(\omega_1) \supseteq \y_2(\omega_2)$ is equivalent to $\y_1(\omega_1) = \y_2(\omega_2)$, because $\y_1(\omega)$ is terminal in $F$. But then $\y_2$ must be a random terminal node as well, whence $\y_1 = \y_2$ and $\omega_1=\omega_2$. Conversely, if these two equalities hold true, then $\y_1(\omega_1) = \y_2(\omega_2)$ follows.
    
    If $k=2$, then $\y_1(\omega_1) \supseteq \y_2(\omega_2)$ is equivalent to the conjunction of these three statements: $\omega_1=\omega_2$, $\omega_2\in D_{\y_1}$, and $w_2\in \y_1(\omega_1)$. By definition of $\ge_\Tr$, this is equivalent to $\y_1 \ge_\Tr \y_2$ and $\omega_1 = \omega_2$.\smallskip

    (Ad ``bijection''):~ Let $(\y_1,\omega_1),(\y_2,\omega_2)\in\Tr\bullet\Omega$ such that $\y_1(\omega_1) = \y_2(\omega_2)$. As the evaluation map is an order embedding, which we have proven just before, we obtain $\y_1 = \y_2$ and $\omega_1 = \omega_2$. Regarding surjectivity, let $y\in F$. If $y$ is terminal, then there is $w\in W$ with $y = \{w\}$ because $F$ is a decision forest on $W$.\footnote{Although this is discussed in \cite{AlosFerrer2005Trees}, we give an argument here for the reader's convenience. Let $f$ be the map from Definition~\ref{1-SDF_AC.def:decision_forest}. If $y$ is terminal, then $(\mc P f)(y) = W(y)$ contains exactly one element, namely $\uparrow y$. As $f$ is injective, $y$ must be a singleton.}
    Hence, $y = \y(\omega)$ for $\omega = \pi(y)$ and the random terminal node $\y = \{(\omega,\{w\})\}$. If $y$ is a move, then, by the definition of $\X$, Property \ref{1-SDF_AC.def:sdf.X.cov}, there is $\x\in\X$ and $\omega\in D_\x$ such that $\x(\omega) = y$.
\end{proof}

\begin{proof}[Proof of Theorem~\ref{1-SDF_AC.thm:Xrm_is_dec_tree}]
    (Ad ``poset''):~ It follows directly from the definitions that $\ge_\Tr$ defines a partial order on $\Tr$.\smallskip
    
    (Ad ``forest''):~ Let $\y_0\in\Tr$ and $\y_1,\y_2\in\uparrow \y_0$. As $D_{\y_0}\neq\emptyset$ by Definition~\ref{1-SDF_AC.def:sdf} and the definition of random terminal nodes, there is $\omega\in D_{\y_0}$, and we have $\omega\in D_{\y_k}$ and $\y_k(\omega) \supseteq \y_0(\omega)$ for both $k=1,2$. As $(F,\supseteq)$ is a forest, $\uparrow \y_0(\omega)$ is a chain, thus $\y_1(\omega) \supseteq \y_2(\omega)$ or $\y_2(\omega) \supseteq \y_1(\omega)$. Hence, by Proposition~\ref{1-SDF_AC.prop:ev_on_Tr_is_iso}, we get $\y_1\ge_\Tr \y_2$ or $\y_2\ge_\Tr \y_1$. We conclude that $(\Tr,\ge_\Tr)$ is a forest. \smallskip

    (Ad ``decision forest''):~ Let $\y_1,\y_2\in\Tr$ be such that $\y_1\neq \y_2$. If they are not comparable by $\ge_\Tr$, that is neither $\y_1 >_\Tr \y_2$ nor $\y_2>_\Tr \y_1$, then any maximal chain $\w$ in $(\Tr,\ge_\Tr)$ with $\y_1\in\w$ satisfies $\y_2\notin\w$. Using the axiom of choice in the form of the Hausdorff maximality principle, there exists indeed such a $\w$. 
    
    For symmetry reasons, it remains to consider the case in that we have $\y_1 >_\Tr \y_2$. Then, by Proposition~\ref{1-SDF_AC.prop:ev_on_Tr_is_iso}, we have for all $\omega^\prime\in D_{\y_2}$, $\y_1(\omega^\prime) \supsetneq \y_2(\omega^\prime)$. As $D_{\y_2} \neq \emptyset$, we can select $\omega\in D_{\y_2}$. We can also select $w\in \y_1(\omega) \setminus \y_2(\omega)$. In particular, $\y_2(\omega) \notin \uparrow \{w\}$. As the latter set is a maximal chain in $(F,\supseteq)$, there is $y_3\in \uparrow \{w\}$ such that $y_3$ and $\y_2(\omega)$ cannot be compared by $\supseteq$. As $\y_1(\omega)$ is also an element of the chain $\uparrow \{w\}$, this implies $\y_1(\omega) \supsetneq y_3$. By Proposition~\ref{1-SDF_AC.prop:ev_on_Tr_is_iso}, there is $\y_3\in\Tr$ such that $\omega\in D_{\y_3}$ and $y_3 = \y_3(\omega)$. This lemma also implies that $\y_2$ and $\y_3$ are not comparable via $\ge_\Tr$. Hence, as we have proven just above, there is a maximal chain $\w$ in $(\Tr,\ge_\Tr)$ with $\y_2\notin\w$ and $\y_3\in\w$. By Proposition~\ref{1-SDF_AC.prop:ev_on_Tr_is_iso}, we obtain $\y_1 \ge_\Tr \y_3$, and using the fact that $(\Tr,\ge_\Tr)$ is a forest, as already proven, we infer that $\y_1\in\w$. Hence there is a maximal chain in $(\Tr,\ge_\Tr)$ separating $\y_1$ and $\y_2$.\smallskip

    (Ad ``rooted tree''):~ In view of the preceding results, it suffices to show that $(\Tr,\ge_\Tr)$ has a maximum. By assumption, the map $\x_0\colon\Omega\to F,\,\omega\to W_\omega$ is a random move, i.e.\ $\x_0\in\X$. Then, we have $\x_0 \ge_\Tr \y$ for all $\y\in\Tr$. Indeed, for all $\y\in\Tr$ and $\omega\in D_\y$, $W_\omega$ is a root of the tree $(T_\omega,\supseteq)$, whence $W_\omega\supseteq \y(\omega)$.\smallskip

    (Ad $\X$ = set of moves):~ Let $\x\in\X$. Then there is $\omega\in D_\x$ and, by Axiom~\ref{1-SDF_AC.def:sdf.X.section}, $\x(\omega)\in X$. Hence, there is $y\in F$ with $\x(\omega) \supsetneq y$. By Proposition~\ref{1-SDF_AC.prop:ev_on_Tr_is_iso}, there is unique $\y\in\Tr$ with $\omega\in D_\y$ and $y = \y(\omega)$, and whence, by the same proposition, $\x >_\Tr \y$. Hence, $\x$ is a move in $(\Tr,\ge_\Tr)$. 

    On the other hand, the elements of $\Tr \setminus \X$, i.e.\ the random terminal nodes, are terminal nodes in $(\Tr,\ge_\Tr)$ directly by definition of $\ge_\Tr$. Hence, $\X$ is the set of moves in $(\Tr,\ge_\Tr)$.
\end{proof}

We continue with the verification of the presented examples of stochastic decision forests.

\begin{proof}[Proof of Lemma~\ref{1-SDF_AC.lemma:simple_sdf1}]
    $(\Omega,\ms E)$ is a measurable space. $F$ is clearly a $W$-poset and the map assigning to any $w\in W$ the chain $\uparrow \{w\}$ is easily seen to be a bijection between $W$ and the set of maximal chains in $(F,\supseteq)$. Moreover, any chain in $(F,\supseteq)$ contains a maximal element, hence, by \cite[Theorem~3]{AlosFerrer2005Trees}, expressed in the language of the present thesis, $F$ a decision forest on $W$ (Axiom~\ref{1-SDF_AC.def:sdf.df}).
   
    The connected components of $(F,\supseteq)$ are easily identified as 
    \[ \{\x_k(\omega) \mid k=0,1,2\} \cup \{\{(\omega,k,m)\} \mid k,m\in\{1,2\}\}, \]
    ranging over $\omega\in\Omega$. $\pi$ is well-defined because all nodes are non-empty subsets of $W$ and its definition does not depend on the choice of the element whose first entry is evaluated. 
    The connected component spelled out above equals $\pi^{-1}(\{\omega\})$ indeed. Hence, Axiom~\ref{1-SDF_AC.def:sdf.conn_comp} is satisfied.

    For all $k=0,1,2$, $\x_k$ is defined on $\Omega$ which is an event. Clearly, $\x_k$ is a section of $\pi$. The union of the images of $\X$'s elements is the set of nodes with at least two elements. As $\{\{w\} \mid w\in W\} \subseteq F$, this union equals $X$. Thus Axioms~\ref{1-SDF_AC.def:sdf.X.section} and \ref{1-SDF_AC.def:sdf.X.cov} are satisfied.

    Furthermore, the relations between the images of all presumed random moves, put aside equalities, amount to
    \[ \forall \omega\in\Omega\forall k=1,2\colon \quad \x_0(\omega) \supseteq \x_k(\omega). \]
    Hence, put aside equalities, $\ge_\X$ is given by $\x_0 \ge_\X \x_k$, for $k=1,2$. Moreover, the roots of $F$ are given by the image of $\x_0$. It follows that $\X$ satisfies Axioms~\ref{1-SDF_AC.def:sdf.X.OC} and \ref{1-SDF_AC.def.sdf.X.surely_NT}.

    As the domains of all $\x\in\X$ are equal to the sample space $\Omega$, the maximality property of Axiom~\ref{1-SDF_AC.def:sdf.X.max} is satisfied. 
\end{proof}

\begin{proof}[Proof of Lemma~\ref{1-SDF_AC.lemma:simple_sdf2}]
    The proof is almost completely analogous to the preceding one, up to small modifications. We only comment on those. Thus, we omit the first two axioms.

    For all $k=0,1$, $\x_k$ is defined on $\Omega$ as in the previous lemma. For $k=2$, it has to be noted that $\{\omega_2\}\in\ms E$. Clearly, $\x'$ is a section of $\pi$ for all $\x'\in\X'$. The union of the images of the elements of $\X'$ is the set of nodes with at least two elements. Again, as $\{\{w'\} \mid w'\in W\} \subseteq F'$, this union equals $X'$. It follows that $\X'$ satisfies Axioms~\ref{1-SDF_AC.def:sdf.X.section} and \ref{1-SDF_AC.def:sdf.X.cov}.

    Furthermore, the relations between the images of all presumed random moves, put aside equalities, amount to
    \[ \forall \omega\in\Omega\colon \quad \x'_0(\omega) \supseteq \x'_1(\omega), \qquad \x'_0(\omega_2) \supseteq \x'_2(\omega_2).\]
    Hence, put aside equalities, $\ge_{\X'}$ is given by $\x'_0 \ge_{\X'} \x'_1$ and $\x'_0 \ge_{\X'} \x'_2$. The roots of $F'$ are given by the image of $\x'_0$. It follows that $\X'$ satisfied Axioms~\ref{1-SDF_AC.def:sdf.X.OC} and \ref{1-SDF_AC.def.sdf.X.surely_NT}.

    The maximality property of Axiom~\ref{1-SDF_AC.def:sdf.X.max} requires only a little bit more justification this time. If $\bar\X'$ is such that $(F,\pi,\bar\X')$ is an \textsc{sdf} and $\X'$ refines $\bar\X'$, then any $\bar\x'\in\bar\X'$ with values in $\im\x'_1 \cup \im\x'_0$, we have $\bar\x'\in\{\x'_1,\x'_0\}$ because the domain of $\x'_1$ and $\x'_0$ is $\Omega$. Hence, there can and must be exactly one other element of $\bar\X'$, namely $\x'_2$. Hence, $\bar\X' = \X'$.
\end{proof}

\begin{proof}[Proof of Lemma~\ref{1-SDF_AC.lemma:absent_minded_driver_Gilboa_sdf}]
    $(F,\pi,\X)$ is easily shown to be an \textsc{sdf}, similarly as above.

    $\rho$ is surjective by assumption. Let $\omega_k\in\rho^{-1}(\{k\})$, for $k=1,2$. Then for both of these $k$, $\x_k(\omega_k) \supsetneq \x_{3-k}(\omega_k)$, but, as a consequence, $\x_k\ngeq_\X \x_{3-k}$. Hence, $(F,\pi,\X)$ is not order consistent.
\end{proof}

\begin{proof}[Proof of Lemma~\ref{1-SDF_AC.lemma:AP_sdf.AssmAP.SDF1}]
    (Ad ``$\Rightarrow$''):~ Suppose \hyperlink{1-SDF_AC.Ass:AP.SDF1}{AP.SDF1} to hold and let $x\in F$ be not a singleton. Then, there is $(t,w)\in \T\times W$ with $x = x_t(w)$. Let $(t',w')\in\T\times W$ such that $x=x_{t'}(w')$. Then, $x_t(w) = x_{t'}(w')$. Hence, $w\in x_{t'}(w')$ and thus $x=x_{t'}(w)$. As $x$ is not a singleton, Assumption~\hyperlink{1-SDF_AC.Ass:AP.SDF1}{AP.SDF1} implies that $t=t'$. Hence, $\T_x = \{t\}$.\smallskip

    (Ad ``$\Leftarrow$'')~ Suppose the right-hand criterion to be satisfied. Let $w\in W$ and $t,u\in\T$ with $t\neq u$ and $x_t(w) = x_u(w)$. Hence, $\T_{x_t(w)}$ is not a singleton, and by our hypothesis $x_t(w)$ must be a singleton. As $w\in x_t(w)$, we get $x_t(w) = \{w\}$.
\end{proof}

\begin{proof}[Proof of Lemma~\ref{1-SDF_AC.lemma:AP_sdf.AssmAP.SDF1_addon}]
    Let $x\in F$ such that $\T_x = \{t\}$ for some non-maximal $t\in\T$, and let $w\in x$. Then there is $u\in\T$ with $t<u$, whence $x = x_t(w) \supsetneq x_u(w)$. As $w\in x_u(w)$, $x$ contains at least two elements.
\end{proof}

\begin{proof}[Proof of Theorem~\ref{1-SDF_AC.thm:AP_sdf}]
    (Ad Axiom~\ref{1-SDF_AC.def:sdf.df}):~ As $F\subseteq \mc P (W)$, $F$ is a $W$-poset. We show that it is a ``bounded'' and ``irreducible'' ``$W$-set tree'', in the language of \cite{AlosFerrer2005Trees}, and that any principal up-set in $(F,\supseteq)$ has a maximal element. From this, using \cite[Theorem~3]{AlosFerrer2005Trees}, we obtain, expressed in the language of the present thesis, that $F$ is a decision forest on $W$.

    A ``$W$-set tree'' is a $W$-poset satisfying ``trivial intersection'' and ``separability'' (see \cite[Definition~3]{AlosFerrer2005Trees}).    
    Regarding ``trivial intersection'' (defined in \cite[(4), p.\ 768]{AlosFerrer2005Trees}), let $x,y\in F$ be such that $x\cap y \neq\emptyset$. Then there is $w\in W$ such that $w\in x \cap y$. Hence, $x$ and $y$ are, respectively, of the form $\{w\}$ or $x_t(w)$, for some $t\in\T$. In each of the four possible cases, we get $x\supseteq y$ or $y\supseteq x$. Hence, the ``trivial intersection'' property is satisfied.
    Regarding ``separability'' (defined in \cite[(8), p.\ 773]{AlosFerrer2005Trees}), let $x,y\in F$ such that $x\supsetneq y$. Then there is $w\in y$, whence $w\in x$. $x$ has at least two elements, thus $x = x_t(w)$ for some $t\in\T$. Let $w'\in x\setminus y$. Then $x \supseteq \{w'\}$ and $y \cap \{w'\} = \emptyset$. Hence, the ``separability'' property is satisfied, and $F$ is a $W$-set tree.

    Regarding ``irreducibility'', let $w_0,w_1\in W$ with $w_0\neq w_1$. Then $\{w_k\}\in F$ satisfies $w_k\in\{w_k\} \setminus \{w_{1-k}\}$ for both $k=0,1$. Whence irreducibility. 
    
    Regarding ``boundedness'', let $c$ be a chain in $(F,\supseteq)$. It suffices to consider the case where $c\neq \emptyset$. First, consider the case that there is $w\in W$ with $\{w\} \in c$. In that case, any $x\in c$ is non-empty and hence the chain property of $c$ implies $w\in x$. Second, if there is no $w\in W$ with $\{w\} \in c$, then there must be a unique $\omega\in\Omega$ such that 
    \[ \T' = \{t'\in\T \mid \exists f_{t'} \in\A^\T\colon x_{t'}(\omega,f_{t'})\in c\} \]
    is non-empty. Uniqueness holds true because if $t_1,t_2\in\T$ and $(\omega_1,f_1), (\omega_2,f_2)\in W$ satisfy $x_{t_1}(\omega_1,f_1) \supseteq x_{t_2}(\omega_2,f_2)$, then, by definition, $\omega_1 = \omega_2$. 
    
    Let $\tilde \T = \{t\in\T \mid \exists t'\in\T'\colon t < t'\}$, a convex subset of $\T$. As $c$ is a chain and by definition of $\T'$, for all $t\in\tilde \T$, there is a unique $a(t)\in\A$ such that for all $t'\in\T'$ with $t < t'$ and $f_{t'}\in \A^\T$ satisfying $x_{t'}(\omega,f_{t'})\in c$ we have $f_{t'}(t) = a(t)$. Let $\tilde f\in\A^\T$ be such that $\tilde f(t) = a(t)$ for all $t\in \tilde \T$. Hence, all $t'\in\T'$ satisfy $x_{t'}(\omega,\tilde f)\in c\subseteq F$. By Assumption~\hyperlink{1-SDF_AC.Ass:AP.SDF2}{AP.SDF2}, there is $f\in\A^\T$ such that $(\omega,f)\in W$ and $x_{t'}(\omega,f) = x_{t'}(\omega,\tilde f)\in c$ for all $t'\in\T'$.

    Hence, as $c$ contains no singleton, we get
    \[ (\ast) \qquad c = \{ x_{t'}(\omega,f) \mid t'\in\T'\}. \]
    Hence, for all $x\in c$, we have $(\omega,f)\in x$. The proof of ``boundedness'' is complete.

    It remains to be shown that any principal up-set contains a maximal element. The crucial point here is that for any $x\in F$, there is $\omega\in\Omega$ such that $x\subseteq (\{\omega\} \times \A^\T)\cap W$. Moreover, for any $w\in x$, $(\{\omega\} \times \A^\T)\cap W = x_0(w)$. Thus $x_0(w)\in \uparrow x$, and if $y\in\uparrow x_0(w)$, then $y = x_0(w)$. Thus, $x_0(w) \in\uparrow x$ is a maximal element for $(F,\supseteq)$.\smallskip

    (Ad Axiom~\ref{1-SDF_AC.def:sdf.conn_comp}):~ Recall from the preceding argument that any $x\in F$ is contained in $(\{\omega\} \times \A^\T)\cap W$ for some $\omega\in\Omega$. Hence, $\pi(x)$ is well-defined and equal to this uniquely determined $\omega$. Let $x,y\in F$. Then $\pi(x) = \pi(y)$ iff there is $\omega\in\Omega$ such that $x,y \subseteq (\{\omega\} \times \A^\T)\cap W$. As $(\{\omega\} \times \A^\T)\cap W$ is a root of $(F,\supseteq)$, as seen just above, this is equivalent to the statement that $x,y$ belong to the same connected component.\smallskip

    (Ad Axiom~\ref{1-SDF_AC.def:sdf.X.section}):~ By definition of the sets $D_{t,f}$, $(t,f)\in\T\times\A^\T$, and in view of the fact that $F$ contains all singletons in $\mc P(W)$, all elements of $\X$ map into the set $X$ of moves of the decision forest $F$ on $W$. Moreover, by Assumption~\hyperlink{1-SDF_AC.Ass:AP.SDF0}{AP.SDF0}, $D_{t,f}\in\ms E$. Furthermore, any $\x\in\X$ is a section of $\pi$. For this represent $\x$ as $\x_t(f)$ for $(t,f)\in\T\times\A^\T$ such that $D_{t,f}\neq\emptyset$, and let $\omega\in D_{t,f}$. Then, 
    \[ \pi(\x_t(f)(\omega)) = \pi(x_t(\omega,f)) = \omega, \]
    because $\omega$ is the first component of $(\omega,f)\in x_t(\omega,f)$.\smallskip

    (Ad Axiom~\ref{1-SDF_AC.def:sdf.X.cov}):~ Let $x\in X$. Then $x$ has at least two elements, because $F$ is a decision forest on $W$. Hence, there are $(\omega,f)\in W$ and $t\in\T$ such that $x = x_t(\omega,f)$. Hence, $\omega\in D_{t,f}$, and $x = \x_t(f)(\omega)$.\smallskip

    (Ad Axiom~\ref{1-SDF_AC.def:sdf.X.OC}):~ Let $\x_1,\x_2\in\X$ such that there is $\omega\in D_{\x_1}\cap D_{\x_2}$ with $\x_1(\omega) \supseteq \x_2(\omega)$. Hence, there are $t_1,t_2\in\T$ and $f\in\A^\T$ such that $\omega \in D_{t_k,f}$ and $\x_k = \x_{t_k}(\omega,f)$, for both $k=1,2$. If we had $t_1 > t_2$, then, by construction of the nodes and by Assumption~\hyperlink{1-SDF_AC.Ass:AP.SDF1}{AP.SDF1}, $\x_1(\omega) = x_{t_1}(\omega,f) \subsetneq x_{t_2}(\omega,f) = \x_2(\omega)$, because they contain at least two elements as moves --- a contradiction. Hence, $t_1\le t_2$.

    Let $\omega'\in D_{t_2,f}$. By definition, $x_{t_2}(\omega',f)$ is a move. As $t_1 \le t_2$, $x_{t_1}(\omega',f) \supseteq x_{t_2}(\omega',f)$, whence $\omega'\in D_{t_1,f}$. In particular, $\x_1(f)(\omega') \supseteq \x_2(f)(\omega')$. Thus, $\x_1 \ge_\X \x_2$.\smallskip

    (Ad Axiom~\ref{1-SDF_AC.def:sdf.X.max}):~ For this part of the proof, suppose that $(I,\A,\T,W)$ is maximal, i.e.\ satisfies Assumption~\hyperlink{1-SDF_AC.Ass:AP.SDF3}{AP.SDF3}. Let $\bar\X$ be a set such that $(F,\pi,\bar\X)$ is an order consistent \textsc{sdf} and that is refined by $\X$. By definition of the latter, for any $\bar\x\in\bar\X$ there is $P_{\bar\x}\subseteq\X$ such that $\bar\x = \bigcup P_{\bar\x}$. 
    As $\X$ and $\bar\X$ induce partitions of $X$, by Proposition~\ref{1-SDF_AC.prop:ev_on_Tr_is_iso}, and as $D_\x\neq \emptyset$ for all $\x\in\X$ by construction of $\X$, we infer that there is a unique map $b\colon\X \to \bar\X, \x\mapsto b(\x)$ such that $D_\x\subseteq D_{b(\x)}$ and $b(\x)|_{D_\x} = \x$, for all $\x\in\X$. 
    For each pair $(t,f)\in\T\times\A^\T$ with $D_{t,f}\neq\emptyset$, denote $\bar \x_t(f) = b({\x_t(f)})$ the image of $\x_t(f)$ under this map.

    Our aim is to show that for all $\bar\x\in\bar\X$, we have $P_{\bar\x} = \{\bar\x\}$. If this were not the case, there would be $\bar\x\in\bar\X$ such that $P_{\bar\x}$ has at least two elements. Hence, there would be $t,u\in\T$ and $f,g\in\A^\T$ with $D_{t,f},D_{u,g}\neq\emptyset$, $\x_t(f) \neq \x_u(g)$, and $\x_t(f),\x_u(g)\in P_{\bar\x}$. This would imply 
    \[ (\dagger)\qquad D_{t,f} \cap D_{u,g} = \emptyset. \] 
    Indeed, otherwise there would be $\omega \in D_{t,f}\cap D_{u,g}$ for that we would obtain $x_t(\omega,f) = \bar\x(\omega) = x_u(\omega,g)$. Hence, for $v = t\wedge u$, we would have $f|_{[0,v)_\T} = g|_{[0,v)_\T}$ and there would be $w\in W$ with $x_t(\omega,f) = x_t(w) = x_u(w) = x_u(\omega,g)$. As these are moves, Assumption~\hyperlink{1-SDF_AC.Ass:AP.SDF1}{AP.SDF1} would imply $t=u=v$, whence $\x_t(f) = \x_u(g)$ which is absurd. Hence, we would necessarily have $D_{t,f} \cap D_{u,g} = \emptyset$.
    
    We can assume, without loss of generality, that $t\le u$. 
    There would be two cases. First, $D_{t,f} \cap D_{t,g} \neq \emptyset$. Then, $f|_{[0,t)_\T} \neq g|_{[0,t)_\T}$, because else $D_{t,f} = D_{t,g}$ whence $D_{u,g} \subseteq D_{t,g} = D_{t,f}$ which contradicts $(\dagger)$. The second case would be $D_{t,f} \cap D_{t,g} = \emptyset$. Then, by Assumption~\hyperlink{1-SDF_AC.Ass:AP.SDF3}{AP.SDF3}, there would be $v\in [0,t)_\T$ such that
    \[ (\circ) \qquad D_{v,f} \cap D_{v,g} \neq \emptyset, \quad \text{and} \quad f|_{[0,v)_\T} \neq g|_{[0,v)_\T}. \]
    We conclude that in both cases, there would be $v\in [0,t]_\T$ satisfying $(\circ)$.

    Let $\omega'\in D_{v,f} \cap D_{v,g}$. Then, $x_v(\omega',f)$ and $x_v(\omega',g)$ would not be comparable in $(F,\supseteq)$ since this would imply $f|_{[0,v)_\T} = g|_{[0,v)_\T}$. Hence, by the definition of $\ge_{\bar\X}$, the same would be true for $\bar\x_v(f)$ and $\bar\x_v(g)$ in $(\bar\X,\ge_{\bar\X})$. But, on the other hand, $\x_v(f) \ge_\X \x_t(f)$ and $\x_v(g) \ge_\X \x_u(g)$, whence by definition of $\ge_\X$ and Axiom~\ref{1-SDF_AC.def:sdf.X.OC} applied to $\bar\X$, we get $\bar\x_v(f),\bar\x_v(g) \in \uparrow \bar \x_t(f)$. As $(\bar\X,\ge_{\bar\X})$ is a forest (which follows from Proposition~\ref{1-SDF_AC.prop:ev_on_Tr_is_iso} as in the proof of Theorem~\ref{1-SDF_AC.thm:Xrm_is_dec_tree}), $\bar\x_v(f)$ and $\bar\x_v(g)$ would have to be comparable, a contradiction. Hence, the initial assumption was false, and we infer that $P_{\bar\x} = \{\bar\x\}$ for all $\bar\x\in\bar\X$. 
    
    This in turn implies that $\x = b(\x)$ for all $\x\in\X$. Hence, we have both $\bar\X\subseteq \X$ and $\X\subseteq\bar\X$, whence $\X = \bar \X$.\smallskip

    (Ad statement about sure non-triviality):~ For all $w\in W$, we have $x_0(w) \supseteq \{w\}$ and $\{w\}\in F$. Hence, for all $\omega\in\Omega$ and $f\in\A^\T$, $x_0(\omega,f)\in X$ iff $\omega\in D_{0,f}$. It is moreover obvious from the definitions that all $\omega\in\Omega$ and $f\in\A^\T$ satisfy $x_0(\omega,f) = W_\omega$.
    Hence, for any $f\in\A^\T$, 
    \[ \{\omega\in\Omega \mid W_\omega\in X\} = D_{0,f} = \big \{\omega\in\Omega \mid \exists f',f''\in\A^\T\colon [f'\neq f'' \text{ and } (\omega,f'),(\omega,f'')\in W]\big \}, \] 
    from which the claimed equivalence directly follows in view of $(\ast)$.   
\end{proof}

\begin{proof}[Proof of Lemma~\ref{1-SDF_AC.lemma:mf_t}]
    Let $x,y\in X$ with $x \supsetneq y$. Then, by definition of $\mf t$ and the construction of nodes in action path \textsc{sdf}, there is $w\in W$ such that $x=x_{\mf t(x)}(w)$ and $y=x_{\mf t(y)}(w)$. Indeed, $y$ is non-empty, and any $w\in y$ satisfies the preceding equalities. If we had $\mf t(x)\ge \mf t(y)$, then $y = x_{\mf t(y)}(w) \supseteq x_{\mf t(x)}(w) = x$, in contradiction to the choice of $x$ and $y$. Hence, by totality of the order on $\T$, $\mf t(x)<\mf t(y)$. 

    Let $\x\in\X$. Then there are $t\in\T$ and $f\in\A^\T$ such that $D_{t,f} \neq \emptyset$ with $\x = \x_t(f)$. Hence, for all $\omega\in D_\x = D_{t,f}$:
    \[\mf t(\x(\omega)) = \mf t(x_t(\omega,f)) = t, \]
    which does not depend on $\omega$.
\end{proof}

\begin{proof}[Proof of Lemma~\ref{1-SDF_AC.lemma:simple_sdf_as_APsdf}]
    Take $\A = \{1,2\}$ and $\T = \{0,1\}$. Recall that $\Omega = \{\omega_1,\omega_2\}$, with $\omega_1 \neq \omega_2$, and $\ms E = \mc P(\Omega)$. 
    
    For the first example, we may set $W = \Omega \times \A^\T$. Any $(\omega,f)\in W$ can be identified with the triple $(\omega,f(0),f(1))$, so that $W$ corresponds to the set denoted by $W$ in the example in Subsection~\ref{1-SDF_AC.subs:simple_sdf}. Under this identification, it is easily verified that $D_{t,f} = \Omega$ for all $f\in\A^\T$ and $t\in\T$, $\x_0 = \x_0(f)$ for all $f\in\A^\T$, and $\x_k = \x_1(f)$ for all $f\in\A^\T$ with $f(0) = k$, for both $k=1,2$. Assumptions \hyperlink{1-SDF_AC.Ass:AP.SDF0}{AP.SDF0} to \hyperlink{1-SDF_AC.Ass:AP.SDF3}{AP.SDF3} are readily verified. Under this identification, the action path construction yields exactly the objects $F$, $\pi$, and $\X$ from the example in Subsection~\ref{1-SDF_AC.subs:simple_sdf}. 

    For the second example, let $W'$ as in Subsection~\ref{1-SDF_AC.subs:simple_sdf}. Interpret every triple $(\omega,k,m)\in W'$ as the pair $(\omega,f)$ where $f\in\A^\T$ is given by $f(0) = k$ and $f(1) = m$. Interpret the pair $(\omega_1,2)$ as the map $f\colon\T\to\{0,1,2\}$ with $f(0) = 2$ and $f(1) = 0$. Then, $W \subseteq {\A'}^\T$ with $\A' = \A \cup \{0\}$, where $0$ is a placeholder for inaction. Furthermore, $D_{t,f} = \Omega$ for all $(t,f)\in\T\times {\A'}^\T$ with $t=0$, or $t=1$ and $f(0) = 1$; and $D_{1,f} = \{\omega_2\}$ for $f\colon\T\to \A'$ with $f(0) = 2$; for all other pairs $(t,f) \in \T\times{\A'}^\T$ we have $D_{t,f} = \emptyset$. Moreover, $\x'_0 = \x_0(f)$ for all $f\in{\A'}^\T$; and $\x'_1 = \x_1(f)$ for all $f\colon\T\to\A'$ with $f(0) = 1$; and $\x'_2 = \x_1(f)$ for $f\colon\T\to \A'$ with $f(0) = 2$. Assumptions \hyperlink{1-SDF_AC.Ass:AP.SDF0}{AP.SDF0} to \hyperlink{1-SDF_AC.Ass:AP.SDF3}{AP.SDF3} are readily verified. Under these identifications, the action path construction yields exactly the objects $F'$, $\pi'$, and $\X'$ from the example in Subsection~\ref{1-SDF_AC.subs:simple_sdf}. 
\end{proof}

\begin{proof}[Proof of the claims in Example~\ref{1-SDF_AC.ex:APsdf}]
    (The case $W = \Omega \times \A^\T$):~ We show that for $W = \Omega\times\A^\T$, $W$ induces an action path \textsc{sdf}. First, we note that for any $\omega\in\Omega$, there is $f\in\A^\T$ such that $(\omega,f)\in W$. 
    
    Clearly, for all $(t,f)\in\T\times\A^\T$, $D_{t,f} = \Omega$, whence \hyperlink{1-SDF_AC.Ass:AP.SDF0}{AP.SDF0}. 
    
    Regarding \hyperlink{1-SDF_AC.Ass:AP.SDF1}{AP.SDF1}, let $w=(\omega,f)\in W$ and $t,u\in\T$ with $t\neq u$ such that $x_t(w) = x_u(w)$. Without loss of generality, assume $t<u$. If $\A$ is a singleton, then $\A^\T$ is a singleton as well. Hence, $x_t(w) = \{w\}$. But, given our assumption on the existence of $w,t,u$ as above, $\A$ must be a singleton, since else there would be $g\in\A^\T$ with $(\omega,g) \in x_t(w)$ and $g(t) \neq f(t)$. Thus, $(\omega,g)\in x_t(w)\setminus x_u(w)$ --- a contradiction. 

    Regarding \hyperlink{1-SDF_AC.Ass:AP.SDF2}{AP.SDF2}, let $\omega\in\Omega$, $\tilde f\in\A^\T$, and $\T'\subseteq\T$ satisfying $x_t(\omega,\tilde f)\in F$ for all $t\in\T'$. Then we already have $(\omega,\tilde f)\in W$.

    \hyperlink{1-SDF_AC.Ass:AP.SDF3}{AP.SDF3} is clearly satisfied because $D_{t,f} = \Omega$ for all $(t,f)\in\T\times\A^\T$.\smallskip

    (The timing problem):~ Clearly, for any $\omega\in\Omega$ there is $f\in\A^\T$ with $(\omega,f)\in W$, e.g.\ the constant map taking only the value $1$ at all times, in all components. 

    Regarding Assumption~\hyperlink{1-SDF_AC.Ass:AP.SDF0}{AP.SDF0}, for all $(t,f)\in\A^\T$, $D_{t,f} = \emptyset$ if $f$ is not decreasing on $[0,t)_\T$ or if $f(t-) = 0$. Else, $D_{t,f} = \Omega$. Hence, \hyperlink{1-SDF_AC.Ass:AP.SDF0}{AP.SDF0} is satisfied.

    Regarding Assumption~\hyperlink{1-SDF_AC.Ass:AP.SDF1}{AP.SDF1}, let $w=(\omega,f)\in W$ and $t,u\in\T$ with $t<u$ such that $x_t(w) = x_u(w)$. Hence, for all decreasing $g\colon \T\to\A$ with $g|_{[0,t)_\T} = f|_{[0,t)_\T}$ we must have $f(v) = g(v)$ for all $v\in [t,u)_\T$. Thus, we must have $f(t-) = 0$. Hence, $x_t(w) = \{(\omega,f)\}$.

    Regarding Assumption~\hyperlink{1-SDF_AC.Ass:AP.SDF2}{AP.SDF2}, let $\omega\in\Omega$, $\tilde f\in \A^\T$, and $\T'\subseteq\T$ such that $x_t(\omega,\tilde f)\in F$ for all $t\in \T'$. Hence, $\tilde f$ is decreasing on $\bigcup_{t\in\T'} [0,t)_\T$. Hence, there is decreasing $f\colon \T\to \A$ such that $f|_{[0,t)_\T} = \tilde f|_{[0,t)_\T}$ for all $t\in\T'$. By construction, $(\omega,f)\in W$.

    Regarding Assumption~\hyperlink{1-SDF_AC.Ass:AP.SDF3}{AP.SDF3}, it is sufficient to note that $D_{t,f}$ equals $\emptyset$ or $\Omega$, for all $(t,f)\in\T\times\A^\T$ which has been established earlier.\smallskip

    (Ad American up-and-out option):~ Let $\A = \{0,1\}$ and $\T=\R_+$. Clearly, for any $\omega\in\Omega$, there is $f\colon\R_+\to\A$ with $(\omega,f)\in W$, namely, the constant path with value $1$.

    Regarding Assumption~\hyperlink{1-SDF_AC.Ass:AP.SDF0}{AP.SDF0}, let $(t,f)\in\T\times\A^{\R_+}$ and $\omega\in\Omega$. Then, we have $\omega\in D_{t,f}$ iff $x_t(\omega,f)$ has at least two elements. This is equivalent to the fact that $f$ is decreasing on $[0,t)$, $f(t-) = 1$, and $P_u(\omega) < 2$ for all $u\in[0,t]$. Whence the claimed representation of $D_{t,f}$ and the fact that $D_{t,f}\in\ms E$, because $P$ has continuous paths so that $\max_{u\in[0,t]} P_u$ is $\ms E$-measurable since $P_u$ is so for all real (and in particular rational) $u\ge 0$.

    Regarding Assumption~\hyperlink{1-SDF_AC.Ass:AP.SDF1}{AP.SDF1}, let $w=(\omega,f)\in W$ and $t,u\in\R_+$ with $t<u$ such that $x_t(w) = x_u(w)$. 
    Hence, for all decreasing $g\colon \R_+\to\{0,1\}$ with $g|_{[0,t)} = f|_{[0,t)}$ and $(\omega,g)\in W$, we must have 
    \[ (\ast) \qquad f(v) = g(v), \quad \text{ for all }~v\in [t,u). \] 
    
    If $f$ is constant to $1$, this implies the existence of $t_0\in[0,t]$ such that $P_{t_0}(\omega) \ge 2$. If such $t_0$ did not exist, the map $g\colon \R_+ \to \{0,1\}$ given by $g(t') = 1\{t' < t\}$ would violate condition $(\ast)$ above. Hence, $x_t(w) = \{w\}$. 
    If $f$ takes the value $0$ and $t_f^\ast < t$, then clearly $x_t(w) = \{w\}$. 
    The remaining case, namely that $f$ takes the value $0$ and $t_f^\ast \ge t$, cannot arise. Indeed, if it did, for both $a\in\{0,1\}$, the map $g_a\colon\R_+ \to \{0,1\}$ given by $g_a(t') = 1$ for $t'<t$, $g_a(t) = a$, and $g_a(t') = 0$ for $t'> t$ would satisfy $t_{g_a}^\ast = t$. Moreover, the fact that $(\omega,f)\in W$ would imply
    \[ \max_{t'\in[0,t]} P_{t'}(\omega) \le \max_{t'\in[0,t_f^\ast]} P_{t'}(\omega) < 2, \]
    hence $(\omega,g_a)\in W$. In particular, $(\ast)$ would imply that $a = g_a(t) = f(t)$, for both $a\in\{0,1\}$ which is impossible. Hence, in any possible case we have $x_t(w) = \{w\}$, thus Assumption~\hyperlink{1-SDF_AC.Ass:AP.SDF1}{AP.SDF1} is satisfied.

    Regarding Assumption~\hyperlink{1-SDF_AC.Ass:AP.SDF2}{AP.SDF2}, let $\omega\in\Omega$, $\tilde f\in \A^{\R_+}$, and $\T'\subseteq\R_+$ such that $x_t(\omega,\tilde f)\in F$ for all $t\in \T'$. Hence, $\tilde f$ is decreasing on $[0,\sup \T')$ (where, in the context of the order on $\R_+$, we have $\sup \emptyset = 0$), and if $\tilde f|_{[0,\sup \T')}$ attains $0$, then $P_u(\omega) < 2$ for all $u\in[0,t_{\tilde f}^\ast]$. Let $f\colon \R_+ \to\{0,1\}$ be given by $f|_{[0,\sup \T')} = \tilde f|_{[0,\sup \T')}$ and, for all $u\in [\sup\T',\infty)$, $f(u) = f(\sup\T'-)$ if $\sup\T' > 0$ and $f(u) = 1$ if $\sup\T' = 0$. By construction, $f|_{[0,t)} = \tilde f|_{[0,t)}$ for all $t\in\T'$. Further, $f$ is decreasing. Moreover, if $f$ attains the value $0$, then it does so on $[0,\sup\T')$ and so does $\tilde f$, and $t_f^\ast = t_{\tilde f}^\ast$. Hence, $(\omega,f)\in W$. We conclude that Assumption~\hyperlink{1-SDF_AC.Ass:AP.SDF2}{AP.SDF2} is satisfied.

    Regarding Assumption~\hyperlink{1-SDF_AC.Ass:AP.SDF3}{AP.SDF3}, let $t\in\R_+$ and $f,g\in\A^\T$ such that $D_{t,f},D_{t,g} \neq\emptyset$. Then, as shown earlier in this proof, $f(t-) = g(t-) = 1$ and
    \[ D_{t,f} = \{\omega\in\Omega \mid \max_{u\in[0,t]} P_u(\omega) < 2\} = D_{t,g}. \]
    Hence, $D_{t,f} \cap D_{t,g} \neq \emptyset$. Thus, Assumption~\hyperlink{1-SDF_AC.Ass:AP.SDF3}{AP.SDF3} is trivially satisfied.
\end{proof}

\subsection{Section~\ref{1-SDF_AC.sec:exogenous_information}}

\begin{proof}[Proof of Lemma~\ref{1-SDF_AC.lemma:simple_sdf1_EIS}]
    There are exactly two ($\sigma$-)algebras on $\Omega$: the discrete and the trivial one, that is, $\mc P(\Omega)$ and $\{\Omega,\emptyset\}$. In the present situation, we have $D_\x = \Omega$ for all $\x\in\X$. Hence, by definition, the set of exogenous information structures $\ms F$ on $(F,\pi,\X)$ admitting recall is given by all families $\ms F = (\ms F_\x)_{\x\in\X}$ of $\sigma$-algebras on $\Omega$ such that $\ms F_{\x_0} \subseteq \ms F_{\x_1} \cap \ms F_{\x_2}$. The claim follows easily from this.
\end{proof}

\begin{proof}[Proof of Lemma~\ref{1-SDF_AC.lemma:simple_sdf2_EIS}]
    There is exactly one ($\sigma$-)algebra on $\{\omega_2\}$, namely $\{\{\omega_2\},\emptyset\}$. On $\Omega$, there are exactly two $\sigma$-algebras, namely $\mc P(\Omega)$ and $\{\Omega,\emptyset\}$, as in the preceding proof. Hence, by definition, the set of exogenous information structures $\ms F'$ on $(F',\pi',\X')$ admitting recall is given by all families $\ms F' = (\ms F'_{\x'})_{\x'\in\X'}$ of $\sigma$-algebras on $\Omega$ such that $\ms F_{\x'_2} =\{\{\omega_2\},\emptyset\}$ and $\ms F_{\x'_0} \subseteq \ms F_{\x'_1}$. The claim follows easily from this.
\end{proof}

\begin{proof}[Proof of Lemma~\ref{1-SDF_AC.lemma:APsdf_EIS_induces_filtration}]
    Suppose that the exogenous information structure $\ms F$ on $\X$ admits recall and let $f\in\A^\T$. Let $t,u\in\T_f$ with $t<u$ and $E\in\ms F_{\x_t(f)}$. Then, $\x_t(f) \ge_\X \x_u(f)$, whence $E\cap D_{u,f} \in \ms F_{\x_u(f)}$ by Definition~\ref{1-SDF_AC.def:EIS}.

    The second claim follows from the first because under its hypothesis $\ms F_{\x_t(f)}$ is a $\sigma$-algebra on $D_{t,f} = \Omega$ for all $t\in\T_f$.
\end{proof}

\begin{proof}[Proof of Theorem~\ref{1-SDF_AC.thm:AP_sdf_EIS}]
    Let $\x\in\X$. Then, by construction, $\ms F_\x$ is a $\sigma$-algebra on $D_\x$ contained in $\ms E$, because $\ms G_{\mf t(x)}$ is a sub-$\sigma$-algebra of $\ms E$ and all $Y_{\x'}$, ranging over $\x'\in\X$, are $\ms E$-measurable.

    Furthermore, let $\x_1,\x_2\in\X$ such that $\x_1\ge_\X\x_2$, and let $E\in \ms F_{\x_1}$. Hence, there is $E'\in \sigma(Y_{\x'} \mid \x'\ge_\X \x_1) \vee \ms G_{\mf t(\x_1)}$ such that $E = E' \cap D_{\x_1}$. Note that
    \[ E'\in \sigma(Y_{\x'} \mid \x'\ge_\X \x_1) \vee \ms G_{\mf t(\x_1)} \subseteq \sigma(Y_{\x'} \mid \x'\ge_\X \x_2) \vee \ms G_{\mf t(\x_2)}. \]
    As $D_{\x_1} \supseteq D_{\x_2}$, we have $ E\cap D_{\x_2} = E' \cap D_{\x_2} $. Hence, $E \cap D_{\x_2} \in \ms F_{\x_2}$.
\end{proof}

\subsection{Section~\ref{1-SDF_AC.sec:adapted_choices}}

\begin{proof}[Proof of Lemma~\ref{1-SDF_AC.lemma:P(c)_compatible_with_conn_comp}]
    As $c\cap W_E = \bigcup_{\omega\in E} (c\cap W_\omega)$ and $\{T_\omega\mid\omega\in\Omega\}$ is a partition of $F$, we infer from Lemma~\ref{1-SDF_AC.lemma:set_forest}, Part~\ref{1-SDF_AC.lemma:set_forest.T=downarrow_V_T}, that the following introductory statement holds true:
    \[ \downarrow (c\cap W_E) = \{x\in F \mid c \cap W_E \supseteq x\} = \{x\in F_E \mid c \supseteq x\} = (\downarrow c) \cap F_E. \]
    
    From this, the claim follows easily: If $x\in P(c\cap W_E)$, then there is $y\in (\downarrow c)\cap F_E$ such that
    \[ (\dagger) \qquad \uparrow x = \uparrow y \setminus ((\downarrow c) \cap F_E). \]
    Hence, $x \in \uparrow y$. There is unique $\omega\in E$ such that $y\in (\downarrow c)\cap T_\omega$. As $T_\omega$ is a connected component of $(F,\supseteq)$ and $y\in T_\omega$, we have $\uparrow y\subseteq T_\omega$, whence $x\in T_\omega$ and
    \[ (\ast) \qquad \uparrow x = \uparrow y \setminus \downarrow c. \]
    If conversely $x\in F_E$ is such that there is $y\in \downarrow c$ satisfying $(\ast)$, then $x\in \uparrow y$, hence $x,y$ belong to the same connected component, whence $y\in F_E$ and $\uparrow y \subseteq F_E$. We infer that $(\dagger)$ holds true, i.e.\ $x\in P(c\cap W_E)$ by the introductory statement.
\end{proof}

\begin{lemma}\label{1-SDF_AC.lemma:simple_sdf1_choices}
    Consider the basic version $(F,\pi,\X)$ of the simple \textsc{sdf} on the exogenous scenario space $(\Omega,\ms E)$ as in Subsection~\ref{1-SDF_AC.subs:simple_sdf_AC}. Let $M$ be the set of maps $\Omega\to \{1,2\}$. 
    Then the following subsets of $W$ define non-redundant and complete choices:
    \begin{itemize}[label=--]
        \item $c_{f\bullet}$, where $f\in M$;
        \item $c_{kg}$, where $k=1,2$ and $g\in M$;
        \item $c_{\bullet g}$, where $g\in M$.
    \end{itemize}
    The corresponding sets of immediate predecessors are given by $P(c_{f\bullet}) = \im\x_0$;  $P(c_{kg}) = \im \x_k$; $P(c_{\bullet g}) = \im \x_1 \cup \im \x_2$.
\end{lemma}

\begin{proof}
    Any set $c$ of the form above is non-empty, and as $F$ contains all singletons in $\mc P(W)$, $c$ is a choice.
    The sets of immediate predecessors are easily shown to be of the claimed form, using Lemma~\ref{1-SDF_AC.lemma:P(c)_compatible_with_conn_comp}. From this, we directly infer the non-redundancy and completeness of all the considered choices.
\end{proof}

\begin{proof}[Proof of Lemma~\ref{1-SDF_AC.lemma:simple_sdf1_RCS}]
    Let $C = \bigcup_{\x\in\X} \ms C_\x$. By Lemma~\ref{1-SDF_AC.lemma:simple_sdf1_choices}, all elements of $C$ are non-redundant and complete choices. The statement about the set of immediate predecessors of first $c_{k \bullet}$ and second $c_{\bullet m}$, $k,m\in\{1,2\}$, stated in Lemma~\ref{1-SDF_AC.lemma:simple_sdf1_choices} shows that the first type of choice is available at $\x_0$, while the second is available at $\x_1$ and $\x_2$.  
\end{proof}

\begin{proof}[Proof of Lemma~\ref{1-SDF_AC.lemma:simple_sdf1_AC}]
    In view of Lemma~\ref{1-SDF_AC.lemma:simple_sdf1_choices}, we only have to verify that for any line and corresponding exogenous information structure (\textsc{eis}) $\ms F$, and any subset $c\subseteq W$ in that line, all $\x\in\X$ that $c$ is available at and $c_\x\in\ms C_\x$, we have
    \[ \x^{-1}(P(c\cap c_\x)) \in \ms F_\x. \]
    For this, one easily verifies that all $k,k',m'\in\{1,2\}$ and $f,g\in M$ satisfy:
    \begin{align*}
        c_{f\bullet} \cap c_{k'\bullet} =&~ c_{k'\bullet} \cap W_{\{f = k'\}}; \\
        c_{kg} \cap c_{\bullet m'} =&~ c_{k m'} \cap W_{\{g = m'\}}; \\
        c_{\bullet g} \cap c_{\bullet m'} =&~  c_{\bullet m'} \cap W_{\{g = m'\}}.
    \end{align*}
    As shown in Lemma~\ref{1-SDF_AC.lemma:simple_sdf1_choices}, we have
    \[ P(c_{k'\bullet} ) = \im \x_0, \quad P(c_{km'} ) = \im \x_k, \quad P(c_{\bullet m'} ) = \im \x_1 \cup \im \x_2.\]
    Then, applying Lemma~\ref{1-SDF_AC.lemma:P(c)_compatible_with_conn_comp} and using the definition of $\ms F$, in each of the five cases respectively, completes the proof.
\end{proof}

\begin{lemma}\label{1-SDF_AC.lemma:simple_sdf2_choices}
    Consider the variant $(F',\pi',\X')$ of the simple \textsc{sdf} on the exogenous scenario space $(\Omega,\ms E)$ as in Subsection~\ref{1-SDF_AC.subs:simple_sdf_AC}. Let $M$ be the set of maps $\Omega\to \{1,2\}$. 
    Then the following subsets of $W'$ define non-redundant and complete choices:
    \begin{itemize}[label=--]
        \item $c'_{f\bullet}$, where $f\in M$;
        \item $c'_{kg}$, where $k=1,2$ and $g\in M$;
        \item $c'_{\bullet g}$, where $g\in M$.
    \end{itemize}
    The corresponding sets of immediate predecessors are given by $P(c'_{f\bullet}) = \im\x'_0$; $P(c'_{kg}) = \im\x'_k$; $P(c'_{\bullet g}) = \im\x'_1\cup \im\x'_2$.
\end{lemma}

\begin{proof}
    Any set $c'$ of the form above is non-empty, and as $F'$ contains all singletons in $\mc P(W')$, any such $c'$ is a choice.
    The sets of immediate predecessors are easily shown to be of the claimed form, using Lemma~\ref{1-SDF_AC.lemma:P(c)_compatible_with_conn_comp}. From this, we directly infer the non-redundancy and completeness of all the considered choices.
\end{proof}

\begin{proof}[Proof of Lemma~\ref{1-SDF_AC.lemma:simple_sdf2_RCS}]
    Let $C' = \bigcup_{\x'\in\X'} \ms C'_{\x'}$. By Lemma~\ref{1-SDF_AC.lemma:simple_sdf2_choices}, all elements of $C'$ are non-redundant and complete choices. The statement about the set of immediate predecessors of first $c'_{k \bullet}$ and second $c'_{\bullet m}$, $k,m\in\{1,2\}$, stated in Lemma~\ref{1-SDF_AC.lemma:simple_sdf2_choices} shows that the first type of choice is available at $\x'_0$, while the second is available at $\x'_1$ and $\x'_2$.  
\end{proof}

\begin{proof}[Proof of Lemma~\ref{1-SDF_AC.lemma:simple_sdf2_AC}]
    In view of Lemma~\ref{1-SDF_AC.lemma:simple_sdf2_choices}, we only have to verify that for any line and corresponding exogenous information structure (\textsc{eis}) $\ms F'$, and any subset $c'\subseteq W$ in that line, all $\x'\in\X'$ that $c'$ is available at and $c'_{\x'}\in\ms C'_{\x'}$, we have
    \[ \x'^{-1}(P(c'\cap c'_{\x'})) \in \ms F'_{\x'}. \]
    For this, one easily verifies that all $k,k',m'\in\{1,2\}$ and $f,g\in M$ satisfy:
    \begin{align*}
        c'_{f\bullet} \cap c'_{k'\bullet} =&~ c'_{k'\bullet} \cap W_{\{f = k'\}}; \\
        c'_{kg} \cap c'_{\bullet m'} =&~ c'_{k m'} \cap W_{\{g = m'\}}; \\
        c'_{\bullet g} \cap c'_{\bullet m'} =&~  c'_{\bullet m'} \cap W_{\{g = m'\}}.
    \end{align*}
    As shown in Lemma~\ref{1-SDF_AC.lemma:simple_sdf2_choices}, we have
    \[ P(c'_{k'\bullet} ) = \im \x'_0, \quad P(c'_{km'} ) = \im \x'_k, \quad P(c'_{\bullet m'} ) = \im \x'_1 \cup \im \x'_2.\]
    Then applying Lemma~\ref{1-SDF_AC.lemma:P(c)_compatible_with_conn_comp} and using the definition of $\ms F'$, in each of the three cases respectively, completes the proof.
\end{proof}

\begin{proof}[Proof of Lemma~\ref{1-SDF_AC.lemma:AP_downarrow_c}]
    Let $t\in\T$ and $c\in\ms C_t$. Let $x\in F$. 
    
    If $x = \{w\}$ for some $w\in W$, then clearly $x\in \downarrow c$ iff $w\in c$.
    It remains to consider the case $x = x_u(w)$ for some $w=(\omega,f)\in W$ and $u\in\T$. 
    
    If $t<u$ and $w\in c$, then any $w' = (\omega',f')\in x$ satisfies $\omega' = \omega$ and $f'|_{[0,u)_\T} = f|_{[0,u)_\T}$, in particular $f'|_{[0,t]_\T} = f|_{[0,t]_\T}$, hence $w'\in c$. We conclude for this case that $x\subseteq c$. 

    If $t<u$ and $w\in c$ do not both hold true, then either $w\notin c$, or $w\in c$ and $u\le t$. If $w\notin c$, then $w\in x \setminus c$, hence $x\nsubseteq c$. If $w\in c$ and $u\le t$, then, by Assumption~\hyperlink{1-SDF_AC.Ass:AP.C1}{AP.C1}, there is $w' \in x_t(w) \setminus c \subseteq x_u(w) \setminus c = x\setminus c$. Whence $x\nsubseteq c$.
\end{proof}

\begin{lemma}\label{1-SDF_AC.lemma:AP_uparrow_x}
    Consider action path \textsc{sdf} data $(I,\A,\T,W)$ on an exogenous scenario space $(\Omega,\ms E)$ and the induced action path \textsc{sdf} $(F,\pi,\X)$. Let $t\in\T$ and $w\in W$. Then,
    \[ \uparrow x_t(w) = \{ x_u(w) \mid u\in \T\colon u\le t\}. \]
\end{lemma}

\begin{proof}
    The inclusion $\supseteq$ is clear from the definition of the nodes of the action path \textsc{sdf}. For the converse inclusion, let $x\in\uparrow x_t(w)$. If $x$ is a singleton, then $x = x_t(w)$. If $x$ is no singleton, then it is a move and there are $u\in \T$ and $w'\in W$ with $x =  x_u(w')$. As $w\in x_t(w) \subseteq x$, we even have $x = x_u(w)$. By means of Lemma~\ref{1-SDF_AC.lemma:mf_t}, we infer $u = \mf t(x) \le \mf t(x_t(w)) = t$.
\end{proof}

\begin{proof}[Proof of Lemma~\ref{1-SDF_AC.lemma:AP_P(c)}]
    Let $t\in\T$ and $c\in\ms C_t$. Let $x\in F$. By definition, $x\in P(c)$ is equivalent to the existence of $y\in \downarrow c$ satisfying
    \[ (\ast)\qquad \uparrow x = \uparrow y \setminus \downarrow c. \]

    If $x = x_t(w)$ for $w\in c$, then, by Assumption~\hyperlink{1-SDF_AC.Ass:AP.C1}{AP.C1}, there is $w'\in x_t(w) \setminus c$. Thus, $x$ has at least two elements. In view of Lemmata \ref{1-SDF_AC.lemma:AP_downarrow_c} and \ref{1-SDF_AC.lemma:AP_uparrow_x} and Assumption~\hyperlink{1-SDF_AC.Ass:AP.SDF1}{AP.SDF1}, $(\ast)$ is satisfied for $y = \{w\}$.

    If, conversely, $(\ast)$ is satisfied for some $y\in \downarrow c$, then $x$ is a move, whence the existence of $w_0\in W$ and $t_0\in\T$ with $x = x_{t_0}(w_0)$. Moreover, there is $w\in y$, and as $y\in \downarrow c$, we obtain $w\in c$. By $(\ast)$, we know that $x\in\uparrow y$, hence also $w\in x$. Thus, $x = x_{t_0}(w)$. From representation $(\ast)$ and Lemma~\ref{1-SDF_AC.lemma:AP_downarrow_c}, we immediately get $t_0 \le t$. Moreover, by Assumption~\hyperlink{1-SDF_AC.Ass:AP.C1}{AP.C1} on $c\in\ms C_t$, $x_t(w)$ has at least two elements. Indeed, as $w\in c$, there is $w'\in x_t(w)\setminus c$. Hence, by Lemma~\ref{1-SDF_AC.lemma:AP_sdf.AssmAP.SDF1}, $\T_{\x_t(w)} = \{t\}$. Hence, $x_t(w) \in \uparrow y \setminus \downarrow c = \uparrow x$, again by representation $(\ast)$, the fact that $y\in\downarrow c$ and Lemma~\ref{1-SDF_AC.lemma:AP_downarrow_c}. Thus, by Lemma~\ref{1-SDF_AC.lemma:mf_t}, $t = \mf t(x_t(w)) \le \mf t(x) = t_0$. Hence, $t_0 = t$ and $x = x_t(w)$ with $w\in c$.\smallskip
\end{proof}

\begin{proof}[Proof of Lemma~\ref{1-SDF_AC.lemma:C_t_non-redundant_complete}]
    By Assumption~\hyperlink{1-SDF_AC.Ass:AP.C0}{AP.C0}, $c$ is a non-empty union of singletons in $\mc P(W)$, which are elements of $F$ by construction. Hence, it is a choice.

    Concerning non-redundancy, let $\omega\in\Omega$ and suppose there is $w\in c \cap W_\omega$. Then, $x_t(w)\in P(c) \cap T_\omega$, by Lemma~\ref{1-SDF_AC.lemma:AP_P(c)}. Hence, by contraposition, if $P(c) \cap T_\omega = \emptyset$, then $c\cap W_\omega = \emptyset$ as well.

    Concerning completeness, let $\x\in\X$ such that there is $\omega\in D_\x$ with $\x(\omega) \in P(c)$. By Lemma~\ref{1-SDF_AC.lemma:AP_P(c)}, there is $f\in\A^\T$ such that $(\omega,f)\in c$ and $\x(\omega) = x_t(\omega,f)$. First, we infer that $D_{t,f}\neq\emptyset$, whence $D_\x = D_{t,f}$ and $\x = \x_t(f)$, by Proposition~\ref{1-SDF_AC.prop:ev_on_Tr_is_iso}. Second, we infer that $x_t(\omega,f)\cap c \neq\emptyset$. By Assumption~\hyperlink{1-SDF_AC.Ass:AP.C2}{AP.C2}, for any $\omega'\in D_{t,f}$ there is $w' \in x_t(\omega',f)\cap c$. In particular, $\x(\omega') = x_t(\omega',f) = x_t(w')$. Hence, $\x(\omega') \in P(c)$, by Lemma~\ref{1-SDF_AC.lemma:AP_P(c)}. We conclude that $\x^{-1}(P(c)) = D_\x$.\smallskip

    Regarding the proof of the second sentence, let $\x\in\X$ such that $c$ is available at $\x$. There is $\omega\in D_\x$. Then, $\x(\omega)\in P(c)$. As in the proof of completeness above, we infer the existence of $f\in\A^\T$ such that $(\omega,f)\in c$, $\omega\in D_{t,f} = D_\x$ and $\x = \x_t(f)$.
\end{proof}

\begin{proof}[Proof of Proposition~\ref{1-SDF_AC.prop:APsdf_RCS}]
    Let $\x\in\tilde\X^i$, $t=\mf t(\x)$, and $c\in \ms C_\x^i$. In particular, $c\in\ms C_t$, and, by Lemma~\ref{1-SDF_AC.lemma:C_t_non-redundant_complete}, $c$ is a non-redundant and complete, in particular $\tilde\X^i$-complete choice. Let $\omega\in D_\x$. Then, by definition of $\ms C_\x^i$, there is $w\in\x(\omega)\cap c$. Hence, $\x(\omega) = x_{\mf t(\x)}(w) = x_t(w)$, and, by Lemma~\ref{1-SDF_AC.lemma:AP_P(c)}, $\x(\omega)\in P(c)$. We conclude that $c$ is available at $\x$.
\end{proof}

\begin{proof}[Proof of Theorem~\ref{1-SDF_AC.thm:APsdf_AC}]
    Let $c = c(A_{<t},i,g)$.
    \smallskip

    (Ad \ref{1-SDF_AC.thm:APsdf_AC.non_red_and_compl}):~ By Lemma~\ref{1-SDF_AC.lemma:C_t_non-redundant_complete}, $c$ is a non-redundant and complete, in particular $\tilde\X^i$-complete choice which proves the first claim. \smallskip

    (Ad \ref{1-SDF_AC.thm:APsdf_AC.Dx_subset_D}):~ 
    Let $\x\in\tilde\X^i$ be such that $c$ is available at $\x$. Let $\omega\in D_\x$. Then $\x(\omega) \in P(c)$, hence, by Lemma~\ref{1-SDF_AC.lemma:AP_P(c)}, there is $w\in c$ with $\x(\omega) = x_t(w)$. There is $f\in\A^\T$ such that $w=(\omega,f)$, and as $w\in c$, we have $f(t) \in A^{i,g}_{t,\omega}$. Hence, $\omega\in D$. We conclude that $D_\x\subseteq D$.\smallskip
    
    (Helpful statements for $c'\in\ms C^i_\x$ and $\x$ that $c$ is available at):~ Let $\x\in\tilde\X^i$ be such that $c$ is available at $\x$. By definition of $\X$, there is $f_0\in\A^\T$ with $D_{t,f_0} \neq\emptyset$ such that $\x = \x_t(f_0)$. 
    
    Let $c'\in\ms C^i_\x$. We compute the set $P(c\cap c')$ and its preimage under $\x$. By definition of $\ms C^i_\x$, there are $A'_{<t}\subseteq\A^{[0,t)_\T}$ and $A_t^{\prime i}\in\ms B(\A^i)$ such that, with $A'_t = (A'_{t,\omega})_{\omega\in\Omega}$ and $A'_{t,\omega} = (p^i)^{-1}(A_t^{\prime i})$ for all $\omega\in D_\x$, and $A'_{t,\omega}=\emptyset$ for all $\omega\notin D_\x$, we have $c' = c(A'_{<t},A'_t)$, $c'\in\ms C_t$, and, for all $\omega\in D_\x$, $\x(\omega) \cap c' \neq\emptyset$.
    
    Let $c_0 = c(A_{<t}\cap A'_{<t},i,g)$. By definition of $c$ and $c'$, we have
    \begin{align*} 
        &~c\cap c' \\
        =&~ \{(\omega,f)\in W_{D_\x} \mid f|_{[0,t)_\T} \in A_{<t} \cap A'_{<t},~ p^i \circ f(t) = g(\omega)\in A_t^{\prime i}\}\\
        =&~ c(A_{<t}\cap A'_{<t},i,g) \cap \{(\omega,f) \in W_{D_\x} \mid g(\omega) \in A_t^{\prime i} \} \\
        =&~ c_0 \cap W_{g|_{D_\x}^{-1}(A_t^{\prime i})}.
    \end{align*}
    Using Lemma~\ref{1-SDF_AC.lemma:P(c)_compatible_with_conn_comp}, we infer
    \[ (\ast)\qquad P(c\cap c') = P(c_0) \cap F_{g|_{D_\x}^{-1}(A_t^{\prime i})}. \]

    Next, we show that
    \[ (\circ)\qquad \forall \omega\in D_\x\exists f\in \A^\T\colon \quad (\omega,f) \in c_0,~ \x(\omega) = x_t(\omega,f).  \]
    For the proof of $(\circ)$, let $\omega\in D_\x$. As $\x$ is available both at $c$ and $c'$, we have $\x(\omega) = x_t(\omega,f_0)\in P(c)\cap P(c')$. Hence, by Lemma~\ref{1-SDF_AC.lemma:AP_P(c)} applied to $c,c'\in\ms C_t$, there are $f,f'\in\A^\T$ with $(\omega,f)\in c$ and $(\omega,f')\in c'$ such that 
    \[ f'|_{[0,t)_\T} = f_0|_{[0,t)_\T} = f|_{[0,t)_\T}. \]
    Thus, $f|_{[0,t)_\T} \in A_{<t}\cap A'_{<t}$. By definition of $f$ and $f_0$, we must have $p^i \circ f(t) = g(\omega)$. Hence, $(\omega,f)\in c_0$. By definition of $f$, we also have $\x(\omega) = x_t(\omega,f)$.

    We infer that $c_0 \in \ms C_t$.
    Indeed, Assumption~\hyperlink{1-SDF_AC.Ass:AP.C0}{AP.C0} is satisfied by $(\circ)$. Concerning \hyperlink{1-SDF_AC.Ass:AP.C1}{AP.C1}, note that $c_0\subseteq c$. Let $w\in c_0$. Thus $w\in c$. Hence, by \hyperlink{1-SDF_AC.Ass:AP.C1}{AP.C1} applied to $c$, there is $w'\in x_t(w) \setminus c \subseteq x_t(w) \setminus c_0$. Regarding \hyperlink{1-SDF_AC.Ass:AP.C2}{AP.C2}, let $f\in\A^\T$ with $f|_{[0,t)_\T} \in A_{<t}\cap A'_{<t}$ such that there is $\omega\in D_{t,f}$ satisfying
    \[ x_t(\omega,f) \cap c_0 \neq \emptyset. \]
    Let $\omega'\in D_{t,f}$. As $c_0 \subseteq c$, we infer that $x_t(\omega,f) \cap c \neq\emptyset$. Hence, by \hyperlink{1-SDF_AC.Ass:AP.C2}{AP.C2} applied to $c$, there exists $w'\in x_t(\omega',f) \cap c$. Let $f'\in\A^\T$ such that $w' = (\omega',f')$. Then, by definition of $w'$ and $f'$, we get $f'|_{[0,t)_\T} = f|_{[0,t)_\T}\in A_{<t}\cap A'_{<t}$, $\omega'\in D$, and $p^i(f'(t)) = g(\omega')$. Hence, $(\omega',f') \in x_t(\omega',f) \cap c_0$. We conclude that $c_0\in\ms C_t$.

    Hence, by Lemma~\ref{1-SDF_AC.lemma:C_t_non-redundant_complete}, $c_0$ defines a non-redundant and complete choice. Moreover, by $(\circ)$ and Lemma~\ref{1-SDF_AC.lemma:AP_P(c)}, $c_0$ is available at $\x$. 
    In particular, we have $\x^{-1}(P(c_0)) = D_\x$. By $(\ast)$, we get:
    \begin{equation*}
    (\dagger)\qquad
    \begin{aligned} 
        \x^{-1}(P(c\cap c')) = \x^{-1}(P(c_0)) \cap \x^{-1}(F_{g|_{D_\x}^{-1}(A_t^{\prime i})}) 
        = D_{\x} \cap g|_{D_\x}^{-1}(A_t^{\prime i}) = g|_{D_\x}^{-1}(A_t^{\prime i}). 
    \end{aligned}
    \end{equation*}
    
    (Ad \ref{1-SDF_AC.thm:APsdf_AC.Fx_mb_=>_adapted}):~ Let $\x\in\tilde\X^i$ be such that $c$ is available at $\x$. If $g|_{D_\x}$ is $\ms F^i_\x$-measurable, then, by $(\dagger)$, $\x^{-1}(P(c\cap c'))\in\ms F^i_\x$ for all $c'\in\ms C^i_\x$. \smallskip
    
    (Ad \ref{1-SDF_AC.thm:APsdf_AC.adapted_=>_Fx_mb}):~ Suppose conversely that $\x^{-1}(P(c\cap c'))\in\ms F^i_\x$ for all $\x\in\tilde\X^i$ that $c$ is available at and all $c'\in\ms C^i_\x$, and that Assumption~\hyperlink{1-SDF_AC.Ass:AP.C3}{AP.C3} is satisfied for $(A_{<t},i,g)$ and $\ms C^i$. 
    
    Let $\x\in\tilde\X^i$ be such that $c$ is available at it. By Assumption~\hyperlink{1-SDF_AC.Ass:AP.C3}{AP.C3}, there is a generator $\ms G(\A^i)$ of $\ms B(\A^i)$, stable under non-trivial intersections, such that for all $G\in\ms G(\A^i)$, we have $c(A_{<t},A^{i,G}_t)\in\ms C^i_\x$. Let $G\in\ms G(\A^i)$ and $c' = c(A_{<t},A^{i,G}_t)$.
    
    Represent $c'$ as in the beginning of the ``Helpful statements'' part above, with $A_t^{\prime i} = G$ and $A'_{<t} = A_{<t}$. Let $c_0$ be defined as in the ``Helpful statements'' part.
    
    Then, we can use these helpful statements to infer that $c_0$ is a non-redundant and complete choice available at $\x$.
    Hence, $\x^{-1}(P(c_0)) = D_\x$, and thus $(\dagger)$ and the hypothesis imply that
    \[ (g|_{D_\x})^{-1}(G) = \x^{-1}(P(c\cap c')) \in \ms F^i_\x. \]
   
    We conclude that $(g|_{D_\x})^{-1}(G)\in\ms F^i_\x$ for all $G\in\ms G(\A^i)$. Trivially, we have $(g|_{D_\x})^{-1}(\emptyset) = \emptyset \in\ms F^i_\x$. As $\ms G(\A^i)\cup \{\emptyset\}$ is a generator of $\ms B(\A^i)$ stable under intersections, in view of the $\pi$-$\lambda$ theorem, $g|_{D_\x}$ is $\ms F^i_\x$-measurable.
\end{proof}

\begin{proof}[Proofs for Example~\ref{1-SDF_AC.ex:APsdf_AC}]
    (Ad $W=\Omega\times\A^\T$):~ We show the following more general statement: If $t\in\T$ and $A_{<t}\subseteq\A^{[0,t)_\T}$ is non-empty, and moreover $A_t = (A_{t,\omega})_{\omega\in\Omega}\in \mc P(\A)^\Omega$ is such that 
    \[ (\ast) \qquad \forall \omega\in\Omega \colon\quad\emptyset \subsetneq A_{t,\omega} \subsetneq \A, \]
    then $c(A_{<t},A_t) \in \ms C_t$. This includes the two following cases: A) $A_t = A_t^{i,g}$ for given $i\in I$ and a map $g\colon \Omega\to\A^i$ because $p^i$ is surjective and $\A^i$ has at least two elements; and B) $A_{t,\omega} = (p^i)^{-1}(G)$ for some $i\in I$ and some set $\emptyset\subsetneq G\subsetneq \A^i$, again because $p^i$ is surjective.
    
    As $A_{<t}\neq\emptyset$ and $(\ast)$ is assumed, there is $w=(\omega,f) \in W=\Omega\times\A^\T$ such that $f|_{[0,t)_\T}\in A_{<t}$ and $f(t) \in A_{t,\omega}$. Hence, $w\in c(A_{<t},A_t)$. Thus, \hyperlink{1-SDF_AC.Ass:AP.C0}{AP.C0} is satisfied.\smallskip

    Regarding \hyperlink{1-SDF_AC.Ass:AP.C1}{AP.C1}, let $w=(\omega,f)\in c(A_{<t},A_t)$. There is $f'\in\A^\T$ with $f'|_{[0,t)_\T} = f|_{[0,t)_\T}$ and $f'(t) \notin A_{t,\omega}$ (by $(\ast)$). Then $w'=(\omega,f')\in \Omega\times\A^\T = W$, hence $w'\in x_t(w)$, but $w'\notin c(A_{<t},A_t)$.\smallskip

    Regarding \hyperlink{1-SDF_AC.Ass:AP.C2}{AP.C2}, let $f\in\A^\T$ with $f|_{[0,t)_\T}\in A_{<t}$ and $\omega\in D_{t,f}$. There is $f'\in\A^\T$ with $f'|_{[0,t)_\T} = f|_{[0,t)_\T}$ and $f'(t) \in A_{t,\omega}$, by $(\ast)$. As $(\omega,f') \in \Omega\times\A^\T = W$, we infer
    \[ (\omega,f') \in x_t(\omega,f) \cap c(A_{<t},A_t). \]

    Hence, $c(A_{<t},A_t)\in\ms C_t$. The general statement above is proven. By considering the case A), we infer that for all $i\in I$ and $g\colon \Omega\to\A^i$, $c(A_{<t},i,g)\in\ms C_t$.\smallskip

    Regarding \hyperlink{1-SDF_AC.Ass:AP.C3}{AP.C3}, suppose that for all $\x\in\tilde\X^i$ with $\mf t(\x)=t$, $\ms C^i_\x$ contains all $c(A_{<t},A_t)$ ranging over all $A_t$ satisfying (AP-RCS.$k$), $k=2,3,4$. Let $\x\in\tilde\X^i$ be such that $c(A_{<t},i,g)$ is available at $\x$.
    By Lemma~\ref{1-SDF_AC.lemma:C_t_non-redundant_complete}, there is $(\omega_0,f_0)\in c(A_{<t},i,g)$ such that $\x = \x_t(f_0)$ and $\omega_0\in D_\x$.
    
    Let $\ms G(\A^i) = \ms B(\A^i)\setminus \{\A^i,\emptyset\}$ which is obviously a generator of $\ms B(\A^i)$ stable under non-trivial intersections. Let $G\in\ms G(\A^i)$. Then, the general statement $(\ast)$ above in case B) applies, hence $c(A_{<t},A_t^{i,G})\in\ms C_t$. To complete the proof that $c(A_{<t},A_t^{i,G})\in\ms C_\x^i$, in view of our additional assumption on $\ms C^i_\x$, it remains to prove Property (AP-RCS.\ref{1-SDF_AC.def:msC.4}) for $c(A_{<t},A_t^{i,G})$, (AP-RCS.\ref{1-SDF_AC.def:msC.3}) having been proven just before and (AP-RCS.\ref{1-SDF_AC.def:msC.2}) being evident. For this, let $\omega'\in D_\x$. Then there is $f'\in\A^\T$ such that $f'|_{[0,t)_\T} = f_0|_{[0,t)_\T}\in A_{<t}$ and $p^i \circ f'(t) \in G$, because $G$ is non-empty and $p^i$ surjective. Then, $w' = (\omega',f')\in \Omega\times \A^\T = W$, and thus $w'\in \x_t(f_0)(\omega') \cap c(A_{<t},A_t^{i,G})$. Property (AP-RCS.\ref{1-SDF_AC.def:msC.4}) therefore holds true for $c(A_{<t},A_t^{i,G})$ and the proof of the statement $c(A'_{<t},A_t^{i,G})\in\ms C_\x^i\cup\{\emptyset\}$ is complete.\medskip

    (Ad timing problem):~ Let $t\in\T$. Further, let $A_{<t}\subseteq \A^{[0,t)_\T}$ be a non-empty set of componentwise decreasing paths $f_t$ such that $p^i\circ f_t = 1_{[0,t)_\T}$, and $g\colon\Omega\to\{0,1\} = \A^i$. Let $c = c(A_{<t},i,g)$. We are going to prove that $c\in\ms C_t$. 

    Regarding \hyperlink{1-SDF_AC.Ass:AP.C0}{AP.C0}, let $\omega\in\Omega$. By assumption on $A_{<t}$, there is componentwise decreasing $f\colon\T \to \A$ such that $f|_{[0,t)_\T}\in A_{<t}$ and $p^i \circ f(t) = g(\omega)$. Hence, $(\omega,f)\in W$, and even $(\omega,f)\in c$. Thus, $c\neq\emptyset$.\smallskip

    Regarding \hyperlink{1-SDF_AC.Ass:AP.C1}{AP.C1}, let $w=(\omega,f)\in c$. There is componentwise decreasing $f'\in\A^\T$ with $f'|_{[0,t)_\T} = f|_{[0,t)_\T}$ and $p^i\circ f'(t)\neq g(\omega)$, because $p^i\circ f|_{[0,t)_\T} = 1_{[0,t)_\T}$. Then $w'=(\omega,f')\in W$, hence $w'\in x_t(w)$, but $w'\notin c$.\smallskip

    Regarding \hyperlink{1-SDF_AC.Ass:AP.C2}{AP.C2}, let $f\in\A^\T$ with $f|_{[0,t)_\T}\in A_{<t}$ and $\omega\in D_{t,f}$. There is componentwise decreasing $f'\in\A^\T$ with $f'|_{[0,t)_\T} = f|_{[0,t)_\T}$ and $p^i \circ f'(t) = g(\omega)$, by assumption on $A_{<t}$. As $(\omega,f') \in W$, we infer
    \[ (\omega,f') \in x_t(\omega,f) \cap c. \]
    We conclude that $c\in \ms C_t$.\smallskip

    Regarding Assumption~\hyperlink{1-SDF_AC.Ass:AP.C3}{AP.C3}, suppose that for all $\x\in\tilde\X^i$ with $\mf t(\x)=t$, $\ms C^i_\x$ contains all $c(A_{<t},A_t)$ ranging over all $A_t$ satisfying (AP-RCS.$k$), $k=2,3,4$. Let $\x\in\tilde\X^i$ be such that $c(A_{<t},i,g)$ is available at $\x$. By Lemma~\ref{1-SDF_AC.lemma:C_t_non-redundant_complete}, there is $(\omega_0,f_0)\in c(A_{<t},i,g)$ such that $\omega_0 \in D_{t,f_0}=D_\x$ and $\x = \x_t(f_0)$. As $c(A_{<t},i,g)\in\ms C_t$, Assumption~\hyperlink{1-SDF_AC.Ass:AP.C1}{AP.C1} implies, that $p^i \circ f_0|_{[0,t)_\T}$ is constant with value $1$. Indeed, there is $f_1\in\A^\T$ with $(\omega,f_1)\in x_t(\omega,f_0)\setminus c(A_{<t},i,g)$. In particular, $f_1$ is componentwise decreasing and takes the same values on $[0,t)_\T$ as $f_0$. If $p^i \circ f_0$ took the value $0$ at some point in $[0,t)_\T$, the monotonicity would imply $p^i\circ f_1(t) = 0 = p^i\circ f_0(t) = g(\omega)$, whence $(\omega,f_1)\in c(A_{<t},i,g)$, in contradiction to the choice of $f_1$.

    Further, let $\ms G(\A^i) = \ms B(\A^i)\setminus \{\A^i,\emptyset\}$ which is obviously a generator of $\ms B(\A^i)$ stable under non-trivial intersections. As $\A^i = \{0,1\}$, $\ms G(\A^i) = \{\{0\},\{1\}\}$. Let $G\in\ms G(\A^i)$. Thus, $G$ is a singleton and $c(A_{<t},A_t^{i,G}) = c(A_{<t},i,g_G)$ for the constant map $g_G$ with value given by the unique element of $G$. We have shown just beforehand that $c(A_{<t},i,g_G)\in\ms C_t$. To complete the proof of the fact that $c(A_{<t},A_t^{i,G})\in\ms C_\x^i$, in view of our additional assumption on $\ms C^i_\x$, it thus remains to show Axiom (AP-RCS.\ref{1-SDF_AC.def:msC.4}) in the definition of $\ms C_\x$, because (AP-RCS.\ref{1-SDF_AC.def:msC.3}) has just been proven and (AP-RCS.\ref{1-SDF_AC.def:msC.2}) is evident by construction. For this, let $\omega'\in D_\x$. As $p^i\circ f_0$ only takes the value $1$ on $[0,t)_\T$, there is componentwise decreasing $f' \colon \T\to\A$ such that $f'|_{[0,t)_\T} = f_0|_{[0,t)_\T}$ and $p^i\circ f'(t) = g_G(\omega')$. Hence, $(\omega',f')\in \x_t(f_0)(\omega')\cap c(A_{<t},i,g_G) = \x(\omega') \cap c(A_{<t},A_t^{i,G})$. This completes the proofs for the timing game example.\medskip

    (Ad up-and-out option exercise example):~ Let $t\in\R_+$, $A_{<t} = \{1\}^{[0,t)}$ and $D$ be the set of $\omega\in\Omega$ such that $\max_{u\in [0,t]} P_u(\omega) < 2$. We suppose that $D\neq\emptyset$. Let $g\colon D\to\{0,1\}$ be a map and let $c = c(A_{<t},i,g)$. We are first going to prove that $c\in\ms C_t$. 

    We start with the proof of \hyperlink{1-SDF_AC.Ass:AP.C0}{AP.C0}. There is $\omega_0\in D$. As $P$ is continuous, there is $\e>0$ such that $\max_{u\in [0,t+\e]} P_u(\omega_0) < 2$. Then, regardless of the value of $g(\omega_0)$, there is decreasing $f\colon{\R_+} \to \A$ such that $f|_{[0,t)} = 1_{[0,t)}$, $f(t) = g(\omega_0)$, and $f(t+\e) = 0$. Hence, $(\omega_0,f)\in W$, and even $(\omega_0,f)\in c$. Thus, $c\neq\emptyset$.\smallskip

    Regarding \hyperlink{1-SDF_AC.Ass:AP.C1}{AP.C1}, let $w=(\omega,f)\in c$. Then, $\max_{u\in [0,t]} P_u(\omega) < 2$, and by continuity of $P$, there is $\e>0$ such that $\max_{u\in [0,t+\e]} P_u(\omega) < 2$. Hence, regardless of the value of $g(\omega)$, there is decreasing $f'\in\A^{\R_+}$ with $f'|_{[0,t)} = f|_{[0,t)} = 1_{[0,t)}$, $f'(t)\neq g(\omega)$, and $f'(t+\e) = 0$. Then $w'=(\omega,f')\in W$, hence $w'\in x_t(w)$, but $w'\notin c$.\smallskip

    Regarding \hyperlink{1-SDF_AC.Ass:AP.C2}{AP.C2}, let $f\in\A^{\R_+}$ with $f|_{[0,t)}\in A_{<t}$ and $\omega\in D_{t,f}$. Then there is decreasing $\tilde f\colon\R_+\to\A$ with $(\omega,\tilde f)\in x_t(\omega,f)$, that is, $\tilde f|_{[0,t)} = f|_{[0,t)}$ and $\max_{u\in [0,t]} P_u(\omega) < 2$. By continuity of $P$, there is $\e>0$ such that $\max_{u\in [0,t+\e]} P_u(\omega) < 2$. Hence, regardless of the value of $g(\omega)$, there is decreasing $f'\in\A^{\R_+}$ with $f'|_{[0,t)} = f|_{[0,t)} = 1_{[0,t)}$, $f'(t) = g(\omega)$, and $f'(t+\e) = 0$. Hence, $(\omega,f') \in W$, and we infer
    \[ (\omega,f') \in x_t(\omega,f) \cap c. \]
    We conclude that $c\in \ms C_t$.\smallskip

    Regarding Assumption~\hyperlink{1-SDF_AC.Ass:AP.C3}{AP.C3}, suppose that for all $\x\in\tilde\X^i$ with $\mf t(\x)=t$, $\ms C^i_\x$ contains all $c(A_{<t},A_t)$ ranging over all $A_t$ satisfying (AP-RCS.$k$), $k=2,3,4$. Let $\x\in\tilde\X^i$ be such that $c(A_{<t},i,g)$ is available at $\x$. By Lemma~\ref{1-SDF_AC.lemma:C_t_non-redundant_complete}, there is $(\omega_0,f_0)\in c(A_{<t},i,g)$ such that $\x = \x_t(f_0)$ and $\omega_0\in D_\x$. As $c(A_{<t},i,g)\in\ms C_t$, Assumption~\hyperlink{1-SDF_AC.Ass:AP.C1}{AP.C1} implies, that $f_0|_{[0,t)}$ is constant with value $1$. Indeed, there is $f_1\in\A^\T$ with $(\omega,f_1)\in x_t(\omega,f_0)\setminus c(A_{<t},i,g)$. In particular, $f_1$ is componentwise decreasing and takes the same values on $[0,t)_\T$ as $f_0$. If $f_0$ took the value $0$ at some point in $[0,t)$, the monotonicity would imply $f_1(t) = 0 = f_0(t) = g(\omega)$, whence $(\omega,f_1)\in c(A_{<t},i,g)$, in contradiction to the choice of $f_1$.

    Let $\ms G(\{0,1\}) = \ms B(\{0,1\})\setminus \{\{0,1\},\emptyset\}$ which is obviously a generator of $\ms B(\{0,1\})$ stable under non-trivial intersections. Note that $\ms G(\{0,1\}) = \{\{0\},\{1\}\}$. Let $G\in\ms G(\{0,1\})$.
    Thus, $G$ is a singleton and $c(A_{<t},A_t^{i,G}) = c(A_{<t},i,g_G)$ for the constant map $g_G$ with value given by the unique element of $G$.

    We have shown just above that $c(A_{<t},i,g_G)\in\ms C_t$. To complete the proof of the fact that $c(A_{<t},i,g_G)\in\ms C_\x^i$, in view of our additional assumption on $\ms C^i_\x$, it remains to show Axiom (AP-RCS.\ref{1-SDF_AC.def:msC.4}), because (AP-RCS.\ref{1-SDF_AC.def:msC.3}) has just been proven and (AP-RCS.\ref{1-SDF_AC.def:msC.2}) is evident by construction. For this, let $\omega'\in D_\x = D_{t,f_0}$. In particular, $\max_{u\in [0,t]} P_u(\omega') < 2$. By continuity of $P$, there is $\e>0$ such that $\max_{u\in [0,t+\e]} P_u(\omega') < 2$. Then, regardless of the value of $g(\omega)$, and as $f_0$ only takes the value $1$ on $[0,t)$, there is componentwise decreasing $f' \colon {\R_+}\to\A$ such that $f'|_{[0,t)} = f_0|_{[0,t)}$, $f'(t) = g_G(\omega')$, and $f'(t+\e) = 0$. Hence, $(\omega',f')\in \x(\omega')\cap c(A_{<t},i,g_G)$. This completes the proofs for the up-and-out option exercise problem example.
\end{proof}

\section{Chapter~\ref{chap:2-SEF_G}}
\subsection{Section~\ref{2-SEF_G.sec:Stochastic extensive forms}}\label{2-SEF_G.subsec:appendix.proofs.1}

\begin{lemma}\label{2-SEF_G.lemma:Xi_random_moves}
    Let $\F$ be a stochastic pseudo-extensive form on some exogenous scenario space $(\Omega,\ms E)$, and $i\in I$. Then, 
    \[ X^i = \{ \x(\omega) \mid \x\in \X^i,~\omega\in D_\x\}. \]
\end{lemma}

\begin{proof}
    Let $x\in X^i$. Then, by Axiom~\ref{1-SDF_AC.def:sdf}.\ref{1-SDF_AC.def:sdf.X.cov}, there is $\x\in\X$ and $\omega\in D_\x$ with $x=\x(\omega)$. Hence, $A^i(x)\subseteq A^i(\x)$, by definition of $A^i(.)$. As $A^i(x)\neq\emptyset$ by assumption, $A^i(\x)\neq\emptyset$, whence $\x\in\X^i$.

    Conversely, let $\x\in\X^i$ and $\omega\in D_\x$. Then, $A^i(\x(\omega)) = A^i(\x) \neq\emptyset$, by $\X^i$-completeness of all elements of $C^i$. Hence, $\x(\omega)\in X^i$.
\end{proof}

Letting $\X^i \bullet \Omega = \{(\x,\omega) \in \X\bullet\Omega \mid \x\in\X^i\}$, the lemma states that $X^i = \mc P\ev(\X^i \bullet \Omega)$, just as $X = \mc P\ev(\X \bullet \Omega)$ and $F = \mc P\ev(\Tr \bullet \Omega)$.

\begin{proof}[Proof of Proposition~\ref{2-SEF_G.prop:information_sets}]
    Let $\F = (F,\pi,\X,I,\ms F,\ms F,C)$ be a tuple as in Definition~\ref{2-SEF_G.def:SEF} satisfying Axioms~\ref{2-SEF_G.def:SEF}.$k$, $k=1,\dots,5$, but not necessarily Axiom~\ref{2-SEF_G.def:SEF}.\ref{2-SEF_G.def:SEF.choice_completeness}.\smallskip

    (Ad \ref{2-SEF_G.prop:information_sets.P(c)_partition}):~ First, any $c\in C^i$ is non-empty. Hence, there is $\omega\in\Omega$ with $c\cap W_\omega\neq\emptyset$. By non-redundancy, $P(c)\cap T_\omega\neq\emptyset$, whence $P(c)\neq\emptyset$. 
    
    Next, let $x\in X^i$. Hence, $A^i(x) \neq \emptyset$ so that there is $c\in C^i$ such that $x\in P(c)$. 

    Then, let $c,c'\in C^i$ such that $P(c) \cap P(c') \neq \emptyset$. Then, by Axiom~\ref{2-SEF_G.def:SEF}.\ref{2-SEF_G.def:SEF.P(c)}, $P(c) = P(c')$.\smallskip

    (Ad \ref{2-SEF_G.prop:information_sets.A(x)_partition}):~ First, note that $A^i(x)\neq\emptyset$ for all $x\in X^i$, by definition.
    
    As $A^i(\x(\omega)) = A^i(\x)$ for all $\x\in\X^i$ and $\omega\in D_\x$, Lemma~\ref{2-SEF_G.lemma:Xi_random_moves} easily implies that 
    \[ \{A^i(x) \mid x\in X^i\} = \{A^i(\x) \mid \x\in\X^i\}. \]

    Let $c\in C^i$. As $c$ is non-empty, there is $\omega\in\Omega$ with $c\cap W_\omega \neq \emptyset$. By non-redundancy, $P(c) \cap T_\omega \neq \emptyset$. Hence, there is $x\in T_\omega \subseteq F$ with $x\in P(c)$. It follows easily from the definition of $P(c)$ that $x\in X$, whence $c\in A^i(x)$, $i\in J(x)$, and $x\in X^i$.

    Let $x,x'\in X^i$ such that $A^i(x) \cap A^i(x') \neq \emptyset$. Let $c\in A^i(x) \cap A^i(x')$. Hence $x,x'\in P(c)$. Let $c'\in A^i(x)$. Then $x\in P(c) \cap P(c')$. Hence, using Axiom~\ref{2-SEF_G.def:SEF}.\ref{2-SEF_G.def:SEF.P(c)} we obtain $x'\in P(c) = P(c')$. Thus, $c'\in A^i(x')$. We conclude that $A^i(x) \subseteq A^i(x')$. The same argument can be repeated with the roles of $x$ and $x'$ reversed, whence in total $A^i(x) = A^i(x')$.\smallskip

    (Ad \ref{2-SEF_G.prop:information_sets.P(c)=P(c')}):~ Let $c,c'\in C^i$. By Axiom~\ref{2-SEF_G.def:SEF}.\ref{2-SEF_G.def:SEF.P(c)}, $P(c) = P(c')$ is true iff there is $x\in P(c) \cap P(c')$. Such an $x$ must be a move by definition of immediate predecessors. Hence, this statement is equivalent to saying that $c,c'\in A^i(x)$.\smallskip

    (Ad \ref{2-SEF_G.prop:information_sets.A(x)=A(x')}):~ Let $x,x'\in X^i$. Using Part~\ref{2-SEF_G.prop:information_sets.A(x)_partition} just proved before, $A^i(x) = A^i(x')$ is equivalent to the existence of $c \in A^i(x) \cap A^i(x')$, or put equivalently, $x,x'\in P(c)$ for some $c\in C^i$.\smallskip

    (Ad \ref{2-SEF_G.prop:information_sets.exists_mfP}):~ It suffices to show that there is a unique equivalence relation $\sim$ on $\X^i$ satisfying $\x\sim\x'$ iff $A^i(\x) = A^i(\x')$, for all $\x,\x'\in \X^i$. Uniqueness is trivial. Concerning existence, there clearly is a binary relation $\sim$ with the preceding property. Moreover, reflexivity, symmetry, and transitivity are trivial. Hence, $\sim$ is an equivalence relation.\smallskip

    (Ad \ref{2-SEF_G.prop:information_sets.Bij_mfP_P(c)}):~ Let $\Phi$ be the map with domain $\mf P^i$ given by
    \[ \forall \mf p \in \mf P^i\colon\qquad \Phi(\mf p) = \bigcup_{\x\in\mf p} \im \x.\]
    
    Concerning the claim about both the codomain and the image of $\Phi$, we prove the following helpful statement: 
    \[ (\ast) \qquad \forall \mf p\in\mf P^i\forall \x_0\in\mf p\forall c_0\in A^i(\x_0)\colon ~\Phi(\mf p) = P(c_0) . \]

    For the proof, let $\mf p \in \mf P^i$, $\x_0\in\mf p$, and $c_0\in A^i(\x_0)$. 
    By definition of $\mf P^i$, $A^i(\x) = A^i(\x') \neq \emptyset$ for all $\x,\x'\in\mf p$. By Parts~\ref{2-SEF_G.prop:information_sets.P(c)=P(c')} and~\ref{2-SEF_G.prop:information_sets.A(x)_partition}, proven above, we obtain that 
    \[ (\dagger) \qquad \forall \x,\x'\in \mf p \forall c\in A^i(\x) \forall c'\in A^i(\x')\colon ~ P(c) = P(c') . \]

    First, let $x\in \Phi(\mf p)$. Take $\x\in\mf p$ and $\omega\in D_\x$ such that $x = \x(\omega)$. As $\x\in\mf p$, there is $c\in A^i(\x)$. By $(\dagger)$, $P(c) = P(c_0)$, whence $x\in P(c) = P(c_0)$. This shows that $\Phi(\mf p)\subseteq P(c_0)$.

    Second, let $x\in P(c_0)$. By Axiom~\ref{1-SDF_AC.def:sdf}.\ref{1-SDF_AC.def:sdf.X.cov}, there is $\x\in\X$ and $\omega\in D_\x$ such that $x = \x(\omega)$. Thus $c_0\in A^i(\x)$ and $\x\in\X^i$. By Part~\ref{2-SEF_G.prop:information_sets.A(x)=A(x')}, proven before, $A^i(\x) = A^i(\x_0)$ must hold true. As $\x_0\in\mf p$, we infer that $\x\in\mf p$ as well. This shows that $P(c_0) \subseteq \Phi(\mf p)$. We conclude that $\Phi(\mf p)=P(c_0)$, and the proof of $(\ast)$ is complete.

    Now, regarding the codomain of $\Phi$, let $\mf p\in\mf P^i$. As $\mf P^i$ is a partition, $\mf p$ is non-empty. Hence, we can choose $\x_0\in\mf p$. As $\x_0\in\X^i$, there is $c_0\in A^i(\x_0)$. By $(\ast)$, $\Phi(\mf p) = P(c_0)$. Hence, the codomain is of the claimed form.

    Next, we determine the image of $\Phi$. Let $c_0\in C^i$. By Part~\ref{2-SEF_G.prop:information_sets.P(c)_partition}, proven before, there is $x_0\in X^i$ such that $x_0\in P(c_0)$. By Lemma~\ref{2-SEF_G.lemma:Xi_random_moves}, there are $\x_0\in\X^i$ and $\omega\in D_\x$ with $x_0 = \x_0(\omega)$. In particular, $c_0\in A^i(\x_0)$. By construction of $\mf P^i$, there is $\mf p\in\mf P^i$ such that $\x_0\in\mf p$. Then, statement $(\ast)$ implies that $\Phi(\mf p) = P(c_0)$. Hence, the image of $\Phi$ is given by the set of all $P(c)$, $c\in C^i$.

    It remains to prove injectivity. Let $\mf p,\mf p'\in \mf P^i$ such that $\Phi(\mf p) = \Phi(\mf p')$. As shown just before, there is $c\in C^i$ such that $\Phi(\mf p) = P(c) = \Phi(\mf p')$. By Part~\ref{2-SEF_G.prop:information_sets.P(c)_partition}, there is $x\in P(c)$. Hence, there are representatives $\x\in\mf p$ and $\x'\in\mf p'$ of both endogenous information sets and $\omega \in D_\x \cap D_{\x'}$ such that $\x(\omega) = x = \x'(\omega)$. By Parts~\ref{2-SEF_G.prop:information_sets.A(x)_partition} and~\ref{2-SEF_G.prop:information_sets.A(x)=A(x')}, we infer that $A^i(\x) = A^i(\x')$, whence $\mf p = \mf p'$. We conclude that $\Phi$ is injective.\smallskip

    (Ad \ref{2-SEF_G.prop:information_sets.msC_msF_const_on_mfP}):~ Let $\mf p\in\mf P^i$ and $\x,\x'\in \mf p$. Then, by definition of $\mf P^i$, $A^i(\x) = A^i(\x')\neq \emptyset$. Hence, by Axiom~\ref{2-SEF_G.def:SEF}.\ref{2-SEF_G.def:SEF.endo_exo_compatible}, $\ms F_\x^i = \ms F_{\x'}^i$ and $\ms C_\x^i=\ms C_{\x'}^i$. As a consequence,
    $ D_\x = \bigcup \ms F_\x^i = \bigcup \ms F_{\x'}^i = D_{\x'}$.
\end{proof}

\begin{lemma}\label{2-SEF_G.lemma:perfect_endo_information_implies_perfect_endo_recall}
    Let $\F$ be a stochastic pseudo-extensive form on some exogenous scenario space $(\Omega,\ms E)$ and $i\in I$ an agent. If $i$ has perfect endogenous (exogenous) information, then $i$ admits perfect endogenous (exogenous, respectively) recall.
\end{lemma}

\begin{proof}[Proof of Lemma~\ref{2-SEF_G.lemma:perfect_endo_information_implies_perfect_endo_recall}]
    Let $\F$ be a $\psi$-\textsc{sef}, and $i\in I$ an agent.\smallskip
    
    (Ad endogenous case):~ Suppose that $i$ has perfect endogenous information. Let $c,c'\in C^i$ and $\omega\in\Omega$ such that $c\cap c'\cap W_\omega\neq \emptyset$. By non-redundancy and Lemma~\ref{1-SDF_AC.lemma:P(c)_compatible_with_conn_comp}, we get $P(c\cap W_\omega) = P(c) \cap T_\omega \neq \emptyset$, and similarly $P(c'\cap W_\omega) = P(c') \cap T_\omega \neq \emptyset$. Proposition~\ref{2-SEF_G.prop:information_sets}, Part~\ref{2-SEF_G.prop:information_sets.Bij_mfP_P(c)}, and the fact that $i$ has perfect information, we infer the existence of $x,x'\in X^i$ such that
    \[ P(c\cap W_\omega) =  P(c) \cap T_\omega = \{x\}, \qquad P(c'\cap W_\omega) = P(c') \cap T_\omega = \{x'\}. \]
    Directly from the definition of immediate predecessors, we infer that $x\supseteq c\cap W_\omega$ and $x'\supseteq c'\cap W_\omega$. Hence, $x\cap x' \neq \emptyset$. By the representation by decision paths inherent in the definition of decision forests (compare Definition~\ref{1-SDF_AC.def:decision_forest}), we have $x\supseteq x'$ or $x'\supseteq x$. 

    Without loss of generality, we assume that $x\supseteq x'$. If $x=x'$, then $P(c) \cap P(c') \neq \emptyset$, whence by Axiom~\ref{2-SEF_G.def:SEF}.\ref{2-SEF_G.def:SEF.P(c)} $P(c) = P(c')$ and $c \cap W_\omega = c' \cap W_\omega$. 
    
    If $x\supsetneq x'$, then Axiom~\ref{2-SEF_G.def:SEF}.\ref{2-SEF_G.def:SEF.enough_choices} implies the existence of $\tilde c\in A^i(x)$ such that $\tilde c \supseteq x'$. As $c\in A^i(x)$, Proposition~\ref{2-SEF_G.prop:information_sets}, Part~\ref{2-SEF_G.prop:information_sets.P(c)=P(c')} implies $P(c) = P(\tilde c)$. Moreover, $\tilde c \supseteq x' \supseteq c'\cap W_\omega$, whence
    \[ c\cap \tilde c \cap W_\omega \supseteq c \cap c' \cap W_\omega \neq \emptyset. \]
    Axiom~\ref{2-SEF_G.def:SEF}.\ref{2-SEF_G.def:SEF.P(c)} yields $c\cap W_\omega = \tilde c \cap W_\omega$. We conclude that $c\cap W_\omega \supseteq c'\cap W_\omega$.\smallskip

    (Ad exogenous case):~ Suppose that $i$ has perfect exogenous information. Then, for any $\x,\x'\in \X^i$ with $\x\ge_\X \x'$ and any $E\in{\ms F}_\x^i$, we have $E\in\ms E$ and, hence, $E\cap D_{\x'} \in \ms E|_{D_{\x'}} = {\ms F}^i_{\x'}$.    
\end{proof}

\begin{proof}
    [Proof of Lemma~\ref{2-SEF_G.lemma:Heraclitus_property}]
    The proof is completely analogous to the proofs of \cite[Proposition 13]{AlosFerrer2005Trees} and \cite[Proposition 4.1]{AlosFerrer2016Theory}. Nevertheless, we give a proof here, both because of the different formal setting and for the reader's convenience.\smallskip

    (Ad \ref{2-SEF_G.lemma:Heraclitus_property.X}):~ Let $x,x'\in X$ such that $A^i(x) \cap A^i(x') \neq \emptyset$ and $x\supseteq x'$. In particular, $i\in J(x) \cap J(x')$.
    If $x\neq x'$, then, by Axiom~\ref{2-SEF_G.def:SEF.enough_choices} there would be $c'\in A^i(x)$ such that $c'\supseteq x'$.    
    By Proposition~\ref{2-SEF_G.prop:information_sets}, Part~\ref{2-SEF_G.prop:information_sets.A(x)_partition}, $A^i(x) = A^i(x')$, and thus $c'\in A^i(x')$ as well. In other words, $x'\in P(c')$. Hence, by definition of the immediate predecessor operator, there would be $y'\in\downarrow c'$ such that
     \[ \uparrow x' = \uparrow y' \setminus \downarrow c'. \]
    As $x'\in\downarrow c'$, this is a contradiction. Hence, the assumption was false and we conclude that $x=x'$.\smallskip

    (Ad \ref{2-SEF_G.lemma:Heraclitus_property.rmX}):~ Let $\x,\x'\in \X$ such that $A^i(\x)\cap A^i(\x') \neq \emptyset$ and $\x\ge_\X \x'$. In particular, $\x,\x'\in \X^i$. 
    Then, there is $\omega\in D_{\x'}$, and we have $\x(\omega) \supseteq \x'(\omega)$ and $A^i(\x(\omega)) \cap A^i(\x'(\omega))\neq \emptyset$. By the statement of the first part, just proven before, we get $\x(\omega) = \x'(\omega)$. Since $\x,\x'\in\X^i$ and evaluation $\X^i\bullet \Omega\to X$ is injective, we obtain $\x = \x'$.
\end{proof}

\begin{proof}
    [Proof of Lemma~\ref{2-SEF_G.lemma:completeness}]
    Let $\F = (F,\pi,\X,I,\ms F,\ms C,C)$ be a tuple satisfying the conditions defining a stochastic pseudo-extensive form on some exogenous scenario space $(\Omega,\ms E)$ possibly except Axiom~\ref{2-SEF_G.def:SEF.choice_completeness}, according to Definition~\ref{2-SEF_G.def:SEF}. For any $i\in I$, let $\hat C^i$ be as defined in the lemma's statement.\smallskip
    
    (Helpful statement \ref{2-SEF_G.lemma:completeness.def:hatC.iii}):~ To start, let us prove that Properties~\ref{2-SEF_G.lemma:completeness}.\ref{2-SEF_G.lemma:completeness.def:hatC.i} and~\ref{2-SEF_G.lemma:completeness}.\ref{2-SEF_G.lemma:completeness.def:hatC.ii} can be strengthened. Namely:
    \begin{enumerate}[label=(iii)]
        \item\label{2-SEF_G.lemma:completeness.def:hatC.iii} For all $i\in I$, all $\hat c\in \hat C^i$ and $\omega\in\Omega$ with $\hat c \cap W_\omega \neq \emptyset$, there is $c\in C^i$ such that $\hat c\cap W_\omega = c \cap W_\omega$ and $P(\hat c) = P(c)$.
    \end{enumerate}
    For the proof, let $i\in I$, $\hat c\in \hat C^i$ and $\omega\in\Omega$ be such that $\hat c \cap W_\omega \neq \emptyset$. By definition of $\hat C^i$, Property~\ref{2-SEF_G.lemma:completeness.def:hatC.i}, there is $c \in C^i$ such that $\hat c \cap W_\omega = c \cap W_\omega$. By Lemma~\ref{1-SDF_AC.lemma:P(c)_compatible_with_conn_comp}, we have
    \[ P(c) \cap T_\omega = P(c\cap W_\omega) = P(\hat c \cap W_\omega) = P(\hat c) \cap T_\omega, \]
    and by non-redundancy of $\hat c$ these sets are non-empty. Let $\mf P^i$ the partition of $\X^i$ according to Remark~\ref{2-SEF_G.rmk:prop_information_sets} and Proposition~\ref{2-SEF_G.prop:information_sets}, Part~\ref{2-SEF_G.prop:information_sets.exists_mfP}. Then, there is $\mf p\in\mf P^i$ such that
    \[ P(c) \cap T_\omega = \{\x(\omega) \mid \x\in\mf p\colon~ \omega \in D_\x\}. \]
    As $\hat c$ is an $\X^i$-complete choice and $\omega\in D_\x = D_\mf p$ for all $\x\in\mf p$ by Remark~\ref{2-SEF_G.rmk:prop_information_sets} and Proposition~\ref{2-SEF_G.prop:information_sets}, Part~\ref{2-SEF_G.prop:information_sets.msC_msF_const_on_mfP}, we obtain that
    \[ P(\hat c) \supseteq \{\x(\omega') \mid \x\in\mf p, ~\omega'\in D_\x\} = P(c). \]
    By definition of $\hat C^i$, Property~\ref{2-SEF_G.lemma:completeness.def:hatC.ii}, there is $c_1\in C^i$ such that $P(\hat c) \subseteq P(c_1)$. Hence, $\emptyset \neq P(c) \subseteq P(c_1)$ and thus, by Axiom~\ref{2-SEF_G.def:SEF}.\ref{2-SEF_G.def:SEF.P(c)}, $P(c) = P(c_1)$, whence $P(c) = P(\hat c)$ which proves \ref{2-SEF_G.lemma:completeness.def:hatC.iii}.\smallskip
    
    (Ad Properties~\ref{2-SEF_G.lemma:completeness}.\ref{2-SEF_G.lemma:completeness.property_1} and~\ref{2-SEF_G.lemma:completeness}.\ref{2-SEF_G.lemma:completeness.property_2}):~
    Both properties follow from the trivial inclusion $C^i \subseteq \hat C^i$ for all $i\in I$ and Property~\ref{2-SEF_G.lemma:completeness.def:hatC.iii}. \smallskip

    (Ad basic properties in Definition~\ref{2-SEF_G.def:SEF})~ For each $i\in I$, $\hat C^i$ is a set of choices, by construction. 
    
    For the remainder of the proof, the operators and sets associated to $\hat C$ and $(F,\pi,\X)$ will be denoted with a hat on top. That is, for $\hat C$ and $(F,\pi,\X)$, denote the set of choices in $\hat C^i$ available at a move $x\in X$ or random move $\x\in\X$ by $\hat A^i(x)$ or $\hat A^i(\x)$, respectively. Furthermore, for $x\in X$ and $\x\in\X$, let
    \[ \hat J(x) = \{i\in I \mid \hat A^i(x) \neq\emptyset\}, \qquad \hat J(\x) = \{i\in I \mid \hat A^i(\x) \neq\emptyset\}, \]
    and, for $i\in I$, let
    \[ \hat X^i = \{x\in X \mid i\in \hat J(x)\}, \qquad \hat\X^i = \{\x\in\X \mid i\in \hat J(\x)\}, \qquad \hat\X^i\bullet \Omega = \{(\x,\omega)\in\hat\X^i\times\Omega \mid \omega\in D_\x\}. \]
    Then, clearly, all $x\in X$, $\x\in\X$, and $i\in I$ satisfy $A^i(x) \subseteq \hat A^i(x)$ and $A^i(\x) \subseteq \hat A^i(\x)$, $J(x) = \hat J(x)$ and $J(\x) = \hat J(\x)$. In particular, $X^i = \hat X^i$ and $\X^i = \hat\X^i$.

    As a consequence, $\ms F$ is a family of exogenous information structures on $\hat\X^i$, $i\in I$, and $\ms C$ is a family of reference choice structures on $\hat\X^i$, $i\in I$. By hypothesis, for any $i\in I$, any element of $\hat C^i$ is $\ms F^i$-$\ms C^i$-adapted, and the evaluation $\hat\X^i \bullet \Omega = \X^i \bullet \Omega \to X$ is injective.\smallskip   
    
    (Ad Axiom~\ref{2-SEF_G.def:SEF}.\ref{2-SEF_G.def:SEF.P(c)}):~ Let $i\in I$ and $\hat c,\hat c'\in \hat C^i$ such that $P(\hat c) \cap P(\hat c') \neq \emptyset$. Then, by Property~\ref{2-SEF_G.lemma:completeness.def:hatC.iii}, there are $c,c'\in C^i$ with $P(c) = P(\hat c)$ and $P(c') = P(\hat c')$. Hence, by Axiom~\ref{2-SEF_G.def:SEF}.\ref{2-SEF_G.def:SEF.P(c)} applied to $\F$, 
    \[ P(\hat c) = P(c) = P(c') = P(\hat c'). \]
    Furthermore, let $\omega\in\Omega$ such that $\hat c \cap \hat c' \cap W_\omega \neq \emptyset$. According to Property~\ref{2-SEF_G.lemma:completeness.def:hatC.iii}, $c$ and $c'$ can be chosen such that $\hat c \cap W_\omega = c \cap W_\omega$ and $\hat c' \cap W_\omega = c' \cap W_\omega$. In particular we get $c\cap c' \cap W_\omega \neq \emptyset$. As $P(c) = P(c')$, Axiom~\ref{2-SEF_G.def:SEF}.\ref{2-SEF_G.def:SEF.P(c)} applied to $\F$ yields 
    \[ \hat c \cap W_\omega = c\cap W_\omega = c' \cap W_\omega = \hat c'\cap W_\omega. \]

    (Ad Axiom~\ref{2-SEF_G.def:SEF}.\ref{2-SEF_G.def:SEF.outcomes_faithful}):~ Let $x\in X$ and $(\hat c^i)_{i\in J(x)} \in \bigtimes_{i\in J(x)} \hat C^i$. Let $\omega = \pi(x)$. Then, by definition of $\hat C$, there is $(c^i)_{i\in J(x)} \in \bigtimes_{i\in J(x)} C^i$ such that for all $i\in J(x)$ we have $\hat c^i \cap W_\omega = c^i \cap W_\omega$. As $x \in T_\omega$, we have $W_\omega \supseteq x$, whence
    \[ x \cap \bigcap_{i\in J(x)} \hat c^i = x \cap \bigcap_{i\in J(x)} (\hat c^i \cap W_\omega) = x \cap \bigcap_{i\in J(x)} (c^i \cap W_\omega) = x \cap \bigcap_{i\in J(x)} c^i \neq \emptyset, \]
    because $\F$ satisfies \ref{2-SEF_G.def:SEF}.\ref{2-SEF_G.def:SEF.outcomes_faithful}.

    (Ad Axiom~\ref{2-SEF_G.def:SEF}.\ref{2-SEF_G.def:SEF.weak_separation}):~ Let $y,y'\in F$ with $\pi(y) = \pi(y')$ and $y\cap y' = \emptyset$. Let $\omega = \pi(y)$. Then, by Axiom~\ref{2-SEF_G.def:SEF}.\ref{2-SEF_G.def:SEF.weak_separation} applied to $\F$, there are $i\in I$ and $c,c'\in C^i$ with $c \supseteq y$, $c' \supseteq y'$ and $c\cap c' \cap W_\omega = \emptyset$. By Property~\ref{2-SEF_G.lemma:completeness.def:hatC.iii} of $\hat C^i$, there are $\hat c, \hat c'\in \hat C^i$ such that $\hat c\cap W_\omega = c\cap W_\omega$ and $\hat c' \cap W_\omega = c'\cap W_\omega$ and $P(\hat c) = P(c)$ as well as $P(\hat c') = P(c')$. Hence,
    \[ \hat c \supseteq \hat c \cap W_\omega = c \cap W_\omega \supseteq y \cap W_\omega = y, \]
    and similarly, $\hat c' \supseteq y'$. Moreover, 
    \[ \hat c \cap \hat c' \cap W_\omega = c \cap c' \cap W_\omega = \emptyset. \]

    (Ad Axiom~\ref{2-SEF_G.def:SEF}.\ref{2-SEF_G.def:SEF.separation}):~ For the proof of this axiom's validity, suppose that $\F$ satisfies Axiom~\ref{2-SEF_G.def:SEF}.\ref{2-SEF_G.def:SEF.separation}. We ought to show that $\hat\F$ does as well. 
    
    Let $y,y'\in F$ with $\pi(y) = \pi(y')$ and $y\cap y' = \emptyset$. Let $\omega = \pi(y)$. Then, by Axiom~\ref{2-SEF_G.def:SEF}.\ref{2-SEF_G.def:SEF.separation} applied to $\F$, there are $x\in X$, $i\in I$ and $c,c'\in C^i$ with $x\cap c \supseteq y$, $x\cap c' \supseteq y'$, $c\cap c' \cap W_\omega = \emptyset$ and $x\in P(c) \cap P(c) \cap T_\omega$. By Property~\ref{2-SEF_G.lemma:completeness.def:hatC.iii} of $\hat C^i$, there are $\hat c, \hat c'\in \hat C^i$ such that $\hat c\cap W_\omega = c\cap W_\omega$ and $\hat c' \cap W_\omega = c'\cap W_\omega$, and $P(\hat c) = P(c)$ as well as $P(\hat c') = P(c')$. Hence,
    \[ x\cap \hat c = x \cap \hat c \cap W_\omega = x \cap c \cap W_\omega = x \cap c \supseteq y, \]
    and similarly, $x \cap \hat c' \supseteq y'$. Moreover, 
    \[ \hat c \cap \hat c' \cap W_\omega = c \cap c' \cap W_\omega = \emptyset, \]
    and $x \in P(c) \cap P(c') \cap T_\omega = P(\hat c) \cap P(\hat c') \cap T_\omega$.\smallskip

    (Ad Axiom~\ref{2-SEF_G.def:SEF}.\ref{2-SEF_G.def:SEF.enough_choices}):~ Let $x\in X$, $i\in J(x)$ and $y\in \downarrow x \setminus \{x\}$. By Axiom~\ref{2-SEF_G.def:SEF}.\ref{2-SEF_G.def:SEF.enough_choices} applied to $\F$, we obtain $c\in A^i(x)$ with $c\supseteq y$. Let $\omega = \pi(x)$. Property~\ref{2-SEF_G.lemma:completeness.def:hatC.iii} of $\hat C$ ensures the existence of $\hat c\in\hat C^i$ such that $\hat c \cap W_\omega = c\cap W_\omega$ and $P(\hat c) = P(c)$. Hence, $x\in P(\hat c)$ alias $\hat c\in \hat A^i(x)$. Moreover, 
    \[ \hat c \cap W = c \cap W_\omega \supseteq y \cap W_\omega = y. \]

    (Ad Axiom~\ref{2-SEF_G.def:SEF}.\ref{2-SEF_G.def:SEF.endo_exo_compatible}):~ Let $i\in I$ and $\x,\x'\in\X$ such that $\hat A^i(\x) \cap \hat A^i(\x')\neq\emptyset$. Hence, there is $\hat c\in \hat C^i$ such that $\x(\omega),\x'(\omega') \in P(\hat c)$ for some (and any) $\omega\in D_\x$ and $\omega'\in D_{\x'}$. Following Property~\ref{2-SEF_G.lemma:completeness.def:hatC.iii} of $\hat C$, there is $c\in C^i$ such that $P(\hat c) = P(c)$. Hence, $c\in A^i(\x) \cap A^i(\x')$, whence $\ms F^i_\x = \ms F^i_{\x'}$ and $\ms C^i_\x = \ms C^i_{\x'}$. \smallskip

    (Ad Axiom~\ref{2-SEF_G.def:SEF}.\ref{2-SEF_G.def:SEF.choice_completeness}):~ Let $i\in I$ and $\hat c'$ an $\ms F^i$-$\ms C^i$-adapted choice satisfying Properties~\ref{2-SEF_G.def:SEF}.\ref{2-SEF_G.def:SEF.choice_completeness.i} and~\ref{2-SEF_G.def:SEF}.\ref{2-SEF_G.def:SEF.choice_completeness.ii}, that is:
    \begin{enumerate}[label=(\roman*)]
        \item any $\omega\in\Omega$ with $\hat c'\cap W_\omega \neq \emptyset$ admits $\hat c\in \hat C^i$ with $\hat c'\cap W_\omega = \hat c \cap W_\omega$,
        \item and there is $\hat c\in \hat C^i$ with $P(\hat c') = P(\hat c)$,
    \end{enumerate}
    Let $\omega\in \Omega$ be such that $\hat c'\cap W_\omega$. Take $\hat c\in\hat C^i$ with $\hat c'\cap W_\omega = \hat c \cap W_\omega$. By definition of $\hat C$, there is $c\in C^i$ with $\hat c\cap W_\omega = c\cap W_\omega$, whence $\hat c' \cap W_\omega = c\cap W_\omega$. 

    Further, take $\hat c\in\hat C^i$ such that $P(\hat c') = P(\hat c)$. By Property~\ref{2-SEF_G.lemma:completeness.def:hatC.iii} of $\hat C$ there is $c\in C^i$ such that $P(\hat c) = P(c)$, implying $P(\hat c') \subseteq P(c)$. 

    We conclude that $\hat c'\in\hat C^i$ which shows that the axiom is satisfied by $\F'$.
\end{proof}

\begin{proof}
    [Proof of Proposition~\ref{2-SEF_G.prop:strategies}]
    Let $i\in I$. Denote the surjections $X^i \surj \X^i$ and $\X^i \to \mf P^i$ by $p_{X,\X;i}$ and $p_{\X,\mf P;i}$ respectively. By Proposition~\ref{2-SEF_G.prop:information_sets}, we have for all $x\in X^i$ and $\x\in\X^i$:
    \[ A^i(x) = A^i(p_{X,\X;i}(x)), \qquad A^i(\x) = A^i(p_{\X,\mf P;i}(\x)). \]
    Moreover, for all $x,x'\in X^i$ we have
    \[ (p_{\X,\mf P;i}\circ p_{X,\X;i})(x) = (p_{\X,\mf P;i}\circ p_{X,\X;i})(x') \qquad \Longleftrightarrow \qquad A^i(x) = A^i(x'), \]
    and similarly, for all $\x,\x'\in \X^i$ we have
    \[ p_{\X,\mf P;i}(\x) = p_{\X,\mf P;i}(\x') \qquad \Longleftrightarrow \qquad A^i(\x) = A^i(\x'). \]
    The claim follows from this using the universal property of the quotient in the category of sets.
\end{proof}

\begin{proof}
    [Proof of Lemma~\ref{2-SEF_G.lemma:simple_sef1}]
    Let $(F,\pi,\X,I,\ms F,\ms C,C)$ be a tuple satisfying the hypothesis of the lemma. Let $i$ be the unique element of $I$. Recall the table defining $C$.
    
    According to Lemma~\ref{1-SDF_AC.lemma:simple_sdf1}, $(F,\pi,\X)$ defines a stochastic decision forest on $(\Omega,\ms E)$. By Lemma~\ref{1-SDF_AC.lemma:simple_sdf1_EIS}, $\ms F^i$ defines an exogenous information structure for it. By Lemma~\ref{1-SDF_AC.lemma:simple_sdf1_RCS}, $\ms C^i$ defines a reference choice structure for it, and by Lemma~\ref{1-SDF_AC.lemma:simple_sdf1_AC}, the elements of $C^i$ are $\ms F^i$-$\ms C^i$-adapted choices.

    (Ad Axiom~\ref{2-SEF_G.def:SEF}.\ref{2-SEF_G.def:SEF.P(c)}):~ The immediate predecessor sets of all choices in $C^i$ have been explicitly calculated in Lemma~\ref{1-SDF_AC.lemma:simple_sdf1_choices}. It follows from the results of that lemma that for all $c,c'\in C^i$, $P(c)$ and $P(c')$ can only non-trivially intersect if $c$ and $c'$ are denoted in the same column of the table. Given this, closer inspection of the table reveals that if $P(c)$ and $P(c')$ non-trivially intersect, then $P(c) = P(c')$, and $c\cap W_\omega$ and $c'\cap W_\omega$ are either disjoint or equal, for all $\omega\in\Omega$.\smallskip

    (Ad Axiom~\ref{2-SEF_G.def:SEF}.\ref{2-SEF_G.def:SEF.outcomes_faithful}):~ There is only one agent $i\in I$ here, hence this axiom is trivially satisfied. Indeed, if $x\in X$ and $c\in C^i$ such that $i\in J(x)$, then $x\in P(c)$. Hence, there is $y\in \downarrow c$ such that $\uparrow x = \uparrow y \setminus \downarrow c$. Hence, $x\cap c \supseteq y \cap c = y \neq \emptyset$.\smallskip

    (Ad Axiom~\ref{2-SEF_G.def:SEF}.\ref{2-SEF_G.def:SEF.separation}):~ Let $y,y'\in T_\omega$ for some $\omega\in\Omega$ with $y\cap y' = \emptyset$. Then, $y$ and $y'$ are moves at time $1$ or terminal nodes. As $T_\omega$ is a finite tree, there is a $\supseteq$-minimal $x\in F$ with $x\supseteq y\cup y'$. In particular, $x$ must be a move with $\pi(x) = \omega$. Let $i\in I$. For any possible values of $x$ and $y$, there are $c,c'\in A^i(x)$ such that $x\cap c \supseteq y$ and $x\cap c' \supseteq y'$, as evident from the table and Lemma~\ref{1-SDF_AC.lemma:simple_sdf1_choices}.\smallskip

    (Ad Axiom~\ref{2-SEF_G.def:SEF}.\ref{2-SEF_G.def:SEF.enough_choices}):~ Let $x\in X$, $i\in J(x)$, and $y\in\downarrow x \setminus \{x\}$. As evidenced by the table and Lemma~\ref{1-SDF_AC.lemma:simple_sdf1_choices}, there is $c\in A^i(x)$ such that $c \supseteq y$.\smallskip

    (Ad Axiom~\ref{2-SEF_G.def:SEF}.\ref{2-SEF_G.def:SEF.endo_exo_compatible}):~ Upon consulting the table, we infer using Lemma~\ref{1-SDF_AC.lemma:simple_sdf1_choices} that $\x_1$ and $\x_2$ are the only two random moves sharing an available choice $c$, and this only in the cases given by th second, fourth, and eighth line of the table. In these cases, we have indeed $\ms F^i_{\x_1} = \ms F^i_{\x_2}$ (\textsc{eis} = 1, 2(a), or 3). We also have $\ms C^i_{\x_1} = \ms C^i_{\x_2}$.\smallskip

    (Ad Axiom~\ref{2-SEF_G.def:SEF}.\ref{2-SEF_G.def:SEF.choice_completeness}):~ Let $c'$ be an $\ms F^i$-$\ms C^i$-adapted choice satisfying \ref{2-SEF_G.def:SEF}.\ref{2-SEF_G.def:SEF.choice_completeness.i} and~\ref{2-SEF_G.def:SEF}.\ref{2-SEF_G.def:SEF.choice_completeness.ii}. From the latter two properties and Lemma~\ref{1-SDF_AC.lemma:simple_sdf1_choices}, we infer that 1) $P(c') = \im \x_0$ and $c' = c_{f\bullet}$ for some $f\in M$, or 2a) we are in line two, four, or eight and $P(c') = \im\x_1 \cup \im\x_2$ and $c' = c_{\bullet g}$ for some $g\in M$, or 2b) we are in line one, three, five, six, or seven and $P(c') \in \{\im\x_1,\im\x_2\}$ and $c' = c_{kg}$ for some $k=1,2$ and $g\in M$.
    Given this, the $\ms F^i$-$\ms C^i$-adaptedness implies that $c'$ has to be one of the entries in the given line of the table, thus an element of $C^i$.
\end{proof}
    
\begin{proof}
    [Proof of Lemma~\ref{2-SEF_G.lemma:simple_sef2}]
    The proof of this lemma is highly analogous to that of Lemma~\ref{2-SEF_G.lemma:simple_sef1} just above. Let $(F',\pi',\X',I',\ms F',\ms C',C')$ be a tuple satisfying the hypothesis of the lemma. Let $i$ be the unique element of $I'$. Recall the table defining $C'$.
    
    According to Lemma~\ref{1-SDF_AC.lemma:simple_sdf1}, $(F',\pi',\X')$ defines a stochastic decision forest on $(\Omega,\ms E)$. By Lemma~\ref{1-SDF_AC.lemma:simple_sdf1_EIS}, $\ms F^{\prime i}$ defines an exogenous information structure for it. By Lemma~\ref{1-SDF_AC.lemma:simple_sdf1_RCS}, $\ms C^{\prime i}$ defines a reference choice structure for it, and by Lemma~\ref{1-SDF_AC.lemma:simple_sdf1_AC}, the elements of $C^{\prime i}$ are $\ms F^{\prime i}$-$\ms C^{\prime i}$-adapted choices.

    (Ad Axiom~\ref{2-SEF_G.def:SEF}.\ref{2-SEF_G.def:SEF.P(c)}):~ The immediate predecessor sets of all choices in $C^{\prime i}$ have been explicitly calculated in Lemma~\ref{1-SDF_AC.lemma:simple_sdf2_choices}. It follows from the results of that lemma that for all $c,c'\in C^{\prime i}$, $P(c)$ and $P(c')$ can only non-trivially intersect if $c$ and $c'$ are denoted in the same column of the table. Given this, closer inspection of the table reveals that if $P(c)$ and $P(c')$ non-trivially intersect, then $P(c) = P(c')$, and $c\cap W_\omega$ and $c'\cap W_\omega$ are either disjoint or equal, for all $\omega\in\Omega$.\smallskip

    (Ad Axiom~\ref{2-SEF_G.def:SEF}.\ref{2-SEF_G.def:SEF.outcomes_faithful}):~ There is only one agent $i\in I'$ here, hence this axiom is trivially satisfied. Exactly the same abstract argument can be used as in the proof of Lemma~\ref{2-SEF_G.lemma:simple_sef1}, Part~\ref{2-SEF_G.def:SEF}.\ref{2-SEF_G.def:SEF.outcomes_faithful}, just above.\smallskip

    (Ad Axiom~\ref{2-SEF_G.def:SEF}.\ref{2-SEF_G.def:SEF.separation}):~ Let $y,y'\in T'_\omega$ for some $\omega\in\Omega$ with $y\cap y' = \emptyset$. Then, $y$ and $y'$ are moves at time $1$ or terminal nodes. As $T_\omega$ is a finite tree, there is a $\supseteq$-minimal $x\in F$ with $x\supseteq y\cup y'$. In particular, $x$ must be a move with $\pi(x) = \omega$. Let $i\in I$. For any possible values of $x$ and $y$, there are $c,c'\in A^i(x)$ such that $x\cap c \supseteq y$ and $x\cap c' \supseteq y'$, as evident from the table and Lemma~\ref{1-SDF_AC.lemma:simple_sdf2_choices}.\smallskip

    (Ad Axiom~\ref{2-SEF_G.def:SEF}.\ref{2-SEF_G.def:SEF.enough_choices}):~ Let $x\in X$, $i\in J(x)$, and $y\in\downarrow x \setminus \{x\}$. As evidenced by the table and Lemma~\ref{1-SDF_AC.lemma:simple_sdf2_choices}, there is $c\in A^i(x)$ such that $c \supseteq y$.\smallskip

    (Ad Axiom~\ref{2-SEF_G.def:SEF}.\ref{2-SEF_G.def:SEF.endo_exo_compatible}):~ Upon consulting the table, we infer using Lemma~\ref{1-SDF_AC.lemma:simple_sdf2_choices} that there are no two random moves sharing an available choice $c$. Hence, this axiom is trivially satisfied.\smallskip

    (Ad Axiom~\ref{2-SEF_G.def:SEF}.\ref{2-SEF_G.def:SEF.choice_completeness}):~ Let $c'$ be an $\ms F^{\prime i}$-$\ms C^{\prime i}$-adapted choice satisfying \ref{2-SEF_G.def:SEF}.\ref{2-SEF_G.def:SEF.choice_completeness.i} and~\ref{2-SEF_G.def:SEF}.\ref{2-SEF_G.def:SEF.choice_completeness.ii}. From the latter two properties and Lemma~\ref{1-SDF_AC.lemma:simple_sdf2_choices}, we infer that 1) $P(c') = \im \x'_0$ and $c' = c'_{f\bullet}$ for some $f\in M$, or 2) $P(c') \in \{\im\x'_1,\im\x'_2\}$ and $c' = c'_{kg}$ for some $k=1,2$ and $g\in M$.
    Given this, the $\ms F^{\prime i}$-$\ms C^{\prime i}$-adaptedness implies that $c'$ has to be one of the entries in the given line of the table, thus an element of $C^{\prime i}$.
\end{proof}

\begin{proof}[Proof of Theorem~\ref{2-SEF_G.thm:absent_minded_driver_Gilboa_sef}]
    It has been demonstrated in the ``Simple examples'' subsections in Chapter~\ref{chap:1-SDF_AC} that $(F,\pi,\X)$ is a stochastic decision forest, that $\ms F^i$ is an exogenous information structure, that $\ms C^i$ is a reference choice structure, and that $C^i$ is a set of $\ms F^i$-$\ms C^i$-adapted choices, for both $i\in I$.\smallskip

    (Ad Axiom~\ref{2-SEF_G.def:SEF}.\ref{2-SEF_G.def:SEF.P(c)}):~ Recall that, for all $i\in I=\{1,2\}$, $E\in\ms F^i_{\x_i}$, we have
    \[ P(c_i(E)) = \im \x_i, \]
    and $\{\im \x_1, \im\x_2\}$ defines a (though not ``order consistent'') partition of $X$.
    Hence, for $c,c'\in C^i$, $P(c)$ and $P(c')$ can only non-trivially intersect if $c$ and $c'$ are available at the same random move, and in that case, $P(c) = P(c')$. If this is the case, then for all $\omega\in\Omega$, $c\cap W_\omega$ and $c'\cap W_\omega$ can both only equal ``exit'' or ``continue'' in scenario $\omega$, i.e.\ $\op{Ex}_i \cap W_\omega$ or $\op{Ct}_i \cap W_\omega$, and thus are either equal or disjoint.\smallskip

    (Ad Axiom~\ref{2-SEF_G.def:SEF}.\ref{2-SEF_G.def:SEF.outcomes_faithful}):~ At any move, there is exactly one active agent $i\in I$, hence this axiom is trivially satisfied. Exactly the same abstract argument can be used as in the proof of Lemma~\ref{2-SEF_G.lemma:simple_sef2}, Part~\ref{2-SEF_G.def:SEF}.\ref{2-SEF_G.def:SEF.outcomes_faithful}, just above.\smallskip

    (Ad Axiom~\ref{2-SEF_G.def:SEF}.\ref{2-SEF_G.def:SEF.separation}):~ Let $y,y'\in T_\omega$ for some $\omega\in\Omega$ with $y\cap y' = \emptyset$. Then, $\{(\omega,D)\}\in\{y,y'\}$, or, $\{\{(\omega,H)\},\{(\omega,M)\}\} = \{y,y'\}$. In the first case, take $x = \x_{\rho(\omega)}(\omega)$, and in the second, take $x=\x_{3-\rho(\omega)}(\omega)$. In both cases, we can take the two choices $c$ and $c'$ to ``continue'' and to ``exit'', or conversely, that are available at $x$, i.e.\ $x\in P(c)\cap P(c') \cap W_\omega$, such that $c\cap x \supseteq y$ and $c'\cap x \supseteq y'$. By construction, $c\cap c' = \emptyset$.\smallskip

    (Ad Axiom~\ref{2-SEF_G.def:SEF}.\ref{2-SEF_G.def:SEF.enough_choices}):~ Let $x\in X$, $i\in J(x)$, and $y\in\downarrow x \setminus \{x\}$. Then $x=\x_i(\omega)$, for some $\omega\in\Omega$. If $\rho(\omega) = i$, $y$ can be any terminal node in $T_\omega$; else, $y\in\{\{(\omega,H)\},\{(\omega,M)\}\}$. By appropriately choosing $c$ to be either ``continue'' or ``exit'', we clearly get $c\supseteq y$ in any of these cases.\smallskip

    (Ad Axiom~\ref{2-SEF_G.def:SEF}.\ref{2-SEF_G.def:SEF.endo_exo_compatible}):~ As $\X^1$ and $\X^2$ are singletons, this condition is trivially satisfied.\smallskip

    (Ad Axiom~\ref{2-SEF_G.def:SEF}.\ref{2-SEF_G.def:SEF.choice_completeness}):~ Let $i\in I=\{1,2\}$ and $c'$ be an $\ms F^{i}$-$\ms C^{i}$-adapted choice satisfying \ref{2-SEF_G.def:SEF}.\ref{2-SEF_G.def:SEF.choice_completeness.i} and~\ref{2-SEF_G.def:SEF}.\ref{2-SEF_G.def:SEF.choice_completeness.ii}. We infer that $P(c') = \im\x_i$ from the latter one. By the first condition, $c'$ must consist in ```continue'' or ``exit'' in every scenario. By adaptedness, and the definition of $\ms C^i$, both ``continue'' and ``exit'' must be chosen on $\ms F^i_{\x_i}$-measurable events. Hence, $c' = c_i(E)$ for some $E\in\ms F_{\x_i}^i$. Thus, $c'\in C^i$.
\end{proof}

\subsection{Section~\ref{2-SEF_G.sec:AP_SEF}}\label{2-SEF_G.subsec:appendix.proofs.2}

As a preparation for proving Theorem~\ref{2-SEF_G.thm:AP_sef}, we consider the following lemmata. The first lemma relates the domain $D$ of the ``random action'' $g$ to the domain of those random moves that the corresponding choice is available at in view of the second lemma. 

\begin{lemma}\label{2-SEF_G.lemma:AP_sef_compute_D}
    Let $\D$ be action path $\psi$-\textsc{sef} data on an exogenous scenario space $(\Omega,\ms E)$, $i\in I$, and $c\in C^i$. Let $t\in\T$, $A_{<t} \in \mc H_t^i$, $D\in\ms E$, $g\colon D\to\A^i$ such that $c = c(A_{<t},i,g)$. Then, for any $f\in \A^\T$ with $f|_{[0,t)_\T}\in A_{<t}$, we have
    \[ D = D_{t,f}. \]
\end{lemma}

\begin{proof}
    Let $f\in\A^\T$ be such that $f|_{[0,t)_\T}\in A_{<t}$. 

    Let $\omega\in D$. By the definition of $C^i$ and the fact that $c\in C^i$, there is $\tilde f\in\A^\T$ such that $(\omega,\tilde f)\in c$ and $\tilde f|_{[0,t)_\T} = f|_{[0,t)_\T}$. As $c\in\ms C_t$, Assumption \hyperlink{2-SEF_G.Ass:AP.C1}{AP.C1} implies that $\omega\in D_{t,\tilde f} = D_{t,f}$. We conclude that $D\subseteq D_{t,f}$. 
    
    As $c\neq\emptyset$, according to Assumption \hyperlink{2-SEF_G.Ass:AP.C0}{AP.C0}, we infer that $D\neq\emptyset$. Let $\omega\in D$ as above, whence $\omega \in D_{t,f}$. Then, as above,
    \[ (\omega,\tilde f) \in x_t(\omega,f) \cap c. \]
    By Assumption \hyperlink{2-SEF_G.Ass:AP.C2}{AP.C2}, all $\omega'\in D_{t,f}$ satisfy $x_t(\omega',f) \cap c\neq \emptyset$ and must therefore be elements of $D$. Hence, $D_{t,f} \subseteq D$.
\end{proof}

\begin{lemma}\label{2-SEF_G.lemma:AP_sef_compute_P(c)}
    Let $\D$ be action path $\psi$-\textsc{sef} data on an exogenous scenario space $(\Omega,\ms E)$, $i\in I$, and $c\in C^i$. Let $t\in\T$, $A_{<t} \in \mc H_t^i$, $D\in\ms E$, $g\colon D\to \A^i$ be such that $c = c(A_{<t},i,g)$. Then,
    \begin{equation*}
    \begin{aligned}
        P(c) =&~\{ x_t(\omega,f) \mid (\omega,f)\in D \times \A^\T\colon~ f|_{[0,t)_\T} \in A_{<t}\}\\
        =&~\{ x_t(\omega,f) \mid (\omega,f)\in \Omega \times \A^\T\colon~ \omega\in D_{t,f},~ f|_{[0,t)_\T} \in A_{<t}\}.
    \end{aligned}
    \end{equation*} 
\end{lemma}

\begin{proof}
    It suffices to prove the first equality because the second one follows from it in view of Lemma~\ref{2-SEF_G.lemma:AP_sef_compute_D}.

    As $c\in\ms C_t$ by assumption, we have 
    \[ P(c) = \{x_t(w) \mid w \in c\}, \]
    by Lemma~\ref{1-SDF_AC.lemma:AP_P(c)}. Now, let $w = (\omega,f)\in \Omega\times \A^\T$. Then, by the definition of $c$, $w\in c$ implies that $f|_{[0,t)_\T} \in A_{<t}$ and $\omega\in D$. Conversely, if $f|_{[0,t)_\T} \in A_{<t}$ and $\omega\in D$, then there is $\tilde f\in\A^\T$ such that $\tilde w = (\omega,\tilde f)\in c$ and $\tilde f|_{[0,t)_\T} = f|_{[0,t)_\T}$ since $c\in C^i$. Hence, $x_t(w) = x_t(\tilde w) \in P(c)$.
\end{proof}

\begin{proof}
    [Proof of Proposition~\ref{2-SEF_G.prop:Dfti}]
    (Ad \ref{2-SEF_G.prop:Dfti.hatDft=unionDfti}):~ This follows directly from the fact that for $a',a''\in\A$ we have $a'=a''$ iff for all $i\in I$, $p^i(a') = p^i(a'')$.\smallskip

    (Ad \ref{2-SEF_G.prop:Dfti.hatDft_subseteq_Dft}):~ Let $t\in\T$, $f\in\A^\T$ and $\omega\in \hat D_{t,f}$. Then there are  $f',f'' \in\A^\T$ such that $(\omega,f'), (\omega,f'')\in W$, $f'|_{[0,t)_\T} = f|_{[0,t)_\T} = f''|_{[0,t)_\T}$ and $f'(t) \neq f''(t)$. Hence, $(\omega,f')$ and $(\omega,f'')$ provide two distinct elements of $x_t(\omega,f)$, whence $\omega\in D_{t,f}$.\smallskip

    (Ad \ref{2-SEF_G.prop:Dfti.Dfti_nonempty}):~ Let $t\in\T$, $f\in\A^\T$ and $i\in I$. Then the assertion that $D_{t,f} \neq \emptyset$ and $\x_t(f) \in \X^i$ hold true is equivalent to the one that there are $\omega\in \Omega$ and $c\in C^i$ such that $x_t(\omega,f)\in P(c)$.

    To show the implication ``$\Rightarrow$'', suppose that there are such $\omega$ and $c$. Then, by Lemmata \ref{2-SEF_G.lemma:AP_sef_compute_P(c)} and~\ref{2-SEF_G.lemma:AP_sef_compute_D}, we can represent $c$ using $A_{<t}\in\mc H^i_t$ and $g\colon D_{t,f} \to \A^i$ as $c = c(A_{<t},i,g)$. As $c\in\ms C_t$ by hypothesis, by Assumption \hyperlink{2-SEF_G.Ass:AP.C1}{AP.C1}, there is $f'\in\A^\T$ such that $(\omega,f')\in W$, $f'|_{[0,t)_\T} = f|_{[0,t)_\T}$ and $p^i \circ f'(t) \neq g(\omega)$. On the other hand, by Lemma~\ref{2-SEF_G.lemma:AP_sef_compute_P(c)}, the fact that $x_t(\omega,f)\in P(c)$ implies that there is $f''\in\A^\T$ satisfying $(\omega,f'')\in W$, $f''|_{[0,t)_\T} = f|_{[0,t)_\T}$ and $p^i \circ f''(t) = g(\omega)$. Hence, $\omega\in D_{t,f}^i$.

    As the preceding argument can be made for any $\omega\in D_{t,f}$ it follows that under the assumption ``$D_{t,f}\neq\emptyset$ and $\x_t(f)\in \X^i$'', we have $D_{t,f} \subseteq D_{t,f}^i$. In view of Parts~\ref{2-SEF_G.prop:Dfti.hatDft=unionDfti} and~\ref{2-SEF_G.prop:Dfti.hatDft_subseteq_Dft}, we obtain $D_{t,f} = D_{t,f}^i$.

    To show the other implication ``$\Leftarrow$'', suppose that there is $\omega\in D_{t,f}^i$. Hence, there is $f'\in\A^\T$ with $f'|_{[0,t)_\T}=f|_{[0,t)_\T}$ and $(\omega,f')\in W$. Hence, $\omega\in D_{t,f}^i = D_{t,f'}^i$. Since $\omega\in D_{t,f'}^i \subseteq D_{t,f'}$, Assumption \hyperlink{2-SEF_G.Ass:AP.SEF2}{AP.SEF2} combined with Lemma~\ref{2-SEF_G.lemma:AP_sef_compute_P(c)} directly yields the existence of $c\in C^i$ with $x_t(\omega,f) = x_t(\omega,f')\in P(c)$.
\end{proof}

\begin{example}\label{2-SEF_G.ex:SEF_with_Dtf_neq_hatDtf}
    Consider singleton $\Omega$, $\T = \R_+$ with standard order, singleton $I$, let $\A = \R$ and $W$ be the set of pairs $(\omega,f)\in\Omega\times\A^\T$ such that $f(0) = 0$ and $f$ is right-constant, that is, for all $t\in\T$, there is $\e>0$ such that $f|_{[t,t+\e)}$ is single-valued. Then, $(I,\A,\T,W)$ clearly satisfies the Assumptions \hyperlink{2-SEF_G.Ass:AP.SDF0}{AP.SDF$k$}, $k=0,\dots,3$.
    
    Further, let $i\in I$, $\ms F^i_\x = \ms E = \mc P\Omega$ for all $\x\in\tilde\X^i$, $\ms F^i = (\ms F^i_\x)_{\x\in\tilde\X^i}$, $\ms F=(\ms F^i)_{i\in I}$, and $\mc H^i_t = \big\{\{f\in \A^{[0,t)} \mid D_{t,f}^i\neq \emptyset\}\big\}$, for the unique action index $i\in I$ and all $t\in\T$, $\mc H^i = (\mc H^i_t)_{t\in\T}$, $\mc H = (\mc H^i)_{i\in I}$. Then, $(I,\A,\T,W,\ms F,\mc H)$ clearly satisfies the Assumptions \hyperlink{2-SEF_G.Ass:AP.SEF0}{AP.SEF$k$}, $k=0,\dots,3$.
    
    However, we clearly have $D_{0,f} = \Omega \neq \emptyset = \hat D_{0,f}$, for all $f\in\A^\T$.
\end{example}

\begin{proof}
    [Proof of Theorem~\ref{2-SEF_G.thm:AP_sef}]
    Let $\D$ action path $\psi$-\textsc{sef} data on an exogenous scenario space $(\Omega,\ms E)$, and let $\F$ be the induced \textsc{sef} candidate.\smallskip
    
    (Ad basic properties in Definition~\ref{2-SEF_G.def:SEF}):~ According to Theorem~\ref{1-SDF_AC.thm:AP_sdf}, $(F,\pi,\X)$ defines an order consistent stochastic decision forest on $(\Omega,\ms E)$. Moreover, by Corollary \ref{2-SEF_G.cor:tildeXi=Xi}, for all $i\in I$, $\tilde\X^i = \X^i$. Hence, $\ms F$ defines a family of exogenous information structures on $\X^i$, $i\in I$, and, by Proposition~\ref{1-SDF_AC.prop:APsdf_RCS}, $\ms C$ defines a family of reference choice structures on $\X^i$, $i\in I$. By construction, for each $i\in I$, the elements of $C^i$ are $\ms F^i$-$\ms C^i$-adapted choices. Moreover, by order consistency and \ref{1-SDF_AC.prop:ev_on_Tr_is_iso}, evaluation $\X^i\bullet\Omega\to X$ is injective.\smallskip

    (Ad Axiom~\ref{2-SEF_G.def:SEF}.\ref{2-SEF_G.def:SEF.P(c)}):~ Let $i\in I$ and $c,c'\in C^i$ such that $P(c) \cap P(c') \neq \emptyset$. Represent $c$ and $c'$ using $t,t'\in\T$, $A_{<t}\in \mc H^i_t$, $A'_{<t'}\in\mc H^i_{t'}$, $D,D'\in\ms E$, $g\colon D\to \A^i$ and $g'\colon D'\to \A^i$ such that
    \[ c = c(A_{<t},i,g), \qquad c' = c(A'_{<t'},i,g').\]
    In view of Lemma~\ref{2-SEF_G.lemma:AP_sef_compute_P(c)}, there are $(\omega,f)\in D\times\A^\T$ and $(\omega',f')\in D'\times\A^\T$ such that $f|_{[0,t)_\T} \in A_{<t}$, $f'|_{[0,t')_\T} \in A'_{<t'}$, $p^i\circ f(t) = g(\omega)$, $p^i\circ f'(t) = g'(\omega')$, and
    \[ x_t(\omega,f) = x_{t'}(\omega',f'). \]
    By applying $\pi$, we obtain $\omega = \omega'$. By applying the ``time'' map $\mf t$, we obtain $t = t'$. By the definition of nodes in action path \textsc{sdf}, we get $f|_{[0,t)_\T} = f'|_{[0,t)_\T}$. Hence, $A_{<t} \cap A'_{<t} \neq\emptyset$. By Assumption \hyperlink{2-SEF_G.Ass:H1}{AP.H1}, $A_{<t} = A'_{<t}$. By Lemma~\ref{2-SEF_G.lemma:AP_sef_compute_D} we get $D=D'$, and by Lemma~\ref{2-SEF_G.lemma:AP_sef_compute_P(c)}, we obtain $P(c) = P(c')$. 

    We remain in the situation where $P(c) = P(c')$ and consider $\omega\in\Omega$ such that $c\cap c'\cap W_\omega \neq \emptyset$. Then, $\omega\in D$ and $g(\omega) = g'(\omega)$, which implies $c\cap W_\omega = c'\cap W_\omega$.\smallskip

    (Ad Axiom~\ref{2-SEF_G.def:SEF}.\ref{2-SEF_G.def:SEF.outcomes_faithful}):~ Let $x\in X$ and $(c^i)_{i\in J(x)} \in \bigtimes_{i\in J(x)} C^i$. Represent $x$ using $t = \mf t(x) \in \T$ and $w = (\omega,f)\in W$, as $x = x_t(w)$. For any $i\in J(x)$, represent $c^i$ using $A^i_{<t}\in\mc H^i_t$, $D^i\in\ms E$, and $g^i\colon D^i\to \A^i$, as $c = c(A^i_{<t},i,g^i)$. 

    For all $i\in J(x)$, we have $x\in P(c^i)$. By Lemma~\ref{2-SEF_G.lemma:AP_sef_compute_P(c)}, this implies that, for all $i\in J(x)$, $(\omega,f|_{[0,t)_\T})\in D^i\times A^i_{<t}$. Hence, for any $i\in J(x)$ there is $f_i\in\A^\T$ with $(\omega,f_i)\in c^i$ and $f_i|_{[0,t)_\T} = f|_{[0,t)_\T}\in A^i_{<t}$.
    
    Then, by Assumption \hyperlink{2-SEF_G.Ass:AP.SEF0}{AP.SEF0} there is $\tilde f\in\A^\T$ such that for all $i\in J(x)$, $p^i\circ \tilde f(t) = p^i \circ f_i(t) = g^i(\omega)$ and $(\omega,\tilde f)\in x_t(\omega,f_i) = x_t(\omega,f)$. We conclude that
    \[ (\omega,\tilde f) \in x \cap \bigcap_{i\in J(x)} c^i. \]

    (Ad Axiom~\ref{2-SEF_G.def:SEF}.\ref{2-SEF_G.def:SEF.weak_separation}):~ Let $y,y'\in F$ with $y\cap y' = \emptyset$ and $\pi(y) = \pi(y')$. Denote $\omega = \pi(y)$. There are $f,f'\in\A^\T$ and $t_0\in\T$ such that a) $(\omega,f)\in y$, $(\omega,f')\in y'$, and $f(t_0)\neq f'(t_0)$ and b) $t_0 < \mf t(y) \wedge \mf t(y')$.
    
    By Assumption \hyperlink{2-SEF_G.Ass:AP.psi-SEF3}{AP.$\psi$-SEF3}, there are $t\in\T$ and $i\in I$ such that $p^i \circ f(t) \neq p^i \circ f'(t)$, $\omega\in D_{t,f}^i \cap D_{t,f'}^i$, and $t\le t_0$. There is a unique pair $A_{<t},A'_{<t}\in\mc H^i_t$ such that $f|_{[0,t)_\T} \in A_{<t}$ and $f'|_{[0,t)_\T} \in A'_{<t}$. By Assumption \hyperlink{2-SEF_G.Ass:AP.SEF2}{AP.SEF2} and definition of $C$, there are $g,g'\colon D_{t,f} \to \A^i$ such that, with $c=c(A_{<t},i,g)$ and $c'=c(A'_{<t},i,g')$, we have $(\omega,f)\in c\in C^i$ and $(\omega,f')\in c'\in C^i$. In particular, $c\cap c'\cap W_\omega = \emptyset$. Moreover, $\omega\in D_{t,f}\cap D_{t,f'}$, and thus Lemma~\ref{2-SEF_G.lemma:AP_sef_compute_P(c)} implies that $c$ ($c'$) is available to agent $i$ at the random moves $\x_t(f)$ ($\x_t(f')$, respectively).

    It remains to prove that $c \supseteq y$ and $c' \supseteq y'$. In view of the problem's symmetry, it suffices to give a proof of the first inclusion.
    The inclusion $c\supseteq y$ holds true by construction if $y$ is a singleton. Else, there is $u\in\T$ such that $y = x_u(\omega,f)$. By construction, $t\le t_0 <\mf t(y) = u$, whence $t<u$. Hence, we can argue as follows. If $w\in y$, then there is $\tilde f\in\A^\T$ with $\tilde f|_{[0,u)_\T} = f|_{[0,u)_\T}$ such that $w = (\omega,\tilde f)$. In particular, $\tilde f|_{[0,t)_\T} = f|_{[0,t)_\T}$ and $\tilde f(t) = f(t)$, so that $w = (\omega,\tilde f)\in c$ as well. We conclude that $c \supseteq y$.\smallskip
    
    (Ad Axiom~\ref{2-SEF_G.def:SEF}.\ref{2-SEF_G.def:SEF.separation}):~ To show this property, we assume that $\D$ is \textsc{sef} data, so that even Assumption \hyperlink{2-SEF_G.Ass:AP.SEF3}{AP.SEF3} is satisfied. 
    
    Let $y,y'\in F$ with $y\cap y' = \emptyset$ and $\pi(y) = \pi(y')$. Denote $\omega = \pi(y)$. There are $f,f'\in\A^\T$ such that $(\omega,f)\in y$ and $(\omega,f')\in y'$, and $f\neq f'$. Hence, by Assumption \hyperlink{2-SEF_G.Ass:AP.SEF3}{AP.SEF3} there is $t\in\T$ such that $f|_{[0,t)_\T} = f'|_{[0,t)_\T}$ and $f(t) \neq f'(t)$. There is $i\in I$ such that $p^i \circ f(t) \neq p^i \circ f'(t)$. Hence, $\omega\in D_{t,f}^i$. There is unique $A_{<t}\in\mc H^i_t$ such that $f|_{[0,t)_\T} \in A_{<t}$. By Assumption \hyperlink{2-SEF_G.Ass:AP.SEF2}{AP.SEF2} and definition of $C$, there are $g,g'\colon D_{t,f} \to \A^i$ such that, with $c=c(A_{<t},i,g)$ and $c'=c(A_{<t},i,g')$, we have $(\omega,f)\in c\in C^i$ and $(\omega,f')\in c'\in C^i$. In particular, $c\cap c'\cap W_\omega = \emptyset$. Moreover, $\omega\in D_{t,f}$, and thus Lemma~\ref{2-SEF_G.lemma:AP_sef_compute_P(c)} implies that $c$ and $c'$ are available to agent $i$ at the random move $\x_t(f)$.

    It remains to prove that $x_t(\omega,f) \cap c \supseteq y$ and $x_t(\omega,f) \cap c' \supseteq y'$. By symmetry of the problem, it suffices to give a proof for the first inclusion. As $x_t(\omega,f) \cap y\neq \emptyset$, both nodes are contained in some decision path alias maximal chain in $(F,\supseteq)$ and are thus comparable. The same holds true for $x_t(\omega,f)$ and $y'$. If we had $y\supseteq x_t(\omega,f)$, then $y$ and $y'$ would be comparable since in $(F,\supseteq)$ principal up-sets are chains. But this would imply $y\cap y'\neq\emptyset$, a contradiction. Hence, $x_t(\omega,f) \supsetneq y$. The second part of the inclusion, namely $c\supseteq y$, holds true by construction if $y$ is a singleton. Else, there is $u\in\T$ such that $y = x_u(\omega,f)$. As $x_t(\omega,f) \supsetneq y$, the strict monotonicity of $\mf t$ implies that $t<u$. Hence, we can argue as follows. If $w\in y$, then there is $\tilde f\in\A^\T$ with $\tilde f|_{[0,u)_\T} = f|_{[0,u)_\T}$ such that $w = (\omega,\tilde f)$. In particular, $\tilde f|_{[0,t)_\T} = f|_{[0,t)_\T}$ and $\tilde f(t) = f(t)$, so that $w = (\omega,\tilde f)\in c$ as well. We conclude that $x_t(\omega,f) \cap c \supseteq y$.\smallskip

    (Ad Axiom~\ref{2-SEF_G.def:SEF}.\ref{2-SEF_G.def:SEF.enough_choices}):~ Let $x\in X$, $i\in J(x)$ and $y\in \downarrow x \setminus \{x\}$. Let $\omega = \pi(x)$. There is $f\in\A^\T$ such that $(\omega,f)\in y$. Then, there is $t\in\T$ such that $x = x_t(\omega,f)$. As $i\in J(x)$, we have $\x_t(f)\in\X^i$ and $\omega\in D_{t,f}$. By Proposition~\ref{2-SEF_G.prop:Dfti}, we infer that $\omega\in D_{t,f}^i$. Hence, by Assumption \hyperlink{2-SEF_G.Ass:AP.SEF2}{AP.SEF2}, there are $g\colon D_{t,f}\to\A^i$ and $A_{<t}\in\mc H_t^i$ with $f|_{[0,t)_\T}\in A_{<t}$ such that $c=c(A_{<t},i,g)$ satisfies $(\omega,f)\in c \in C^i$. By Lemma~\ref{2-SEF_G.lemma:AP_sef_compute_P(c)}, $x = x_t(\omega,f)\in P(c)$, whence $c\in A^i(x)$.

    Let $w\in y$. Then there is $\tilde f\in\A^\T$ such that $w=(\omega,\tilde f)$. If $y$ is a singleton, then $\tilde f = f$, whence $w\in c$. If $y$ is not a singleton, then there is $u\in\T$ such that $y = x_u(\omega,f)$. By strict monotonicity of $\mf t$, we have $t<u$. In the very same way as we did in the proof of the previous axiom, we infer that $\tilde f|_{[0,t)_\T} = f|_{[0,t)_\T}$ and $\tilde f(t) = f(t)$, whence $(\omega,\tilde f)\in c$. To conclude, we have shown that $c\supseteq y$.\smallskip

    (Ad Axiom~\ref{2-SEF_G.def:SEF}.\ref{2-SEF_G.def:SEF.endo_exo_compatible}):~ Let $i\in I$ and $\x,\x'\in\X$ such that $A^i(\x) \cap A^i(\x') \neq \emptyset$. Hence, there are $\omega\in D_\x$, $\omega'\in D_{\x'}$ and $c\in C^i$ such that $\x(\omega),\x'(\omega)\in P(c)$. There are $A_{<t}\in\mc H^i_t$, $D\in\ms E$, $g\colon D\to \A^i$ such that $c = c(A_{<t},i,g)$. By Lemma~\ref{2-SEF_G.lemma:AP_sef_compute_P(c)}, there are $f,f'\in\A^\T$ with $f|_{[0,t)_\T},f'|_{[0,t)_\T}\in A_{<t}$ such that $\x(\omega) = x_t(\omega,f)$ and $\x'(\omega') = x_t(\omega',f')$. As the left-hand sides are moves, $D_{t,f}$ and $D_{t,f'}$ are non-empty, and by Definition~\ref{1-SDF_AC.def:sdf}.\ref{1-SDF_AC.def:sdf.X.OC}, we get $\x = \x_t(f)$ and $\x' = \x_t(f')$. 
    By Definition~\ref{2-SEF_G.def:mcH}.\ref{2-SEF_G.def:mcH.msF_compatible}, we obtain 
    \[ \ms F^i_\x = \ms F^i_{\x_t(f)} = \ms F^i_{\x_t(f')} = \ms F^i_{\x'}. \]

    Note that this implies
    \[ (\ast) \qquad D_\x = D_{\x'}. \]

    It remains to show that $\ms C^i_{\x} = \ms C^i_{\x'}$. In view of the problem's symmetry, it suffices to show the inclusion ``$\subseteq$''. Let $c_1\in\ms C^i_{\x_t(f)}$. Then there are $\tilde A_{<t}\in \mc H_t^i$ and $A_t^i\in\ms B(\A^i)$ such that with $A_t$ as in ($\ms C_{\x}^i$.\ref{2-SEF_G.def:msC.2}), we have $c_1 = c(\tilde A_{<t},A_t)$, $c_1\in\ms C_t$, and for all $\tilde\omega\in D_\x$, $\x(\tilde\omega) \cap c_1 \neq\emptyset$. Hence, $f|_{[0,t)_\T}\in \tilde A_{<t}\cap A_{<t}$, and thus $\tilde A_{<t} = A_{<t}$ by Definition~\ref{2-SEF_G.def:mcH}.\ref{2-SEF_G.def:mcH.partition}.

    In view of $(\ast)$, it remains to show Property (AP-RCS*.\ref{2-SEF_G.def:msC.4}) for $\x'$, that is, for all $\tilde\omega\in D_{\x'}$, we have $\x'(\tilde\omega) \cap c_1 \neq \emptyset$. Let $\tilde\omega\in D_{\x'}$. Then $\tilde\omega\in D_\x$. Hence, there is $\tilde f\in\A^\T$ such that $(\tilde\omega,\tilde f)\in\x(\tilde\omega) \cap c_1$. By Assumption \hyperlink{2-SEF_G.Ass:AP.C1}{AP.C1}, there $\tilde f_1\in\A^\T$ such that $(\tilde\omega,\tilde f_1) \in \x(\tilde\omega) \setminus c_1$. Hence, $\tilde\omega\in D_{t,\tilde f}^i$, and thus, by Assumption \hyperlink{2-SEF_G.Ass:AP.SEF2}{AP.SEF2}, there is $c_2\in C^i$ such that a) $(\tilde\omega,\tilde f)\in c_2$, and b) there is $\tilde f'\in\A^\T$ satisfying $(\tilde\omega,\tilde f') \in c_2$ and $\tilde f'|_{[0,t)_\T} = f'|_{[0,t)_\T}$. As a consequence, $(\tilde\omega,\tilde f')\in \x'(\tilde\omega)$, $\tilde f'|_{[0,t)_\T}\in A_{<t}$, and $p^i \circ \tilde f'(t) = p^i \circ \tilde f(t)$, whence $(\tilde\omega,\tilde f')\in c_1$. We have thus shown the existence of an element of $\x'(\tilde\omega) \cap c_1$.\smallskip

    (Ad Axiom~\ref{2-SEF_G.def:SEF}.\ref{2-SEF_G.def:SEF.choice_completeness}):~ Let $i\in I$ and $c'$ an $\ms F^i$-$\ms C^i$-adapted choice satisfying Properties~\ref{2-SEF_G.def:SEF.choice_completeness.i} and~\ref{2-SEF_G.def:SEF.choice_completeness.ii} in Definition~\ref{2-SEF_G.def:SEF}. 

    By the latter one, there is $c\in C^i$ with $P(c) = P(c')$. There is $t\in\T$, $A_{<t}\in\mc H^i_t$, $D\in\ms E$, and $g\colon D \to \A^i$ such that $c = c(A_{<t},i,g)$. 
    
    Note that the non-redundancy of both $c$ and $c'$ and Lemma~\ref{1-SDF_AC.lemma:P(c)_compatible_with_conn_comp} imply that
    \begin{equation*}
    (\star)\qquad
    \begin{aligned} 
        D =&~ \{\omega\in \Omega \mid c\cap W_\omega \neq \emptyset\} = \{\omega\in\Omega \mid P(c) \cap T_\omega \neq \emptyset\} \\
        =&~ \{\omega\in\Omega \mid P(c') \cap T_\omega \neq \emptyset\} = \{\omega\in\Omega \mid c'\cap W_\omega \neq \emptyset\}.
    \end{aligned}
    \end{equation*}
    
    Let us define a function $g'\colon D \to \A^i$ as follows. Let $\omega\in D$. Then, in view of $(\star)$ and Property~\ref{2-SEF_G.def:SEF.choice_completeness.i}, there is $c_\omega\in C^i$ with $c'\cap W_\omega = c_\omega\cap W_\omega$. Such $c_\omega$ satisfies
    \[ P(c) \cap T_\omega = P(c') \cap T_\omega = P(c_\omega) \cap T_\omega \neq \emptyset,\] 
    by Lemma~\ref{1-SDF_AC.lemma:P(c)_compatible_with_conn_comp} and non-redundancy of the choices involved. We obtain, as shown shortly afterwards, that any $c_\omega\in C^i$ with $c'\cap W_\omega = c_\omega \cap W_\omega$ must admit a representation
    \[ (\ast) \qquad c_\omega = c(A_{<t},i,g_\omega) \] 
    for some $g_\omega\colon D \to \A^i$ such that for all $(\omega',f'_{<t})\in D\times A_{<t}$ there is $f'\in\A^\T$ with $f'|_{[0,t)_\T} = f'_{<t}$ and $(\omega',f')\in c_\omega$. As $c_\omega\cap W_\omega\neq\emptyset$,  $g_\omega(\omega)$ is independent of the choice of $c_\omega\in C^i$ such that $c'\cap W_\omega = c_\omega \cap W_\omega$. Hence, we can and do define $g'(\omega)$ by the equation
    \[ g'(\omega) = g_\omega(\omega). \]
    
    To see the existence of a representation as in $(\ast)$, represent $c_\omega$ --- which is an element of $C^i$ by assumption --- as $c(A_{<t;\omega},i,g_\omega)$ for some $A_{<t;\omega}\in\mc H^i_t$, $D_\omega\in\ms E$ with $\omega\in D_\omega$, and $g_\omega\colon D_\omega \to \A^i$, such that for all $(\omega',f'_{<t})\in D_\omega\times A_{<t;\omega}$ there is $f'\in\A^\T$ with $f'|_{[0,t)_\T} = f'_{<t}$ and $(\omega',f')\in c_\omega$. Then, there is an element $x\in P(c) \cap T_\omega = P(c_\omega)\cap T_\omega$, which by Lemma~\ref{2-SEF_G.lemma:AP_sef_compute_P(c)} applied to both $c$ and $c_\omega$, can be represented as $x = x_t(\omega,f)$ with $f\in \A^\T$ such that $f|_{[0,t)_\T}\in A_{<t}\cap A_{<t;\omega}$. Hence, by Definition~\ref{2-SEF_G.def:mcH}.\ref{2-SEF_G.def:mcH.partition}, $A_{<t} = A_{<t;\omega}$. Moreover, by Lemma~\ref{2-SEF_G.lemma:AP_sef_compute_D} applied to $c$ and $c_\omega$, we get $D = D_{t,f} = D_\omega$. This proves the existence of the representation $(\ast)$. In particular, $g'$ is well-defined.

    We now claim that 
    \[ (\dagger)\qquad c' = c(A_{<t},i,g'). \]
    It suffices to show that for all $\omega\in\Omega$, we have
    \[ c'\cap W_\omega = c(A_{<t},i,g') \cap W_\omega. \]
    This is clear for $\omega\in D^\complement$, by $(\star)$ and by definition of $c(A_{<t},i,g)$. If on the other hand $\omega\in D$, then there is $c_\omega\in C^i$ with $c_\omega\cap W_\omega = c' \cap W_\omega$ that can be represented as in $(\ast)$, and we have
    \[ c'\cap W_\omega = c_\omega\cap W_\omega = c(A_{<t},i,g_\omega) \cap W_\omega = c(A_{<t},i,g') \cap W_\omega,\]
    by $(\ast)$ and the definition of $g'$. This proves $(\dagger)$.

    It remains to show that 
    \begin{enumerate}[label=(\alph*),ref=(\alph*)]
        \item\label{2-SEF_G.thm:AP_SEF.proof.choice_compl.claim.i} $c'\in \ms C_t$, and
        \item\label{2-SEF_G.thm:AP_SEF.proof.choice_compl.claim.ii} for all $(\omega,f_{<t})\in D\times A_{<t}$ there is $f\in\A^\T$ with $(\omega,f)\in c'$ and $f|_{[0,t)_\T} = f_{<t}$.
    \end{enumerate}
    Regarding \ref{2-SEF_G.thm:AP_SEF.proof.choice_compl.claim.i}, Assumption \hyperlink{2-SEF_G.Ass:AP.C0}{AP.C0} is satisfied for $c'$ because it is satisfied for $c$. Indeed, as $c\neq \emptyset$ by assumption, there is $\omega\in\Omega$ such that $c\cap W_\omega \neq\emptyset$. Hence, there is $\omega\in \Omega$ satisfying $P(c) \cap T_\omega \neq \emptyset$ by non-redundancy. Thus, 
    \[ P(c') \cap T_\omega = P(c) \cap T_\omega \neq \emptyset,\]
    whence $c' \neq \emptyset$.

    For Assumption \hyperlink{2-SEF_G.Ass:AP.C1}{AP.C1}, let $w\in c'$ and let $\omega\in\Omega$ be such that $w\in W_\omega$. Let $c_\omega\in C^i$ such that $c_\omega \cap W_\omega = c' \cap W_\omega$. Then, $w\in c'\cap W_\omega = c_\omega\cap W_\omega$. As $c_\omega\in \ms C_t$, there is 
    \[ w'\in x_t(w) \setminus c_\omega = (x_t(w) \cap W_\omega) \setminus c_\omega = (x_t(w) \cap W_\omega) \setminus c' \subseteq x_t(w) \setminus c'. \]

    For Assumption \hyperlink{2-SEF_G.Ass:AP.C2}{AP.C2}, let $f\in\A^\T$ with $f|_{[0,t)_\T}\in A_{<t}$ such that $x_t(\omega_0,f) \cap c' \neq \emptyset$ for some $\omega_0 \in D_{t,f}$. Hence, $c' \cap W_{\omega_0} \neq \emptyset$, which implies $\omega_0 \in D$. Then, there is $c_{\omega_0}\in C^i$ such that $c_{\omega_0} = c(A_{<t},i,g_{\omega_0})$, for some $g_{\omega_0}\colon D\to \A^i$ as in $(\ast)$, $c_{\omega_0} \cap W_{\omega_0} = c' \cap W_{\omega_0}$ and $x_t(\omega_0,f) \cap c_{\omega_0} \neq\emptyset$. As $c_{\omega_0}\in C^i$, Lemma~\ref{2-SEF_G.lemma:AP_sef_compute_D} implies that $D = D_{t,f}$, and Lemma~\ref{2-SEF_G.lemma:AP_sef_compute_P(c)} implies that $x_t(\omega_0,f)\in P(c_{\omega_0})$. As a consequence,
    \[ x_t(\omega_0,f) \in P(c_{\omega_0}) \cap T_{\omega_0} = P(c') \cap T_{\omega_0} = P(c) \cap T_{\omega_0}, \]
    where the first equality follows from Lemma~\ref{1-SDF_AC.lemma:P(c)_compatible_with_conn_comp} and the fact that $c_{\omega_0} \cap W_{\omega_0} = c' \cap W_{\omega_0}$. Hence, $x_t(\omega_0,f)\in X^i$ and $\x_t(f)\in\X^i$. By completeness of $c$, we get that $c$ is available at the random move $\x_t(f)$.
    
    Let $\omega\in D_{t,f}$. Then, $\omega\in D$. Thus, there is $c_\omega\in C^i$ as in $(\ast)$ such that $c_\omega \cap W_\omega = c' \cap W_\omega$ and 
    \[ x_t(\omega,f)\in P(c) \cap T_\omega = P(c') \cap T_\omega = P(c_\omega) \cap T_\omega, \]
    which we can show using the same argument as above. By Lemma~\ref{1-SDF_AC.lemma:AP_P(c)}, we have $P(c_\omega) = \{x_t(w) \mid w\in c_\omega\}$. Hence, there is $\tilde f\in\A^\T$ with $\tilde f|_{[0,t)_\T} = f|_{[0,t)_\T}$ such that $(\omega,\tilde f) \in c_\omega$. Fix such an $\tilde f$. We infer that 
    \[ (\omega,\tilde f) \in c_\omega \cap W_\omega = c' \cap W_\omega. \]
    Hence, $(\omega,\tilde f) \in x_t(\omega,f) \cap c'$. We have thus completely proven that $c'\in\ms C_t$.

    Regarding \ref{2-SEF_G.thm:AP_SEF.proof.choice_compl.claim.ii}, let $(\omega,f_{<t})\in D\times A_{<t}$. Then, by $(\ast)$, there is $c_\omega\in C^i$ such that there is $f\in\A^\T$ with $f|_{[0,t)_\T} = f_{<t}$ and
    \[ (\omega,f) \in c_\omega \cap W_\omega = c' \cap W_\omega \subseteq c'. \]

    We conclude that $c'\in C^i$, and the proof is complete.
\end{proof}

\begin{proof}
    [Proof of Proposition~\ref{2-SEF_G.prop:compute_Ai(x)_mfp}]
    (Ad \ref{2-SEF_G.prop:compute_Ai(x)_mfp.Ai(x)}):~ Let $t\in\T$ and $f\in\A^\T$ such that $D_{t,f}\neq\emptyset$. Let $\x = \x_t(f)$. Furthermore, let $c\in C^i$, and represent it by $u\in\T$, $A_{<u}\in \mc H^i_u$, $D\in\ms E$, and $g\colon D\to\A^i$ via $c = c(A_{<u},i,g)$.

    Then, $c\in A^i(\x)$ iff $\x(\omega) \in P(c)$ for some and all $\omega\in D_\x$. By Lemmata \ref{2-SEF_G.lemma:AP_sef_compute_D} and~\ref{2-SEF_G.lemma:AP_sef_compute_P(c)}, this is equivalent to $t=u$, $f|_{[0,t)_\T}\in A_{<u}=A_{<t}$ and $D=D_{t,f}$. \smallskip

    (Ad \ref{2-SEF_G.prop:compute_Ai(x)_mfp.mfp}):~ Call the presumed map introduced in the claim $\p$.
    
    To see that $\p$ defines a map indeed, it suffices to show that for any pair $(t,A_{<t})$ as in the proposition and any $f\in\A^\T$ with $f|_{[0,t)_\T}\in A_{<t}$ we have $D_{t,f} \neq \emptyset$. Indeed, by \ref{2-SEF_G.def:mcH}.\ref{2-SEF_G.def:mcH.partition}, $D_{t,f}^i \neq \emptyset$, and thus, by Proposition~\ref{2-SEF_G.prop:Dfti}, $D_{t,f}\neq\emptyset$.
    
    Regarding injectivity of $\p$, we show the following stronger statement: For any two pairs $(t,A_{<t})$ and $(u,A'_{<u})$, as in the statement, such that the respective values $\mf p = \p(t,A_{<t})$ and $\mf p' = \p(u,A'_{<u})$ non-trivially intersect, it must hold true that $t=u$ and $A_{<t}=A'_{<u}$. Indeed, if $\mf p\cap\mf p' \neq\emptyset$, then both $A_{<t}$ and $A'_{<u}$ are non-empty, there are $f,f'\in \A^\T$ with $f|_{[0,t)_\T} \in A_{<t}$, $f'|_{[0,u)_\T}\in A'_{<u}$ with $\x_t(f) = \x_u(f')$. Applying the ``time'' map $\mf t$ to both sides yields $t=u$, whence $A_{<t} \cap A'_{<u}\neq\emptyset$. Thus, by Definition~\ref{2-SEF_G.def:mcH}.\ref{2-SEF_G.def:mcH.partition}, $A_{<t} = A'_{<u}$.

    Regarding the claim about the image, we recall that according to the preceding step, the values of $\p$ are pairwise disjoint. As elements of $\mc H^i_t$ are non-empty for all $t\in\T$, the values of $\p$ are non-empty. Moreover, using Proposition~\ref{2-SEF_G.prop:Dfti}.\ref{2-SEF_G.prop:Dfti.Dfti_nonempty} and Definition~\ref{2-SEF_G.def:mcH}.\ref{2-SEF_G.def:mcH.partition}, we infer that the image $\im\p$ of $\p$ defines a partition of $\X^i$. 
    
    It remains to show that this partition equals $\mf P^i$. For showing this, let $\x,\x'\in\X^i$. $\x$ and $\x'$ belong to the same element of $\im\p$ iff there are $t\in\T$, $A_{<t}\in\mc H^i_t$, $f,f'\in\A^\T$ such that $f|_{[0,t)_\T},f'|_{[0,t)_\T}\in A_{<t}$, $\x=\x_t(f)$ and $\x'=\x_t(f')$. Note that this implies that $D_{t,f}^i,D_{t,f'}^i \ne \emptyset$. By definition of the latter events, for all $\omega\in D_{t,f}^i$ and $\omega'\in D_{t,f}^i$, $f,f'$ can even be chosen such that $(\omega,f),(\omega',f')\in W$.
    
    Hence, in view of Assumption \hyperlink{2-SEF_G.Ass:AP.SEF2}{AP.SEF2} and the definition of $C^i$, this is equivalent to the existence of $c\in C^i$ that can be represented as $c=c(A_{<t},i,g)$ by $A_{<t}\in \mc H^i_t$ and $g\colon D\to\A^i$, for some $t\in\T$ and $D\in\ms E$, such that there are $f,f'\in\A^\T$ satisfying $f|_{[0,t)_\T},f'|_{[0,t)_\T}\in A_{<t}$, $\x=\x_t(f)$, $\x'=\x_t(f')$, as well as $D_{t,f} = D = D_{t,f'}$. The latter is implied by Lemma~\ref{2-SEF_G.lemma:AP_sef_compute_D}.
    
    By Part~\ref{2-SEF_G.prop:compute_Ai(x)_mfp.Ai(x)}, this statement is equivalent to $A^i(\x) \cap A^i(\x') \neq \emptyset$. By Proposition~\ref{2-SEF_G.prop:information_sets}.\ref{2-SEF_G.prop:information_sets.A(x)_partition}, this is equivalent to $A^i(\x) = A^i(\x')$ which, by definition, is equivalent to $\x,\x'\in\mf p$ for some $\mf p\in\mf P^i$. We conclude that $\mf P^i = \im \p$.
\end{proof}

\begin{proof}
    [Proof of Theorem~\ref{2-SEF_G.thm:adapted_choices_mb_functions}]
    Let the data be given as in the proposition's statement, that is, let $(\Omega,\ms E)$ be an exogenous scenario space, $\D$ be action path $\psi$-\textsc{sef} data on it, $\F$ be the induced action path $\psi$-\textsc{sef}, $i\in I$, $t\in\T$, $A_{<t}\in\mc H^i_t$, $D\in\ms E$ and $g\colon D\to \A^i$ be a map such that $c = c(A_{<t},i,g)\in\ms C_t$.
    
    Then Parts~\ref{2-SEF_G.thm:adapted_choices_mb_functions.non_red_compl}, \ref{2-SEF_G.thm:adapted_choices_mb_functions.Dx_subseteq_D}, and~\ref{2-SEF_G.thm:adapted_choices_mb_functions.mb}.($\Leftarrow$) follow directly from Parts~\ref{1-SDF_AC.thm:APsdf_AC.non_red_and_compl}, \ref{1-SDF_AC.thm:APsdf_AC.Dx_subset_D}, and~\ref{1-SDF_AC.thm:APsdf_AC.Fx_mb_=>_adapted} in Theorem~\ref{1-SDF_AC.thm:APsdf_AC}. It thus remains to prove Part~\ref{2-SEF_G.thm:adapted_choices_mb_functions.mb}.($\Rightarrow$). To prove this, it suffices to show that \hyperlink{1-SDF_AC.Ass:AP.C3}{Assumption~AP.C3} is satisfied for $(A_{<t},i,g)$ and $\ms C^i$, in view of Part~\ref{1-SDF_AC.thm:APsdf_AC.adapted_=>_Fx_mb} in Theorem~\ref{1-SDF_AC.thm:APsdf_AC}.

    To show this, let $\x\in\X^i$ be such that $c$ is available at $\x$. There is $(t,f)\in\T\times\A^\T$ with $\x = \x_t(f)$ and $D_{t,f} \neq\emptyset$. As $\x\in\X^i$, we have, by Proposition \ref{2-SEF_G.prop:Dfti}, $D_{t,f}^i \neq\emptyset$.

    Thus, by Assumption \hyperlink{2-SEF_G.Ass:AP.SEF1}{AP.SEF1} there is a generator $\ms G(\A^i)$ of $\ms B(\A^i)$, stable under non-trivial intersections, such that for all $G\in\ms G(\A^i)$, we have $c(A_{<t},A_t^{i,G}) \in\ms C_\x^i$ (with the notation from that assumption). This proves that \hyperlink{1-SDF_AC.Ass:AP.C3}{Assumption~AP.C3} is satisfied, thus completing the proof.  
\end{proof}

\begin{proof}
    [Proof of Theorem~\ref{2-SEF_G.thm:link_endogenous_information_H}]
    Let $(\Omega,\ms E)$ be an exogenous scenario space, $\D$ be action path $\psi$-\textsc{sef} data on it, $\F$ be the induced action path $\psi$-\textsc{sef}, and $i\in I$.\smallskip
    
    (Ad \ref{2-SEF_G.thm:link_endogenous_information_H.perfect_recall}.$\Rightarrow$):~ Suppose that $i$ admits perfect endogenous recall, and let $t,u\in\T$ with $t<u$, $A_{<t}\in\mc H^i_t$ and $A_{<u}\in\mc H_u^i$. Note first that $A_{<u}\neq\emptyset$, by definition (\ref{2-SEF_G.def:mcH}.\ref{2-SEF_G.def:mcH.partition}). Hence, $(\mc P p_{u,t})(A_{<u})\neq \emptyset$. Thus, we cannot have $(\mc P p_{u,t})(A_{<u}) \cap A_{<t} = \emptyset$ and $(\mc P p_{u,t})(A_{<u}) \subseteq A_{<t}$ at the same time. It therefore remains to show the following assertion:
    \[ (\ast) \quad \Bigg[  (\mc P p_{u,t})(A_{<u}) \cap A_{<t} \neq \emptyset \quad \Longrightarrow \quad \Big[ (\mc P p_{u,t})(A_{<u}) \subseteq A_{<t}, ~ \forall f,f'\in A_{<u}\colon p^i\circ f(t) = p^i\circ f'(t) \Big]\Bigg]. \]
    Let us therefore consider $A_{<u}\in\mc H_u^i$ and $A_{<t}\in\mc H^i_t$ such that  $(\mc P p_{u,t})(A_{<u}) \cap A_{<t} \neq \emptyset$.
    
    From this, we infer that there is $f\in\A^\T$ such that $f|_{[0,u)_\T} \in A_{<u}$ and $f|_{[0,t)_\T} \in A_{<t}$. As a consequence, by Definition~\ref{2-SEF_G.def:mcH}.\ref{2-SEF_G.def:mcH.partition}, both $D_{t,f}^i\neq\emptyset$ and $D_{f,u}^i \neq\emptyset$. By Proposition~\ref{2-SEF_G.prop:Dfti}.\ref{2-SEF_G.prop:Dfti.Dfti_nonempty}, we obtain $D_{t,f}^i = D_{t,f}$ and $D_{f,u}^i = D_{f,u}$. In particular, we obtain
    \[ D_{f,u}^i = D_{f,u} \subseteq D_{t,f} = D_{t,f}^i. \]
    
    Let $\omega\in D_{f,u}^i$. Upon modifying $f$ at time $u$ and later times, we can assume that $(\omega,f)\in W$. By Assumption \hyperlink{2-SEF_G.Ass:AP.SEF2}{AP.SEF2} and the definition of $C$, there are $g_t^\omega\colon D_{t,f} \to \A^i$ and $g_u^\omega\colon D_{f,u} \to \A^i$ such that $p^i \circ f(t) = g_t^\omega(\omega)$, $p^i\circ f(u) = g_u^\omega(\omega)$, and, with $c_t^\omega = c(A_{<t},i,g_t^\omega)$ and $c_u^\omega = c(A_{<u},i,g_u^\omega)$, $c_t^\omega,c_u^\omega \in C^i$. In particular, 
    $ (\omega,f) \in c_t^\omega\cap c_u^\omega\cap W_\omega$. 
    Hence, as $i$ admits perfect endogenous recall, we have 
    $c_t^\omega \cap W_\omega \supseteq c_u^\omega \cap W_\omega$ or $c_t^\omega \cap W_\omega \subseteq c_u^\omega \cap W_\omega$. As $\omega\in D_{f,u}^i$ and $(\omega,f)\in W$, there is $f'\in\A^\T$ with $(\omega,f')\in x_u(\omega,f)$ such that $p^i \circ f'(u) \neq p^i \circ f(u) = g_u^\omega(\omega)$. Hence, $(\omega,f') \in (c_t^\omega \cap W_\omega)\setminus (c_u^\omega \cap W_\omega)$. 
    We conclude that all $\omega\in D_{f,u}^i$ satisfy
    \[ (\dagger)\qquad c_t^\omega \cap W_\omega \supseteq c_u^\omega \cap W_\omega. \]

    We now show that $(\mc P p_{u,t})(A_{<u}) \subseteq A_{<t}$. For this, let $f_{<u}\in A_{<u}$. There is an $\omega\in D_{f,u}$, and thus, as $c_u^\omega\in C^i$, there is $f'\in\A^\T$ with $(\omega,f')\in c_u^\omega$ and $f'|_{[0,u)_\T} = f_{<u}$. Then, by $(\dagger)$, $(\omega,f')\in c_t^\omega$. In particular, $p_{u,t}(f_{<u}) = f'|_{[0,t)_\T} \in A_{<t}$. This shows the claimed inclusion.
    
    Next, let $f_{<u},f'_{<u}\in A_{<u}$. Again, there is $\omega\in D_{f,u}$, and as $c_u^\omega\in C^i$, there are $f,f'\in \A^\T$ with $(\omega,f),(\omega,f')\in c_u^\omega$, $f|_{[0,u)_\T} = f_{<u}$, and $f'_{[0,u)_\T} = f'_{<u}$. By $(\dagger)$, we have
    \[ p^i \circ f_{<u}(t) = p^i \circ f(t) = g_t^\omega(\omega) = p^i\circ f'(t) = p^i \circ f'_{<u}(t). \]

    (Ad \ref{2-SEF_G.thm:link_endogenous_information_H.perfect_recall}.$\Leftarrow$):~ Suppose the right-hand condition on $\mc H^i$ to be satisfied. 
    Let $c,c'\in C^i$ and $\omega\in\Omega$ such that $c\cap c' \cap W_\omega \neq \emptyset$. We have to show that $c\cap W_\omega$ and $c'\cap W_\omega$ can be compared by set inclusion.

    Represent $c,c'\in C^i$ using $t,u\in \T$, $A_{<t}\in \mc H^i_t$, $A'_{<u}\in\mc H^u_t$, $D_t,D_u\in \ms E$, and $g_t \colon D_t \to \A^i$, $g_u\colon D_u \to \A^i$ such that $c = c(A_{<t},i,g_t)$ and $c' = c(A'_{<u},i,g_u)$. By hypothesis, there is $f\in \A^\T$ with $(\omega,f) \in c \cap c'$. 
    
    Without loss of generality, we can assume that $t\le u$. First, consider the case ``$t=u$''. 
    Then, $f|_{[0,t)_\T} \in A_{<t} \cap A'_{<u}$. Hence, by Definition~\ref{2-SEF_G.def:mcH}.\ref{2-SEF_G.def:mcH.partition}, $A_{<t} = A'_{<u}$. Moreover, $g_t(\omega) = p^i \circ f(t) = p^i \circ f(u) = g_u(\omega)$. Hence, $c\cap W_\omega = c'\cap W_\omega$.

    In the other case ``$t<u$'', $ f|_{[0,t)_\T} \in (\mc P p_{u,t})(A'_{<u}) \cap A_{<t}$. Thus, by hypothesis, $(\mc P p_{u,t})(A'_{<u}) \subseteq A_{<t}$. We claim that $c\cap W_\omega \supseteq c' \cap W_\omega$. To show this, let $w'\in c'\cap W_\omega$. Represent $w' = (\omega,f')$ for some $f'\in \A^\T$. In particular, $f'|_{[0,u)_\T} \in A'_{<u}$. We infer that $f'|_{[0,t)_\T} = p_{u,t}(f'|_{[0,u)_\T}) \in A_{<t}$, and, since $f|_{[0,u)_\T}\in A'_{<u}$, $p^i \circ f'(t) = p^i \circ f(t) = g_t(\omega)$. Hence, since $w'\in W$, we get $w'\in c \cap W_\omega$.\smallskip

    (Ad \ref{2-SEF_G.thm:link_endogenous_information_H.perfect_information}):~ By definition, $i$ has perfect endogenous information iff a) all $\mf p\in\mf P^i$ are singletons and b) for all $j\in I\setminus\{i\}$ we have $\X^i \cap \X^j = \emptyset$. By Proposition~\ref{2-SEF_G.prop:compute_Ai(x)_mfp}.\ref{2-SEF_G.prop:compute_Ai(x)_mfp.mfp}, a) is equivalent to the statement that for all $t\in\T$, all $A\in \mc H_t^i$ are singletons. By Proposition~\ref{2-SEF_G.prop:compute_Ai(x)_mfp}.\ref{2-SEF_G.prop:compute_Ai(x)_mfp.Ai(x)}, and in view of Assumption~\hyperlink{2-SEF_G.Ass:AP.SEF2}{AP.SEF2}, b) is equivalent to the statement that for all $t\in\T$, all $j\in I\setminus \{i\}$, all $A\in\mc H^i_t$ and all $A'\in\mc H^j_t$, we have $A\cap A' = \emptyset$.
\end{proof}

\subsection{Section~\ref{2-SEF_G.sec:well-posedness_equilibrium}}\label{2-SEF_G.subsec:appendix.proofs.3}

\begin{proof}[Proof of Lemma~\ref{2-SEF_G.lemma:closed_history}]
    Let $(F,\ge)$ be a decision forest and $h\in H$ be a history. Let $W(h)$ the set of all maximal chains $w$ with $h\subseteq w$, which is non-empty by the Hausdorff maximality principle. \smallskip
    
    (Ad \ref{2-SEF_G.lemma:closed_history.1}):~ Let $h_0 = \bigcap W(h)$, i.e.\ the intersection of all maximal chains $w$ with $h\subseteq w$. As $h\subseteq h_0$, $h_0 \neq\emptyset$. As an intersection of chains, $h_0$ is a chain. Concerning upward-closure, let $x\in h_0$ and let $w\in W(h)$. Then $x\in w$. As $(F,\ge)$ is a forest, $\uparrow x$ is a chain. As $w$ is a chain containing $x$, $\uparrow x \cup w$ is a chain as well. As $w$ is a maximal chain, $\uparrow x \subseteq w$. Hence, $\uparrow x \subseteq h_0$.
    
    Furthermore, by construction, $h_0$ clearly satisfies Property a). Regarding Property b), let $h_1$ be a history satisfying a), that is, $h_1 \subseteq w$ for all maximal chains $w$ with $h\subseteq w$. Then, $h_1\subseteq h_0$ by definition of $h_0$.
    
    Regarding uniqueness, let $\overline h$ an upward closed chain satisfying a) and b). Then, $h_0 \subseteq \overline h$ by b) applied to $\overline h$ and a) applied to $h_0$; and $\overline h\subseteq h_0$ by b) applied to $h_0$ and a) applied to $\overline h$. Hence, $h_0 = \overline h$.\smallskip

    (Ad \ref{2-SEF_G.lemma:closed_history.2}):~ For every $w\in W(h)$, there is $x\in w$ with $h\subseteq \uparrow x$, because $h$ is non-maximal and upward closed. As $\uparrow x \subseteq w$, we infer 
    \[ \bigcap\{\uparrow x \mid x\in F\colon h\subseteq \uparrow x\} \subseteq \bigcap W(h). \]
    On the other hand, for every $x\in F$ with $h\subseteq \uparrow x$ there is a maximal chain $w$ with $\uparrow x \subseteq w$, because $(F,\ge)$ is a forest and by the Hausdorff maximality principle. In particular, $w\in W(h)$. Moreover, for any $y\in w \setminus \uparrow x$, we have $W(x) \neq W(y)$, because $(F,\ge)$ is a decision forest. As $x\ge y$ and $h\subseteq \uparrow x$, we then have $W(y) \subsetneq W(x) \subseteq W(h)$. Hence, for such $y$, there is $w'\in W(h)$ with $y\notin w'$. Thus $\bigcap W(h) = w\cap \bigcap W(h) \subseteq \uparrow x$. We conclude that 
    \[ \bigcap W(h)\subseteq \bigcap\{\uparrow x \mid x\in F\colon h\subseteq \uparrow x\}. \]
    From Part~\ref{2-SEF_G.lemma:closed_history.1}, it follows directly that $\overline h = \bigcap W(h)$ which then implies the claim.\smallskip

    (Ad \ref{2-SEF_G.lemma:closed_history.3}):~ Note that the right-hand condition is equivalent to $W(h) = W(h')$. If this is true, then, in view of Part \ref{2-SEF_G.lemma:closed_history.1} just proven, $\overline h = \bigcap W(h) = \bigcap W(h') = \overline{h'}$. Conversely, if $\overline{h} = \overline{h'}$, and $w\in W(h)$, then $h' \subseteq\overline{h'}=\overline h\subseteq w$, hence $w\in W(h')$. Thus, $W(h)\subseteq W(h')$. Repeating the argument with $h$ and $h'$ swapped, yields $W(h) = W(h')$.\smallskip

    (Ad \ref{2-SEF_G.lemma:closed_history.4}):~ Let $B_h$ the set of lower bounds of $h$, i.e.\
    \[ B_h = \{ y\in F\mid \forall x\in h\colon x\ge y\}. \]
    The statement is equivalent to saying that $\overline h=h$ if $B_h$ has no maximum, and $\overline h = h \cup \{\max B_h\}$ otherwise. It is this latter statement that is proven in the following.

    First, suppose that $B_h$ has no maximum. It suffices to show that $F\setminus h \subseteq F\setminus \overline h$, because $h\subseteq \overline h$ by construction. Let $x\in F\setminus h$. If $x\notin B_h$, then $x\notin\overline h$, because $\overline h$ is a chain containing $h$. It remains to consider the case where $x\in B_h \setminus h$. By hypothesis, there is $x'\in B_h$ with $x\ngeq x'$. By Part~\ref{2-SEF_G.lemma:closed_history.2}, $x\notin \overline h$.

    Second, suppose that $B_h$ has a maximum. Let $x\in F$ such that $h\subseteq \uparrow x$. Then, $x\in B_h$, whence $\max B_h\in\uparrow x$. Hence, by Part~\ref{2-SEF_G.lemma:closed_history.2}, $\max B_h\in\overline h$. As $h\subseteq\overline h$ by construction, we obtain $\overline h \supseteq h\cup \{\max B_h\}$. Conversely, if $x\in\overline h \setminus h$, then $x\in w\setminus h$ for any $w\in W(h)$. As $W(h)\neq\emptyset$ and $h$ is an upward closed chain, $x\in B_h$. Furthermore, for any $y\in B_h$, $h\subseteq \uparrow y$, by definition. Thus, by Part~\ref{2-SEF_G.lemma:closed_history.2} just proven, $x\in\uparrow y$, i.e.\ $x\ge y$. Thus, $x=\max B_h$. We conclude that $\overline h = h\cup \{\max B_h\}$.
\end{proof}

\begin{proof}[Proof of Lemma~\ref{2-SEF_G.lemma:histories_canonical_inj}]
    (Ad range):~ Let $x\in X$. Then $\uparrow x$ is a non-empty, upward closed chain, and it is non-maximal because $x$ is not terminal. Clearly, $x = \inf \uparrow x$ and $x\in\uparrow x$, so that $\uparrow x$ is closed by \ref{2-SEF_G.lemma:closed_history}. Further, let $\x\in\X$ and $(F,\pi,\X)$ be order consistent. If there is $\x'\in\X$ admitting $\omega\in D_\x \cap D_{\x'}$ with $\x'(\omega) \supseteq \x(\omega)$, then by order consistency, $D_{\x'}\supseteq D_\x$ and $\x'(\omega') \supseteq \x(\omega')$ for all $\omega'\in D_\x$.\smallskip

    (Ad injectivity):~ First, let $x,x'\in X$ be such that $\uparrow x = \uparrow x'$. Then, $x\supseteq x' \supseteq x$, whence $x=x'$. Second, let $\x,\x'\in\X$ be such that $D_{\x} = D_{\x'}$ and $\uparrow \x(\omega) = \uparrow \x'(\omega)$ for all $\omega\in D_\x$. Then, by the first part of the proof, $\x(\omega) = \x'(\omega)$ for all $\omega\in D_\x$, whence $\x = \x'$.
\end{proof}

\begin{lemma}\label{2-SEF_G.lemma:random_histories_h_f(h)}
    Let $(F,\pi,\X)$ be an order consistent, surely non-trivial, and maximal stochastic decision forest on an exogenous scenario space $(\Omega,\ms E)$. Let $\h$ be a random history with domain $D_\h$, $\omega\in D_\h$, and $f(\h) =  \{\x\in\X \mid \exists \omega'\in D_\x\cap D_\h\colon \x(\omega')\in\h(\omega')\}$. Then, we have:
    \[ \h(\omega) = \{ \x(\omega) \mid \x\in f(\h)\}. \]
\end{lemma}

\begin{proof}
    Given the data $(F,\pi,\X)$, $\h$, $\omega$, $f$ from the lemma, let $\x\in\X$. Then $\x\in f(\h)$ iff there is $\omega'\in D_\x\cap D_\h$ such that $\x(\omega')\in\h(\omega')$. As $\h$ is a random history, this latter statement is equivalent to saying that $D_\x\supseteq D_\h$ and for all $\omega'\in D_\h$ we have $\x'(\omega') \in \h(\omega')$. By the same argument, though, this latter statement is equivalent to saying that $D_\h\subseteq D_\x$ and $\x(\omega)\in \h(\omega)$.
\end{proof}

\begin{proof}[Proof of Proposition~\ref{2-SEF_G.prop:random_histories}]
    (Ad \ref{2-SEF_G.prop:random_histories.image_H_Tr}):~ Let $\h\in\H$. Recall that, by definition, $f(\h) = \{\x\in\X \mid \exists \omega\in D_\x\cap D_\h\colon \x(\omega)\in\h(\omega)\}$. 

    There is $\omega\in D_\h$ and $\h(\omega)$ is non-empty. There is, thus, $\x\in\X$ with $\omega\in D_\x$ and $\x(\omega)\in\h(\omega)$ because $\X$ covers $X$. Hence, $\x\in f(\h)$. We therefore see that $f(\h)$ is non-empty.
    
    Let $\x,\x'\in f(\h)$. Then, $D_\x\cap D_{\x'} \supseteq D_\h$ which is non-empty. Hence, there is $\omega\in D_\h$ with $\x(\omega),\x'(\omega)\in\h(\omega)$. As $\h(\omega)$ is a chain, $\x(\omega)$ and $\x'(\omega)$ are related via $\supseteq$. Up to changing the roles of $\x$ and $\x'$, we obtain $\x(\omega)\supseteq \x'(\omega)$, and thus, by order consistency, $\x\ge_\X \x'$. Thus, $f(\h)$ is a chain.

    Let $\x\in f(\h)$ and $\x'\in\X$ such that $\x'\ge_\X \x$. Then, $D_{\x'}\supseteq D_\x\supseteq D_\h$, and for all $\omega\in D_\h$ we have $\x'(\omega) \supseteq \x(\omega)$. As $\h(\omega)$ is upward closed, $\x'(\omega)\in\h(\omega)$ for all $\omega\in D_\h$. Hence, $\x'\in f(\h)$, and $f(\h)$ is upward closed.

    Regarding non-maximality, we first note that for any $\omega\in D_\h$, there is $y\in F\setminus \h(\omega)$ with $x \supsetneq y$ for all $x\in\h(\omega)$, by non-maximality of $\h(\omega)$. In that case we infer that, for all $\x\in f(\h)$, $\omega\in D_\x$ and $\x(\omega)\in\h(\omega)$, whence $\x(\omega) \supsetneq y$. 
    
    If a) there is $\omega\in D_\h$ and $y\in X\setminus \h(\omega)$ with $x \supsetneq y$ for all $x\in\h(\omega)$, then there is $\x'\in\X$ with $y = \x'(\omega)$, whence $\x>_\X \x'$. Then, we conclude that $f(\h)$ is a non-maximal chain in $(\X,\ge_\X)$ and that $f(\h) \in H_\Tr$.
    
    If, however, b) there is no such pair $(\omega,y)$, then $f(\h)$ is a maximal chain in $(\X,\ge_\X)$ and for any $\omega\in D_\h$, there is a terminal node $y\in F \setminus \h(\omega)$ with $x \supsetneq y$ for all $x\in\h(\omega)$. We infer that there is a random terminal node $\y\colon \{\omega\} \to \{y\}$, and $\x >_\Tr\y = \{(\omega,y)\}$. Hence, $D_\h \subseteq \bigcap_{\x\in h_\Tr} D_\x$ and for any $\omega\in D_\h$, there is $w_\omega\in W_\omega$ with $\y(\omega) = \{w_\omega\}$ whence $w_\omega \in \bigcap_{\x\in h_\Tr} \x(\omega)$. Moreover, $D_\h\neq \emptyset$, which implies that $f(\h)$ is not a maximal chain in $(\Tr,\ge_\Tr)$. $D_\h \in\ms E$ by hypothesis. We have shown that $f(\h)$ is contained in the set described in the claim by the disjunction of conditions a) and b). \smallskip

    For the converse statement, let $h_\Tr$ be an element of the set described in the claim by the disjunction of conditions a) and b). First, suppose a) that $h_\Tr\in H_\Tr$ is a non-maximal chain in $(\X,\ge_\X)$. Then, there is $\x_1\in\X$ with $\x>_\X \x_1$ for all $\x\in h_\Tr$. Let $D = D_{\x_1}$. Second, suppose b) that $h_\Tr$ is a maximal chain in $(\X,\ge_\X)$ admitting non-empty $D\in\ms E\setminus\{\emptyset\}$ with $D\subseteq \bigcap_{\x\in h_\Tr} D_\x$ such that for any $\omega\in D$ there is $w\in W_\omega$ with $w\in \bigcap_{\x\in h_\Tr} \x(\omega)$.

    In both cases, we have a non-empty event $D$ and we define a map $\h$ with domain $D$ by letting 
    \[ (\ast)\qquad\h(\omega) = \{\x(\omega) \mid \x\in h_\Tr\}, \]
    for all $\omega\in D$. We claim that $\h\in\H$ and that $f(\h) = h_\Tr$.
    
    Regarding the first of these claims, the domain $D$ is by construction a non-empty event. Moreover, for any $\omega\in D$, $\h(\omega)\in H$. Indeed, let $\omega\in D$. Then, as $h_\Tr$ is a non-empty chain, $\h(\omega)$ is so, too, by definition of $\ge_\Tr$. Moreover, let $x\in\h(\omega)$ and $x'\in\uparrow x$. Then, there is $\x\in h_\Tr$ with $x=\x(\omega)$, and $x'\in X$. Accordingly, there is $\x'\in\X$ with $x' = \x'(\omega)$. Thus, by order consistency, $\x'\ge_\X \x$. As $h_\Tr$ is upward closed, $\x'\in h_\Tr$, whence $x'=\x'(\omega)\in h(\omega)$. Hence, $\h(\omega)$ is upward closed. Furthermore, $\h(\omega)$ is not maximal as a chain. To show this, we distinguish the two cases from before. In case a), $\x>_\Tr \x_1$ by construction, in particular, $\x(\omega) \supsetneq \x_1(\omega)$ for all $\x\in h_\Tr$. In case b), $\x(\omega) \supsetneq w_\omega$ for all $\x\in h_\Tr$. Hence, $\h(\omega)$ is not maximal as a chain in $(F,\supseteq)$. This shows that for any $\omega\in D$, $\h(\omega)\in H$. Further, let $\x\in\X$ admit $\omega\in D_\x\cap D$ with $\x(\omega)\in\h(\omega)$. Then, by order consistency and Proposition~\ref{1-SDF_AC.prop:ev_on_Tr_is_iso}, $\x\in h_\Tr$. Hence, $D_\x\supseteq D$ and $\x(\omega')\in\h(\omega')$ for all $\omega'\in D$. It is thus proven that $\h\in\H$.

    Regarding the second of the two claims above, namely that $f(\h) = h_\Tr$, let $\x\in\X$. By definition of $f$, $\x\in f(\h)$ holds true iff there is $\omega\in D_\x\cap D_\h$ with $\x(\omega)\in\h(\omega)$. By definition of $\h$, the latter is equivalent to saying that there is $\omega\in D_\x\cap D$ with $\x(\omega) = \x'(\omega)$ for some $\x'\in h_\Tr$. By order consistency and Proposition~\ref{1-SDF_AC.prop:ev_on_Tr_is_iso}, and by definition of $D$, this is equivalent to $\x\in h_\Tr$. Thus, $f(\h) = h_\Tr$.    
    \smallskip

    (Ad \ref{2-SEF_G.prop:random_histories.faithful}):~ Let $\h_1,\h_2\in\H$ such that $f(\h_1) = f(\h_2)$. Let $\omega\in D_{\h_1}\cap D_{\h_2}$. \\Then, by Lemma~\ref{2-SEF_G.lemma:random_histories_h_f(h)}, 
    \[ \h_1(\omega) = \{\x(\omega) \mid \x\in f(\h_1)\} = \{\x(\omega) \mid \x\in f(\h_2)\} = \h_2(\omega). \]
    Hence, there is a map $\h$ on $D = D_{\h_1} \cup D_{\h_2}$ satisfying $\h|_{D_{\h_k}} = \h_k$ for both $k=1,2$. By construction, $D$ is a non-empty event and $\h(\omega)$ is a history in $(F,\supseteq)$ for any $\omega\in D$. Moreover, we clearly have $f(\h_1) = f(\h) = f(\h_2)$. Thus, if there are $\x\in\X$ and $\omega\in D_\x\cap D_\h$ with $\x(\omega)\in\h(\omega)$, then $\x\in f(\h) = f(\h_1) = f(\h_2)$, and thus, $D_\x\supseteq D_{\h_1} \cup D_{\h_2} = D_\h$ and for all $\omega'\in D_\h$ we have $\x(\omega')\in \h(\omega')$, since $\h_1,\h_2\in\H$.\smallskip

    (Ad \ref{2-SEF_G.prop:random_histories.closed}):~ Let $\h\in\H$ be a closed random history. We have to show that $\overline{f(\h)} = f(\h)$. In view of Lemma~\ref{2-SEF_G.lemma:closed_history} it remains to show that, if $f(\h)$ admits an infimum in $(\Tr,\ge_\Tr)$, then $\inf f(\h)\in f(\h)$. Suppose that $f(\h)$ has an infimum. There is $\omega\in D_\h$. From order consistency and Lemma~\ref{2-SEF_G.lemma:random_histories_h_f(h)}, it follows directly that $(\inf f(\h))(\omega)$ is an (and the) infimum of $\h(\omega)$ in $(F,\supseteq)$. As $\h(\omega)$ is a closed history by hypothesis, we get $(\inf f(\h))(\omega)\in\h(\omega)$. Thus $\inf f(\h)\in\X$, $D_\h\subseteq D_{\inf f(\h)}$, and $\inf f(\h) \in f(\h)$.
\end{proof}

\begin{proof}[Proof of Lemma~\ref{2-SEF_G.lemma:R_continuous_in_h}]
    Let $\F$ be a stochastic pseudo-extensive form on an exogenous scenario space $(\Omega,\ms E)$, $s\in S$, $h\in H$ and $w\in W$. It suffices to show that for all $x\in X$, we have
    \[ x\subseteq \bigcap h \qquad \Longleftrightarrow \qquad x\subseteq \bigcap \overline h. \]
    This is trivial if $h=\overline h$. Therefore, suppose that the latter is not the case. By Lemma~\ref{2-SEF_G.lemma:closed_history}, $h$ has an infimum and $\overline h = h \cup \{\inf h\}$. As $h\subseteq \overline h$, the implication ``$\Leftarrow$'' is evident. For the converse one, suppose that $x\subseteq \bigcap h$. In other words, $x\subseteq y$ for all $y\in h$. By definition of the infimum, we infer $x\subseteq \inf h = \bigcap \overline h$.
\end{proof}

\begin{lemma}\label{2-SEF_G.lemma:extension_of_X_strategies}
    Let $\F$ be a stochastic pseudo-extensive form, $i\in I$, and $s^i_0 \colon X^i_0 \to C^i$ be a map on some set of $i$'s moves $X^i_0 \subseteq X^i$. Then $s^i_0$ is the restriction of an $X$-strategy iff 
    \begin{enumerate}
        \item\label{2-SEF_G.lemma:extension_of_X_strategies.1} for all $x\in X^i_0$, we have $s^i(x) \in A^i(x)$;
        \item\label{2-SEF_G.lemma:extension_of_X_strategies.2} for all $x,x'\in X^i_0$ with $A^i(x) = A^i(x')$, we have $s^i(x) = s^i(x')$.
    \end{enumerate}
\end{lemma}

\begin{proof}
    $S^i$ is non-empty because $A^i(\mf p) \neq \emptyset$ for all $\mf p\in \mf P^i$, by definition. Hence, in view of Proposition~\ref{2-SEF_G.prop:strategies} we can choose an $X$-strategy $s_1^i$. Then, define $s^i$ as follows. Let $x\in X^i$. If there is $x_0\in X^i_0$ with $A^i(x) = A^i(x_0)$, let $s^i(x) = s^i_0(x_0)$. Else, let $s^i(x) = s_1^i(x)$. By Property~\ref{2-SEF_G.lemma:extension_of_X_strategies.2}, $s^i$ is well-defined.

    By Property~\ref{2-SEF_G.lemma:extension_of_X_strategies.1} and the fact that $s^i_1$ is an $X$-strategy, we clearly have $s^i(x)\in A^i(x)$ for all $x\in X^i$. Moreover, if $x,x'\in X^i$ satisfy $A^i(x) = A^i(x')$, there are two cases. First, if there is $x_0\in X^i_0$ with $A^i(x) = A^i(x_0)$, then we also have $A^i(x') = A^i(x_0)$, hence $s^i(x) = s^i_0(x_0) = s^i(x')$. Else, there is no such $x_0$. Then, neither is there $x_0\in X_0^i$ with $A^i(x') = A^i(x_0)$. Hence, $s^i(x) = s_1^i(x) = s_1^i(x') = s^i(x')$, because $s_1^i$ is an $X$-strategy.
\end{proof}

\begin{proof}[Proof of Theorem~\ref{2-SEF_G.thm:well_posed.onto_outcomes}]
    Let $h\in H$ and $w\in\bigcap h$. Let $i\in I$ and $X_0^i = \{x\in X^i \mid w\in x\subseteq \bigcap h\}$. 
    
    As $F$ is a decision forest, $\uparrow \{w\} = \{ y\in F \mid w\in y\}$ is a maximal chain. Let $x\in X_0^i$. Then, $x\in X$ and, thus, $\uparrow x$ is not a maximal chain. As a consequence, there is $y\in\downarrow x\setminus\{x\}$ with $w\in y$, and by Axiom~\ref{2-SEF_G.def:SEF}.\ref{2-SEF_G.def:SEF.enough_choices}, there is $c\in A^i(x)$ with $c\supseteq y$. Let $s_0^i(x) = c$. 
    
    This defines a map $s^i_0\colon X^i_0 \to C^i$ with $s^i_0(x) \in A^i(x)$ for all $x\in X_0^i$. By the Heraclitus property from Lemma~\ref{2-SEF_G.lemma:Heraclitus_property}, $x,x'\in X^i_0$ with $x\neq x'$ necessarily satisfy $A^i(x) \cap A^i(x') = \emptyset$. Hence, by Lemma~\ref{2-SEF_G.lemma:extension_of_X_strategies}, $s_0^i$ can be extended to an $X$-strategy $s^i$, which uniquely corresponds to a strategy by Proposition~\ref{2-SEF_G.prop:strategies}. Letting $s = (s^i)_{i\in I}$, we have $s\in S$ and, by construction, $w\in R(s,w\mid h)$.
\end{proof}

\begin{proof}[Proof of Proposition~\ref{2-SEF_G.prop:scwise_SEF}]
    Let $\F$ be a stochastic pseudo-extensive form on an exogenous scenario space $(\Omega,\ms E)$, and let $\omega\in\Omega$. Let $C_\omega$ be defined as in the claim. Let $\Omega_\omega = \{\omega\}$ and $\ms E_\omega = \mc P\Omega_\omega$.\smallskip 

    (Ad ``$T_\omega$ induces a stochastic decision forest''):~ $T_\omega$ is a connected component of the decision forest $F$. Hence, by Theorem~\ref{1-SDF_AC.thm:decision_forest=forest_of_decision_trees}, it is a decision tree over $W_\omega$. With $\pi_\omega = \pi|_{T_\omega}$ and ${\X_\omega}$ being the set of maps $\Omega_\omega \to X$, $(T_\omega,\pi_\omega,{\X_\omega})$ is a stochastic decision forest on $(\Omega_\omega,\ms E_\omega)$.\smallskip

    (Ad ``$C_\omega$ is a family of sets of $\X^i_\omega$-complete choices and evaluation maps are injective''):~ Let $c\in C^i$ such that $c\cap W_\omega \neq \emptyset$. There is a non-empty set $F_c\subseteq F$ of nodes such that $c = \bigcup F_c$. As $c\cap W_\omega = \bigcup (F_c \cap T_\omega)$, $F_c \cap T_\omega$ is non-empty. Moreover, $F_c \cap T_\omega$ is a set of nodes in $T_\omega$. Hence, $c\cap W_\omega$ is a choice in $(T_\omega,\pi_\omega,{\X_\omega})$.
    
    Let ${X_\omega}$, $P_\omega(.)$, $A^i_\omega(.)$, $J_\omega(.)$, ${\X^i_\omega}$, etc.\ be associated to $({T_\omega},\pi_\omega,{\X_\omega})$ and $C_\omega$ as $X$, $P(.)$, $A^i(.)$, $J(.)$, $\X^i$ etc.\ are associated to $(F,\pi,\X)$ and $C$. We infer that for any $c\subseteq W$ we have
    \begin{equation}\label{2-SEF_G.eq:P_omega(c_omega)}
        P_\omega(c\cap W_\omega) = P(c\cap W_\omega) = P(c) \cap T_\omega,
    \end{equation}
    using the definition of $P$ and $P_\omega$, as well as Lemma~\ref{1-SDF_AC.lemma:P(c)_compatible_with_conn_comp}. 
    Furthermore, note that for all $x\in {X_\omega}$ and $\x\in {\X_\omega}$ we have
    \begin{equation}\label{2-SEF_G.eq:Ai'(x)}
        A^i_\omega(x) = \{ c \cap W_\omega \mid c\in A^i(x)\}, \qquad A^i_\omega(\x) = A^i_\omega(\x(\omega)). 
    \end{equation}
    As a consequence, we have
    \begin{equation}\label{2-SEF_G.eq:Xi'}
        {X^i_\omega} = X^i\cap T_\omega, \qquad {\X^i_\omega} =  (X^i\cap T_\omega)^{\Omega_\omega}.
    \end{equation}
    Moreover, ${\X^i_\omega}\bullet\Omega_\omega = {\X^i_\omega} \times \Omega_\omega$, and the evaluation map from that set to ${X_\omega}$ is clearly injective.
    \smallskip

    For any $i\in I$ and $\x\in{\X^i_\omega}$, let $(\ms F_\omega)^{i}_{\x} = \ms E_\omega$ and $(\ms C_\omega)^{i}_{\x} = \emptyset$. Let $\ms F_\omega = ((\ms F_\omega)^i)_{i\in I}$ and $\ms C_\omega = ((\ms C_\omega)^{i})_{i\in I}$. Let
    \[ \F_\omega = ({T_\omega},\pi_\omega,{\X_\omega},I,\ms F_\omega,\ms C_\omega,C_\omega). \]

    (Ad ``$\ms F_\omega$ and $\ms C_\omega$ define adequate \textsc{eis} and reference choice structures'' and Axioms~\ref{2-SEF_G.def:SEF.endo_exo_compatible}, \ref{2-SEF_G.def:SEF.choice_completeness}):~ For any $i\in I$, $(\ms F_\omega)^{i}$ defines an exogenous information structure on ${\X_\omega^i}$ and $(\ms C_\omega)^{i}$ defines a reference choice structure on ${\X_\omega^i}$. Trivially, any $c\in C^i_\omega$ is $(\ms F_\omega)^{i}$-$(\ms C_\omega)^{i}$-adapted. Axioms  \ref{2-SEF_G.def:SEF.endo_exo_compatible} and~\ref{2-SEF_G.def:SEF.choice_completeness} are trivially satisfied.\smallskip

    (Ad Axiom~\ref{2-SEF_G.def:SEF.P(c)}):~ Let $i\in I$ and $c_\omega,c'_\omega\in C^i_\omega$ such that $P_\omega(c_\omega) \cap P_\omega(c'_\omega) \neq \emptyset$. There are $c,c'\in C^i$ with $c_\omega = c\cap W_\omega$ and $c'_\omega = c'\cap W_\omega$. Hence, using Equation~\ref{2-SEF_G.eq:P_omega(c_omega)}, we infer $P(c) \cap P(c') \neq \emptyset$. Thus, by Axiom~\ref{2-SEF_G.def:SEF.P(c)} applied to $\F$, we get $P(c) = P(c')$. Using Equation~\ref{2-SEF_G.eq:P_omega(c_omega)}, we obtain
    \[ P_\omega(c_\omega) = P(c) \cap T_\omega = P(c') \cap T_\omega = P_\omega(c'_\omega). \]
    Furthermore, Axiom~\ref{2-SEF_G.def:SEF.P(c)} applied to $c$, $c'$, and $\omega$ implies that $c_\omega = c'_\omega$ or $c_\omega \cap c'_\omega = \emptyset$. Hence, $\F_\omega$ satisfies Axiom~\ref{2-SEF_G.def:SEF.P(c)}.\smallskip

    (Ad Axiom~\ref{2-SEF_G.def:SEF.outcomes_faithful}):~ Let $x\in T_\omega$ and $(c^i_\omega)_{i\in J_\omega(x)} \in \bigtimes_{i\in J_\omega(x)} C^i_\omega$. Using Equation~\ref{2-SEF_G.eq:Ai'(x)}, we get $J_\omega(x) = J(x)$. Hence, there is $(c^i)_{i\in J(x)} \in \bigtimes_{i\in J(x)} C^i$ such that $c^i_\omega = c^i \cap W_\omega$ for all $i\in J(x)$. Then, Axiom~\ref{2-SEF_G.def:SEF.outcomes_faithful} applied to $x$ and $(c^i)_i$ yields:
    \[ x\cap \bigcap_{i\in J_\omega(x)} c^i_\omega = x\cap \bigcap_{i\in J(x)} c^i \neq \emptyset. \]

    (Ad Axiom~\ref{2-SEF_G.def:SEF.weak_separation}):~ Let $y,y'\in T_\omega$ disjoint. By Axiom~\ref{2-SEF_G.def:SEF.weak_separation} applied to $\F$, there are $i\in I$ and $c,c'\in C^i$ such that $y\subseteq c$, $y'\subseteq c'$, and $c\cap c'\cap W_\omega = \emptyset$. Then, $c_\omega = c\cap W_\omega\neq \emptyset$ and $c'_\omega = c'\cap W_\omega \neq \emptyset$, hence, $c_\omega,c'_\omega\in C^i_\omega$, and $y\subseteq c_\omega$, $y'\subseteq c'_\omega$, and $c_\omega\cap c'_\omega = \emptyset$.\smallskip

    (Ad Axiom~\ref{2-SEF_G.def:SEF.separation}):~ For this Axiom, suppose that $\F$ is even a stochastic extensive form. Let $y,y'\in T_\omega$ disjoint. By Axiom~\ref{2-SEF_G.def:SEF.separation} applied to $\F$, there are $x\in X$, $i\in I$ and $c,c'\in C^i$ such that $y\subseteq x\cap c$, $y'\subseteq x\cap c'$, $c\cap c'\cap W_\omega = \emptyset$, and $x\in P(c) \cap P(c') \cap W_\omega$. Then, $c_\omega = c\cap W_\omega\neq \emptyset$ and $c'_\omega = c'\cap W_\omega \neq \emptyset$, hence, $c_\omega,c'_\omega\in C^i_\omega$, and $y\subseteq x\cap c_\omega$, $y'\subseteq x\cap c'_\omega$, $c_\omega\cap c'_\omega = \emptyset$, and $x\in P_\omega(c) \cap P_\omega(c')$, in view of Equation~\ref{2-SEF_G.eq:P_omega(c_omega)}.\smallskip

    (Ad Axiom~\ref{2-SEF_G.def:SEF.enough_choices}):~ Let $x\in T_\omega$, $i\in J_\omega(x)$, and $y\in \downarrow x \setminus \{x\}$. Using Equation~\ref{2-SEF_G.eq:Ai'(x)}, we get $J_\omega(x) = J(x)$. Hence, by Axiom~\ref{2-SEF_G.def:SEF.enough_choices} applied to $\F$, there is $c\in A^i(x)$ with $c\supseteq y$. Hence, by Equation~\ref{2-SEF_G.eq:Ai'(x)}, $c_\omega = c\cap W_\omega\in A^i_\omega(x)$ and $c_\omega \supseteq y$.
\end{proof}

\begin{proof}[Proof of Theorem~\ref{2-SEF_G.thm:SEF_well-posed}]
    The claimed equivalences follow directly from the following statements:
    \begin{enumerate}
        \item\label{2-SEF_G.thm:SEF_well-posed.proof.H} $H = \bigcup_{\omega\in\Omega} H_\omega$, where $H_\omega$ is the set of histories in $(T_\omega,\supseteq)$;
        \item\label{2-SEF_G.thm:SEF_well-posed.proof.Homega_Womega_comp} for all $\omega\in\Omega$ and $h\in H_\omega$, we have $\bigcap h \subseteq W_\omega$;
        \item\label{2-SEF_G.thm:SEF_well-posed.proof.si_restriction} if $s^i\in S^i$ is an $X$-strategy for agent $i\in I$ and $\omega\in\Omega$, then the map $s^i_\omega$ with domain $X^i\cap T_\omega$ defined by the assignment $x \mapsto s^i(x) \cap W_\omega $ defines an $X$-strategy in $(T_\omega,I,C_\omega)$;
        \item\label{2-SEF_G.thm:SEF_well-posed.proof.siomega_extension} if conversely $s^i_\omega$ is an $X$-strategy for an agent $i\in I$ in $(T_\omega,I,C_\omega)$, for some $\omega\in\Omega$, then there is an $X$-strategy $s^i$ for $i$ in $\F$ such that $s^i_\omega(x) = s^i(x) \cap W_\omega$ for all $x\in X^i\cap T_\omega$;
        \item\label{2-SEF_G.thm:SEF_well-posed.proof.R=R_omega} for all $s\in S$, $\omega\in\Omega$, $h\in H_\omega$ and $w\in \bigcap h$ we have
        \[ R(w,s \mid h) \cap W_\omega = R_\omega(w,s_\omega \mid h), \]
        where $s_\omega = (s^i_\omega)_{i\in I}$, $s^i_\omega$ is the restriction of $s^i$ according to \ref{2-SEF_G.thm:SEF_well-posed.proof.si_restriction} for each $i\in I$, and $R_\omega$ is the map $R(\F_\omega)$ associated to the $\psi$-\textsc{sef} $\F_\omega$ according to Definition~\ref{2-SEF_G.def:sef_well-posed}.
    \end{enumerate}

   (Ad \ref{2-SEF_G.thm:SEF_well-posed.proof.H}):~ For $\omega\in\Omega$, let $H_\omega$ be the set of histories in $(T_\omega,\supseteq)$. Then, similarly, $H$ is the disjoint union of all $H_\omega$, because histories are chains, and the connected components of $(F,\supseteq)$ are given by the collection of all $T_\omega$, $\omega\in\Omega$.\smallskip

   (Ad \ref{2-SEF_G.thm:SEF_well-posed.proof.Homega_Womega_comp}):~ For all $\omega\in\Omega$, $W_\omega$ is the root of $(T_\omega,\supseteq)$, whence the claim using Part~\ref{2-SEF_G.thm:SEF_well-posed.proof.H}.\smallskip

    (Ad \ref{2-SEF_G.thm:SEF_well-posed.proof.si_restriction}):~ Let $s^i\in S^i$ is an $X$-strategy for agent $i\in I$ and $\omega\in\Omega$. By Equation~\ref{2-SEF_G.eq:Xi'} established in the proof of Proposition~\ref{2-SEF_G.prop:scwise_SEF} we have ${X^i_\omega} = X^i \cap T_\omega$. Moreover, if $x\in X^i\cap T_\omega$, then $s^i(x) \in A^i(x)$ by definition of $s^i$. Hence, by Equation~\ref{2-SEF_G.eq:Ai'(x)} established in the proof of Proposition~\ref{2-SEF_G.prop:scwise_SEF}, we get $s^i(x) \cap W_\omega\in A^i_\omega(x)$. 
    
    Furthermore, for all $x,x'\in {X^i_\omega}$ with $A^i_\omega(x) = A^i_\omega(x')$ we have $A^i(x) = A^i(x')$. Indeed, if $x,x'\in {X^i_\omega}$ satisfy $A^i_\omega(x) = A^i_\omega(x')$, then there is $c\in C^i$ with $x,x'\in P_\omega(c\cap W_\omega) = P(c\cap T_\omega)\subseteq P(c)$, by Proposition~\ref{2-SEF_G.prop:information_sets} and Equation~\ref{2-SEF_G.eq:P_omega(c_omega)}. Thus, by Proposition~\ref{2-SEF_G.prop:information_sets}, $A^i(x) = A^i(x')$. As $s^i$ is an $X$-strategy, we obtain $s^i(x) = s^i(x')$. As a consequence, $s^i(x) \cap W_\omega = s^i(x') \cap W_\omega$.\smallskip

    (Ad \ref{2-SEF_G.thm:SEF_well-posed.proof.siomega_extension}):~ Conversely, let $s^i_\omega$ be an $X$-strategy for an agent $i\in I$ in $(T_\omega,I,C_\omega)$. Let $X_0^i$ be a representative system of the partition $\{P_\omega(c_\omega) \mid c_\omega\in C^i_\omega\}$, see Proposition~\ref{2-SEF_G.prop:information_sets}, Part~\ref{2-SEF_G.prop:information_sets.P(c)_partition}. Then, by definition of $C^i_\omega$ and by Equation~\ref{2-SEF_G.eq:Ai'(x)}, there is a map $\tilde s^i_0 \colon X^i_0 \to C^i$ with $\tilde s^i_0(x) \cap W_\omega = s^i_\omega(x)$ and $\tilde s^i_0(x) \in A^i(x)$, for all $x\in X^i_0$. For all $x,x'\in X^i_0$ with $A^i(x) = A^i(x')$, we have $A^i_\omega(x) = A^i_\omega(x')$, by Equation~\ref{2-SEF_G.eq:Ai'(x)}. Thus, $x,x'\in P_\omega(c_\omega)$ for some $c_\omega\in C^i_\omega$, by Proposition~\ref{2-SEF_G.prop:information_sets}, Part~\ref{2-SEF_G.prop:information_sets.A(x)=A(x')}, applied to $\F_\omega$. As $X^i_0$ is a representative system, we infer $x=x'$, whence $\tilde s^i_0(x) = \tilde s^i_0(x')$.

    Hence, by Lemma~\ref{2-SEF_G.lemma:extension_of_X_strategies} $\tilde s^i_0$ can be extended to an $X$-strategy $s^i$ for $i$ on $\F$. Let $x\in X^i \cap T_\omega$. Then, by Equation~\ref{2-SEF_G.eq:Xi'}, $x\in {X^i_\omega}$ and consequently there is $c_\omega\in C^i_\omega$ with $x\in P_\omega(c_\omega)$. There are $x_0\in X^i_0$ with $x_0\in P_\omega(c_\omega)$ and $c\in C^i$ with $c_\omega = c\cap W_\omega$. Hence, by Equation~\ref{2-SEF_G.eq:P_omega(c_omega)}, $x,x_0\in P(c)$ and $x,x_0 \in P_\omega(c_\omega)$, which implies both $A^i(x) = A^i(x_0)$ and $A^i_\omega(x) = A^i_\omega(x_0)$, by Proposition~\ref{2-SEF_G.prop:information_sets}, Part~\ref{2-SEF_G.prop:information_sets.A(x)=A(x')}.
    As $s^i_\omega$ and $s^i$ are $X$-strategies, we infer that
    \[ s^i_\omega(x) = s^i_\omega(x_0) = \tilde s^i(x_0) \cap W_\omega = s^i(x_0) \cap W_\omega = s^i(x) \cap W_\omega. \]

    (Ad \ref{2-SEF_G.thm:SEF_well-posed.proof.R=R_omega}):~ Let $s\in S$, $\omega\in\Omega$, $h\in H_\omega$ and $w\in \bigcap h$. 
    Then, 
    \begin{align*}
        R(w,s\mid h) \cap W_\omega =&\, \bigcap \Big\{ s^i(x) \cap W_\omega \mid x \in X,\,i\in J(x)\colon~ w\in x \subseteq \bigcap h \Big\} \\
        =&\, \bigcap \Big\{ s^i_\omega(x) \mid x \in X_\omega,\,i\in J_\omega(x)\colon~ w\in x \subseteq \bigcap h \Big\} \\
        =&\, R_\omega(w,s \mid h).
    \end{align*}
    Indeed, $x\in X$ and $i\in J(x)$ satisfy $w\in x\subseteq \bigcap h$ iff $x\in X^i \cap T_\omega$ and $w\in x\subseteq \bigcap h$, because the collection of $W_{\omega'}$, $\omega'\in\Omega$, equals the set of roots of the decision forest $F$, and is in particular a partition of $W$ (Theorem~\ref{1-SDF_AC.thm:decision_forest=forest_of_decision_trees}). But $X^i\cap T_\omega = X^i_\omega$ by Equation~\ref{2-SEF_G.eq:Xi'}. Hence, $x\in X^i\cap T_\omega$ is equivalent to $x\in X_\omega$ and $i\in J_\omega(x)$.
\end{proof}

\begin{proof}[Proof of Corollary \ref{2-SEF_G.cor:SEF_existence=~order-theoretic_properties}]
    This is a direct consequence of Theorem~\ref{2-SEF_G.thm:SEF_well-posed} and \cite[Theorem~2]{AlosFerrer2008Trees} (alias \cite[Theorem~5.2]{AlosFerrer2016Theory}). 
    
    Every $\psi$-\textsc{sef} with decision forest $F$ satisfies Property~\ref{2-SEF_G.def:sef_well-posed}.\ref{2-SEF_G.def:sef_well-posed.well_posed.existence} iff for every $\psi$-sef $\F$ with decision forest $F$ on an exogenous scenario space $(\Omega,\ms E)$, the induced classical $\psi$-\textsc{sef}s $(T_\omega,I,C_\omega)$ do so for all $\omega\in\Omega$, by Theorem~\ref{2-SEF_G.thm:SEF_well-posed}. This is the case iff for all connected components $T$ of $F$, all classical $\psi$-\textsc{sef}s with decision tree $T$ satisfy Property~\ref{2-SEF_G.def:sef_well-posed}.\ref{2-SEF_G.def:sef_well-posed.well_posed.existence}. Indeed, the ``if'' part is clear, and the ``only if'' part is shown using the exogenous scenario space $\Omega$ given by the set of connected components of $(F,\supseteq)$ with $\ms E = \mc P\Omega$. 
    
    By \cite[Theorem~2]{AlosFerrer2008Trees}, for all connected components $T$ of $F$, all classical $\psi$-\textsc{sef}s with decision tree $T$ satisfy Property~\ref{2-SEF_G.def:sef_well-posed}.\ref{2-SEF_G.def:sef_well-posed.well_posed.existence} (are ``playable everywhere'' in the language of \cite{AlosFerrer2008Trees}) iff all connected components $T$ of $F$ are weakly up-discrete and coherent with respect to $\supseteq$. But this is clearly equivalent to $(F,\supseteq)$ being weakly up-discrete and coherent.
\end{proof}

\begin{proof}[Proof of Corollary \ref{2-SEF_G.cor:SEF_well-posedness=~order-theoretic_properties}]
    This is a direct consequence of Theorem~\ref{2-SEF_G.thm:SEF_well-posed}, and \cite[Theorem~6]{AlosFerrer2011Comment} and \cite[Corollary 5]{AlosFerrer2011Comment} (alias \cite[Theorem~5.5 and Corollary 5.4]{AlosFerrer2016Theory}). For the remainder of the proof, let $\F$ be a stochastic extensive form on an exogenous scenario space $(\Omega,\ms E)$.\smallskip

    (Ad equivalence of \ref{2-SEF_G.cor:SEF_well-posedness=~order-theoretic_properties.well-posed} and~\ref{2-SEF_G.cor:SEF_well-posedness=~order-theoretic_properties.order_properties}):~ $\F$ is well-posed iff for all $\omega\in\Omega$, $(T_\omega,I,C_\omega)$ is well-posed, by Theorem~\ref{2-SEF_G.thm:SEF_well-posed}. By \cite[Theorem~6]{AlosFerrer2011Comment}, the latter is equivalent to the statement that for all $\omega\in\Omega$, $(T_\omega,\supseteq)$ is regular, weakly up-discrete, and coherent. Clearly, this is equivalent to $(F,\supseteq)$ having these three properties.\smallskip

    (Ad equivalence of \ref{2-SEF_G.cor:SEF_well-posedness=~order-theoretic_properties.well-posed} and~\ref{2-SEF_G.cor:SEF_well-posedness=~order-theoretic_properties.order_properties2}):~ $\F$ is well-posed iff for all $\omega\in\Omega$, $(T_\omega,I,C_\omega)$ is well-posed, by Theorem~\ref{2-SEF_G.thm:SEF_well-posed}. By \cite[Corollary 5]{AlosFerrer2011Comment}, the latter is equivalent to the statement that for all $\omega\in\Omega$, $(T_\omega,\supseteq)$ is regular, and up-discrete. Again, this is clearly equivalent to $(F,\supseteq)$ having these two properties.
\end{proof}

\begin{proof}[Proof of Theorem~\ref{2-SEF_G.thm:AP_sef_well-posed}]
    Let $\D$ be action path stochastic extensive form data on an exogenous scenario space $(\Omega,\ms E)$ with time $\T$ and let $\F$ be the induced stochastic extensive form.\medskip

    \emph{First implication}: Suppose $\T$ to be well-ordered.     
    In view of Theorem~\ref{2-SEF_G.thm:SEF_well-posed}, it suffices to show that the underlying decision forest $F$ is up-discrete and regular with respect to $\supseteq$.\smallskip

    (Ad up-discreteness):~ Let $c$ be a non-empty chain in $(F,\supseteq)$. Then, as $F$ is a decision forest, $\bigcap c\neq\emptyset$. Indeed, $c$ is contained in some maximal chain of the form $\uparrow \{w\}$ for some $w\in W$, see Definition~\ref{1-SDF_AC.def:decision_forest} and Proposition~\ref{1-SDF_AC.prop:f(v)=uparrow v}. Therefore, there is $w=(\omega,f)\in \bigcap c$, which is an element of $W$. Hence, there is $\T'\subseteq \T$ with
    \[ c \cup \{\{w\}\} = \{ x_t(\omega,f) \mid t\in\T'\} \cup \{\{w\}\}. \]
    If $\T'$ is empty, then $w = \max c$. If $\T'$ is non-empty, then, by hypothesis, it has a minimum $t_0$ with respect to the well-order $\le$ on $\T$. Then, by the very definition of $x_t(\omega,f)$ for $t\in\T'$ and the fact that they all contain $w$, we directly obtain $x_{t_0}(\omega,f) = \max c$.\smallskip

    (Ad regularity):~ Let $x\in F$ be non-maximal. If $\uparrow x \setminus \{x\}$ has a minimum, then it also has an infimum. It therefore remains to consider the case where $\uparrow x \setminus \{x\}$ has no minimum. For that proof, let
    \[ B_x = \{ y\in F \mid \forall z\in\uparrow x \setminus \{x\} \colon z\supseteq y\} \]
    denote the set of lower bounds of $\uparrow x \setminus \{x\}$, a set that clearly contains $x$.
    
    If $x$ is terminal, then $x = \{w\}$ for some $w = (\omega,f)\in W$. Let $y\in B_x$ and $w'\in y$. Then $w'=(\omega,f')$ for some $f'\in\A^\T$ because $\uparrow x\setminus \{x\}$ is non-empty by non-maximality of $x$, implying that $x$ and $y$ are elements of the same connected component. If we had $f\neq f'$, then by hypothesis there would be a minimal $t\in\T$ with $f(t) \neq f'(t)$. Thus, $x_t(\omega,f)\in X$, and as $\uparrow x \setminus \{x\}$ is assumed to have no minimum, there would be $u\in\T$ with $t<u$ and $x_u(\omega,f)\in \uparrow x \setminus \{x\}$. But then $w'\notin x_u(\omega,f)$, in contradiction to $w'\in y \subseteq x_u(\omega,f)$, as $y\in B_x$. Hence, the assumption $f\neq f'$ was false and we must have $f=f'$. Thus, $w'=w$ and $y=\{w\}=x$. We conclude that $B_x=\{x\}$. Hence, $B_x$ has a maximum and $\uparrow x \setminus \{x\}$ an infimum.

    If $x$ is not terminal, then $x\in X$ and $x = x_{\mf t(x)}(w)$ for some $w = (\omega,f)\in W$. Let $y\in B_x$ and let $w'\in y$. Then $w' = (\omega,f')$ for some $f'\in \A^\T$, for similar reasons as above. Then $f'=f$ or $f'\neq f$. In the latter case, the hypothesis implies the existence of minimal $t\in\T$ with $f(t) \neq f'(t)$. If we had $t<\mf t(x)$, then there would be $u\in \T$ with $t<u<\mf t(x)$, since otherwise $x_t(\omega,f)$ would be a minimum of $\uparrow x\setminus \{x\}$ which does not exist by assumption. But then, just as above, $w'\notin x_u(\omega,f)$, in contradiction to $w'\in y \subseteq x_u(\omega,f)$, as $y\in B_x$. Hence, $\mf t(x) \le t$ which implies that $f'|_{[0,\mf t(x))_\T} = f|_{[0,\mf t(x))_\T}$. Hence, whether $f=f'$ or not, it follows that $w'=(\omega,f')\in x$. We infer that $y\subseteq x$. Thus, $x$ is a maximum of $B_x$ and an infimum of $\uparrow x \setminus \{x\}$.    \medskip

    \emph{Second implication}: Suppose $\F$ to be well-posed and suppose that for any $t\in\T$ there is $f\in\A^\T$ with $D_{t,f}\neq\emptyset$. Then, by Corollary~\ref{2-SEF_G.cor:SEF_well-posedness=~order-theoretic_properties}, $(F,\supseteq)$ is up-discrete.

    Let $T\subseteq\T$ be a non-empty set of time. Let $t\in T$. Then, there is $f\in\A^\T$ with $D_{t,f}\neq\emptyset$. Hence, there is $w\in W$ such that $x_t(w)\in X$, i.e.\ is a move. By up-discreteness, the non-empty chain
    \[ \{ x_u(w) \mid u\in T\} \]
    has a maximum with respect to ``$\supseteq$''. This maximum can be written as $x_{u^\ast}(w)$ for some $u^\ast\in T$. We now show that $u^\ast$ is a (and the) minimum of $T$.

    If not, there would be $u\in T$ with $u<u^\ast$, because $\T$ is totally ordered. Note that $x_{u^\ast}(w)$ has at least two elements because it is a superset of the move $x_t(w)$. Hence, $x_u(w) \supsetneq x_{u^\ast}(w)$, by \hyperlink{1-SDF_AC.Ass:AP.SDF1}{{Assumption~AP.SDF1}} --- a contradiction to the definition of $u^\ast$. Hence, $u^\ast = \min T$. We conclude that $\T$ is well-ordered.    
\end{proof}

\begin{lemma}\label{2-SEF_G.lemma:closed_histories_in_coherent_regular_forests}
    Let $(F,\ge)$ be a coherent and regular decision forest and $h\in H$ a closed history. Then, there is $x\in F$ with $h = \uparrow x$.
\end{lemma}

\begin{proof}
    Let $h\in H$ be a closed history. It suffices to show that $h$ has a minimum $x$ since then $h = \uparrow x$, because $h$ is upward closed. 
    
    If $h$ had no minimum, then by coherence there would be a continuation $c$ with a maximum $x$. Then $h = \uparrow x \setminus \{x\}$. Indeed, if $y\in h$, then $y\supsetneq x$ because $h\cup c$ is a chain, $h$ is upward closed, and $x\notin h$. Conversely, let $z\in \uparrow x \setminus \{x\}$. For any $y\in h$, we would have $y\supseteq z$ or $z\supseteq y$ because $\uparrow x$ would be a chain containing $h$; and for any $y\in c$ we would have $z\supsetneq y$ because $x = \max c$. Hence $h\cup c \cup \{z\}$ would be a chain. By maximality of $h\cup c$ and $z\notin c$, we would get $z\in h$. 
    
    Having established that $h = \uparrow x \setminus \{x\}$, we would infer from regularity, that $h = \uparrow x \setminus \{x\}$ would have an infimum. As $h$ is supposed to be closed, it would contain its infimum and thus $h$ would have a minimum --- a contradiction.
\end{proof}

\begin{proof}[Proof of Proposition~\ref{2-SEF_G.prop:random_histories=~random_moves}]
    Let $\F$ be a well-posed stochastic extensive form on an exogenous scenario space $(\Omega,\ms E)$. The first claim follows directly from Corollary~\ref{2-SEF_G.cor:SEF_well-posedness=~order-theoretic_properties} and Lemma~\ref{2-SEF_G.lemma:closed_histories_in_coherent_regular_forests}. Next, let $\h\in\H$ be a closed random history and suppose that $(F,\pi,\X)$ is order consistent. By the first part proven just above, for any $\omega\in D_\h$, there is $x_\omega\in T_\omega$ with $\h(\omega) = \uparrow x_\omega$. For any $\omega\in D_\h$, $x_\omega \in X$ because $\h(\omega)$ is not a maximal chain, but upward closed. For every $\omega\in D_\h$, there is $\x_\omega\in\X$ with $\omega\in D_{\x_\omega}$ and $x_\omega = \x_\omega(\omega)$. 

    Let $\omega,\omega'\in D_\h$. Then, as $\h$ is a random history, $D_\h \subseteq D_{\x_\omega}\cap D_{\x_{\omega'}}$ and $\x_\omega(\omega')\in\h(\omega')$. Thus, $\x_\omega(\omega') \supseteq \x_{\omega'}(\omega')$. By order consistency, then, $\x_\omega \ge_\X \x_{\omega'}$. Repeating the argument after having permuted $\omega$ and $\omega'$ yields the converse inequality whence $\x_\omega = \x_{\omega'}$. 
\end{proof}

\begin{proof}[Proof of Theorem~\ref{2-SEF_G.thm:absent_minded_driver_Gilboa_sef_well-posed}]
    Let $p\in[0,1]$, $s\in S$ and $\Pr$ be a suitable \textsc{eu} preference structure. Let $i,j\in\{1,2\}$ with $i\neq j$ and let $\tilde s\in S$ with $\tilde s^i = s^i$. As above, $\tilde s^j$ can be identified with an exit event $\tilde E_j\in\ms F^j_{\x_j}$.

    Note that $E_{\neg i,j} = \{\rho = j\}$, $E_{i,j} = \{\rho = i\}$, and $E^{s}_{i,j} = \{\rho=i\}\cap E_i^\complement$. Hence, using the $\P$-independence of $\rho$ and $\ms F^i_{\x_i}$,
    \[ \P(E_{\neg i,j} \cup E^{s}_{i,j}) = \P(\rho = j) + \P(\rho = i)(1-\P(E_i)) = \frac 12 (1+p). \]
    Thus, for every $E\in\ms E$, dynamic consistency of $(\Pi,s)$ implies that
    \[ \P_{j,\{\x_j\}}(E) = 2\,\frac{\P((\{\rho = j\} \cup [\{\rho=i\}\cap E_i^\complement])\cap E)}{1+p}. \]
    Easy computations using the $\P$-independence of $\rho$, $\ms F^i_{\x_i}$ and $\ms F^j_{\x_j}$ yield:
    \begin{align*}
        \P_{j,\{\x_j\}}(\rho=j,\tilde E_j^\complement,E_i\mid \ms F^j_{\x_j}) =&\, \frac{(1-p)1_{\tilde E_j^\complement}}{1+p}, \\
        \P_{j,\{\x_j\}}(\rho=i,\tilde E_j\mid \ms F^j_{\x_j}) =&\, \frac{p1_{\tilde E_j}}{1+p}, \\
        \P_{j,\{\x_j\}}(\rho=j,\tilde E_j^\complement,E_i^\complement\mid \ms F^j_{\x_j}) =&\, \frac{p1_{\tilde E_j^\complement} }{1+p}, \\
        \P_{j,\{\x_j\}}(\rho=i,\tilde E_j^\complement\mid \ms F^j_{\x_j}) =&\, \frac{p1_{\tilde E_j^\complement}}{1+p}. \\
    \end{align*}
    
    Then,
    \begin{align*}
        \pi_{j,\{\x_j\}} =&\, \E_{j,\{\x_j\}}\Big[4 \cdot \Big(1\{\rho=j,\tilde E_j^\complement,E_i\} + 1\{\rho=i,\tilde E_j\}\Big) \\
        & \qquad\quad + 1 \cdot \Big(1\{\rho=j,\tilde E_j^\complement,E_i^\complement\} + 1\{\rho=i,\tilde E_j^\complement\}\Big) \\
        & \qquad\quad + 0 \cdot 1\{\rho=j,\tilde E_j\} \Big] \\
        =&\, \frac{1}{1+p} \Big(1_{\tilde E_j^\complement} \big(4(1-p)+p+p\big) + 1_{\tilde E_j} 4p\Big)\\
        =&\, \frac{1}{1+p} \Big(1_{\tilde E_j^\complement} \big(4-2p\big) + 1_{\tilde E_j} 4p\Big).\\
    \end{align*}
    
    For $p<\frac 23$, this is maximised $\P_{j,\{\x_j\}}$-almost surely by all $\tilde E_j\in\ms F^j_{\x_j}$ with $\P_{j,\{\x_j\}}(\tilde E_j) = 0$. If $(s,\Pr)$ were in equilibrium, then, $p = 1 - \P(E_j) = 1$, a contradiction. 

    For $p>\frac 23$, conversely, the above expression is maximised $\P_{j,\{\x_j\}}$-almost surely by all $\tilde E_j\in\ms F^j_{\x_j}$ with $\P_{j,\{\x_j\}}(\tilde E_j) = 1$. If $(s,\Pr)$ were in equilibrium, then, $p = 1 - \P(E_j) = 0$, a contradiction. 

    Hence, if $(s,\Pr)$ is in equilibrium, then $p = \frac 23$. Conversely, suppose that $p = \frac 23$. Then, by the computation above, $\pi_{j,\{\x_j\}} = \frac 85$, which is independent on the chosen strategy $\tilde s^j$ alias $\tilde E_j$. In particular, $\tilde s^j$ is a best response to $s^i$. Switching the roles of $i$ and $j$ also shows that $\pi_{i,\{\x_i\}} = \frac 85$ so that $(s,\Pr)$ is in equilibrium.
\end{proof}

\section{Chapter~\ref{chap:3-SPF_VECT}}
\subsection{Section~\ref{3-SPF_VECT.sec:vERT}}

\subsubsection{The complete total order $\ovT$}

\begin{proof}[Proof of Lemma~\ref{3-SPF_VECT.lemma:suprema_and_infima_in_ovT}]
    Let $\alpha\in{\mf w_1}+1$ and $S\subseteq \ovTa$ be some subset, $a=\inf\mc Pp(S)$, $b=\sup\mc Pp(S)$ in $\bRp$.
    
    If $a \in \mc Pp(S)$, then $\mc P\pi(S \cap [\{a\} \times (\sup\alpha+1)])\neq\emptyset$. Hence, this set has a minimum $\gamma$ in $\sup\alpha+1$. Then, $(a,\gamma)$ defines a minimum of $S$. If $a\in\R_+\setminus\mc Pp(S)$, then $(a,\sup\alpha)$ is an infimum of $S$ in $\ovTa$. Else, $a = \infty$ and $S=\emptyset$. Then, $\infty$ is an infimum of $S$ in $\ovT$.

    If $b\in\mc Pp(S)$, then $\mc P\pi(S \cap [\{b\} \times (\sup\alpha+1)])\neq\emptyset$. Thus, this set has a supremum $\gamma$ in $\sup\alpha+1$. Then, $(b,\gamma)$ defines a supremum of $S$. If $b\notin\mc Pp(S)$, then $(b,0)$ is a supremum of $S$.
\end{proof}

\begin{proof}[Proof of Proposition~\ref{3-SPF_VECT.prop:ovT_small_completion_of_T}]
    $\ovTa$ is a complete lattice by Lemma~\ref{3-SPF_VECT.lemma:suprema_and_infima_in_ovT}. It then suffices to show that, via set inclusion, $\ovTa$ defines a dense completion of $\Ta$ (see Corollary~\ref{3-SPF_VECT.cor:dense=>small}). 
    
    For this, note that, in case $\alpha>0$, for any $t\in\R_+$, we have $\sup A_t = (t,\sup\alpha) = \inf B_t$, where
    \[ A_t = p^{-1}([0,t]_{\R_+})\cap \Ta, \qquad B_t = p^{-1}((t,\infty)_{\R_+})\cap \Ta, \]
    by Lemma~\ref{3-SPF_VECT.lemma:suprema_and_infima_in_ovT}. Moreover, the same lemma implies $\inf \emptyset = \infty = \sup \T$. In view of Equation~\ref{3-SPF_VECT.eq:ovT=union(T,tops)}, $\ovTa$ is a dense completion of $\Ta$.  
\end{proof}

\subsubsection{Topology and $\sigma$-algebras on $\ovT$}

\begin{proof}[Proof of Lemma~\ref{3-SPF_VECT.lemma:msO_ovT.subbase_base}]
    (Ad ``$\ms G_\ovT(\T)$ is a subbase of $\ms O_\ovT$''):~
    It suffices to show that elements of $\ms G_\ovT(\ovT)$ are unions of subsets of $\ms G_\ovT(\T)$. Let $t\in\ovT$. Then, there are $A_t,B_t\subseteq \T$ with $\sup A_t = t = \inf B_t$. We infer that
    \[ [0,t)_\ovT = \bigcup_{u\in A_t} [0,u)_\ovT, \qquad (t,\infty]_\ovT = \bigcup_{u\in B_t} (u,\infty]_\ovT. \]

    (Ad ``$\ms U_\ovT$ is a base of $\ms O_\ovT$''):~ We have just seen that $\ms G_\ovT(\T)$ is a subbase of $\ms O_\ovT$.
    It is evident that $\ms U_\ovT \cup \{\ovT\}$ is the set of intersections of finite subsets of $\ms G_\ovT(\T)$,\footnote{... the empty intersection being equal to $\ovT$.} and, by basic topology, a base of $\ms O_\ovT$. Moreover, $\ovT = [0,1)_\ovT \cup (0,\infty]_\ovT$. Hence, $\ms U_\ovT$ is a base, too.
\end{proof}

\begin{proof}[Proof of Proposition~\ref{3-SPF_VECT.prop:msK_compact_class_generating_msI}]    
    (Ad compactness of $[t,u]_\ovT$):~ It is well-known that complete totally ordered lattices are compact. We nevertheless give a proof for the reader's convenience.
    
    Let $t,u\in\ovT$ and $\ms C$ be an open covering of $[t,u]_\ovT$ in $\ovT$, i.e.\ $\ms C \subseteq\ms O_\ovT$ and $[t,u]_\ovT \subseteq\bigcup\ms C$. We have to show that $\ms C$ admits a finite subcovering. By Lemma~\ref{3-SPF_VECT.lemma:msO_ovT.subbase_base} and basic topology (Alexander subbase theorem), it suffices to consider the case where $\ms C\subseteq \ms G_\ovT$. 
    Let, in the complete lattice $\ovT$,
    \[ a = \inf \{ t'\in \T \mid (t',\infty]_\ovT \in \ms C\}, \qquad b = \sup \{ u'\in \T \mid [0,u')_\ovT \in \ms C\}. \]
    If $a<t$, then there is $t'\in\T$ with $t'<t$ such that $(t',\infty]_\ovT\in\ms C$, whence the finite subcovering $[t,u]_\ovT \subseteq (t',\infty]_\ovT$. Similarly, if $u<b$, we get the a finite subcovering $[t,u]_\ovT \subseteq [0,u')_\ovT\in\ms C$ for some $u'\in\T$. 
    
    It remains to consider the case $t\le a$ and $b\le u$. We claim that $a<b$. As $b\le u$ and $\ms C$ covers $[t,u]_\ovT$, there is $t'\in\T$ such that $u\in(t',\infty]_\ovT\in\ms C$, whence $a < u$. Thus $a\in[t,u]_\ovT$, whence we infer -- using the definition of $a$ and the covering property of $\ms C$ -- the existence of $u'\in\T$ such that $a\in [0,u')_\ovT\in\ms C$. Hence, $a<b$. As a consequence, there are $t',u'\in\T$ with $a\le t'<u'\le b$ such that $[0,u')_\ovT,(t',\infty]_\ovT\in\ms C$. We then have $[t,u]_\ovT \subseteq \ovT = [0,u')_\ovT \cup (t',\infty]_\ovT$.\smallskip

    (Ad ``$\ms K_\ovT(\T)$ is an intersection-stable compact class''):~ As a total order, $\T$ is a lattice. Hence, $\ms K_\ovT(\T)$ is intersection-stable. It is a compact class, by basic topology, because its elements are compact with respect to the fixed topology $\ms O_\ovT$.\smallskip
    
    (Ad $\ms I_\ovT(\T) = \sigma(\ms G_\ovT(\T)) = \sigma(\ms K_\ovT(\T))$):~ The first equality is the definition. For the second one, let $u,t\in\T$. Then, $\pi(t)$ is countable. Thus,
    \[ (u,\infty]_\ovT = \ovT \setminus [0,u]_\ovT, \qquad [0,t)_\ovT = \bigcup_{\gamma\in\pi(t)} \big[0,(p(t),\gamma)\big]_\ovT, \]
    are elements of $\sigma(\ms K_\ovT(\T))$. Conversely,
    \[ [t,u]_\ovT = \ovT \setminus \big([0,t)_\ovT \cup (u,\infty]_\ovT\big),\]
    which is an element of $\sigma(\ms G_\ovT(\T))$. We conclude that $\sigma(\ms G_\ovT(\T))=\sigma(\ms K_\ovT(\T))$.
\end{proof}

\begin{proof}[Proof of Corollary~\ref{3-SPF_VECT.cor:msI_is_generated_by_all_kinds_of_principal_up/down-sets}]
    We denote the four sets in the claim, ordered from left to right, by $\mc M_i$, $i=1,\dots,4$. Clearly, all these four sets are intersection-stable and, by the preceding Proposition~\ref{3-SPF_VECT.prop:msK_compact_class_generating_msI} and complement-stability, contained in $\ms I_\ovT(\T)$. It is also clear that $\ms G_\ovT(\T) = \mc M_2 \cup \mc M_4$ and $\ms K_\ovT(\T) \subseteq \sigma(\mc M_1 \cup \mc M_3)$. We conclude that 
    \[ \ms I_\ovT(\T) = \sigma(\mc M_1 \cup \mc M_3) = \sigma(\mc M_2 \cup \mc M_4). \]
    
    It remains to show that $\sigma(\mc M_i) = \sigma(\mc M_{i+2})$, for both $i=1,2$. As $\sigma(\mc M_1) = \sigma(\mc M_4)$ and $\sigma(\mc M_2) = \sigma(\mc M_3)$, this is equivalent to proving $\sigma(\mc M_1) = \sigma(\mc M_2)$.    
    
    Let $u\in\T$. 
    Note that
    \[ [0,u]_\ovT = \big[0,(p(u),\pi(u)+1)\big)_\ovT.\]
    This is an element of $\mc M_2$ since $\pi(u)+1<\mf w_1$. 
    Hence, $\mc M_1\subseteq\mc M_2$, and thus $\sigma(\mc M_1) \subseteq \sigma(\mc M_2)$.

    Regarding the converse inclusion, if $u = 0$, then $[0,u)_\ovT = \emptyset$. If $u\in\R_+\setminus\{0\}$, then there is a sequence $(u_n)_{n\in\N}$ valued in $[0,u)_{\R_+}$ converging to $u$, and we have
    \[ [0,u)_\ovT = \bigcup_{n\in\N} [0,u_n]_\ovT. \]
    Else, $u\in\T\setminus \R_+$. Then, $\pi(u) > 0$ and
    \[ [0,u)_\ovT = \bigcup_{\gamma\in\pi(u)} \big[0,(p(u),\gamma)\big]_\ovT, \]
    a countable union. In all three cases, $[0,u)_{\ovT} \in\sigma(\mc M_1)$. We conclude that $\sigma(\mc M_1) = \sigma(\mc M_2)$, completing the proof.
\end{proof}

\begin{proof}[Proof of Lemma~\ref{3-SPF_VECT.lemma:generator_msP}]
    (Ad Part~\ref{3-SPF_VECT.lemma:generator_msP.alpha}):~ This property is the subject of a classical exercise on product $\sigma$-algebras; its proof is sketched for the reader's convenience. Let $\alpha\in\mf w_1$ and $\ms A^\alpha$ denote the $\sigma$-algebra on $\ovT$ generated by $\ms G_{\ovT,\times}^\alpha$. 

    First, for fixed $C\in\ms G_{\alpha+1}$, consider the set $\ms D_\bRp^\alpha$ of $D\in\ms B_\bRp$ such that $(\rho^\alpha)^{-1}(D\times C)\in\ms A^\alpha$. By construction, $\ms D_\bRp$ contains the intersection-stable generator $\ms G_\bRp$ of $\ms B_\bRp$. Moreover, it is a Dynkin system on $\bRp$. Indeed, there is a countable subset $\ms C_{\bRp}\subseteq \ms G_{\bRp}$ with $\bigcup \ms C_{\bRp} = \bRp$, whence 
    \[ (\rho^\alpha)^{-1}\big(\bRp\times C\big) = \bigcup_{B\in\ms C_{\bRp}} (\rho^\alpha)^{-1}\big(B\times C\big)~ \in \ms A^\alpha. \]
    Stability under complements and countable (disjoint) unions is easily verified. 
    Hence, by Dynkin's $\pi$-$\lambda$-theorem, $\ms D_\bRp^\alpha = \ms B_\bRp$.
    
    Second, for fixed $B\in\ms B_\bRp$ consider the set $\ms D_{\alpha+1}$ of $D\in\ms B_{\alpha+1}$ such that $(\rho^\alpha)^{-1}(B\times D)\in\ms A^\alpha$. By the first step, $\ms D_{\alpha+1}$ contains the intersection-stable generator $\ms G_{\alpha+1}$ of $\ms B_{\alpha+1}$. Moreover, it is a Dynkin system on $\alpha+1$. Indeed, there is a countable subset $\ms C_{\alpha+1}\subseteq \ms G_{\alpha+1}$ with $\bigcup \ms C_{\alpha+1} = \alpha+1$, whence 
    \[ (\rho^\alpha)^{-1}\big(B\times (\alpha+1)\big) = \bigcup_{C\in\ms C_{\alpha+1}} (\rho^\alpha)^{-1}\big(B\times C\big)~ \in \ms A^\alpha. \]
    Stability under complements and countable (disjoint) unions is also easily verified. 
    Hence, by Dynkin's $\pi$-$\lambda$-theorem, $\ms D_{\alpha+1} = \ms B_{\alpha+1}$.

    As the set of products $B\times C$ ranging over $B\in\ms B_\bRp$ and $C\in\ms B_{\alpha+1}$ generates $\ms B_{\bRp\times(\alpha+1)}$, we infer that $\rho^\alpha$ is $\ms A^\alpha$-$\ms B_{\bRp\times(\alpha+1)}$-measurable. By definition of $\ms P_\ovT^\alpha$, we infer that $\ms P_\ovT^\alpha \subseteq\ms A^\alpha$. On the other hand, we have $\ms G_{\ovT,\times}^\alpha\subseteq \ms P_\ovT^\alpha$, whence $\ms A^\alpha \subseteq \ms P_\ovT^\alpha$. As $\ms G_{\ovT,\times}^\alpha$ inherits intersection-stability from $\ms G_\bRp$ and $\ms G_{\alpha+1}$, Claim~\ref{3-SPF_VECT.lemma:generator_msP.alpha} obtains.\smallskip

    (Ad Part~\ref{3-SPF_VECT.lemma:generator_msP.mfw1}):~ Let $\ms A$ denote the $\sigma$-algebra on $\ovT$ generated by $\ms G_{\ovT,\times}$. By construction and Part~\ref{3-SPF_VECT.lemma:generator_msP.alpha}, we have $\ms P_\ovT^\alpha\subseteq\ms A$ for all $\alpha\in\mf w_1$. Hence, by Part~\ref{3-SPF_VECT.lemma:generator_msP.alpha}, for any $\alpha\in\mf w_1$, $\rho^\alpha$ is $\ms A$-$\ms B_{\bRp\times(\alpha+1)}$-measurable. By definition of $\ms P_\ovT$, therefore, $\ms P_\ovT\subseteq\ms A$. On the other hand, $\ms G_{\ovT,\times}\subseteq\ms P_\ovT$, whence the converse inclusion $\ms A \subseteq \ms P_\ovT$.

    Regarding the second sentence, suppose that for all $\alpha,\beta\in\mf w_1$ with $\alpha < \beta$ and all $C\in\ms G_{\alpha+1}$, we have either $\alpha\in C$ and $C\cup (\alpha,\beta]_{\mf w_1} \in \ms G_{\beta+1}$, or $\alpha\notin C$ and $C\in\ms G_{\beta+1}$. Then, we claim that for all $\alpha,\beta\in\mf w_1$ with $\alpha < \beta$, $\ms G_{\ovT,\times}^\alpha\subseteq \ms G_{\ovT,\times}^\beta$ which then implies the claim, because $\ms G_{\ovT,\times}^\beta$ is intersection-stable by Part~\ref{3-SPF_VECT.lemma:generator_msP.alpha}. For proving the claim just made, let $\alpha,\beta\in\mf w_1$ be such that $\alpha < \beta$ and let $B\in\ms G_\bRp$ and $C\in\ms G_{\alpha+1}$. If $\alpha\notin C$, then $C\in\ms G_{\beta+1}$ and, thus $(\rho^\alpha)^{-1}(B\times C) = (\rho^\beta)^{-1}(B\times C)\in \ms G_{\ovT,\times}^\beta$. If $\alpha\in C$, then $C\cup (\alpha,\beta]_{\mf w_1} \in \ms G_{\beta+1}$, and, thus, $(\rho^\alpha)^{-1}(B\times C) = (\rho^\beta)^{-1}(B\times (C\cup (\alpha,\beta]_{\mf w_1}))\in \ms G_{\ovT,\times}^\beta$. This proves the claim.\smallskip

    (Ad Part~\ref{3-SPF_VECT.lemma:generator_msP.compact_class}):~ Suppose that $\ms G_Y$ consists of compact sets in $Y$ for all $Y\in\{\bRp\} \cup {\mf w_1}$ and the only element $B$ of $\ms G_\bRp$ with $\infty\in B$ is $B = \{\infty\}$. To start, we recall that the product of two compact topological spaces is compact.\footnote{See, e.g., \cite[Appendix~A1]{Rudin1991Functional}.} Hence, if $B\subseteq\bRp$ and $C\subseteq\alpha+1$ are compact, for some $\alpha\in\mf w_1$, then $B\times C$, equipped with product topology, is compact as well. 

    Let $B\in\ms G_\bRp$, $\alpha\in\mf w_1$, and $C\in\ms G_{\alpha+1}$. If $\infty\notin B$, then
    \[ (\rho^\alpha)^{-1}(B\times C) = \begin{cases} B\times C, &\text{if } \alpha\notin C, \\ B\times(C\cup[\alpha+1,\mf w_1]_{\mf w_1+1}), &\text{else,}\end{cases} \]
    which is compact with respect to the product topology. If $\infty\in B$, then $B = \{\infty\}$, whence
    \[ (\rho^\alpha)^{-1}(B\times C) = \begin{cases} B\times \{0\}, &\text{if } 0\in C, \\ \emptyset, &\text{else,}\end{cases} \]
    which is equally compact. Hence, under the hypotheses made, all elements of $\ms G_{\ovT,\times}$ can be seen as compact subsets of $\bRp\times(\mf w_1+1)$ equipped with product topology. Therefore, $\ms G_{\ovT,\times}$, and also its subsets $\ms G_{\ovT,\times}^\alpha$, $\alpha\in\mf w_1$, all yield compact classes.
\end{proof}

\begin{proof}[Proof of Lemma~\ref{3-SPF_VECT.lemma:sets_in_msPovT^alpha_otimes_msE_are_inactive_beyond_alpha}]
    Let
    \[ \ms M_\alpha = \Big\{ M\in\ms P_\ovT \otimes \ms E \mid \forall x,y\in\ovT \setminus \Ta\forall\omega\in\Omega\colon \big[(x,\omega)\in M \text{ and } p(x) = p(y) ~ \Rightarrow ~ (y,\omega) \in M\big]\Big\}. \]
    We have to show that $\ms P_\ovT^\alpha\otimes\ms E \subseteq \ms M_\alpha$. In view of the definition of $\ms P_\ovT^\alpha$ and basic measure theory, it suffices to show that:
    \begin{enumerate}
        \item\label{3-SPF_VECT.cor:sets_in_msTovT^alpha_otimes_msE_are_inactive_beyond_alpha.msMa_contains_generator} for all $B\in\ms B_{\bRp\times(\alpha+1)}$ and $E\in\ms E$,  $(\rho^\alpha)^{-1}(B)\times E\in\ms M_\alpha$,
        \item\label{3-SPF_VECT.cor:sets_in_msTovT^alpha_otimes_msE_are_inactive_beyond_alpha.msMa_sigma-algebra} $\ms M_\alpha$ is a $\sigma$-algebra on $\ovT\times\Omega$.
    \end{enumerate}
    
    (Ad Statement~\ref{3-SPF_VECT.cor:sets_in_msTovT^alpha_otimes_msE_are_inactive_beyond_alpha.msMa_contains_generator}):~ Let $B\in\ms B_{\bRp\times(\alpha+1)}$ and $E\in\ms E$. Let $x,y\in\ovT\setminus\Ta$ and $\omega\in\Omega$ such that $(x,\omega)\in (\rho^\alpha)^{-1}(B)\times E$ and $p(x) = p(y)$. As $\pi(x),\pi(y) \ge \alpha$, 
    \[ \rho^\alpha(x) = (p(x),\alpha) = (p(y),\alpha) = \rho^\alpha(y). \]
    Hence $(y,\omega)\in (\rho^\alpha)^{-1}(B)\times E$.\smallskip

    (Ad Statement~\ref{3-SPF_VECT.cor:sets_in_msTovT^alpha_otimes_msE_are_inactive_beyond_alpha.msMa_sigma-algebra}):~ Clearly, $\emptyset\in\ms M_\alpha$. Next, let $M\in\ms M_\alpha$. Then, $M^\complement\in\ms P_\ovT\otimes\ms E$. For the proof, let $x,y\in\ovT\setminus\Ta$ and $\omega\in\Omega$ such that $(y,\omega)\notin M^\complement$. Thus, $(y,\omega)\in M$. If $p(x) = p(y)$, then $(x,\omega)\in M$, whence $(x,\omega)\notin M^\complement$. By contraposition, $M^\complement\in\ms M_\alpha$. Finally, let $(M_n)_{n\in\N}$ be an $\ms M_\alpha$-valued sequence and $M = \bigcup_{n\in\N} M_n$. Let $x,y\in\ovT\setminus\Ta$ and $\omega\in\Omega$ such that $(x,\omega)\in M$ and $p(x) = p(y)$. Then, there is $n\in\N$ such that $(x,\omega)\in M_n$. Hence, $(y,\omega)\in M_n$. Therefore, $(y,\omega)\in M$. The proof is complete. 
\end{proof}

\begin{proof}[Proof of Proposition~\ref{3-SPF_VECT.prop:exhaust_msP_ovT_otimes_msE}]
    The second equality a direct consequence of the definition of $\ms P_\ovT^\alpha$, $\alpha\in\mf w_1$, and the product $\sigma$-algebra. We therefore focus on the proof of the first equality.\smallskip

    (Ad ``$\supseteq$'' and monotonicity of the union):~ Let $\alpha,\beta\in{\mf w_1}$ be such that $\alpha\le\beta$. Then, by Lemma~\ref{3-SPF_VECT.lemma:generator_msP} and Example~\ref{3-SPF_VECT.ex:generator_msP}, we find generators $\ms G^\gamma_\ovT$ of $\ms P^\gamma_\ovT$, $\gamma\in\mf w_1$, such that $\ms G^\alpha_\ovT\subseteq\ms G^\beta_\ovT$. Hence, $\ms P_\ovT^\alpha\subseteq\ms P_\ovT^\beta$. By construction, $\ms P_\ovT^\beta\subseteq\ms P_\ovT$. Then, it follows directly from the definition of the product $\sigma$-algebra -- i.e.\ the smallest one making the set-theoretic projections measurable -- that
    \[ \ms P_\ovT\otimes\ms E \supseteq \ms P_\ovT^\beta \otimes\ms E \supseteq \ms P_\ovT^\alpha \otimes\ms E. \]

    (Ad ``$\subseteq$''):~ Let $\ms A = \bigcup_{\alpha\in{\mf w_1}}\ms P_\ovT^\alpha\otimes\ms E$. In view of Equation~\ref{3-SPF_VECT.eq:msP_ovT^alpha_def}, and by basic measure theory, $\ms A$ contains a generator of $\ms P_\ovT\otimes\ms E$, namely
    \[ \{ (\rho^\alpha)^{-1}(B) \times E \mid \alpha\in\mf w_1,\, B\in\ms B_{\bRp\times(\alpha+1)},\, E\in\ms E\}. \]
    It therefore suffices to show that $\ms A$ is a $\sigma$-algebra on $\ovT\times\Omega$. As $\ms P_\ovT^0\otimes\ms E$ is a $\sigma$-algebra on $\ovT\times\Omega$, $\emptyset\in\ms A$. If $A\in\ms A$, then there is $\alpha\in{\mf w_1}$ such that $A\in\ms P_\ovT^\alpha\otimes\ms E$. Hence, $A^\complement\in\ms P_\ovT^\alpha\otimes\ms E \subseteq \ms A$. Finally, let $(A_n)_{n\in\N}$ be an $\ms A$-valued sequence. For any $n\in\N$, there is minimal $\alpha_n\in{\mf w_1}$ such that $A_n\in\ms P_\ovT^{\alpha_n}\otimes\ms E$. Let $\alpha = \sup_{n\in\N} \alpha_n$. As the supremum of a countable set of countable ordinals, $\alpha$ is a countable ordinal as well, i.e.\ $\alpha\in{\mf w_1}$. Using monotonicity of the union, proven in the first step above, we infer that, for all $n\in\N$, $A_n\in\ms P_\ovT^\alpha\otimes\ms E$. Hence, $\bigcup_{n\in\N} A_n\in\ms P_\ovT^\alpha\otimes\ms E\subseteq\ms A$. We conclude that $\ms A$ is a $\sigma$-algebra, thereby completing the proof.
\end{proof}

\begin{proof}[Proof of Corollary~\ref{3-SPF_VECT.cor:msP_ovT_quasi_subsetneq_msB_ovT}]
    (Ad inclusion):~  
    Let $\alpha\in\mf w_1$. Let, in addition, $B = [0,t)_\bRp$ and $C = \beta \setminus \{0\}$ for $t\in\R_+$ and $\beta\in\alpha+1$. Then
    \[ (\rho^\alpha)^{-1}(B\times C) = B \times C = \bigcup_{x\in B} ((x,0),(x,\beta))_\ovT~\in\ms O_\ovT\subseteq\ms B_\ovT. \]
    Furthermore, for open real intervals $B$ as above, we have
    \[ (\rho^\alpha)^{-1}(B\times (\alpha+1)) = B \times (\mf w_1+1) = \bigcap_{n\in\N} [0,t+2^{-n})_\ovT~\in\ms B_\ovT. \]
    Moreover, 
    \[ (\rho^\alpha)^{-1}(B\times\{0\}) = \begin{cases} (\rho^\alpha)^{-1}(B\times(\alpha+1)), &\text{if } \alpha = 0, \\ B\times\{0\}, &\text{else.} \end{cases} \]
    Hence, for $B$ as above and any $\beta\in\alpha+2$, we get 
    \[ (\rho^\alpha)^{-1}(B\times\beta) \in\ms B_\ovT\vee \sigma(B'\times\{0\} \mid B'\in\ms B_\bRp). \]
    Moreover, for $B = \{\infty\}$ and $C \in \alpha+2$, we have
    \[ (\rho^\alpha)^{-1}(B\times C) = \begin{cases} \emptyset, &\text{if } C\cap\{0\} = \emptyset, \\ \{\infty\}, &\text{else,} \end{cases}\quad \in\ms B_\ovT. \]
    
    Let $\ms G_\bRp$ be the set consisting of a) all intervals $[0,t)_\bRp$ running over $t\in\R_+$ and b) the set $\{\infty\}$, and $\ms G_{\alpha+1} = \alpha+2$, for any $\alpha\in\mf w_1$. With the notations from Lemma~\ref{3-SPF_VECT.lemma:generator_msP}, we have $\ms G_{\ovT,\times}\subseteq \ms B_\ovT\vee \sigma(B'\times\{0\} \mid B'\in\ms B_\bRp)$. Hence, by Lemma~\ref{3-SPF_VECT.lemma:generator_msP}, the claimed inclusion obtains.\smallskip

    (Ad inequality):~ Let $V\subseteq\R_+$ a non-Lebesgue-measurable set, and $U = V \times \{1\} \subseteq\ovT$. Then, 
    \[ U = \bigcup_{t\in V} (t,(t,2))_{\ovT} ~\in\ms O_\ovT\subseteq\ms B_\ovT. \]
    If we had $U\in\ms P_\ovT$, then there would be $\alpha\in\mf w_1$ with $U\in\ms P_\ovT^\alpha$, by Equation~\ref{3-SPF_VECT.eq:msP_ovT} following Proposition~\ref{3-SPF_VECT.prop:exhaust_msP_ovT_otimes_msE}. More precisely, $U = (\rho^\alpha)^{-1}(B)$ for some $B\in\ms B_{\bRp\times(\alpha+1)}$. Then, $\alpha\ge 2$, and $B = U = V\times\{1\}$. As $\alpha+1$ is Polish, measurable projection onto $\R_+$ would imply that $V \in (\ms B_\bRp)^{\mathrm u}$ --- which is false. We conclude that $U\notin\mc P_\ovT$.
\end{proof}

\begin{proof}[Proof of Lemma~\ref{3-SPF_VECT.lemma:msI_ovT_subseteq_msP_ovT}]
    It suffices to show that $\ms G_\ovT(\T)\subseteq\ms P_\ovT$. Let $t\in\T$ and $\alpha = \pi(t)$. Then,
    \[ [0,t)_\ovT = (\rho^0)^{-1}([0,p(t))_\bRp\times 1) \cup (\rho^\alpha)^{-1}(\{p(t)\} \times \alpha), \]
    which is an element of $\ms P_\ovT^\alpha\subseteq\ms P_\ovT$. Similarly, 
    \[ (t,\infty]_\ovT = (\rho^0)^{-1}((p(t),\infty]_\bRp\times 1) \cup (\rho^{\alpha+1})^{-1}(\{p(t)\} \times \{\alpha+1\}), \]
    which is an element of $\ms P_\ovT^{\alpha+1}\subseteq\ms P_\ovT$.
\end{proof}

\begin{proof}[Proof of Lemma~\ref{3-SPF_VECT.lemma:iota_alpha_&_p_mb}]
    (Ad $\iota_\alpha$):~ Let $\alpha,\beta\in{\mf w_1}$ and $B\in\ms B_{\bRp}$, $C\subseteq\beta+1$. Then,
    \[ \iota_\alpha^{-1}((\rho^\beta)^{-1}(B\times C)) = (\rho^\beta\circ\iota_\alpha)^{-1}(B\times C) = \begin{cases} B \times (C\cap(\alpha+1)), &\text{if } \alpha\le\beta, \\ B \times C, &\text{if } \alpha > \beta,\, \beta \notin C, \\ B\times (C\cup (\beta,\alpha]_{\mf w_1}), &\text{if }\alpha>\beta,\, \beta\in C. \end{cases} \]
    In view of the definition of $\ms P_\ovT$, we infer that $\iota_\alpha$ is $\ms B_{\bRp\times(\alpha+1)}|_{\ovT_{\alpha+1}}$-$\ms P_\ovT$-measurable.\smallskip

    (Ad $p$):~ Let $c\in\R_+$ with $c>0$. Then, for all $\alpha\in\mf w_1$,
    \[ p^{-1}([0,c)_\bRp) = \bigcup_{n\in\N} [0,c(1-2^{-n})]_\ovT = (\rho^\alpha)^{-1}([0,c)_\bRp\times(\alpha+1)), \]
    which is thus an element of both $\ms I_\ovT(\T)$ and $\ms P_\ovT^\alpha$. Hence, $p$ is measurable with respect to both $\sigma$-algebras.
\end{proof}

\subsubsection{Continuous functions on $\ovT$}

\begin{proof}[Proof of Lemma~\ref{3-SPF_VECT.lemma:left/right-limit_points_in_ovT}]
    If $t\in\ovT$ is such that $\pi(t) = \beta+1$ for some $\beta\in\On$, then for $u = (p(t),\beta)$, we have $(u,\infty]_\ovT \cap [0,t)_\ovT = \emptyset$. Thus, $t$ is not a left-limit point. Moreover, $\ovT \cap [0,0)_\ovT = \emptyset$, hence $0$ is not a left-limit point either.

    Let $t\in\ovT\setminus\{0\}$. Then, every neighbourhood $U$ of $t$ contains an open interval $(t_0,u_0)_\ovT$, with $t_0 < t < u_0$, $t_0,u_0\in\T$, by Lemma~\ref{3-SPF_VECT.lemma:msO_ovT.subbase_base}.
    If $\pi(t) = 0$, then $p(t_0) < p(t)$. Hence, there is $x\in\R_+$ with $p(t_0) < x < p(t)$, whence $x\in U \cap [0,t)_\ovT$. Thus, $t$ is a left-limit point. 
    If $\pi(t)$ is a limit ordinal, then $u = (p(t_0),\pi(t_0) + 1) \in \T$ satisfies $t_0 <u<t$. Thus, $t$ is a left-limit point.\smallskip

    $\infty$ is clearly not a right-limit point, because $\ovT\cap(\infty,\infty]_\ovT = \emptyset$. Let $t\in\ovT\setminus\{\infty\}$. If $\pi(t) < {\mf w_1}$, then $u = (p(t),\pi(t) + 1) \in\T$. Thus, $[0,u)_\ovT \cap (t,\infty]_\ovT = \emptyset$, and $t$ is therefore not a right-limit point. If $\pi(t) = {\mf w_1}$, then, again, every neighbourhood $U$ of $t$ contains an open interval $(t_0,u_0)_\ovT$, with $t_0 < t < u_0$, $t_0,u_0\in\T$, by Lemma~\ref{3-SPF_VECT.lemma:msO_ovT.subbase_base}. Hence, $p(t) < p(u_0)$. Therefore, there is $x\in\R$ with $p(t) < x < p(u_0)$. Hence, $x\in U \cap (t,\infty]_\ovT$. Therefore, $t$ is a right-limit point.
\end{proof}

\begin{proof}[Proof of Lemma~\ref{3-SPF_VECT.lemma:Q1.mb_fcts_are_left-cont_on_pi=mfw1}]
    Let $t\in\R_+$ and $y = f(t,\mf w_1)$. As $Y$ is metrisable, $\{y\}\in \ms B_Y$ and $f^{-1}(\{y\})\in\ms P_\ovT$, by hypothesis. By Proposition~\ref{3-SPF_VECT.prop:exhaust_msP_ovT_otimes_msE} applied to singleton $\Omega$, there is $\alpha\in\mf w_1$ with $f^{-1}(\{y\})\in\ms P_\ovT^\alpha$. By Lemma~\ref{3-SPF_VECT.lemma:sets_in_msPovT^alpha_otimes_msE_are_inactive_beyond_alpha} applied to singleton $\Omega$, $f(t,\beta) \in \{y\}$ for all $\beta\in[\alpha,\mf w_1]_{\mf w_1+1}$, and, in particular, $f(u) = y$ for all $u\in ((t,\alpha),(t,\mf w_1)]_\ovT$. Thus, $f$ is left-constant at $(t,\mf w_1)$.
    
    In particular, $f$ is left-continuous at $(t,\mf w_1)$.
\end{proof}

Our proof of Proposition~\ref{3-SPF_VECT.prop:Q1.cont_fcts_are_mb} is based on several lemmata. As explained in the proof below, it suffices to perform the main analysis under the assumption that $Y=\R$. The first lemma is about the cardinality of the set of left-jumps of làg functions, which generalises its well-known counterpart in classical theory. For this, let us note that for any làg function $f\colon\ovT\to\R$, there is a function $\Delta_\nearrow f\colon\ovT \to \R$ uniquely defined by $\Delta_\nearrow f(0) = 0$ and for $t\in\ovT\setminus \{0\}$:
\[ 
    \Delta_\nearrow f(t) = 
    \begin{cases} 
        f(t) - f(p(t),\beta), & \text{if }\pi(t) = \beta + 1 \text{ for some }\beta\in{\mf w_1}, \\
        f(t) - \lim_{u\nearrow t} f(u), & \text{else.}  
    \end{cases} 
\]

\begin{lemma}\label{3-SPF_VECT.lemma:lag_fct_have_countable_jumps}
    Let $f\colon\ovT\to\R$ be làg. Then, the set
    \begin{equation}\label{3-SPF_VECT.eq:N=t|Delta_-f(t)neq0}
        N = \{t\in\ovT \mid \Delta_\nearrow f(t) \neq 0\} 
    \end{equation}
    is countable. Moreover, for all $t,u\in\ovT$ with $t\le u$ and $(t,u]_\ovT \subseteq \ovT\setminus N$, we have
    \[ p(t) = p(u) \qquad \Longrightarrow\qquad f(t) = f(u). \]
\end{lemma} 

\begin{proof}[Proof of Lemma~\ref{3-SPF_VECT.lemma:lag_fct_have_countable_jumps}]
    (Ad ``$N$ is countable''):~ Let $\e>0$ and
    \[ N_\e = \{t\in\ovT \mid |\Delta_\nearrow f(t)| \ge 2\e\}. \]
    If $N_\e$ were infinite, the axiom of choice would yield a strictly increasing sequence $(t_n)_{n\in\N}$ valued in $N_\e$ and such that $(t_n,t_{n+1})_\ovT\neq \emptyset$ for all $n\in\N$. Then, the axiom of choice would yield a sequence $(u_n)_{n\in\Nast}$ with $t_n < u_{n+1} < t_{n+1}$ for all $n\in\N$ such that $|f(t_n) - f(u_n)|\ge\e$ for all $n\in\Nast$, because $(t_n)_{n\in\N}$ would be $N_\e$-valued.

    As both sequences would be increasing and $\ovT$ is a complete lattice, they would converge in $\ovT$, and, by choice of $(u_n)_{n\in\N}$, their limits coincide. Let $t = \lim_{n\to\infty} t_n = \lim_{n\to\infty} u_n$. $\pi(t)$ could not be a successor ordinal of the form $\beta + 1$, $\beta\in\On$, since then, for all but finitely many $n\in\N$, we would have $t_n \in ((p(t),\beta),\infty]_\ovT$, that is, $t\le t_n$, which contradicts the strict monotonicity of $(t_n)_{n\in\N}$. Moreover, by strict monotonicity of the considered sequences, $t>0$.

    Hence, by Lemma~\ref{3-SPF_VECT.lemma:left/right-limit_points_in_ovT}, $t$ would be a left-limit point. By hypothesis, $f$ would then have left-limits in $t$. In particular, for all but finitely many $n\in\N$, $|f(t_n) - f(u_n)|<\e$ -- in contradiction to the choice of $(u_n)_{n\in\N}$. Thus, $N_\e$ must be finite.

    We conclude that
    \[ N = \bigcup_{n\in\N} N_{2^{-n}} \]
    is countable.

    (Ad second claim):~ Let $t,u\in\ovT$ satisfy $t\le u$ and $p(t) = p(u)$. We show that if $f(t) \neq f(u)$, then $(t,u]_\ovT \cap N\neq\emptyset$.
    
    If $f(t) \neq f(u)$, then the set 
    \[ S = \{ \alpha\in(\pi(t),\pi(u)]_{{\mf w_1}+1} \mid f(t) \neq f(p(t),\alpha) \} \]
    contains $\pi(u)$. Therefore, it has a minimum, which we denote by $\alpha$. Clearly, $\pi(t) < \alpha$, and, for all $\beta\in [\pi(t),\alpha)_{{\mf w_1}+1}$, we have $f(p(t),\beta) = f(t)$. We infer that $\Delta_\nearrow f(p(t),\alpha) = f(p(t),\alpha) - f(t) \neq 0$. Therefore, $(t,u]_\ovT \cap N\neq\emptyset$.
\end{proof}

For the next lemma, let us note that for any function $f\colon\ovT\to\R$, $N$ as in the preceding Lemma~\ref{3-SPF_VECT.lemma:lag_fct_have_countable_jumps}, and $M = \mc Pp(N)$, there is a map $\gamma\colon M \times ({\mf w_1}+1)\to({\mf w_1}+1)$ satisfying, for any $t\in M$:
\begin{itemize}
    \item $\gamma(t,0) = 0$,
    \item for any ordinal $\alpha\in{\mf w_1}$, $\gamma(t,\alpha+1) = \inf N(t,\alpha)$ is the infimum of 
    \[ N(t,\alpha) = \{\beta\in (\gamma(t,\alpha),{\mf w_1}]_{{\mf w_1}+1} \mid (t,\beta)\in N\} \]
    in ${\mf w_1}+1$,
    \item for any limit ordinal $\alpha\in{\mf w_1}$:
    \[ \gamma(t,\alpha) = \sup_{\beta\in\alpha} \gamma(t,\beta). \]
\end{itemize}
For $t\in M$, let $D_t = \{\alpha\in{\mf w_1}+1 \mid \gamma(t,\alpha) < {\mf w_1}\}$. The aim of $\gamma$ is to count the jumps of $f$. 

\begin{lemma}\label{3-SPF_VECT.lemma:gamma_monotone,Dt_countable_succ_ordinal}
    Let $f\colon\ovT\to\R$ be làg, and left-continuous at $(t,{\mf w_1})$, for any $t\in\R_+$. Let $N$, $M$, $\gamma$, and $(D_t)_{t\in M}$ be defined as just before. Then, for all $t\in M$:
    \begin{enumerate}
        \item\label{3-SPF_VECT.lemma:gamma_monotone,Dt_countable_succ_ordinal.gamma} for all $\alpha,\beta\in{\mf w_1}+1$ with $\alpha < \beta$ we have $\gamma(t,\alpha) \le \gamma(t,\beta)$ with strict inequality iff $\alpha\in D_t$,
        \item\label{3-SPF_VECT.lemma:gamma_monotone,Dt_countable_succ_ordinal.Dt_countable} $D_t$ is countable,
        \item\label{3-SPF_VECT.lemma:gamma_monotone,Dt_countable_succ_ordinal.Dt_countable_succ_ordinal} there is $\delta_t\in{\mf w_1}$ with $D_t = \delta_t+1$.
    \end{enumerate}
\end{lemma}

\begin{proof}
    Let $t\in M$.\smallskip
    
    (Ad Part~\ref{3-SPF_VECT.lemma:gamma_monotone,Dt_countable_succ_ordinal.gamma}):~ For the proof of monotonicity, we use transfinite induction. Let $S = \{\beta\in{\mf w_1}+1 \mid \forall\alpha\in\beta\colon \gamma(t,\alpha)\le \gamma(t,\beta)\}$. Clearly, $0\in S$. If $\beta\in S$, then $\beta+1\in S$, because $\gamma(t,\beta) \le \gamma(t,\beta+1)$ by definition. If $\beta\in{\mf w_1}+1$ is a limit ordinal such that $\beta\subseteq S$, and $\alpha\in\beta$, then $\gamma(t,\alpha) \le \gamma(t,\alpha+1) \le \gamma(t,\beta)$, because $\alpha+1<\beta$ and by definition of $\gamma$. Hence, $S={\mf w_1}+1$.\footnote{Indeed, the principle of transfinite induction could be reformulated here by saying that: if this were not the case, then $({\mf w_1}+1) \setminus S$ would have a minimum, a contradiction to the afore-said.} 
    
    For the claim about strict inequality, let $\alpha,\beta\in{\mf w_1}+1$ with $\alpha < \beta$. If $\gamma(t,\alpha) < {\mf w_1}$, then, by definition of $\gamma$, $\gamma(t,\alpha) < \gamma(t,\alpha+1) \le \gamma(t,\beta)$. If $\gamma(t,\alpha) = {\mf w_1}$, then, by monotonicity, $\gamma(t,\alpha) = {\mf w_1} = \gamma(t,\beta)$.\smallskip

    (Ad Part~\ref{3-SPF_VECT.lemma:gamma_monotone,Dt_countable_succ_ordinal.Dt_countable}):~ Let $\gamma^\ast_t = \sup \mc P\pi(N\cap p^{-1}(t))$, which is countable. Let $\alpha\in D_t$. If $\alpha$ is not a limit ordinal, then clearly $\gamma(t,\alpha) \in \gamma^\ast_t+1$. If $\alpha$ is a limit ordinal, then $\gamma(t,\alpha) = \sup \{ \gamma(t,\beta+1) \mid \beta\in\alpha \}$ -- because, by definition of a limit ordinal, $\beta\in\alpha$ implies $\beta+1\in\alpha$ and $\gamma(t,.)$ is monotone. Hence, $\gamma(t,\alpha) \le \gamma^\ast_t$, i.e.\ $\gamma(t,\alpha)\in \gamma^\ast_t+1$. Therefore, and by Part~\ref{3-SPF_VECT.lemma:gamma_monotone,Dt_countable_succ_ordinal.gamma}, $\gamma(t,.)|_{D_t}$ defines an injection of $D_t$ into the countable set $\gamma^\ast_t+1$. We conclude that $D_t$ is countable. \smallskip

    (Ad Part~\ref{3-SPF_VECT.lemma:gamma_monotone,Dt_countable_succ_ordinal.Dt_countable_succ_ordinal}):~ Note that, by monotonicity of $\gamma(t,.)$, $D_t$ is a downward closed subset of the ordinal ${\mf w_1}$, hence an ordinal itself. As $D_t$ is countable, therefore, $D_t\in{\mf w_1}$. If $D_t$ were a limit ordinal, then $\gamma(t,D_t) = \sup_{\beta\in D_t} \gamma(t,\beta)$ would be countable as the supremum of a countable set of countable ordinals, whence the contradiction $D_t\in D_t$.\footnote{Recall that $D_t \in D_t$ is equivalent to $D_t < D_t$, and that in ZFC, a set cannot contain itself as an element.} Therefore, as $0\in D_t$, $D_t$ is a successor ordinal. Because $D_t$ is countable, its predecessor $\delta_t$ is countable as well.
\end{proof}

Given a function $f\colon\ovT\to\R$, $N$, $M$, $\gamma$, $(D_t)_{t\in M}$, and $(\delta_t)_{t\in M}$ as defined before Lemma~\ref{3-SPF_VECT.lemma:gamma_monotone,Dt_countable_succ_ordinal}, let $D = \bigcup_{t\in M} \{t\} \times D_t$, let $g\colon D\to\T$ be given by $g(t,\alpha) = (t,\gamma(t,\alpha))$, and define the set 
\begin{equation}\label{3-SPF_VECT.eq:def_mcN}
    \mc N = \big\{ [g(t,\alpha),g(t,\alpha+1))_{\ovT} \mid t\in M,\, \alpha\in\delta_t\big\} \cup\big\{[g(t,\delta_t),(t,{\mf w_1})]_{\ovT} \mid t\in M \big\}. 
\end{equation}
So, $g$ simply represents the jump counter $\gamma$ in terms of the time half-axis $\ovT$ (or more precisely, its subset $\T$).

\begin{lemma}\label{3-SPF_VECT.lemma:mcN_measurable_partition,f_vertically_constant}
    Let $f\colon\ovT\to\R$ be làg, and left-continuous at $(t,{\mf w_1})$, for any $t\in\R_+$. Let $N$, $M$, and $\mc N$ be defined as just before. Then, the following statements hold true:
    \begin{enumerate}
        \item\label{3-SPF_VECT.lemma:mcN_measurable_partition,f_vertically_constant.mcN_mb} $\mc N\subseteq\ms I_\ovT(\T)$,
        \item\label{3-SPF_VECT.lemma:mcN_measurable_partition,f_vertically_constant.mcN_countable} $\mc N$ is countable,
        \item\label{3-SPF_VECT.lemma:mcN_measurable_partition,f_vertically_constant.bigcup_mcN_mb} ${\bigcup \mc N}\in\ms I_\ovT(\T)$,
        \item\label{3-SPF_VECT.lemma:mcN_measurable_partition,f_vertically_constant.mcN_partition} $\mc N$ is a partition of $\bigcup_{t\in M} p^{-1}(t)$, i.e.\ $\emptyset\notin\mc N$, $\mc N$ is disjoint and ${\bigcup \mc N} = \bigcup_{t\in M} p^{-1}(t)$,
        \item\label{3-SPF_VECT.lemma:mcN_measurable_partition,f_vertically_constant.f_outside_mcN} for all $t\in\ovT\setminus {\bigcup \mc N}$, $f(t) = f(p(t))$,
        \item\label{3-SPF_VECT.lemma:mcN_measurable_partition,f_vertically_constant.f_on_mcN} for any $S\in\mc N$ and all $t,u\in S$, $f(t) = f(u)$.
    \end{enumerate}
\end{lemma}

\begin{proof}
    For the proof, let $f\colon\ovT\to\R$ be làg, and left-continuous at $(t,{\mf w_1})$, for any $t\in\R_+$, and let $N$, $M$, $\gamma$, $(D_t)_{t\in M}$, $D$, $g$, and $\mc N$ be defined as described above Lemmata~\ref{3-SPF_VECT.lemma:gamma_monotone,Dt_countable_succ_ordinal} and~\ref{3-SPF_VECT.lemma:mcN_measurable_partition,f_vertically_constant}. We use Lemma~\ref{3-SPF_VECT.lemma:gamma_monotone,Dt_countable_succ_ordinal} throughout the whole proof. \smallskip
    
    (Ad Part~\ref{3-SPF_VECT.lemma:mcN_measurable_partition,f_vertically_constant.mcN_mb}):~ It suffices to note that, by definition, $g$ maps to $\T$ and that for any $t\in\T$ and $u\in\ovT$ with $\pi(u) = {\mf w_1}$, we have
    \[ [t,u]_\ovT = \bigcap_{n\in\N} [t,p(u) + 2^{-n})_\ovT \quad \in\ms I_\ovT. \]

    (Ad Part~\ref{3-SPF_VECT.lemma:mcN_measurable_partition,f_vertically_constant.mcN_countable}):~ $N$ and $D_t$, for all $t\in M$, are countable. Thus, $M = \mc Pp(N)$ is countable as well, and so is $D$. In particular, $\mc N$ is countable.\smallskip

    (Ad Part~\ref{3-SPF_VECT.lemma:mcN_measurable_partition,f_vertically_constant.bigcup_mcN_mb}):~ This statement follows from the preceding Parts~\ref{3-SPF_VECT.lemma:mcN_measurable_partition,f_vertically_constant.mcN_mb} and~\ref{3-SPF_VECT.lemma:mcN_measurable_partition,f_vertically_constant.mcN_countable}.\smallskip

    (Ad Part~\ref{3-SPF_VECT.lemma:mcN_measurable_partition,f_vertically_constant.mcN_partition}):~ For any $t\in M$, $g(t,.)$ is strictly increasing and maps into $\T$. Hence, $\emptyset\notin\mc N$. It is moreover clear that ${\bigcup \mc N} \subseteq \bigcup_{t\in M} p^{-1}(t)$. On the other hand, let $t\in M$ and $u\in p^{-1}(t)$. Let $\alpha = \{\beta\in D_t \mid 0 \le \gamma(t,\beta) \le \pi(u)\}$. By monotonicity of $\gamma(t,.)$, $\alpha$ is a downward closed subset of the ordinal $D_t$ and thus an ordinal itself. We have $0\in\alpha$ alias $\alpha > 0$. $\alpha$ cannot be a limit ordinal, because if it were, then $\gamma(t,\alpha) = \sup_{\beta\in\alpha} \gamma(t,\beta) \le \pi(u)$, whence the contradiction $\alpha\in\alpha$. Hence, $\alpha$ is a successor ordinal. In particular, $\alpha$ has a maximum $\beta$ and $\gamma(t,\beta) \le \pi(u) < \gamma(t,\beta+1)$. Thus, $u\in[g(t,\beta),g(t,\beta+1))_\ovT$ if $\beta < \delta_t$, and $u\in [g(t,\delta_t),(t,{\mf w_1})]_\ovT$ if $\beta = \delta_t$. $\mc N$ is disjoint because $g(t,.) = (t,\gamma(t,.))$ is strictly monotone on $D_t$, and the fibres of $p$ are disjoint.\smallskip

    (Ad Parts~\ref{3-SPF_VECT.lemma:mcN_measurable_partition,f_vertically_constant.f_outside_mcN} and~\ref{3-SPF_VECT.lemma:mcN_measurable_partition,f_vertically_constant.f_on_mcN}):~ These parts follow from the previous one by a direct application of Lemma~\ref{3-SPF_VECT.lemma:lag_fct_have_countable_jumps}. Regarding Part \ref{3-SPF_VECT.lemma:mcN_measurable_partition,f_vertically_constant.f_outside_mcN}, let $t\in \ovT \setminus \bigcup\mc N$. For all $u\in (p(t),t]_\ovT$, we have $p(u) = p(t)\notin M$, by Part~\ref{3-SPF_VECT.lemma:mcN_measurable_partition,f_vertically_constant.mcN_partition}, and, in particular, $u\notin N$. Thus, $(p(t),t]_\ovT\cap N = \emptyset$, whence $f(u) = f(t)$ for all $u\in (p(t),t]_\ovT$, by the mentioned lemma. Regarding Part~\ref{3-SPF_VECT.lemma:mcN_measurable_partition,f_vertically_constant.f_on_mcN}, let $S\in\mc N$ and $t,u\in S$ with $t\le u$, without loss of generality. By construction of $S$ (based on the definition of $g$ and $\gamma$), we have $p(t) = p(u)$ and there is no $x\in N$ with $t < x \le u$, that is, $(t,u]_\ovT\cap N = \emptyset$. Hence, by the mentioned lemma, $f(t) = f(u)$.
\end{proof}

\begin{proof}[Proof of Proposition~\ref{3-SPF_VECT.prop:Q1.cont_fcts_are_mb}]
    \emph{The main case $Y=\R$}: We first assume that $Y=\R$. 
    Let $f\colon\ovT\to\R$ be làg, and left-continuous at $(t,{\mf w_1})$ for any $t\in\R_+$. Furthermore, let $N$, $M$, $\gamma$, $(D_t)_{t\in M}$, $D$, $g$, and $\mc N$ be defined as described above Lemmata~\ref{3-SPF_VECT.lemma:gamma_monotone,Dt_countable_succ_ordinal} and~\ref{3-SPF_VECT.lemma:mcN_measurable_partition,f_vertically_constant}.\smallskip

    (First argument):~ Let $h\colon \bRp \to \R$ be defined by $h(0) = f(0)$ and, for all $t\in\bRp\setminus\{0\}$, $h(t) = \lim_{u\nearrow t} f(u)$. Then, by adapting classical methods, it follows that $h$ is left-continuous. Indeed, let $\e>0$. Then, for any $t\in\bRp\setminus\{0\}$, there is $a_t\in\ovT$ with $a_t<t$ such that, for all $x\in (a_t,t)_\ovT$, we have $|f(x) - h(t)| < \e/2$. As $t\in\bRp$, $p(a_t) < p(t)$, and hence we can, upon making $a_t$ larger, assume $a_t\in\bRp$. Let us fix some $t\in\bRp\setminus\{0\}$ and take arbitrary $x\in (a_t,t)_\bRp$. If $x\notin N$, we have $h(x) = f(x)$ and thus $|h(x)-h(t)|<\e/2<\e$. If $x\in N$, then there is $y\in [(a_x,x)_\bRp \cap (a_t,t)_\bRp] \setminus N$, because $N$ is countable and $a_x,a_t\in\R_+$. Therefore, $|h(x)-h(t)|\le |f(y)-h(x)|+|f(y)-h(t)|< \e/2+\e/2=\e$. We conclude that $h$ is left-continuous.\smallskip

    (Second argument):~ As a left-continuous function $\bRp\to\R$, $h$ is $\ms B_\bRp$-$\ms B_\R$-measurable. Hence, by Lemma~\ref{3-SPF_VECT.lemma:iota_alpha_&_p_mb}, $h\circ p$ is $\ms I_\ovT(\T)$-$\ms B_\R$-measurable. By Lemma~\ref{3-SPF_VECT.lemma:mcN_measurable_partition,f_vertically_constant}, Parts~\ref{3-SPF_VECT.lemma:mcN_measurable_partition,f_vertically_constant.bigcup_mcN_mb}, \ref{3-SPF_VECT.lemma:mcN_measurable_partition,f_vertically_constant.mcN_partition}, and~\ref{3-SPF_VECT.lemma:mcN_measurable_partition,f_vertically_constant.f_outside_mcN}, $h\circ p(t) = f\circ p(t) = f(t)$ for all $t\in \ovT\setminus \bigcup\mc N$, and $\ovT\setminus \bigcup\mc N\in\ms I_\ovT(\T)$. Hence, \[f\, 1_{\ovT\setminus \bigcup\mc N} = h\circ p\,  1_{\ovT\setminus \bigcup\mc N}\] is $\ms I_\ovT(\T)$-$\ms B_\R$-measurable.\smallskip

    (Third argument):~ $f$ is constant on any $S\in\mc N$, by Lemma~\ref{3-SPF_VECT.lemma:mcN_measurable_partition,f_vertically_constant}, Part~\ref{3-SPF_VECT.lemma:mcN_measurable_partition,f_vertically_constant.f_on_mcN}. Hence, by Part~\ref{3-SPF_VECT.lemma:mcN_measurable_partition,f_vertically_constant.mcN_mb} of the same lemma, $f1_S$ is $\ms I_\ovT(\T)$-$\ms B_\R$-measurable for any $S\in\mc N$. Hence, using that $\mc N$ is a countable partition (Parts~\ref{3-SPF_VECT.lemma:mcN_measurable_partition,f_vertically_constant.mcN_countable} and~\ref{3-SPF_VECT.lemma:mcN_measurable_partition,f_vertically_constant.mcN_partition} of the aforementioned lemma), we infer that \[ f\,1_{\bigcup\mc N} = \sum_{S\in\mc N} f\,1_S \] is $\ms I_\ovT(\T)$-$\ms B_\R$-measurable.\smallskip

    Combining the results of the second and third arguments, we infer that \[f = f\,1_{\ovT\setminus \bigcup\mc N} + f\, 1_{\bigcup \mc N}\] is $\ms I_\ovT(\T)$-$\ms B_\R$-measurable.\medskip

    \emph{The case of general metrisable $Y$}: Let $Y$ be a metrisable topological space and $f\colon\ovT\to Y$ be làg, and left-continuous at $(t,{\mf w_1})$ for any $t\in\R_+$. Let $d_Y$ be a metric generating the topology on $Y$, let $\e>0$ and $y\in Y$. It suffices to show that $f^{-1}(B_\e(y))\in\ms I_\ovT(\T)$. For this, define
    \[ \tilde f\colon\ovT\to\R,\, t\mapsto d_Y(f(t),y). \]
    By the triangular inequality, any $t\in\ovT$ and $z\in Y$ satisfy
    \[ d_Y(f(t),y) \le d_Y(f(t),z) + d_Y(z,y), \]
    whence $|\tilde f(t)-d_Y(y,z)| \le d_Y(f(t),z)$. We infer that $\tilde f$ is làg, and left-continuous at $(t,{\mf w_1})$ for any $t\in\R_+$. Thus, by the main case studied just above, $\tilde f$ is $\ms I_\ovT(\T)$-$\ms B_\R$-measurable. Therefore, 
    \[ f^{-1}(B_\e(y)) = \tilde f^{-1}(B_\e(0))\, \in \ms I_\ovT(\T). \]
    This completes the proof.
\end{proof}

\subsubsection{Measurable projection and section}

\begin{proof}[Proof of Proposition~\ref{3-SPF_VECT.prop:Q2.random_sets_of_time}]
    Let $M\in\ms P_\ovT\otimes\ms E$. By Proposition~\ref{3-SPF_VECT.prop:exhaust_msP_ovT_otimes_msE}, there is $\alpha\in\mf w_1$ and $M_\alpha\in\ms B_{\bRp\times(\alpha+1)}\otimes\ms E$ with $M = (\rho^\alpha\times\id_\Omega)^{-1}(M_\alpha)$. As $\ovT_{\alpha+1} = \T_{\alpha+1}\cup\{\infty\}  \in \ms B_{\bRp\times(\alpha+1)}$ and $\im\rho^\alpha\subseteq \ovl\T_{\alpha+1}$, we can assume without loss of generality that $M_\alpha\subseteq \ovT_{\alpha+1}\times\Omega$. As $\rho^\alpha(t) = t$ for all $t\in\ovT_{\alpha+1}$, we obtain $M_\alpha = M \cap (\ovT_{\alpha+1}\times\Omega)$.

    As $M_\alpha\subseteq M$, it is clear that $\mc P\prj_\Omega(M_\alpha) \subseteq \mc P\prj_\Omega(M)$. For the converse inclusion, let $\omega\in\mc P\prj_\Omega(M)$. Then, there is $t\in\ovT$ such that $(t,\omega)\in M$. If $\pi(t) \le \alpha$, then $(t,\omega)\in M_\alpha$, whence $\omega\in \mc P\prj_\Omega(M_\alpha)$. Else $\pi(t) > \alpha$. Then $t\in\ovT\setminus\Ta$ and $p(t) < \infty$. By Lemma~\ref{3-SPF_VECT.lemma:sets_in_msPovT^alpha_otimes_msE_are_inactive_beyond_alpha}, we infer that $u = (p(t),\alpha)$ satisfies $(u,\omega) \in M$. As $p(u) < \infty$ and $\pi(u)\le \alpha$, we get $(u,\omega)\in M_\alpha$. Hence, $\omega\in \mc P\prj_\Omega(M_\alpha)$. We conclude that $\mc P\prj_\Omega(M) = \mc P\prj_\Omega(M_\alpha)$.
\end{proof}

\begin{proof}[Proof of Theorem~\ref{3-SPF_VECT.thm:mb_proj_section}]
    Let $M\in\ms P_\ovT\otimes\ms E$. By Proposition~\ref{3-SPF_VECT.prop:Q2.random_sets_of_time}, there is $\alpha\in{\mf w_1}$  with 
    \[ (\ast) \qquad M_\alpha = M \cap (\ovT_{\alpha+1}\times\Omega)~ \in \ms B_{\bRp\times(\alpha+1)}\otimes\ms E, \quad\text{and}\quad  \mc P\prj_\Omega(M) = \mc P\prj_\Omega(M_\alpha). \]
    
    As $\bRp\times(\alpha+1)$ is a Polish space, the classical theorems on measurable projection and section apply to $M_\alpha$ and the canonical projection $(\bRp\times(\alpha+1))\times\Omega\to\Omega$, which is the restriction of $\prj_\Omega\colon\ovT\times\Omega\to\Omega$. Thus, $\mc P\prj_\Omega(M) = \mc P\prj_\Omega(M_\alpha)\in\ms E^{\mathrm u}$, and there is $\ms E^{\mathrm u}|_{\mc P\prj_\Omega(M_\alpha)}$-$\ms B_{\bRp\times(\alpha+1)}$-measurable $\sigma_\alpha\colon\mc P\prj_\Omega(M_\alpha)\to\bRp\times(\alpha+1)$ such that $[\![\sigma_\alpha]\!]\subseteq M_\alpha$. In particular, $\sigma_\alpha$ is $\ovT_{\alpha+1}$-valued.
    
    Let $\sigma = \iota_\alpha\circ\sigma_\alpha$. Then, by Lemma~\ref{3-SPF_VECT.lemma:iota_alpha_&_p_mb}, and another application of $(\ast)$, $\sigma$ is $\ms E^{\mathrm u}|_{\mc P\prj_\Omega(M)}$-$\ms P_\ovT$-measurable and $[\![\sigma]\!]\subseteq M$. This completes the proof.
\end{proof}

\subsection{Section~\ref{3-SPF_VECT.sec:sto_proc_in_vERT}}

\subsubsection{Augmentation and right-limits of exogenous information flow}

\begin{proof}[Proof of Equation~\ref{3-SPF_VECT.eq:(ovl_msF)_+=ovl_(msF_+)}] 
    This is a classical argument in stochastic analysis, added here for the reader's convenience only. By definition, $\ovl{\ms F_{\infty\Plus}} = \ovl{\ms F_{\infty}} =  \ovl{\ms F}_{\infty} = \ovl{\ms F}_{\infty\Plus}$. It remains to show that, for all $t\in\ovT\setminus\{\infty\}$, we have 
    \[ \bigcap_{\P\in\mf P_{\ms E}}\Big[\Big(\bigcap_{\R\ni u>t} \ms F_u\Big) \vee \ms N_\P\Big] = \bigcap_{\R\ni u>t}\bigcap_{\P\in\mf P_{\ms E}} \Big(\ms F_u \vee \ms N_\P\Big). \] 
    To start, note that $(\bigcap_{\R\ni u>t} \ms F_u) \vee \ms N_\P \subseteq \ms F_v \vee \ms N_\P$ for all real $v>t$ and all $\P\in\mf P_{\ms E}$. Whence the inclusion ``$\subseteq$''. For the other inclusion's proof, let $E\in\bigcap_{\R\ni u>t}\bigcap_{\P\in\mf P_{\ms E}} (\ms F_u\vee\ms N_\P)$ and $\P\in\mf P_{\ms E}$. For $n\in\N$, let $u_n = p(t) + 2^{-n}$. Then, for any $n\in\N$, there is $E_n^\P\in\ms F_{u_n}$ such that $\P(E\Delta E_n^\P) = 0$. Let $E^\P = \limsup_{n\to\infty} E^\P_{u_n}$. Then, $E^\P\in \bigcap_{\R\ni u>t} \ms F_u$. As \[ \bigcap_{n\in\N} E^\P_{u_n} \subseteq E^\P \subseteq \bigcup_{n\in\N} E^\P_{u_n}, \] $\sigma$-subadditivity yields \[ \P(E\Delta E^\P) \le \sum_{n\in\N} \P(E\Delta E^\P_{u_n}) = 0. \] 
    Hence, $E\in (\bigcap_{\R\ni u>t} \ms F_u) \vee \ms N_\P$. 
    As this holds true for any $\P\in\mf P_{\ms E}$, we infer that \[ E\in\bigcap_{\P\in\mf P_{\ms E}}\Big[\Big(\bigcap_{\R\ni u>t} \ms F_u\Big) \vee \ms N_\P\Big]. \]
\end{proof}

\subsubsection{Progressively measurable processes}

\begin{proof}[Sketch of a proof of Remark~\ref{3-SPF_VECT.rmk:tau_&_msF_tau}]
    Most claims can be proven by just copying the standard arguments from the classical theory of stochastic processes (as exposed, for example, in \cite{Kallenberg2021Foundations}). So we limit ourselves to explaining those points that require some additional thought.\smallskip

    Regarding Properties~\ref{3-SPF_VECT.rmk:tau_&_msF_tau.sup_of_tau_n} and~\ref{3-SPF_VECT.rmk:tau_&_msF_tau.inf_of_tau_n}, the standard argument can directly by applied. Concerning Property~\ref{3-SPF_VECT.rmk:tau_&_msF_tau.tau<t_in_msG_t}, note that for any $t\in\ovT$ with $\pi(t) < \mf w_1$, $\pi(t)$ is countable. Hence, with $Q_t = [0,t)_\ovT \cap \Q$, we have
    \[ \{\tau < t\} = \bigcup_{q\in Q_t} \{\tau \le q\} \cup \bigcup_{\beta\in\pi(t)} \{\tau \le (p(t),\beta)\}\, \in\ms F_t. \]
    As a consequence, $\{\tau = t\} = \{\tau \le t\} \setminus \{\tau < t\} \in\ms F_t$.\smallskip

    Next, consider Claim~\ref{3-SPF_VECT.rmk:tau_&_msF_tau.F_tau_sigma-alg}. The fact that $\ms F_\tau$ is a sub-$\sigma$-algebra of $\ms E$ follows exactly as in the classical case. The claim on augmentedness has also nothing special in the vertically extended case, but it seems less standard; hence, we give the proof for the reader's convenience. It suffices to show that $\ovl{\ms F_\tau}\subseteq\ms F_\tau$. Indeed, we have --- using universal completeness of $\ms E$ and universal augmentedness of $\ms F$ in $\ms E$ ---
    \begin{align*}
        \ovl{\ms F_\tau} =&~ \bigcap_{\P\in\mf P_{\ms E}} \big\{F\in\ms E \mid \forall t\in\ovT\colon F \cap \{\tau\le t\}\in\ms F_t\big\} \vee \ms N_\P \\
        =&~ \bigcap_{\P\in\mf P_{\ms E}} \Big\{E\in\ms E \vee \ms N_\P \mid \exists F\in\ms E \colon \big[\P(F\Delta E) = 0,\, \forall t\in\ovT\colon F \cap \{\tau\le t\}\in\ms F_t \big] \Big\} \\
        =&~ \Big\{ E\in\underbrace{\ms E^{\mathrm u}}_{=\ms E} \mid \forall \P\in\mf P_{\ms E}\exists F\in\ms E\colon \big[\P(F\Delta E) = 0,\, \forall t\in\ovT\colon F \cap \{\tau\le t\}\in\ms F_t \big] \Big\} \\
        \subseteq&~ \big\{ E\in \ms E \mid \forall \P\in\mf P_{\ms E} \forall t\in\ovT\colon E\cap \{\tau\le t\} \in \ms F_t\vee\ms N_\P\big\} \\
        =&~ \big\{ E\in \ms E \mid \forall t\in\ovT\colon E\cap \{\tau\le t\} \in \underbrace{\bigcap_{\P\in\mf P_{\ms E}} (\ms F_t\vee\ms N_\P)}_{=\ms F_t}\big\} \\
        =&~ \ms F_\tau.
    \end{align*}

    Properties~\ref{3-SPF_VECT.rmk:tau_&_msF_tau.msF_tau=msG_t_if_tau=t} and~\ref{3-SPF_VECT.rmk:tau_&_msF_tau.F_tau_subseteq_F_sigma_if_tau_leq_sigma} are shown as in the classical case. Property~\ref{3-SPF_VECT.rmk:tau_&_msF_tau.tau_is_msF_tau_msI_mb} follows as in the classical case, using the fact that $\ms I_\ovT(\T)$ is generated by all sets of the form $[0,t]_{\ovT}$, $t\in\T$, according to Corollary~\ref{3-SPF_VECT.cor:msI_is_generated_by_all_kinds_of_principal_up/down-sets}.
\end{proof}

\begin{proof}[Proof of Example~\ref{3-SPF_VECT.ex:counterex_stopping_times}]
    (Ad Part~\ref{3-SPF_VECT.ex:counterex_stopping_times.stopping_time}):~  Let $t\in\R_+$. Then,
    \[ \qquad \{\tau\le t\} = \{\sigma < t\} \cup (\{\sigma = t\} \cap V) \in \ms F_t, \]
    because $\{\sigma = t\} \cap V \subseteq \{\sigma=t\}$ and $\ms F$ and $\ms E$ are $\P$-complete. Moreover, for any $\alpha\in\mf w_1 \setminus \{0\}$:
    \[ \{\tau \le (t,\alpha)\} = \{\sigma\le t\} \in \ms F_t = \ms F_{(t,\alpha)}. \]
    Hence, $\tau$ is a $\ms F$-stopping time in $\ovT$.\smallskip
    
    (Ad Part~\ref{3-SPF_VECT.ex:counterex_stopping_times.distribution}):~ 
    From the preceding equations, we infer that $\P(\tau \le t) = \P(\sigma \le t)$ for all $t\in\T$. Hence, $\P_\tau = \P_\sigma$, by Corollary~\ref{3-SPF_VECT.cor:msI_is_generated_by_all_kinds_of_principal_up/down-sets}.\smallskip
    
    (Ad Part~\ref{3-SPF_VECT.ex:counterex_stopping_times.order}): This third claim is evident from the definition of $\tau$.\smallskip

    (Ad Part~\ref{3-SPF_VECT.ex:counterex_stopping_times.V_non_Lebesgue}):~ Suppose that $V\notin \ms E$. Then
    \[ \{\pi\circ \tau = 0\} = V \,\notin\ms E, \]
    and, in particular, $\{\pi\circ\tau=0\}\notin\ms F_\tau$. As $\bRp\times 1\in\ms P_\ovT$, $\tau$ is not $\ms F_\tau$-$\ms P_\ovT$-measurable. This proves Claim~\ref{3-SPF_VECT.ex:counterex_stopping_times.V_non_Lebesgue.tau_is_not_msF_tau_msT_mb}.

    Regarding Claim~\ref{3-SPF_VECT.ex:counterex_stopping_times.V_non_Lebesgue.[[tau]]}, note that
    \[ V = \prj_\Omega([\![\tau]\!]\cap (\rho^1)^{-1}(\bRp\times \{0\})). \]
    As $V\notin\ms E$ and $\ms E$ is $\P$-complete, by the Measurable Projection Theorem~\ref{3-SPF_VECT.thm:mb_proj_section}, we obtain $[\![\tau]\!]\cap (\rho^1)^{-1}(\bRp\times \{0\})\notin\ms P_\ovT\otimes\ms E$. Hence, $[\![\tau]\!] \notin \ms P_\ovT\otimes\ms E$, and, in particular, $[\![\tau]\!]$ is not $\ms F$-progressively measurable.

    Moreover, by classical theory, it follows that $[\![\sigma]\!]$ is $\ms F$-progressively measurable. Indeed, for every $x\in\bRp$ and every $\alpha\in\mf w_1$,
    \begin{equation*} 
    \begin{aligned}
        (\ast) \qquad &~\{(\sigma(\omega),\alpha,\omega) \mid \omega\in\Omega\colon \sigma(\omega)\le x\} \\
        =&~ ([0,x]_\bRp\times\{\alpha\}\times\Omega) \setminus \Big( \bigcup_{\Q\ni q < x} \big[([0,q)_\bRp\times\{\alpha\}\times \{q < \sigma\}) \cup ((q,x]_\bRp\times\{\alpha\} \times \{\sigma \le q\})\big]\Big) \\
        &~~\in \ms B_{\bRp\times(\alpha+2)}\times\ms F_x. 
    \end{aligned}
    \end{equation*}
    Taking the preimage of this under $\rho^2\times\id_\Omega$, for every $t\in\ovT$, with $x=p(t)$ and $\alpha=0$, yields:
    \[ [\![\sigma]\!]\cap [\![0,t]\!] = (\rho^2\times\id_\Omega)^{-1}\Big(\{(\sigma(\omega),0,\omega) \mid \omega\in\Omega\colon \sigma(\omega)\le x\}\Big)\, \in \ms P_\ovT\times\ms F_x\subseteq\ms P_\ovT\otimes\ms F_t. \]
    Furthermore,
    \[ V^\complement = \prj_\Omega([\![0,\tau)\!)\cap [\![\sigma]\!]). \]
    As $V^\complement\notin\ms E$ and $\ms E$ is $\P$-complete, by the Measurable Projection Theorem~\ref{3-SPF_VECT.thm:mb_proj_section}, we infer $[\![0,\tau)\!)\cap[\![\sigma]\!]\notin\ms P_\ovT\otimes\ms E$. Hence, $[\![0,\tau)\!) \notin \ms P_\ovT\otimes\ms E$, and, in particular, $[\![0,\tau)\!)$ is not $\ms F$-progressively measurable.

    Regarding $(\!(\tau,\infty]\!]$ a similar argument can be used. Let $\sigma'\colon\Omega\to\ovT,\, \omega\mapsto(\sigma(\omega),1)$. Taking the preimage of expression $(\ast)$ under $\rho^2\times\id_\Omega$ for any $t\in\ovT$, $x=p(t)$, and $\alpha=1$, we get, if $\pi(t) > 0$:
    \[ [\![\sigma']\!]\cap [\![0,t]\!] = (\rho^3\times\id_\Omega)^{-1}\Big(\{(\sigma(\omega),1,\omega) \mid \omega\in\Omega\colon \sigma(\omega)\le x\}\Big)\, \in \ms P_\ovT\times\ms F_x\subseteq\ms P_\ovT\otimes\ms F_t. \]
    If $\pi(t) = 0$, then 
    \[ [\![\sigma']\!]\cap [\![0,t]\!] = \bigcup_{\Q \ni q < t} [\![\sigma']\!] \cap [\![0,(q,1)]\!]\, \in \ms P_\ovT\otimes\ms F_t. \]
    Then, note that
    \[ V = \prj_\Omega((\!(\tau,\infty]\!]\cap [\![\sigma']\!]). \]
    We conclude as above, using the Measurable Projection Theorem~\ref{3-SPF_VECT.thm:mb_proj_section}, that $(\!(\tau,\infty]\!]$ is not $\ms F$-progressively measurable.
\end{proof}

\subsubsection{Optional times}

We prepare the proof of Theorem~\ref{3-SPF_VECT.thm:optional_times} with a basic lemma. It is based on the classical argument showing progressive measurability of the converse graph of stopping times in standard real time. Let us introduce the following notation. If $\tau\colon\Omega\to\ovT$ is a map and $\alpha\in{\mf w_1}+1$, then let $\tau_\alpha$ be the map $\Omega\to\ovT$ satisfying $p\circ\tau = p\circ\tau_\alpha$ and $\pi(\tau_\alpha(\omega)) = \alpha$ for all $\omega\in \{\tau < \infty\}$.

\begin{lemma}\label{3-SPF_VECT.lemma:[0,tau_omega1]_is_prog_mb_if_tau<t_is_msF_t}
    Let $\tau\colon\Omega\to\ovT$ be a map satisfying $\{\tau < t\}\in\ms F_t$ for any $t\in\bRp$.\footnote{By Remark~\ref{3-SPF_VECT.rmk:tau_&_msF_tau}, Item~\ref{3-SPF_VECT.rmk:tau_&_msF_tau.tau<t_in_msG_t}, any $\ms F$-stopping time satisfies this condition.} The sets
    $[\![0,\tau_{{\mf w_1}}]\!]$ and $(\!(\tau_{{\mf w_1}},\infty]\!]$ 
    are $\ms F$-progressively measurable with respect to the interval $\sigma$-algebra $\ms I_\ovT(\T)$.
\end{lemma}

\begin{proof}
    Let $t\in\ovT$, and $Q_t = [0,p(t))_\bRp\cap \Q$. Then, using the fact that $Q_t$ is countable, and by Lemma~\ref{3-SPF_VECT.cor:msI_is_generated_by_all_kinds_of_principal_up/down-sets}:
    \[ (\!(\tau_{{\mf w_1}},\infty]\!] \cap ([0,t]_\ovT\times\Omega) = \bigcup_{q\in Q_t} (q,t]_\ovT \times \{\tau < q\} ~\in \ms I_\ovT(\T)\otimes\ms F_t. \]
    Hence, $(\!(\tau_{{\mf w_1}},\infty]\!]\in \Prg(\ms I_\ovT(\T),\ms F)$. As a consequence, by Remark~\ref{3-SPF_VECT.rmk:prog_mb_basic_properties}, Item~\ref{3-SPF_VECT.rmk:prog_mb_basic_properties.sigma_algebra_prog_mb},
    \[ [\![0,\tau_{{\mf w_1}}]\!] = (\ovT\times\Omega) \setminus (\!(\tau_{{\mf w_1}},\infty]\!]~\in \Prg(\ms I_\ovT(\T),\ms F) \]
    as well.
\end{proof}

\begin{proof}[Proof of Theorem~\ref{3-SPF_VECT.thm:optional_times}]
    For a plan of the proof, see Figure~\ref{3-SPF_VECT.fig:thm.optional_times.proof.plan}. The completeness assumption is only made in the proof of implications ``$k\,\Rightarrow\,6$'' for $k\in\{1,2,4\}$.
    \begin{figure}[h]
        \centering
        \[
            \begin{tikzcd}[row sep=2cm, column sep=2cm]
                & 1 \arrow[rd, Rightarrow, bend right=12] & & \\
                3 \arrow[r, Leftrightarrow, bend right=12] & 2 \arrow[r, Rightarrow, bend right=12] & 6 \arrow[r, Rightarrow, bend right=12] \arrow[lu, Rightarrow, bend right=12] \arrow[ll, Rightarrow, bend right=12] \arrow[ld, Rightarrow, bend right=12] & 7 \arrow[l, Rightarrow, bend right=12] \\ 
                5 \arrow[r, Leftrightarrow] & 4 \arrow[ru, Rightarrow, bend right=12] & & 
            \end{tikzcd}
        \]
        \caption{Plan of the proof of Theorem~\ref{3-SPF_VECT.thm:optional_times}}
        \label{3-SPF_VECT.fig:thm.optional_times.proof.plan}
    \end{figure}
    \smallskip
    
    (Ad ``\ref{3-SPF_VECT.thm:optional_times.[0,tau)} $\Leftrightarrow$ \ref{3-SPF_VECT.thm:optional_times.[tau,infty]}'' and ``\ref{3-SPF_VECT.thm:optional_times.(tau,infty]} $\Leftrightarrow$ \ref{3-SPF_VECT.thm:optional_times.[0,tau]}''):~ This follows directly from the fact that $\Prg(\ms P_\ovT,\ms F)$ is stable under complements (see Remark~\ref{3-SPF_VECT.rmk:prog_mb_basic_properties}, Item~\ref{3-SPF_VECT.rmk:prog_mb_basic_properties.sigma_algebra_prog_mb}).\smallskip

    (Ad \ref{3-SPF_VECT.thm:optional_times.[tau]} $\Rightarrow$ \ref{3-SPF_VECT.thm:optional_times.pi_bounded,pi_tau_tau-mb}):~ Suppose that $(\Omega,\ms E,\ms F)$ is universally complete and that $[\![\tau]\!]$ is $\ms F$-progressively measurable. Then, as $\ms F_\infty\subseteq\ms E$, applying Relation~\ref{3-SPF_VECT.eq:prog_mb} with $t=\infty$, basic measure theory yields $[\![\tau]\!]\in\ms P_\ovT\otimes\ms E$. Hence, by Proposition~\ref{3-SPF_VECT.prop:exhaust_msP_ovT_otimes_msE}, $[\![\tau]\!]\in\ms P_\ovT^\alpha\otimes\ms E$ for some $\alpha\in{\mf w_1}$. Then, $\pi\circ\tau\le \alpha$. Indeed, if there existed $\omega\in\Omega$ with $\pi\circ\tau(\omega) > \alpha$, Lemma~\ref{3-SPF_VECT.lemma:sets_in_msPovT^alpha_otimes_msE_are_inactive_beyond_alpha} would imply that $(u,\omega)\in[\![\tau]\!]$ for all $u\in\ovT\setminus\Ta$ with $p(u) = p(\tau(\omega))$. $(\ovT\setminus \Ta) \cap p^{-1}(p(\tau(\omega))$ is infinite, but, as $\tau$ is a set-theoretic function, $[\![\tau]\!] \cap (\ovT\times\{\omega\})$ is a singleton -- whence a contradiction.

    For the remaining part of the claim, let $\beta\in\alpha+1$ and $t\in\ovT$. Let $M_t = [\![\tau]\!] \cap ([0,t]_\ovT \times \Omega)$. Then, by progressive measurability, $M_t \in \ms P_\ovT\otimes\ms F_t$. As a consequence,
    \[ M^\beta_t = M_t \cap [(\rho^{\beta+1})^{-1}(\bRp\times\{\beta\}) \times\Omega] \in\ms P_\ovT \otimes\ms F_t. \]
    By Theorem~\ref{3-SPF_VECT.thm:mb_proj_section}, and universal completeness of $\ms F_t$,
    \[ \{\pi\circ\tau = \beta,\, \tau\le t\} = \mc P\prj_\Omega(M^\beta_t) \in \ms F_t. \]

    (Ad \ref{3-SPF_VECT.thm:optional_times.[0,tau)} $\Rightarrow$ \ref{3-SPF_VECT.thm:optional_times.pi_bounded,pi_tau_tau-mb}):~ Suppose that $(\Omega,\ms E,\ms F)$ is universally complete and that $[\![0,\tau)\!)$ is $\ms F$-progressively measurable. As in the proof of the implication (\ref{3-SPF_VECT.thm:optional_times.[tau]} $\Rightarrow$ \ref{3-SPF_VECT.thm:optional_times.pi_bounded,pi_tau_tau-mb}) we get that $[\![0,\tau)\!)\in\ms P_\ovT^\alpha\otimes\ms E$ for some $\alpha\in{\mf w_1}$. We infer that $\pi\circ\tau\le \alpha$. Indeed, if there existed $\omega\in\Omega$ with $\pi\circ\tau(\omega) > \alpha$, then, with $t = (p(\tau(\omega)),\alpha)$, we would have $(t,\omega)\in[\![0,\tau)\!)$. Hence, Lemma~\ref{3-SPF_VECT.lemma:sets_in_msPovT^alpha_otimes_msE_are_inactive_beyond_alpha} would imply that, for $u = (p(\tau(\omega)),{\mf w_1})$, $(u,\omega)\in [\![0,\tau)\!)$, which is absurd.

    Note that as a direct consequence of the hypothesis, $[\![\tau,\infty]\!]$ is also $\ms F$-progressively measurable. By Theorem~\ref{3-SPF_VECT.thm:mb_proj_section}, we infer that, for any $t\in\ovT$:
    \[ \{\tau \le t\} = \mc P \prj_\Omega\big([\![\tau,\infty]\!]\cap([0,t]_\ovT\times\Omega)\big) \in \ms F_t. \]
    Thus, $\tau$ is an $\ms F$-stopping time, and satisfies the hypothesis of Lemma~\ref{3-SPF_VECT.lemma:[0,tau_omega1]_is_prog_mb_if_tau<t_is_msF_t}, by Remark~\ref{3-SPF_VECT.rmk:tau_&_msF_tau}, Item~\ref{3-SPF_VECT.rmk:tau_&_msF_tau.tau<t_in_msG_t}.

    For the remaining part of the claim, let $\beta\in\alpha+1$ and $t\in\ovT$. Let $M_t = [\![\tau,\tau_{{\mf w_1}}]\!] \cap ([0,t]_\ovT \times\Omega)$. By hypothesis, and Lemma~\ref{3-SPF_VECT.lemma:[0,tau_omega1]_is_prog_mb_if_tau<t_is_msF_t}, $M_t\in\ms P_\ovT\otimes\ms F_t$. In particular,
    \[ M_t^\beta = M_t \cap [(\rho^{\beta+1})^{-1}(\bRp\times\{\beta\}) \times\Omega] \in\ms P_\ovT \otimes\ms F_t. \]
    Hence, by Theorem~\ref{3-SPF_VECT.thm:mb_proj_section}, and universal completeness of $\ms F_t$, 
    \[ \{\pi\circ\tau \le \beta,\, \tau_\beta \le t\} = \mc P\prj_\Omega(M_t^\beta)\in\ms F_t. \]
    If $\beta \le \pi(t)$, then 
    \[ \{\pi\circ\tau \le \beta,\, \tau\le t\} = \{\pi\circ\tau \le \beta,\, \tau_\beta \le t\} \in \ms F_t. \]
    Else, $\pi(t) < \beta$, so that $\pi(t) + 1$ is countable. Let $Q_t = [0,p(t))_\bRp \cap \Q$, a countable set as well. As $\beta \le {\mf w_1}$, Remark~\ref{3-SPF_VECT.rmk:tau_&_msF_tau}, Item~\ref{3-SPF_VECT.rmk:tau_&_msF_tau.tau<t_in_msG_t}, and the previous case yield:
    \[ \{\pi\circ\tau \le \beta,\, \tau\le t\} = \bigcup_{\gamma\in\pi(t)+1} \{\tau = (p(t),\gamma)\}\cup \bigcup_{q\in Q_t} \{\pi\circ\tau\le \beta,\, \tau\le(q,{\mf w_1})\}\, \in\ms F_t. \]
    We conclude that
    \[ \{\pi\circ\tau = \beta,\,\tau\le t\} = \{\pi\circ\tau \le \beta,\,\tau\le t\} \setminus \bigcup_{\gamma\in\beta} \{\pi\circ\tau \le \gamma,\,\tau\le t\}\, \in \ms F_t, \]
    because $\beta$ is countable.\smallskip

    (Ad \ref{3-SPF_VECT.thm:optional_times.(tau,infty]} $\Rightarrow$ \ref{3-SPF_VECT.thm:optional_times.pi_bounded,pi_tau_tau-mb}):~ Suppose that $(\Omega,\ms E,\ms F)$ is universally complete and that $(\!(\tau,\infty]\!]$, or equivalently $[\![0,\tau]\!]$, is $\ms F$-progressively measurable and $\pi\circ\tau < {\mf w_1}$. As in the proof of the implication (\ref{3-SPF_VECT.thm:optional_times.[tau]} $\Rightarrow$ \ref{3-SPF_VECT.thm:optional_times.pi_bounded,pi_tau_tau-mb}) we infer that $[\![0,\tau]\!]\in\ms P_\ovT^\alpha\otimes\ms E$ for some $\alpha\in{\mf w_1}$. We infer that $\pi\circ\tau\le \alpha$. Indeed, if there existed $\omega\in\Omega$ with $\pi\circ\tau(\omega) > \alpha$, then, with $t = (p(\tau(\omega)),\alpha)$, we would have $(t,\omega)\in[\![0,\tau]\!]$. Hence, Lemma~\ref{3-SPF_VECT.lemma:sets_in_msPovT^alpha_otimes_msE_are_inactive_beyond_alpha} would imply that, for $u = (p(\tau(\omega)),{\mf w_1})$, $(u,\omega)\in [\![0,\tau]\!]$, whence $\pi\circ\tau(\omega) = {\mf w_1}$ -- in contradiction to the second part of the hypothesis.

    By Theorem~\ref{3-SPF_VECT.thm:mb_proj_section}, we infer from the hypothesis that, for any $t\in\bRp$:
    \[ \{\tau < t\} = \mc P \prj_\Omega\big((\!(\tau,\infty]\!]\cap([0,t]_\ovT\times\Omega)\big) \in \ms F_t. \]
    Thus, $\tau$ satisfies the hypothesis of Lemma~\ref{3-SPF_VECT.lemma:[0,tau_omega1]_is_prog_mb_if_tau<t_is_msF_t}.

    For the remaining part of the claim, let $\beta\in{\mf w_1}$, $t\in\ovT$, and $M_t = (\!(\tau,\tau_{{\mf w_1}}]\!] \cap ([0,t]_\ovT \times\Omega)$. By hypothesis, and Lemma~\ref{3-SPF_VECT.lemma:[0,tau_omega1]_is_prog_mb_if_tau<t_is_msF_t}, $M_t\in\ms P_\ovT\otimes\ms F_t$. In particular,
    \[ M_t^\beta = M_t \cap [(\rho^{\beta+1})^{-1}(\bRp\times\{\beta\}) \times\Omega] \in\ms P_\ovT \otimes\ms F_t. \]
    Hence, by Theorem~\ref{3-SPF_VECT.thm:mb_proj_section}, and universal completeness of $\ms F_t$, 
    \[ \{\pi\circ\tau < \beta,\, \tau_\beta \le t\} = \mc P\prj_\Omega(M_t^\beta)\in\ms F_t. \]
    If $\beta \le \pi(t)$, then 
    \[ \{\pi\circ\tau < \beta,\, \tau\le t\} = \{\pi\circ\tau < \beta,\, \tau_\beta \le t\} \in \ms F_t. \]
    Else, $\pi(t) < \beta$, so that $\pi(t) + 1$ is countable. Let $Q_t = [0,p(t))_\bRp \cap \Q$, a countable set as well. As $\beta \le {\mf w_1}$, Remark~\ref{3-SPF_VECT.rmk:tau_&_msF_tau}, Item~\ref{3-SPF_VECT.rmk:tau_&_msF_tau.tau<t_in_msG_t}, and the previous case yield:
    \[ \{\pi\circ\tau < \beta,\, \tau\le t\} = \bigcup_{\gamma\in\pi(t)+1} \{\tau = (p(t),\gamma)\}\cup \bigcup_{q\in Q_t} \{\pi\circ\tau < \beta,\, \tau\le(q,{\mf w_1})\}\, \in \ms F_t. \]
    We conclude that
    \[ \{\pi\circ\tau = \beta,\,\tau\le t\} = \{\pi\circ\tau < \beta+1,\,\tau\le t\} \setminus \bigcup_{\gamma\in\beta+1} \{\pi\circ\tau < \gamma,\,\tau\le t\}\, \in \ms F_t, \]
    because $\beta+1\in{\mf w_1}$. In particular, this relation holds true if $\beta\in\alpha+1$.\smallskip

    (Ad \ref{3-SPF_VECT.thm:optional_times.stopping_times,pi_bounded,tau_tau-mb} $\Rightarrow$ \ref{3-SPF_VECT.thm:optional_times.pi_bounded,pi_tau_tau-mb}):~ Suppose Condition \ref{3-SPF_VECT.thm:optional_times.stopping_times,pi_bounded,tau_tau-mb} to hold true. Let $\alpha\in{\mf w_1}$ such that $\pi\circ\tau\le \alpha$. Hence, for all $\beta\in\alpha+1$ and $t\in\ovT$, we have, by definition of $\ms F_\tau$ and $\ms P_\ovT$,
    \[ \{\pi\circ\tau = \beta,\, \tau\le t\} = \big\{\rho^{\alpha+1}\circ\tau \in \bRp\times\{\beta\}\big\} \cap \{\tau\le t\} \, \in \ms F_t. \]
    
    (Ad \ref{3-SPF_VECT.thm:optional_times.pi_bounded,pi_tau_tau-mb} $\Rightarrow$ \ref{3-SPF_VECT.thm:optional_times.stopping_times,pi_bounded,tau_tau-mb}):~ Suppose Condition \ref{3-SPF_VECT.thm:optional_times.pi_bounded,pi_tau_tau-mb} to hold true. Let $\alpha\in{\mf w_1}$ such that $\pi\circ\tau\le \alpha$. Then, $\alpha+1$ is countable, hence, for any $t\in\ovT$, we have:
    \[ \{\tau\le t\} = \bigcup_{\beta\in\alpha+1} \{\pi\circ\tau = \beta,\, \tau\le t\} \, \in\ms F_t. \]
    Hence, $\tau$ is an $\ms F$-stopping time. Note that, for all $\gamma\in{\mf w_1}$, all $\beta\in\gamma+1$, and all $s\in\bRp$, we have
    \begin{equation}\label{3-SPF_VECT.eq:rho^gamma_circ_tau}
    \begin{aligned}
        \big\{\rho^\gamma\circ\tau \in [0,s]_\bRp\times\{\beta\}\big\} =& \begin{cases} \{\pi\circ\tau = \beta,\, p\circ\tau\le s\}, & \text{if } \beta < \gamma, \\ \{\beta \le \pi\circ\tau \le\alpha,\, p\circ\tau \le s\}, & \text{else} \end{cases} \\
        =& \begin{cases} \{\pi\circ\tau = \beta,\, \tau\le (s,{\mf w_1})\}, & \text{if } \beta < \gamma, \\ \{\beta \le \pi\circ\tau \le\alpha,\, \tau \le (s,{\mf w_1})\}, & \text{else.} \end{cases}
    \end{aligned}
    \end{equation}
    Hence, for all $t\in\ovT$, using that $\alpha$ is countable, we obtain:
    \[ \big\{\rho^\gamma\circ\tau \in [0,s]_\ovT\times\{\beta\}\big\} \cap \{\tau\le t\} = \begin{cases} \{\pi\circ\tau = \beta,\, \tau\le (s,{\mf w_1}) \wedge t\}, & \text{if } \beta < \gamma, \\ \{\beta \le \pi\circ\tau \le\alpha,\, \tau \le (s,{\mf w_1}) \wedge t\}, & \text{else} \end{cases} \, \in \ms F_t. \]
    By Lemma~\ref{3-SPF_VECT.lemma:generator_msP}, and the hypothesis, it follows easily that $\tau$ is $\ms F_\tau$-$\ms P_\ovT$-measurable.\medskip

    (Ad \ref{3-SPF_VECT.thm:optional_times.pi_bounded,pi_tau_tau-mb} $\Rightarrow$ [\ref{3-SPF_VECT.thm:optional_times.[tau]}, \ref{3-SPF_VECT.thm:optional_times.[tau,infty]}, and \ref{3-SPF_VECT.thm:optional_times.(tau,infty]}]):~ Suppose that Property~\ref{3-SPF_VECT.thm:optional_times.pi_bounded,pi_tau_tau-mb} is satisfied. By what has already been proven, Statement~\ref{3-SPF_VECT.thm:optional_times.stopping_times,pi_bounded,tau_tau-mb} then also holds. Let $\alpha\in{\mf w_1}$ be such that $\pi\circ\tau\le\alpha$. Then, in particular, $\pi\circ\tau < {\mf w_1}$. \smallskip
    
    In a \emph{first step}, we show the following intermediate result.

    \begin{lemma}\label{3-SPF_VECT.lemma:tau^beta_msF^beta_stopping_time}
        Let $\beta\in\alpha+1$, $\Omega^\beta = \{\pi\circ\tau = \beta\}$, $\ms F^\beta = (\ms F^\beta_t)_{t\in\bRp}$ be given by $\ms F^\beta_t = \ms F_{(t,\beta)}|_{\Omega^\beta}$ for $t\in\R_+$ and $\ms F^\beta_\infty = \ms F_\infty|_{\Omega^\beta}$, and $\tau^\beta = p\circ\tau|_{\Omega^\beta}$. Then, $\Omega^\beta\in\ms E$, $\ms F^\beta$ defines a filtration on $(\Omega^\beta,\ms E|_{\Omega^\beta})$ with classical real time half-axis $\bRp$, and $\tau^\beta$ is an $\ms F^\beta$-stopping time in the classical sense, that is, for all $t\in\R_+$, we have:
        \[ \{\omega\in\Omega^\beta \mid \tau^\beta(\omega) \le t\} \,\in \ms F_t^\beta. \]
    \end{lemma}

    \begin{proof}[Proof of Lemma~\ref{3-SPF_VECT.lemma:tau^beta_msF^beta_stopping_time}]
        First, note that, by Property~\ref{3-SPF_VECT.thm:optional_times.pi_bounded,pi_tau_tau-mb}:
        \[ \Omega^\beta = \{\pi\circ\tau=\beta,\, \tau\le\infty\} \,\in\ms F_\infty\subseteq\ms E. \]
        As $\ms F^\beta$ clearly inherits the monotonicity from $\ms F$, we infer that $\ms F^\beta$ defines a filtration on $(\Omega^\beta,\ms E|_{\Omega^\beta})$ with classical real time half-axis $\bRp$. Concerning the proof of the stopping time property, let $t\in\R_+$. Then,
        \[ \{\omega\in\Omega^\beta \mid \tau^\beta(\omega) \le t\} = \{\pi\circ\tau=\beta,\,\tau\le (t,\beta)\}\,\in\ms F_{(t,\beta)}|_{\Omega^\beta} = \ms F^\beta_t, \]
        by Property~\ref{3-SPF_VECT.thm:optional_times.pi_bounded,pi_tau_tau-mb}.
    \end{proof}

    For the \emph{second step} of Part ``\ref{3-SPF_VECT.thm:optional_times.pi_bounded,pi_tau_tau-mb} $\Rightarrow$ [\ref{3-SPF_VECT.thm:optional_times.[tau]}, \ref{3-SPF_VECT.thm:optional_times.[tau,infty]}, and \ref{3-SPF_VECT.thm:optional_times.(tau,infty]}]'' in the proof of Theorem~\ref{3-SPF_VECT.thm:optional_times}, let $\beta\in\alpha+1$, and $t\in\ovT$. Let $x = p(t)$, $Q_x = [0,x)_\bRp\cap \Q$, and extend $\ms F$ so that $\ms F_{(x,\beta)} = \ms F_\infty$ in case $x = \infty$. Let us define the following sets, describing the graph, epigraph, strict epigraph of $\tau^\beta$ below $x$:
    \begin{equation*}
        \begin{aligned}
             G_{x}^\beta =&~ \{(u,\omega)\in\bRp\times\Omega^\beta \mid u = \tau^\beta(\omega) \} \cap ([0,x]_\bRp\times\Omega^\beta), \\
             E_{x}^\beta =&~ \{(u,\omega)\in\bRp\times\Omega^\beta \mid  \tau^\beta(\omega) \le u \} \cap ([0,x]_\bRp\times\Omega^\beta), \\
             \mathrm{s}E_{x}^\beta =&~ \{(u,\omega)\in\bRp\times\Omega^\beta \mid  \tau^\beta(\omega) < u \} \cap ([0,x]_\bRp\times\Omega^\beta).
        \end{aligned}
    \end{equation*}
    In this step, we wish to analyse measurability properties of $\tau^\beta$ and of these sets in particular. By Lemma~\ref{3-SPF_VECT.lemma:tau^beta_msF^beta_stopping_time}, $\tau^\beta$ is an $\ms F^\beta$-stopping time in the classical sense. In particular,
    \begin{equation}\label{3-SPF_VECT.eq:tau^beta<x_cap_Omega^beta_in_msFx}
        \{\omega\in \Omega^\beta \mid \tau^\beta(\omega) < x\} = \bigcup_{q\in Q_x} \{\omega\in \Omega^\beta \mid \tau^\beta(\omega) \le q\} ~ \in \bigvee_{q\in Q_x} \ms F_q^\beta \subseteq \ms F_x^\beta. 
    \end{equation}    
    Moreover, classical theory (or Lemma~\ref{3-SPF_VECT.lemma:[0,tau_omega1]_is_prog_mb_if_tau<t_is_msF_t}) implies that 
    \[ G_{x}^\beta, \, E_{x}^\beta, \, \mathrm{s}E_{x}^\beta~\in\ms B_\bRp\otimes\ms F^\beta_{x}.  \]
    
    By basic measure theory, we have:
    \[ \ms B_\bRp\otimes\ms F^\beta_{x} = \ms B_\bRp\otimes\ms F_{(x,\beta)}|_{\Omega^\beta} \subseteq \big(\ms B_\bRp\otimes\ms F_{(x,\beta)}\big)|_{\bRp\times\Omega^\beta}. \]
    Hence, there are $\tilde G_{x}^\beta,\tilde E_{x}^\beta,\widetilde{\mathrm{s}E}_{x}^\beta\in \ms B_\bRp\otimes\ms F_{(x,\beta)}$ with 
    \[ G_{x}^\beta = \tilde G_{x}^\beta \cap (\bRp\times\Omega^\beta), \qquad E_{x}^\beta = \tilde E_{x}^\beta \cap (\bRp\times\Omega^\beta), \qquad \mathrm{s}E_{x}^\beta = \widetilde {\mathrm{s}E}_{x}^\beta \cap (\bRp\times\Omega^\beta). \]
    Then, by hypothesis,
    \[ G_{x}^\beta = G_{x}^\beta \cap [\bRp\times\{\tau\le (x,\beta)\}] = \tilde G_{x}^\beta \cap[\bRp\times \{\pi\circ\tau = \beta,\,\tau\le (x,\beta)\}] ~ \in \ms B_\bRp\otimes\ms F_{(x,\beta)}, \]
    and similarly, 
    \[ E_{x}^\beta = E_{x}^\beta \cap [\bRp\times\{\tau\le (x,\beta)\}] = \tilde E_{x}^\beta \cap[\bRp\times\{\pi\circ\tau=\beta,\,\tau\le (x,\beta)\}] ~ \in \ms B_\bRp\otimes\ms F_{(x,\beta)}, \]
    \[ \mathrm{s}E_{x}^\beta = \mathrm{s}E_{x}^\beta \cap [\bRp\times\{\tau\le (x,\beta)\}] = \widetilde {\mathrm{s}E}_{x}^\beta \cap[\bRp\times\{\pi\circ\tau=\beta,\,\tau\le (x,\beta)\}] ~ \in \ms B_\bRp\otimes\ms F_{(x,\beta)}. \]
    In particular, for any $\gamma\in{\mf w_1}$ and any $S\in \ms B_{\gamma+1}$, we get that
    \[ S \times G_{x}^\beta, \, S\times E_{x}^\beta,\, S\times \mathrm{s}E_{x}^\beta\in \ms B_{\gamma+1}\otimes\ms B_{\bRp}\otimes\ms F_{(x,\beta)}. \]
    For any $\gamma\in{\mf w_1}$, the map $f_\gamma\colon\ovT\times\Omega \to (\gamma+1)\times\bRp\times\Omega,\, (u,\omega)\mapsto (\pi\circ\rho^{\gamma}(u),p\circ\rho^{\gamma}(u),\omega)$ is a composition of suitably measurable transformations, and is therefore $\ms P_\ovT\otimes\ms F_{(x,\beta)}$-$\ms B_{\gamma+1}\otimes\ms B_{\bRp}\otimes\ms F_{(x,\beta)}$-measurable. Hence, for any $\gamma\in{\mf w_1}$ and any $S\in \ms B_{\gamma+1}$,
    \begin{equation}\label{3-SPF_VECT.eq:f_gamma^-1(S_times_graph)_is_mb}
        f_{\gamma}^{-1}(S \times G_{x}^\beta),\, f_{\gamma}^{-1}(S \times E_{x}^\beta),\, f_\gamma^{-1}( S\times \mathrm{s}E_{x}^\beta)~ \in \ms P_\ovT\otimes\ms F_{(x,\beta)}.
    \end{equation}
    
    In a \emph{third step}, we show that, for all $t\in\ovT$, $x=p(t)$, and $Q_x = [0,x)_\bRp\cap \Q$, we have the following two decompositions:
    \begin{equation}\label{3-SPF_VECT.eq:[[tau]]_cap_[[0,t]]_decomposition}
        [\![\tau]\!] \cap ([0,t]_\ovT\times\Omega) = \bigcup_{\substack{\beta\in\alpha+1\colon\\ \beta > \pi(t)}}\bigcup_{q\in Q_x} f_{\alpha+1}^{-1}(\{\beta\} \times G_{q}^\beta) \cup \bigcup_{\substack{\beta\in\alpha+1\colon\\ \beta \le \pi(t)}}  f_{\alpha+1}^{-1}(\{\beta\} \times G_{x}^\beta),
    \end{equation}
    and
    \begin{equation}\label{3-SPF_VECT.eq:[[tau,t]]_decomposition}
        \begin{aligned}
            &~[\![\tau,\infty]\!] \cap ([0,t]_\ovT\times\Omega)\\
            =&~ \bigcup_{\substack{\beta\in\alpha+1\colon\\ \beta > \pi(t)}}\Bigg[\bigcup_{q\in Q_x} \Big[ f_{\alpha}^{-1}([(\alpha+1)\setminus \beta] \times E_{q}^\beta) \cup f_{\alpha}^{-1}(\beta \times \mathrm{s}E_{q}^\beta)\Big] \\
            &\qquad\qquad\quad \cup \Big [ [x,t]_\ovT \times \{\omega\in\Omega^\beta \mid \tau^\beta(\omega) < x\}\Big ] \Bigg] \\ 
            &~\cup\bigcup_{\substack{\beta\in\alpha+1\colon\\ \beta \le \pi(t)}} \Bigg[\Big[ f_{\alpha}^{-1}([(\alpha+1) \setminus \beta] \times E_{x}^\beta) \cap ([0,t]_\ovT\times\Omega)\Big]\cup f_{\alpha}^{-1}(\beta \times \mathrm{s}E_{x}^\beta)\Bigg] 
        \end{aligned}
    \end{equation}
    For the proof of these two decompositions, let $t\in\ovT$ and $(u,\omega)\in\ovT\times\Omega$. We first prove Decomposition~\ref{3-SPF_VECT.eq:[[tau]]_cap_[[0,t]]_decomposition}. By definition, $(u,\omega) \in [\![\tau]\!] \cap ([0,t]_\ovT\times\Omega)$ iff $u = \tau(\omega) \le t$. This latter statement splits into two disjunct cases, generated by the alternative ``$\pi(\tau(\omega)) > \pi(t)$'' or ``$\pi(\tau(\omega)) \le \pi(t)$''. 
    \begin{enumerate}
        \item On the one hand, $u = \tau(\omega) \le t$ and $\pi(\tau(\omega)) > \pi(t)$ both hold true iff there is $q\in\Q$ with $u=\tau(\omega) \le q < p(t)$ and $\pi(u) > \pi(t)$. This is equivalent to the relation $\pi(t) < \pi(u) \le \alpha$ and the existence of $q\in Q_x$ with $\omega\in\Omega^{\pi(u)}$ and $p(u) = \tau^{\pi(u)}(\omega) \le q$, i.e.\ $(p(u),\omega)\in G_q^{\pi(u)}$. This in turn is equivalent to the existence of $q\in Q_x$ and $\beta\in\alpha+1$ with $\beta > \pi(t)$ such that $(u,\omega)\in f_{\alpha+1}^{-1}(\{\beta\} \times G_{q}^\beta)$. 
        \item On the other hand, $u = \tau(\omega) \le t$ and $\pi(\tau(\omega)) \le \pi(t)$ both hold true iff $p(u) = p\circ\tau(\omega) \le p(t)$ and $\pi(u) = \pi\circ\tau(\omega) \le \pi(t)$. This is equivalent to the conjunction of the relation $\pi(u) \le \alpha \wedge \pi(t)$ and the statement that $\omega\in\Omega^{\pi(u)}$ and $p(u) = \tau^{\pi(u)}(\omega) \le x$ hold true, i.e.\ $(p(u),\omega) \in G_x^{\pi(u)}$. This in turn is equivalent to the existence of $\beta\in\alpha+1$ with $\beta\le\pi(t)$ such that  $(u,\omega) \in f_{\alpha+1}^{-1}(\{\beta\} \times G_{x}^\beta)$. 
    \end{enumerate}
    The first decomposition, Equation~\ref{3-SPF_VECT.eq:[[tau]]_cap_[[0,t]]_decomposition}, is proven.

    We continue with proving the second decomposition, Equation~\ref{3-SPF_VECT.eq:[[tau,t]]_decomposition}. By definition, $(u,\omega) \in [\![\tau,\infty]\!] \cap ([0,t]_\ovT\times\Omega)$ iff $\tau(\omega) \le u \le t$. This latter statement splits into four disjunct cases, generated by the following two alternatives: ``$\pi(\tau(\omega)) > \pi(t)$'' or ``$\pi(\tau(\omega))\le\pi(t)$''; ``$\pi(\tau(\omega))\le \pi(u)$'' or ``$\pi(\tau(\omega)) > \pi(u)$''. 
    \begin{enumerate}
        \item First, $\tau(\omega)\le u \le t$, $\pi(\tau(\omega)) > \pi(t)$, and $\pi(\tau(\omega))\le \pi(u)$ all hold true iff $p\circ\tau(\omega) \le p(u) < p(t)$ and $\pi(u) \ge \pi(\tau(\omega)) > \pi(t)$; a condition which is satisfied iff there is $q\in\Q$ with $p\circ\tau(\omega)\le p(u) \le  q <  x$ and $\pi(u) \ge \pi \circ \tau(\omega) > \pi(t)$. This is equivalent to the existence of $q\in Q_x$ and $\beta\in\alpha+1$ with $\beta > \pi(t)$ such that $(p(u),\omega)\in E_q^\beta$ and $\pi(u)\ge\beta$, i.e.\ the fact that $(u,\omega)\in f_{\alpha}^{-1}([(\alpha+1)\setminus \beta] \times E_{q}^\beta)$.
        \item Second, $\tau(\omega)\le u \le t$, $\pi(\tau(\omega)) > \pi(t)$, and $\pi(\tau(\omega)) > \pi(u)$ all hold true iff a) $p\circ\tau(\omega) < p(u) < p(t)$ and $\pi(u) \vee \pi(t) < \pi(\tau(\omega))$, or b) $p\circ\tau(\omega) < p(u) = p(t)$ and $\pi(u) \le \pi(t) < \pi(\tau(\omega))$. Condition a) is equivalent to the existence of $q\in Q_x$ and $\beta\in\alpha+1$ with $\beta > \pi(t)$ such that $(p(u),\omega)\in\mathrm{s}E_q^\beta$ and $\pi(u) < \beta$, i.e.\ $f_\alpha(u,\omega)\in \beta\times\mathrm{s}E_q^\beta$, while condition b) is equivalent to $x\le u \le t$ and the existence of $\beta\in\alpha+1$ with $\beta > \pi(t)$ such that $\omega\in\Omega^\beta$ and $\tau^\beta(\omega) < x$.
        \item Third, $\tau(\omega)\le u \le t$, $\pi(\tau(\omega)) \le \pi(t)$, and $\pi(\tau(\omega))\le \pi(u)$ all hold true iff a) $p\circ\tau(\omega)\le p(u) < p(t)$ and $\pi\circ\tau(\omega) \le \pi(t) \wedge \pi(u)$, or b) $p\circ\tau(\omega)\le p(u) = p(t)$ and $\pi\circ\tau(\omega) \le \pi(u) \le \pi(t)$. a) is equivalent to the existence of $q\in Q_x$ and $\beta\in\alpha+1$ with $\beta \le \pi(t)$ such that $(p(u),\omega)\in E_q^\beta$ and $\pi(u) \ge \beta$, while b) is equivalent to the existence of $\beta\in \alpha+1$ with $\beta \le \pi(t)$ such that $(p(u),\omega)\in E_x^\beta$, $\pi(u) \ge \beta$, and $u\in [x,t]_\ovT$. Hence, the conjunction of a) and b) is equivalent to the existence of $\beta\in\alpha+1$ with $\beta\le\pi(t)$ such that $f_\alpha(u,\omega) \in [(\alpha+1)\setminus \beta]\times E_x^\beta$ and $u\le t$.
        \item Fourth, $\tau(\omega)\le u \le t$, $\pi(\tau(\omega)) \le \pi(t)$, and $\pi(\tau(\omega)) > \pi(u)$ all hold true iff $p\circ\tau(\omega) < p(u) \le p(t)$, $\pi(u) < \pi\circ\tau(\omega) \le \pi(t)$. This is equivalent to the existence of $\beta\in\alpha+1$ with $\beta \le \pi(t)$ such that $(p(u),\omega)\in \mathrm{s}E_x^\beta$ and $\pi(u) < \beta$, i.e.\ $f_\alpha(u,\omega) \in \beta \times \mathrm{s}E_x^\beta$.
    \end{enumerate}
    \smallskip
    
    In a \emph{fourth step}, we combine the second and third steps, namely the Statements~\ref{3-SPF_VECT.eq:[[tau]]_cap_[[0,t]]_decomposition}, \ref{3-SPF_VECT.eq:[[tau,t]]_decomposition}, \ref{3-SPF_VECT.eq:f_gamma^-1(S_times_graph)_is_mb}, and~\ref{3-SPF_VECT.eq:tau^beta<x_cap_Omega^beta_in_msFx}, recall that, with the notation from the third step, 
    \[ \{\omega\in\Omega^\beta \mid \tau^\beta(\omega) < x\} = \{\pi\circ\tau = \beta,\, \tau < x\} = \bigcup_{\Q\ni q < x} \{\pi\circ\tau = \beta,\, \tau\le q\}\, \in\ms F_x, \]
    and infer that $[\![\tau]\!]$ and $[\![\tau,\infty]\!]$ are $\ms F$-progressively measurable. By complement-stability of $\Prg(\ms P_\ovT,\ms F)$, then also 
    \[ (\!(\tau,\infty]\!] = [\![\tau,\infty]\!] \setminus [\![\tau]\!] \]
    is $\ms F$-progressively measurable. The proof is complete.
\end{proof}

\begin{proof}[Proof of Corollary~\ref{3-SPF_VECT.cor:xi_tau_is_F_tau_mb}]
    Under the completeness hypothesis, the map $f\colon\Omega\to\ovT\times\Omega,\, \omega\mapsto (\tau(\omega),\omega)$ is $\ms F_\tau$-$\Prg(\ms F)$-measurable. Indeed, for any $t\in\ovT$ and $M\in\Prg(\ms F)$, we then have $[\![\tau]\!]\cap M \cap [\![0,t]\!] \in \ms P_\ovT\otimes\ms F_t$, by Theorem~\ref{3-SPF_VECT.thm:optional_times}, and, by Theorem~\ref{3-SPF_VECT.thm:mb_proj_section} --- now using completeness ---
    \[ f^{-1}(M) \cap \{\tau\le t\} = \prj_\Omega([\![\tau]\!]\cap M \cap [\![0,t]\!])\, \in \ms F_t. \]
    Hence, $\xi_\tau = \xi \circ f$ is $\ms F_\tau$-$\ms Y$-measurable.
\end{proof}

\begin{proof}[Proof of Proposition~\ref{3-SPF_VECT.prop:basic_properties_of_optional_times}]
    (Ad~\ref{3-SPF_VECT.prop:basic_properties_of_optional_times.translated_stopping_times}):~ Let $\sigma = p\circ\tau + \e$. Then, $\pi\circ\sigma = 0$. Moreover, we have
    \[ \{\pi\circ\sigma = 0,\, \sigma\le \infty\} = \Omega~\in\ms F_\infty = \ms F_{\infty+}, \]
    and, for all $t\in\ovT\setminus\{\infty\}$:
    \begin{equation*}
        \begin{aligned}
            \{\pi\circ\sigma = 0,\, \sigma\le t\} =&~ \{p\circ\tau + \e \le t\} \\
            =&~ \{p\circ\tau + \e \le p(t) \} \\ 
            =&~  \{p\circ\tau \le p(t) - \e \} \\
            =&~ \{ \tau \le (p(t)-\e,{\mf w_1}) \} ~\in \ms F_{(p(t) - \e,{\mf w_1})}.    
        \end{aligned}
    \end{equation*}
    Note that $\ms F_{(p(t),{\mf w_1})} \subseteq \ms F_u$ for all real $u>t$. Hence, if $\e=0$, $\{\pi\circ\sigma = 0,\, \sigma\le t\}\in\ms F_{t\Plus}$. If $\e>0$, then $\ms F_{(p(t) - \e,{\mf w_1})}\subseteq \ms F_t$.\smallskip

    (Ad~\ref{3-SPF_VECT.prop:basic_properties_of_optional_times.lim}):~ Let $C$ be the set of $\omega\in\Omega$ such that $(\tau_n(\omega))_{n\in\N}$ converges in $\bRp\times(\alpha+1)$. As $\pi|_{\bRp\times(\alpha+1)}$ is continuous as a map $\bRp\times(\alpha+1)\to(\alpha+1)$, we have $\pi\circ\tau(\omega) = \lim_{n\to\infty} \pi\circ\tau_n(\omega) \le \alpha$ for all $\omega\in C$. For all $\omega\in C^\complement$, we have $\pi\circ\tau(\omega) = \pi(\infty) = 0\le\alpha$. 
    
    In view of Theorem~\ref{3-SPF_VECT.thm:optional_times}, it remains to consider arbitrary $\beta\in\alpha+1$ and $t\in\ovT$ and check the $\ms F_{t\Plus}$-measurability of $\{\pi\circ \tau=\beta,\, \tau\le t\}$.
    In doing so, we use the following facts: a) convergence of a sequence in $\bRp\times(\alpha+1)$ is equivalent to the convergence of the component sequences; b) convergence of a sequence $(x_n)_{n\in\N}$ in $\bRp$ is equivalent to $(x_n)_{n\in\N}$ being Cauchy or the set $\{n\in\N \mid x_n \le \ell\}$ being finite for all $\ell\in\N$; c) for all $n\in\N$, $p\circ\tau_n$ is an $\ms F_\Plus$-optional time (by Part~\ref{3-SPF_VECT.prop:basic_properties_of_optional_times.translated_stopping_times} just proven before), and thus, by classical theory, $\{p\circ\tau_n \le p\circ\tau_m + \kappa\} \in (\ms F_\Plus)_{p\circ\tau_n}$, for all $m,n\in\N$, $\kappa\in\R_+$.
    
    i) We first focus on the case $\beta\le\pi(t)$ and $t<\infty$. In this case, using Properties a), b), and c), we infer that all $m\in\N$ satisfy:
    \begin{equation*}
        \begin{aligned}
            &~\{\pi\circ \tau=\beta,\, \tau\le t\} = \{\pi\circ\tau = \beta,\, p\circ\tau \le p(t)\} \\
            =&~ 
            \begin{cases}
                &\bigcap_{\gamma\in\beta}\bigcap_{\ell=m}^\infty \bigcup_{k\in\N} \bigcap_{n,m=k}^\infty \Big(\{\pi\circ\tau_n\in (\gamma,\beta],\,\tau_n \le p(t)+2^{-\ell}\} \\
                &\qquad\qquad \cap\{p\circ\tau_n\le p\circ\tau_m + 2^{-\ell},\,p\circ\tau_n \le p(t)+2^{-\ell}\} \\
                &\qquad\qquad \cap \{p\circ\tau_m\le p\circ \tau_n + 2^{-\ell},\, p\circ\tau_m \le p(t)+2^{-\ell}\}\Big),\\
                &\qquad\qquad\qquad\qquad\qquad   \text{if } \beta \text{ is a limit ordinal,} \\
                &\bigcap_{\ell=m}^\infty \bigcup_{k\in\N} \bigcap_{n,m=k}^\infty \Big(\{\pi\circ\tau_n=\beta,\,\tau_n \le p(t)+2^{-\ell}\} \\
                &\qquad\qquad \cap\{p\circ\tau_n\le p\circ\tau_m + 2^{-\ell},\,p\circ\tau_n \le p(t)+2^{-\ell}\} \\
                &\qquad\qquad \cap \{p\circ\tau_m\le p\circ \tau_n + 2^{-\ell},\, p\circ\tau_m \le p(t)+2^{-\ell}\}\Big),\\
                &\qquad\qquad\qquad\qquad\qquad  \text{else}
            \end{cases}
            \\
            &~\in\ms F_{(p(t) + 2^{-m})\Plus}.
        \end{aligned}
    \end{equation*} 
    As this holds true for all $m\in\N$, we obtain $\{\pi\circ \tau=\beta,\, \tau\le t\}\in \ms F_{t\Plus}$. 

    ii) Second, we consider the case $\pi(t) < \beta$ and $t<\infty$. Then, by case i) just studied, we get
    \begin{align*}
        &~\{\pi\circ\tau = \beta,\, \tau\le t\} = \{\pi\circ\tau = \beta,\, \tau < p(t) \} \\
        =&~ \bigcup_{\substack{q\in\Q\colon\\ q < p(t)}} \{\pi\circ\tau = \beta,\, \tau \le (q,\beta)\} ~\in \bigvee_{\substack{q\in\Q\colon\\ q < p(t)}} \ms F_{(q,\beta)\Plus} \subseteq \ms F_{p(t)} \subseteq \ms F_{t\Plus}.
    \end{align*}

    iii) Finally, we consider the case $t=\infty$. Then, if $\beta > 0$, using results i) and ii) yield
    \[ \{\pi\circ \tau=\beta,\, \tau\le t\} \cap C = \{\pi\circ \tau=\beta,\,\tau < \infty\} \cap C = \bigcup_{n\in\N} \{\pi\circ \tau=\beta,\, \tau\le n\}~\in\ms F_\infty; \]
    and
    \[ \{\pi\circ \tau=\beta,\, \tau\le t\} \setminus C = \{\pi\circ\tau = \beta,\, \tau = \infty\} \setminus C = \emptyset~\in\ms F_\infty, \]
    whence $\{\pi\circ \tau=\beta,\, \tau\le t\}\in\ms F_\infty = \ms F_{\infty\Plus}$.    
    If $\beta = 0$, then, again by i), 
    \begin{equation*}
        \begin{aligned}
            &~\{\pi\circ \tau=\beta,\, \tau\le t\} \cap C \\
            =&~ \bigcup_{n\in\N} \{\pi\circ \tau=0,\, \tau\le n\} \cup \Big(\liminf_{n\to\infty} \{\pi\circ\tau_n = 0\} \cap \bigcap_{k\in\N}\liminf_{n\to\infty} \{\tau_n > k\}\Big) ~ \in \ms F_\infty;
        \end{aligned}
    \end{equation*}
    and
    \begin{equation*}
        \begin{aligned}
            &~\{\pi\circ \tau=\beta,\, \tau\le t\} \setminus C \\
            =&~ C^\complement \\
            =&~ \Big[\bigcap_{\gamma\in(\alpha+1)\setminus\{0\}} \bigcap_{\delta\in\gamma} \big(\liminf_{n\to\infty} \{\delta < \pi\circ \tau_n \le \gamma\}\big)^\complement \cap \big(\liminf_{n\to\infty}\{\pi\circ\tau_n=0\}\big)^\complement\Big] \\
            & \cup \Bigg[\Big(\bigcup_{\ell\in\N} \bigcap_{k\in\N} \bigcup_{n,m=k}^\infty \{p\circ\tau_n\le p\circ\tau_m + 2^{-\ell}\}^\complement \cup \{p\circ\tau_m\le p\circ \tau_n + 2^{-\ell}\}^\complement \Big) \\
            &\qquad\cap \Big(\bigcup_{\ell\in\N} \limsup_{n\to\infty} \{ \tau_n \le \ell\}\Big)\Bigg] \quad \in \ms F_{\infty\Plus},
        \end{aligned}
    \end{equation*}
    whence again $\{\pi\circ \tau=\beta,\, \tau\le t\}\in\ms F_{\infty\Plus}$.
    \smallskip
    
    (Ad~\ref{3-SPF_VECT.prop:basic_properties_of_optional_times.comparable}):~ Let $t\in\ovT$ and $Q_t = [0,t)_\ovT\cap\Q$. Note that
    \[ \{p\circ\tau_0 < p\circ\tau_1,\ \tau_1 \le t\} = \bigcup_{q\in Q_t} \{\tau_0 < q < \tau_1 \le t\} ~\in\ms F_t, \]
    and that
    \[ \{p\circ\tau_0 = p\circ\tau_1,\, \tau_0\le t,\, \tau_1\le t\} = \Big[\bigcap_{q\in Q_t} \{ \tau_0 < q < \tau_1\}^\complement \cap \{\tau_1 < q < \tau_0\}^\complement \Big] \cap \{\tau_0\le t,\, \tau_1\le t\} \, \in \ms F_t. \]
    As both $\tau_0$ and $\tau_1$ are optional times, there is $\alpha\in\mf w_1$ such that $\pi\circ\tau_k \le \alpha$ for both $k\in\{0,1\}$. Hence,
    \begin{align*}
        &~ \{\tau_0\le\tau_1\} \cap \{\tau_1\le t\} \\
        =&~ \{p\circ\tau_0 < p\circ\tau_1,\, \tau_1 \le t\} \cup \Big( \{p\circ\tau_0 = p\circ\tau_1,\, \tau_0\le t,\, \tau_1\le t\} \cap \{ \pi\circ\tau_0 \le \pi\circ\tau_1,\, \tau_0\le t,\, \tau_1\le t\}\Big) \\
        =&~ \{p\circ\tau_0 < p\circ\tau_1,\, \tau_1 \le t\} \cup \Big( \{p\circ\tau_0 = p\circ\tau_1,\, \tau_0\le t,\, \tau_1\le t\} \\
        &\qquad\qquad\qquad\qquad \cap \bigcup_{\beta\in\alpha+1}\bigcup_{\gamma\in\beta+1}\{ \pi\circ\tau_0 =\gamma,\, \tau_0\le t\} \cap \{ \pi\circ\tau_1 =\beta,\, \tau_1\le t\}\Big)~\in\ms F_t.
    \end{align*}
    Thus, $\{\tau_0\le\tau_1\}\in\ms F_{\tau_1}$. 
    Similarly, we get
    \begin{align*}
        &~ \{\tau_0<\tau_1\} \cap \{\tau_1\le t\} \\
        =&~ \{p\circ\tau_0 < p\circ\tau_1,\, \tau_1 \le t\} \cup \Big( \{p\circ\tau_0 = p\circ\tau_1,\, \tau_0\le t,\, \tau_1\le t\} \cap \{ \pi\circ\tau_0 < \pi\circ\tau_1,\, \tau_0\le t,\, \tau_1\le t\}\Big) \\
        =&~ \{p\circ\tau_0 < p\circ\tau_1,\, \tau_1 \le t\} \cup \Big( \{p\circ\tau_0 = p\circ\tau_1,\, \tau_0\le t,\, \tau_1\le t\} \\
        &\qquad\qquad\qquad\qquad \cap \bigcup_{\beta\in\alpha+1}\bigcup_{\gamma\in\beta}\{ \pi\circ\tau_0 =\gamma,\, \tau_0\le t\} \cap \{ \pi\circ\tau_1 =\beta,\, \tau_1\le t\}\Big)~\in\ms F_t.
    \end{align*}
    Hence, $\{\tau_0 < \tau_1\}\in\ms F_{\tau_1}$. By complement-stability, $\{\tau_1\le\tau_0\}\in\ms F_{\tau_1}$. Upon switching the roles of $\tau_0$ and $\tau_1$, we infer $\{\tau_0\le\tau_1\}\in\ms F_{\tau_0}$.\smallskip

    (Ad~\ref{3-SPF_VECT.prop:basic_properties_of_optional_times.max_min}):~ For $k\in\{0,1\}$, let $\alpha_k\in\mf w_1$ be such that $\pi\circ\tau_k \le \alpha_k$. Let $\alpha = \alpha_0 \vee \alpha_1$. Hence, $\pi \circ (\tau_0 \wedge \tau_1)\le \alpha$ and $\pi \circ (\tau_0\vee\tau_1) \le \alpha$. 

    Furthermore, let $\beta\in\alpha+1$ and $t\in\ovT$. Then, using Part~\ref{3-SPF_VECT.prop:basic_properties_of_optional_times.comparable},
    \begin{align*}
        &~ \{\pi\circ(\tau_0\wedge\tau_1)=\beta,\, \tau_0\wedge\tau_1\le t\} \\
        =&~ \big(\{\tau_0\le\tau_1,\, \tau_0\le t\} \cap \{\pi\circ\tau_0=\beta,\, \tau_0\le t\}\big) \cup \big(\{\tau_1\le\tau_0,\, \tau_1\le t\} \cap \{\pi\circ\tau_1=\beta,\, \tau_1\le t\}\big)\, \in\ms F_t.
    \end{align*}
    Similarly, using Part~\ref{3-SPF_VECT.prop:basic_properties_of_optional_times.comparable} again, we show that
    \begin{align*}
        &~ \{\pi\circ(\tau_0\vee\tau_1)=\beta,\, \tau_0\vee\tau_1\le t\} \\
        =&~ \big(\{\tau_0\le\tau_1,\, \tau_1\le t\} \cap \{\pi\circ\tau_1=\beta,\, \tau_1\le t\}\big) \cup \big(\{\tau_1\le\tau_0,\, \tau_0\le t\} \cap \{\pi\circ\tau_0=\beta,\, \tau_0\le t\}\big)\, \in\ms F_t.
    \end{align*}
\end{proof}

\begin{remark}\label{3-SPF_VECT.rmk:basic_properties_of_optional_times.proofs_using_universal_completeness}
    If $(\Omega,\ms E,\ms F)$ is universally complete, some of the proofs above can be simplified, by using Theorem~\ref{3-SPF_VECT.thm:optional_times}. For example, Property~\ref{3-SPF_VECT.prop:basic_properties_of_optional_times.comparable} can be proven as follows.

    \begin{proof}[Proof of Property~\ref{3-SPF_VECT.prop:basic_properties_of_optional_times.comparable} under universal completeness]
    Note that, for any $t\in\ovT$:
    \begin{equation*}
        \begin{aligned}
            \{\tau_0\le\tau_1\} \cap \{\tau_1\le t\} =&~ \prj_\Omega([\![\tau_1]\!] \cap [\![\tau_0,t]\!]), \\
            \{\tau_1 < \tau_0\} \cap \{\tau_0\le t\} =&~ \prj_\Omega([\![\tau_0]\!] \cap (\!(\tau_1,t]\!]). \\
        \end{aligned}
    \end{equation*}
    As $\tau_0$ and $\tau_1$ are $\ms F$-optional times, $[\![\tau_1]\!] \cap [\![\tau_0,\infty]\!]$ and $[\![\tau_0]\!] \cap (\!(\tau_1,t]\!]$ are $\ms F$-progressively measurable, by Theorem~\ref{3-SPF_VECT.thm:optional_times}. Hence, their respective intersections with $[\![0,t]\!]$ are elements of $\ms P_\ovT\otimes\ms F_t$. Applying the Measurable Projection Theorem~\ref{3-SPF_VECT.thm:mb_proj_section} and using the universal completeness assumption, we obtain that the left-hand side of the two equations above are elements of $\ms F_t$. 
    
    As this holds true for any $t\in\ovT$, we infer $\{\tau_0\le\tau_1\} \in \ms F_{\tau_1}$ and $\{\tau_0\le\tau_1\} = \{\tau_1 < \tau_0\}^\complement\in\ms F_{\tau_0}$. 
    \end{proof}
\end{remark}

In addition, the following claims can be easily shown using universal completeness.
\begin{proposition}\label{3-SPF_VECT.prop:basic_properties_of_optional_times.addon_under_completeness}
    Suppose that $(\Omega,\ms E,\ms F)$ is a universally complete and let $(\tau_n)_{n\in\N}$ be a sequence of $\ms F$-optional times. Then, the following statements holds true:
    \begin{enumerate}[start=5]
        \item\label{3-SPF_VECT.prop:basic_properties_of_optional_times.sup} The scenariowise supremum $\sup_{n\in\N} \tau_n$ is an $\ms F$-optional time.
        \item\label{3-SPF_VECT.prop:basic_properties_of_optional_times.inf} The scenariowise infimum $\sigma = \inf_{n\in\N} \tau_n$ is an $\ms F$-optional time iff $\bigcup_{n\in\N} \{\sigma = \tau_n\} = \Omega$.
    \end{enumerate}
\end{proposition}
\begin{proof}
    (Ad~\ref{3-SPF_VECT.prop:basic_properties_of_optional_times.sup}):~ We assume that $\tau = \sup_{n\in\N} \tau_n$. By Theorem~\ref{3-SPF_VECT.thm:optional_times}, $[\![0,\tau_n)\!)$ is $\ms F$-progressively measurable for any $n\in\N$. As
    \[ [\![0,\tau)\!) = \bigcup_{n\in\N} [\![0,\tau_n)\!), \]
    $[\![0,\tau)\!)$ is so, too. Using the completeness assumption and applying Theorem~\ref{3-SPF_VECT.thm:optional_times} again, we infer that $\tau$ is an $\ms F$-optional time.\smallskip

    (Ad~\ref{3-SPF_VECT.prop:basic_properties_of_optional_times.inf}):~ By Theorem~\ref{3-SPF_VECT.thm:optional_times}, $(\!(\tau_n,\infty]\!]$ is $\ms F$-progressively measurable for any $n\in\N$. As
    \[ (\!(\sigma,\infty]\!] = \bigcup_{n\in\N} (\!(\tau_n,\infty]\!], \]
    $(\!(\sigma,\infty]\!]$ is so, too. If $\pi\circ\sigma < {\mf w_1}$, then, using the completeness assumption and applying Theorem~\ref{3-SPF_VECT.thm:optional_times} again, we infer that $\sigma$ is an $\ms F$-optional time. Conversely, if $\sigma$ is an optional time, then $\pi\circ\sigma<{\mf w_1}$, by the same theorem.
    
    It remains to prove that 
    \[ \bigcup_{n\in\N} \{\sigma = \tau_n\} = \{\pi\circ\sigma < {\mf w_1}\}. \]
    As $\pi\circ\tau_n < {\mf w_1}$ for any $n\in\N$, the inclusion ``$\subseteq$'' obtains. For the proof of the inclusion ``$\supseteq$'', let $\omega\in\Omega$ satisfy $\pi\circ\sigma < {\mf w_1}$. Then, $\{p\circ\tau_n(\omega) \mid n\in\N\}$ has a minimum $x$ in $\bRp$. Hence, $\{\pi\circ \tau_n(\omega) \mid n\in\N\colon p\circ\tau_n(\omega) = x\}$ has a minimum because ${\mf w_1}$ is well-ordered. If $n$ denotes an element of $\N$ that this minimum is attained at, then $\sigma(\omega) = \tau_n(\omega)$.
\end{proof}

\begin{proof}[Proof of Proposition~\ref{3-SPF_VECT.prop:any_ovl_msF-optional_time_has_msF-optional_time_version}]
    Let $\P\in\mf P_{\ms E}$ and $\ovl\tau$ be an $\ovl{\ms F}$-optional time. Then, there is $\alpha\in{\mf w_1}$ such that $\pi\circ\ovl\tau\le\alpha$. 
    For $n\in\N$, define
    \[ \ovl\tau_n = \big(\inf\{k\in\N \mid \ovl\tau \le k 2^{-n}\}\cdot 2^{-n}, \pi\circ\ovl\tau\big). \]
    Then, $\pi\circ\ovl\tau_n\le \alpha$. Moreover, for any $\beta\in\alpha+1$ and $t\in\ovT$, we have
    \[ \{\pi\circ\ovl\tau_n = \beta,\, \ovl\tau_n \le t\} = \bigcup_{\substack{k\in\N\colon\\ (k2^{-n},\beta)\le t}} \{\pi\circ\ovl\tau=\beta,\, \ovl\tau\le k2^{-n}\} \,\in\ovl{\ms F}_t. \]
    Hence, $(\ovl\tau_n)_{n\in\N}$ is a sequence of $\ms F$-optional times.

    For $k,n\in\N$ and $\beta\in\alpha+1$, let
    \[ \ovl E_{k,\beta}^n = \{\ovl\tau_n = (k2^{-n},\beta)\}. \]
    $\ovl E_{k,\beta}^n\in\ovl{\ms F}_{(k2^{-n},\beta)}$, because $\ovl\tau_n$ is an $\ovl{\ms F}$-optional time. Hence, there is a family $(E_{k,\beta}^n)_{k,n\in\N,\,\beta\in\alpha+1}$ of events such that, for all $k,n\in\N$ and $\beta\in\alpha+1$:
    \[ E_{k,\beta}^n \in\ms F_{(k2^{-n},\beta)}, \qquad \P(E_{k,\beta}^n\Delta\ovl E_{k,\beta}^n) = 0. \]
    For each $n\in\N$, define
    \[ M^n = \bigcup_{k\in\N} \bigcup_{\beta\in\alpha+1} \{(k2^{-n},\beta)\}\times E_{k,\beta}^n, \qquad \tau_n = D_{M^n}. \]
    We clearly have $\pi\circ\tau_n \le \alpha$. Moreover, for any $k\in\N$ and $\beta\in\alpha+1$, we have
    \[ \{\tau_n = (k2^{-n},\beta)\} = E_{k,\beta}^n \setminus \Big(\bigcup_{\substack{\ell\in\N,\,\gamma\in\alpha+1\colon\\(\ell 2^{-n},\gamma) < (k 2^{-n},\beta)}} E_{\ell,\gamma}^n  \Big), \]
    which is an element of $\ms F_{(k2^{-n},\beta)}$. As a consequence,
    \[ \{\tau_n = \infty\} = \bigcap_{k\in\N,\,\beta\in\alpha+1} \{\tau_n = (k2^{-n},\beta)\}^\complement~ \in\ms F_\infty. \]
    Hence, for any $\beta\in\alpha+1$ and all $t\in\ovT$, we have --- with the understanding that $\ovl\N = \N\cup\{\infty\}$ and $\infty \cdot 2^{-n} = \infty$ ---
    \[ \{\pi\circ\tau_n=\beta,\, \tau_n \le t\} =  \bigcup_{\substack{k\in\ovl\N\colon\\(k2^{-n},\beta)\le t}}\{\tau_n = (k2^{-n},\beta)\} ~\in \ms F_t. \]
    Hence, by Theorem~\ref{3-SPF_VECT.thm:optional_times} (the claims without completeness assumption), $\tau_n$ is an $\ms F$-optional time. Moreover,
    \begin{equation*}
        \begin{aligned}
            \P(\tau_n \neq \ovl\tau_n,\, \tau_n < \infty) \le&~ \sum_{k\in\N} \sum_{\beta\in\alpha+1} \P(\{\tau_n=(k2^{-n},\beta)\}\setminus \ovl E^n_{k,\beta})  \\
            \le&~ \sum_{k\in\N} \sum_{\beta\in\alpha+1} \P(E^n_{k,\beta}\setminus \ovl E^n_{k,\beta}).
        \end{aligned}
    \end{equation*}
    Noting that 
    \[ \{\tau_n < \infty\} = \bigcup_{k\in\N,\,\beta\in\alpha+1} E_{k,\beta}^n, \qquad \{\ovl\tau_n < \infty\} = \bigcup_{k\in\N,\,\beta\in\alpha+1} \ovl E_{k,\beta}^n, \] 
    we obtain
    \begin{equation*}
        \begin{aligned}
            \P(\tau_n \neq \ovl\tau_n,\, \tau_n = \infty) =&~ \P(\ovl\tau_n < \infty,\,\tau_n = \infty) \\
            \le&~ \sum_{k\in\N} \sum_{\beta\in\alpha+1} \P(\ovl E^n_{k,\beta} \cap \{\tau_n = \infty\}) \\
            \le&~ \sum_{k\in\N} \sum_{\beta\in\alpha+1} \P(\ovl E^n_{k,\beta} \setminus E^n_{k,\beta}).
        \end{aligned}
    \end{equation*}
    Hence,
    \[ \P(\tau_n \neq \ovl\tau_n) \le \sum_{k\in\N} \sum_{\beta\in\alpha+1} \P(\ovl E^n_{k,\beta} \Delta E^n_{k,\beta} ) = 0. \]
    Thus, $\P(\tau_n = \ovl\tau_n) = 1$. Let $\tau$ be the map satisfying $\tau(\omega) = \lim_{n\to\infty} \tau_n(\omega)$ for all $\omega\in\Omega$ that this limit exists for, and $\tau(\omega) = \infty$ otherwise. By Proposition~\ref{3-SPF_VECT.prop:basic_properties_of_optional_times}, $\tau$ is an $\ms F_\Plus$-optional time. Since $\ovl\tau_n \to \ovl\tau$ as $n\to\infty$ pointwise in $\bRp\times(\alpha+1)$, we have:
    \[ \{\tau \neq \ovl\tau\} \subseteq \bigcup_{n\in\N} \{\tau_n \neq \ovl\tau_n\}. \]
    This leads to 
    \[ \P(\tau \neq \ovl\tau) \le \sum_{n\in\N} \P(\tau_n\neq\ovl\tau_n) = 0, \]
    whence $\P(\tau = \ovl\tau) = 1$.
\end{proof}

\begin{proof}[Proof of Theorem~\ref{3-SPF_VECT.thm:debut}]
    (Ad first statement):~ Let $t\in\ovT$. Then, by progressive measurability, $M\cap[\![0,t]\!]\in \ms P_\ovT\otimes\ms F_t$. 
    Hence, if $t$ is not a right-limit point, or equivalently, $\pi(t) < {\mf w_1}$ (Lemma~\ref{3-SPF_VECT.lemma:left/right-limit_points_in_ovT}), measurable projection (Theorem~\ref{3-SPF_VECT.thm:mb_proj_section}) yields
    \begin{equation}\label{3-SPF_VECT.eq:DM<=t_for_t_not_right-limit_point}
        \{D_M \le t\} = \mc P\prj_\Omega\Big(M\cap[\![0,t]\!]\Big)~\in (\ms F_t)^{\mathrm u} \subseteq \ovl{\ms F}_t = \ovl{\ms F}_{t+},
    \end{equation}
    because the $\pi$-fibres of $\ovT$ are well-ordered and $\ms F_t$ is universally complete. 
    As a consequence, if $t$ is a right-limit point, or equivalently, $\pi(t) = {\mf w_1}$, i.e.\ $t = (p(t),{\mf w_1})$, then 
    \[ \{D_M \le t\} = \bigcap_{k\in\N} \bigcap_{\ell=k}^\infty \{D_M \le p(t) + 2^{-\ell}\} ~ \in \bigcap_{k\in\N} \ovl{\ms F}_{p(t)+2^{-k}} = \ovl{\ms F}_{t+}, \]
    because $p(t) + 2^{-\ell} \in \R_+$ for all $\ell\in\N$, which is therefore not a right-limit point (Lemma~\ref{3-SPF_VECT.lemma:left/right-limit_points_in_ovT}), so that Equation~\ref{3-SPF_VECT.eq:DM<=t_for_t_not_right-limit_point} applies.\medskip

    (Ad second statement):~ As $M$ is $\ms F$-progressively measurable, there is $\alpha\in{\mf w_1}$ such that $M\in \ms P_\ovT^\alpha\otimes\ms E$. Therefore, for all $t\in\ovT$ with $\pi(t) > \alpha$ we have:
    \begin{equation*}
        (\ast)\qquad (t,\omega)\in M ~ \Longrightarrow ~ (p(t),\alpha,\omega) \in M,
    \end{equation*}
    by Lemma~\ref{3-SPF_VECT.lemma:sets_in_msPovT^alpha_otimes_msE_are_inactive_beyond_alpha}.\smallskip
    
    Regarding the implication ``\ref{3-SPF_VECT.thm:debut.DM_optional_time} $\Rightarrow$ \ref{3-SPF_VECT.thm:debut.pi_circ_DM<omega1}'' we simply note that, if $D_M$ is an optional time, then $\pi\circ D_M < {\mf w_1}$ (see Theorem~\ref{3-SPF_VECT.thm:optional_times}). \smallskip
    
    For the implication ``\ref{3-SPF_VECT.thm:debut.pi_circ_DM<omega1} $\Rightarrow$ \ref{3-SPF_VECT.thm:debut.[DM]_cap_[0,infty)_subseteq_M}'', suppose that $\pi\circ D_M < {\mf w_1}$ and let $\omega\in\Omega$ such that $D_M(\omega) < \infty$. Let $x = p\circ D_M(\omega)$ and $S = \{\gamma\in{\mf w_1}+1 \mid (x,\gamma,\omega)\in M\}$. If $S$ were empty, then $D_M(\omega) = (x,{\mf w_1})$ which is excluded by hypothesis. Hence, $S$ has a minimum $\gamma_\ast$ and, in particular, $D_M(\omega) = (x,\gamma_\ast)$. Therefore, $(D_M(\omega),\omega)\in M$.

    For the proof of the remaining implication ``\ref{3-SPF_VECT.thm:debut.[DM]_cap_[0,infty)_subseteq_M} $\Rightarrow$ \ref{3-SPF_VECT.thm:debut.DM_optional_time}'', suppose that $[\![D_M]\!]\cap[\![0,\infty)\!)\subseteq M$. First, we get that $\pi\circ D_M \le \alpha$. Indeed, if we had $\pi\circ D_M(\omega) > \alpha$, then $t = D_M(\omega)$ and $u = (p(t),\alpha)$ would satisfy $(t,\omega)\in M$ and $u<t$. By $(\ast)$, we would obtain $(u,\omega)\in M$. Hence, $D_M(\omega) = t \le u$ --- a contradiction. So we infer that $\pi\circ D_M \le \alpha$.

    For the remainder of the proof, define, for any $\beta\in\alpha+1$, 
    \[ M^\beta = M \cap (\rho^{\alpha+1}\times\id_\Omega)^{-1}(\bRp\times\{\beta\}\times\Omega). \]
    This set is $\ms F$-progressively measurable. Hence, by the first statement proven above, $D_{M^\beta}$ is an $\ovl{\ms F}_+$-stopping time for any $\beta\in\alpha+1$. Thus, for any $t\in\bRp$, we have, with $Q_t = [0,t)_\ovT\cap\Q$,
    \[ \{D_{M^\beta} < t\} = \bigcup_{q\in Q_t} \{D_{M^\beta} \le q\}\, \in\bigvee_{q\in Q_t} \ovl{\ms F}_{q+} \subseteq \ovl{\ms F}_t. \]
    By Lemma~\ref{3-SPF_VECT.lemma:[0,tau_omega1]_is_prog_mb_if_tau<t_is_msF_t}, then, for any $\beta\in\alpha+1$, the set 
    \[ N^\beta = [\![0,(D_{M^\beta})_{\mf w_1}]\!]\]
    is $\ovl{\ms F}$-progressively measurable.\footnote{We recall that $(D_{M^\beta})_{\mf w_1}$ is the map $\Omega\to\ovT$ with $p\circ (D_{M^\beta})_{\mf w_1} = p\circ D_{M^\beta}$ and, for all $\omega\in \{D_{M^\beta} < \infty\}$, $\pi\circ (D_{M^\beta})_{\mf w_1} = \mf w_1$.}

    \emph{First intermediate claim}:~ Then, we claim that for any $\beta\in\alpha+1$ and $t\in\ovT$, we have
    \begin{equation}\label{3-SPF_VECT.eq:Debut.intermediate_claim_1}
    \begin{aligned}
        \prj_\Omega(N^\beta\cap M^\beta\cap [\![0,t]\!]) =&~ \{\omega\in\Omega \mid (D_{M^\beta}(\omega),\omega)\in M^\beta\cap [\![0,t]\!]\} \\
        =&~ \{\pi\circ D_{M^\beta} = \beta,\, D_{M^\beta} \le t\}.
    \end{aligned}
    \end{equation}
    For the \emph{proof}, let $\beta\in\alpha+1$ and $t\in\ovT$. Regarding the first equality, the inclusion ``$\supseteq$'' is evident. Concerning the proof of inclusion ``$\subseteq$'', let $(u,\omega)\in N^\beta\cap M^\beta\cap[\![0,t]\!]$. First, we infer directly that, by the definition of the début, we have $D_{M^\beta}(\omega) \le u \le t$. Second, as $N^\beta\cap M^\beta$ is $\ovl{\ms F}$-progressively measurable, there is $\gamma\in\mf w_1$ such that $N^\beta\cap M^\beta\in \ms P_\ovT^\gamma\otimes{\ms E}^{\mathrm u}$. If $\pi(u) \ge \gamma$, then, by Lemma~\ref{3-SPF_VECT.lemma:sets_in_msPovT^alpha_otimes_msE_are_inactive_beyond_alpha}, all $v\in\ovT$ with $p(v) = p(u)$ and $\pi(v) \ge \gamma$ satisfy $(v,\omega)\in N^\beta\cap M^\beta$ as well. Hence, $D_{M^\beta}(\omega) \le (p(u),\gamma)$. Moreover, as $(u,\omega)\in N^\beta$, we have $p(u) \le p\circ D_{M^\beta}(\omega)$. These both relations imply $\pi\circ D_{M^\beta}(\omega) \le \gamma$. As $\mf w_1$ is well-ordered, we infer that $(D_{M^\beta}(\omega),\omega)\in M^\beta$. Indeed, otherwise the definition of the début would imply the impossible statement $D_{M^\beta}(\omega) \ge (p\circ D_{M^\beta}(\omega), \pi\circ D_{M^\beta}(\omega) + 1) > D_{M^\beta}(\omega)$.

    Regarding the second equality, the inclusion ``$\subseteq$'' is clear that time. Inclusion ``$\supseteq$'' follows again from the well-ordering on $\mf w_1$. Indeed, let $\omega\in\{\pi\circ D_{M^\beta} = \beta\}$. If we had $(D_{M^\beta}(\omega),\omega)\notin M^\beta$, then, by definition of the début, we would again obtain the impossible statement $D_{M^\beta}(\omega) \ge (p\circ D_{M^\beta}(\omega), \beta+ 1) > D_{M^\beta}(\omega)$. This completes the proof of the first intermediate claim.

    \emph{Second intermediate claim}:~ For any $\beta\in\alpha+1$ and $t\in\ovT$, we have
    \begin{equation}\label{3-SPF_VECT.eq:Debut.intermediate_claim_2}
        \{\pi\circ D_{M^\beta} = \beta,\, D_{M^\beta} \le t\} \, \in\ovl{\ms F}_t.
    \end{equation}
    For the \emph{proof}, note that the set under scrutiny equals $\prj_\Omega(N^\beta\cap M^\beta\cap [\![0,t]\!])$, by the first intermediate claim, see Equation~\ref{3-SPF_VECT.eq:Debut.intermediate_claim_1}. The $\in$-relation follows from the fact that $N^\beta\cap M^\beta$ is $\ovl{\ms F}$-progressively measurable, the fact that 
    \[ (\ovl{\ms F}_t)^{\mathrm u} \subseteq \ovl{\ovl{\ms F}_t} = \ovl{\ms F_t} = \ovl{\ms F}_t \] 
    (when augmenting in $\ms E$), and the Measurable Projection Theorem~\ref{3-SPF_VECT.thm:mb_proj_section}.

    \emph{Third intermediate claim}:~ For any $\beta\in\alpha+1$, the map
    \begin{equation}\label{3-SPF_VECT.eq:Debut.intermediate_claim_3}
        \sigma_\beta\colon \Omega\to\ovT,\, \omega\mapsto\begin{cases} D_{M^\beta}(\omega), &\text{if } \pi\circ D_{M^\beta}(\omega) = \beta, \\ \infty, &\text{else,} \end{cases}
    \end{equation}
    is an $\ovl{\ms F}$-optional time.

    For the \emph{proof}, let $\beta\in\alpha+1$. We clearly have $\pi\circ\sigma_\beta\le \beta$. Next, we study the measurability of $\{\pi\circ\sigma_\beta,\, \sigma_\beta\le t\}$ for all $\gamma\in\beta+1$ and $t\in\ovT$. If $\gamma = \beta > 0$, we obtain 
    \[ \{\pi\circ\sigma_\beta = \beta,\, \sigma_\beta\le t\} = \{\pi\circ D_{M^\beta} = \beta,\, D_{M^\beta} \le t\}\, \in \ovl{\ms F}_t, \]
    by the second intermediate claim, see Equation~\ref{3-SPF_VECT.eq:Debut.intermediate_claim_2}. If $\beta > 0 = \gamma$, then we obtain
    \[ \{\pi\circ\sigma_\beta = 0,\, \sigma_\beta\le t\} = \begin{cases} \emptyset, &\text{if } t < \infty, \\\{\pi\circ D_{M^\beta}\}^\complement, &\text{else,}  \end{cases}\quad \in \ovl{\ms F}_t, \]
    again, by the second intermediate claim applied to $t=\infty$, see Equation~\ref{3-SPF_VECT.eq:Debut.intermediate_claim_2}. If $\gamma \in(\beta+1)\setminus\{0,\beta\}$, then
    \[ \{\pi\circ\sigma_\beta = 0,\, \sigma_\beta\le t\} = \emptyset\, \in\ovl{\ms F}_t. \]
    Finally, if $\beta = \gamma = 0$, then, by looking separately at the cases $t < \infty$ and $t=\infty$ we infer that
    \[ \{\pi\circ\sigma_\beta = 0,\, \sigma_\beta\le t\} = \{D_{M^0} \le p(t)\}\, \in\ovl{\ms F}_{p(t)+} = \ovl{\ms F}_{p(t)} \subseteq \ovl{\ms F}_t, \]
    since $D_{M^0}$ is an $\ovl{\ms F}_+$-stopping time. This completes the proof of the third intermediate claim.

    \emph{Fourth intermediate claim}:~ We have
    \begin{equation}\label{3-SPF_VECT.eq:Debut.intermediate_claim_4}
        D_M = \inf_{\beta\in\alpha+1} \sigma_\beta, \qquad \Omega = \bigcup_{\beta\in\alpha+1} \{D_M = \sigma_\beta\}. 
    \end{equation}
    For the \emph{proof}, note that, for any $\beta\in\alpha+1$, the definitions of $M^\beta$ and $\sigma_\beta$ directly imply that
    \[ (\dagger) \qquad D_M \le D_{M^\beta} \le \sigma_\beta, \]
    whence $D_M \le \inf_{\beta\in\alpha+1} \sigma_\beta$. 
    For the remainder of the proof of this intermediate claim, let $\omega\in\Omega$. Then, $D_M(\omega) = \infty$ or $D_M(\omega) < \infty$. In the first case, by $(\dagger)$, $D_M(\omega) = \sigma_\beta(\Omega)$ for all $\beta\in\alpha+1$. In the second case, we have $(D_M(\omega),\omega)\in M$ by assumption (Property~\ref{3-SPF_VECT.thm:debut.[DM]_cap_[0,infty)_subseteq_M}). As $\pi\circ D_M(\omega) \le \alpha$, there is $\beta\in\alpha+1$ such that $(D_M(\omega),\omega) \in M^\beta$. Hence, $\pi\circ D_{M^\beta}(\omega) = \beta$ and $D_{M^\beta}(\omega)\le D_M(\omega)$. In view of the definition of $\sigma_\beta$, we infer that equality holds true in $(\dagger)$. In total, we conclude that $D_M(\omega) = \inf_{\beta\in\alpha+1} \sigma_\beta(\omega)$ and $\Omega = \bigcup_{\beta\in\alpha+1} \{D_M = \sigma_\beta\}$.

    \emph{Conclusion}:~ By the third intermediate claim (cf.\ Equation~\ref{3-SPF_VECT.eq:Debut.intermediate_claim_3}), $\sigma_\beta$ is an $\ovl{\ms F}$-optional time for any $\beta\in\alpha+1$. Hence, by the fourth intermediate claim (cf.\ Equation~\ref{3-SPF_VECT.eq:Debut.intermediate_claim_4}) and Proposition~\ref{3-SPF_VECT.prop:basic_properties_of_optional_times.addon_under_completeness}, Part~\ref{3-SPF_VECT.prop:basic_properties_of_optional_times.inf}, $D_M$ is an $\ovl{\ms F}$-optional time --- because $\alpha+1$ is countable and $(\Omega,\ms E^{\mathrm u},\ovl{\ms F})$ is universally complete. This completes the proof of the theorem.
\end{proof}

\subsubsection{Optional processes}

\begin{proof}[Proof of Lemma~\ref{3-SPF_VECT.lemma:Prd_subseteq_Opt}]
    The second inclusion follows directly from Definition~\ref{3-SPF_VECT.def:prog_mb}, Remark~\ref{3-SPF_VECT.rmk:prog_mb_basic_properties}, and Theorem~\ref{3-SPF_VECT.thm:optional_times}. Regarding the first inclusion, let $E\in\ms F_0$. Then, $M = (\{0\}\times E) \cup (\{(0,1)\}\times E^\complement)$ is clearly $\ms F$-progressively measurable since $\ms F_0 \subseteq \ms F_{(0,1)}$. Moreover, one easily sees (or otherwise applies Theorem~\ref{3-SPF_VECT.thm:debut} to see) that $D_M$ is an $\ms F$-optional time. Hence, 
    \[ \{0\}\times E = [\![0,D_M)\!) \in \Opt(\ms F). \]
    Furthermore, let $\tau$ be an $\ms F$-optional time. Then, the $E = \{\tau = \infty\}$ satisfies $E\in\ms F_\infty$, by Remark~\ref{3-SPF_VECT.rmk:tau_&_msF_tau}, Part~Remark~\ref{3-SPF_VECT.rmk:tau_&_msF_tau.tau<t_in_msG_t}. Moreover, the map $\sigma\colon\Omega\to\ovT$ such that, first, $p\circ\sigma = p\circ\tau$ and, second, for all $\omega\in E^\complement$, $\pi\circ\sigma(\omega) = \pi\circ\tau(\omega)+1$, is an $\ms F$-optional time. 
    
    Indeed, there is $\alpha\in\mf w_1$ with $\pi\circ\tau\le\alpha$. Thus, $\pi\circ\sigma \le \pi\circ\tau +1\le \alpha+1$. On $E^\complement$, $\pi\circ\sigma$ is valued in the set of successor ordinals in $\mf w_1$, and equal to zero on $E$. Moreover, for all $\beta\in\mf w_1$ and $t\in\ovT$, we have
    \[ \{\pi\circ\sigma = \beta+1,\, \sigma\le t\} = \bigcup_{\Q\ni q < t}\{\pi\circ\tau = \beta, \, \tau \le q\} \cup \bigcup_{\gamma\in \pi(t)\cap (\alpha+1)}\{\pi\circ\tau = \beta, \, \tau \le (p(t),\gamma)\}\, \in\ms F_{t}, \]
    and 
    \[ \{\pi\circ\sigma = 0,\, \sigma\le t\} = \begin{cases}\emptyset, &\text{if } t < \infty, \\ E, &\text{else,} \end{cases} ~\in\ms F_\infty. \]
    This shows that $\sigma$ is an $\ms F$-optional time.
    
    Hence,
    \[ [\![0,\tau]\!] = [\![0,\sigma)\!) \cup (\{\infty\}\times E)~\in \Opt(\ms F). \]
\end{proof}

\begin{proof}[Proof of Lemma~\ref{3-SPF_VECT.lemma:Opt_proc_generated_by_xi*1[tau,sigma)}]
    Let $\ms M$ be a $\sigma$-algebra as in the lemma. Then, $\ms M$ must contain $[\![0,\tau)\!)$ for any $\ms F$-optional time $\tau$, since we can take $\alpha=2$, $\xi_0 = 1$, $\xi_1 = \xi_2 = 0$, $\tau_1=\tau$ in Equation~\ref{3-SPF_VECT.eq:Opt_proc_generated_by_xi*1[tau,sigma)}. Moreover, let $E\in\ms F_\infty$. Then, taking $\alpha=1$, $\xi_0=0$ and $\xi_1=1_E$ in Equation~\ref{3-SPF_VECT.eq:Opt_proc_generated_by_xi*1[tau,sigma)}, we infer that $\{\infty\}\times E\in \ms M$. Hence, $\Opt(\ms F) \subseteq \ms M$.

    On the other hand, all processes as in Equation~\ref{3-SPF_VECT.eq:Opt_proc_generated_by_xi*1[tau,sigma)} are $\Opt(\ms F)$-measurable. Indeed, let $\alpha\in{\mf w_1}$, $(\tau_\beta)_{\beta\in\alpha+1}$ and $(\xi^\beta)_{\beta\in\alpha+1}$ be given as in the statement of the lemma. Then, $[\![\tau_\alpha]\!] = \{\infty\}\times\Omega\in\Opt(\ms F)$ and, for $\beta\in\alpha$, 
    \[ [\![\tau_{\beta},\tau_{\beta+1})\!) = [\![0,\tau_{\beta+1})\!) \setminus [\![0,\tau_{\beta})\!) ~ \in \Opt(\ms F). \]
    In a first step, suppose that, for all $\beta\in\alpha+1$, 
    \[ (\ast) \qquad \xi^\beta = 1(E_\beta)\]
    for some $E_\beta\in\ms F_{\tau_\beta}$. We have
    \[ (\ovT\times E_\alpha) \cap [\![\tau_\alpha]\!] = \{\infty\}\times E_\alpha ~\in \Opt(\ms F). \]
    In addition, for any $\beta\in\alpha$, we see, using Theorem~\ref{3-SPF_VECT.thm:optional_times}, Property~\ref{3-SPF_VECT.thm:optional_times.pi_bounded,pi_tau_tau-mb}, that (with the usual measure-theoretic convention $\infty\cdot 0 = 0$)
    \[ \sigma_{\beta} = \tau_{\beta}\, 1_{E_{\beta}} + \infty\, 1_{E_{\beta}^\complement} \]
    is an $\ms F$-optional time. Hence,
    \[  (\ovT\times E_{\beta}) \cap [\![\tau_{\beta},\tau_{\beta+1})\!) = [\![\sigma_{\beta},\tau_{\beta+1})\!) = [\![0,\tau_{\beta+1})\!) \setminus [\![0,\sigma_{\beta})\!) ~ \in \Opt(\ms F).\]
    As $\alpha$ is countable, this implies that under Assumption~$(\ast)$, the process in Equation~\ref{3-SPF_VECT.eq:Opt_proc_generated_by_xi*1[tau,sigma)} is $\Opt(\ms F)$-measurable. 
    In a second step, we directly infer that the same result obtains under the weaker hypothesis that for all $\beta\in\alpha+1$, 
    \[ \xi^\beta = \sum_{\ell = 1}^{N_\beta} x_{\ell,\beta} 1_{E_{\ell,\beta}}, \]
    for some integer $N_\beta$, some reals $x_{1,\beta},\dots,x_{N_\beta,\beta}\in\R$, and some $E_{1,\beta},\dots,E_{N_\beta,\beta}\in\ms F_{\tau_{\beta}}$, i.e.\ $\xi^\beta$ is a simple function with respect to $\ms F_{\tau_\beta}$.  
    In the third and final step, we consider the general case. Then, for any $\beta\in\alpha+1$, there is a sequence $(\xi^{\beta,m})_{m\in\N}$ of simple functions with respect to $\ms F_{\tau_\beta}$ converging pointwise to $\xi^\beta$. As measurability is stable under pointwise convergence of real-valued functions, we infer that
    \[ \xi^\alpha \circ \prj_\Omega\, 1[\![\tau_\alpha]\!] + \sum_{\beta\in\alpha} \xi^{\beta}\circ\prj_\Omega\, 1[\![\tau_{\beta},\tau_{\beta+1})\!) = \lim_{m\to\infty}\Big( \xi^{\alpha,m} \circ \prj_\Omega\, 1[\![\tau_\alpha]\!] + \sum_{\beta\in\alpha} \xi^{\beta,m}\circ\prj_\Omega\, 1[\![\tau_{\beta},\tau_{\beta+1})\!)\Big)\]
    is $\Opt(\ms F)$-measurable.
\end{proof}

\begin{proof}[Proof of Lemma~\ref{3-SPF_VECT.lemma:Opt_proc_generated_by_xi*1[tau,sigma)_2}]
    Let $\mc S$ be a set of maps $\ovT\times\Omega\to\R$ satisfying conditions a), b), and c). 
    In a \emph{first step}, let $\ms D$ be the set of $M\in\Opt(\ms F)$ such that $1_M\in\mc S$. 
    
    We first show that $\ms D$ is a Dynkin system, and then infer that $\ms D = \Opt(\ms F)$. Taking $\tau=\infty$ and $E=\Omega$ in a), we get that $1_\Omega\in\mc S$, whence $\Omega\in\ms D$. Moreover, for all $M\in\ms D$, $1_{M^\complement} = 1_{\Omega} - 1_M\in\mc S$ by b), whence $M^\complement\in\ms D$. Moreover, if $(D_n)_{n\in\N}$ is a pairwise disjoint $\ms D$-valued sequence, then $D = \bigcup_{n\in\N} D_n$ satisfies
    \[ 1_D = \sum_{k=0}^\infty 1_{D_k} = \lim_{n\to\infty} \sum_{k=0}^n 1_{D_k}, \]
    whence $1_D\in\mc S$ by b) and c), thus $D\in\ms D$.

    Next, we note that $\ms D$ contains an intersection-stable generator of $\Opt(\ms F)$. Namely, a) implies that, for any $\ms F$-optional time, $[\![0,\tau)\!)\in\ms D$, and, for any $E\in\ms F_\infty$, $\{\infty\}\times E\in\ms D$. Hence, 
    \begin{equation}\label{3-SPF_VECT.eq:generator_of_Opt}
        \ms G = \Big\{\{\infty\}\times E\mid E\in\ms F_\infty\Big\} \cup \Big\{ [\![0,\tau)\!) \mid \tau~\ms F\text{-optional time}\Big\} 
    \end{equation}
    satisfies $\ms G \subseteq\ms D$. By definition, $\ms G$ generates the $\sigma$-algebra $\Opt(\ms F)$. Proposition~\ref{3-SPF_VECT.prop:basic_properties_of_optional_times}, Part~\ref{3-SPF_VECT.prop:basic_properties_of_optional_times.max_min}, implies that the pointwise minimum $\sigma\wedge\tau$ of two $\ms F$-optional times $\sigma,\tau$ is again an $\ms F$-optional time. As $[\![0,\sigma)\!) \cap [\![0,\tau)\!) = [\![0,\sigma\wedge\tau)\!)$, we readily infer that $\ms G$ intersection-stable. 

    Hence, by Dynkin's $\pi$-$\lambda$-theorem, $\ms D = \Opt(\ms F)$.
    In a \emph{second step}, we show that all $\ms F$-optional processes are elements of $\mc S$. Let $\xi$ be an $\ms F$-optional process. Then, there is a sequence $(\xi_n)_{n\in\N}$ of simple functions with respect to $\Opt(\ms F)$ that converges pointwise to $\xi$. By the first step and Property~b), $\xi_n\in\mc S$ for any $n\in\N$. Therefore, by Property~c), $\xi = \lim_{n\to\infty} \xi_n \in \mc S$ as well.\smallskip

    It remains to show that the set of $\ms F$-optional processes satisfies properties a), b), and c). Property~a) is satisfied by construction. Properties b) and c) follow from the fact that the real-valued $\ms F$-optional processes are exactly the $\Opt(\ms F)$-$\ms B_\R$-measurable functions, combined with basic measure theory.
\end{proof}

\begin{proof}[Proof of Proposition~\ref{3-SPF_VECT.prop:optional_processes=mcV_omega1}]
    Denote the set of real-valued $\ms F$-optional processes by $L^0(\Opt(\ms F);\R)$. We have to show that
    \[ L^0(\Opt(\ms F);\R) = \mc V_{{\mf w_1}}. \]

    (Step 1):~ We first show that for all $\alpha\in{\mf w_1}+1$, we have $\mc V_{\alpha}\subseteq L^0(\Opt(\ms F);\R)$. For $\alpha = {\mf w_1}$, this yields the inclusion ``$\supseteq$''.

    If the claim did not hold true, then it would fail for at least one $\alpha\in{\mf w_1}+1$. Hence, by the well-order property of ordinals, there would be a smallest one that fails. Let us call it $\alpha^\ast$. By definition of $\mc V_0$ and since $\R$-linear combinations of real-valued measurable functions are measurable, we have $\alpha^\ast > 0$. If $\alpha^\ast$ were a limit ordinal, we would get
    \[ \mc V_{\alpha^\ast} = \bigcup_{\beta\in\alpha^\ast} \mc V_\beta \subseteq \bigcup_{\beta\in\alpha^\ast} L^0(\Opt(\ms F);\R) = L^0(\Opt(\ms F);\R), \]
    which contradicts the definition of $\alpha^\ast$. If $\alpha^\ast$ were a successor ordinal, there would be $\gamma\in{\mf w_1}$ with $\alpha^\ast = \gamma + 1$. Hence, if $\xi\in\mc V_{\alpha^\ast}$, then there is a $\mc V_\gamma$-valued sequence $(\xi_n)_{n\in\N}$ with $\xi = \lim_{n\to\infty} \xi_n$ pointwise as $n\to\infty$. By definition of $\alpha^\ast$ and $\gamma$, $\xi_n\in L^0(\Opt(\ms F);\R)$ for all $n\in\N$. As pointwise limits of real-valued measurable functions are measurable, $\xi$ would be measurable again. This would prove that $\mc V_{\alpha^\ast}\subseteq L^0(\Opt(\ms F);\R)$, contradicting the definition of $\alpha^\ast$. As non-zero ordinals are either successors or limits, we conclude that the claim must be correct.\smallskip

    (Step 2):~ Let 
    \[ \ms D = \{ M\in\Opt(\ms F) \mid 1_M \in \mc V_{{\mf w_1}} \}. \]
    Using Dynkin's $\pi$-$\lambda$-theorem, we show that $\ms D = \Opt(\ms F)$. 

    First, we show that $\ms D$ is a Dynkin system. Note that
    \[ 1(\ovT\times\Omega) = 1{[\![0,\infty)\!)} + 1(\{\infty\}\times\Omega) ~\in\mc V_0\subseteq\mc V_{{\mf w_1}}, \]
    whence $\ovT\times\Omega\in\ms D$. Moreover, if $M_0,M_1\in\ms D$ with $M_0\subseteq M_1$, then there are $\alpha_0,\alpha_1\in{\mf w_1}$ such that $1(M_k)\in\mc V_{\alpha_k}$ for $k=0,1$. Without loss of generality, $\alpha_0 \le \alpha_1$. Hence, $M_0,M_1\in \mc V_{\alpha_1}$. As $\mc V_{\alpha_1}$ is an $\R$-vector space, we infer
    \[ 1(M_1 \setminus M_0) = 1({M_1}) - 1({M_0})~ \in\mc V_{\alpha_1}\subseteq \mc V_{{\mf w_1}}, \]
    whence $M_1\setminus M_0\in\ms D$. Next, let $(M_n)_{n\in\N}$ be an increasing sequence valued in $\ms D$ and $M = \bigcup_{n\in\N} M_n$. Then, there is a sequence $(\alpha_n)_{n\in\N}$ valued in ${\mf w_1}$ such that $1(M_n)\in\mc V_{\alpha_n}$, for all $n\in\N$. Let $\alpha = \sup_{n\in\N} \alpha_n$, an element of ${\mf w_1}$ again. Then, $1(M_n)\in\mc V_{\alpha}$ for all $n\in\N$. Hence,
    \[ 1(M) = \lim_{n\to\infty} 1(M_n)~\in \mc V_{\alpha+1} \subseteq \mc V_{{\mf w_1}}, \]
    whence $M\in\ms D$. We have proven that $\ms D$ is a Dynkin system.

    Second, we show that $\ms D$ contains an intersection-stable generator of $\Opt(\ms F)$. By definition, $\mc V_0$ contains $1_M$, for any $M\in\ms G$, where $\ms G$ is the intersection-stable generator of $\Opt(\ms F)$ from Equation~\ref{3-SPF_VECT.eq:generator_of_Opt}.\footnote{Recall that the only non-trivial assertion to verify in order to prove intersection-stability of $\ms G$ is that the pointwise minimum of two $\ms F$-optional times $\sigma,\tau$ is an $\ms F$-optional time again, because $[\![0,\sigma)\!) \cap [\![0,\tau)\!) = [\![0,\sigma\wedge\tau)\!)$. But this follows from Proposition~\ref{3-SPF_VECT.prop:basic_properties_of_optional_times}, Part~\ref{3-SPF_VECT.prop:basic_properties_of_optional_times.max_min}.}
    We infer that $\ms G \subseteq \ms D$. Thus, $\ms D$ contains an intersection-stable generator of $\Opt(\ms F)$.

    Third, combining these two intermediate results implies that $\ms D = \Opt(\ms F)$, by Dynkin's $\pi$-$\lambda$-theorem.\smallskip

    (Step 3):~ We now show the inclusion ``$\subseteq$''. Let $\xi\in L^0(\Opt(\ms F);\R)$. Then, by basic measure theory, there is a $\Opt(\ms F)$-valued family $(M_{n,k})_{n\in\N,\,k\in\{0,\dots,n\}}$ and a real-valued family $(x_{n,k})_{n\in\N,\,k\in\{0,\dots,n\}}$ such that with
    \[ \xi_n = \sum_{k=0}^n x_{n,k}\, 1(M_{n,k}), \qquad n\in\N, \]
    we have $\xi_n \to \xi$ pointwise as $n\to\infty$. For each pair $(n,k)\in\N^2$ with $k\le n$, there is $\alpha_{n,k}\in{\mf w_1}$ such that $1(M_{n,k})\in \mc V_{\alpha_{n,k}}$, by Step~2. Let $\alpha = \sup_{(n,k)\in\N^2\colon k\le n} \alpha_{n,k}$, which is an element of ${\mf w_1}$. Hence, for all $n\in\N$, $\xi_n\in V_\alpha$. Therefore, $\xi\in \mc V_{\alpha+1}\subseteq\mc V_{{\mf w_1}}$. 
\end{proof}

\begin{proof}[Proof of Corollary~\ref{3-SPF_VECT.cor:Opt.uniform_upper_bound_on_vertical_activity}]
    Without loss of generality, we may assume that $Y=\R$. We say that a map $\xi\colon\ovT\times\Omega\to\R$ has Property $P$ iff there is $\alpha\in{\mf w_1}$ such that for all $t\in\ovT$ with $\pi(t) \ge \alpha$ and all $\omega\in\Omega$, $\xi(t,\omega) = \xi((p(t),\alpha),\omega)$ holds true. Now, for any $\gamma\in{\mf w_1}+1$, we make the following claim $C(\gamma)$: any $\xi\in\mc V_\gamma$ has Property $P$. Using Proposition~\ref{3-SPF_VECT.prop:optional_processes=mcV_omega1}, the corollary is just the special case of $C({{\mf w_1}})$.

    If $C(\gamma)$ were not correct for all $\gamma\in{\mf w_1}+1$, then there would be a minimal $\gamma\in{\mf w_1}+1$ such that $C(\gamma)$ is incorrect. Denote this hypothetical minimum by $\gamma^\ast$.\smallskip
    
    (Step~1):~ We claim that if $\xi^1,\xi^2\colon\ovT\times\Omega\to\R$ are maps having Property $P$, respectively, then for any $x_1,x_2\in\R$, the linear combination $\xi = x_1\xi^1+x_2\xi^2$ does so, too. Indeed, let $\alpha_1,\alpha_2\in{\mf w_1}$ be such that for both $k=0,1$, all $t\in\ovT$ with $\pi(t) \ge \alpha_k$ and all $\omega\in\Omega$, $\xi^k(t,\omega) = \xi^k((p(t),\alpha_k),\omega)$. Then, let $\alpha = \alpha_1 \vee \alpha_2$, which is an element of ${\mf w_1}$. Then, for all $t\in\ovT$ with $\pi(t) \ge \alpha$ and all $\omega\in\Omega$, we have
    \[ \xi(t,\omega) = x_1\xi^1(t,\omega)+x_2\xi^2(t,\omega) = x_1\xi^1((p(t),\alpha),\omega)+x_2\xi^2((p(t),\alpha),\omega) = \xi((p(t),\alpha_k),\omega). \]
    Hence, $\xi$ has Property $P$.\smallskip

    (Step~2):~ We claim that if $(\xi^n)_{n\in\N}$ is a family of maps $\ovT\times\Omega\to\R$ having Property $P$ and converging pointwise to a map $\xi$, then $\xi$ has Property $P$ as well. Indeed, let $(\alpha_n)_{n\in\N}$ be a ${\mf w_1}$-valued sequence such that for all $n\in\N$, all $t\in\ovT$ with $\pi(t) \ge \alpha_n$ and all $\omega\in\Omega$, $\xi^n(t,\omega) = \xi^n((p(t),\alpha_n),\omega)$ holds true. Let $\alpha = \sup_{n\in\N} \alpha_n$, which is an element of ${\mf w_1}$ again. Then, for all $t\in\ovT$ with $\pi(t) \ge \alpha$ and all $\omega\in\Omega$, we get:
    \[ \xi(t,\omega) = \lim_{n\to\infty} \xi^n(t,\omega) = \lim_{n\to\infty} \xi^n((p(t),\alpha),\omega) = \xi((p(t),\alpha),\omega). \]
    Thus, $\xi$ has Property $P$.\smallskip

    (Step~3):~ $C(0)$ is correct, so that $\gamma^\ast > 0$. Indeed, if $\xi = 1[\![0,\tau)\!)$ for an $\ms F$-optional time $\tau$, then $\pi\circ\tau\le\alpha$ for some $\alpha\in{\mf w_1}$. Now, for $(t,\omega)\in\ovT\times\Omega$ we have
    \[ t < \tau(\omega) \qquad \Longleftrightarrow \qquad \Big[ p(t) < p\circ\tau(\omega) \text{ or } \Big( p(t) = p\circ\tau(\omega) \text{ and } \pi(t) < \pi\circ \tau(\omega)\Big)\Big]. \]
    If $\pi(t) \ge \alpha$, then the right-hand side is equivalent to $p(t) < p\circ\tau(\omega)$. Hence, if $\pi(t)\ge\alpha$, $t<\tau(\omega)$ iff $(p(t),\alpha) < \tau(\omega)$; in other words:
    \[ \xi(t,\omega) = \xi((p(t),\alpha),\omega). \]
    If $\xi = 1(\{\infty\}\times E)$ for $E\in\ms F_\infty$, then $\xi(t,\omega) = \xi(p(t),\omega)$ for all $t\in\ovT$; here, $\alpha = 0$ does the job already. Moreover, by Step 1, the set $S$ of maps $\ovT\times\Omega\to\R$ having Property $P$ is an $\R$-vector space. As $S$ contains all maps of the form given in Equation~\ref{3-SPF_VECT.eq:def.mcV0}, $S$ contains the $\R$-vector space generated by them, namely $\mc V_0$. Hence, $C(0)$ is correct and $\gamma^\ast > 0$.\smallskip

    (Step~4):~ If $\gamma^\ast$ were a limit ordinal, then for any $\xi\in\mc V_{\gamma^\ast}$, there would be $\beta\in\gamma^\ast$ with $\xi\in V_\beta$. As $C(\beta)$ is assumed to hold true, $\xi$ would have Property $P$. Thus, any element of $V_{\gamma^\ast}$ would have Property $P$, i.e.\ $C(\gamma^\ast)$ would be correct --- a contradiction to the definition of $\gamma^\ast$. If $\gamma^\ast$ were a successor ordinal, there would be $\beta\in\gamma^\ast$ with $\gamma^\ast = \beta+1$. Then, for any $\xi\in\mc V_{\gamma^\ast}$, there would be a $\mc V_{\beta}$-valued sequence $(\xi^n)_{n\in\N}$ converging pointwise to $\xi$, by construction of the hierarchy. As $\beta < \gamma^\ast$, all members of that sequence would have Property $P$. Hence, by Step~2, $\xi$ would also have Property $P$. Thus, all elements of $V_{\gamma^\ast}$ would have Property $P$, i.e.\ $C(\gamma^\ast)$ would be correct --- contradicting again the definition of $\gamma^\ast$. As a non-zero ordinal is either a limit or a successor, we conclude that Claim $C(\gamma)$ is correct for all $\gamma\in{\mf w_1}+1$.
\end{proof}

\subsubsection{Tilting convergence}

\begin{proof}[Proof of Lemma~\ref{3-SPF_VECT.lemma:simple_opt_proc_induced_grid}]
    (Ad grid property, last sentence):~ Represent $\xi$ as in Equation~\ref{3-SPF_VECT.eq:Opt_proc_generated_by_xi*1[tau,sigma)}, for suitable $\alpha\in\mf w_1$, $\tau_\beta$ and $\xi^\beta$, $\beta\in\alpha+1$. Define $G\colon(\alpha+1)\times\Omega\to\ovT$ by $G(0,.) = 0$, Equation~\ref{3-SPF_VECT.eq:compatibility_simple_proc_grid} for all $\beta\in\alpha$ and $\omega\in\Omega$, and $G(\gamma,\omega) = \sup_{\beta\in\gamma} G(\beta,\omega)$ for all limit ordinals $\gamma\in\alpha+1$ and $\omega\in\Omega$.

    Then, by transfinite induction, we infer using completeness, the Début Theorem~\ref{3-SPF_VECT.thm:debut}, and Proposition~\ref{3-SPF_VECT.prop:basic_properties_of_optional_times} (Part~\ref{3-SPF_VECT.prop:basic_properties_of_optional_times.sup}), that $G(\beta,.)$ is an $\ms F$-optional time, for any $\beta\in\alpha+1$. By construction, $G(\beta,\omega) < G(\gamma,\omega)$ holds true for all $\beta,\gamma\in\alpha+1$ and $\omega\in\Omega$ with $G(\beta,\omega) < \infty$ because $\xi$ has locally right-constant paths. Moreover, using transfinite induction, again, we see that, for all $\beta\in\alpha+1$, we have $\tau_\beta \le G(\beta,.)$. In particular, $\infty = \tau_\alpha \le G(\alpha,.)$. Thus, $\tau^G_\alpha = G(\alpha,.) = \infty$. Hence, $G$ is an $\ms F$-adapted grid. It follows directly from the definition, using transfinite induction, that $G$ is classical if $\xi$ is so.\medskip 

    (Ad representation):~ Let
    \[ \xi' = \xi_{\tau_\alpha^G}\circ\prj_\Omega\, 1[\![\tau_\alpha^G]\!] + \sum_{\beta\in\alpha} \xi_{\tau_\beta^G}\circ\prj_\Omega\,1[\![\tau_\beta^G,\tau_{\beta+1}^G)\!), \]
    and let $M = \{\xi \neq \xi'\}$, an $\ms F$-optional set, by Corollary~\ref{3-SPF_VECT.cor:xi_tau_is_F_tau_mb} and completeness. As both $\xi$ and $\xi'$ have locally right-constant paths, the Début Theorem~\ref{3-SPF_VECT.thm:debut} implies, together with completeness, that $D_M$ is an $\ms F$-optional time. We show that $D_M = \infty$. We do so by showing the claim (C1) $D_M \ge \tau_\beta^G$ for all $\beta\in\alpha+1$, using transfinite induction. Inserting $\beta = \alpha$ into (C1) then yields $D_M \ge \infty$, whence $D_M = \infty$. (C1) holds true for $\beta = 0$ because $\tau_0^G = 0$. 
    
    Suppose that (C1) holds true for $\beta\in\alpha$, and let $\omega\in\Omega$.     
    If we had $D_M(\omega) < \tau_{\beta+1}^G(\omega)$, then, using local right-constancy of paths, the induction hypothesis, and the definition of $\xi'$, $\xi_{D_M}(\omega) \neq {\xi'}_{D_M}(\omega) = \xi_{\tau_\beta^G}(\omega)$. Thus, $\tau_\beta^G(\omega) < D_M(\omega) < \tau_{\beta+1}^G(\omega)$. As $\xi'(\omega)$ is constant on $[\tau_\beta^G(\omega),\tau_{\beta+1}^G(\omega))_\ovT$, this would imply that $\xi_{D_M}(\omega)\neq {\xi'}_{D_M}(\omega) = {\xi'}_{\tau_\beta^G}(\omega) = \xi_{\tau_\beta^G}(\omega)$, whence the contradiction $\tau^G_{\beta+1}(\omega) = G(\beta+1,\omega) \le D_M(\omega) < \tau_{\beta+1}(\omega)$. Therefore, $D_M(\omega) \ge \tau_{\beta+1}^G(\omega)$.

    For all limit ordinals $\gamma\in\alpha+1$ such that (C1) holds true for all $\beta\in\gamma$, we have $D_M \ge \tau_\beta^G$ for all $\beta\in\gamma$, and thus $D_M \ge \sup_{\beta\in\gamma} \tau_\beta^G = \tau_\gamma^G$, because $G$ is an $\ms F$-adapted grid. The proof of claim (C1) for all $\beta\in\alpha+1$ is complete, and we infer $D_M = \infty$.

    By definition of $\xi'$, we have ${\xi'}_\infty = \xi_{\tau_\alpha^G} = \xi_\infty$. Thus, $D_M = \infty$ implies that $M=\emptyset$ which is equivalent to $\xi = \xi'$.
\end{proof}

\begin{proof}[Proof of Lemma~\ref{3-SPF_VECT.lemma:refining,convergent_grid->psi,delta,gamma}]
    (Ad Part~\ref{3-SPF_VECT.lemma:refining,convergent_grid->psi,delta,gamma.exists_delta^n_and_psi^n}):~ As a subset of a well-order, $\{ \beta\in\alpha_n \mid G_n(\beta,\omega) \ge t \}$ is itself well-ordered, and therefore, by basic well-order theory (see, e.g., \cite[Lemma~10.1]{Ziegler2017Mathematische}), there is a unique ordinal $\delta^n(t,\omega)$ admitting an order isomorphism 
    \[ \tilde\psi^n(t,\omega)\colon\delta^n(t,\omega) \to \{ \beta\in\alpha_n \mid G_n(\beta,\omega) \ge t \} \]
    and $\tilde\psi^n(t,\omega)$ is uniquely determined by the requirement that, for all $\beta'\in\delta^n(t,\omega)$: 
    \[ \tilde\psi^n(t,\omega)(\beta') = \min \{ \beta\in\alpha_n \mid G_n(\beta,\omega) \ge t \}\setminus [\mc P\tilde\psi^n(t,\omega)](\beta')\Big);\] 
    that is, $\tilde\psi^n(t,\omega)(0) = \min\{ \beta\in\alpha_n \mid G_n(\beta,\omega) \ge t \}$, $\tilde\psi^n(t,\omega)(\beta'+1) = \tilde\psi^n(t,\omega)(\beta') +1$ for all $\beta'\in\delta^n(t,\omega)$ with $\beta'\in\delta^n(t,\omega)$, and $\tilde\psi^n(t,\omega)(\gamma') = \sup_{\beta'\in\gamma'} \tilde\psi^n(t,\omega)(\beta')$ for all limit ordinals $\gamma'\in\delta^n(t,\omega)$. $\tilde\psi^n(t,\omega)$ can be extended to $\delta^n(t,\omega)+1$ by letting $\psi^n(t,\omega)(\delta^n(t,\omega)) = \alpha_n$. As $G_n(\alpha_n,\omega) = \infty \ge t$, this yields an order isomorphism $\psi^n(t,\omega)$ as claimed. From the recursion above, we infer that $\psi^n(t,\omega)(\beta') = \psi^n(t,\omega)(0) + \beta'$ for all $\beta'\in\delta^n(t,\omega)+1$.

    Uniqueness of the isomorphism follows directly from what has been shown before, by basic well-order theory. We give an argument here for the reader's convenience only. For this, consider an arbitrary order isomorphism $f\colon \delta^n(t,\omega)+1 \to \{ \beta\in\alpha_n+1 \mid G_n(\beta,\omega) \ge t \}$, and let $S = \{\beta'\in\delta^n(t,\omega)+1 \mid f(\beta') \neq \psi^n(t,\omega)(0) + \beta'\}$. If $S$ were non-empty, it would have a minimum $\beta_0$. There would be $\beta_1,\beta_2\in\delta^n(t,\omega)+1$ with $f(\beta_0) = \psi^n(t,\omega)(0) + \beta_1$ and $\psi^n(t,\omega)(0) + \beta_0 = f(\beta_2)$. In particular, $\beta_1,\beta_2 > \beta_0$, and
    \[ f(\beta_0) = \psi^n(t,\omega)(0) + \beta_1 > \psi^n(t,\omega)(0) + \beta_0 = f(\beta_2),\]
    implying the contradiction $\beta_0 > \beta_2$.\medskip

    (Ad Part~\ref{3-SPF_VECT.lemma:refining,convergent_grid->psi,delta,gamma.delta^n_incr_in_n}):~ Let $n\in\N$. There is an order-embedding $j\colon\alpha_n+1 \inj \alpha_{n+1}+1$ such that $G_n = G_{n+1} \circ (j\times\id_\Omega)$. Hence, if $(\beta,\omega)\in (\alpha_n+1)\times\Omega$ is such that $G_n(\beta,\omega) \ge t$, then $G_{n+1}(j(\beta),\omega)\ge t$. Therefore, for any $\omega\in\Omega$, $\{\beta\in\alpha_n+1 \mid G_n(\beta,\omega)\ge t\}$ can be order-embedded into $\{\beta\in\alpha_{n+1}+1 \mid G_{n+1}(\beta,\omega)\ge t\}$. Via the order isomorphism from Part~\ref{3-SPF_VECT.lemma:refining,convergent_grid->psi,delta,gamma.exists_delta^n_and_psi^n}, $\delta^n(t,\omega)+1$ can be embedded in to $\delta^{n+1}(t,\omega)+1$, whence $\delta^n(t,\omega) + 1\le \delta^{n+1}(t,\omega) + 1$ which implies $\delta^n(t,\omega) \le \delta^{n+1}(t,\omega)$.
    
    Regarding the second claim, as a supremum of countably many countable, non-zero ordinals, $\delta(t,\omega)$ is countable and non-zero.
\end{proof}

\begin{proof}[Proof of Lemma~\ref{3-SPF_VECT.lemma:tilting_conv_as_ptw_conv?}]
    Let $n\in\N$, $(t,\beta,\omega)\in\ovT\times\Omega$ with $\beta\in\delta^n(t,\omega)+1$. 
    For all $\beta_0\in\alpha_n+1$ we infer, using that $G_n$ is an $\ms F$-adapted grid:
    \begin{align*}
        &\psi^n(t,\omega)(0) = \beta_0 \\
        \Leftrightarrow~&\Big( \tau^{G_n}_{\beta_0}(\omega) \ge t, ~\forall\beta'\in\beta_0\colon \tau^{G_n}_{\beta'}(\omega) < t \Big) \\
        \Leftrightarrow~&\begin{cases}
            t = 0, &\text{if } \beta_0 = 0, \\
            \tau^{G_n}_{\beta_0}(\omega) \ge t > \tau^{G_n}_{\beta'_0}(\omega), &\text{if } \beta_0 = \beta'_0 + 1 \text{ for some }\beta'_0\in\mf w_1, \\
            \tau^{G_n}_{\beta_0}(\omega) = t, & \text{if } \beta_0\in \mathrm{L}(\On).
        \end{cases}
    \end{align*}
    Hence, in view of Lemma~\ref{3-SPF_VECT.lemma:refining,convergent_grid->psi,delta,gamma}, we get:
    \begin{align*}
        &\xi^n\Big(G_n(\psi^n(t,\omega)(\beta),\omega),\omega\Big) \\
        =~&\xi^n\Big(G_n(\psi^n(t,\omega)(0)+\beta,\omega),\omega\Big) \\
        =~&\sum_{\beta_0\in\alpha_n+1} \xi^n_{\tau^{G_n}_{\beta_0+\beta}}(\omega)\, 1\{\psi^n(t,\omega)(0) = \beta_0\} \\
        =~& \xi^n_{\tau^{G_n}_{\beta}}(\omega)\,  1[\![0]\!](t,0,\omega) \\
        & + \sum_{\beta_0\in\alpha_n+1} \xi^n_{\tau^{G_n}_{\beta_0+1+\beta}}(\omega)\, 1(\!(\tau^{G_n}_{\beta_0},\tau^{G_n}_{\beta_0+1}]\!](t,0,\omega) \\
        & + \sum_{\beta_0\in(\alpha_n+1)\cap \mathrm{L}(\On)} \xi^n_{\tau^{G_n}_{\beta_0+\beta}}(\omega)\, 1[\![\tau^{G_n}_{\beta_0}]\!](t,0,\omega).
    \end{align*}
\end{proof}

\begin{proof}[Proof of Proposition~\ref{3-SPF_VECT.prop:optional_times_are_tilting_limits_of_classical_very_simple_optional_procs}]
    Let $\tau$ be an $\ms F$-optional time. By definition, there is $\alpha\in\mf w_1$ with $\pi\circ\tau < \alpha$. We choose $\alpha$ to be a limit ordinal. This is possible because, if necessary, we could replace $\alpha$ with the ordinal $\alpha' = \sup \{\alpha+\beta \mid \beta\in\mf w\}$. Clearly, the updated bound then satisfies $\alpha'\in \mf w_1$ and $\pi\circ\tau < \alpha'$. \smallskip

    (Construction of $(G_n)_{n\in\N}$):~ 
    For any $a,b\in\bRp$ with $a<b$, fix a continuous order embedding $h_{a,b}\colon\alpha\inj[a,b)_\bRp$ such that:
    \begin{itemize}
        \item $h_{a,b}(0) = a$;
        \item $\sup_{\beta\in\alpha} h_{a,b}(\beta) = b$;
        \item $\sup_{\beta\in\alpha} \big(h_{a,b}(\beta+1) - h_{a,b}(\beta)\big) \le \frac{b-a}{2} \wedge 1$.
    \end{itemize}
    
    For any $n\in\N$, define $\alpha_n\in\mf w_1$ and an order embedding $g_n\colon(\alpha_n+1)\to\bRp$ recursively as follows. 
    Let $\alpha_0 = \alpha$, and $g_0$ be given by $g_0(\beta) = h_{0,\infty}(\beta)$ for $\beta\in\alpha$ and $g_0(\alpha) = \infty$. Further, let $n\in\N$ and suppose that $\alpha_n\in \mf w_1$ and $g_n\colon(\alpha_n+1)\to\bRp$ is an order embedding. Define $g_{n+1}$ as follows. Let $\tilde g_{n+1}\colon\alpha_n\times \alpha\to \R_+$ be given by
    \[ \tilde g_{n+1}(\beta_0,\beta) = h_{g_n(\beta_0),g_n(\beta_0+1)}(\beta). \]
    Equip $\alpha_n\times\alpha$ with lexicographic order. Then, $\tilde g_{n+1}$ defines a continuous order embedding and $\alpha_n\times\alpha$ defines a well-order. Therefore, the latter is order-isomorphic to a unique ordinal $\alpha_{n+1}$, and it is countable, thus $\alpha_{n+1}\in\mf w_1$. As a consequence, $\tilde g_{n+1}$ induces a continuous order embedding $\alpha_{n+1}\to\R_+$ which we extend to an order embedding $g_{n+1}\colon(\alpha_{n+1}+1)\to\bRp$ by letting $g_{n+1}(\alpha_{n+1}) = \infty$. It is clear from this construction that the sequence $(G_n)_{n\in\N}$ given by 
    \[ G_n \colon (\alpha_n+1)\times\Omega\to\ovT,\, (\beta,\omega)\mapsto g_n(\beta), \qquad n\in\N, \]
    defines a refining sequence of classical, deterministic grids. Moreover, it is convergent, since --- by choice of the family $(h_{a,b})_{a,b}$ --- we have $\Delta(G_n,.) \le 2^{-n}$ for all $n\in\Nast$.\smallskip

    (Construction of $(\xi^n)_{n\in\N}$):~ Let $n\in\N$ and, for any $\omega\in\Omega$, 
    \[ \sigma_n(\omega) = \inf\{\beta\in\alpha_n+1 \mid p\circ\tau(\omega) \le g_n(\beta)\} + \pi\circ\tau(\omega). \]
    Upon extending $g_n$ to $\mf w_1$ by $g_n(\beta) = \infty$ for $\beta\in (\alpha_n,\mf w_1)_{\mf w_1}$, let $\tau_n = g_n \circ \sigma_n$, and $\xi^n = 1[\![0,\tau_n)\!)$.

    We show that $\tau_n$ is an $\ms F$-optional time. As $\pi\circ\tau_n = 0$, $\im\tau_n\subseteq \im g_n$, and $g_n$ has countable image, it suffices to show that $\{\tau_n = t\} \in \ms F_t$ for all $t\in\im g_n \setminus \{\infty\}$. This follows once we show that $\{\sigma_n = \beta^\star\}\in\ms F_{g_n(\beta^\star)}$ for all $\beta^\star\in\alpha_n$.

    Let $\beta_1,\beta_2\in\mf w_1$. Then, we have the following equivalences:
    \begin{align*}
        &\quad \inf \{\beta\in\alpha_n+1 \mid p\circ\tau(\omega) \le g_n(\beta) \} = \beta_1, \, \pi\circ\tau(\omega) = \beta_2 \\
        \Leftrightarrow&\quad \forall \beta\in\beta_1\colon \Big( (g_n(\beta),\beta_2) < \tau(\omega) \le (g_n(\beta_1),\beta_2)\Big),\, \pi\circ\tau(\omega) = \beta_2.
    \end{align*}
    Hence, for $\beta^\star\in\alpha_n$,
    \begin{align*}
        &~\{\sigma_n = \beta^\star\} \\
        =&~\bigcup_{\substack{\beta_1,\beta_2\in\mf w_1\colon\\ \beta_1+\beta_2 = \beta^\star}} \bigcap_{\beta\in\beta_1} \{ (g_n(\beta),\beta_2) < \tau(\omega) \le (g_n(\beta_1),\beta_2) \} \cap \{\pi\circ\tau = \beta_2\} \quad \in \ms F_{g_n(\beta^\star)},
    \end{align*}
    because $(g_n(\beta_1),\beta_2) \le g_n(\beta_1+\beta_2)$ in $\ovT$ for all $\beta_1,\beta_2\in\mf w_1$, and $\tau$ is an $\ms F$-optional time. Thus, $\tau_n$ is an $\ms F$-optional time.

    As $\tau_n$ is an $\ms F$-optional time, $\xi^n$ is $\ms F$-optional. As $\tau_n$ is $\bRp$-valued, $\xi^n$ is classical and very simple, and, by construction, the grid $G_n$ is compatible with $\xi^n$. It remains to show that $(\xi^n\mid G_n)\convT 1[\![0,\tau)\!)$ as $n\to\infty$.\smallskip

    (Ad Convergence):~ Let $(t,\beta,\omega)\in\ovT\times\Omega$ with $\beta\in\gamma(t,\omega)$, and let $n\in\N$ be large enough such that $\beta\in\delta^n(t,\omega)+1$. Then, $\xi^n\Big(G_n(\psi^n(t,\omega)(\beta),\omega),\omega\Big) = 1$ iff $G_n(\psi^n(t,\omega)(\beta),\omega) < \tau^n(\omega)$. By definition of $G_n$ and $\tau^n$, the latter is equivalent to $g_n(\psi^n(t,\omega)(\beta)) < g_n(\sigma_n(\omega))$. By definition of $g_n$, and because $\psi^n(t,\omega)$ maps into $\alpha_n+1$, this is equivalent to $\psi^n(t,\omega)(\beta) < \sigma_n(\omega)$. By Lemma~\ref{3-SPF_VECT.lemma:refining,convergent_grid->psi,delta,gamma}, and by definition of $\sigma_n$, this is equivalent to
    \begin{equation}\label{3-SPF_VECT.eq:tilting_approx_of_optional_times}
        \inf\{\beta'\in\alpha_n+1 \mid g_n(\beta')\ge t\} + \beta < \inf\{\beta'\in\alpha_n+1 \mid g_n(\beta')\ge  p\circ\tau(\omega)\} + \pi\circ\tau(\omega). 
    \end{equation}
    If $(t,\beta)\ge \tau(\omega)$, then, by construction of $(G_k)_{k\in\N}$ and the fact that $\pi\circ\tau<\alpha$, there is $N\in\N$ such that for all integers $n\ge N$ Inequality~\ref{3-SPF_VECT.eq:tilting_approx_of_optional_times} is not satisfied. Thus, in that case, 
    \[ \xi^n\Big(G_n(\psi^n(t,\omega)(\beta),\omega),\omega\Big) \to 0 = 1[\![0,\tau)\!)(t,\beta,\omega), \qquad \text{as }n\to\infty. \] 
    If, conversely, $(t,\beta) < \tau(\omega)$, then, again by construction of $(G_k)_{k\in\N}$ and the inequality $\pi\circ\tau<\alpha$, there is $N\in\N$ such that for all integers $n\ge N$ Inequality~\ref{3-SPF_VECT.eq:tilting_approx_of_optional_times} is satisfied. Thus, in that case, 
    \[ \xi^n\Big(G_n(\psi^n(t,\omega)(\beta),\omega),\omega\Big) \to 1 = 1[\![0,\tau)\!)(t,\beta,\omega), \qquad \text{as }n\to\infty.\] 

    Furthermore, note that $\gamma(t,\omega) \ge \alpha$ for all $(t,\omega)\in\R_+\times\Omega$, by construction of $(G_k)_{k\in\N}$, and $\gamma(\infty,\omega) > 0$ for all $\omega\in\Omega$ by the general construction of $\gamma$. Hence, for $(t,\beta,\omega)\in\ovT\times\Omega$ with $\beta\notin\gamma(t,\omega)$, we have $\beta\ge\alpha$. As a consequence, $1[\![0,\tau)\!)(t,\beta,\omega) =\lim_{\beta'\nearrow \gamma(t,\omega)} 1[\![0,\tau)\!)(t,\beta',\omega)$, since $\gamma(t,\omega)$ is a limit ordinal and $\gamma(t,\omega) \ge \alpha > \pi\circ\tau(\omega)$.

    We conclude that $(\xi^n\mid G_n)\convT 1[\![0,\tau)\!)$ as $n\to\infty$.
\end{proof}

\begin{proof}[Proof of Theorem~\ref{3-SPF_VECT.thm:density_classical_very_simple_opt_proc}]
    As any Polish space $Y$ can be measure-theoretically embedded into $\R$,\footnote{That is, there is a measurable injection $\p\colon Y\inj \R$ with measurable image, and a measurable inverse $\im\p \to Y$.} we can suppose without loss of generality that $Y = \R$. For any two $\ms F$-optional processes $\xi,\xi'$ valued in $\R$, the process $\tilde\xi\colon\ovT\times\Omega\to\R^2,\, (t,\omega)\mapsto (\xi_t(\omega),\xi'_t(\omega))$ is again $\ms F$-optional because for all $B,B'\in\ms B_\R$, we have
    \[ \tilde\xi^{-1}(B\times B') = \xi^{-1}(B) \cap {\xi'}^{-1}(B') \, \in\Opt(\ms F).\]
    Thus, for any continuous --- and \emph{a fortiori} Borel-measurable --- $f\colon\R^2\to\R$, $\xi'' = f\circ\tilde\xi$ is $\ms F$-optional.

    Furthermore, pointwise addition and scalar multiplication with $\lambda\in\R$ can be described by the continuous map $\R^2\to\R,\, (x,y) \mapsto \lambda x + y$. With this, the theorem is a direct consequence of Lemma~\ref{3-SPF_VECT.lemma:Opt_proc_generated_by_xi*1[tau,sigma)_2} and Proposition~\ref{3-SPF_VECT.prop:optional_times_are_tilting_limits_of_classical_very_simple_optional_procs}, together with the fact that, for any $\ms F_\infty$-measurable, real-valued random variable $\xi^\infty$, the process $\xi^\infty\circ\prj_\Omega\, 1[\![\infty]\!]$ is a classical, very simple $\ms F$-optional process.
\end{proof}

\subsection{Section~\ref{3-SPF_VECT.sec:SPF}}

\subsubsection{Information sets, counterfactuals, and equilibrium}
\begin{proof}[Proof of Proposition~\ref{3-SPF_VECT.prop:information_sets}]
    (Ad Part~\ref{3-SPF_VECT.prop:information_sets.f(t,omega,h)=f(t,omega,h')}):~ Let $V$ be the set of $(\ms P_\ovT\otimes\ms E\otimes \{\emptyset,\B^\ovT\}) \vee \Prd(\ms H^i)$-measurable $f\colon\ovT\times W\to \R$ such that for $t\in\ovT$, $\omega\in\Omega$, and $h,h'\in\B^\ovT$ with $h|_{[0,t)_\ovT} = h'|_{[0,t)_\ovT}$, we have $f(t,\omega,h) = f(t,\omega,h')$. Let $V_b\subseteq V$ be the subset of bounded $f\in V$.

    $V_b$ is clearly stable under pointwise addition, multiplication, and real scalar multiplication and contains all constant functions. It remains to show the Claim \textbf{(CL1)} that $V_b$ contains the functions $1_M$ for all $M\in\mc G_k$, $k=1,2$, for an intersection-stable generator $\mc G_1$ of $\ms P_\ovT\otimes\ms E\otimes \{\emptyset,\B^\ovT\}$ and an intersection-stable generator $\mc G_2$ of $\Prd(\ms H^i)$. Indeed, if such $\mc G_k$, $k=1,2$, exist, then we may assume that they contain $\ovT\times\Omega\times\B^\ovT$. Under this assumption, $\mc G = \{M\cap N\mid M\in\mc G_1,\,N\in \mc G_2\}$ defines an intersection-stable generator of $(\ms P_\ovT\otimes\ms E\otimes \{\emptyset,\B^\ovT\}) \vee \Prd(\ms H^i)$. Then, as $V_b$ is stable under multiplication, $1_G\in V_b$ for all $G\in\mc G$, and products of such indicators are again in $V_b$. Moreover, it is clear that $V$ is closed under pointwise convergence and the limit of a $V_b$-valued pointwise converging sequence $(f_n)_{n\in\N}$ such that $0\le f_n\le C$ for some real constant $C>0$ is again bounded. Hence, using the functional monotone class theorem and Claim (CL1), we infer that $V$ equals the set of all $(\ms P_\ovT\otimes\ms E\otimes \{\emptyset,\B^\ovT\}) \vee \Prd(\ms H^i)$-measurable $f\colon\ovT\times W\to \R$. As $\Opt(\ms E\otimes\{\emptyset,\B^\ovT\})\subseteq \ms P_\ovT\otimes\ms E\otimes \{\emptyset,\B^\ovT\}$, and as we make Assumption~\ref{3-SPF_VECT.prop:information_sets.Ass.msMi_endogenously_predictable}, the claim of the proposition's first part follows.

    We prove Claim (CL1). 
    For this, let $t\in\ovT$, $\omega\in\Omega$, and $h,h'\in\B^\ovT$. 
    First, let $T\in\ms P_\ovT$, $E\in\ms E$, and $f = 1(T\times E\times\B^{\ovT})$. Then, for all $t,\omega,h,h'$ as above, we have $f(t,\omega,h) = f(t,\omega,h')$. Thus, $f\in V_b$.
    Next, let $H\in\ms H^i_0$ and $f = 1(\{0\}\times H)$. By Assumption~\ref{3-SPF_VECT.prop:information_sets.Ass.Hi0_open-loop}, there is $E\in\ms E$ such that $H = E\times\B^\ovT$. Then, again, for all $t,\omega,h,h'$ as above, we have $f(t,\omega,h) = f(t,\omega,h')$. Hence, $f\in V_b$. Next, take an $\ms H^i$-optional time $\sigma$ and let $f = 1[\![0,\sigma]\!]$. If $t=0$, then $f(t,\omega,h) = 1 = f(t,\omega,h')$. Suppose that $t>0$. We have $\{f_t = 0\} = \{\sigma < t\}$. As $\pi\circ\sigma$ is bounded above by some fixed countable ordinal, there is a countable subset $Q\subseteq[0,t)_\ovT$ such that 
    \[\{f_t = 0\} = \{\sigma < t\} = \bigcup_{u\in Q} \{\sigma \le u\}. \]
    Let $u\in Q$. Then, we have $\{\sigma\le u\}\in \ms H^i_u$. Hence, by \textsc{spf} Axiom~\ref{3-SPF_VECT.def:spf.msH_non_anticipative}, using the notation from Definition~\ref{3-SPF_VECT.def:spf}, there is $H_u \in\tilde{\ms H}^i_u$ with
    \[ 1\{\sigma\le u\}(\omega,h) = 1(H_u)(\omega,\op{proj}_{[0,u]_\ovT}(h)) = 1(H_u)(\omega,\op{proj}_{[0,u]_\ovT}(h')) = 1\{\sigma\le u\}(\omega,h') ,\]
    because $u<t$. 
    Thus, $f_t(\omega,h) = 1-\sup_{u\in Q} 1\{\sigma\le u\}(\omega,h) = 1-\sup_{u\in Q} 1\{\sigma\le u\}(\omega,h') = f_t(\omega,h')$. Thus, $f\in V_b$. 
    Furthermore, if $\tau$ is another $\ms H^i$-optional time, the minimum $\tau\wedge\sigma$ is as well by Proposition~\ref{3-SPF_VECT.prop:basic_properties_of_optional_times}, and as $V_b$ is stable under differences, we also get that $f = 1(\!(\sigma,\tau]\!]\in V_b$. This proves Claim (CL1).\smallskip

    (Ad Part~\ref{3-SPF_VECT.prop:information_sets.chi|<taui=chi|<taui_=>_chi_=~_chi'}):~ Let $\tau^i$ be an optional time for $i$ and let $\chi,\chi'$ be state processes such that there is $N\in\ms N$ satisfying, for all $\omega\in\Omega\setminus N$, 
    \[ \chi|_{[0,\tau^i(\omega,\chi(\omega))_\ovT} = \chi'|_{[0,\tau^i(\omega,\chi(\omega))_\ovT}. \]
    Hence, by Part~\ref{3-SPF_VECT.prop:information_sets.f(t,omega,h)=f(t,omega,h')}, any $\omega\in\Omega\setminus N$, any $\ms M^i$-measurable $f\colon\ovT\times W\to \R$ and any $t\in[0,\tau^i(\omega,\chi(\omega))]_\ovT$ satisfy
    \[ (\ast) \qquad f(t,\omega,\chi(\omega)) = f(t,\omega,\chi'(\omega)). \]
    Whence, $\chi\approx_{i,\tau^i} \chi'$.\smallskip

    (Ad Part~\ref{3-SPF_VECT.prop:information_sets.chi_=~_chi'_=>_chi|<taui=chi|<taui}):~ Let $\tau^i$ be an optional time for $i$ and let $\chi,\chi'$ be state processes that are left-continuous at all $u\in\ovT$ with $\pi(u) = \mf w_1$ such that $\chi\approx_{i,\tau^i} \chi'$. We make Assumption~\ref{3-SPF_VECT.prop:information_sets.Ass.Hit_closed-loop}. There is $N\in\ms N$ such that for all $\omega\in\Omega\setminus N$, all $t\in[0,\tau^i(\omega,\chi(\omega))]_\ovT$, and all $\ms M^i$-measurable $f\colon\T\times W\to \R$, Equation~$(\ast)$ is satisfied. 
    
    As $\Prd(\ms H^i) \subseteq \ms M^i$, it suffices to select some embedding of measurable spaces $\p\colon \B \inj [0,1]$,\footnote{That is, $\p$ is Borel measurable, injective, has Borel-measurable image, and Borel measurable inverse $\im\p\to\B$.} and to show the intermediate claim that for all $u\in\ovT$ with $\pi(u) < \mf w_1$, the map
    \[ f^u\colon\T\times W \to \R,\, (t,\omega,h) \mapsto \begin{cases} \p\circ h(u), &\text{if } u<t, \\ -1, &\text{else,} \end{cases} \]
    is $\ms H^i$-predictable. Indeed, then we infer that, for all $\omega\in\Omega\setminus N$ and $u\in [0,\tau^i(\omega,\chi(\omega)))_\ovT$ with $\pi(u)<\mf w_1$, it holds true that $\chi(u,\omega) = \chi'(u,\omega)$. By the left-continuity assumption, this extends to all $u\in [0,\tau^i(\omega,\chi(\omega)))_\ovT$.

    For the proof of the intermediate claim above, fix $u\in\ovT$ with $\pi(u) < \mf w_1$. Then, $\{f^u = -1\} = [\![0,u]\!] \in \Prd(\ms H^i)$. Moreover, for any Borel set $B\in\ms B_\R$ with $-1\notin B$, 
    \[ \sigma^u\colon W\to\ovT,\, (\omega,h) \mapsto \begin{cases} u, &\text{if } \p\circ h(u)\in B, \\ \infty, &\text{else,} \end{cases} \]
    is clearly an $\ms H^i$-optional time, in view of Assumption~\ref{3-SPF_VECT.prop:information_sets.Ass.Hit_closed-loop}. As
    \[ \{f^u \in B\} = (\!(\sigma^u,\infty]\!] \, \in\Prd(\ms H^i), \]
    $f^u$ is $\ms H^i$-predictable, and the proof of this part of the proposition is complete.\smallskip

    The final statement of the proposition follows directly from the last two parts.
\end{proof}

\subsubsection{Timing games}

\begin{proof}[Proof of Lemma~\ref{3-SPF_VECT.lemma:z_optional}]
    (Ad~$\tau_{\bm b}$):~ Let $\bm b\in\B$. As $z$ has locally right-constant paths, we have, for all $(t,\omega,h)\in[\![0,\tau_{\bm b})\!)$, $h(\tau_{\bm b}(\omega,h)) = z_{\tau_{\bm b}}(\omega,h)\le\bm b$ and $h(t) = z_t(\omega,h) > \bm b$. Hence, by definition of $W$ and the upper vertical level, $\pi\circ\tau_{\bm b}(\omega,h) < \alpha$. Moreover, for all $t,u\in\ovT$ with $u\le t$, and $\beta\in\mf w_1$, we have
    \begin{align*}
        &~\{\tau_{\bm b} \le u,\, \pi\circ\tau_{\bm b} = \beta\} \\
        =&~\Big(\Omega\times \{h\in\B^\ovT \mid \exists v\in[0,u]_\ovT\colon ( h(v) \le \bm b, \forall v'\in [0,v)_\ovT\colon h(v') > \bm b,\pi(v) = \beta)\}\Big) \cap W \\
        =&~ (\id_\Omega\times\op{proj}_{[0,t]_\ovT})^{-1}(\tilde H)\cap W,
    \end{align*}
    for some subset $\tilde H\subseteq\Omega\times\B^{[0,t]_\ovT}$. Hence, by definition of $\ms H^i$, $\tau_{\bm b}$ is an $\ms H^i$-optional time.\smallskip

    (Ad~$\tau_{\bm b}^-$):~ We clearly have $\tau_{\bm 1} = \tau_{\bm 1}^-$. In the following, we suppose that $\bm b\neq\bm 1$. 
    Let $w\in W$. Then, $z_{\tau_{\bm b}}(w) \le \bm b < z_{\tau_{\bm b}^-}(w)$, because $z$ has locally right-constant paths. Hence, if $\tau_{\bm b}(w) = (t,\beta)$ for some $t\in\bRp$ and $\beta\in\mf w_1$, then $\tau_{\bm b}^-(w) = (t,\beta+1)$. As $\pi\circ\tau_{\bm b}(w) < \alpha$, we infer that $\tau_{\bm b}^-(w) = (p\circ\tau_{\bm b}(w),\pi\circ\tau_{\bm b}(w) + 1)$. Hence, $[\![0,\tau_{\bm b}^-)\!) = [\![0,\tau_{\bm b}]\!]$, and $\pi\circ\tau_{\bm b}^- = \pi\circ\tau_{\bm b} + 1 \le \alpha$. Let $t\in\ovT$ and $\beta\in\mf w_1$. We have
    $\{\tau_{\bm b}^- \le t,\, \pi\circ\tau_{\bm b}^- = \beta\} = \emptyset$, if $\beta$ is not a successor ordinal or if $t = 0$. Else, there is $\gamma\in\mf w_1$ with $\beta = \gamma +1$ and $t>0$. If $\pi(t)$ is a successor ordinal as well, there is $\delta\in\mf w_1$ with $\pi(t) = \delta+1$, whence
    \[ \{\tau_{\bm b}^- \le t,\, \pi\circ\tau_{\bm b}^- = \beta\} = \{\tau_{\bm b} \le (p(t),\delta),\, \pi\circ\tau_{\bm b} = \gamma\}\, \in \ms H^i_{(p(t),\delta)} \subseteq \ms H^i_t. \]
    If $\pi(t)$ is not a successor ordinal and $t>0$, then there is a strictly increasing sequence $(u_n)_{n\in\N}\in \ovT^\N$ with $u_n \nearrow t$. Hence,
    \[ \{\tau_{\bm b}^- \le t,\, \pi\circ\tau_{\bm b}^- = \beta\} = \bigcup_{n\in\N} \{\tau_{\bm b} \le u_n,\, \pi\circ\tau_{\bm b} = \gamma\}\, \in \bigvee_{n\in\N}\ms H^i_{u_n} \subseteq \ms H^i_t. \]
    Thus, $\tau_{\bm b}^-$ is an $\ms H^i$-optional time with the claimed properties.\smallskip

    (Intermediate result):~ Let $\bm b\in\B$ and
    \[ \tau_{<\bm b} = \inf\{t\in\ovT \mid z_u < \bm b\}. \]
    As $z$ is decreasing, we have, with $\downarrow \bm b = \{\bm b'\in\B \mid \bm b' \le \bm b\}$,
    \[ \tau_{<\bm b} = \inf_{\bm b'\in \downarrow\bm b \setminus\{\bm b\}} \tau_{\bm b'}. \]
    As an infimum of a finite number of $\ms H^i$-optional times, this is again an $\ms H^i$-optional time by an application of Proposition~\ref{3-SPF_VECT.prop:basic_properties_of_optional_times}, Part~\ref{3-SPF_VECT.prop:basic_properties_of_optional_times}.\smallskip
    
    (Ad~$z$):~ Then, as $z$ has locally right-constant and componentwise decreasing paths, for any $\bm b\in\B$, we obtain the representations
    \[ (\ast) \qquad
    \{ z = \bm b \} = \begin{cases} [\![\tau_{\bm b},\infty]\!], &\text{if }\bm b = \bm 0,\\  [\![\tau_{\bm b},\tau_{<\bm b})\!), &\text{else,} \end{cases} \qquad
    \{ z_-= \bm b \} = \begin{cases} [\![0,\tau_{<\bm b}]\!], &\text{if }\bm b = \bm 1,\\  (\!(\tau_{\bm b},\tau_{<\bm b}]\!], &\text{else.} \end{cases}
    \]
    Hence, $z$ is $\ms H^i$-optional and has upper vertical level smaller than or equal to $\alpha$, and $z_-$ is $\ms H^i$-predictable and has (upper) vertical level smaller than or equal to $\alpha$ (if $\alpha$ is a limit ordinal, respectively). Let $\beta\in\alpha$ and $h\colon\ovT\to\B$ be given by $h(t) = \bm 1$ for $t\in[0,(0,\beta))_\ovT$, and $h(t) = \bm 0$ for $t\in [(0,\beta),\infty]_\ovT$. Then, for any $\omega\in\Omega$, we have $(\omega,h)\in W$. Plugging in $(\omega,h)$ into $z$ and $z_-$ for all $\beta\in\alpha$ shows that $z$ has upper vertical level $\alpha$ and that $z_-$ has (upper) vertical level $\alpha$ (if $\alpha$ is a limit ordinal).
\end{proof}

\begin{proof}[Proof of Lemma~\ref{3-SPF_VECT.lemma:fsharp_preserves_optionality}]
    The map $f^\#$ is well-defined because for any $(\omega,h)\in W$, $(\omega,f(\omega,h))\in W$. Using Theorem~\ref{3-SPF_VECT.thm:optional_times}, Corollary~\ref{3-SPF_VECT.cor:xi_tau_is_F_tau_mb}, Theorem~\ref{3-SPF_VECT.thm:debut}, and the completeness assumption on the data $\ms H^\vee$, we infer that $f$ is a simple $\ms H^\vee$-optional process of the form
    \[ f_t(w) = \begin{cases} f_{\tau_k}(w), &\text{if } (t,w)\in [\![\tau_k,\tau_{k+1})\!) \text{ for } k = 0,\dots,|I|, \\ \bm 0, &\text{if } t = \infty, \end{cases} \]
    for $\ms H^\vee$-optional times $\tau_0,\dots,\tau_{|I|}$ with $0 = \tau_0 \le \dots \le \tau_{|I|+1} = \infty$.\smallskip

    (First step):~ Let $\id_\Omega\star f$ denote the map $W\to W,\, (\omega,h) \mapsto (\omega,f(\omega,h))$. We show that, for any $t\in\ovT$, $\id_\Omega\star f$ is $\ms H^\vee_t$-$\ms H^\vee_t$-measurable. By basic measure theory, using universal completeness of $(W,\ms H^\vee_\infty,\ms H^\vee)$, it suffices to show that it is $\ms H^{\vee}_t$-$\ms H^{i,0}_t$-measurable for all $i\in I$, with the notation from the definition of the data $\mathbf F$.
    
    Let $i\in I$, $E\in\ms F^i_0$. Then, $(\id_\Omega\star f)^{-1}(E\times\B^\ovT) = E\times\B^\ovT\in\ms H^{\vee}_0$.     
    Next, let $t\in\ovT\setminus\{0\}$, $i\in I$, $E\in\ms F^i_t$, $n\in\Nast$, for $\ell=1,\dots,n$, $(u_\ell,\bm b_\ell)\in [0,t]_\ovT\times\B$, and $B = \{h\in\B^\ovT \mid \forall \ell = 1,\dots,n\colon h(u_\ell) = \bm b_\ell\}$. Then, for $H_t = E\times B$, 
    \begin{align*} 
        (\id_\Omega\star f)^{-1}(H_t) =&~ (E\times\B^\ovT) \cap \bigcap_{\ell=1}^n \Big(\bigcup_{k=0}^{|I|} \{ w\in W \mid \tau_k(w) \le u_\ell < \tau_{k+1}(w),\, f_{\tau_k}(w) = \bm b_\ell\}\\
        &\qquad\qquad\qquad\qquad\qquad\qquad \cup \{w\in W \mid u_\ell = \infty,\, \bm b_\ell = \bm 0\} \Big)
    \end{align*}
    is an element of $\ms H^{\vee}_t$. Finally, let again $t\in\ovT\setminus\{0\}$, $\beta\in\mf w_1$, and $\bm b\in\B$. Let 
    \[ \tilde f\colon \ovT\times W \to \B,\, (t,w)\mapsto \begin{cases} \bm 1, &\text{if } t=0, \\ f_t(w), &\text{else,} \end{cases} 
    \]
    and 
    $\sigma_{\bm b} = \inf\{u\in\ovT \mid \tilde f_u \le \bm b\}$. Then, Theorem~\ref{3-SPF_VECT.thm:debut}, local right-constancy and optionality of $\tilde f$ as well as the completeness assumption on $\ms H^\vee$ imply that $\sigma_{\bm b}$ is an $\ms H^\vee$-optional time. For $H_t = \{\tau_{\bm b}\le t,\, \pi\circ \tau_{\bm b}=\beta\}$, we infer that
    \begin{align*}
        (\id_\Omega\star f)^{-1}(H_t) =&~ \{\sigma_{\bm b} \le t,\, \pi\circ\sigma_{\bm b}=\beta\}\, \in \ms H^\vee_t.
    \end{align*}
    We conclude that $\id_\Omega\star f$ is $\ms H^\vee_t$-$\ms H^\vee_t$-measurable.\smallskip

    ($\Opt(\ms H^\vee)$-$\Opt(\ms H^\vee)$-measurability):~ Now let $\tau$ be an $\ms H^\vee$-optional time and let $\tau^f = \tau \circ (\id_\Omega\star f)$. There is $\alpha\in\mf w_1$ with $\pi\circ\tau\le \alpha$, whence $\pi\circ\tau^f\le \alpha$. Moreover, for any $t\in\ovT$ and $\beta\in\mf w_1$, we have, using the first step's result,
    \[ \{\tau^f\le t,\, \pi\circ\tau^f = \beta\} = (\id_\Omega\star f)^{-1}(\{\tau\le t,\, \pi\circ\tau = \beta\}) \, \in \ms H^\vee_t. \]
    Hence, 
    \[ (f^\#)^{-1}([\![0,\tau)\!)) = [\![0,\tau^f)\!)\,\in\Opt(\ms H^\vee). \]
    Moreover, for any $E\in\ms H^\vee_\infty$, we have $(\id_\Omega\star)^{-1}(E)\in\ms H^\vee_\infty$ by the first step, whence
    \[ (f^\#)^{-1}(\{\infty\}\times E) = \{\infty\} \times (\id_\Omega\star f)^{-1}(E)\, \in \Opt(\ms H^\vee). \]
    This completes the proof of $\Opt(\ms H^\vee)$-$\Opt(\ms H^\vee)$-measurability.\smallskip

    ($\Prg(\ms H^\vee)$-$\Prg(\ms H^\vee)$-measurability):~ Let $M\in\Prg(\ms H^\vee)$ and $t\in\ovT$. Then, $M \cap [\![0,t]\!] \in\ms P_\ovT\otimes\ms H^\vee_t$, whence
    \begin{align*}
        (f^\#)^{-1}(M) \cap [\![0,t]\!] =&~ [\id_\ovT\times(\id_\Omega\star f)]^{-1}(M) \cap [\![0,t]\!] \\
        =&~ [\id_\ovT\times(\id_\Omega\star f)]^{-1}(M \cap [\![0,t]\!]) \,\in\ms P_\ovT\otimes\ms H^\vee_t,
    \end{align*}
    because $\id_\ovT \times (\id_\Omega\star f)$ is $\ms P_\ovT\otimes\ms H^\vee_t$-$\ms P_\ovT\otimes\ms H^\vee_t$-measurable by the first step and basic measure theory. This proves $\Prg(\ms H^\vee)$-$\Prg(\ms H^\vee)$-measurability.
\end{proof}

\begin{proof}[Proof of Lemma~\ref{3-SPF_VECT.lemma:eta_circ(id_Omega_star_chi)_inherits_optionality}]
    (First step):~ As $\chi$ is $\ms F^\vee$-optional, it is $\ms F^\vee$-adapted (this follows from Corollary~\ref{3-SPF_VECT.cor:Opt.uniform_upper_bound_on_vertical_activity}). Using this and the fact that $\ms F^\vee$ is universally augmented in $\ms E$, applying methods similar to those employed in the proof of Lemma~\ref{3-SPF_VECT.lemma:fsharp_preserves_optionality}, one proves that $\id_\Omega\star\chi$ is $\ms F^\vee_t$-$\ms H^{i,0}_t$-measurable for any $i\in I$ and $t\in\ovT$. Hence, it is $\ms F^\vee_t$-$\ms H^\vee_t$-measurable for any $t\in\ovT$, by basic measure theory and the universal completeness of $(\Omega,\ms E,\ms F^\vee)$.\smallskip

    (Second step):~ Let $\tau$ be an $\ms H^\vee$-optional time, and $\tau^\chi = \tau\circ(\id_\Omega\star\chi)$. Then, it follows from the first step, in a way completely analogous to the proof of Lemma~\ref{3-SPF_VECT.lemma:fsharp_preserves_optionality}, second step, that $\tau^\chi$ is an $\ms F^\vee$-optional time. Hence,
    \[ [\id_\ovT\times(\id_\Omega\star\chi)]^{-1}([\![0,\tau)\!)) = [\![0,\tau^\chi)\!) \, \in\Opt(\ms F^\vee). \]
    Moreover, for any $E\in\ms H^\vee_\infty$, we have $(\id_\Omega\star\chi)^{-1}(E)\in\ms F^\vee_\infty$, by the first step, whence
    \[ [\id_\ovT\times(\id_\Omega\star\chi)]^{-1}(\{\infty\}\times E) = \{\infty\}\times (\id_\Omega\star\chi)^{-1}(E)\, \in \Opt(\ms F^\vee). \]
    Thus, $[\id_\ovT\times(\id_\Omega\star\chi)]$ is $\Opt(\ms F^\vee)$-$\Opt(\ms H^\vee)$-measurable. As a consequence, the composition $\eta\circ [\id_\ovT\times(\id_\Omega\star\chi)]$ is $\Opt(\ms F^\vee)$-$\ms B_\B$-measurable, i.e.\ $\ms F^\vee$-optional. 
\end{proof}

\begin{proof}[Proof of Theorem~\ref{3-SPF_VECT.thm:timing_SPF_well-posed}]
    (Ad ``$\mathbf F$ is \textsc{spf}'': basic properties of the data):~ Any $(\xi,\chi)\in\mc W$ satisfies $(\omega,\chi(\omega))\in W$ for all $\omega\in\Omega$. Moreover, for all $i\in I$ and $t,u\in\ovT$ with $t\le u$, it follows readily from the definition that $\ms H^i_t$ is a $\sigma$-algebra on $W$ and that we have $\ms H^i_t \subseteq \ms H^i_u$, which proves that $\ms H^i$ is a filtration. Moreover, $\ms F^i_t \otimes \{\emptyset,\B^\ovT\}\subseteq\ms H^i_t$, for all $t\in\ovT$; hence, $\Prd(\ms H^i) \subseteq\ms M^i\subseteq\Opt(\ms H^i)$. 
    It remains to show the Axioms in Definition~\ref{3-SPF_VECT.def:spf}.\smallskip
    
    (Ad Axiom~\ref{3-SPF_VECT.def:spf.msH_non_anticipative}):~ This axiom is satisfied by construction. \smallskip

    (Ad Axiom~\ref{3-SPF_VECT.def:spf.msF^i_zeta_subseteq_msE}):~ 
    Let $\chi$ be such that there is $\xi$ with $(\xi,\chi)\in \mc W$ and $i\in I$. $\chi$ is $\ms F^\vee$-optional, hence -adapted. Thus, for any $i\in I$, all
    \[ H_t \in \ms F^i_\infty\otimes\ms B_\B^{[0,t]_\ovT}\otimes\{\emptyset,\B^{(t,\infty]_\ovT}\}|_W \]
    satisfy $(\id_\Omega\star\chi)^{-1}(H_t)\in\ms F^\vee_\infty\subseteq\ms E$. 
    
    Let $i\in I$, $t\in\ovT$, $\beta\in\mf w_1$. For any state process $\chi$, let 
    \[ \tilde\chi\colon\ovT\times\Omega\to\B,\, (t,\omega) \mapsto \begin{cases} \bm 1, &\text{if } t=0, \\ \chi_t(\omega), &\text{else,} \end{cases} \]
    and $M^{\bm b}_\chi = \{(u,\omega)\in\ovT\times\Omega \mid \tilde \chi(u) \le \bm b\}$ and $D_{M^{\bm b}_\chi}$ be the début of the set $M^{\bm b}_\chi$. By construction, $\tilde\chi$ is left-constant at all $u\in\ovT$ with $\pi(u) = \mf w_1$. Moreover, $\tilde\chi$ is also right-constant at all those $u$. Thus, as $\ms F^\vee$ is universally augmented in the universally complete $\sigma$-algebra $\ms E$, $D_{M^{\bm b}_\chi}$ is an $\ms F^\vee$-optional time, by Theorem~\ref{3-SPF_VECT.thm:debut}. Hence,
    \[ (\id_\Omega\star\chi)^{-1}(\{\tau_{\bm b}\le t,\, \pi\circ\tau_{\bm b} = \beta\}) = \{D_{M^{\bm b}_\chi} \le t,\, \pi\circ D_{M^{\bm b}_\chi} = \beta\}\, \in\ms F^\vee_t\subseteq\ms E. \]
    
    We conclude that for $i\in I$, $\id_\Omega\star\chi$ is $\ms E$-$\ms H^{i,0}_\infty$-measurable, with $\ms H^{i,0}_\infty$ as in the definition of the data $\mathbf F$. As $\ms E$ is universally complete, basic measure theory implies that $\id_\Omega\star\chi$ is also $\ms E$-$\ms H^{i,1}_\infty$-measurable, because $\ms H^{i,1}_\infty = [\ms H^{i,0}_\infty]^{\mathrm u}$, as in the definition of the data $\mathbf F$. As $\ms H^i_\infty\subseteq \ms H^{i,1}_\infty$, we infer that, in particular, $\id_\Omega\star\chi$ is $\ms E$-$\ms H^i_\infty$-measurable.
    \smallskip    

    (Ad Axiom~\ref{3-SPF_VECT.def:spf.nu_Obs(xi)}):~ This is evident because $\xi = \chi$ for all $(\xi,\chi)\in\mc W$. We conclude that $\mathbf F$ is an \textsc{spf}. \smallskip

    (Ad Axiom~\ref{3-SPF_VECT.def:spf.msMi-measurability}):~ This axiom is satisfied by construction.\smallskip
    
    (Ad well-posedness):~ Let $s\in\mc S$, $i\in I$, $\tau^i$ be an optional time for $i$, and $\tilde\chi$ be a state process. 
    
    We now define two sequences $(\sigma_n)_{n\in\N}$ of 
    ${\ms H}^\vee$-optional times 
    $\sigma_n\colon W\to\ovT$ with $\pi\circ\sigma_n<\alpha$ for all $n\in\N$
    and $(\eta^n)_{n\in\N}$ of right-continuous, decreasing
    ${\ms H}^\vee$-optional processes 
    $\eta^n\colon\ovT\times W\to \B$ with upper vertical level smaller than or equal to $\alpha$ and $\eta^n_\infty = \bm 0$, satisfying the following ``extra properties'', for all $n\in\N$:
    \begin{enumerate}
        \item\label{3-SPF_VECT.thm:timing_SPF_well-posed.construction.1} If $n>0$, then $\sigma_{n-1}\le\sigma_n$ and, for all $w \in \{\sigma_{n-1} < \infty\}$, we have $\sigma_{n-1}(w) < \sigma_{n}(w)$.
        \item\label{3-SPF_VECT.thm:timing_SPF_well-posed.construction.2} If $n>0$, then $s_{\sigma_n(\omega,h)}(\omega,\eta^n(\omega,h)) < \eta^n_{\sigma_n(\omega,h)}(\omega,h)$ for all $(\omega,h)\in W$ with $\sigma_n(\omega,h) < \infty$.
        \item\label{3-SPF_VECT.thm:timing_SPF_well-posed.construction.3} If $n>0$, then for all $t,u\in\ovT$ and $w\in W$ with $(t,w), (u,w)\in[\![\sigma_{n-1},\sigma_n)\!)$, we have $\eta^n_t(w) = \eta^n_u(w)$ and $|\eta^n_t(w)|_1 \le |I|-n+1$.\footnote{Here, $|.|_1$ denotes the $\L^1$-norm. That is, for $\bm b = (b^i)_{i\in I}\in\B$, we have $|\bm b|_1 = \sum_{i\in I} |b^i| = \sum_{i\in I} b^i$ which is the number of components with value one, or game-theoretically speaking, the number of active agents.}
        \item\label{3-SPF_VECT.thm:timing_SPF_well-posed.construction.4} If $n>0$, then we have $\eta^{n-1}|_{[\![0,\sigma_{n-1})\!)} = \eta^{n}|_{[\![0,\sigma_{n-1})\!)}$.
        \item\label{3-SPF_VECT.thm:timing_SPF_well-posed.construction.5} $\eta^n|_{[\![0,\sigma_0)\!)} = \eta^0|_{[\![0,\sigma_0)\!)}$.
        \item\label{3-SPF_VECT.thm:timing_SPF_well-posed.construction.6} For all $(t,\omega,h)\in [\![\sigma_0,\sigma_n)\!)$ we have
        \[ \eta^n_t(\omega,h) = s_t(\omega,\eta^{n}(\omega,h)). \]
    \end{enumerate}
    
    Let $\sigma_0 = \tau^i$ and $\eta^0 = z$. 
    By construction, $\sigma_0$ is an $\ms H^\vee$-optional time, and $\eta^0$ is right-continuous, decreasing and 
    ${\ms H}^\vee$-optional. Moreover, $\eta^0_\infty = \bm 0$ by assumption. The extra properties above are clearly satisfied for $n=0$.
    
    Now let $k\in\N$ and $\sigma_n$ and $\eta^n$ with the claimed properties be given for all $n = 0,\dots, k$. Let, for any $t\in\ovT$ and $(\omega,h)\in W$:
    \begin{gather*}
        \eta^{k+1}_t(\omega,h) = \begin{cases} \eta^k_t(\omega,h), &\text{if } (t,\omega,h)\in [\![0,\sigma_{k})\!), \\ s_{\sigma_{k}(\omega,h)}(\omega,\eta^k(\omega,h)), &\text{if } (t,\omega,h)\in[\![\sigma_{k},\infty)\!), \\ \bm 0, &\text{else,} \end{cases} \\
        \sigma_{k+1}(\omega,h) = \inf \{t \ge \sigma_k(\omega,h) \mid s_t(\omega,\eta^{k+1}(\omega,h)) \neq \eta^{k+1}_{t}(\omega,h)\}.
    \end{gather*}
    With the notation from Lemma~\ref{3-SPF_VECT.lemma:fsharp_preserves_optionality}, $\ovT\times W \to \B,\, (t,\omega,h)\mapsto s_t(\omega,\eta^k(\omega,h))$ equals the composition $s\circ (\eta^k)^\#$, and, by that lemma and the induction hypothesis, it is $\ms H^\vee$-progressively measurable. Therefore, using again the induction hypothesis, completeness, and Corollary~\ref{3-SPF_VECT.cor:xi_tau_is_F_tau_mb}, we obtain that $[s\circ(\eta^k)^\#]_{\sigma_{k}}$ is $\ms H^\vee_{\sigma_{k}}$-measurable and $\eta^{k+1}$ is a (simple) $\ms H^\vee$-optional process. Moreover, $\eta^{k+1}$ clearly has right-continuous and decreasing paths, because $s_t(\omega,h) \le h(t-)$ for all $(t,\omega,h)\in\ovT\times W$. Moreover, $\eta^{k+1}_\infty = \bm 0$ by definition. As $\pi\circ\sigma_k < \alpha$ and as $s$ has upper vertical level smaller than or equal to $\alpha$, $\eta^{k+1}$ has upper vertical level smaller than or equal to $\alpha$.

    Now we show that $\sigma_{k+1}$ is an $\ms H^\vee$-optional time with $\pi\circ\sigma_{k+1}<\alpha$. As $\eta^{k+1}$ is $\ms H^\vee$-optional, $s\circ (\eta^{k+1})^\#$ is $\ms H^\vee$-progressively measurable, by Lemma~\ref{3-SPF_VECT.lemma:fsharp_preserves_optionality}. 
    In particular, both $s\circ(\eta^{k+1})^\#$ and $\eta^{k+1}$ are $\ms H^\vee$-progressively measurable. $\eta^{k+1}$ is left- and right-constant and $s$ is left-constant at all $u\in\ovT$ with $\pi(u) = \mf w_1$. If we had $\pi \circ \sigma_{k+1}(\omega,h) = \mf w_1$ for some $(\omega,h)\in W$, then $\sigma_{k+1}(\omega,h) < \infty$ and $(s\circ(\eta^{k+1})^\#)_{\sigma_{k+1}}(\omega,h) = \eta^{k+1}_{\sigma_{k+1}}(\omega,h)$. Using local right-constanty of $\eta^{k+1}$ and the fact that the paths of $s$ are lower semicontinuous from the right, we see that there would be $i\in I$ such that, for any $t\in\ovT$ with $\sigma_{k+1}(\omega,h)<t$ and infinitely many $u\in (\sigma_{k+1}(\omega,h),t)_\ovT$, we have $(s^i\circ(\eta^{k+1})^\#)_{\sigma_{k+1}}(\omega,h) = 0$ and $(s^i\circ(\eta^{k+1})^\#)_{u}(\omega,h) = 1$. Then, choosing $t$ sufficiently small so that $\eta^{k+1}$ is constant on $[\sigma_{k+1}(\omega,h),t)_\ovT$, there would be such $u$ with
    \[ 1 = (s^i\circ(\eta^{k+1})^\#)_{u}(\omega,h) \le \eta^{k+1}_{u-}(\omega,h) = \eta^{k+1}_{\sigma_{k+1}}(\omega,h) = (s^i\circ(\eta^{k+1})^\#)_{\sigma_{k+1}}(\omega,h) = 0 \]
    --- a contradiction. Hence, $\pi\circ\sigma_{k+1} < \mf w_1$. Thus, by augmentedness of $\ms H^\vee$ and Theorem~\ref{3-SPF_VECT.thm:debut}, $\sigma_{k+1}$ is an $\ms H^\vee$-optional time. Moreover, as $s$ and $\eta^{k+1}$ have upper vertical level inferior or equal to $\alpha$, the preceding inequality even implies $\pi\circ\sigma_{k+1} < \alpha$.

    It remains to show the extra properties for $n=k+1$. Property~\ref{3-SPF_VECT.thm:timing_SPF_well-posed.construction.1} follows directly from the definitions of $\eta^{k+1}$ and $\sigma_{k+1}$; the fact that $\pi\circ\sigma_k < \mf w_1$ ($\sigma_k$ is an optional time by induction hypothesis); and the equality
    \[ (\dagger) \qquad s_{\sigma_{k}(\omega,h)}(\omega,\eta^k(\omega,h)) = s_{\sigma_{k}(\omega,h)}(\omega,\eta^{k+1}(\omega,h)), \qquad (\omega,h)\in W. \]
    We briefly prove $(\dagger)$. For this, fix $(\omega,h)\in W$ and let $t = \sigma_k(\omega,h)$, a deterministic optional time. By Axiom~\ref{3-SPF_VECT.def:spf.msMi-measurability} applied to $t$ and $\beta = \pi(t)$, for all $i\in I$, there is $\ms M^i$-measurable $\tilde s^i$ with $\tilde s^i_t = s^i_t$. By definition, we have $\eta^k(\omega,h)|_{[0,t)_\ovT} = \eta^{k+1}(\omega,h)|_{[0,t)_\ovT}$. Hence, by Proposition~\ref{3-SPF_VECT.prop:information_sets}, Part~\ref{3-SPF_VECT.prop:information_sets.f(t,omega,h)=f(t,omega,h')}, all $i\in I$ satisfy
    \[ s^i_{\sigma_{k}(\omega,h)}(\omega,\eta^k(\omega,h)) = \tilde s^i_{t}(\omega,\eta^k(\omega,h)) = \tilde s^i_{t}(\omega,\eta^{k+1}(\omega,h)) = s^i_{\sigma_{k}(\omega,h)}(\omega,\eta^{k+1}(\omega,h)), \]
    which proves $(\dagger)$.
    
    Regarding Property~\ref{3-SPF_VECT.thm:timing_SPF_well-posed.construction.2}, let $(\omega,h)\in W$ such that $\sigma_{k+1}(\omega,h) < \infty$. By construction, $\eta^{k+1}(\omega,h)$ is constant on $[\sigma_k(\omega,h),\infty)_\ovT$. By Property~\ref{3-SPF_VECT.thm:timing_SPF_well-posed.construction.1}, just shown above, $[\sigma_k(\omega,h),\sigma_{k+1}(\omega,h))_\ovT$ is non-empty. Hence,
    \[ s_{\sigma_{k+1}(\omega,h)}(\omega,\eta^{k+1}(\omega,h)) \le \eta^{k+1}_{\sigma_{k+1}(\omega,h)-}(\omega,h) = \eta^{k+1}_{\sigma_{k+1}(\omega,h)}(\omega,h). \]
    Note that
    \[ (\ast) \qquad s_{\sigma_{k+1}(\omega,h)}(\omega,\eta^{k+1}(\omega,h)) \neq \eta^{k+1}_{\sigma_{k+1}(\omega,h)}(\omega,h), \]
    because $\pi\circ\sigma_{k+1} < \mf w_1$ as shown above.  
    Thus, we get $s_{\sigma_{k+1}(\omega,h)}(\omega,\eta^{k+1}(\omega,h)) < \eta^{k+1}_{\sigma_{k+1}(\omega,h)}(\omega,h)$ as claimed.

    Regarding Property~\ref{3-SPF_VECT.thm:timing_SPF_well-posed.construction.3}, by construction $\eta^{k+1}$ is scenariowise constant on $[\![\sigma_{k},\sigma_{k+1})\!)$. Let $(t,w)\in[\![\sigma_{k},\sigma_{k+1})\!)$. If $k=0$, then $|\eta^{k+1}_t(w)| \le |I| = |I|-(k+1)+1$. If $k>0$, then, by the induction hypothesis on Property~\ref{3-SPF_VECT.thm:timing_SPF_well-posed.construction.2} (by the choice of $(t,w)$, $\sigma_k(w) < \infty$), and by monotonicity of $\eta^k$, we have 
    \[ |\eta^{k+1}_t(w)|_1 = |s_{\sigma_{k}(\omega,h)}(\omega,\eta^{k}(\omega,h))|_1 < |\eta^{k}_{\sigma_{k}(\omega,h)}(\omega,h)|_1 \le |\eta^{k}_{\sigma_{k-1}(\omega,h)}(\omega,h)|_1\le |I|-k+1. \]
    Hence, $|\eta^{k+1}_t(w)|_1 \le |I|-k+1-1 = |I|-(k+1)+1$.

    Property~\ref{3-SPF_VECT.thm:timing_SPF_well-posed.construction.4} holds true for $n=k+1$ by definition. By Property~\ref{3-SPF_VECT.thm:timing_SPF_well-posed.construction.4} and the induction hypothesis for Properties~\ref{3-SPF_VECT.thm:timing_SPF_well-posed.construction.1} and~\ref{3-SPF_VECT.thm:timing_SPF_well-posed.construction.5}, we infer Property~\ref{3-SPF_VECT.thm:timing_SPF_well-posed.construction.5} for $n=k+1$.

    Regarding Property~\ref{3-SPF_VECT.thm:timing_SPF_well-posed.construction.6}, first let $(t,\omega,h)\in [\![\sigma_0,\sigma_k)\!)$. Then, if $\pi(t) < \mf w_1$, using the induction hypothesis, Proposition~\ref{3-SPF_VECT.prop:information_sets} (Part~\ref{3-SPF_VECT.prop:information_sets.f(t,omega,h)=f(t,omega,h')}) combined with Axiom~\ref{3-SPF_VECT.def:spf.msMi-measurability} applied to the optional time $t$ and $\beta=\pi(t)$, and Property~\ref{3-SPF_VECT.thm:timing_SPF_well-posed.construction.4} for $n=k+1$ which we have proven above, we get
    \[ \eta^{k+1}_t(\omega,h) = \eta^k_t(\omega,h) = s_t(\omega,\eta^k(\omega,h)) = s_t(\omega,\eta^{k+1}(\omega,h)). \]
    If $\pi(t) = \mf w_1$, both $\eta^{k+1}(\omega,h)$ and $s(\omega,\eta^{k+1}(\omega,h))$ are left-constant at $t$ by progressive measurability of $\eta^{k+1}$ and $s$ (Remark~\ref{3-SPF_VECT.rmk:prog_mb_basic_properties}, Part~\ref{3-SPF_VECT.rmk:prog_mb_basic_properties.scwise_left-constant_at_(t,mfw1)}). Whence, using the fact that $\pi\circ\sigma_0 < \mf w_1$ and what has been shown just before, we infer $\eta^{k+1}_t(\omega,h) = s_t(\omega,\eta^{k+1}(\omega,h))$. 
    Second, let $(t,\omega,h)\in[\![\sigma_k,\sigma_{k+1})\!)$. Then, by definition of $\sigma_{k+1}$, we obtain
    \[ \eta^{k+1}_t(\omega,h) = s_t(\omega,\eta^{k+1}(\omega,h)). \]
    The construction is complete.
    
    Next, let $n_\ast = |I| + 1$ and $\eta = \eta^{n_\ast}$. We claim that:
    \begin{enumerate}[start=7]
        \item\label{3-SPF_VECT.thm:timing_SPF_well-posed.construction.final1} $\sigma_{n_\ast} = \infty$.
        \item\label{3-SPF_VECT.thm:timing_SPF_well-posed.construction.final2} $\eta|_{[\![0,\tau^i)\!)} = z|_{[\![0,\tau^i)\!)}$, and $s_t(\omega,\eta(\omega,h)) = \eta_t(\omega,h)$ for all $(t,\omega,h)$ with $\tau^i(\omega,h) \le t$.
    \end{enumerate}
    Suppose that there is $(\omega,h)\in W$ with $\sigma_{n_\ast}(\omega,h) < \infty$. Then, by Properties~\ref{3-SPF_VECT.thm:timing_SPF_well-posed.construction.1} and~\ref{3-SPF_VECT.thm:timing_SPF_well-posed.construction.3}, $\eta^{n_\ast}_{\sigma_{n_\ast}-} = \bm 0$. Property~\ref{3-SPF_VECT.thm:timing_SPF_well-posed.construction.2} then yields the contradiction $s_{\sigma_{n_\ast}(\omega,h)}(\omega,\sigma_{n_\ast}(\omega,h)) < \bm 0$. Hence, $\sigma_{n_\ast} = \infty$. The second claim now follows from the initial values of $\sigma_0$ and $\eta^0$, Properties~\ref{3-SPF_VECT.thm:timing_SPF_well-posed.construction.5} and~\ref{3-SPF_VECT.thm:timing_SPF_well-posed.construction.6}, as well as the convention that $s_\infty = \bm 0 = \eta_\infty$.

    Let $\chi = \eta\circ(\id_\Omega\star\tilde\chi)$. By Lemma~\ref{3-SPF_VECT.lemma:eta_circ(id_Omega_star_chi)_inherits_optionality}, $\chi$ is $\ms F^\vee$-optional. It is decreasing and right-continuous, and satisfies $\chi_\infty = 0$. $\eta = \eta^{n_\ast}$ has upper vertical level inferior or equal to $\alpha$, and hence the same is true of $\chi$. Hence, $(\chi,\chi)\in\mc W$. By construction of $\eta$, for all $\omega\in \Omega$, we have:
    \begin{gather*}
        \chi|_{[0,\tau^i(\omega,\tilde\chi(\omega)))_\ovT} = \tilde\chi|_{[0,\tau^i(\omega,\tilde\chi(\omega)))_\ovT}, \\
        s\llcorner \chi|_{[\tau^i(\omega,\tilde\chi(\omega))\infty]_\ovT} = \chi|_{[\tau^i(\omega,\tilde\chi(\omega))\infty]_\ovT},
    \end{gather*}
    where we used Property~\ref{3-SPF_VECT.thm:timing_SPF_well-posed.construction.final2}. We infer using Proposition~\ref{3-SPF_VECT.prop:information_sets} that $\chi\approx_{i,\tau^i} \tilde\chi$.

    Let $\chi'$ be another state process such that $\chi'\approx_{i,\tau^i} \tilde\chi$ satisfying $s\llcorner \chi'(t,\omega) = \chi'(t,\omega)$ for all $(t,\omega)\in [\![\tau^i\circ(\id_\Omega\star\chi(\omega),\infty]\!]$. As both $\chi'$ and $\chi$ are $\ms F^\vee$-optional with locally right-constant paths,
    \[ \sigma = \inf\{t\in\ovT \mid \chi'_t \neq \chi_t \} \]
    defines an $\ms F^\vee$-optional time with $\chi'_\sigma \neq \chi_\sigma$ on $\{\sigma < \infty\}$. Hence, for all $\omega\in\{\sigma<\infty\}$, we have
    \[ \chi'_\sigma = (s\llcorner \chi')_\sigma = (s\llcorner \chi)_\sigma = \chi_\sigma, \]
    again by a scenario- and componentwise application of \textsc{spf} Axiom~\ref{3-SPF_VECT.def:spf.msMi-measurability} combined with Proposition~\ref{3-SPF_VECT.prop:information_sets}, Part~\ref{3-SPF_VECT.prop:information_sets.f(t,omega,h)=f(t,omega,h')}.\footnote{Namely, for fixed $\omega\in\Omega$, $\sigma(\omega)$ is a deterministic optional time. For any $i\in I$, there is $\ms M^i$-measurable $\tilde s^i$ with $s^i_{\sigma(\omega)} = \tilde s^i_{\sigma(\omega)}$. Now, plug in both $(\omega,\chi(\omega))$ and $(\omega,\chi'(\omega))$ and apply Proposition~\ref{3-SPF_VECT.prop:information_sets}, Part~\ref{3-SPF_VECT.prop:information_sets.f(t,omega,h)=f(t,omega,h')}.} Thus, $\{\sigma < \infty\} = \emptyset$. As $\chi'_\infty = \bm 0 = \chi_\infty$, we infer that $\chi' = \chi$. The proof of well-posedness is complete.
\end{proof}

\begin{proof}[Proof of Theorem~\ref{3-SPF_VECT.thm:timing_game-Out(s)_is_tilting_limit}]
    Note that, being an action process in $\mathbf F$, every component of $\xi$ is an $\ms F^\vee$-optional decreasing process valued in $\{0,1\}$ taking the value zero in time $\infty$. Hence, any component is of the form $1[\![0,\sigma)\!)$ for some $\ms F^\vee$-decision time $\sigma$, by Theorem~\ref{3-SPF_VECT.thm:debut} and universal completeness of $(\Omega,\ms E,\ms F^\vee)$. Then, the theorem follows directly from Proposition~\ref{3-SPF_VECT.prop:optional_times_are_tilting_limits_of_classical_very_simple_optional_procs} in combination with Remark~\ref{3-SPF_VECT.rmk:basic_properties_tilting_convergence}.
\end{proof}

\begin{proof}[Proof of Theorem~\ref{3-SPF_VECT.thm:Riedel-Steg_equilibrium}]
    ($s$ is a strategy profile):~ Let $i\in I$. We start with proving all properties not concerning measurability. Regarding lower semicontinuity from the right, let $t\in\ovT$ be a right-limit point, i.e.\ $\pi(t) = \mf w_1$, and let $(\omega,h)\in W$ such that $s^i_t(\omega,h) = 1$. Then, $t<\tau(\omega)$. Hence, there is $u\in\R_+$ such that $t < u < \tau(\omega)$, so that $s^i(\omega,h)$ is constant with value $1$ on $[t,u)_\ovT$. Thus, $s^i$ is lower semicontinuous from the right. 
    Moreover, $s^i_t(\omega,h) = 0$ for all $(t,\omega,h)\in\ovT\times W$ with $\pi(t) \ge \mf w$. Hence, $s^i$ has upper vertical level inferior or equal to $\alpha = \mf w+1$. By construction, we have $s^i_t(\omega,h) \le h^i(t-)$ for all $(t,\omega,h)\in\ovT\times W$ and $s^i_\infty = 0$.

    It remains to verify $\ms H^i$-progressive measurability and ``local'' $\ms M^i$-measurability. We first show that the map
    \[ {f^i}\colon\ovT\times W\to\R,\, (t,\omega,h) \mapsto \begin{cases} \upsilon^{i,\pi(t)}(\omega), &\text{if }\pi(t) < \mf w, \\ -1, &\text{else,} \end{cases} \]
    is $\ms H^i$-progressively measurable. As ${f^i}$ is valued in $\{-1\}\cup[0,1]_{\R_+}$, it suffices to show that, for all $b\in\R$ with $b>-1$ and $u\in\ovT$, we have $\{{f^i} > b\} \in \ms M^i$. Let such $b$ and $u$ be given. Then,
    we have
    \begin{align*}
        &~\{{f^i} > b\}  \\
        =&~ \bigcup_{\beta\in\mf w} (\rho^{\mf w})^{-1}(\bRp\times\{\beta\}) \times \{\upsilon^{i,\beta} > b\} \times \B^\ovT \, \in\ms P_\ovT\otimes \ms F_0^i\otimes\{\emptyset,\B^\ovT\}.
    \end{align*}
    Now, we have 
    \[ \ms P_\ovT\otimes \ms F_0^i\otimes\{\emptyset,\B^\ovT\} \subseteq \Prg(\ms F^i \otimes \{\emptyset,\B^\ovT\}) \subseteq \Prg(\ms H^i) \]
    by Remark~\ref{3-SPF_VECT.rmk:prog_mb_basic_properties}, Part~\ref{3-SPF_VECT.rmk:prog_mb_basic_properties.examples_prog_mb_sets}.

    The map $\tilde\tau\colon W \to \ovT,\, (\omega,h) \mapsto \tau(\omega)$, defines an $\ms F^i\otimes \{\emptyset,\B^\ovT\}$-optional time because $\tau$ is an $\ms F^i$-optional time. In particular, $[\![0,\tilde\tau)\!) \in \ms M^i$. 
    
    Moreover, with $\B_{i,0} = \{\bm b = (b^i)_{i\in I}\in\B \mid b^i = 0\}$, we have
    \[ \tau_{i,0} = \inf\{t\in\ovT \mid z^i_{t-} = 0\} = \inf_{\bm b\in\B_{i,0}} \tau^-_{\bm b}. \]
    For any $\bm b\in\B_{i,0}$, $\tau^-_{\bm b}$ is an $\ms H^i$-optional time with $[\![0,\tau_{\bm b}^-)\!) \in \Prd(\ms H^i) \subseteq \ms M^i$, in view of Lemma~\ref{3-SPF_VECT.lemma:z_optional}. By Proposition~\ref{3-SPF_VECT.prop:basic_properties_of_optional_times}, Part~\ref{3-SPF_VECT.prop:basic_properties_of_optional_times.max_min}, infima of finitely many optional times are optional times. Thus, $\tau_{i,0}$ is an $\ms H^i$-optional time. Moreover,
    \[ [\![0,\tau_{i,0})\!) = \bigcap_{\bm b\in\B_{i,0}} [\![0,\tau^-_{\bm b})\!) \, \in\ms M^i. \]
    
    Let us extend $\eta^j$ to a process 
    \[ \tilde\eta^j\colon\ovT\times W \to (0,\infty)_{\R_+},\, (t,\omega,h) \mapsto \begin{cases} \eta^j_{p(t)}(\omega), &\text{if } t < \infty, \\ 1, &\text{else.} \end{cases} \] As $\eta^j$, seen as a classical process with time axis $\R_+$, is $\ms G^i$-progressively measurable in the classical sense, $\tilde\eta^j$ is $\ms F^i\otimes\{\emptyset,\B^\ovT\}$-optional and thus $\ms M^i$-measurable in the sense of the present text. 
    With this, we get the representation
    \[ s^i = 1[\![0,\tau_{i,0})\!) \cdot \Big(1[\![0,\tilde\tau)\!) + 1[\![\tilde\tau,\infty)\!)\cdot 1\Big\{{f^i}\ge\frac{\tilde\eta^j}{1+\tilde\eta^j}\Big\}
    \Big), \]
    from which we infer that $s^i$ is $\ms H^i$-progressively measurable.

    It remains to show that $s^i$ is ``locally'' $\ms M^i$-measurable, for any $i\in I$. For this let $i\in I$, $\beta\in\mf w_1$, and $\tau^i$ be an optional time for $i$. Let
    \[ \tilde s^i = 1[\![0,\tau_{i,0})\!) \cdot \Big(1[\![0,\tilde\tau)\!) + 1[\![\tilde\tau,\infty)\!)\cdot 1\Big\{\upsilon^{i,\beta} \ge \frac{\tilde\eta^j_{\tau^i}}{1+\tilde\eta^j_{\tau^i}},\,\beta\in\mf w\Big\} \Big). \]
    As $\upsilon^{i,\beta}$, seen as a process constant in time, is $\ms F^i\otimes\{\emptyset,\B^\ovT\}$-optional and thus $\ms M^i$-measurable, combining with the measurability properties of the other relevant processes in the definition of $\tilde s^i$, we infer that $\tilde s^i$ is $\ms M^i$-measurable. Moreover, we have $s^i_{\tau^i} = \tilde s^i_{\tau^i}$ on $\{\pi\circ\tau^i=\beta\}$, as wanted.
    \smallskip

    (Construction of $\Pr$):~ Let $\Pi = (\ms P^{i,\tau^i},\kappa^{i,\tau^i},p_{i,\mf p}, \ms P_{i,\mf p})_{\mf p=(\tau^i,\mf x)\in\mf P^i,\, i\in I}$ be given as follows: 
    \begin{itemize}
        \item for any $i\in I$ and any optional time $\tau^i$ for $i$, $\ms P^{i,\tau^i}$  
        is the coarsest (i.e.\ two-element) $\sigma$-algebra on $\mf P^i(\tau^i)$; 
        \item for any $i\in I$, any optional time $\tau^i$ for $i$, any $\mf p\in \mf P^i(\tau^i)$ and $E\in\ms E$, let $\kappa^{i,\tau^i}(E,\mf p) = \P(E)$;
        \item for any $i\in I$, any optional time $\tau^i$ for $i$, any $\mf p = (\tau^i,\mf x) \in\mf P^i(\tau^i)$, let $\ms P_{i,\mf p}$ be the coarsest (i.e.\ two-element) $\sigma$-algebra on $\mf x$; 
        \item for any $i\in I$, any optional time $\tau^i$ for $i$, any $\mf p = (\tau^i,\mf x) \in\mf P^i(\tau^i)$, and any $\omega\in\Omega$, let $p_{i,\mf p}(\omega) = \Out^\star(s \mid \tau^i,\tilde\chi)$ for $\tilde\chi\in\mf x$.
    \end{itemize}
    $\Pi$ clearly defines a belief system and $U = (u_{i,\mf p})_{i\in I,\, \mf p\in\mf P^i}$ with $u_{i,\mf p} = u_i$ obviously defines a taste system in the sense of Definition~\ref{3-SPF_VECT.def:equilibrium}. $\ms W$ is a $\sigma$-algebra on $W$.

    We show that $\Pr$ is an \textsc{eu} preference structure. For this, let $i\in I$, $\tau^i$ be an optional time for $i$ and $\mf p=(\tau^i,\mf x)\in\mf P^i(\tau^i)$ be a corresponding information set. Regarding Property~\ref{3-SPF_VECT.def:EU_pref_str.uip_mb}, the map $u_i$ is bounded and $\ms W$-$\ms B_\R$-measurable, because, first, the maps $W\ni (\omega,h) \mapsto \sigma_i(h)$ are measurable functions of the family $(\tau_{\bm b})_{\bm b\in\B}$. And, second, for any $(\omega,h)\in W$,  
    \[ a_i(h) = 1\{\prj_{\mf w+1} \circ \rho^{\mf w} \circ \sigma_i(h) < \mf w\}, \]
    where $\prj_{\mf w+1}\colon \bRp\times(\mf w+1)\to(\mf w+1)$ is the canonical projection. Hence, $W\ni(\omega,h)\mapsto a_i(h)$ is measurable as well. 
    Furthermore, for the proof of Property~\ref{3-SPF_VECT.def:EU_pref_str.Out_mb}, let $s'\in\mc S$ and denote the state processes induced by $s'$ and $s$ given the information set $\mf p$ --- or more precisely, given arbitrary $\tilde\chi\in\mf x$ and time $\tau^i$ ---  by $\chi'$ and $\chi$, respectively. This makes sense because information sets are small here and characterised by ``equality on $[\![0,\tau^i)\!)$'', see Proposition~\ref{3-SPF_VECT.prop:information_sets}. Then, $p_{i,\mf p}$ is constant with value $\chi$, by construction. Moreover, as $\chi\in\mf x$, we get, for all $\omega\in\Omega$, 
    \[ \Out^{s'}_{i,\mf p}(\omega) = (\omega,\chi'(\omega)) = (\id_\Omega\star\chi')(\omega). \]
    By Axiom~\ref{3-SPF_VECT.def:spf.msF^i_zeta_subseteq_msE}, $(\id_\Omega\star\chi')$ is $\ms E$-$\ms H^j_\infty$-measurable for all $j\in I$. By universal completeness and basic measure theory, then, it is also $\ms E$-$\ms H^\vee_\infty$-measurable. Property~\ref{3-SPF_VECT.def:EU_pref_str.phi_mb} is trivially satisfied because $p_{i,\mf p}$ is constant. The constancy of $p_{i,\mf p}$ also implies that Property~\ref{3-SPF_VECT.def:EU_pref_str.chi_i,mfp_state-proc} holds true. Hence, $\Pr$ is an expected utility preference structure.\smallskip
    
    (Dynamic consistency of $(\Pr,s)$):~ Property~\ref{3-SPF_VECT.def:EU_pref_str.consistency.ui=uip} in the definition of dynamic consistency is satisfied by construction. For the proof of Property~\ref{3-SPF_VECT.def:EU_pref_str.consistency.p}, let $i\in I$, $\tau^i,\sigma^i$ be optional times for $i$ with $\tau^i\le\sigma^i$, $\mf p=(\tau^i,\mf x)\in\mf P^i(\tau^i)$ an information set for $i$ with time $\tau^i$ and $\omega\in\Omega$. Using Proposition~\ref{3-SPF_VECT.prop:information_sets}, let $\chi$ denote the state process induced by $s$ given $\mf p$.\footnote{More precisely, by the proposition, the induced state process does not depend on the choice of $\tilde\chi\in\mf x$.} Then, by construction, $p_{i,\mf p}(\omega) = \chi$ and $\Out^\star(s \mid \tau^i,p_{i,\mf p}(\omega)) = \chi$. Moreover, $\p^s_{i,\mf p,\sigma^i}(\chi)$ equals the information set $\mf p' = (\sigma^i,\mf x')$ at time $\sigma^i$ with $\chi\in\mf x'$. Hence, by definition of $p_{i,\mf p'}$, $p_{i,\p^s_{i,\mf p,\sigma^i}(p_{i,\mf p}(\omega))}(\omega)$ equals the state process induced by $s$ given $\mf p'$ alias $\chi$ at time $\sigma^i$. It follows, then, directly from the definition of ``state process induced by $s$'' that $p_{i,\p^s_{i,\mf p,\sigma^i}(p_{i,\mf p}(\omega))}(\omega)=\chi$. Hence, Property~\ref{3-SPF_VECT.def:EU_pref_str.consistency.p} is satisfied as well. For the proof of Property~\ref{3-SPF_VECT.def:EU_pref_str.consistency.P}, let $i\in I$, $\tau^i,\sigma^i$ be optional times for $i$ with $\tau^i\le\sigma^i$, $\mf p = (\tau^i,\mf x)\in\mf P^i(\tau^i)$ and $E\in\ms E$. Note that $p_{i,\mf p}$ is constant, whence $\kappa^{i,\sigma^i}(E,\p^s_{i,\mf p,\sigma^i}\circ p_{i,\mf p}) = \P(E) = \P(E \mid \p^s_{i,\mf p,\sigma^i}\circ p_{i,\mf p})$. 
    We conclude that $(\Pr,s)$ is dynamically consistent.\smallskip

    (Ad dynamic rationality):~ Let $i,j\in I$ with $i\neq j$, $\mf p = (\tau^i,\mf x)\in\mf P^i$ such that there is $\ms F^i$-progressively measurable $\chi\in\mf x$, and $\tilde s\in \mc S$ such that $\tilde s^j = s^j$. Without loss of generality, we can assume that $\chi$ has no jumps on $[\![\tau^i\circ(\id_\Omega\star\chi)]\!]$, that is, for any $\omega\in\Omega$, $\chi$ is constant on $[\tau^i(\omega,\chi(\omega)),\infty]_\ovT$ and, if $\tau^i(\omega,\chi(\omega)) > 0$, there is $u\in [0,\tau^i(\omega,\chi(\omega)))_\ovT$ such that $\chi$ is even constant on $[u,\infty]_\ovT$.
    
    Let $\tilde\chi$ denote the state process induced by $\tilde s$ given $(\tau^i,\chi_{i,\mf p})$. Recall that $\chi_{i,\mf p}$, by construction, is the state process induced by $s$ given $(\tau^i,\chi_{i,\mf p})$, that is $\chi_{i,\mf p} = \Out^s_{i,\mf p}$. By Proposition~\ref{3-SPF_VECT.prop:information_sets}, $\chi$, $\chi_{i,\mf p}$, and $\tilde\chi$ cannot be distinguished from another until $\tau^i$. In particular, we have, for all $\omega\in\Omega$,
    \[ \tau^i(\omega,\chi(\omega)) = \tau^i(\omega,\chi_{i,\mf p}(\omega)) = \tau^i(\omega,\tilde\chi(\omega)). \]
    Let $\hat\tau^i = \tau^i(\omega,\chi(\omega))$, an $\ms F^i$-optional time. Indeed, this can be shown as in Lemma~\ref{3-SPF_VECT.lemma:eta_circ(id_Omega_star_chi)_inherits_optionality}, using the additional assumption that $\chi$ is $\ms F^i$-progressively measurable and thus even $\ms F^i$-optional, because it is componentwise decreasing, locally right-constant, and $(\Omega,\ms E,\ms F^i)$ is universally complete so that Theorem~\ref{3-SPF_VECT.thm:debut} can be applied. Furthermore, let $\tilde\sigma_k = \sigma_k\circ\tilde\chi$ and $\hat\sigma_k = \sigma_k\circ\chi_{i,\mf p}$ for both $k=1,2$, which are $\ms F^\vee$-optional times.

    First note that, as $\chi_{i,\mf p}$ and $\tilde\chi$ coincide on $[\![0,\hat\tau^i)\!)$,
    \begin{align*}
        \E[u_i\circ\Out^{\tilde s}_{i,\mf p}\, 1\{\tilde\sigma_i < \hat\tau^i\} \mid \ms F^\ast_{i,\mf p}] = \E[u_i\circ\Out^s_{i,\mf p}\, 1\{\hat\sigma_i < \hat\tau^i\} \mid \ms F^\ast_{i,\mf p}].
    \end{align*}
    Similarly, we get
    \begin{align*}
        \E[u_i\circ\Out^{\tilde s}_{i,\mf p}\, 1\{\tilde\sigma_j<\hat\tau^i\} \mid \ms F^\ast_{i,\mf p}] = \E[u_i\circ\Out^{s}_{i,\mf p}\, 1\{\hat\sigma_j<\hat\tau^i\} \mid \ms F^\ast_{i,\mf p}]
    \end{align*}
    Furthermore, we have
    \begin{align*}
         \E[u_i\circ\Out^{\tilde s}_{i,\mf p}\, 1\{\hat\tau^i\le\tilde\sigma_i\wedge\tilde\sigma_j<\tau\} \mid \ms F^\ast_{i,\mf p}] \le 0,
    \end{align*}
    with equality if $\tilde s = s$, because $s$ avoids stopping before $\tau$.

    It remains to consider the most central event $E =\{\hat\tau^i\vee\tau\le\tilde\sigma_i\wedge\tilde\sigma_j\}$. Note that $E$ can be decomposed into the three cases: a) $\hat\tau^i\vee\tau\le\tilde\sigma_i < \tilde\sigma_j$, b) $\hat\tau^i\vee\tau\le\tilde\sigma_i = \tilde\sigma_j$, c) $\hat\tau^i\vee\tau\le\tilde\sigma_j < \tilde\sigma_i$. By definition of $s^j$, we have $p\circ\hat\tau^i = p\circ\tilde\sigma_j$ and $\pi\circ\tilde\sigma_j<\mf w$ $\P$-almost surely on $E\cap \{\pi\circ\hat\tau^i < \mf w\}$. Now, for any optional time $\sigma$ and $\alpha\in\mf w$, let $\sigma\oplus\alpha$ denote the optional time given by $p\circ(\sigma\oplus\alpha) = p\circ\sigma$ and, on $\{\sigma < \infty\}$, $\pi\circ(\sigma\oplus\alpha) = \pi\circ\sigma + \alpha$.\footnote{We have already used this construction several times in case $\alpha=1$, for instance, in the proof of Lemma~\ref{3-SPF_VECT.lemma:Prd_subseteq_Opt}. The general construction obtains by iterating this process finitely many times.} Also, let use recall that we extend $\eta^k$ to a map $\ovT\times\Omega\to\R$, via the requirement $\eta^k = \eta^k \circ (p\times\id_\Omega)$, an $\ms I_\ovT(\T)\otimes\ms F^i_0$-measurable map. 
    
    Then, we get --- with conditional probability denoting conditional expectation of the corresponding indicator ---:
    \begin{align*}
        &~\E\big[u_i\circ\Out^s_{i,\mf p}\, 1_E \mid \ms F^\ast_{i,\mf p}\big]\\
        =&~\E\big[\eta^i_{\tilde\sigma_i}\, 1\{\hat\tau^i\vee\tau\le\tilde\sigma_i < \tilde\sigma_j,\, \pi\circ\hat\tau^i < \mf w\} - 1\{\hat\tau^i\vee\tau\le\tilde\sigma_i = \tilde\sigma_j,\,\pi\circ\hat\tau^i < \mf w\} \mid \ms F^\ast_{i,\mf p}\big] \\
        =&~ \sum_{\beta,\gamma\in\mf w\colon \beta<\gamma} \eta^i_{\hat\tau^i}\, \P\big(\tilde\sigma_i=\hat\tau^i\oplus\beta,\, \tilde\sigma_j = \hat\tau^i\oplus\gamma,\, \tau\le\hat\tau^i\oplus\beta,\, \pi\circ\hat\tau^i < \mf w \mid \ms F^\ast_{i,\mf p}\big) \\
        &~\quad - \sum_{\beta\in\mf w} \P\big(\tilde\sigma_i = \hat\tau^i\oplus\beta = \tilde\sigma_j \ge \tau, \, \pi\circ\hat\tau^i < \mf w \mid \ms F^\ast_{i,\mf p}\big).
    \end{align*}
    We have used that $\tilde\eta^i$ is $\ms M^i$-measurable, whence the $\ms F^\ast_{i,\mf p}$-measurability of $\eta^i_{\hat\tau^i}$.

    Now, we have, for $\beta,\gamma\in\mf w$ with $\beta < \gamma$:
    \begin{align*}
        &~ \big\{ \tilde\sigma_i=\hat\tau^i\oplus\beta,\, \tilde\sigma_j = \hat\tau^i\oplus\gamma,\, \tau\le\hat\tau^i\oplus\beta,\, \pi\circ\hat\tau^i < \mf w \big\} \\
        =&~ \Big\{\forall\delta\in[0,\beta)_{\mf w}\colon \tilde s^i_{\hat\tau^i\oplus\delta} \circ (\id_\Omega\star\chi)=1, \, \tilde s^i_{\hat\tau^i\oplus\beta} \circ (\id_\Omega\star\chi)=0, \\
        &\qquad\tau\le\hat\tau^i\oplus\beta,\, \pi\circ\hat\tau^i<\mf w, \\
        &\qquad \forall \delta\in [0,\gamma)_{\mf w}\colon f^j_{\hat\tau^i+\delta} \ge \frac{\eta_{\hat\tau^i}^i}{1+\eta_{\hat\tau^i}^i},\, f^j_{\hat\tau^i+\gamma} < \frac{\eta_{\hat\tau^i}^i}{1+\eta_{\hat\tau^i}^i} \Big\}. 
    \end{align*}
    We have crucially used that $\tilde\sigma_i=\hat\tau^i\oplus\beta$ and $\tilde\sigma_j = \hat\tau^i\oplus\gamma$ imply that $\chi$ and $\tilde\chi$ coincide on $[\![0,\hat\tau^i\oplus\beta)\!)$. Now, as $\chi$ is $\ms F^i$-optional, $\tilde s^i$ is $\ms H^i$-progressively measurable, $\tau$ is an $\ms F^i$-optional time, the first two lines of the set on the right-hand side of the equation are $\ms F^i_\infty$-measurable conditions. The third line on the right-hand side, however, can be written as
    \[ \Big\{\omega\in\Omega \mid h\big(\hat\tau^i,\frac{\eta_{\hat\tau^i}^i}{1+\eta_{\hat\tau^i}^i},\upsilon^j\big) = 0\Big\}, \]
    for some $\ms P_\ovT\otimes\ms B_\R\otimes(\ms B_\R|_{[0,1]_\R})$-measurable map $h$. As $\ms F^\ast_{i,\mf p} \subseteq\ms F^i_\infty$, and as the family $(\upsilon^{j,\ell})_{\ell\in\N}$ is $\P$-independent and $\P$-independent from $\ms F^i$, an application of the power rule and the basic behaviour of conditional expectation on functions with independent arguments, we get that
    \begin{align*}
        &~\P\big(\tilde\sigma_i=\hat\tau^i\oplus\beta,\, \tilde\sigma_j = \hat\tau^i\oplus\gamma,\, \tau\le\hat\tau^i\oplus\beta,\, \pi\circ\hat\tau^i < \mf w \mid \ms F^\ast_{i,\mf p}\big)  \\
        =&~ \lambda^\ast_{i,\mf p} \cdot \frac{\tilde \eta^i_{\tau^i}}{(1+\tilde \eta^i_{\tau^i})^{\gamma+1}}.
    \end{align*}
    where
    \[ 
        \lambda^\ast_{i,\mf p} = \P\big(\forall\delta\in[0,\beta)_{\mf w}\colon \tilde s^i_{\hat\tau^i\oplus\delta} \circ (\id_\Omega\star\chi)=1, \, \tilde s^i_{\hat\tau^i\oplus\beta} \circ (\id_\Omega\star\chi)=0, 
        \tau\le\hat\tau^i\oplus\beta,\, \pi\circ\hat\tau^i<\mf w \mid \ms F^\ast_{i,\mf p}\big). 
    \]
    Using a completely analogous argument, we also obtain 
    \begin{align*}
        &~\P\big(\tilde\sigma_i = \hat\tau^i\oplus\beta = \tilde\sigma_j \ge \tau, \, \pi\circ\hat\tau^i < \mf w \mid \ms F^\ast_{i,\mf p}\big) \\
        =&~ \lambda^\ast_{i,\mf p}\cdot \frac{\tilde \eta^i_{\tau^i}}{(1+\tilde \eta^i_{\tau^i})^{\beta+1}}
    \end{align*}

    Hence, we conclude that:\footnote{These computations generalise and are related to the well-known verifications for the classical discrete-time grab-the-dollar game as well as the stochastic variant in \cite{Riedel2017Subgame}.}
    \begin{align*}
        &~\E\big[u_i\circ\Out^s_{i,\mf p}\, 1_E \mid \ms F^\ast_{i,\mf p}\big]\\ 
        =&~  \sum_{\beta\in\mf w}  \lambda^\ast_{i,\mf p}\, \Big(\eta^i_{\tau^i} \sum_{\gamma=\beta+1}^\infty \frac{\tilde \eta^i_{\tau^i}}{(1+\tilde \eta^i_{\tau^i})^{\gamma+1}} - \frac{\tilde \eta^i_{\tau^i}}{(1+\tilde \eta^i_{\tau^i})^{\beta+1}} \Big) \\
        =&~ \sum_{\beta\in\mf w}  \lambda^\ast_{i,\mf p}\,  (1-1) \frac{\eta^i_{\tau^i}}{(1+\eta^i_{\tau^i})^{\beta+1}} \\
        =&~ 0.        
    \end{align*}

    Putting all pieces together, we infer that
    \[ \pi_{i,\mf p}(\tilde s) \le \pi_{i,\mf p}(s), \]
    with equality if $\tilde s = s$. This completes the proof of dynamic rationality on $\tilde{\mf P}$. We conclude that $(s,\Pr)$ is in equilibrium on $\tilde{\mf P}$.
\end{proof}

\subsubsection{Stochastic differential games and control}

\begin{proof}[Proof of Proposition~\ref{3-SPF_VECT.prop:SDG_well-posed_SPF}]
    Axioms~\ref{3-SPF_VECT.def:spf.msH_non_anticipative} and~\ref{3-SPF_VECT.def:spf.msMi-measurability} are satisfied by assumption. Axiom~\ref{3-SPF_VECT.def:spf.msF^i_zeta_subseteq_msE} is satisfied automatically, because $\chi$ is a stochastic process by assumption and $\ms H^i_\infty \subseteq [\ms E\otimes (\ms B_{\B})^{\otimes\ovT}|_W]^{\mathrm u}$ for all $i\in I$. Axiom~\ref{3-SPF_VECT.def:spf.nu_Obs(xi)} is also satisfied, as proved next.
    
    Let $(\xi,\chi)$ and $(\xi',\chi')$ be elements of $\mc W$, $i\in I$, $\tau^i$ be an $\ms H^i$-optional time such that $[\![0,\tau^i)\!)\in\ms M^i$ and $\hat\tau^i = \tau^i \circ (\id\star\chi)$ be the induced optional time on $\Omega$. Suppose that $\xi|_{[\![0,\hat\tau^i]\!]} \cong_\P \xi'|_{[\![0,\hat\tau^i]\!]}$. $\chi$ (resp.\ $\chi'$) is the up to $\P$-indistinguishability, unique solution to System~\ref{3-SPF_VECT.eq:stochastic_differential_games.1} associated to $\xi$ (resp.\ $\xi'$) and the initial data $(0,\hat\chi^0)$, by the third axiom defining $\mc W$. 
    Then, by the fourth axiom, 
    there is an, up to $\P$-indistinguishability unique, solution $\tilde\chi$ to System~\ref{3-SPF_VECT.eq:stochastic_differential_games.1} for $\xi$ with initial data $(\hat\tau^i,\chi'_{[\![0,\hat\tau^i]\!]})$ satisfying $(\xi,\tilde\chi)\in\mc W$. As $\chi'_0 = \hat \chi^0$ by definition, and $\xi|_{[\![0,\hat\tau^i]\!]} \cong_\P \xi'|_{[\![0,\hat\tau^i]\!]}$ by hypothesis, \hyperlink{3-SPF_VECT.Ass:SDG}{Assumption SDG}, applied to $(\xi,\tilde\chi)$ and $(\xi',\chi')$, implies that $\tilde\chi$ then also solves System~\ref{3-SPF_VECT.eq:stochastic_differential_games.1} for $\xi$ with initial data $(0,\hat\chi^0)$. Thus, by the third axiom, $\chi$ and $\tilde\chi$ are $\P$-indistinguishable. Whence, $\chi|_{[\![0,\hat\tau^i]\!]} \cong_\P \tilde \chi|_{[\![0,\hat\tau^i]\!]} \cong_\P\chi'|_{[\![0,\hat\tau^i]\!]}$.

    The claim on well-posedness follows directly from the construction of $\mc S$.
\end{proof}

\chapter{Supplementary material}\label{chap:App2-Additional_mat}
\section{The smallest completion of posets}\label{3-SPF_VECT.sec:Dedekind-MacNeille_completion}
In this note, we characterise the Dedekind-MacNeille completion as the smallest completion of a poset using elementary methods. Although the results shown in the following are slightly stronger than those the author could find in the literature (e.g.\ in \cite{Davey2002Introduction}), it is to be assumed that the results, tools, and proofs that follow in this subsection are indeed classical and well-known, and not new. In any case, the aim of the present development is to provide a focused and self-contained overview on the mentioned characterisation as simply and clearly as possible.\smallskip

\subsection{Basic definitions and notation}

A \emph{partially ordered set}, in short \emph{poset}, is a pair $(P,\le)$ consisting of a set $P$ and a \emph{partial order} $\le$ on $P$, that is, a binary relation that is reflexive, antisymmetric, and transitive. For pragmatic reasons, we write $P$ instead of $(P,\le)$ and do not indicate the dependence of $\le$ on $P$, unless strictly necessary. For any poset $P$ with partial order $\le$, there is a dual poset on $P$ whose partial order is given by $x\ge y$ iff $x\le y$, for all $x,y\in P$. Moreover, a binary relation $<$ is induced by letting $x<y$ iff $x\le y$ and $x\neq y$, for all $x,y\in P$. The corresponding dual relation is denoted by $>$. A \emph{totally ordered set}, or \emph{chain}, is a poset $T$ such that for all $x,y\in T$, $x\le y$ or $y\le x$. Given two posets $P$ and $Q$, a set-theoretic map $f\colon P \to Q$ is said \emph{monotone} iff for all $x,y\in P$ with $x\le y$, we have $f(x) \le f(y)$. An \emph{embedding} between two posets $P$ and $Q$, or of $P$ into $Q$, is a set-theoretic map $f\colon P \to Q$ such that for all $x,y\in P$, $x\le y$ iff $f(x) \le f(y)$.

Posets define the objects and monotone set-theoretic maps between posets the morphisms of a category, denoted by $\Pos$. It is easily checked that $\Pos$-isomorphisms are exactly the surjective embeddings. For a fixed poset $P$, there is a further category, denoted by $\PPos$: its objects are given by the $\Pos$-morphisms with domain $P$ and the morphisms between two objects $\p\colon P \to Q$ and $\psi\colon P \to R$ are given by all $\Pos$-morphisms $f\colon Q \to R$ such that $\psi = f \circ \p$. By definition, a \emph{$\PPos$-embedding} is a $\PPos$-morphism that is an embedding. It is easily checked that $\PPos$-isomorphisms are exactly the $\PPos$-morphisms that are $\Pos$-isomorphisms, or equivalently, the surjective $\PPos$-embeddings.

For any poset $P$ and any subset $A\subseteq P$, the sets of lower and upper bounds are defined by
\[ A^\ell = \{x\in P \mid \forall a\in A \colon x \le a \}, \qquad A^u = \{x\in P \mid \forall a\in A \colon x\ge a\} . \]
In case $A = \{a\}$ is a singleton, then we write $\downarrow a = A^\ell$, the \emph{principal down-set of $a$}, and $\uparrow a = A^u$, the \emph{principal up-set of $a$}. $A$ is \emph{downward closed} iff for all $x\in A$, $\downarrow x \subseteq A$. $A$ is \emph{upward closed} iff for all $x\in A$, $\uparrow x \subseteq A$. For $n\in\N$ and symbols $i_1,\dots,i_n = \ell,u$, we write $A^{i_n i_{n-1} \dots i_2 i_1} = ((\dots((A^{i_n})^{i_{n-1}})\dots)^{i_2})^{i_1}$. An \emph{infimum} of $A$ is an element of $A^\ell \cap A^{\ell u}$. If it exists, it is unique and denoted by $\inf A$. Conversely, a \emph{supremum} of $A$ is an element of $A^u \cap A^{u\ell}$. If it exists, it is unique and denoted by $\sup A$. For $\Pos$-isomorphisms $f$, $f$ or $\mc P f$, respectively, commutes with both $.^u$ and $.^\ell$, and in particular with the existence and, if applicable, the values of suprema and infima. Note also that when passing from the partial order to its dual $\ge$, lower bounds, principal down-sets, downward closed sets, and infima become upper bounds, principal up-sets, upward closed sets, and suprema, and vice versa.

We collect some basic observations whose proofs are direct and therefore omitted (see \cite[Section 7.37 and Lemma 7.39]{Davey2002Introduction} for a discussion of the first four points from that the others follow). If $P$ is a poset and $A,B\subseteq P$ are subsets, then:
\begin{enumerate}
    \item\label{3-SPF_VECT.obs:A_subset_Aul} $A\subseteq A^{u\ell}$ and $B\subseteq B^{\ell u}$;
    \item\label{3-SPF_VECT.obs:ul_antitone} if $A\subseteq B$, then $A^u\supseteq B^u$ and $A^\ell\supseteq B^\ell$;
    \item\label{3-SPF_VECT.obs:Au=Aulu} $A^u = A^{u\ell u}$ and $B^\ell = B^{\ell u \ell}$;
    \item\label{3-SPF_VECT.obs:downarrowu=uparrow} for all $x\in P$, $(\downarrow x)^u = \uparrow x$ and $(\uparrow x)^\ell = \downarrow x$;
    \item\label{3-SPF_VECT.obs:supA=infAu} $A$ admits a supremum iff $A^u$ admits an infimum, and in that case $\sup A = \inf A^u$;
    \item\label{3-SPF_VECT.obs:infB=supBl} $B$ admits an infimum iff $B^\ell$ admits a supremum, and in that case $\inf B = \sup B^\ell$;
    \item\label{3-SPF_VECT.obs:sup_monotone} if $A,B$ admit suprema and $A\subseteq B$, then $\sup A \le \sup B$;
    \item\label{3-SPF_VECT.obs:inf_antitone} if $A,B$ admit infima and $A\subseteq B$, then $\inf A \ge \inf B$.
\end{enumerate}

A \emph{lattice} is a poset $L$ such that for all $x,y\in L$, $\{x,y\}$ admits both an infimum and a supremum. This terminology is reasonable because such $L$ defines an algebraic lattice, whose meet and join operations $\wedge$ and $\vee$ are given by $\inf$ and $\sup$, and conversely, any algebraic lattice $(L,\wedge,\vee)$ naturally defines a poset by letting $x\le y$ iff $x\wedge y = x$, for all $x,y\in L$, and both operations are essentially inverse to each other. A lattice $L$ is \emph{complete} iff any subset of $L$ admits both an infimum and a supremum. A subset $D$ of $L$ is said \emph{meet-dense} iff any $x\in L$ admits a subset $S\subseteq D$ with $x = \sup S$. A subset $D$ of $L$ is said \emph{join-dense} iff any $x\in L$ admits a subset $S\subseteq D$ with $x = \inf S$, or equivalently, iff $D$ is meet-dense with respect to the dual order.

\subsection{Dense completions}
A \emph{completion} of a poset $P$ is a pair $(\p,L)$, given by a complete lattice $L$ and an embedding $\p\colon P \inj L$ of $P$ into $L$. However, by slight abuse, we may refer to $\p$ or $L$ as the completion if the other datum is clear from the context. Note that any completion of a poset $P$ is an object of $\PPos$. Many possible completions exist and are of interest in the literature. Here, we are interested in the following notions, for which we introduce names.
\begin{definition}
    Let $P$ be a poset and $\p\colon P \inj L$ be a completion. We call $\p$ \emph{dense} iff $\im\p$ is both join- and meet-dense in $L$. We call $\p$ \emph{small} iff for any completion $\psi\colon P \inj M$, there is a $\PPos$-embedding $(\p,L)\inj (\psi,M)$.
\end{definition}

These notions are invariant under isomorphisms in $\PPos$.

\begin{lemma}\label{3-SPF_VECT.lemma:isomorphisms_dense_small}
    Let $P$ be a poset and $\p\colon P\inj L$, $\psi\colon P\inj M$ be two $\PPos$-isomorphic completions. Then, $(\p,L)$ is dense (small) iff $(\psi,M)$ is dense (small, respectively). 
\end{lemma}

\begin{proof}
    Let $f\colon (\p,L)\to (\psi,M)$ be a $\PPos$-isomorphism and $g$ its inverse. For symmetry reasons, it suffices to show that if $(\psi,M)$ has the relevant property, then $(\p,L)$ does so, too.\smallskip
    
    (Ad ``dense''):~ Let $(\psi,M)$ be dense and $x\in L$. Then, there are $A,B\subseteq P$ such that
    \[ \sup\mc P\psi(A) = f(x) = \inf\mc P\psi(B). \]
    As $g$ is a $\Pos$-isomorphism with $\p = g\circ \psi$ and $\mc P$ a functor, we infer that
    \begin{align*} 
        &\sup\mc P\p(A) = \sup\mc P(g\circ\psi)(A) = g(\sup\mc P\psi(A)) \\
        =~& x = g(\inf\mc P\psi(B)) = \inf \mc P(g\circ\psi)(B)= \inf \mc P\p(B). 
    \end{align*}
    Hence, $(\p,L)$ is dense.\smallskip

    (Ad ``small''):~ Let $(\psi,M)$ be small and $\rho\colon P \inj N$ be a completion of $P$. Then, there is a $\PPos$-embedding $h\colon(\psi,M)\inj(\rho,N)$. Then, $h\circ f$ is a $\PPos$-embedding $(\p,L)\inj(\rho,N)$. Hence, $(\p,L)$ is small.
\end{proof}

In \cite[Theorem~7.4.1]{Davey2002Introduction}, it is shown that for any poset $P$ there is an up to $\PPos$-isomorphism unique dense completion, and a representative of it is constructed explicitly, namely the Dedekind-MacNeille completion.
In this note, we show the following stronger result: a completion is dense iff it is small, for any poset $P$ there is an up to unique $\PPos$-isomorphism unique small completion,  and it can be represented by the Dedekind-MacNeille completion. Moreover, our method of proof is more elementary.\smallskip

The building block of the proof will be the following lemma. Given a (dense) completion $\p\colon P \inj L$ of a poset $P$, we try to find a representative system for the expression of elements $x\in L$ as suprema and infima over sets in $\im\p$.

\begin{lemma}\label{3-SPF_VECT.lemma:x=sup=inf_std_representatives}
    Let $P$ be a poset, $\p\colon P \inj L$ be a completion of $P$, $x\in L$, and $A,B\subseteq P$ such that
    \[ \sup \mc P\p(A) = x = \inf \mc P\p(B). \]
    Then,
    \begin{equation}\label{3-SPF_VECT.lemma:x=sup=inf_std_representatives:representatives}
        A^{u\ell} = \{y\in P \mid \p(y) \le x\}, \qquad B^{\ell u} = \{y\in P \mid \p(y) \ge x\}, 
     \end{equation}
    and 
    \begin{equation}\label{3-SPF_VECT.lemma:x=sup=inf_std_representatives:representation}
        \sup \mc P\p(A^{u\ell}) = x = \inf\mc P\p(B^{\ell u}).    
    \end{equation}
\end{lemma}

\begin{proof}
    In view of duality, it suffices to show the left-hand equalities in Equations~\ref{3-SPF_VECT.lemma:x=sup=inf_std_representatives:representatives} and~\ref{3-SPF_VECT.lemma:x=sup=inf_std_representatives:representation}. We start with a preliminary observation, namely that 
    \[ (\ast) \qquad \mc P\p(B^\ell) \subseteq [\mc P\p(B)]^\ell, \qquad \mc P\p(A^u) \subseteq [\mc P\p(A)]^u. \]
    Indeed, if $c\in P$ satisfies with $c\le b$ for all $b\in B$, then $\p(c) \le \p(b)$ for all $b\in B$, which shows the left-hand statement. The right-hand one follows from the left-hand one by duality.
    \smallskip

    (Ad Equation~\ref{3-SPF_VECT.lemma:x=sup=inf_std_representatives:representation}):~ Let $a\in A$ and $b\in B$. Then, $\p(a) \le x \le \p(b)$, whence $a\le b$. Thus, $B\subseteq A^u$, and, by Observation~\ref{3-SPF_VECT.obs:ul_antitone}, 
    \[ (\dagger) \qquad A^{u\ell}\subseteq B^\ell. \]
    Hence, using $(\ast)$ and Observation~\ref{3-SPF_VECT.obs:A_subset_Aul}, we get
    \[ \mc P\p(A) \subseteq \mc P\p(A^{u\ell}) \subseteq \mc P\p(B^\ell) \subseteq [\mc P\p(B)]^\ell. \]
    By Observations~\ref{3-SPF_VECT.obs:infB=supBl} and~\ref{3-SPF_VECT.obs:sup_monotone}, this implies
    \[ x = \sup\mc P\p(A) \le \sup\mc P\p(A^{u\ell}) \le \sup[\mc P\p(B)]^\ell = \inf\mc P\p(B) = x, \]
    whence $x = \sup\mc P(A^{u\ell})$.\smallskip

    (Ad Equation~\ref{3-SPF_VECT.lemma:x=sup=inf_std_representatives:representatives}):~ Regarding ``$\supseteq$'', let $y\in P$ be such that $\p(y) \le x$. 
    Let $z\in A^u$. 
    Then, by Observations $(\ast)$, \ref{3-SPF_VECT.obs:supA=infAu}, and~\ref{3-SPF_VECT.obs:inf_antitone}
    \[ x = \sup \mc P\p(A) = \inf [\mc P\p(A)]^u \le \inf \mc P\p(A^u) \le \p(z). \]
    Hence, $\p(y) \le \p(z)$. As $\p$ is an embedding, $y\le z$. We have shown that $y\in A^{u\ell}$.

    Regarding ``$\subseteq$'', let $y\in A^{u\ell}$. By $(\dagger)$, $y\in B^\ell$. In view of $(\ast)$, thus, $\p(y) \in [\mc P\p(B)]^\ell$, and, by Observation~\ref{3-SPF_VECT.obs:infB=supBl},
    \[ \p(y) \le \sup [\mc P\p(B)]^\ell = \inf \mc P\p(B) = x. \]
\end{proof}

As a consequence, if $\p$ is a dense completion, then for any $x\in L$ there are $A,B\subseteq P$ with $A = A^{u\ell}$ and $B = B^{\ell u}$ -- namely the sets on the right-hand sides of the equations in Equation~\ref{3-SPF_VECT.lemma:x=sup=inf_std_representatives:representatives} -- such that Equation~\ref{3-SPF_VECT.lemma:x=sup=inf_std_representatives:representation} holds true. Together with Observation~\ref{3-SPF_VECT.obs:downarrowu=uparrow}, this motivates the following, classical construction. For any poset $P$, let
\[ \DM(P) = \{A\subseteq P \mid A^{u\ell} = A\}, \quad \p_\DM\colon P \to \DM(P),\, x \mapsto \downarrow x, \]
and equip $\DM(P)$ with the partial order ``$\subseteq$''. This is well-defined by Observation~\ref{3-SPF_VECT.obs:downarrowu=uparrow}.
\begin{thm}\label{3-SPF_VECT.thm:DM_is_dense_completion}
    Let $P$ be a poset. Then, $\p_\DM\colon P \to \DM(P)$ is a dense completion of $P$.
\end{thm}

This classical theorem can be found, for instance, in \cite[Theorems~7.40, 7.41]{Davey2002Introduction}. $\p_\DM$ is called the \emph{Dedekind-MacNeille completion of $P$}. It generalises the construction of the extended real line out of the rationals by Dedekind cuts.

\begin{proof}
    It is clear that $\DM(P)$ is a poset and that $\p_\DM$ is an embedding. To show that $\DM(P)$ is a complete lattice, let $S\subseteq\DM(P)$. In $P$, define the sets $A=[\bigcup S]^{u\ell}$ and $B=\bigcap S$. By Observation~\ref{3-SPF_VECT.obs:Au=Aulu}, $A\in\DM(P)$. Moreover, for any $C\in S$, $B\subseteq C$. Applying Observation~\ref{3-SPF_VECT.obs:ul_antitone} twice, we get $B^{ul} \subseteq C$. Hence, $B^{ul}\subseteq \bigcap S = B$. Using Observation~\ref{3-SPF_VECT.obs:A_subset_Aul}, we infer $B\in\DM(P)$. We claim that
    \[ \sup S = A, \qquad \inf S = B.\]
    
    Regarding the first equality, we have to show that $A\in S^u \cap S^{u\ell}$. For this, note first that, for all $C\in \DM(P)$, we have $C\in S^u$ iff $\bigcup S \subseteq C$. By Observation~\ref{3-SPF_VECT.obs:A_subset_Aul}, $\bigcup S\subseteq [\bigcup S]^{u\ell} = A$, whence $A\in S^u$. Further, let $C\in S^u$. Then $\bigcup S \subseteq C$. Applying Observation~\ref{3-SPF_VECT.obs:ul_antitone} twice yields $A = [\bigcup S]^{u\ell} \subseteq C$. Thus, $A\in S^{u\ell}$.

    Regarding the second equality, we have to show that $B\in S^\ell \cap S^{\ell u}$. For this, note first that, for all $C\in\DM(P)$, we have $C\in S^\ell$ iff $C\subseteq \bigcap S$. For $C=B$ the latter condition is evidently satisfied, whence $B\in S^\ell$. Further, let $C\in S^\ell$. Then, $C\subseteq \bigcap S = B$. Hence, $B\in S^{\ell u}$. We have shown that $\DM(P)$ is a complete lattice.\smallskip

    It remains to show that $\p_\DM$ is dense. For this, let $A\in\DM(P)$. Let $B = A^u$. Then, we claim that
    \[ \sup \mc P\p_\DM(A) = A = \inf \mc P\p_\DM(B). \]
    Regarding the first equality, let $x\in A$. Then, clearly $\p_\DM(x) = \downarrow x \subseteq A$, because $A = A^{u \ell}$ is downward closed. Thus, $A\in [\mc P\p_\DM(A)]^u$. Let $C\in [\mc P\p_\DM(A)]^u$. Then, 
    \[ C\supseteq \bigcup [\mc P\p_\DM(A)] = \{ x\in P \mid \exists a\in A\colon x \le a\} \supseteq A. \]
    Hence, $A\in [\mc P\p_\DM(A)]^{u\ell}$. We conclude that $\sup \mc P\p_\DM(A) = A$.

    Regarding the second equality, let $x\in A$ and $y\in B = A^u$. Then, $x\le y$, whence $x\in \downarrow y = \p_\DM(y)$. Thus, $A\subseteq \p_\DM(y)$. We infer that $A\in[\mc P\p_\DM(B)]^\ell$. Next, let $C\in[\mc P\p_\DM(B)]^\ell$. That is,
    \[ C \subseteq \bigcap [\mc P\p_\DM(B)] = \{ x\in P \mid \forall b\in B\colon x\le b\} = B^\ell = A^{u\ell} = A. \]
    Thus, $A\in [\mc P\p_\DM(B)]^{\ell u}$. We conclude that $\inf \mc P\p_\DM(B) = A$.
\end{proof}

\subsection{Small completions}
We begin with analysing the extension of morphisms into complete lattices onto a dense completion of the domain, thereby establishing that any dense completion is small.

\begin{proposition}\label{3-SPF_VECT.prop:extension_of_morphisms_into_complete_lattices}
    Let $P$ be a poset, $\p\colon P \inj L$ be a dense completion and $f\colon P \to M$ be an object in $\PPos$ for some complete lattice $M$. Then, the set-theoretic map $f_L\colon L \to M$ given by
    \[ f_L(x) = \sup \{ f(y) \mid y\in P \colon \p(y) \le x\}, \qquad x\in L, \]
    is a $\PPos$-morphism $(\p,L) \to (f,M)$. If $f$ is an embedding, then $f_L$ is so as well.
\end{proposition}

Combining the two statements from this proposition, we directly obtain:

\begin{corollary}\label{3-SPF_VECT.cor:dense=>small}
    Any dense completion $\p\colon P\inj L$ of a poset $P$ is small.\qed
\end{corollary}

Corollary~\ref{3-SPF_VECT.cor:dense=>small} and Theorem~\ref{3-SPF_VECT.thm:DM_is_dense_completion} directly imply the following.

\begin{corollary}\label{3-SPF_VECT.cor:exists_small_completion}
    For any poset $P$, there is a small completion, namely $\DM(P)$.\qed
\end{corollary}

\begin{proof}[Proof of the proposition]
    Regarding the first claim, let $x,x'\in L$ be such that $x\le x'$. Then, all $y\in P$ with $\p(y) \le x$ satisfy $\p(y) \le x'$. Hence, by Observation~\ref{3-SPF_VECT.obs:sup_monotone}, $f_L(x)\le f_L(x')$.

    Further, let $z\in P$. Then, for all $y\in P$ with $\p(y) \le \p(z)$, we have $y\le z$, because $\p$ is an embedding, whence $f(y) \le f(z)$. Furthermore, if $w\in M$ is such that $f(y) \le w$ for all $y\in P$ with $\p(y) \le \p(z)$, then, in particular, $f(z)\le w$. Thus \[ f(z) = \sup\{ f(y) \mid y\in P \colon \p(y) \le \p(z)\} = f_L(\p(z)). \]

    Regarding the second claim, suppose that $f$ is an embedding. Let $x,x'\in L$ with $f_L(x) \le f_L(x')$. We show the auxiliary \hypertarget{3-SPF_VECT.clm:1}{Claim 1} that for all $y\in P$ with $\p(y) \le x$ we also have $\p(y) \le x'$. For this, let $y\in P$ with $\p(y) \le x$. By assumption on $x$ and $x'$, we have $f(y) \le f_L(x) \le f_L(x')$, whence by transitivity
    \[ (\ast) \qquad f(y) \in [\mc Pf(\{y'\in P \mid \p(y') \le x'\})]^{u\ell}. \]
    We infer that 
    \[ (\dagger)\qquad f(y) \in \mc Pf(\{y'\in P \mid \p(y') \le x'\}^{u\ell}) = \mc Pf(\{y'\in P \mid \p(y') \le x'\}). \]
    For the proof, note that the equality follows from Lemma~\ref{3-SPF_VECT.lemma:x=sup=inf_std_representatives}. Indeed, there is $A\subseteq P$ with $\sup \mc P\p(A) = x'$ because $\p$ is dense. Then, by the lemma, $\sup \mc P\p(A^{u\ell}) = x'$ and $A^{u\ell} = \{y'\in P \mid \p(y') \le x'\}$. Hence, by Observation~\ref{3-SPF_VECT.obs:Au=Aulu}, 
    \[ \{y'\in P \mid \p(y') \le x'\}^{u\ell} = A^{u\ell u\ell} = A^{u\ell} = \{y'\in P \mid \p(y') \le x'\}, \]
    which implies the equality. 
    For the proof of the $\in$-relation, let $z\in \{y'\in P \mid \p(y') \le x'\}^{u}$. As $f$ is monotone, we infer that $f(z) \in [\mc Pf(\{y'\in P \mid \p(y') \le x'\})]^{u}$. Hence, by $(\ast)$, $f(y) \le f(z)$. As $f$ is an embedding, $y\le z$. Hence, $y\in \{y'\in P \mid \p(y') \le x'\}^{u\ell}$, whence $f(y)\in \mc Pf(\{y'\in P \mid \p(y') \le x'\}^{u\ell})$, as claimed.

    As an embedding, $f$ is injective. By $(\dagger)$, then, $\p(y) \le x'$. This shows our auxiliary \hyperlink{3-SPF_VECT.clm:1}{Claim~1}. Hence, by Observation~\ref{3-SPF_VECT.obs:sup_monotone} and Lemma~\ref{3-SPF_VECT.lemma:x=sup=inf_std_representatives}, using density of $\p$, we infer
    \[ x = \sup \mc P\p(\{y\in P \mid \p(y) \le x\}) \le \sup \mc P\p(\{y'\in P \mid \p(y') \le x'\}) = x'. \]
    This shows that $f_L$ is an embedding.
\end{proof}

Next, we analyse the uniqueness of dense completions. From this, all remaining open claims follow easily, as shown afterwards.

\begin{proposition}\label{3-SPF_VECT.prop:P-morphisms_between_dense_completions}
    Let $P$ be a poset and $\p\colon P \inj L$, $\psi\colon P\inj M$ be dense completions. Then, there is a unique $\PPos$-morphism $f\colon(\p,L) \to (\psi,M)$, and $f$ is a $\PPos$-isomorphism.
\end{proposition}

\begin{proof}
    By Proposition~\ref{3-SPF_VECT.prop:extension_of_morphisms_into_complete_lattices}, there are $\PPos$-embeddings $\psi_L\colon (\p,L) \inj (\psi,M)$ and $\p_M\colon (\psi,M)\inj(\p,L)$. For symmetry reasons, it remains to show that any $\PPos$-morphism $f\colon (\p,L)\inj(\psi,M)$ is equal to $\psi_L$ and that $\p_M$ is surjective.

    To show this, let $f\colon(\p,L)\inj(\psi,M)$ be a $\PPos$-morphism and $x\in L$. By density, there are $A,B\subseteq P$ such that
    \[ \sup \mc P\p(A) = x = \inf \mc P\p(B). \]
    As $f$ is a $\PPos$-morphism and $\mc P$ is a functor, we infer
    \[ \sup \mc P \psi(A) = \sup \mc P(f\circ\p)(A) \le f(x) \le \inf\mc P(f\circ \p)(B) = \inf\mc P\psi(B). \]
    As a consequence, the fact that $\p_M$ is a $\PPos$-morphism and $\mc P$ a functor, implies
    \[ x = \sup\mc P\p(A) = \sup\mc P(\p_M\circ \psi)(A) \le \p_M\circ f(x) \le \inf\mc P(\p_M\circ\psi)(B) = \inf\mc P\p(B) = x. \]
    Hence, $\p_M \circ f(x) = x$. Thus, $\p_M$ is surjective. As $\psi_L$ is a $\PPos$-morphism $(\p,L)\inj(\psi,M)$, this result can be applied to $\psi_L$ (that is, we can plug in $\psi_L$ for $f$). Then, we get $\p_M(f(x)) = x = \p_M(\psi_L(x))$, whence $f(x) = \psi_L(x)$ because $\p_M$ is an embedding.
\end{proof}

\begin{corollary}\label{3-SPF_VECT.cor:dense=small}
    For any poset $P$, a completion of $P$ is dense iff it is small.
\end{corollary}

\begin{proof}
    Let $(\p,L)$ be a completion of a poset $P$. If it is dense, then it is small, by Corollary~\ref{3-SPF_VECT.cor:dense=>small}. For the converse implication, suppose it to be small. Then, $\DM(P)$ is a small completion of $P$ as well, by Corollary~\ref{3-SPF_VECT.cor:exists_small_completion}. Hence, there are $\PPos$-embeddings $f\colon (\p,L) \inj (\p_\DM,\DM(P))$ and $g\colon (\p_\DM,\DM(P)) \inj (\p,L)$. Thus, $f\circ g$ defines a $\PPos$-embedding of $\DM(P)$ into itself. As $\DM(P)$ is dense, by Theorem~\ref{3-SPF_VECT.thm:DM_is_dense_completion}, Proposition~\ref{3-SPF_VECT.prop:P-morphisms_between_dense_completions} implies that $f\circ g = \id_{\DM(P)}$. Hence, $f$ is surjective and, thus, a $\PPos$-isomorphism with inverse $g$. By Lemma~\ref{3-SPF_VECT.lemma:isomorphisms_dense_small}, $(\p,L)$ is dense.
\end{proof}

\begin{thm}\label{3-SPF_VECT.thm:exists_unique_small_completion=DM}
    For any poset $P$, there is an up to unique $\PPos$-isomorphism unique small completion $(\p,L)$, given by the Dedekind-MacNeille completion $\DM(P)$.
\end{thm}

\begin{proof}
    Let $P$ be a poset. By Corollary~\ref{3-SPF_VECT.cor:exists_small_completion}, $\DM(P)$ is a small completion of $P$. By Proposition~\ref{3-SPF_VECT.prop:P-morphisms_between_dense_completions} and Corollary~\ref{3-SPF_VECT.cor:dense=small}, between any to small completions there is a unique $\PPos$-isomorphism.
\end{proof}

In that sense, the Dedekind-MacNeille completion $\DM(P)$ is the smallest completion of a poset $P$, the use of the definite article being completely specified. 

\subsection{Further results}

We have discussed small and dense completions through general embeddings, and the Dedekind-MacNeille completion is given by a specific embedding. If some complete lattice $L$ is fixed as a reference, and all posets under consideration are embedded into $L$, then we may wish to construct the small completion as a subset of $L$. This is discussed next.

\begin{proposition}\label{3-SPF_VECT.prop:universal_embedding_of_small_completions}
    Let $L$ be a complete lattice and $(P_i)_{i\in I}$ be a family of subsets of $L$. Then, there is a family $(L_i)_{i\in I}$ of subsets of $L$ such that for each $i\in I$, set-theoretic inclusion $P_i \inj L_i$ is a small completion of $P_i$. This statement also holds true if the property ``small'' is replaced with ``dense'' or with ``$P_i$-$\Pos$-isomorphic to $\DM(P_i)$''. The completions satisfy
    \begin{equation}\label{3-SPF_VECT.eq:L_i}
        \{ x\in L \mid \exists A,B\subseteq P_i \colon \sup A = x = \inf B\}\subseteq L_i,\qquad i\in I.
    \end{equation}
\end{proposition}

\begin{proof}
    For any $i\in I$, set-theoretic inclusion $\iota_i \colon P_i\inj L$ defines a completion. Denote the Dedekind-MacNeille completion of $P_i$ by $\p_\DM^i\colon P_i\inj \DM(P_i)$. Then, by Theorem~\ref{3-SPF_VECT.thm:exists_unique_small_completion=DM}, there is a $P_i$-$\Pos$-embedding $f_i\colon (\p_\DM^i,\DM(P_i))\inj(\iota_i,L)$. Let $L_i = \im f_i$. Then, $(\iota_i,L_i)$ is $P_i$-$\Pos$-isomorphic to the Dedekind-MacNeille completion $(\p_\DM^i,\DM(P_i))$ of $P_i$ -- in particular, it is a small and dense completion of $P_i$, by Lemma~\ref{3-SPF_VECT.lemma:isomorphisms_dense_small}, Theorem~\ref{3-SPF_VECT.thm:DM_is_dense_completion}, and Corollary~\ref{3-SPF_VECT.cor:dense=>small}.
    
    For the proof of ``$\subseteq$'' in Equation~\ref{3-SPF_VECT.eq:L_i}, let $x\in L$ be such that there are $A,B\subseteq P_i$ with $\sup A = x = \inf B$. Then, $A,B \subseteq L_i$, and, as $L_i$ is a complete lattice, $A$ has a supremum and $B$ an infimum in $L_i$, denoted by $\sup^{L_i} A$ and $\inf^{L_i} B$, respectively. As $L_i\subseteq L$, $\inf^{L_i} B\le x\le\sup^{L_i} A$. But for all $a\in A$ and $b\in B$, we have $a\le x\le b$. Whence $a\le \inf^{L_i} B$ for all $a\in A$ and thus $\sup^{L_i} A \le \inf^{L_i} B$. Hence, $\inf^{L_i} B = x =\sup^{L_i} A$ and $x\in L_i$.
\end{proof}

\begin{remark}
    The embedded completions $L_i$, $i\in I$, need not be unique and the inclusion in Equation~\ref{3-SPF_VECT.eq:L_i} can be strict. To see this, consider the real interval $L = [0,3]_\R$, a complete lattice, and $P_i = [0,1)_\R\cup (2,3]_\R$. Then, there exists an uncountable set of small completions embedded into $L$, namely $L_i^x = P_i \cup \{x\}$, for any $x\in[1,2]_\R$. Obviously, in any of the cases, the inclusion in Equation~\ref{3-SPF_VECT.eq:L_i} is strict.
\end{remark}

The small completion restricts to the subcategory of chains, as the following result implies.

\begin{proposition}\label{3-SPF_VECT.prop:DM(T)_chain_if_T_chain}
    For any chain $T$, its Dedekind-MacNeille completion $\DM(T)$ is a chain as well.
\end{proposition}

\begin{proof}
    We have to show that for all $A,B\in\DM(T)$ with $B\nsubseteq A$, we have $A\subseteq B$.

    Let $A,B\in\DM(T)$ such that there is $b\in B\setminus A$, and let $a\in A$. As $A = A^{u\ell}$ und $B = B^{u\ell}$, $A$ and $B$ are downward closed. Hence, $b\nleq a$, and thus, by the assumption on $T$, $a\le b$. Thus, $a\in B$.
\end{proof}

\end{document}